\documentclass[12pt,a4paper,twoside]{report} 
\usepackage{t1enc}
\usepackage{mathrsfs}
\usepackage[utf8]{inputenc}
\usepackage[dvips]{graphicx}
\usepackage{afterpage}
\usepackage{amsmath}
\usepackage{amssymb}
\usepackage{bbm}
\usepackage{color}
\definecolor{czerwony}{rgb}{0,0,0}
\definecolor{green}{rgb}{0,0,0}

\usepackage[linktocpage]{hyperref}

\begin{document}

\newpage
\thispagestyle{empty}

\begin{center}
{\large
Faculty of Physics \\
Adam Mickiewicz University \\
Pozna\'n, Poland \\
}
\end{center}

\vspace{2cm}

\begin{center}
{\bf {\large Ph.D. Thesis}}
\end{center}

\vspace{1cm}

\begin{center}
{\LARGE {\bf The influence of magnetic field on the superconducting properties and the BCS-BEC crossover in systems with local fermion pairing} }
\end{center}

\vspace{1.5cm}

\begin{center}
{\large Agnieszka Cichy}
\end{center}

\vspace{1.5cm}

\vspace{0.5cm}

\begin{flushright}
{\large
{\bf Supervisor} \\
Prof. dr hab. Roman Micnas \\
Adam Mickiewicz University \\
Solid State Theory Division \\
}
\end{flushright}

\vspace{3cm}

\begin{center}
Pozna\'n 2012
\end{center}

\newpage
\thispagestyle{empty}
\mbox{}
\newpage

\vspace*{17cm}
\begin{flushright}
 \emph{\large To my Parents}
\end{flushright}

\newpage
\thispagestyle{empty}
\mbox{}
\tableofcontents

\chapter*{Acknowledgements}
\addcontentsline{toc}{chapter}{Acknowledgements}
I would like to thank my supervisor Roman Micnas, who introduced me to condensed matter physics and was always patient in answering all my questions and sharing his great experience. 

I thank Stanisław Robaszkiewicz for many insightful discussions and constant support in my work.

I would like to thank all the people from the Solid State Theory division for useful discussions and very friendly atmosphere during my PhD thesis work.   

I thank Piotr Tomczak who contributed a lot to my education in numerical methods, which was of utmost importance for this work.

I am very grateful to Karl Jansen thanks to whom I could learn the principles of Lattice QCD and do some work in this fascinating field. I also thank DESY for hospitality during my stays in Zeuthen.   

I acknowledge useful discussions with: Maciej Bąk, \textcolor{green}{Ravindra Chhajlany}, Krzysztof Cichy (who is my husband now), Przemysław Grzybowski, Andrzej Koper and Tomasz Polak. 

This work was partly financed from Ministry of Science and Higher Education grant nr. N N0305/B/H03/2011/40. 

Last but not least, I thank my family and friends who supported me over many years -- especially my husband Krzysztof and my parents. My son Filip provided great motivation in the final months of writing this thesis. \textcolor{green}{I also thank Monika Rudnicka, Alina Piwowarczyk, Gosia and Maciej Niemir, Małgorzata Mrowińska, Ania and Marcin Makowscy, Asia and Dominik Niedzielscy, Wojtek Kowalewski, Ania Dyrdał, Piotr Trocha for their support and care.}

\chapter*{List of abbreviations}
\addcontentsline{toc}{chapter}{List of abbreviations}
\begin{footnotesize}
\begin{tabular}[c]{ll}
AAHM & Asymmetric Attractive Hubbard Model\\
AHM & Attractive Hubbard Model\\
BCC & Body-Centered Cubic\\
BCS & Bardeen-Cooper-Schrieffer (theory)\\
BEC & Bose-Einstein Condensation\\
BP & Breached Pair\\
BP-1 & Breached Pair with 1 Fermi surface\\
BP-2 & Breached Pair with 2 Fermi surfaces\\
CC & Chandrasekhar-Clogston (limit)\\
CDW & Charge Density Wave\\
CO & Charge Ordering\\
DMRG & Density Matrix Renormalization Group\\
DOS & Density Of States\\
FCC & Face-Centered Cubic\\
FFLO & Fulde-Ferrell-Larkin-Ovchinnikov\\
FS & Fermi Surface\\
GS & Gapless\\
HFA & Hartree-Fock Approximation\\
HTSC & High-T$_c$ superconductivity\\
KM & Kadanoff-Martin\\
KT & Kosterlitz-Thouless\\
LP & Local Pair\\
LPU & Large-Positive-$U$ \\
MF & Mean-Field\\
NO & NOrmal (phase)\\
NO-I & partially polarized NOrmal (phase)\\
NO-II & fully polarized NOrmal (phase)\\
NSR & Nozi\'eres--Schmitt-Rink\\
PG & Pseudo-Gap\\
PS & Phase Separation\\
PS-I & partially polarized Phase Separation (SC$_0$+NO-I)\\
PS-II & fully polarized Phase Separation (SC$_0$+NO-II)\\
PS-III & Phase Separation (SC$_M$+NO-II)\\
QCP & Quantum Critical Point\\
QMC & Quantum Monte Carlo\\
RT & Reentrant transition\\
SC & SuperConducting\\
SC$_0$ & unpolarized SuperConducting (phase)\\
SC$_M$ & homogeneous magnetized SuperConducting (phase)\\ 
SDW & Spin Density Wave\\
TCP & TriCritical Point\\
\end{tabular}
\end{footnotesize}

\chapter*{Introduction}
\addcontentsline{toc}{chapter}{Introduction}

The immense development of experimental techniques in cold atomic Fermi
gases in the last years has allowed investigation of not only strongly
correlated
systems in condensed matter physics, but also in high energy physics and even
astrophysics (neutron stars) systems. The ability to control the interactions
via the Feshbach resonance in ultracold fermionic gases sets new
perspectives
for experimental realization and study of many different unconventional
systems, including: spin-polarized superfluidity (with
population imbalance), superconductivity with nontrivial Cooper pairing,
Bose-Fermi mixtures, mixtures of fermions with unequal masses and mixtures of
fermions with three hyperfine states (in analogy to quantum chromodynamics). The
possibility to control the population imbalance has also motivated the attempts
to understand the BCS-BEC crossover phase diagrams at zero and finite
temperatures \textcolor{czerwony}{in the presence of spin polarization}. 

The aim of the thesis was to investigate superconducting properties in the
presence of Zeeman magnetic field in systems with local fermion pairing
on the lattice. The study also concerned the evolution from the weak
coupling
(BCS-like) limit to the strong coupling limit of tightly bound local pairs (BEC)
with increasing attraction, both in the ground state and at finite temperatures,
within the spin-polarized extended Hubbard model. The analysis was also extended
to the case of spin dependent hopping integrals (mass
imbalance), with special attention paid to the BCS-BEC crossover physics in the
ground state. The methods used included: the mean field approximation
(BCS-Stoner \textcolor{czerwony}{type}) and the estimation of the phase coherence temperature within
the Kosterlitz-Thouless \textcolor{czerwony}{(KT)} scenario in two dimensions.
The BCS-BEC crossover was also analyzed in three
dimensions, at finite temperatures, within the spin polarized Attractive Hubbard
Model (AHM), going beyond the mean field approximation. In this case, the
critical temperatures of the superconducting transition were determined within
the self-consistent T-matrix method. The strong coupling expansion was applied
to map the spin polarized AHM onto
the model of hard-core bosons on the lattice.   
  
The thesis is organized as follows.

In \textbf{chapter 1}, the theoretical principles of superconductivity and
superfluidity are reviewed, beginning with a historical
introduction. The fundamentals of the BCS-BEC crossover physics are given,
including the
characteristic features of the BCS and BEC limits. We also show new experimental
and theoretical possibilities of studying cold atomic Fermi gases.
	
\textbf{Chapter 2} gives a discussion of the spin-polarized extended Hubbard
model. The Hamiltonian is introduced and its symmetries are discussed.
Next, applying the broken symmetry Hartree approximation, the set of
self-consistent equations for the superconducting gap parameter, the number of
particles and magnetization are obtained. From the partition function, the
grand canonical potential of the superconducting and normal state are
determined. 

In \textbf{chapter 3}, some theoretical aspects of the KT
transition in 2D systems are outlined, followed by a brief discussion of the
two-dimensional XY model.
This model is a good example of a system which does not have the long
range order \textcolor{czerwony}{at finite temperatures but it exhibits a phase transition}. Finally, the superfluid density in a general case is calculated and further used to
determine the KT
temperature from a universal relation.
   
In \textbf{chapter 4}, the influence of the Zeeman
term on the superfluid characteristics of the attractive Hubbard model with spin
independent hopping integrals is analyzed, for the square and simple cubic
lattices. Within
the mean-field approach, we construct phase diagrams in two ways: by fixing the
chemical potential or the electron concentration, and show the relevant
differences resulting from these possibilities. The importance of the Hartree
term in the broken symmetry Hartree-Fock approximation is indicated. For the
two-dimensional case, the KT transition in the weak coupling
regime is investigated. Finally, the influence of the pure d-wave
pairing symmetry on the superconducting phases stability in a 2D system is
analyzed.

In \textbf{chapter 5}, the influence of magnetic field on the
BCS-BEC crossover in the ground state for the square and simple
cubic lattices is analyzed,
within the spin-polarized \textcolor{czerwony}{AHM}. The development of
experimental techniques in cold atomic Fermi gases with tunable attractive
interactions (through the Feshbach resonance) allowed a study of the
BCS-BEC crossover and the properties of exotic states in these systems.

In \textbf{chapter 6}, the BCS-BEC crossover analysis is extended to finite \textcolor{czerwony}{temperatures} in 2D
within the Kosterlitz-Thouless scenario. \textcolor{czerwony}{We also investigate} the effects of pairing fluctuations \textcolor{czerwony}{using} self-consistent
T-matrix \textcolor{czerwony}{approach} \textcolor{czerwony}{to study the BCS-BEC crossover in 3D}. The T-matrix approach goes beyond the standard mean field, since
it includes the effects of non-condensed pairs and allows a description
of
the BEC side of the crossover. The crossover diagrams include the
pseudogap state. We make a comparison of the results obtained within the
so-called $(GG_0)G_0$ and $(GG)G_0$ schemes, both for the 3D continuum model
with contact attraction and for the AHM, for \textcolor{czerwony}{a} simple cubic lattice. We also
discuss a generalization of the $T_c$ equations in non-zero Zeeman magnetic
field case and \textcolor{czerwony}{show numerical solutions}. 

\textbf{Chapter 7} presents the results concerning the effects of
spin-dependent hopping integrals ($t^{\uparrow}\neq t^{\downarrow}$) on the
stability of the superfluid phases.
We study the evolution from the weak to strong coupling limit,
within the \textcolor{czerwony}{AHM} in magnetic field with spin-dependent
hopping integrals, for square and simple
cubic lattices. We also construct the BCS-BEC crossover phase diagrams in finite
temperatures in 2D, taking into account the Kosterlitz-Thouless transition. The
strong coupling expansion is applied to map the spin polarized AHM onto the
model of
hard-core bosons on the lattice. We show the occurrence of competition between
charge ordering and superconducting phases for any
particle concentration in the spin dependent hopping integrals case. This
mapping is specific to the lattice fermion model.  

\textcolor{czerwony}{Appendices A-E contain additional material.}







\newpage
\thispagestyle{empty}
\mbox{}
\chapter{Principles of superconductivity and superfluidity}
\section{Historical perspective}

\textbf{Superconductivity}, \textbf{superfluidity} and \textbf{Bose-Einstein
condensation} (BEC) are among the most interesting phenomena of condensed
matter physics. They have been a scientific mystery for many years of
the 20th century and they still provide many theoretical and experimental
challenges. The properties of superconductors and superfluids are seemingly very
different, but they are similar in many respects, e.g. both evince frictionless
flow and contain quantized vortices. The appearance of these effects
depends upon whether the system in question is electrically neutral
(superfluids) or charged (superconductors).    

The phenomenon of superconductivity, i.e. the ability of some materials to
conduct the electric current without resistance, was discovered in 1911 by
Kamer\-lingh-Onnes \cite{onnes} in a mercury (Hg) sample which was cooled to
$4.2$ K. Two years after this spectacular discovery, superconductivity was found
in lead (Pb) with critical temperature $T_c=7.2$ K and in 1930 in niobium (Nb)
at $T_c=9.2$ K. 

In 1933 Meissner and Ochsenfeld \cite{meisner} discovered that magnetic flux is
canceled from the inside of a sample when it is cooled below its
superconducting transition temperature in a weak external magnetic field.
Therefore, a characteristic feature of superconducting materials is perfect
diamagnetism. Twenty years after this discovery, \textcolor{czerwony}{with advent of the Ginzburg-Landau theory}, 
superconductors were
proposed to be divided into two classes -- those from which magnetic field is
fully canceled were named \textbf{type-I superconductors}, while those
from which strong magnetic field is only partly canceled were named
\textbf{type-II superconductors}. 

At around the same time (1924), Indian physicist Satyendra Bose sent to
Albert Einstein a paper in which he derived the Planck black-body radiation
formula with the use of exclusively statistical arguments \cite{castin}. Bose
was a relatively unknown scientist and his earlier papers had been ignored.
However, Einstein realized that there were spectacular novel ideas in Bose's
letter and helped him to publish the results in \emph{Zeitschrift f\"ur Physik}
\cite{bose}. He was also interested in this topic and wrote two papers in which
a complete theory of bosonic particles was included. The statistics that governs
the behavior of such particles has been named \textbf{Bose-Einstein statistics}.

In one of his papers, Einstein noticed \cite{einstein} that if the number of
particles is conserved, the system of non-interacting particles undergoes a
thermodynamic phase transition at sufficiently low temperatures, called the
\textbf{Bose-Einstein condensation (BEC)}. The condensation discovered by
Einstein results from the fact that the total number of states at decreasing
energy becomes very low in the thermodynamic limit, and there is no space for
all particles when temperature decreases. Thus, the system can accumulate
particles only in the ground state.

However, for a long time no physical system was known to actually have
shown Bose-Einstein condensation. In 1938 F. London suggested
\cite{london} that the superfluidity phenomenon which was discovered earlier in
$^{4}$He by Allen, Misener and Kapitza \cite{allen, kapica} can be a manifestation of bosonic
condensation. The absence of superfluidity in  $^{3}$He (the fermionic helium
isotope), whose nuclei are fermions, supported this thesis. However, much
later $^{3}$He turned out to be superfluid at a yet much lower temperature than
$^{4}$He (by Osheroff, Richardson, Lee) \cite{osheroff}. For many years, the nature of relation between the
superfluid properties and Bose-Einstein condensation was unclear. The
breakthrough took place in the 1950s when O. Penrose and L. Onsager found the
connection between superfluidity and the long-range order present in
highly-correlated bosonic systems \cite{penrose}. It permitted an estimation of
the number of condensed atoms in the liquid $^{4}$He as 8$\%$. The number is relatively
small in contrast to the BEC theory which describes the case of an ideal gas and neglects any
interactions between the particles. However, the interactions are strong in
liquid helium and hence can not be neglected.

Strong interactions in  $^{4}$He made it impossible to carry out the
\emph{ab initio} calculations of its properties. However, superfluidity itself
(the lack of viscosity in a liquid) was explained within the Landau
phenomenological theory in 1941 \cite{landau1, landau2}. In this theory,
superfluidity results from the fact that when the accessible energy is very low,
only long-wavelength phonons can be excited. In 1947 N. Bogoliubov extended
this approach by introducing the low-energetic spectrum of phonons and
assuming that the dynamics of the system is dominated by the atoms which
constitute the condensate \cite{bogoliubov1, bogoliubov2}. In this paper, the
Bogoliubov canonical transformation was introduced, which turned out to be
very useful for the description of many phenomena, such as:
\emph{superconductivity}, \emph{atomic condensates} and also in \emph{nuclear
physics}. The weakly interacting systems were studied also in perturbation
theory by K. Huang in the 1950s \cite{huang}.

In 1957, the mechanism of superconductivity was finally explained in the
framework of quantum mechanics by J. Bardeen, L. Cooper and R. Schrieffer
\cite{cooper}-\cite{barden2}. \textbf{The BCS theory} clarifies that the
interaction of electrons with the crystalline lattice (electron-phonon coupling)
in a superconductor leads to an attractive interaction and the creation of pairs
of electrons with opposite momenta and spins (Cooper pairs). Cooper has shown
that even a very weak attractive interaction causes a Fermi sea instability.
Therefore, the formation of Cooper pairs is more energetically favorable for
superconducting systems than the occupation of allowed single particle states. The BCS
theory based on the above background explains consistently many physical
properties of phonon superconductors. The theory predicted (among other things)
the formation of a minimum excitation energy (or energy gap) in a superconductor
below the critical temperature ($T_c$). The energy gap increases gradually
with decreasing temperature. The Cooper pairs are well-defined objects in
momentum space. The BCS theory describes conventional superconductors, for which
the critical temperature is of order 10$^4$ times weaker than the Fermi energy
($E_F$) (strictly speaking $k_B T_c \sim 10^{-4}$).

In 1986 Bednorz and M\"uller \cite{bednorz} obtained a ceramic material
consisting of lanthanum (La), barium (Ba), copper (Cu) and oxygen (O) (LaBaCuO),
which exhibits the phase transition from the superconducting to the normal state
at $T_c=35$ K. 
From this time on, many materials have been discovered with yet higher
transition temperatures. These compounds have been named High-T$_c$
superconductors (HTSC) and this phenomenon -- High-T$_c$ superconductivity.
High-T$_c$ superconductors are anisotropic ionic crystals which are, depending
on doping, antiferromagnetic insulators or superconductors with very untypical
properties in the normal state. They have critical temperatures $T_c\sim 100$ K,
i.e. an order of magnitude higher than conventional superconductors. They are
also extreme II-type superconductors, i.e. they have considerably shorter
coherence length ($\xi \sim 10$ {\AA}) than conventional superconductors and
large penetration depth ($\lambda \sim 3000$ {\AA}). A very important feature of
unconventional superconductors is a linear dependence
between the superconducting transition temperature and the inverse square of
penetration depth (extrapolated to $T=0 K$). The plot of this dependence has been
named Uemura plot \cite{uemura}. 

Although it is known that conventional phonon mechanism plays some role, there
is still no complete microscopic theory to describe unconventional
superconductivity. Therefore, many hypotheses on the source of occurrence of
this phenomenon have been put forward. 

As mentioned before, in the weak coupling limit the fermion superfluid or
superconducting (charged fermion superfluid) system is described within the
Bardeen-Cooper-Schrieffer theory. With increasing attraction, the system can
evolve to the Bose-Einstein condensation limit of preformed fermionic pairs
(pairing in real space). Already in 1980 A. Leggett showed that when
attraction between two fermions with opposite spins is weak, the BCS superfluid
is stable, but when the attraction becomes strong, the system is on the BEC side
at $T=0$ \cite{leggett}. Leggett considered the zero-range attractive potential
(the contact interaction). In 1985 Leggett's analysis was extended to non-zero
temperatures (especially the critical temperature for the emergence of
superfluidity) by P. Nozi\'eres and S. Schmitt-Rink (NSR) \cite{nozieres}. They
used a diagrammatic method for a model of fermions with a finite-range
attractive interaction.

The most interesting and promising ideas assume that the properties of
High-T$_c$ superconductors (and also other unconventional superconductors)
place them between two regimes: BCS and BEC \cite{Robaszkiewicz}-\cite{chen}.
\textbf{The BCS-BEC crossover phenomenon} is also used: (a) to describe
superfluidity in ultracold atomic gases, (b) to describe the
insulator-superconductor transition in disordered systems, (c) to study Bose
condensation of excitons, (d) in high-energy physics (quantum chromodynamics on
a lattice (lattice QCD)), (e) in astrophysics (neutron stars). 

The physics of the BCS-BEC crossover will be described in section
\ref{sec:bcs-bec}.


\section{Basics of the BCS-BEC crossover physics}
\label{sec:bcs-bec}    

As mentioned before, there has been a broad agreement that the physics of
BCS-BEC crossover is crucial for unconventional superconductivity. 

The properties of a fermionic system with attractive interaction are very
different in the extreme BCS and BEC limits, especially in the normal state. 

The characteristic features of \textbf{the BCS limit} are the following:
\begin{itemize}
\item pairs form and condense at the same temperature ($T_c$),
\item  pairing takes place in momentum space and only a small number of
electrons in the vicinity of the Fermi surface form the Cooper pairs,
\item attractive interaction between fermions ($|U|$) is weak,
\item pair size in the condensate is much larger than the average distance
between them (hence the pairs strongly overlap),
\item the energy gap decreases monotonically with increasing $T$ and vanishes at
$T_c$,
\item the critical temperature and the thermodynamics are determined by single
particle excitations -- broken Cooper pairs with exponentially small gap,
\item above the critical temperature the normal state is described by the Fermi
liquid theory.
\end{itemize}
On the other hand, \textbf{the BEC limit (or the local pair limit (LP))} is
characterized by the following features:
\begin{itemize}
 \item pairs are formed at a temperature $T_p$, much higher than the critical
temperature ($T_p\gg T_c$) in which the long-range phase coherence occurs and
phase transition to the superconducting state takes place,
\item pairs form in real space and all electrons are paired,
\item attractive interaction between fermions is strong,
\item pairs are much smaller than the average distance between them,
\item pair binding energy is proportional to $|U|$, pairs exist above $T_c$ up
to $T_p$,
\item the critical temperature and thermodynamics are determined by collective
modes,
\item the normal state is described by the Bose liquid of tightly bound and
phase incoherent pairs (in the range of $T_c<T<T_p$).
\end{itemize}

For intermediate couplings, the normal state can have a so-called
\emph{pseudo-gap} (PG) in the single particle energy spectrum and deviations
from the standard Landau theory of the Fermi liquids occur.

A schematic phase diagram for systems in which the evolution from the BCS to
BEC limit takes place is presented in Fig. \ref{fig:micnas1}. The solid
lines corresponds to the border between the superconducting and the normal state
($T_c \sim t\exp (-1/N(0)|U|)$, where $N(0)$ -- density of states at Fermi level) -- in the weak coupling regime and $T_c\sim t^2/|U|$
-- in the strong coupling limit). The intermediate region (the dashed part of
the diagram) between the Fermi liquid and the Bose liquid regimes is the area 
in which the pairs are formed ($T_p\sim |U|$). The continuation of $T_c$, determined
within continuum model, is denoted by the dashed line in the phase diagram.

\begin{figure}[h!]
\begin{center}
\includegraphics[width=0.6\textwidth]{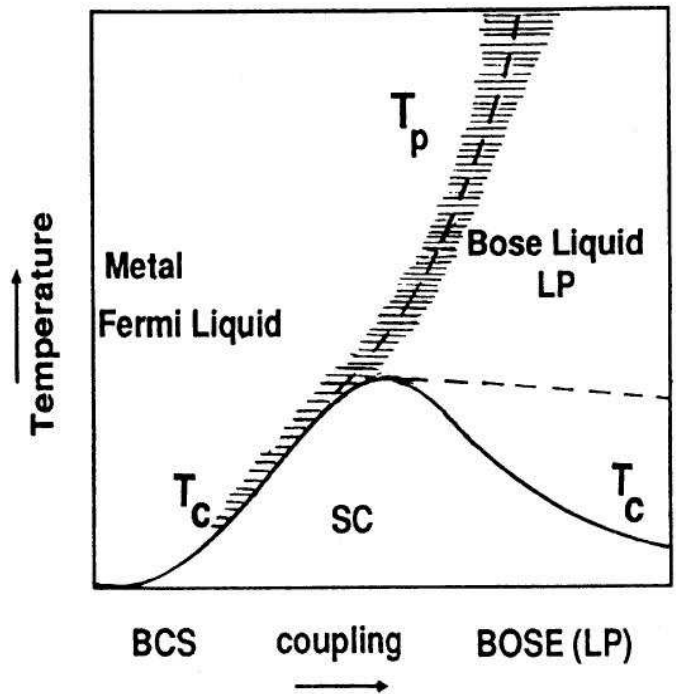}
\caption[{\label{fig:micnas1} Schematic phase diagram for systems in which the
BCS-BEC crossover takes place \textcolor{green}{\cite{Micnas1998}}.}]{\label{fig:micnas1} Schematic phase diagram for systems in which the
BCS-BEC crossover takes place \textcolor{green}{\cite{Micnas1998}} (see also chapter \ref{6.2.2}).}
\end{center}
\end{figure}

The effects of the BCS-BEC crossover are clearly visible in the behavior of
chemical potential $\mu$ at $T=0$. In the weak coupling limit, $\mu = E_F$
($E_F$ -- Fermi energy) and the conventional BCS theory is valid. At a suitably
strong coupling, the chemical potential drops to $\mu =0$ and becomes negative
($\mu <0$) in the BEC limit. The point at which $\mu=0$ is the BCS-BEC
crossover point \cite{leggett}. However, the behavior of the system does not
resemble neither the BCS nor the BEC regime near $\mu =0$, so one can speak of a
whole crossover region, not only of one crossover point. When $\mu \simeq E_F$,
the Fermi surface is present in the system and we deal with a Cooper
pairs condensate. On the other hand, when $\mu$ becomes negative, the Fermi
surface vanishes and the bosonic regime appears.  

\begin{figure}[h!]
\begin{center}
\includegraphics[width=0.6\textwidth]{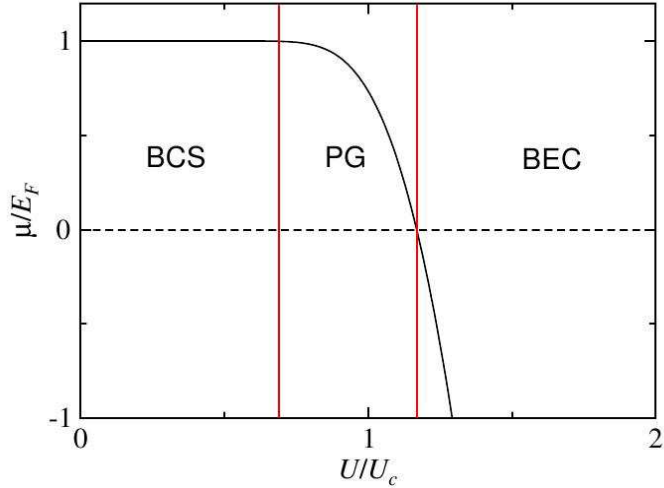}
\caption{\label{fig:levin} Chemical potential behavior in units of the Fermi
energy at $T=0$ in three regimes: BCS, BCS-BEC crossover (PG) and BEC \textcolor{green}{\cite{levin}}.}
\end{center}
\end{figure}  

The evolution from the BCS superfluid to the Bose-Einstein
condensation limit of preformed fermionic pairs cannot be directly examined
experimentally by tuning the interaction strength in superconductors, nuclear
matter or in neutron stars. However, the BCS-BEC crossover in superconductors
can be analyzed experimentally by changing particle concentration (doping). The
tuning of interactions itself, in turn, is feasible in ultracold Fermi
atoms through \textbf{Feshbach resonances} \textcolor{green}{\cite{Fano, Feshbach}}. These
resonances take place when the energy difference $\Delta E$ between a bound
state with energy $E_{res}$ (the closed channel) and the threshold energy
$E_{th}$  of scattering states (the open channel) goes to zero, by using an
external magnetic field $B_0$. The Feshbach resonance is characterized by a
divergence in the scattering length $a_s$. The BCS limit occurs when the
dimensionless scattering parameter: $1/k_F a_s \rightarrow -\infty$, while the
BEC limit corresponds to $1/k_F a_s \rightarrow \infty$, where $k_F$ is the
Fermi momentum. 

In 2003 three experimental groups observed a condensation of pairs of fermionic
atoms of ultracold gases in the region of the BCS-BEC crossover. M. Grimm et al.
\textcolor{green}{\cite{jochim, chin}} and also W Ketterle et al. \cite{ketterle-znow, zwierlein}
used $^{6}$Li atoms. D. S. Jin et al. \cite{Jin, regal} cooled down the trapped
atom gas of $^{40}$K to a temperature of the order of $5\cdot 10^{-8}$ K.
Afterwards, the interactions between atoms were tuned through the
Feshbach resonance. The appearance of the energy gap was shown on the
basis of radiospectroscopic measurements \cite{kinast, bourdel}. Years later the
existence of the superfluid phase was eventually confirmed through the presence
of quantized vortices \cite{schunck}. 

\section{The new possibilities}
\label{sec:new_poss}

Unconventional superconductivity with nontrivial Cooper pairing in strongly
correlated electron systems and spin-polarized superfluidity (in the context of
cold atomic Fermi gases) are currently investigated and also widely discussed in
the leading world literature. They lay down one of the most
investigated directions of studies in the range of condensed matter
physics and ultracold quantum gases. 

Recently, experimental groups from MIT \cite{ketterle}-\cite{ketterle3}
and from the Rice University \cite{li} have begun investigations of quantum
Fermi gases ($^{6}$Li) with unequal numbers of fermions with down ($\downarrow$)
and up ($\uparrow$) spins ($N_{\downarrow}\neq N_{\uparrow}$ -- systems with
\emph{population imbalance}). The possibility to control the population
imbalance has motivated the attempts to understand the BCS-BEC crossover phase
diagrams at zero and finite temperatures.

Experiments have indicated the presence of an unpolarized superfluid core in the
center of the trap and a polarized normal state surrounding this core in the
density profiles of trapped Fermi mixtures with population imbalance. Hence, a
phase separation appears between the unpolarized BCS and the polarized normal
state.

Experiments in which the fermionic (or bosonic) gases are put on optical
lattices have also been carried out \textcolor{green}{\cite{chin, stoferle}}. Both the depth of the
periodic trapped potential and the geometry can be fully controlled. In
this way, strongly correlated systems with different geometries of the
lattice can be investigated. Atomic gases with tunable interactions on optical
lattices allow experimental realization of the Hubbard model.               


The possibility to create the population imbalance in conventional
superconductors is to apply an external magnetic field, but this field is
shielded by the orbital motion of electrons (the Meissner effect). However, a
mixture with arbitrary population ratio can be prepared in atomic Fermi gases.
Hence, the influence of the Zeeman magnetic field on superfluidity can be also
investigated.   


In the presence of the Zeeman magnetic field ($h$), the densities of states are
different for the particles with spin down and spin up. The population imbalance
introduces a mismatch between the Fermi surfaces. At stronger imbalance, in
the weak coupling regime, superfluidity is destroyed and there is a first-order
phase transition to the polarized normal state at a universal value of the
critical magnetic field $h_c=\Delta_{0}/\sqrt{2} \approx 0.707 \Delta_0$ (the
Chandrasekhar-Clogston limit (CC) \cite{chandrasekhar, clogston}), where
$\Delta_0$ is the gap at $T=0$, $h=0$. 
Rather recently a behavior in accordance with the CC limit has been observed in
population imbalanced atomic Fermi gases \cite{CC}.

In the weak coupling limit, at a large difference in the occupation number (or
at a strong magnetic field) states with nontrivial Cooper pairing can exist. An
example of such pairing is the formation of Cooper pairs across the spin-split
Fermi surface with non-zero total momentum ($\vec{k} \uparrow$,
$-\vec{k}+\vec{q} \downarrow$), leading to the so-called Fulde-Ferrell
\cite{Fulde} and Larkin-Ovchinnikov \cite{Larkin} (FFLO) state, which is favored
\cite{Bianchi}-\cite{Mierzejewski} (against the normal state) up to
$h_{c}^{FFLO}=0.754 \Delta_0$ in three dimensions. There are experimental
and theoretical premises that the FFLO state can be found in heavy-fermion
superconductors (e.g. CeCoIn$_5$). However, the observation of such a state is
extremely difficult in superconducting systems because of the very strong
destructive influence of the orbital effect on the superconductivity, as
mentioned above. 

There has been much work on exact numerical studies (Quantum Monte Carlo (QMC)
simulations and density-matrix renormalization group (DMRG)) of the 1D
attractive Hubbard model with fermion population imbalance
\cite{Feiguin}-\cite{Tezuka}. The results seem to suggest that the FFLO state
can be obtained in one-dimensional systems, which is consistent with the
fact that $h_c^{FFLO}$ diverges as $T\rightarrow 0$ in $d=1$.

Another kind of pairing and phase coherence is the spatially homogeneous
spin-polarized superconductivity (breached pair (BP)) which has a gapless
spectrum for the majority spin species \cite{Sheehy, Sheehy3, Sheehy2}. The
coexistence of the superfluid and the normal component in the isotropic state is
characteristic of the BP phase. The state of this type was
originally considered by Sarma \cite{Sarma}. He studied the case of a
superconductor in an external magnetic field within the BCS theory. All orbital
effects were neglected. He showed that self-consistent mean field solutions with
gapless spectrum ($\Delta (h)$) are energetically unstable at $T=0$, in contrary
to the fully gapped BCS solutions. On the other hand, a non-zero temperature can
lead to the stabilization of a spin-polarized state.


\begin{figure}[h!]
\begin{center}
\includegraphics[width=0.7\textwidth]{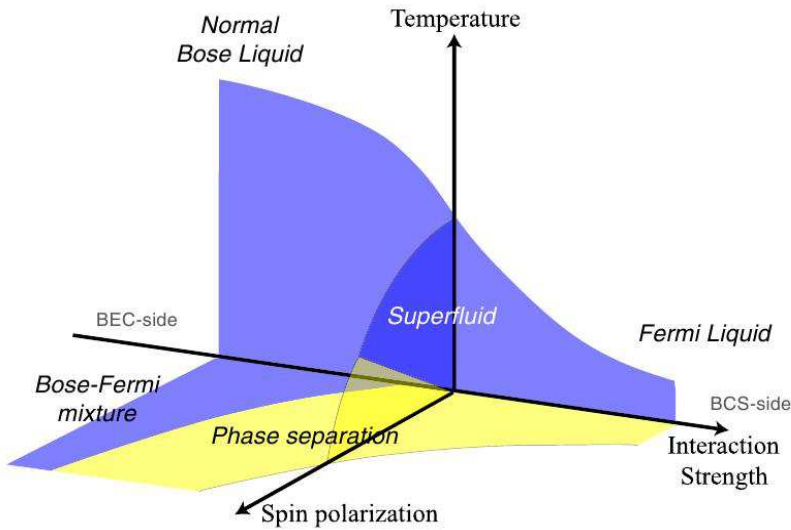}
\caption{\label{fig:ketterle} Schematic diagram (temperature-interaction
strength-spin polarization) in which various possible phases of two-component
Fermi gas are illustrated \cite{Ketterle}.}
\end{center}
\end{figure}    

As shown in Fig. \ref{fig:ketterle}, one can expect at least three phases when
the interaction is changed and the population imbalance is fixed. In the weak
coupling region at $T=0$, the superfluid BCS state with equal particle densities
is spatially separated from the unpaired fermions. On the other hand, at $T=0$,
for strong attraction the two-component Fermi gas with population imbalance
evolves into a coherent mixture of local pairs (LP's) and excess spin-up
fermions (Bose-Fermi mixture). 

New possibilities appear when one investigates mixtures of fermions with unequal
masses. These systems are very interesting not only in the context of atomic
physics, but also in that of condensed-matter physics and color
superconductivity in high energy physics. If the number of hyperfine states is
extended from two to three, there can be three types of $s$-wave pairing. Such
pairs are analogous to quarks which can form pairs in different color states in
the cores of neutron stars \cite{deMello}.  

First theoretical studies of Fermi condensates in systems with spin and mass
imbalances have shown that the BP state can have excess fermions with two Fermi
surfaces (BP-2 or interior gap state) \cite{Wilczek}-\cite{Iskin-2}. However,
the problem of stability of the BP-2 state is still open. According to some
investigations, the interior gap state proposed by Liu and Wilczek
\cite{Wilczek} is always unstable even for large mass ratio and the phase
separation is favorable \nolinebreak{\cite{Parish, Parish2}}. On the other
hand, at strong attraction the homogeneous magnetized superconducting
phase can only have one Fermi surface (BP-1). Details concerning these exotic
phases will be discussed later.


\newpage
\thispagestyle{empty}
\mbox{}
\chapter{The Hubbard model}
\section{Introduction}

The Hubbard model is one of the most important models in solid state physics.
Despite its simplicity, it describes many interesting phenomena, such as:
metal-superconductor transition, antiferromagnetism, ferrimagnetism,
ferromagnetism and superconductivity in strongly co\-rrelated systems
\cite{Fazekas}. The Hubbard model is the simplest many-body Hamiltonian which
allows a description of two opposing tendencies: (a) the kinetic energy
leading to the delocalization of electrons (metal-like behavior), (b) the
Coulomb repulsion leading to the localization of electrons on the lattice
(metal-insulator transition -- the Mott transition).

\section{The Hubbard Hamiltonian}

Let us consider the case of a partially filled $s$-band which contains $n$
electrons per atom. Let $\Psi_{\vec{k}}$ be the Bloch functions of this band and
$\epsilon_{\vec{k}}$ -- the energies corresponding to the Bloch functions. The
wave functions and the energies are calculated in the Hartree-Fock potential
which represents the average $s$-band electrons interaction with the electrons
of other bands and other $s$-band electrons.

The electrons dynamics can be described through the following Hamiltonian
\cite{Hubbard}:

\begin{eqnarray}
\label{hami}
H&=&\sum_{\vec k, \sigma} \epsilon_{\vec k} c^{\dag}_{\vec k \sigma} c_{\vec k
\sigma}+\frac{1}{2}\sum_{\vec k_1, \vec k_2, \vec k'_1, \vec
k'_2}\sum_{\sigma_1, \sigma_2} (\vec k_1 \vec k_2 \arrowvert 1\slash r\arrowvert
\vec k'_1 \vec k'_2)c_{\vec k_1\sigma_1}^{\dag} c_{\vec k_2 \sigma_2}^{\dag}
c_{\vec k'_2 \sigma_2} c_{\vec k'_1 \sigma_1} \nonumber \\
&-&\sum_{\vec k, \vec k'}\sum_{\sigma} \left( 2(\vec k \vec k' \arrowvert
1\slash r \arrowvert \vec k \vec k' ) -(\vec k \vec k' \arrowvert 1\slash r
\arrowvert \vec k \vec k' )\right) \nu_{\vec k'} c_{\vec k \sigma}^{\dag}c_{\vec
k \sigma} ,
\end{eqnarray}
where: 
sum over $\vec k$ from the first Brillouin zone, $\nu_{\vec k}$ -- band filling
of electrons with momentum $\vec{k}$
\begin{equation}
(\vec k_1 \vec k_2 \arrowvert 1\slash r \arrowvert \vec k'_1 \vec k'_2 )= e^2 \int
\frac{\varphi^*_{\vec k_1}(\vec x)\varphi_{\vec k'_1}(\vec x)\varphi^*_{\vec
k_2}(\vec x')\varphi_{\vec k'_2}(\vec x')}{\arrowvert \vec x - \vec x'
\arrowvert} d\vec x d\vec x'.
\end{equation}
1. Hamiltonian term -- the band energ\textcolor{green}{ies} of electrons,
\\
2. Hamiltonian term -- the Coulomb interaction of electrons,
\\
3. Hamiltonian term -- prevents double counting of the electrons interactions
within the band (they were included when $\epsilon_{\vec{k}}$ \textcolor{green}{were} calculated
within the Hartree-Fock approximation (HFA)).

Here, we introduce the Wannier functions:
\begin{equation}
\phi (\vec{x}-\vec{R_i})=\frac{1}{\sqrt{N}} \sum_{\vec k}e^{-i\vec{k} \cdot
\vec{x}} \varphi_{\vec k} (\vec x),
\end{equation}
where $N$ is the number of atoms.
Then, one can show that:
\begin{equation}
\varphi_{\vec k} (\vec x)=\frac{1}{\sqrt{N}}\sum_i e^{i\vec k \vec R_i} \phi
(\vec x - \vec R_i),
\end{equation}
where the sum goes over all positions of atoms $\vec R_i$. Let us introduce
creation and annihilation operators of electrons with spin $\sigma$ in the
Wannier state $\phi (\vec x- \vec R_i)$:
\begin{equation}
c_{\vec k \sigma}= \frac{1}{\sqrt{N}}\sum_i e^{i \vec k \vec R_i} c_{i\sigma},
\end{equation}
\begin{equation}
c_{\vec k \sigma}^{\dag}= \frac{1}{\sqrt{N}}\sum_i e^{-i \vec k \vec R_i}
c_{i\sigma}^{\dag}.
\end{equation}
The Hamiltonian (\textcolor{green}{\ref{hami}}) can be rewritten using creation and annihilation
operators:
\begin{eqnarray}
\label{3.7}
H&=& \sum_{i, j}\sum_{\sigma}t_{ij} c_{i\sigma}^{\dag} c_{j \sigma}
+\frac{1}{2}\sum_{i, j, k, l}\sum_{\sigma, \sigma'} (ij\arrowvert 1\slash r
\arrowvert kl) c^{\dag}_{i \sigma} c^{\dag}_{j\sigma'} c_{l\sigma'}
c_{k\sigma}\nonumber\\
& -& \sum_{i, j, k, l} \sum_{\sigma}\left( 2(ij\arrowvert 1\slash r \arrowvert
kl)-(ij\arrowvert 1\slash r \arrowvert kl)\right)
\nu_{jl}c^{\dag}_{i\sigma}c_{k\sigma},
\end{eqnarray}
where:
\begin{equation}
t_{ij}=\frac{1}{N}\sum_{\vec k} \epsilon_{\vec k} e^{i\vec k (\vec R_i-\vec
R_j)},
\end{equation}
\begin{equation}
\label{3.9}
(ij\arrowvert 1\slash r \arrowvert kl)=e^2\int \frac{\phi^* (\vec x-\vec
R_i)\phi (\vec x-\vec R_k)\phi^* (\vec x'-\vec R_j) \phi (\vec x'-\vec
R_l)}{\arrowvert \vec x-\vec x' \arrowvert}d\vec x d\vec x',
\end{equation}
\label{3.10}
\begin{equation}
\nu_{jl}=\sum_{\vec k}\nu_{\vec k}e^{i\vec k (\vec R_j-\vec R_l)}.
\end{equation}

Let us simplify the Hamiltonian (\ref{3.7}). The Wannier functions resemble the
atomic $s$ functions, since we consider a narrow energy band. Then, the size of
the atomic shell is much smaller than the inter-atomic distances. Hence, the
largest integral among (\ref{3.9}) is $(ii\arrowvert 1\slash r \arrowvert ii)
\equiv U$, suggesting that we can neglect all integrals (\ref{3.9}), except for
the integral U. Then, the Hamiltonian (\ref{3.7}) takes the form:
\begin{equation}
\label{3.11}
H= \sum_{i, j}\sum_{\sigma}t_{ij} c^{\dag}_{i\sigma} c_{j\sigma}+\frac{1}{2} U
\sum_i \sum_{\sigma}n_{i\sigma}n_{i-\sigma}- U\sum_i
\sum_{\sigma}\nu_{ii}n_{i\sigma}.
\end{equation}
We get from eq.~\eqref{3.10}:
\begin{equation}
\nu_{ii}=\frac{1}{N}\sum_{\vec k}\nu_{\vec k}=\frac{1}{2} n
\end{equation}
and the last term in (\ref{3.11}) reduces to a constant so we can neglect it and
obtain final form of the Hamiltonian:
\begin{equation}
H= \sum_{i, j}\sum_{\sigma}t_{ij} c^{\dag}_{i\sigma} c_{j\sigma} +\frac{1}{2}
U\sum_i\sum_{\sigma}n_{i\sigma}n_{i-\sigma}.
\end{equation}

\section{The Spin-Polarized extended Hubbard model}

The Hubbard model can be extended by adding the intersite interaction term to
the Hubbard Hamiltonian. One can also analyze the influence of the pure Zeeman
effect on the superfluid characteristics within a lattice fermion model (the
\textcolor{czerwony}{AHM}). The Hamiltonian of such system takes the form:
\begin{eqnarray}
\label{extham}
H&=&\sum_{ij}\sum_{\sigma}t_{ij}^{\sigma}c_{i\sigma}^{\dag}c_{j\sigma}+\frac{1}{
2}U\sum_{i\sigma}n_{i\sigma}n_{i-\sigma}+\frac{1}{2}\sum_{ij}\sum_{\sigma
\sigma'}W_{ij}n_{i,\sigma}n_{j,\sigma'}\nonumber\\
&-&h\sum_{i}(n_{i\uparrow}-n_{i\downarrow}),
\end{eqnarray}
where: $t_{ij}^{\sigma}$ -- nearest-neighbor spin-dependent hopping;
$\sigma=\uparrow ,\downarrow$ -- spin index;
$n_{i\sigma}=c_{i\sigma}^{\dag}c_{i\sigma}$ -- particle number operator; $U$ --
on-site interaction; $W_{ij}$ -- intersite interaction; $h$ -- Zeeman magnetic
field\footnote{In the case of ultracold Fermi gases, the role of the
Zeeman field is played by the difference in chemical potentials for spin-up and
spin-down fermions, i.e. $h\equiv(\mu_\uparrow-\mu_\downarrow)/2$.}. 

The Hamiltonian (\ref{extham}) (the extended Hubbard model at $h=0$,
$t_{ij}^{\uparrow}=t_{ij}^{\downarrow}=t_{ij}$) has been proposed by Micnas,
Ranninger and Robaszkiewicz while working on the
problem of theoretical models in systems with local fermion pairing
\cite{MicnasModern, MicnasarXiv}. This Hamiltonian describes a system of narrow
band electrons which are strongly coupled with a bosonic field. The electrons
polarize the bosonic field and, in this way, interact effectively with each
other, creating excitations which exhibit a short-range Fr\"ohlich
type-interaction, which competes with the Coulomb interaction. These bosonic
modes can be phonons, excitons, acoustic plasmons etc. The parameters of the
Hamiltonian (\ref{extham}) -- $t_{ij}$, $U$, $W_{ij}$ are effective parameters. 

A typical microscopic mechanism which leads to the effective short-range
attraction is the strong electron-lattice coupling. It causes the formation of
polarons (the anharmonic modes can play an important role in this case) or the
interaction of electrons with excitations -- excitons, plasmons, magnons
\cite{MicnasarXiv}.

Two cases are considered most often within the extended Hubbard model
\cite{Micnas1998}:
\begin{itemize}
 \item $U_{eff}<0$, $W_{eff}>0$ -- the induced local attraction plays a crucial
role and dominates over the on-site repulsion. Therefore, this is the case of
the on-site attraction (\emph{negative-U extended Hubbard model}). In the strong coupling
limit, electron pairs form. One can also consider the weak attraction limit. The
Hubbard model with $U_{eff}<0$ is the simplest lattice model of a superconductor
with a short-range coherence length which exhibits the crossover between the
regime of BCS-type superconductivity and Bose-Einstein condensation of local
pairs. It is also an effective model which accounts for the properties of
superconductivity and the charge ordering in the family of barium bismuthates
($Ba_{1-x}K_x BiO_3$, $BaPb_xBi_{1-x}O_3$), Chevrel phases and fullerenes. It can also be used for discussion of 
general properties (such as a pseudogap) cuprates.
\item $W_{eff}<0$, $U_{eff}>0$ -- this is the case of the intersite attraction.
The induced attraction is strong enough to dominate over the intersite Coulomb
repulsion. It is the model which describes the intersite pairing with various
symmetries. In this case the Hamiltonian accounts for the properties of cuprates
and heavy-fermion superconductors.     
\end{itemize}

As mentioned before, the recent progress in experimental techniques gives new
possibilities to observe various exotic phases appearing in strongly correlated
systems. The properties of two-component Fermi mixtures with arbitrary
population ratio ($n_{\uparrow} \neq n_{\downarrow}$) and spin dependent hopping
integrals $t^{\uparrow}\neq t^{\downarrow}$ (mass imbalance) on optical lattices
can be described by the extended Hubbard model \eqref{extham}.  

In this thesis we focus on the analysis of the Zeeman term influence on the
superfluid characteristics of the attractive Hubbard model ($U<0$) for the
square and simple cubic lattices ($s$-wave pairing). We also study the
properties of the $t^{\sigma}$ -- $U$ -- $h$ model in which the spin-dependent
hopping ($t^{\uparrow}\neq t^{\downarrow}$) is included. We analyze briefly the
$t$ -- $W$ -- $h$ model in which the attractive interaction between the
electrons on the neighboring sites is taken into account. In this way, we study
the stability of various superconducting phases in an external magnetic field,
with $d$-wave pairing symmetry.

\section{The Hartree-Fock approximation}
\label{HFapprox}
In this thesis we deal with the properties of two- and three dimensional systems
in a Zeeman field, within the spin-polarized extended Hubbard model (in grand canonical ensamble):
\begin{eqnarray}
\label{extham'}
H&=&\sum_{i,j,\sigma}(t_{i,j}^{\sigma}-\mu\delta_{i,j})c_{i,\sigma}^{\dag}c_{j,
\sigma}+U\sum_{i}n_{i,\uparrow}n_{i,\downarrow}+\frac{1}{2}\sum_{i,j,\sigma,
\sigma'}W_{ij}n_{i,\sigma}n_{j,\sigma'}\nonumber \\
&-&h\sum_{i}(n_{i\uparrow}-n_{i\downarrow}),
\end{eqnarray} 
where: $\mu$ -- chemical potential. The Zeeman field ($h$) can originate from an
external magnetic field (in $g \mu_B \slash 2$ units) or from population
imbalance in the context of the cold atomic Fermi gases with $\mu
=(\mu_{\uparrow} +\mu_{\downarrow})/2$ and $h=(\mu_{\uparrow} -
\mu_{\downarrow})/2$, where $\mu_{\sigma}$ is the chemical potential of atoms
with (pseudo) spin-$\sigma$. This model allows inclusion of the
influence of the Zeeman field on the superfluid properties not only for the
isotropic pairing symmetry ($s$-wave) but also for the $s_{x^2+y^2}$-wave and
$d_{x^2-y^2}$-wave pairing symmetries, which originate from the $W$ term
\textcolor{green}{\cite{Micnas, Miyake, tobi}}.

One should emphasize that the attractive Hubbard model ($U<0$) in a Zeeman
field, with spin independent hopping integrals ($t$ -- $U$ -- $h$) can be
transformed to a doped repulsive Hubbard model (see Appendix \ref{appendix1}). 


Let us Fourier transform the Hamiltonian (\ref{extham'}) to the reciprocal
space: 
\begin{equation}
c_{j,\sigma}^{\dag}=\frac{1}{\sqrt{N}}\sum_{\vec{k}}c_{\vec{k}\sigma}e^{i\vec{
r_j}\vec{k}}.
\end{equation}
Then, one obtains:
\begin{eqnarray}
\label{ham'}
H&=&\sum_{\vec{k},\sigma}(\epsilon_{\vec{k}}^{\sigma}-\mu)c_{\vec{k}\sigma}^{
\dag}c_{\vec{k}\sigma}+\frac{1}{N}\sum_{\vec{k_1},\vec{k_2},
\vec{q}}\sum_{\sigma , \sigma'}\Big(U\delta_{\sigma
\bar{\sigma}}+\frac{W}{2}\gamma_{\vec{q}}
\Big)c_{\vec{k_1}\sigma}^{\dag}c_{\vec{k_1}-\vec{q}\sigma}c_{\vec
k_2\sigma'}^{\dag}c_{\vec{k_2}+\vec{q}\sigma'}\nonumber \\
&-&h\sum_{\vec{k}}
(c_{\vec{k}\uparrow}^{\dag}c_{\vec{k}\uparrow}-c_{\vec{k}\downarrow}^{\dag}c_{
\vec{k}\downarrow}),
\end{eqnarray}
where: N -- number of lattice sites. The electron dispersion is
$\epsilon_{\vec{k}}^{\sigma}=\sum_{\vec{\delta}}t_{\vec{\delta}}^{\sigma}e^{\vec
{k} \cdot \vec{\delta}}=-2t^{\sigma} \Theta_{\vec{k}}$;
$\Theta_{\vec{k}}=\sum_{l=1}^{d} \cos(k_l a_l)$ (here $d=2,3$ for two- and
three-dimensional lattices, respectively); $a_l$ is the lattice constant in the
$l$-th direction (we set $a_l=1$ in further considerations);
$\gamma_{\vec{q}}=2\Theta_{\vec{q}}$, $\gamma_{\vec{q}}=\gamma_{-\vec{q}}$,
$\sum_{\vec{q}} \gamma_{\vec{q}}=0$.

We will not consider alternative phases such as FFLO states, in which the
$\vec{q}\neq 0$ pairs can occur. 

Now, we use the Hartree-Fock approximation by separating the two-particle
operators in such a way that:
\begin{equation}
A=\langle A \rangle +\delta A.
\end{equation}
Hence:
\begin{eqnarray}
AB&=&(\langle A \rangle + \delta A)(\langle B \rangle +\delta B)\simeq  \langle
A \rangle \langle B\rangle +\delta A\langle B \rangle + \delta B \langle A
\rangle = \nonumber \\
&=& \langle A \rangle \langle B\rangle+A\langle B \rangle - \langle A \rangle
\langle B\rangle+B\langle A \rangle - \langle A \rangle \langle B\rangle =
\nonumber \\
&=&A\langle B \rangle +B\langle A \rangle - \langle A \rangle \langle B\rangle.
\end{eqnarray}
For example (if $\vec{k_1}=-\vec{k_2}$), we get:
\begin{eqnarray}
&\,&c_{\vec{k_1}\sigma}^{\dag}c_{\vec{k_1}-\vec{q}\sigma}c_{\vec{k_2}\sigma'}^{
\dag}c_{\vec{k_2}+\vec{q}\sigma'}^{\dag}=c_{\vec{k_1}\sigma}^{\dag}c_{\vec{k_1}
-\vec{q}\sigma}c_{-\vec{k_1}\sigma'}^{\dag}c_{-\vec{k_1}+\vec{q}\sigma'}
=\nonumber\\
&=&-c_{\vec{k_1}\sigma}^{\dag}c_{-\vec{k_1}\sigma'}^{\dag}c_{\vec{k_1}-\vec{q}
\sigma}c_{-\vec{k_1}+\vec{q}\sigma'}=-\langle
c_{\vec{k_1}\sigma}^{\dag}c_{-\vec{k_1}\sigma'}^{\dag}\rangle
c_{\vec{k_1}-\vec{q}\sigma}c_{-\vec{k_1}+\vec{q}\sigma'} \nonumber \\ 
&-&\langle c_{\vec{k_1}-\vec{q}\sigma}c_{-\vec{k_1}+\vec{q}\sigma'}\rangle
c_{\vec{k_1}\sigma}^{\dag}c_{-\vec{k_1}\sigma'}^{\dag}+\langle
c_{\vec{k_1}\sigma}^{\dag}c_{-\vec{k_1}\sigma'}^{\dag}\rangle \langle
c_{\vec{k_1}-\vec{q}\sigma}c_{-\vec{k_1}+\vec{q}\sigma'}\rangle,
\end{eqnarray}
while if $\vec{k_2}=\vec{k_1}-\vec{q}$ then:
\begin{eqnarray}
&\,&c_{\vec{k_1}\sigma}^{\dag}c_{\vec{k_1}-\vec{q}\sigma}c_{\vec{k_2}\sigma'}^{
\dag}c_{\vec{k_2}+\vec{q}\sigma'}^{\dag}=c_{\vec{k_1}\sigma}^{\dag}c_{\vec{k_1}
-\vec{q}\sigma}c_{\vec{k_1}-\vec{q}\sigma'}^{\dag}c_{\vec{k_1}\sigma'}=\nonumber
\\
&=&-c_{\vec{k_1}\sigma}^{\dag}c_{\vec{k_1}\sigma'}c_{\vec{k_1}-\vec{q}\sigma'}^{
\dag}c_{\vec{k_1}-\vec{q}\sigma}=
-\langle c_{\vec{k_1}\sigma}^{\dag}c_{\vec{k_1}\sigma'}\rangle
c_{\vec{k_1}-\vec{q}\sigma'}^{\dag}c_{\vec{k_1}-\vec{q}\sigma}\nonumber\\
&-& \langle c_{\vec{k_1}-\vec{q}\sigma'}^{\dag}c_{\vec{k_1}-\vec{q}\sigma}
\rangle c_{\vec{k_1}\sigma}^{\dag}c_{\vec{k_1}\sigma'} + \langle
c_{\vec{k_1}\sigma}^{\dag}c_{\vec{k_1}\sigma'}\rangle  \langle
c_{\vec{k_1}-\vec{q}\sigma'}^{\dag}c_{\vec{k_1}-\vec{q}\sigma} \rangle.
\end{eqnarray}
The anomalous averages have also been taken into account apart from the normal
averages. The averages which describe the charge ordering (CO) and spin density
waves (SDW ordering) have been neglected at this stage of the work. As a result,
we obtain the following Hamiltonian in a bilinear form:
\begin{equation}
\label{ham''}
\textcolor{czerwony}{H=\sum_{\vec{k}\sigma}\bar{\epsilon}_{\vec{k}}^{\sigma}c_{\vec k\sigma}^{\dag}
c_{\vec k\sigma}-\sum_{\vec k}(\Delta_{\vec k} c_{\vec k\uparrow}^{\dag}c_{-\vec
k\downarrow}^{\dag}+H.c.) 
-\frac{1}{2}\sum_{\vec k\sigma} (\Delta_t^{\sigma} (\vec k)c_{\vec
k\sigma}^{\dag}c_{-\vec k\sigma}^{\dag} +H.c.)+C,}
\end{equation}
where: 
\begin{equation}
\label{barepsilon2}
\bar{\epsilon}_{\vec{k}}^{\uparrow}=\epsilon_{\vec{k}}^{\uparrow}-p_{\uparrow}W
\gamma_{\vec{k}}\slash \gamma_0 -\bar{\mu}_{\uparrow}-h,
\end{equation}
\begin{equation}
\label{barepsilon1}
\bar{\epsilon}_{\vec{k}}^{\downarrow}=\epsilon_{\vec{k}}^{\downarrow}-p_{
\downarrow}W \gamma_{\vec{k}}\slash \gamma_0 -\bar{\mu}_{\downarrow}+h,
\end{equation}
\begin{equation}
\bar{\mu}_{\uparrow}=\mu -Un_{\downarrow}-W\gamma_0 n, 
\end{equation}
\begin{equation}
\bar{\mu}_{\downarrow}=\mu -Un_{\uparrow}-W\gamma_0 n, 
\end{equation}
\begin{equation}
\label{pgora}
p_{\uparrow}=\frac{1}{N}\sum_{\vec{q}}\gamma_{\vec{q}} \langle
c_{\vec{q}\uparrow}^{\dag}c_{\vec{q}\uparrow} \rangle,
\end{equation}
\begin{equation}
\label{pdol}
p_{\downarrow}=\frac{1}{N}\sum_{\vec{q}}\gamma_{\vec{q}} \langle
c_{\vec{q}\downarrow}^{\dag}c_{\vec{q}\downarrow} \rangle,
\end{equation}
\begin{equation}
\Delta_{\vec k}=\frac{1}{N}\sum_{\vec{q}}V_{\vec{k}\vec{q}}^s \langle c_{-\vec
q\downarrow} c_{\vec q\uparrow} \rangle,
\end{equation}
\begin{equation}
V_{\vec k \vec q}^s=-U-W\gamma_{\vec{k}-\vec{q}},
\end{equation}
\begin{equation}
\Delta_t^{\sigma}(\vec k)=\frac{1}{N}\sum_{\vec q} V_{\vec k \vec q}^t \langle
c_{-\vec q\sigma}c_{\vec q\sigma} \rangle,
\end{equation}
\begin{equation}
V_{\vec k \vec
q}^t=\frac{1}{2}W(\gamma_{\vec{k}+\vec{q}}-\gamma_{\vec{k}-\vec{q}}),
\end{equation}
\begin{eqnarray}
C&=&-\frac{U}{N}\sum_{\vec{k}\vec{q}} \langle c_{\vec k\uparrow}^{\dag}c_{-\vec
k\downarrow}^{\dag} \rangle \langle c_{\vec q\downarrow}c_{-\vec q\uparrow}
\rangle +\frac{1}{2N}\sum_{\vec{k}\vec{q}\sigma
\sigma'}W\gamma_{\vec{k}-\vec{q}} \langle c_{\vec
k\sigma}^{\dag}c_{-\vec{k}\sigma'}^{\dag} \rangle \langle c_{-\vec
q\sigma}c_{\vec q\sigma'}\rangle \nonumber\\
&+&\frac{1}{2N}\sum_{\vec{k}\vec{q}\sigma}W\gamma_{\vec{k}-\vec{q}}\langle
c_{\vec k\sigma}^{\dag}c_{\vec k\sigma} \rangle \langle c_{\vec q\sigma}^{\dag}
c_{\vec q\sigma} \rangle -UNn_{\uparrow}n_{\downarrow}-\frac{1}{2}W\gamma_0Nn^2.
\end{eqnarray}
$\Delta_{\vec{k}}$ is the order parameter for the singlet pairing, for which 
\textcolor{czerwony}{$\langle c_{-\vec q \downarrow} c_{\vec q \uparrow} \rangle 
= \langle c_{\vec q \downarrow} c_{-\vec q \uparrow} \rangle $.}
$\Delta_t^{\sigma}(k)$ -- the order parameter for the triplet (equal spin) pairing, which
describes the $p$-wave pairing symmetry. In this work we will restrict our
analysis to the properties of the singlet pairing.

The following Bogoliubov transformation can be used to transform the Hamiltonian
(\ref{ham''}) to a diagonal form:
\begin{equation}
\label{3.42}
c_{\vec{k} \uparrow}= u_{\vec{k}}^*\gamma_{\vec{k}0}+\nu_{\vec{k}}
\gamma_{\vec{k}1}^{\dag},
\end{equation}
\begin{equation}
\label{3.42'}
c^{\dag}_{-\vec{k}\downarrow}=-\nu_{\vec{k}}^*\gamma_{\vec{k}0}
+u_{\vec{k}}\gamma_{\vec{k}1}^{\dag}.
\end{equation} 
The quasiparticle operators $\gamma_{\vec{k}0}$ and $\gamma_{\vec{k}1}$ are also
the fermion operators. Therefore, the anticommutation condition takes the form:
\begin{equation}
\label{3.42''}
\vert u_{\vec{k}}\vert^2 + \vert \nu_{\vec{k}} \vert^2 =1.
\end{equation}
Then, the Hamiltonian (\ref{ham''}) can be rewritten:
\begin{eqnarray}
\label{hamtrans}
H&=&\sum_{\vec{k}}[\bar{\epsilon}_{\vec{k}}^{\uparrow}(|u_{\vec{k}}|^2
\gamma_{\vec{k}0}^{\dag}\gamma_{\vec{k}0}-|\nu_{\vec{k}}|^2
\gamma_{\vec{k}1}^{\dag}\gamma_{\vec{k}1})+\bar{\epsilon}_{\vec{k}}^{\downarrow}
(|\nu_{\vec{k}}|^2 \gamma_{\vec{k}1}^{\dag}\gamma_{\vec{k}1}
-|u_{\vec{k}}|^2\gamma_{\vec{k}0}^{\dag}\gamma_{\vec{k}0})\nonumber \\
&+&(\bar{\epsilon}_{\vec{k}}^{\uparrow}+\bar{\epsilon}_{\vec{k}}^{\downarrow}
)(\nu_{\vec{k}}
u_{\vec{k}}\gamma_{\vec{k}0}^{\dag}\gamma_{\vec{k}1}^{\dag}+\nu_{\vec{k}}^{*}
u_{\vec{k}}^{*}\gamma_{\vec{k}1}\gamma_{\vec{k}0})
+|\nu_{\vec{k}}|^2(\bar{\epsilon}_{\vec{k}}^{\uparrow}+\bar{\epsilon}_{\vec{k}}^
{\downarrow})]\nonumber\\
&+&\sum_{\vec{k}}[(\Delta_{\vec{k}}u_{\vec{k}}\nu_{\vec{k}}^*+\Delta^*_{\vec{k}}
u_{\vec{k}}^*\nu_{\vec{k}})(\gamma_{\vec{k}0}^{\dag}\gamma_{\vec{k}0}
+\gamma_{\vec{k}1}^{\dag}\gamma_{\vec{k}1} -1)\nonumber\\
&+&(\Delta_{\vec{k}}
\nu_{\vec{k}}^{*2}-\Delta_{\vec{k}}^*u_{\vec{k}}^{*2})\gamma_{\vec{k}1}\gamma_{
\vec{k}0}+(\Delta_{\vec{k}}^*\nu_{\vec{k}}^2-\Delta_{\vec{k}}u_{\vec{k}}
^2)\gamma_{\vec{k}0}^{\dag}\gamma_{\vec{k}1}^{\dag}].
\end{eqnarray}
We equate the non-diagonal terms of the Hamiltonian (\ref{hamtrans}) to zero:
 \begin{equation}
(\bar{\epsilon}_{\vec{k}}^{\uparrow}+\bar{\epsilon}_{\vec{k}}^{\downarrow})u_{
\vec{k}}\nu_{\vec{k}}
+\nu_{\vec{k}}^2\Delta_{\vec{k}}^*-u_{\vec{k}}^2\Delta_{\vec{k}}=0.
\end{equation}
Hence:
\begin{equation}
\label{wzor}
\frac{\nu_{\vec{k}}}{u_{\vec{k}}}\Delta_{\vec{k}}^*=E_{\vec{k}\uparrow}-\bar{
\epsilon}_{\vec{k}}^{\uparrow}=E_{\vec{k}\downarrow}-\bar{\epsilon}_{\vec{k}}^{
\downarrow},
\end{equation}
where:
\begin{equation}
E_{\vec{k}\uparrow}=\frac{\bar{\epsilon}_{\vec{k}\uparrow}-\bar{\epsilon}_{\vec{
k}\downarrow}}{2}+\sqrt{\Bigg(
\frac{\bar{\epsilon}_{\vec{k}\uparrow}+\bar{\epsilon}_{\vec{k}\downarrow}}{2}
\Bigg)^2+\vert \Delta_{\vec{k}} \vert^2},
\end{equation}
\begin{equation}
E_{\vec{k}\downarrow}=\frac{\bar{\epsilon}_{\vec{k}\downarrow}-\bar{\epsilon}_{
\vec{k}\uparrow}}{2}+\sqrt{\Bigg(
\frac{\bar{\epsilon}_{\vec{k}\uparrow}+\bar{\epsilon}_{\vec{k}\downarrow}}{2}
\Bigg)^2+\vert \Delta_{\vec{k}} \vert^2}.
\end{equation}
After simple mathematical transformations:
\begin{equation}
\label{energy}
E_{\vec{k}\downarrow, \uparrow}= \pm
(-t^{\downarrow}+t^{\uparrow})\Theta_{\vec{k}}\pm \frac{UM}{2}\pm
\frac{1}{2}W(p_{\uparrow}-p_{\downarrow})\frac{\gamma_{\vec{k}}}{\gamma_0} \pm
h+\omega_{\vec{k}},
\end{equation}
where:
\begin{equation}
\omega_{\vec{k}}=\sqrt{((-t^{\uparrow}-t^{\downarrow})\Theta_{\vec{k}}-\bar{\mu}-\textcolor{czerwony}{p\frac{\gamma_{\vec k}}{\gamma_0}W}
)^2+|\Delta_{\vec{k}}|^2},
\end{equation}
\begin{equation}
\bar{\mu}\equiv \frac{\bar{\mu}_{\uparrow}+\bar{\mu}_{\downarrow}}{2}= \mu
-n(\frac{U}{2}+W\gamma_0), 
\end{equation}

\textcolor{czerwony}{$p \equiv \frac{p_{\uparrow}+p_{\downarrow}}{2}$}, $M=n_{\uparrow}-n_{\downarrow}$ is the spin magnetization.

Defining the ratio $r\equiv t^{\uparrow}/t^{\downarrow}$, the quasiparticle
energies (\ref{energy}) can be rewritten as: 

\begin{equation}
\label{energy2}
E_{\vec{k}\downarrow, \uparrow}= \pm
\Bigg(\frac{1-r}{1+r}\Bigg)\epsilon_{\vec{k}}\pm \frac{UM}{2}\pm
\frac{1}{2}W(p_{\uparrow}-p_{\downarrow})\frac{\gamma_{\vec{k}}}{\gamma_0} \pm
h+\omega_{\vec{k}}.
\end{equation}

The condition (\ref{3.42''}) with Eq. (\ref{wzor}) allow calculation of the
coefficients $u_{\vec{k}}$ and $\nu_{\vec{k}}$:
\begin{equation}
\label{nu}
|\nu_{\vec{k}}|^2=\frac{1}{2}\Bigg(1-\frac{\bar{\epsilon}_{\vec{k}\uparrow}+\bar
{\epsilon}_{\vec{k}\downarrow}}{E_{\vec{k}\uparrow}+E_{\vec{k}\downarrow}}\Bigg)
,
\end{equation}
\begin{equation}
\label{u}
|u_{\vec{k}}|^2=\frac{1}{2}\Bigg(1+\frac{\bar{\epsilon}_{\vec{k}\uparrow}+\bar{
\epsilon}_{\vec{k}\downarrow}}{E_{\vec{k}\uparrow}+E_{\vec{k}\downarrow}}\Bigg).
\end{equation}
The remaining terms of the Hamiltonian  (\ref{hamtrans}) are reduced to:
\begin{equation}
 H=\frac{1}{2}\sum_{\vec{k}}
[(\bar{\epsilon}_{\vec{k}}^{\uparrow}+\bar{\epsilon}_{\vec{k}}^{\downarrow})-(E_
{\vec{k}\uparrow}+E_{\vec{k}\downarrow})]+\sum_{\vec{k}}[E_{\vec{k}\uparrow}
\gamma_{\vec{k}0}^{\dag}\gamma_{\vec{k}0}+E_{\vec{k}\downarrow}\gamma_{\vec{k}1}
^{\dag}\gamma_{\vec{k}1}]+C.
\end{equation}
Knowing the coefficients $|u_{\vec{k}}|^2$ and $|\nu_{\vec{k}}|^2$, one can
calculate the operator averages:
\begin{equation}
 \langle c_{\vec{k}\uparrow}^{\dag}c_{\vec{k}\uparrow} \rangle =
|\nu_{\vec{k}}|^2 + \langle |u_{\vec{k}}|^2 \gamma_{\vec{k}0}^{\dag}
\gamma_{\vec{k}0} - |\nu_{\vec{k}}|^2 \gamma_{\vec{k}1}^{\dag} \gamma_{\vec{k}1}
\rangle.
\end{equation}
Now, we substitute: $\langle \gamma_{\vec{k}0}^{\dag}\gamma_{\vec{k}0}\rangle$
and $\langle \gamma_{\vec{k}1}^{\dag}\gamma_{\vec{k}1}\rangle$ with the
Fermi-Dirac distribution functions: $f(E_{\vec{k}\uparrow})=1/(e^{\beta
E_{\vec{k}\uparrow}}+1)$, $f(E_{\vec{k}\downarrow})=1/(e^{\beta
E_{\vec{k}\downarrow}}+1)$, respectively, where $\beta=1/k_B T$. We obtain:
\begin{equation}
\label{ndol}
 \langle c_{\vec{k}\uparrow}^{\dag}c_{\vec{k}\uparrow} \rangle =
|\nu_{\vec{k}}|^2 f(-E_{\vec{k}\downarrow})+|u_{\vec{k}}|^2
f(E_{\vec{k}\uparrow}).
\end{equation}
In a similar way:
\begin{equation}
\label{ngora}
\langle c_{\vec{k}\downarrow}^{\dag}c_{\vec{k}\downarrow} \rangle =
|u_{\vec{k}}|^2 f(-E_{\vec{k}\uparrow})+|\nu_{\vec{k}}|^2
f(E_{\vec{k}\downarrow}).
\end{equation}
These two expressions (\eqref{ndol}-\eqref{ngora}) can be written as:
\begin{equation}
 n_{\sigma}=\frac{1}{N}\sum_{\vec{k}} \langle
c_{\vec{k}\sigma}^{\dag}c_{\vec{k}\sigma}\rangle = \frac{1}{N} \sum_{\vec{k}}
(|u_{\vec{k}}|^2 f(E_{\vec{k},\sigma})+|\nu_{\vec{k}}|^2
f(-E_{\vec{k},-\sigma})).
\end{equation}

After simple transformations, the particle number equation takes the form:
\begin{equation}
\label{n}
n=1-\frac{1}{2N}\sum_{\vec{k}}
\frac{-(t^{\uparrow}+t^{\downarrow})\Theta_{\vec{k}}-\bar{\mu}-\textcolor{czerwony}{p\frac{\gamma_{\vec k}}{\gamma_0}W}}{\omega_{\vec{k}}
}
\Bigg(\tanh\frac{\beta E_{\vec{k}\uparrow}}{2}+\tanh \frac{\beta
E_{\vec{k}\downarrow}}{2}\Bigg),
\end{equation}
where: $n=n_{\uparrow}+n_{\downarrow}$.

The equation for the magnetization is:
\begin{equation}
\label{M}
 M=\frac{1}{2N}\sum_{\vec{k}}\Bigg(\tanh\frac{\beta
E_{\vec{k}\downarrow}}{2}-\tanh \frac{\beta E_{\vec{k}\uparrow}}{2}\Bigg).
\end{equation}

After calculating the anomalous averages, one obtains the following equation for
the superconducting gap parameter:
\begin{equation}
\label{del}
 \Delta_{\vec{k}}=\frac{1}{N}\sum_{\vec{q}}V_{\vec{k}\vec{q}}^s\Delta_{\vec{q}}
F_{\vec{q}}(T),
\end{equation}
where:
\begin{equation}
 F_{\vec{q}}(T)=({2\omega_{\vec{q}}})^{-1}\frac{1}{2}\Bigg(\tanh\frac{\beta
E_{\vec{q}\uparrow}}{2}+\tanh \frac{\beta E_{\vec{q}\downarrow}}{2}\Bigg).
\end{equation}

The pairing potential $V_{\vec{k}\vec{q}}^s$ can be separated for the
two-dimensional square lattice into:
\begin{equation}
V_{\vec{k}\vec{q}}^s= -U+\frac{\vert W \vert}{4} (\gamma_{\vec{k}}
\gamma_{\vec{q}} + \eta_{\vec{k}} \eta_{\vec{q}})+2\vert W \vert
(\textrm{sin}k_x \textrm{sin}q_x + \textrm{sin}k_y \textrm{sin}q_y),
\end{equation}
where: $\gamma_{\vec{k}}=2(\textrm{cos}k_x + \textrm{cos}k_y)$, while
$\eta_{\vec{k}}=2(\textrm{cos}k_x - \textrm{cos}k_y)$.
\\
Eq.~\eqref{del} is separated into parts corresponding to the $s$, $s_{x^2+y^2}$
and $d_{x^2-y^2}$ pairing symmetry. 
\\
In this way:
\begin{equation}
\label{delta''}
\Delta_{\vec{k}}=\Delta_0+\Delta_{\gamma}\gamma_{\vec k}+\Delta_{\eta}\eta_{\vec
k}.
\end{equation}
The self-consistent equations take the form:
\begin{equation}
\label{delta0}
\Delta_0=-U\phi_1,
\end{equation}
\begin{equation}
\label{deltagamma}
\Delta_{\gamma}=\frac{\vert W \vert}{4}\phi_{\gamma}
\end{equation}
for the $s$-wave pairing symmetry, and:
\begin{equation}
\label{deltaeta}
\Delta_{\eta}=\frac{\vert W \vert}{4}\phi_{\eta}
\end{equation}
for the $d$-wave pairing symmetry,
\\ 
where:
\begin{equation}
\phi_1=\frac{1}{N}\sum_{\vec q} \Delta_{\vec q} F_{\vec q},
\end{equation}
\begin{equation}
\phi_{\gamma}=\frac{1}{N}\sum_{\vec q} \Delta_{\vec q}\gamma_{\vec q} F_{\vec
q},
\end{equation}
\begin{equation}
\phi_{\eta}=\frac{1}{N}\sum_{\vec q} \Delta_{\vec q}\eta_{\vec q} F_{\vec q}.
\end{equation}
Inserting to the self-consistent equations \eqref{delta0}, \eqref{deltagamma}
and \eqref{deltaeta} the expression \eqref{delta''}, we obtain a system of
equations for the $s_{x^2+y^2}$ pairing:
\begin{equation}
\label{sext}
\left(\begin{array}{ccc}
1+U\phi_1(T) & U\phi_2(T) \\
-\frac{\vert W \vert}{4}\phi_2(T) & 1-\frac{\vert W \vert}{4}\phi_{\gamma}(T) \\
\end{array}\right)
\left(\begin{array}{ccc}
\Delta_0 \\
\Delta_{\gamma} \\
\end{array} \right) =0,
\end{equation}
where:
\begin{equation}
\label{phi1}
\phi_1(T)=\frac{1}{N}\sum_{\vec q}F_{\vec q}(T),
\end{equation}
\begin{equation}
\phi_2(T)=\frac{1}{N}\sum_{\vec q}\gamma_{\vec q} F_{\vec q}(T),
\end{equation}
\begin{equation}
\label{phigamma}
\phi_{\gamma}(T)=\frac{1}{N}\sum_{\vec q}\gamma_{\vec q}^2 F_{\vec q}(T) .
\end{equation}
For the $d$-wave pairing:
\begin{equation}
\label{d}
\frac{4}{\vert W \vert}=\frac{1}{N}\sum_{\vec q}\eta_{\vec q}^2F_{\vec q}(T).
\end{equation}
From these equations, the value of the order parameter for different types of
pairing and the transition temperature can be determined. 

As follows from eq.~\eqref{sext}, the solution for the $s_{x^2+y^2}$ pairing
depends both on $U$ and on $\vert W \vert$. However, the solution \eqref{d} for
the $d$-wave pairing is \textcolor{czerwony}{formally} independent of $U$. Substitution of the form
\eqref{delta''} neglects the relative phases of $d$- and $s$-waves order
parameters and therefore it is not the most general. Such a coupling of phases
can be relevant when we consider the mixed phases $s-d$ below the critical
temperature \cite{tobi}.

The equations for the Fock parameters \eqref{pgora}, \eqref{pdol} take the form:
\begin{equation}
p_{\uparrow}=\frac{1}{N} \sum_{\vec{q}} \frac{1}{2} \gamma_{\vec{q}}
\Big(|\nu_{\vec{q}}|^2\tanh \frac{\beta
E_{\vec{q}\downarrow}}{2}-|u_{\vec{q}}|^2\tanh \frac{\beta
E_{\vec{q}\uparrow}}{2}\Big), 
\end{equation}
\begin{equation}
p_{\downarrow}=\frac{1}{N} \sum_{\vec{q}} \frac{1}{2} \gamma_{\vec{q}}
\Big(|\nu_{\vec{q}}|^2\tanh \frac{\beta
E_{\vec{q}\uparrow}}{2}-|u_{\vec{q}}|^2\tanh \frac{\beta
E_{\vec{q}\downarrow}}{2}\Big). 
\end{equation}
If $p\equiv (p_{\uparrow}+p_{\downarrow})/2$, we obtain:
\begin{equation}
p=-\frac{1}{N} \sum_{\vec{q}}\frac{1}{2}\gamma_{\vec{q}} 
\frac{-(t^{\uparrow}+t^{\downarrow})\Theta_{\vec{q}}-\bar{\mu}-\textcolor{czerwony}{p\frac{\gamma_{\vec k}}{\gamma_0}W}}{\omega_{\vec{q}}
} \Bigg(\tanh\frac{\beta E_{\vec{q}\uparrow}}{2}+\tanh \frac{\beta
E_{\vec{q}\downarrow}}{2}\Bigg).
\end{equation}
From the partition function, one can determine the grand canonical potential of
the superconducting state:
\begin{eqnarray}
\label{pot}
\frac{\Omega^{SC}}{N}&=&\frac{1}{4} Un(2-n)-\mu+\frac{1}{4}UM^2+W\gamma_0
n-\frac{1}{2}W\gamma_0 n^2-Wp_{\uparrow}^2/2\gamma_0-Wp_{\downarrow}^2/2\gamma_0
\nonumber \\
&+&\frac{1}{N} \sum_{\vec{k}} \frac{|\Delta_{\vec{k}}|^2}{2\omega_{\vec{k}}}
\frac{1}{2} \Bigg(\tanh\frac{\beta E_{\vec{k}\uparrow}}{2}+\tanh \frac{\beta
E_{\vec{k}\downarrow}}{2}\Bigg) \nonumber \\
&-&\frac{1}{\beta N}\sum_{\vec{k}} \ln \Bigg (2\cosh\frac{\beta
(E_{\vec{k}\uparrow}+E_{\vec{k}\downarrow})}{2}+2\cosh\frac{\beta
(-E_{\vec{k}\uparrow}+E_{\vec{k}\downarrow})}{2}\Bigg),
\end{eqnarray}
and also the free energy: $F^{SC}/N=\Omega^{SC} /N +\mu n$.

The equations for the superconducting gap parameter \eqref{del}, the number of
particles (determining $\mu$) \eqref{n} \textcolor{green}{and the} magnetization \eqref{M} satisfy
the following  extremum conditions: $\frac{\partial F}{\partial \Delta}=0$,
$\frac{\partial F}{\partial \mu}=0$, $M=-\frac{1}{N}\frac{\partial F}{\partial
h}$, respectively.

The above equations take into account the spin polarization in the presence of a
magnetic field and spin-dependent hopping ($t^{\uparrow}\neq t^{\downarrow}$)
\cite{Wilczek3, Cuoco}. This method includes the spin-dependent Hartree term and
can be called the BCS-Stoner approach.

The equations for the normal phase ($\Delta = 0$) have the following form:
\begin{equation}
n=1-\frac{1}{2N}\sum_{\vec{k}}\Bigg(\tanh\frac{\beta
E^{NO}_{\vec{k}\uparrow}}{2}+\tanh \frac{\beta
E^{NO}_{\vec{k}\downarrow}}{2}\Bigg),
\end{equation}
\begin{equation}
 M=\frac{1}{2N}\sum_{\vec{k}}\Bigg(\tanh\frac{\beta
E^{NO}_{\vec{k}\downarrow}}{2}-\tanh \frac{\beta
E^{NO}_{\vec{k}\uparrow}}{2}\Bigg),
\end{equation}
\begin{equation}
p=-\frac{1}{N} \sum_{\vec{k}}\frac{1}{2}\gamma_{\vec{k}} 
\Bigg(\tanh\frac{\beta E^{NO}_{\vec{k}\uparrow}}{2}+\tanh \frac{\beta
E^{NO}_{\vec{k}\downarrow}}{2}\Bigg),
\end{equation}
\begin{eqnarray}
\frac{\Omega^{NO}}{N}&=&\frac{1}{4} Un(2-n)-\mu+\frac{1}{4}UM^2 +W\gamma_0
n-\frac{1}{2}W\gamma_0 n^2\nonumber \\
&-&Wp_{\uparrow}^2/2\gamma_0-Wp_{\downarrow}^2/2\gamma_0 -\frac{1}{\beta
N}\sum_{\vec{k}} \ln \Bigg (2\cosh\frac{\beta
(E^{NO}_{\vec{k}\uparrow}+E^{NO}_{\vec{k}\downarrow})}{2}\Bigg)\nonumber\\
&-&\frac{1}{\beta N}\sum_{\vec{k}} \ln \Bigg( 2\cosh\frac{\beta
(-E^{NO}_{\vec{k}\uparrow}+E^{NO}_{\vec{k}\downarrow})}{2}\Bigg),\nonumber\\
\end{eqnarray}
where:
\begin{equation}
E_{\vec{k}\downarrow, \uparrow}^{NO}= \pm
(-t^{\downarrow}+t^{\uparrow})\Theta_{\vec{k}}\pm \frac{UM}{2} \pm
\frac{1}{2}W(p_{\uparrow}-p_{\downarrow})\frac{\gamma_{\vec{k}}}{\gamma_0}\pm
h+\omega_{\vec{k}}^{NO},
\end{equation}
\begin{equation}
\label{omega}
\omega_{\vec{k}}^{NO}=(-t^{\uparrow}-t^{\downarrow})\Theta_{\vec{k}}-\bar{\mu}-\textcolor{czerwony}{p\frac{\gamma_{\vec k}}{\gamma_{0}}W}.
\end{equation}

The equations for the superconducting gap parameter, the number of particles,
the magnetization, the Fock parameter and the grand canonical potential at
$T=0$, both in the superconducting and in the normal state are presented in
Appendix \ref{appendix2}.


\chapter{Kosterlitz-Thouless Transition}

A two-dimensional superconductor at $h=0$ can be classified in the same
universality class as a 2D superfluid system or the 2D XY model. In these cases,
no phase transition exists above $T=0$, understood as the
\textcolor{czerwony}{dis}appearance of a long-range order \cite{Mermin, Hohenberg}. However, there is
some evidence that a \textbf{topological ordering} can occur in these systems
\cite{KT, KT2}. 

In the 2D XY model, the \textbf{Kosterlitz-Thouless (KT) transition} \cite{KT,
KT2} from the disordered to a topologically ordered system takes place. Below
the KT transition temperature ($T_c^{KT}$), the system has a quasi-long-range
(algebraic) order, which is characterized by a power law decay of the order
parameter correlation function and non-zero superfluid stiffness ($\rho_s$).
Below $T_{c}^{KT}$, bound vortex-antivortex pairs exist. These pairs get
unbound when temperature increases above $T^{KT}_c$. 

\section{The XY model}
\label{sec_XY}

As mentioned before, in statistical physics, the two-dimensional XY model
is a good example of a system which does not have a long-range order
(LRO), because it is unstable due to the low-energy excitations of spin waves at only
finite temperatures \cite{Mermin}. This model is a system of spins which can
rotate in the plane of the lattice and is defined by the Hamiltonian:
\begin{equation}
\label{XYham}
H=-J \sum_{\langle ij \rangle} \vec{S_{i}} \cdot \vec{S_{j}} = -J\sum_{\langle
ij \rangle} \cos (\phi_{i} - \phi_{j}), 
\end{equation}
where: $J>0$ -- spin coupling constant, $\sum_{\langle ij \rangle}$ -- sum over
nearest neighbors. \textcolor{czerwony}{In its clasical version}, the spins $\vec{S}$ are vectors of length $1$ oriented in the
$x-y$ plane, $\phi_{i}$ -- angle between the $i$-th spin and the $x$ axis. The
spins are located in the nodes of the square lattice with lattice spacing $a$. 

The XY model exhibits specific symmetries, which make it possible to study the
ferro- and antiferromagnetic ordering.
This model is also used to describe such systems as: (a) superconducting
materials, (b) thin superfluid helium films \cite{Bishop}, (c) Josephson
junctions \cite{Resnick}, (d) gaseous and liquid crystal systems, (e) melting in
2D (dislocations).

It is worth noting that the Hamiltonian \eqref{XYham} has a rotational symmetry,
i.e.: the transformation $\phi_{i} \rightarrow \phi_{i} + \phi_0$, for all $i$
leaves the Hamiltonian invariant. 

At high temperatures, more spins flip randomly, because the interactions between
the spins are weak compared to the thermal energy. As a consequence, the system
is in a disordered state. At $T=0$, all the spins align along a given direction.
At low temperatures, the spins fluctuate weakly around this direction. If
$|\phi_{i} - \phi_{j}| \ll 2\pi$ (the spin-wave approximation), the Hamiltonian
\eqref{XYham} takes the form:
\begin{equation}
\label{XYmodel'}
 H=-\frac{zNJ}{2} +\frac{1}{2} J\sum_{\langle ij \rangle}
(\phi_{i}-\phi_{j})^2=E_0+\frac{1}{4} J \sum_{\vec{r},\vec{a}} (\phi
(\vec{r}+\vec{a})-\phi (\vec{r}))^2, 
\end{equation}
where: $z$ -- the number of nearest neighbors, $E_0$ -- the ground state energy.
The sum over $\vec{a}$ -- the sum over nearest neighbors of site $\vec{r}$. If
$\phi (\vec{r})$ does not vary a lot from site to site, one can write the
Hamiltonian \eqref{XYmodel'} in the continuum limit:
\begin{equation}
\label{XYmodel2}
 H=E_0+\frac{1}{2}J \int d^2 \vec{r} (\vec{\nabla} \phi (\vec{r}))^2,
\end{equation}
where the finite differences and the sum over lattice sites in \eqref{XYmodel'}
are replaced by derivatives and an integral, respectively. The second term is
the spin-wave energy.

Let us calculate the spin-spin correlation function:
\begin{equation}
 g(r)=\langle \vec{S}(\vec{r}) \cdot \vec{S} (0) \rangle =\langle \exp (i(\phi
(\vec{r}) -\phi (0))) \rangle,
\end{equation}
where $\langle ... \rangle$ is the thermal average.

In high temperatures, the correlation function takes the form: $g(r) \sim \exp
(-r/ \xi (T))$. It decays exponentially to zero at large distances and the
system is disordered. On the other hand, if the correlation function decays to a
constant value at $r\rightarrow \infty$, the system exhibits LRO. 

At low temperatures, by using the spin-wave approximation \eqref{XYmodel2}, we
find the correlation function which decays with the power law:
\begin{equation}
 g(r) = \Big(\frac{a}{\pi r} \Big)^{\eta (T)},
\end{equation}
where: $\eta (T)= k_B T/2 J \pi$. This kind of ordering in the system,
characterized by a power-low decay of the correlation function, is called the
quasi-long-rang order (qLRO).
Therefore, there must exist some excitations in the system which can bring the
system from its low-temperature phase to the high-temperature state. 

In 1972, Kosterlitz and Thouless proposed a scenario in which these new
excitations (called vortices) are unbound above the critical temperature
$T_c^{KT}$. On the other hand, the vortices form tightly bound pairs at
$T<T_c^{KT}$. It is worth emphasizing that the vortices can be positively or
negatively charged (i.e. vortices or antivortices).

Let us consider an isolated vortex. In the continuum approximation $\phi (r,
\theta) =n \theta$, where $\phi (r, \theta)$ -- the spin orientation at a polar
coordinates $r$, $\theta$ and $n$ -- the strength of the vortex. Then: $\int
d\vec{l} \cdot \nabla \phi = 2n\pi$ and $\nabla \phi = \frac{n}{r}$. One can
easily show that the energy of the system with an isolated vortex takes the
form:
\begin{equation}
\label{vortex}
E=\pi J n^2 \ln \Big( \frac{L}{a} \Big), 
\end{equation}
where $L$ is the radius of the system. Therefore, the energy of the vortex
increases logarithmically with the size of the system and is infinite in the
thermodynamic limit. 

In turn, the energy of the vortex-antivortex pair is given by:
\begin{equation}
\label{2vortex}
 E_{pair} (\vec{r}_1, \vec{r}_2)= -2\pi n_1 n_2 \ln
\frac{|\vec{r}_1-\vec{r}_2|}{a} ,
\end{equation}
where: $n_1$, $n_2$ -- the strengths of the vortices. One can easily notice that
the energy of an isolated vortex \eqref{vortex} is higher than the energy of the
vortex-antivortex pair \eqref{2vortex} if the radius of the system is much
larger than the distance between two vortices which are oppositely charged. It
means that the existence of bound vortex pairs is energetically favorable in
low temperatures. However, \textcolor{czerwony}{vortex pair unbinding} is much easier at a higher temperature,
because of the thermal energy.

The entropy of an isolated vortex is given by:
\begin{equation}
 S=k_B \ln \Big( \frac{L}{a} \Big)^2.
\end{equation}
Then, the free energy of the system with an isolated vortex takes the form:
\begin{equation}
 F=E-TS\approx (\pi Jn^2 - 2k_B T )\ln \Big(\frac{L}{a} \Big).
\end{equation}
As we can see, for:
\begin{equation}
\label{KTtemp}
 k_B T_c = \frac{\pi}{2} J
\end{equation}
the free energy changes its sign. Obviously, the free energy becomes more
negative with increasing temperature. For sufficiently high $T$, the
vortex-antivortex pairs are destroyed by the thermal energy. 

\section{The Kosterlitz-Thouless critical temperature determination}
\label{3.2}
In Sec. \ref{sec_XY} it has been shown that $T_c^{KT}$ in the XY model is
expressed by eq.~\eqref{KTtemp}. 

As mentioned before, there is experimental evidence which confirms the nature of
the phase transition in thin films of $^{4}$He. It is worth mentioning that the
experimental situation in trapped gases is different from theoretical
predictions. Because of the presence of a trapping potential, the system is
not uniform. Hence, the Bose-Einstein condensation is possible to occur in 2D
systems \cite{Petrov, Bagnato}. However, the KT transition is also observed in
trapped systems \cite{Prokofev, Sachdev}. Recent works of the experimental group
from Paris \cite{Hadzibabic, Kruger} have reported the observation of the KT
transition in a trapped quantum degenerate gas of $^{87}$Rb.

The attractive Hubbard Model, which describes the isotropic
superconductivity, can be mapped into the XY model by means of the functional
integration method \cite{Alvarez, Paolo}. As mentioned \textcolor{czerwony}{above}, the
two-dimensional superconductor at $h=0$ can be classified in the same
universality class as a 2D superfluid system or the 2D XY model. Therefore, the
KT temperature can be determined in the same way as in the previous case. Then,
the KT critical temperature in the superconducting system takes the form
\cite{Halperin}:
\begin{equation}
\label{KT}
 k_B T_c^{KT} =\frac{\pi {\hbar}^2 n_s^*}{2m^*}= \frac{\pi \rho_s
(T_c^{KT})}{2},
\end{equation}
where: $n_{s}^*=n_s/2$ -- the number of pairs in the condensate, $m^*=2m$ -- the
mass of the pair, $n_s$ -- the density of the superconductor electrons, $m$ --
the mass of the electron, $\rho_{s} (T)$ -- the superfluid density or the phase
stiffness. $\rho_s$ is equivalent of the spin stiffness in the XY model
\cite{Fisher, Ohta}.

The superfluid stiffness (or helicity modulus) can be determined from the change
in the grand canonical potential ($\Omega_{\vec{q}}$) of the system in the
response to the order parameter twist \cite{Denteneer-2, Denteneer, Leeuwen}.
For this purpose, we have to determine $\Omega_{\vec{q}}$ with the phase
twist. 

We assume that the fluctuations of the order parameter are specified by the
formula:
\begin{equation}
 \Delta_l = |\Delta_l| e^{2i\vec{q}\cdot \vec{R}_l},
\end{equation}
where $\vec{q}$ is the twist vector, $\vec{R}_l$ -- the position vector of site
$l$.
Then, the effective Hamiltonian \eqref{ham''} takes the form:
\begin{equation}
\label{hameff}
 H_{eff}=\sum_{\vec{k}\sigma} \xi_{\vec{k}+\vec{q}}^{\sigma}
c_{\vec{k}\sigma}^{\dag} c_{\vec{k}\sigma}-\sum_{\vec{k}}
(|\Delta|c_{\vec{k}\uparrow}^{\dag}c_{-\vec{k}\downarrow}^{\dag}+|\Delta|c_{
-\vec{k}\downarrow}c_{\vec{k}\uparrow})+C,
\end{equation}
where: $\xi_{\vec{k}+\vec{q}}^{\downarrow,
\uparrow}=\epsilon_{\vec{k}+\vec{q}}^{\downarrow, \uparrow}
-\bar{\mu}_{\downarrow,\uparrow} \pm h$. We do not take into account the Fock
term in further calculations.

The above Hamiltonian can be written as:
\begin{equation}
 H_{eff}=\sum_{\vec{k}} \Psi_{\vec{k}}^{\dag} \hat{M}_{\vec{k},\vec{q}}
\Psi_{\vec{k}}+\sum_{\vec{k}}
(\xi_{\vec{q}-\vec{k}}^{\uparrow}+\xi_{\vec{q}-\vec{k}}^{\downarrow}),
\end{equation}
where: the matrix $\hat{M}_{\vec{k},\vec{q}}$:
\begin{equation}
\hat{M}_{\vec{k},\vec{q}}=\left(\begin{array}{cccc}
\xi_{\vec{k}+\vec{q}}^{\uparrow} & -|\Delta| & 0 & 0 \\
-|\Delta|^* & -\xi_{\vec{q}-\vec{k}}^{\downarrow} & 0 & 0 \\
0 & 0 & \xi_{\vec{k}+\vec{q}}^{\downarrow} & |\Delta| \\
0 & 0 & |\Delta| & -\xi_{\vec{q}-\vec{k}}^{\uparrow} \\ 
\end{array}\right),
\end{equation}
the operator $\Psi$:
\begin{equation}
\Psi_{\vec{k}}= \left(\begin{array}{cccc}
c_{\vec{k}\uparrow} \\
c_{-\vec{k}\downarrow}^{\dag} \\
c_{\vec{k}\downarrow} \\
c_{-\vec{k}\uparrow}^{\dag} \\
\end{array} \right).
\end{equation}
$\hat{M}_{\vec{k},\vec{q}}$ is a block diagonal matrix. To determine the
quasiparticle energies (eigenvalues of the matrix $\hat{M}_{\vec{k},\vec{q}}$),
we solve the equations:
\begin{equation}
\det \left(\begin{array}{cccc}
\xi_{\vec{k}+\vec{q}}^{\uparrow}-E & -|\Delta| \\
-|\Delta|^* & -\xi_{\vec{q}-\vec{k}}^{\downarrow}-E\\
\end{array}\right)=0,
\end{equation}
\begin{equation}
\det\left(\begin{array}{cccc}
\xi_{\vec{k}+\vec{q}}^{\downarrow}-E & |\Delta| \\
|\Delta| & -\xi_{\vec{q}-\vec{k}}^{\uparrow}-E \\ 
\end{array}\right)=0.
\end{equation}
As a result, we get:
\begin{equation}
E_{\textcolor{czerwony}{{1,2}}}(\vec k, \textcolor{czerwony}{\vec q})=\frac{\xi_{\vec{k}+\vec{q}}^{\uparrow}-\xi_{\vec{q}-\vec{k}}^{
\downarrow}}{2}\pm
\sqrt{\Bigg(\frac{\xi_{\vec{k}+\vec{q}}^{\uparrow}+\xi_{\vec{q}-\vec{k}}^{
\downarrow}}{2}  \Bigg)^2 +|\Delta|^2},
\end{equation}
\begin{equation}
E_{{\textcolor{czerwony}{{3,4}}}}(\vec k \textcolor{czerwony}, {\vec q})=\frac{\xi_{\vec{k}+\vec{q}}^{\downarrow}-\xi_{\vec{q}-\vec{k}}^{
\uparrow}}{2}\pm
\sqrt{\Bigg(\frac{\xi_{\vec{k}+\vec{q}}^{\uparrow}+\xi_{\vec{q}-\vec{k}}^{
\downarrow}}{2}  \Bigg)^2 +|\Delta|^2},
\end{equation}
It is easy to show that: 
$E_{\textcolor{czerwony}{2}}(-\textcolor{czerwony}{\vec k}, \textcolor{czerwony}{\vec q})=-E_{\textcolor{czerwony}{3}}(\textcolor{czerwony}{\vec k, \vec q})$ and $E_{\textcolor{czerwony}{4}}(-\textcolor{czerwony}{\vec k, \vec q})=-E_{\textcolor{czerwony}{1}}(\textcolor{czerwony}{\vec k, \vec q})$.

After some transformations, the grand canonical potential with the phase twist
takes the form:
\begin{eqnarray}
\frac{\Omega_{\vec{q}}}{N}&=&\frac{1}{N}\sum_{k>0}(\xi_{\vec{q}-\vec{k}}^{
\uparrow}+\xi_{\vec{q}-\vec{k}}^{\downarrow}) -\frac{1}{\beta N}\sum_{\vec{k}}
\ln \Bigg[2\cosh\Bigg( \frac{\beta E_{1}(\vec k, \vec q)}{2}\Bigg)\Bigg]\nonumber \\
&-& \frac{1}{\beta N} \sum_{\vec{k}} \ln \Bigg[2\cosh\Bigg( \frac{\beta
E_3(\vec k, \vec q)}{2}\Bigg)\Bigg] +C,
\end{eqnarray}
where $C$ is a constant.

We Taylor-expand $\Omega_{\vec{q}}$ up to the ${q}^2$ term:
\begin{equation}
\frac{\Omega_{\vec{q}}}{N}=\frac{\Omega_{\vec{q}=0}}{N}+2\rho_s {q}^2 +O({q}^4).
\end{equation}
On the other hand:
\begin{equation}
\label{expantion}
 \Omega_{\vec{q}}=\Omega_{\vec{q}=0}+\frac{1}{2} q_x^2 \frac{\partial^2
\Omega_{\vec{q}}}{\partial q_x^2}+\frac{1}{2} q_y^2 \frac{\partial^2
\Omega_{\vec{q}}}{\partial q_y^2}+O(q^4).
\end{equation}
On the square lattice $\frac{\partial^2\Omega_{\vec{q}}}{\partial
q_x^2}=\frac{\partial^2\Omega_{\vec{q}}}{\partial q_y^2}$ and hence:
\begin{equation}
\label{expantion-2}
 \Omega_{\vec{q}}=\Omega_{\vec{q}=0}+\frac{1}{2} q^2 \frac{\partial^2
\Omega_{\vec{q}}}{\partial q_x^2}+O(q^4).
\end{equation}
Comparing the coefficients of $q^2$ in \eqref{expantion} and
\eqref{expantion-2}, we get:
\begin{equation}
 2\rho_s=\frac{1}{2}\frac{\partial^2\Omega_{\vec{q}}}{\partial q_x^2}
\Bigg|_{\vec{q}=0}.
\end{equation}
Calculating this derivative explicitly yields:
\begin{eqnarray}
\label{ro_s}
\rho_s(T)&=&\frac{1}{4N}\sum_{\vec{k}} \Bigg\{ \frac{\partial
^{2}\epsilon^{+}_{\vec{k}}}{\partial k_x^2} -\frac{1}{2}\Bigg[\frac{\partial
^{2}\epsilon^{-}_{\vec{k}}}{\partial
k_x^2}+\frac{\epsilon^{+}_{\vec{k}}}{\omega_{\vec{k}}}   \Bigg(\frac{\partial
^{2}\epsilon^{+}_{\vec{k}}}{\partial k_x^2}\Bigg) 
+\Bigg(\frac{\partial \epsilon^{-}_{\vec{k}}}{\partial k_x}\Bigg)^2
\frac{|\Delta|^2}{\omega_{\vec{k}}^3} \Bigg] \tanh \Bigg(\frac{\beta
E_{\vec{k}\uparrow}}{2}\Bigg)\nonumber \\
&+&\frac{1}{2}\Bigg[\frac{\partial ^{2}\epsilon^{-}_{\vec{k}}}{\partial
k_x^2}-\frac{\epsilon^{+}_{\vec{k}}}{\omega_{\vec{k}}} \Bigg(\frac{\partial
^{2}\epsilon^{+}_{\vec{k}}}{\partial k_x^2}\Bigg)  
-\Bigg(\frac{\partial \epsilon^{-}_{\vec{k}}}{\partial k_x}\Bigg)^2
\frac{|\Delta|^2}{\omega_{\vec{k}}^3} \Bigg] \tanh \Bigg(\frac{\beta
E_{\vec{k}\downarrow}}{2}\Bigg)\nonumber \\
&+&\Bigg[\frac{\partial \epsilon^{+}_{\vec{k}}}{\partial k_x}+
\frac{\epsilon^{+}_{\vec{k}}}{\omega_{\vec{k}}} \Bigg(\frac{\partial
\epsilon^{-}_{\vec{k}}}{\partial k_x}\Bigg)\Bigg]^2 
\frac{\partial f(E_{\vec{k}\uparrow})}{\partial
E_{\vec{k}\uparrow}}+\Bigg[\frac{\partial \epsilon^{+}_{\vec{k}}}{\partial k_x}-
\frac{\epsilon^{+}_{\vec{k}}}{\omega_{\vec{k}}} \Bigg(\frac{\partial
\epsilon^{-}_{\vec{k}}}{\partial k_x}\Bigg)\Bigg]^2 \frac{\partial
f(E_{\vec{k}\downarrow})}{\partial E_{\vec{k}\downarrow}}  \Bigg\},
\end{eqnarray}
where: $\epsilon_{\vec{k}}^{+}=\frac{\xi_{\vec{k}\uparrow}+\xi_{\vec{k}
\downarrow}}{2}$,
$\epsilon_{\vec{k}}^{-}=\frac{\xi_{\vec{k}\uparrow}-\xi_{\vec{k}
\downarrow}}{2}$, \textcolor{czerwony}{$\omega_{\vec k}=\sqrt{\bar{\epsilon}_{\vec k}^2+|\Delta|^2}$}.
Then, evaluating the derivatives with respect to $k_x$ and using the
trigonometric identities:
\begin{equation}
\tanh \Big(\frac{x}{2}\Big)\pm
\tanh\Big(\frac{y}{2}\Big)=\frac{\sinh\Big(\frac{x\pm
y}{2}\Big)}{\cosh\Big(\frac{x}{2}
\Big)\cosh\Big(\frac{y}{2}\Big)},
\end{equation}
\begin{equation}
 \frac{1}{\cosh^2\Big(\frac{x}{2}\Big)}+\frac{1}{\cosh^2\Big(\frac{y}{2}\Big)}
=\frac{4\Big(1+\cosh\Big(\frac{x+y}{2}\Big)\cosh\Big(\frac{x-y}{2}\Big)\Big)}{
\Big(\cosh\Big(\frac{x+y}{2}\Big)+\cosh\Big(\frac{x-y}{2}\Big)\Big)^2},
\end{equation}
\begin{equation}
\frac{1}{\cosh^2\Big(\frac{x}{2}\Big)}-\frac{1}{\cosh^2\Big(\frac{y}{2}\Big)}
=\frac{4\sinh\Big(\frac{x+y}{2}\Big)\sinh\Big(\frac{y-x}{2}\Big)}{
\Big(\cosh\Big(\frac{x+y}{2}\Big)+\cosh\Big(\frac{x-y}{2}\Big)\Big)^2}, 
\end{equation}

the superfluid stiffness \eqref{ro_s} can be rewritten as:
\begin{eqnarray}
\rho_s(T)&=&\frac{1}{4N}\sum_{\vec{k}} \Bigg((t^{\uparrow} \cos k_x
-t^{\downarrow} \cos k_x)X_{\vec{k}}^{a} 
-\Big(\frac{\xi_{\vec{k} \uparrow}+\xi_{\vec{k}
\downarrow}}{2\omega_{\vec{k}}}(t^{\uparrow} \cos k_x+t^{\downarrow} \cos k_x)
\nonumber \\
&+&(t^{\uparrow} \sin k_x-t^{\downarrow} \sin
k_x)^2\frac{|\Delta|^{2}}{\omega_{\vec{k}}^{3}}\Big)X_{\vec{k}}^{b} \nonumber\\
&-&\Big((t^{\uparrow} \sin k_x+t^{\downarrow} \sin k_x)^2 
+\Big( \frac{\xi_{\vec{k} \uparrow}+\xi_{\vec{k}
\downarrow}}{2\omega_{\vec{k}}}\Big)^2(t^{\uparrow} \sin k_x-t^{\downarrow} \sin
k_x)^2\Big)Y_{\vec{k}}^{a} \nonumber \\ 
&+&2\Big(\frac{\xi_{\vec{k} \uparrow}+\xi_{\vec{k}
\downarrow}}{2\omega_{\vec{k}}}\Big) ((t^{\uparrow} \sin k_x)^2-(t^{\downarrow}
\sin k_x)^2)Y_{\vec{k}}^{b}\Bigg),
\end{eqnarray}
where:
\begin{equation}
 X_{\vec{k}}^{a}=\frac{\sinh(\beta
\omega_{\vec{k}})}{\cosh(\beta((-t^{\downarrow}+t^{\uparrow})\Theta_{\vec{k}}
+h+\frac{UM}{2}))+\cosh(\beta \omega_{\vec{k}})},
\end{equation}
\begin{equation}
X_{\vec{k}}^{b}=\frac{\sinh(\beta((-t^{\downarrow}+t^{\uparrow})\Theta_{\vec{k}}
h+\frac{UM}{2}))}{\cosh(\beta((-t^{\downarrow}+t^{\uparrow})\Theta_{\vec{k}}
+h+\frac{UM}{2}))+\cosh(\beta \omega_{\vec{k}})},
\end{equation}
\begin{equation}
Y_{\vec{k}}^{a}=\beta
\frac{\cosh(\beta((-t^{\downarrow}+t^{\uparrow})\Theta_{\vec{k}}+h+\frac{UM}{2}
))\cosh(\beta
\omega_{\vec{k}})+1}{\big(\cosh(\beta((-t^{\downarrow}+t^{\uparrow})\Theta_{\vec
{k}}+h+\frac{UM}{2}))+\cosh(\beta \omega_{\vec{k}})\big)^2},
\end{equation}
\begin{equation}
Y_{\vec{k}}^{b}=\beta
\frac{\sinh(\beta((-t^{\downarrow}+t^{\uparrow})\Theta_{\vec{k}}+h+\frac{UM}{2}
))\sinh(\beta
\omega_{\vec{k}})}{\big(\cosh(\beta((-t^{\downarrow}+t^{\uparrow})\Theta_{\vec{k
}}+h+\frac{UM}{2}))+\cosh(\beta \omega_{\vec{k}})\big)^2},
\end{equation}
$\Delta$, $M$, $\mu$ are found from a set of self-consistent equations
\eqref{n}-\eqref{del}.

If $t^{\uparrow}=t^{\downarrow}$, eq.~\eqref{ro_s} reduces to:
\begin{eqnarray}
\label{ro_s-2}
\rho_s &=&\frac{1}{4N} \sum_{k} \Bigg\{
\frac{\partial^2\epsilon_{\vec{k}}}{\partial k_x^2}\Bigg[
1-\frac{\bar{\epsilon}_{\vec{k}}}{2\omega_{\vec{k}}}
\Bigg(\tanh\Bigg(\frac{\beta
E_{\vec{k}\uparrow}}{2}\Bigg)+\tanh\Bigg(\frac{\beta
E_{\vec{k}\downarrow}}{2}\Bigg) \Bigg)\Bigg]+\nonumber\\
&+&\Bigg(\frac{\partial \epsilon_{\vec{k}}}{\partial k_x}
\Bigg)^2\Bigg[\frac{\partial f(E_{\vec{k}\uparrow})}{\partial
E_{\vec{k}\uparrow}} +\frac{\partial f(E_{\vec{k}\downarrow})}{\partial
E_{\vec{k}\downarrow}} \Bigg] \Bigg\},
\end{eqnarray}
where: $\bar{\epsilon}_{\vec{k}}=\epsilon_{\vec{k}}-\bar{\mu}$,
$\textcolor{czerwony}{E_{\vec{k}\downarrow, \uparrow}=\pm \frac{UM}{2}\pm
h+\omega_{\vec{k}}}$.

Expression
\eqref{ro_s-2} can be written in a simple form:
\begin{equation}
\rho_s(T)=-\frac{t}{N} \sum_{\vec{k}}
\Bigg(\frac{\epsilon_{\vec{k}}-\bar{\mu}}{2\omega_{\vec{k}}} \cos k_x
X_{\vec{k}}^{a}+t\sin^2k_x Y_{\vec{k}}^{a}\Bigg),
\end{equation}
with $X_{\vec{k}}^{a}$ and $Y_{\vec{k}}^{a}$ for $t^{\uparrow}=t^{\downarrow}$.

Finally, if $h=M=0$, $E_{\vec{k}\uparrow}=E_{\vec{k}\downarrow}=E_{\vec{k}}$, where {\textcolor{czerwony}{$E_{\vec k}$ is given by the standard form,}
$\rho_s$ is:
\begin{equation}
 \rho_s=\frac{1}{2N} \sum_{\vec{k}}\Bigg\{
\frac{1}{2}\frac{\partial^2\epsilon_{\vec{k}}}{\partial
k_x^2}\Bigg[1-\frac{\bar{\epsilon}_{\vec{k}}}{2\omega_{\vec{k}}}\tanh\Bigg(\frac
{\beta E_{\vec{k}}}{2}\Bigg)\Bigg]+\Bigg(\frac{\partial
\epsilon_{\vec{k}}}{\partial k_x} \Bigg)^2\frac{\partial
f(E_{\vec{k}})}{\partial E_{\vec{k}}}\Bigg\}\textcolor{green}{.}
\end{equation}

\newpage
\thispagestyle{empty}
\mbox{}

\chapter{Superconducting properties in the presence of \textcolor{czerwony}{a} Zeeman magnetic field: Weak and intermediate coupling}
\label{chapter4}

In this chapter we focus on the analysis of the influence of the Zeeman term on
the superfluid characteristics of the attractive Hubbard model (AHM) with
spin independent hopping integrals ($t^{\uparrow}=t^{\downarrow}$) for
the square and simple cubic lattices. Within the mean-field (the BCS-Stoner)
approach, we construct phase diagrams in two ways: by fixing the chemical
potential ($\mu$) or the electron concentration ($n$), and show the relevant
differences resulting from these possibilities. The importance of the Hartree
term in the broken symmetry Hartree-Fock approximation is indicated. A reentrant
transition is observed for sufficiently high magnetic fields in the temperature
phase diagrams in the weak coupling regime. However, we also find a region
where the superfluid density ($\rho_s$) becomes negative on these diagrams, both
for $d=2$ and $d=3$. For the two-dimensional case, we investigate the
Kosterlitz-Thouless transition (KT) in the weak coupling regime. Finally, we
also briefly discuss the influence of the pure d-wave pairing symmetry on the
polarized superconducting phase stability in the 2D system. Some selected
results have been published by us in Refs. \textcolor{green}{\cite{Acta, Kujawa, Kujawa2}}. 

Because of the fact that most of this chapter is devoted to the analysis of
the superconducting characteristics of the spin-polarized AHM, we write its
Hamiltonian explicitly. The Hamiltonian \eqref{extham'} is reduced to the form:
\begin{equation}
\label{ham}
H=\sum_{ij\sigma} (t_{ij}-\mu \delta_{ij})c_{i\sigma}^{\dag}c_{j\sigma}+U\sum_{i} n_{i\uparrow}n_{i\downarrow}-h\sum_{i}(n_{i\uparrow}-n_{i\downarrow}),
\end{equation}
where: $n_{i\uparrow}=c_{i\uparrow}^{\dag}c_{i\uparrow}$, $n_{i\downarrow}=c_{i\downarrow}^{\dag}c_{i\downarrow}$, $t_{ij}$ -- spin independent hopping integral.

The superconducting gap parameter for the s-wave pairing symmetry is defined by: $\Delta=-\frac{U}{N}\sum_{\vec{i}} \langle c_{i \downarrow} c_{i \uparrow} \rangle=-\frac{U}{N}\sum_{\vec{k}} \langle c_{-\vec{k} \downarrow} c_{\vec{k} \uparrow} \rangle$.

The system of self-consistent equations for the s-wave pairing symmetry case ($W=0$, $U<0$) in the presence of a Zeeman magnetic field is reduced to:
\begin{equation}
\label{del2}
 \Delta=-\frac{U}{N}\sum_{\vec{k}}\frac{\Delta}{2\omega_{\vec{k}}}\frac{1}{2}\Bigg(\tanh\frac{\beta E_{\vec{k}\uparrow}}{2}+\tanh \frac{\beta E_{\vec{k}\downarrow}}{2}\Bigg),
\end{equation}
where: 
\begin{equation}
\label{quasiparticles}
E_{\vec{k}\downarrow, \uparrow}=\pm \frac{UM}{2}\pm h+\omega_{\vec{k}},
\end{equation} 
$\omega_{\vec{k}}=\sqrt{\epsilon_{\vec{k}}-\bar{\mu})^2+|\Delta|^2}$, $\bar{\mu}=\mu-\frac{Un}{2}$.
\\
The particle number equation takes the form:
\begin{equation}
\label{particle_eq}
n=1-\frac{1}{2N}\sum_{\vec{k}}\frac{\epsilon_{\vec{k}}-\bar{\mu}}{\omega_{\vec{k}}}
\Bigg(\tanh\frac{\beta E_{\vec{k}\uparrow}}{2}+\tanh \frac{\beta E_{\vec{k}\downarrow}}{2}\Bigg).
\end{equation}
The equation for the magnetization is:
\begin{equation}
\label{Magn_s}
 M=\frac{1}{2N}\sum_{\vec{k}}\Bigg(\tanh\frac{\beta E_{\vec{k}\downarrow}}{2}-\tanh \frac{\beta E_{\vec{k}\uparrow}}{2}\Bigg).
\end{equation}
The grand canonical potential of the superconducting state:
\begin{eqnarray}
&\frac{\Omega^{SC}}{N}&=\frac{1}{4} Un(2-n)-\mu+\frac{1}{4}UM^2 -\frac{|\Delta|^2}{U}\nonumber \\
&-&\frac{1}{\beta N}\sum_{\vec{k}} \ln \Bigg (2\cosh\frac{\beta (E_{\vec{k}\uparrow}+E_{\vec{k}\downarrow})}{2}+2\cosh\frac{\beta (-E_{\vec{k}\uparrow}+E_{\vec{k}\downarrow})}{2}\Bigg),
\end{eqnarray}
and the free energy: $F^{SC}/N=\Omega^{SC} /N +\mu n$.

\section{Ground state}
\label{ground_state}
The system of the above self-consistent equations was solved numerically
for the 2D square and 3D simple cubic lattices. The sums over the first
Brillouin zone were performed with the use of the
density of states\footnote{The density of states for the 2D square lattice is
expressed by the complete elliptic integral of the first kind $K$ \cite{bak},
while the form of the density of states for a simple cubic lattice has been
taken from Ref. \cite{rjj} \textcolor{czerwony}{(see Appendix \ref{appendix4})}.}, whenever possible. First, the chemical
potential was fixed. The first order transition lines were determined from
the condition $\Omega^{SC} =\Omega^{NO}$. Then, these results were
mapped onto the case of fixed $n$.

Let us start our analysis with the influence of a Zeeman magnetic field on
the order parameter $\Delta$, at fixed $\mu$. As shown in Fig. \ref{fig1}(a),
there are three different solutions for the order parameter at $T=0$. 

The first type of the solution is $\Delta \neq 0$ which is a constant function
of the magnetic field. It means that $n_{\uparrow}=n_{\downarrow}$ and the spin
polarization $P=0$. Such solution will be marked $\Delta(h=0)\equiv \Delta_0$
in further considerations. This is the unpolarized superconducting ground
state (SC$_0$).

For higher magnetic fields, the densities of states are different for the
particles with spin down and spin up ($n_{\uparrow} \neq n_{\downarrow}$, $P\neq
0$). Then, there appears a lower branch of solutions with $\Delta (h)$
and a gapless spectrum for the majority spin species. This is a
polarized superconducting state with a magnetic field dependent order
parameter. However, as follows from Fig. \ref{fig1}(b), this
branch of $\Delta$ is energetically unstable, i.e. the grand canonical potential
for this branch is higher than the grand canonical potential for the SC$_0$
branch.

At $h=\Delta_0$, there is a jump in the order parameter from a constant value to
zero. Hence, one can distinguish a third solution with $\Delta =0$ and $P\neq
0$. This is the Pauli paramagnetic state (normal phase (NO)).     

\begin{figure}
\hspace*{-0.8cm}
\includegraphics[width=0.38\textwidth,angle=270]{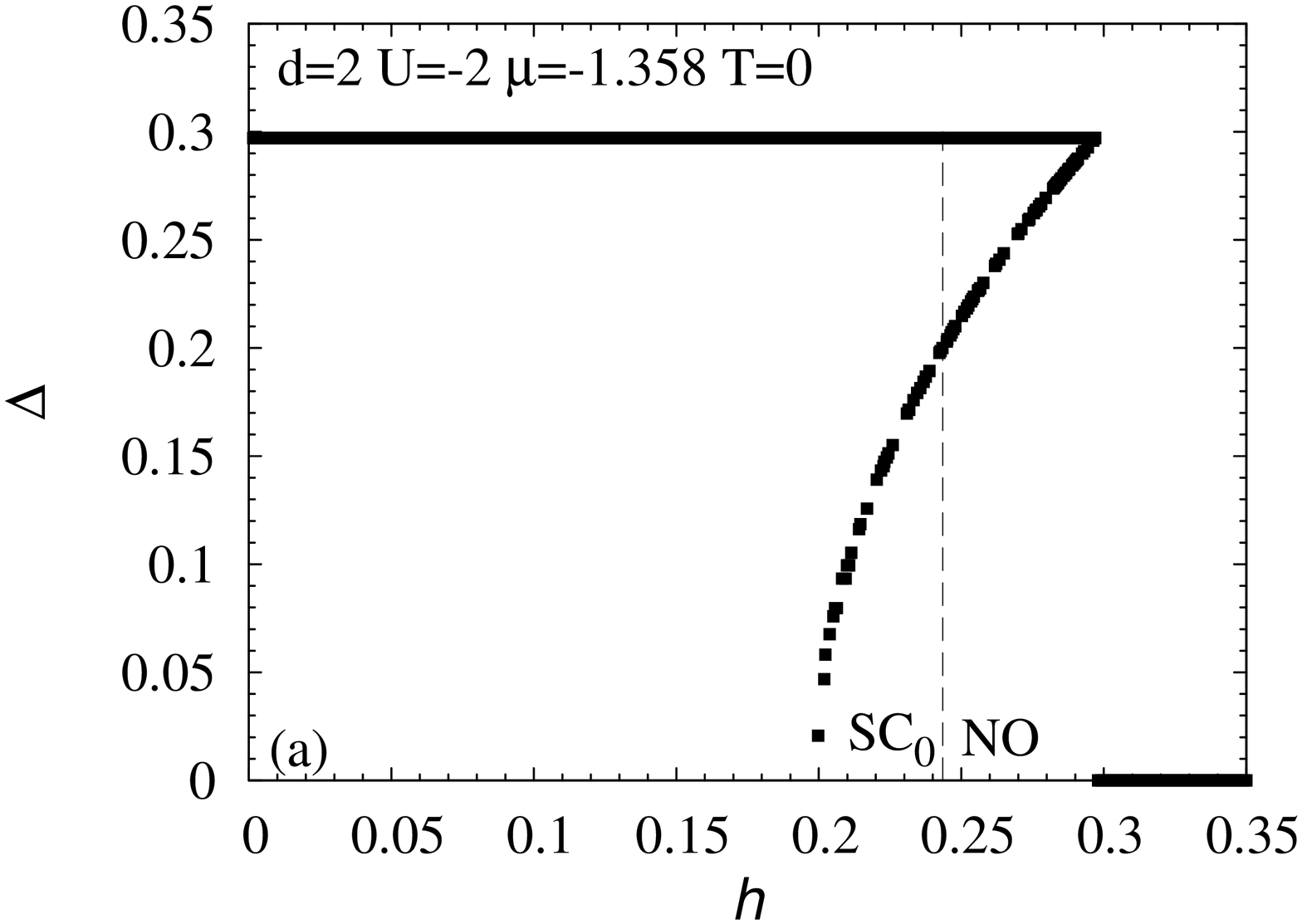}\hspace{-0.2cm}
\hspace*{-0.6cm}
\includegraphics[width=0.38\textwidth,angle=270]{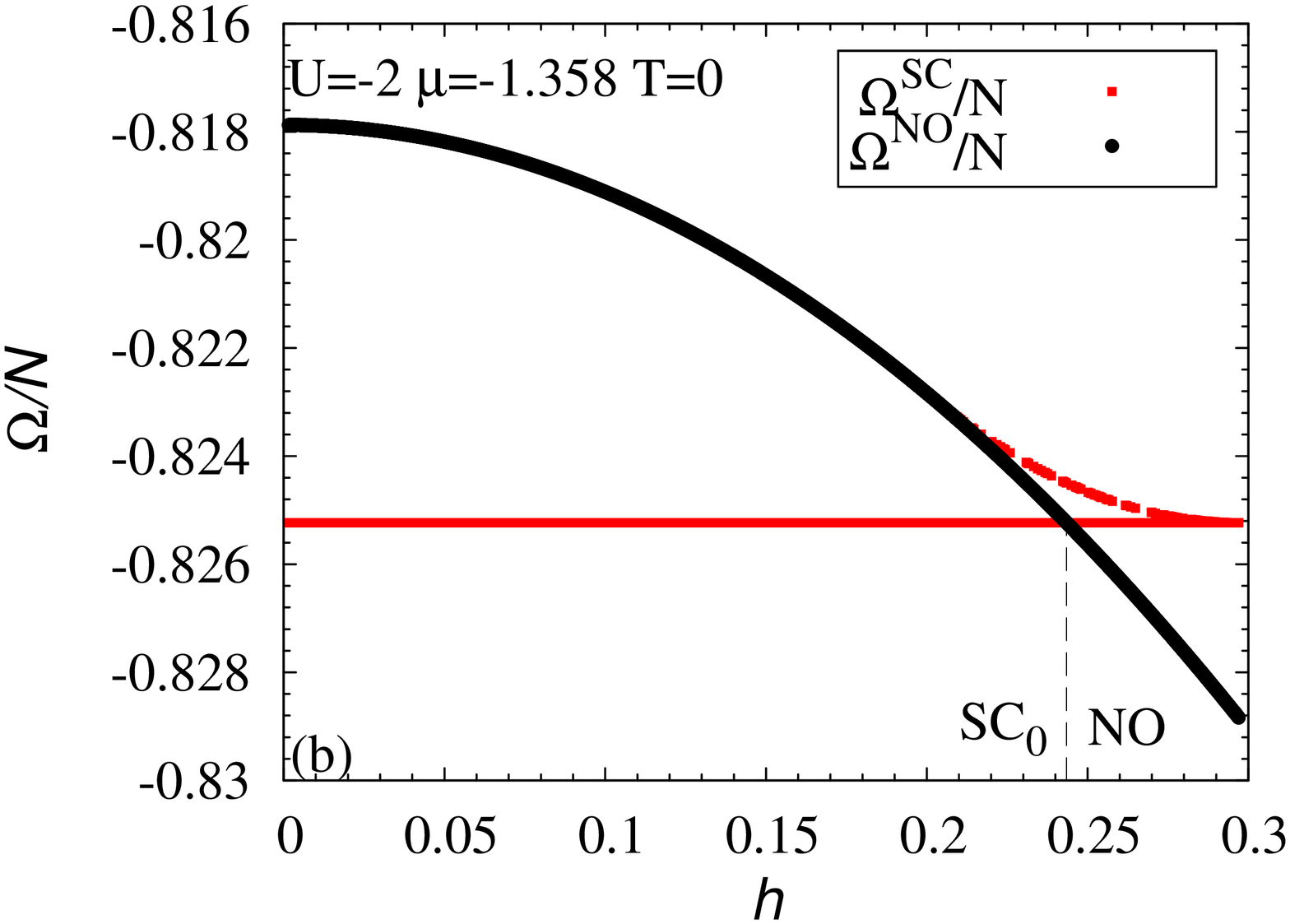}\\
\caption[Dependence of the order parameter (a) and the grand canonical potential
(b) on the magnetic field at $T=0$, $d=2$, $U=-2$, for fixed $\mu \approx
-1.358$.]{\label{fig1} Dependence of the order parameter (a) and the grand
canonical potential (b) on the magnetic field at $T=0$, $d=2$, \textcolor{green}{$U/t=-2$}, for a
fixed $\mu \approx -1.358$. SC$_0$ -- the superconducting phase without the spin
polarization ($n_{\uparrow}=n_{\downarrow}$, $P=0$), NO -- the normal state. In
Fig. (a) the lower branch is unstable (the grand canonical potential for this
branch is higher than the grand canonical potential for the SC$_0$ branch --
Fig. (b) -- red squares). The vertical dashed lines mark the Hartee-Fock phase
transition magnetic field -- the first order phase transition to the normal
state at $T=0$. The chemical potential has been chosen to yield $n \approx0.75$
at $h=0$.}
\end{figure}

If we investigate the behavior of the grand canonical potential vs. the order
parameter (Fig. \ref{fig2}), for fixed $h$ and $T=0$, we find that at $h=0$
(Fig. \ref{fig2}(a)) there is a global minimum for the solution from the upper
branch ($\Delta_0$ with $P=0$), i.e. it is stable.
When the magnetic field is increased, such a situation persists until
$h\approx0.2$. Up to this point, there are two solutions to the
self-consistent
equations ($\Delta=\Delta_0$ and $\Delta=0$).
For $h\in(0.2,0.242)$ (Fig. \ref{fig2}(b)), we still find a global minimum for
the solution from the upper
branch, but there appears a third solution (the lower branch with $\Delta
(h)$, $P\neq 0$), which is a local maximum, indicating that it is unstable. 
If $h$ is larger than approx. 0.242, the global minimum for the solution from
the upper branch becomes a local minimum, whereas the global minimum
corresponds now to $\Delta=0$ -- the system undergoes a first-order phase
transition to the NO state.
Finally, for $h>\Delta_0$, there is no local minimum at $\Delta \neq 0$ and the
self-consistent equations yield only one solution with $\Delta=0$.

\begin{figure}
\hspace*{-0.8cm}
\includegraphics[width=0.38\textwidth,angle=270]{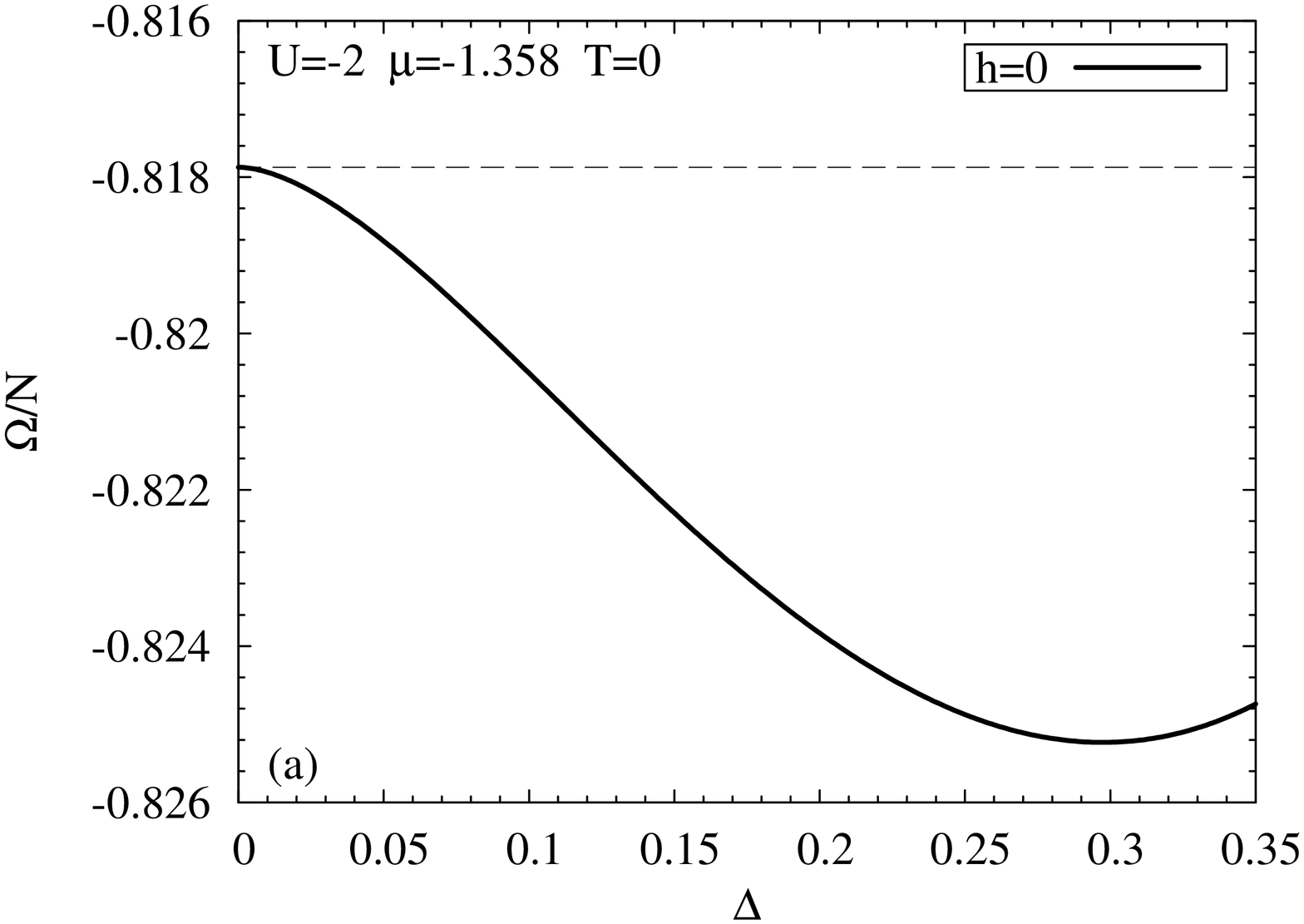}\hspace{-0.2cm}
\hspace*{-0.6cm}
\includegraphics[width=0.38\textwidth,angle=270]
{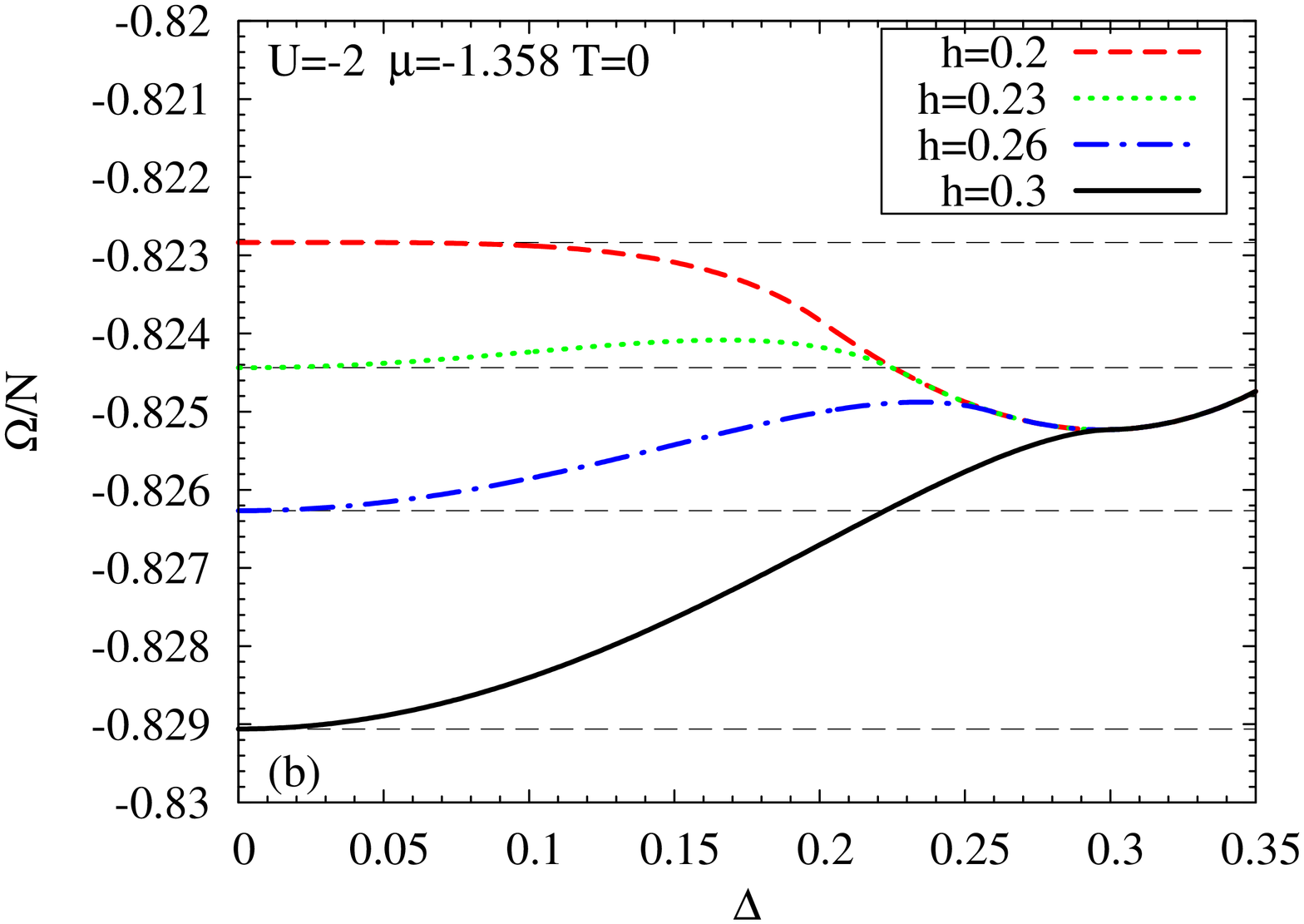}
\caption{\label{fig2} Dependence of the grand canonical potential on the
order parameter at $T=0$ and fixed $\mu\approx -1.358$. (a) $h=0$, (b) $h=0.2$
(red, dashed line), $h=0.23$ (green, dotted line), $h=0.26$ (blue,
dash-dotted line) and $h=0.3$ (black, solid line). \textcolor{green}{$t$ is used as the unit}.}
\end{figure}

The influence of \textcolor{czerwony}{a} Zeeman magnetic field on the superconducting phase is also
manifested by changes in the density of states.

The spin-dependent density of states (DOS) in the superconducting ground state is determined from:
\begin{equation}
\label{DOS}
 g_{\sigma}(E)=\frac{1}{N} \sum_{\vec{k}}\Big[|u_{\vec{k}}|^2\delta
(E-E_{\vec{k}\sigma}) + |\nu_{\vec{k}}|^2\delta (E+E_{\vec{k}\,-\sigma})\Big],
\end{equation}
where: the coefficients $u_{\vec{k}}$ and $\nu_{\vec{k}}$ are given by
Eqs.~\eqref{nu}-\eqref{u}, $E_{\vec{k}\downarrow,\uparrow}$ are expressed in the
general form by Eq.~\eqref{energy} and
$\bar{\epsilon}_{\vec{k}\downarrow,\uparrow}$ --
by Eqs.~\eqref{barepsilon2}-\eqref{barepsilon1}.
The total density of states: $g(E)=g_{\uparrow}(E)+g_{\downarrow}(E)$. 
\textcolor{czerwony}{For numerical calculations, the Dirac delta function is expressed by:  
$\delta (x)=\frac{1}{\pi} lim_{\eta \rightarrow 0}  \frac{\eta}{x^2+\eta^2}$, where $\eta =10^{-3}$.}

Apart from changes in the densities of states, the influence of
\textcolor{czerwony}{a} magnetic field on the superconducting phase is manifested by a change
in the momentum distributions in the Brillouin zone.
The momentum distribution \textcolor{czerwony}{at finite temperatures}, in the Brillouin zone is given
by:
\begin{equation}
 n_{\vec{k}\sigma}=|u_{\vec{k}}|^2 f(E_{\vec{k}\sigma})+|\nu_{\vec{k}}|^2
f(-E_{\vec{k}-\sigma}).
\end{equation}
\textcolor{czerwony}{Above formula at $T=0$ takes the form: $n_{\vec{k}\sigma}=|u_{\vec{k}}|^2 \Theta(-E_{\vec{k}\sigma})+|\nu_{\vec{k}}|^2
\Theta(E_{\vec{k}-\sigma})$, where $\Theta (x)$ is the Heaviside step function: $\Theta(x)=1$ if $x>0$ and zero otherwise.} 

\begin{figure}
\begin{center}
\includegraphics[width=0.44\textwidth,angle=270]{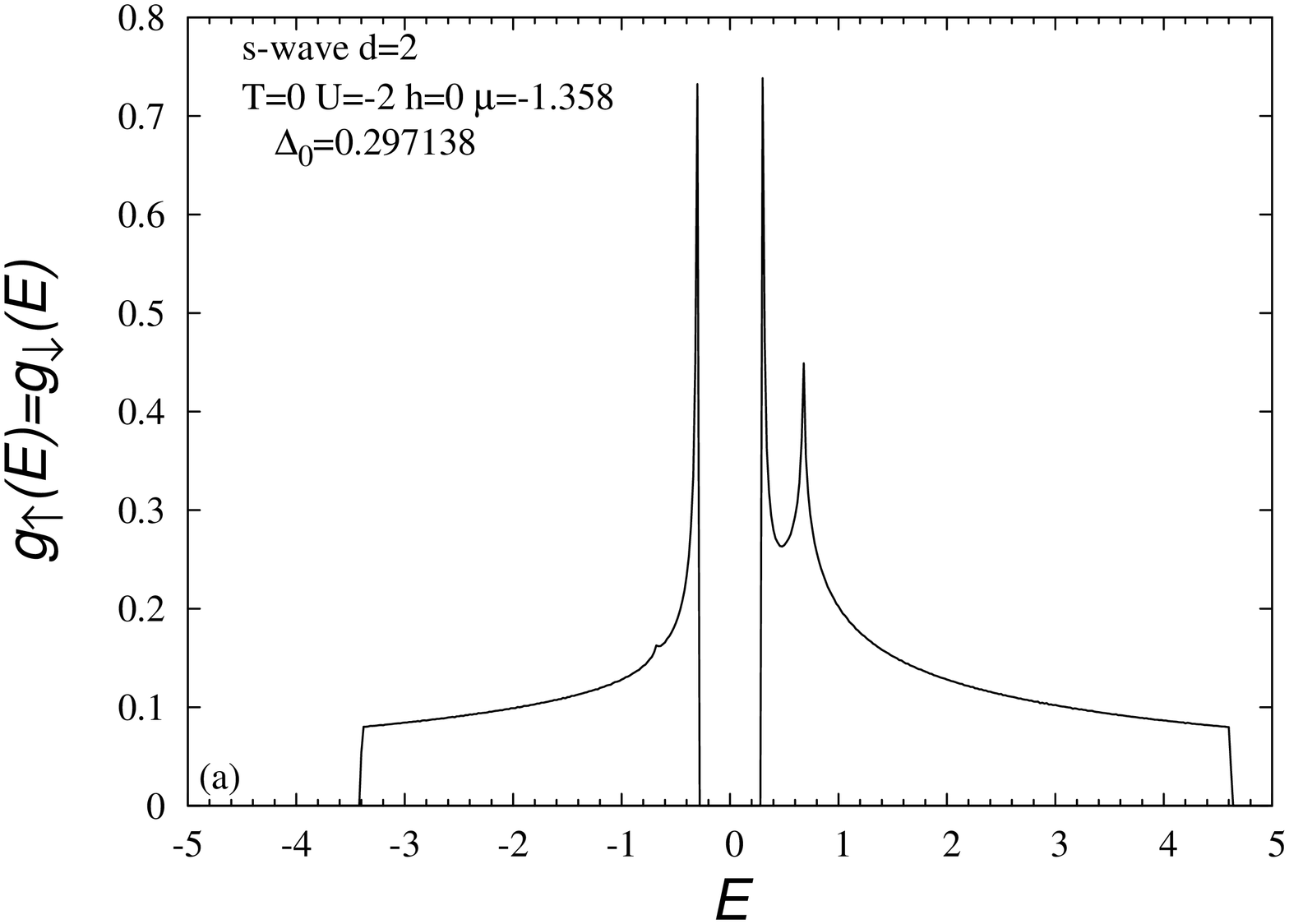}\hspace{-0.2cm}
\includegraphics[width=0.44\textwidth,angle=270]{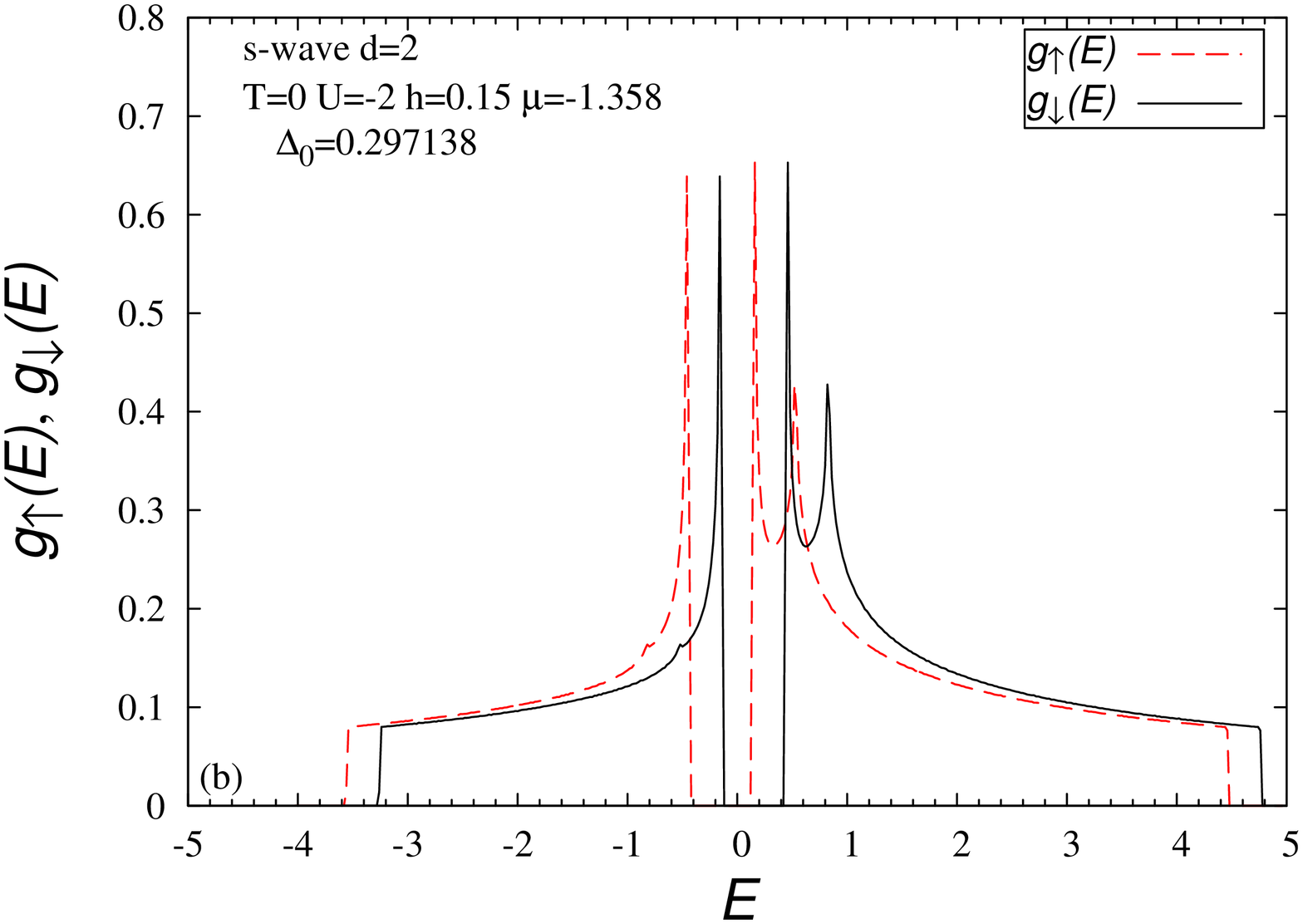}\\
\includegraphics[width=0.44\textwidth,angle=270]{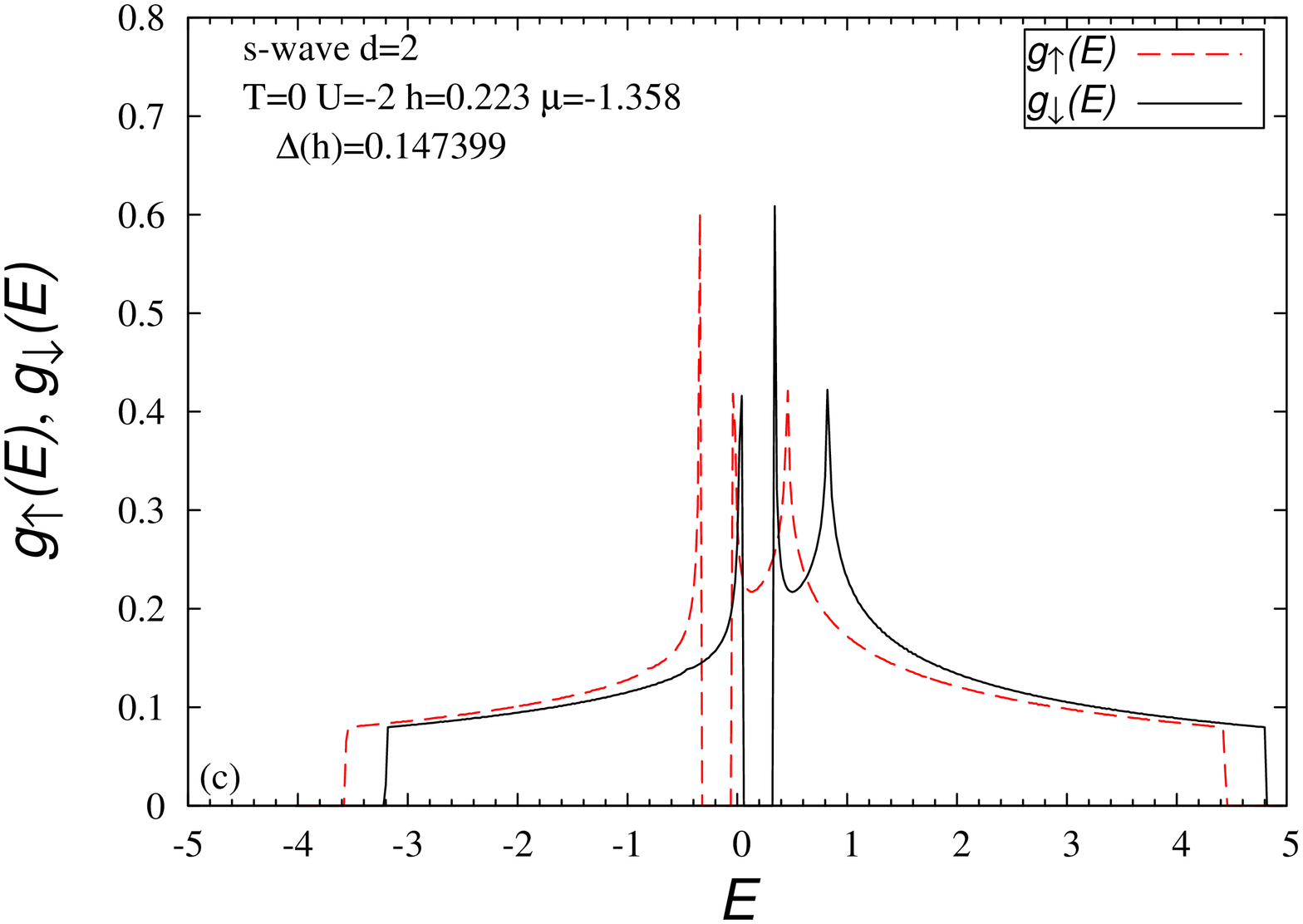}
\caption{\label{fig3} Density of states for two-dimensional s-wave pairing symmetry case, $T=0$, $\mu\approx -1.358$, $U=-2$, (a) $h=0$, (b) $h=0.15$ (the stable branch solution), (c) $h=0.223$ (the unstable branch solution).}
\end{center}
\end{figure}

\begin{figure}
\begin{center}
\includegraphics[width=0.43\textwidth,angle=270]{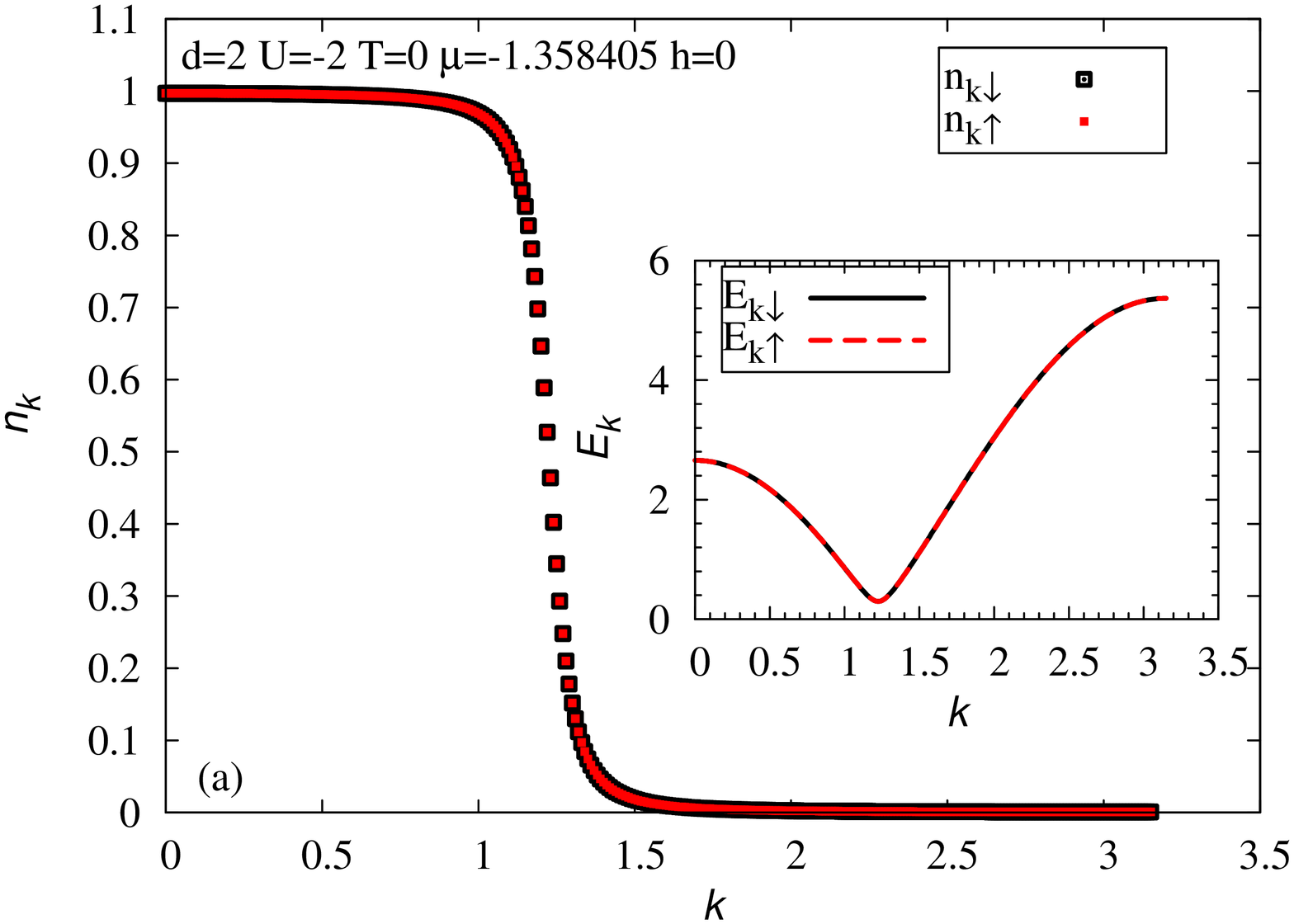}
\includegraphics[width=0.43\textwidth,angle=270]{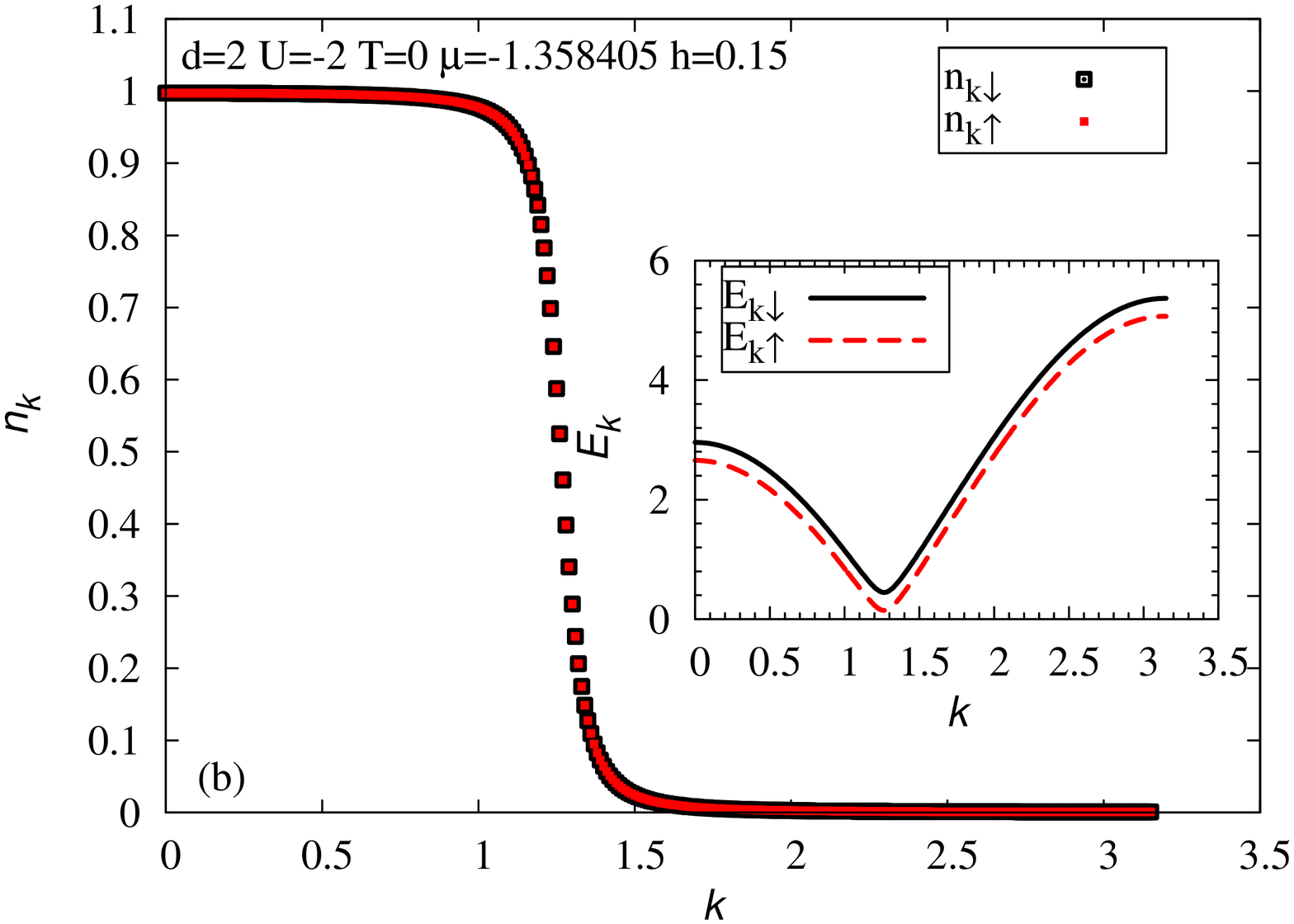}\\
\includegraphics[width=0.43\textwidth,angle=270]{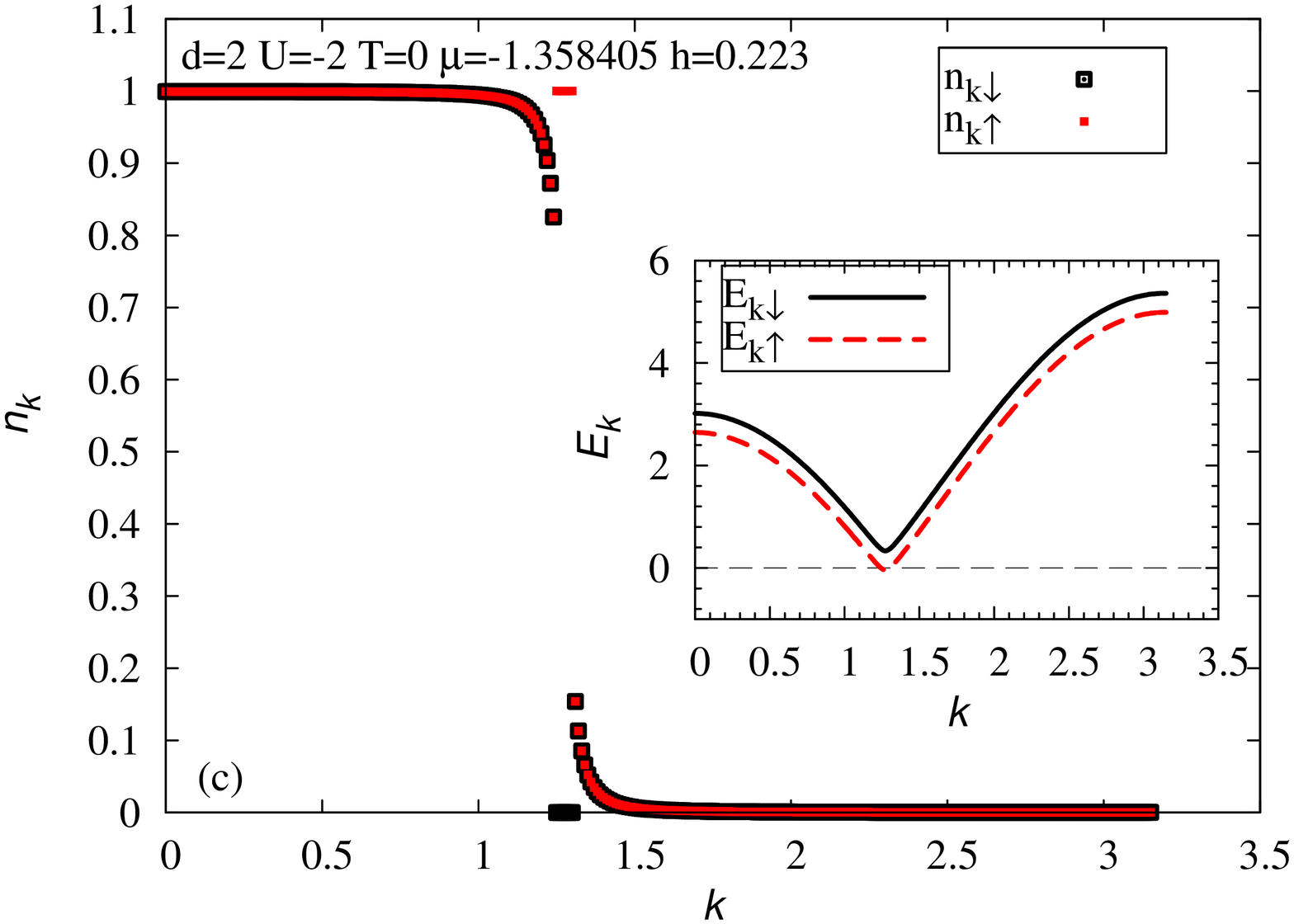}
\caption{\label{n_k} Plots of momentum occupation numbers $n_{\vec{k}\uparrow}$ (red points), $n_{\vec{k}\downarrow}$ (black points) vs. $k\equiv |k_x|=|k_y|$ and corresponding quasiparticle spectra $E_{\vec{k}\uparrow}$, $E_{\vec{k}\downarrow}$ (insets) for $U=-2$, $\mu\approx -1.358$, (a) $h=0$, (b) $h=0.15$, (c) $h=0.223$.}
\end{center}
\end{figure}

Fig. \ref{fig3} shows the densities of states
$g_{\uparrow}(E)$ (red dashed lines) and $g_{\downarrow}(E)$ (black solid
lines), while Fig. \ref{n_k} show momentum occupation numbers
$n_{\vec{k}\uparrow}$ (red points), $n_{\vec{k}\downarrow}$ (black points) vs.
$k\equiv |k_x|=|k_y|$ and the corresponding quasiparticle spectra (insets). The
excitation spectrum of the quasiparticles is given by
Eq.~\eqref{quasiparticles}. For both figures, we give results at \textcolor{czerwony}{a} fixed $\mu
\approx -1.358$,
$h=0$ and two values of the magnetic field. 

If $h=0$, the densities of states are equal for the particles with spin down and
spin up ($g_{\uparrow}(E)=g_{\downarrow}(E)$). The energy gap ($E_g$) occurs in
DOS, for the isotropic s-wave pairing symmetry (Fig. \ref{fig3}(a)). $E_g$ is
bounded by the location of the square root singularities. The width of
the energy gap in the BCS limit is given by: $E_g=2\Delta$.
The energies of the quasiparticles are equal
($E_{\vec{k}\uparrow}=E_{\vec{k}\downarrow} \geq\Delta$),
$n_{\vec{k}\uparrow}=n_{\vec{k}\downarrow}$ and we observe the characteristic
smearing around the Fermi surface (Fig. \ref{n_k}(a)).  

In the presence of the magnetic field, $g_{\uparrow}(E)\neq g_{\downarrow}(E)$.
Let us assume $h>0$. If we take into account the unpolarized solution ($P=0$),
then $g_\uparrow(E)$ moves to the left by $h$ and $g_\downarrow(E)$ to
the right by $h$ (Fig. \ref{fig3}(b)). Since $h<\Delta$, the regions with
$g_\uparrow(E)=0$ and $g_\downarrow(E)=0$ still have an overlap, hence the total
density of states $g(E)$ is gapped.
Accordingly (Fig. \ref{n_k}(b)), $E_{\vec{k}\uparrow}$ moves down by $h$,
yielding a minimum excitation energy for quasiparticles with spin-up of
$\Delta-h>0$ and $E_{\vec{k}\downarrow}$ moves up by $h$, giving a minimum
excitation energy for spin-down of $\Delta+h>0$.
There is still one Fermi surface in the system, since the momentum occupation
numbers $n_{\vec{k}\uparrow}=n_{\vec{k}\downarrow}$ and thus $P=0$.

If $\Delta$ is $h$-dependent and $h>\Delta$, $n_{\uparrow}\neq
n_{\downarrow}$, the shift to the left of $g_\uparrow(E)$ and the shift to the
right of $g_\downarrow(E)$ by $h$ is such that the gapped regions in both
$g_\sigma(E)$ do not overlap any more and thus the total density of states
$g(E)$ becomes gapless (Fig. \ref{fig3}(c)).
Moreover, the excitation energy $E_{\vec{k}\uparrow}=\Delta-h<0$, i.e. the
spectrum for the majority spin species becomes gapless (Fig. \ref{n_k}(c)).
There are two Fermi surfaces in the system, which is manifested in the
momentum occupation numbers by the appearance of a region
with $n_{\vec{k}\uparrow}=1$ and $n_{\vec{k}\downarrow}=0$.
However, as shown above, the spatially homogeneous polarized
superconducting
phase, characterized by the gapless spectrum for the majority spin species is
unstable at $T=0$, on the BCS side and in the s-wave pairing symmetry case.

\subsection{Phase diagrams}

In this subsection, we consider the ground state phase diagrams for the
square and simple cubic lattices. The weak and intermediate couplings are \textcolor{green}{analyzed at $r=1$}. 

As mentioned above, in the absence of a magnetic field, in the weak
coupling \textcolor{czerwony}{regime}, the usual superconducting state (BCS-\textcolor{czerwony}{type} state) is stable at
$T=0$.

The Zeeman magnetic field destroys superfluidity at weak and intermediate
couplings through the paramagnetic effect (or by population imbalance). In
consequence, there is a first order phase transition from the unpolarized
superconducting to the polarized normal state, both for $d=2$ and $d=3$.

\begin{figure}
\begin{center}
\includegraphics[width=0.55\textwidth,angle=270]{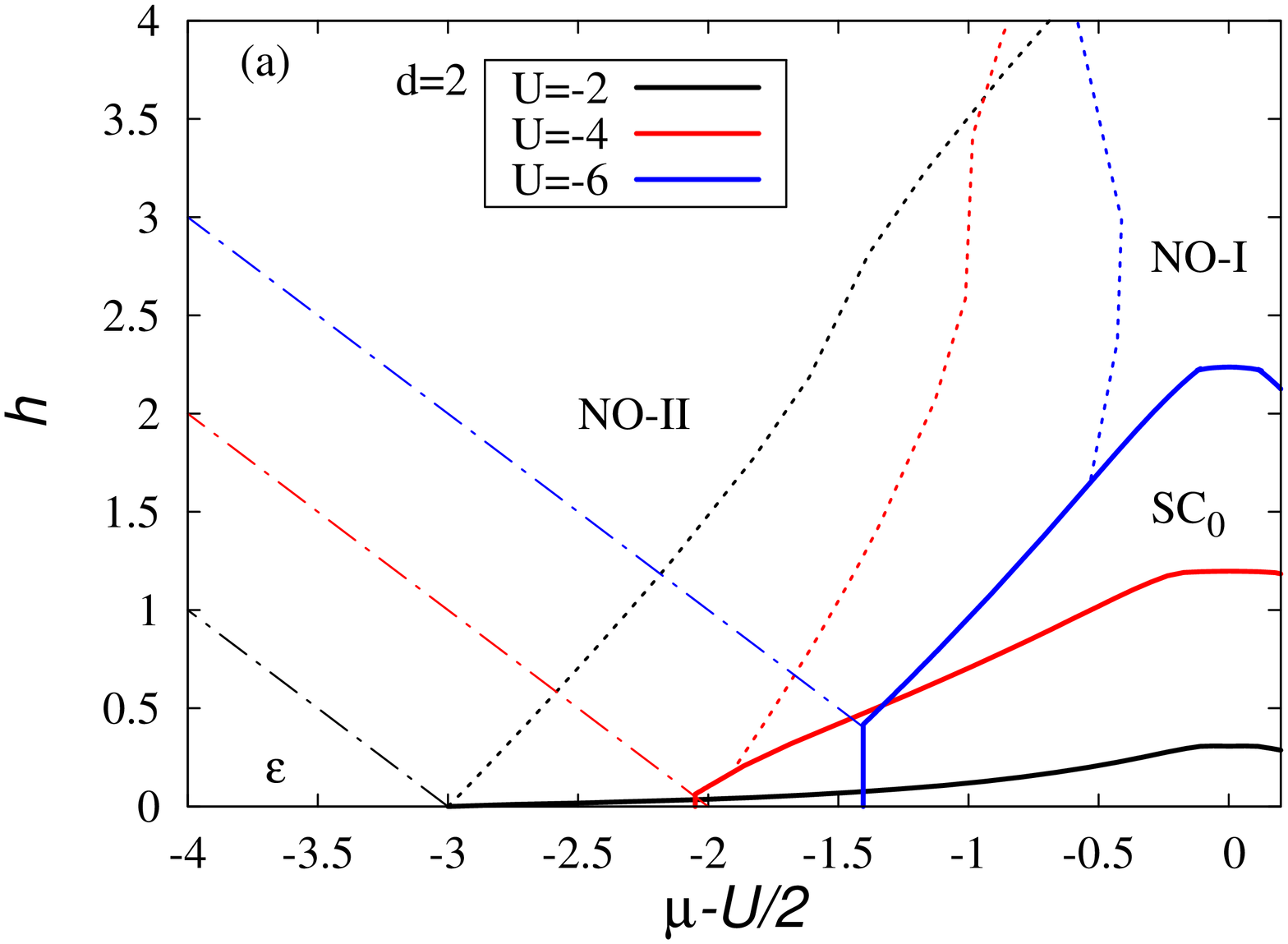}\hspace{-0.2cm}
\includegraphics[width=0.55\textwidth,angle=270]{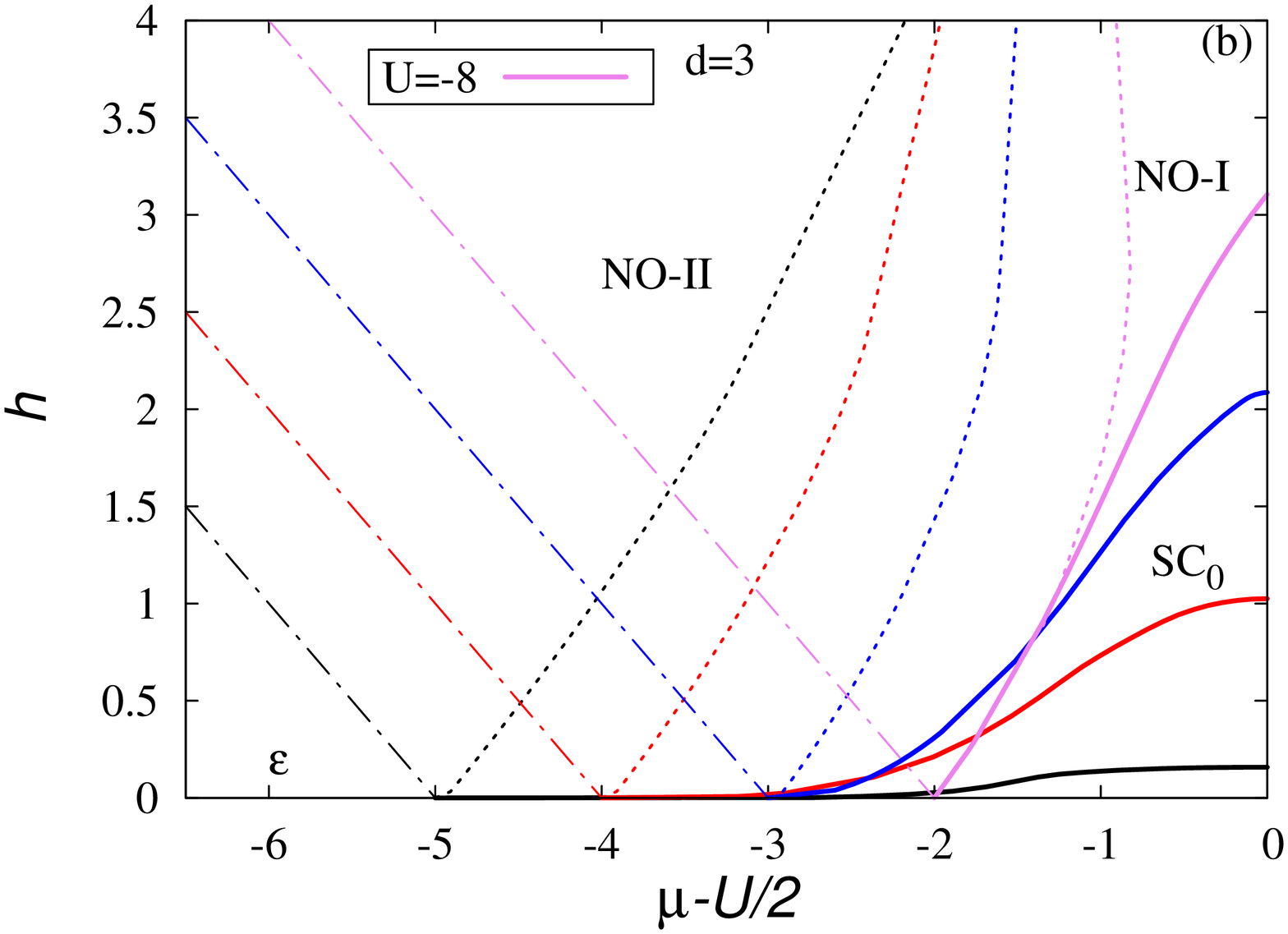}\\
\caption[Critical magnetic field vs. the chemical potential for the first order SC$_0$-NO transition, at $T=0$, three different values of the attractive interaction; (a) square lattice, (b) simple cubic lattice.]{\label{mu_diagram} Critical magnetic field vs. the chemical potential for the first order SC$_0$-NO transition, at $T=0$, three different values of the attractive interaction; (a) square lattice, (b) simple cubic lattice. SC$_0$ -- unpolarized superconducting state with $n_{\uparrow}=n_{\downarrow}$, NO-I -- partially polarized ($P<1$) and NO-II -- fully polarized ($P=1$) normal state, $\varepsilon$ -- empty state. Solid lines denote SC$_0$-NO first order transition, dashed lines separate NO-I and NO-II states, borders between NO-II and empty states are shown with dash-dotted lines. If $|U|$ is higher than $|U_c|^{d=3}$, the behavior of the critical line in $d=3$ becomes similar to that of $d=2$ (\textcolor{green}{$U/t=-8$}).}
\end{center}
\end{figure}

Fig. \ref{mu_diagram} shows the $h-\mu$ phase diagram for the (a) square and (b)
simple cubic lattices, at $T=0$. These diagrams are symmetric with respect to
the sign change of $\mu -U/2$ or $h$. For the sake of clarity, we only show the
range of $\mu-U/2$ from -4 to 0 (for $d=2$), from -6 to 0 (for $d=3$) and for
$h\geq 0$. The solid lines denote the first order phase transition to the NO
phase. The range of stability of the SC$_0$ state depends on the value of the
attractive interaction, i.e. it widens with increasing attraction. For the weak
and intermediate attraction, $\bar{\mu}$ does not drop below the lower band edge
in the three dimensional case, because there exists a threshold value of $|U|$
for which a bound state is formed in the empty lattice. After the transition
from SC$_0$, the system is in the partially polarized normal state (NO-I),
whose polarization is
$P=(n_{\uparrow}-n_{\downarrow})/(n_{\uparrow}+n_{\downarrow})<1$. With further
increase in the magnetic field, the system goes smoothly to the fully
polarized
normal state (NO-II), for which $P=1$ ($n_\downarrow=0$). If the attraction is
increased above the threshold for bound state formation in the empty lattice
($|U_c|^{d=3}/12t=0.659$), a first order transition from SC$_0$ to NO-II takes
place, for some values of the parameters (Fig. \ref{mu_diagram}(b)). It should
be emphasized that the magnetic field modifies the band edge, which is clearly
visible in Fig. \ref{mu_diagram} (the thin dash-dotted lines).
The first order transition lines at $T=0$ were obtained numerically from the
condition $\Omega^{SC}_{T=0} =\Omega^{NO}_{T=0}$ (where $\Omega^{NO}_{T=0}$ and
$\Omega^{SC}_{T=0}$ denote the grand canonical potential of the normal
($\Delta=0$, $P\neq 0$) and the superconducting ($\Delta \neq 0$, $P=0$) state,
respectively).  

\begin{figure}[t!]
\begin{center}
\includegraphics[width=0.55\textwidth,angle=270]{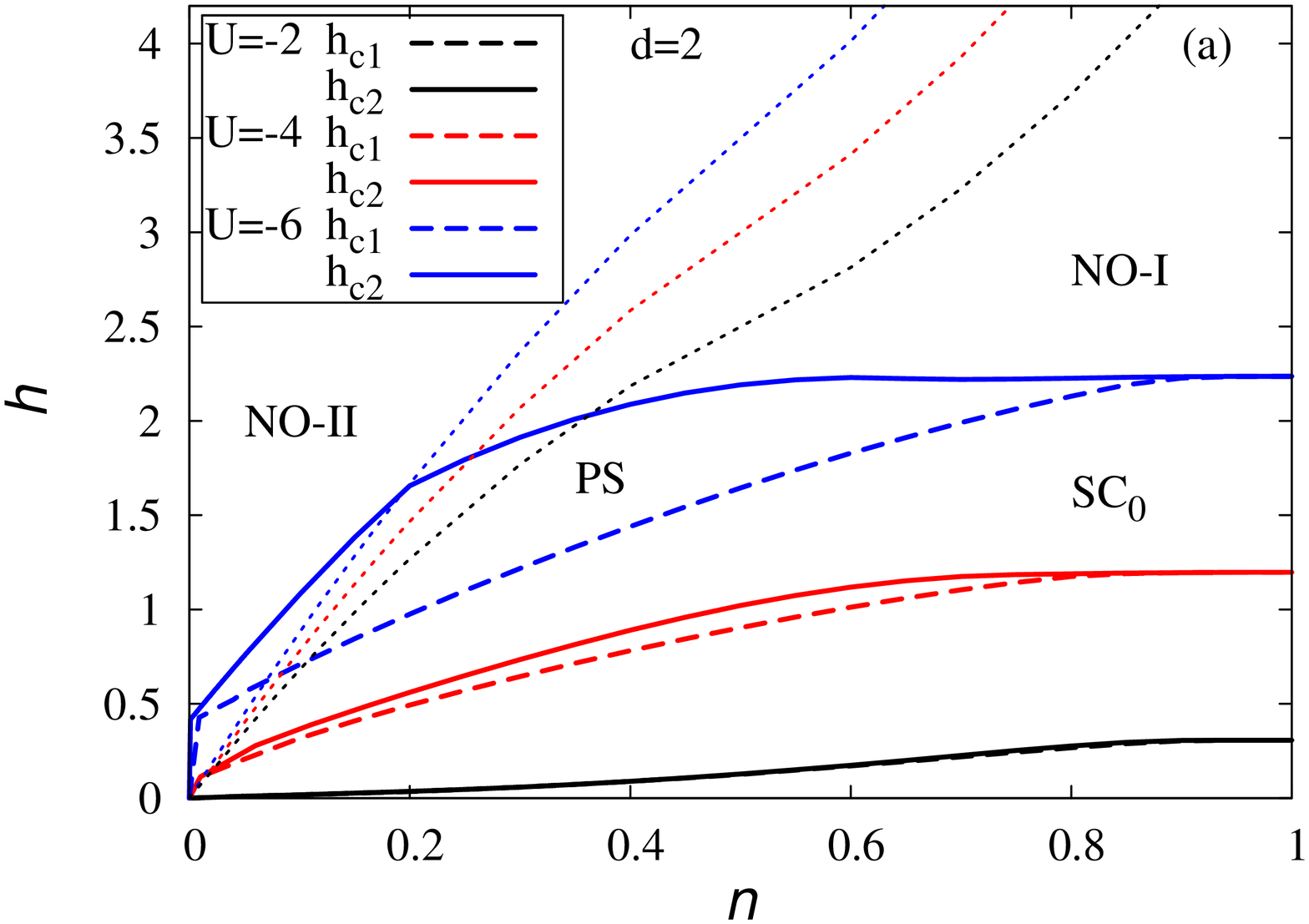}\hspace{-0.2cm}
\includegraphics[width=0.55\textwidth,angle=270]{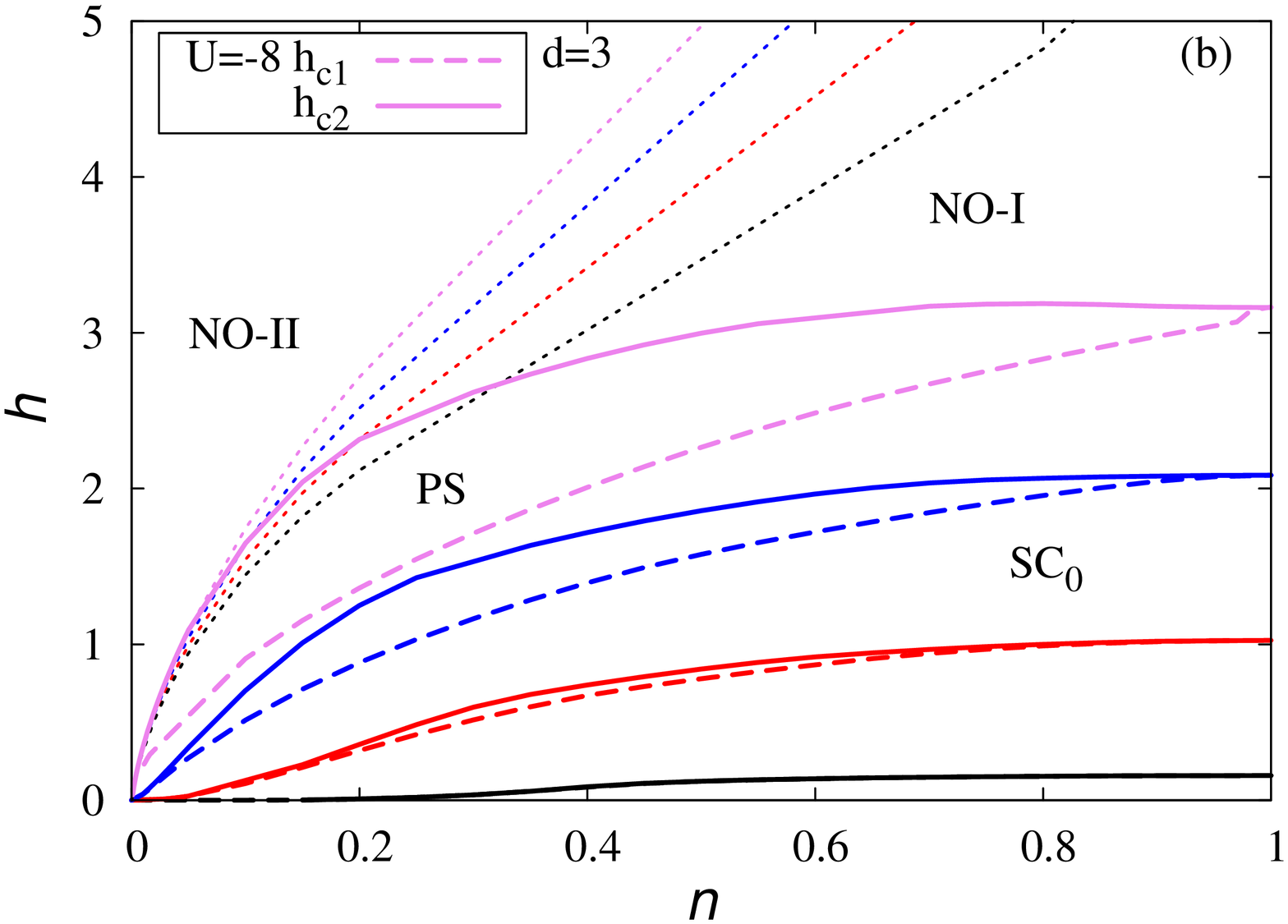}\\
\caption[Critical magnetic field vs. electron concentration for the first order SC$_0$-NO transition, at $T=0$; 
(a) square lattice, (b) simple cubic lattice.]{\label{n_diagram} Critical magnetic field vs. electron concentration 
for the first order SC$_0$-NO transition, at $T=0$; (a) square lattice, (b) simple cubic lattice. PS -- phase separation. 
Thick solid and dashed lines limit the region of PS, thin dashed lines border NO-I and NO-II states. 
\textcolor{czerwony}{Charge density wave (CDW)} state, being degenerated with SC for $h=0$, $n=1$, is not shown.}
\end{center}
\end{figure}

The situation is more complex for the two dimensional case (Fig.
\ref{mu_diagram}(a)). Since a critical interaction value for which a bound state
is formed in the empty lattice does not exist, for any small value of the
attraction, $\bar{\mu}$ can drop below the lower band edge. For $n\rightarrow
0$, $\bar{\mu}=-4t-\frac{1}{2} E_b$ ($E_b$ -- the binding energy in two-body
problem). It is in agreement with a rigorous result of Randeria et al.
\cite{randeria2, randeria3} that in the low density limit and $d=2$, the
presence of the bound state in the two-body problem is a necessary and
sufficient condition for $s$-wave SC$_0$ to take place. Moreover, as opposed to
the $d=3$ case, for some values of the parameters, we find a first order
transition directly from SC$_0$ to NO-II, even for weak and intermediate
attractive interaction (Fig. \ref{mu_diagram}(a)).

There are relevant differences between the phase diagrams obtained for fixed
chemical potential and those for fixed electron concentration. Fig.
\ref{n_diagram} shows the dependence of the critical magnetic fields on the
electron concentration, for three (four) different values of attraction, for
$d=2$ (a) and $d=3$ (b). Here, due to the particle-hole symmetry, we only show
the range of $n$ from 0 to 1. In contrast to the fixed chemical potential case,
if the number of particles is fixed and $n\neq 1$, one obtains two critical
Zeeman fields ($h_{c1}$, $h_{c2}$) in the phase diagrams. The two critical field
lines determine the phase separation region between SC$_0$ with the number of
particles $n_s$ and NO with the number of particles $n_n$. For the case of the
square lattice and for higher values of the attractive interaction or
smaller $n$, the transition from the PS region can be directly to NO-II. For the
case of
\textcolor{czerwony}{a} simple cubic lattice, the situation is different, as long as $|U|$ is
smaller than $|U_c|$ -- the first order transition from PS to NO-I takes
place in the whole range of the parameters and afterwards the system goes
smoothly from NO-I to NO-II. 

\section{Finite temperatures}

This section presents the results concerning the influence of the magnetic field
on superfluidity at finite temperatures, based on the HF-BCS
theory. For the $d=2$ system at $h=0$, the SC-NO transition is of the
\textcolor{czerwony}{KT type}. According to Eq. (\ref{KT}), the KT
transition temperature is found from the intersection point of the straight line
$\frac{2}{\pi} k_B T$ with the curve $\rho_s(T)$ \footnote{It should be added that we use expression for $\rho_s (T)$ \textcolor{czerwony}{derived in chapter \ref{3.2}. Thus, $T_c$ is an upper bound on actual transition temperature.}}. Thus, we take into account
phase fluctuations in our treatment, in 2D. We analyze the
weak coupling regime for the $d=2$ and $d=3$ cases.

The results shown below for chosen values of the parameters $U$, $n$ and $\mu$,
illustrate the typical behavior, but our analysis is general within the Hartree
approximation. We start by investigating the stability of the gap solutions, at
$T\neq 0$, for fixed $\mu \approx -1.358$ in $d=2$.
\begin{figure}
\hspace*{-0.8cm}
\includegraphics[width=0.35\textwidth,angle=270]{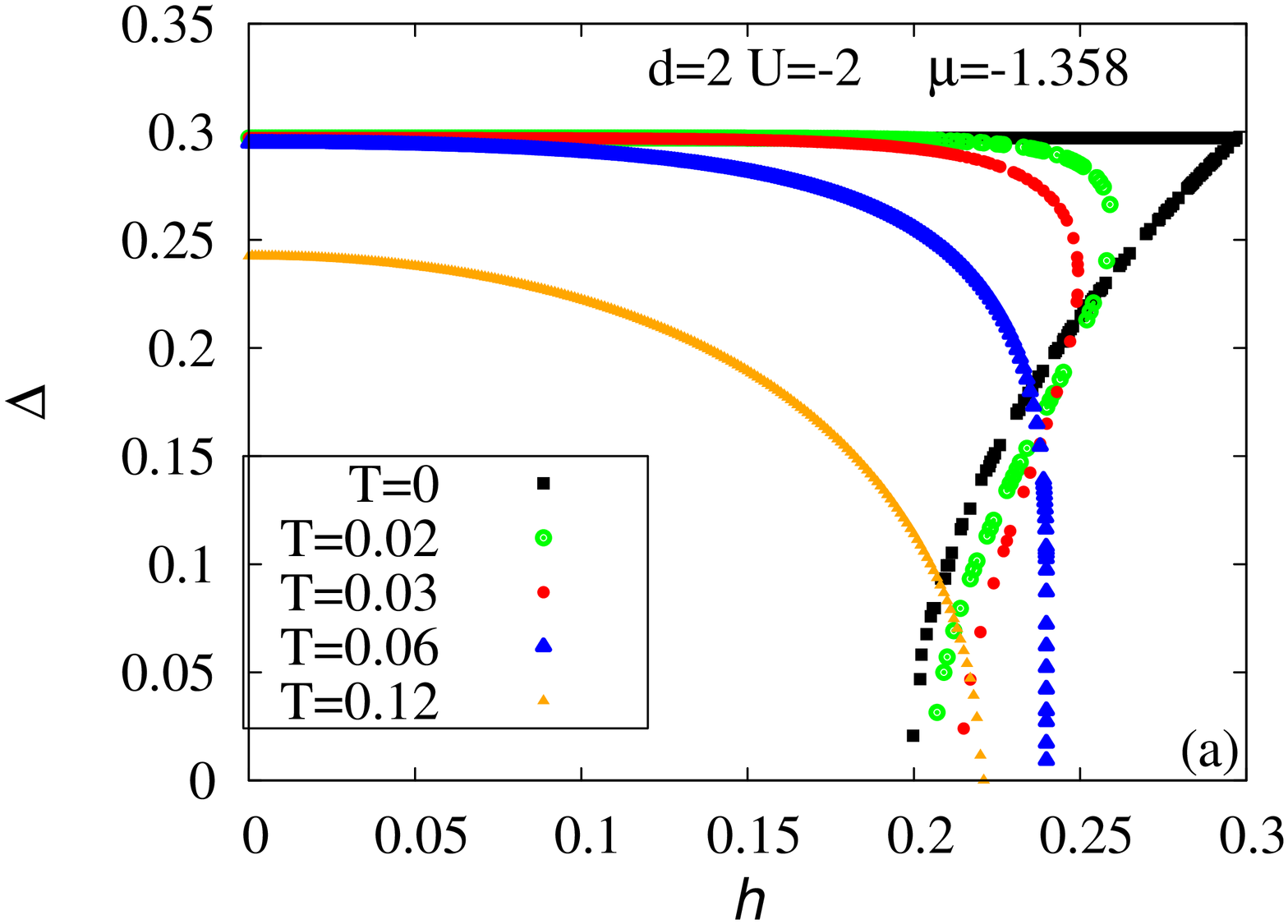}
\hspace*{-0.6cm}
\includegraphics[width=0.35\textwidth,angle=270]{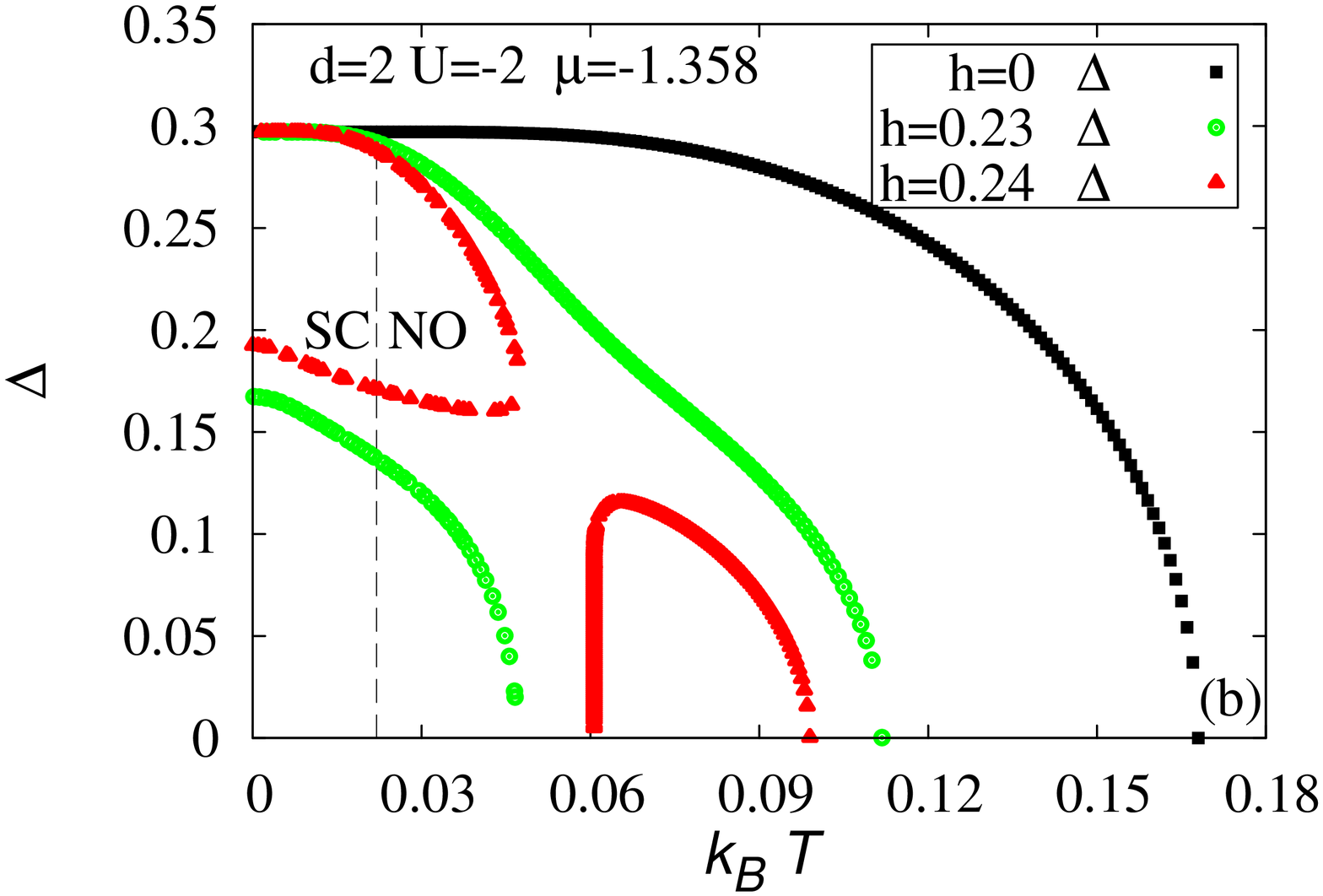}\\
\hspace*{-0.8cm}
\includegraphics[width=0.35\textwidth,angle=270]{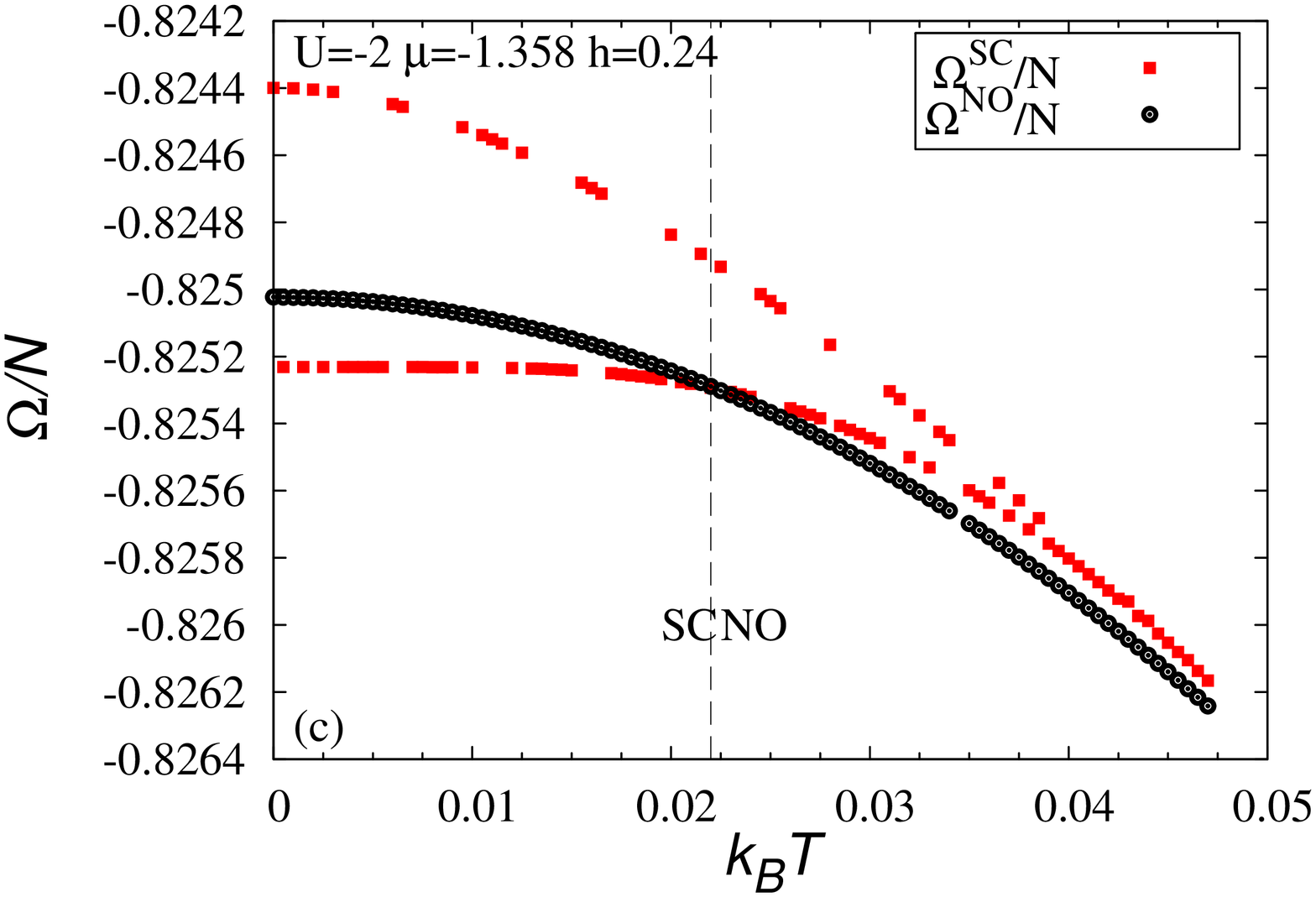}
\hspace*{-0.6cm}
\includegraphics[width=0.35\textwidth,angle=270]{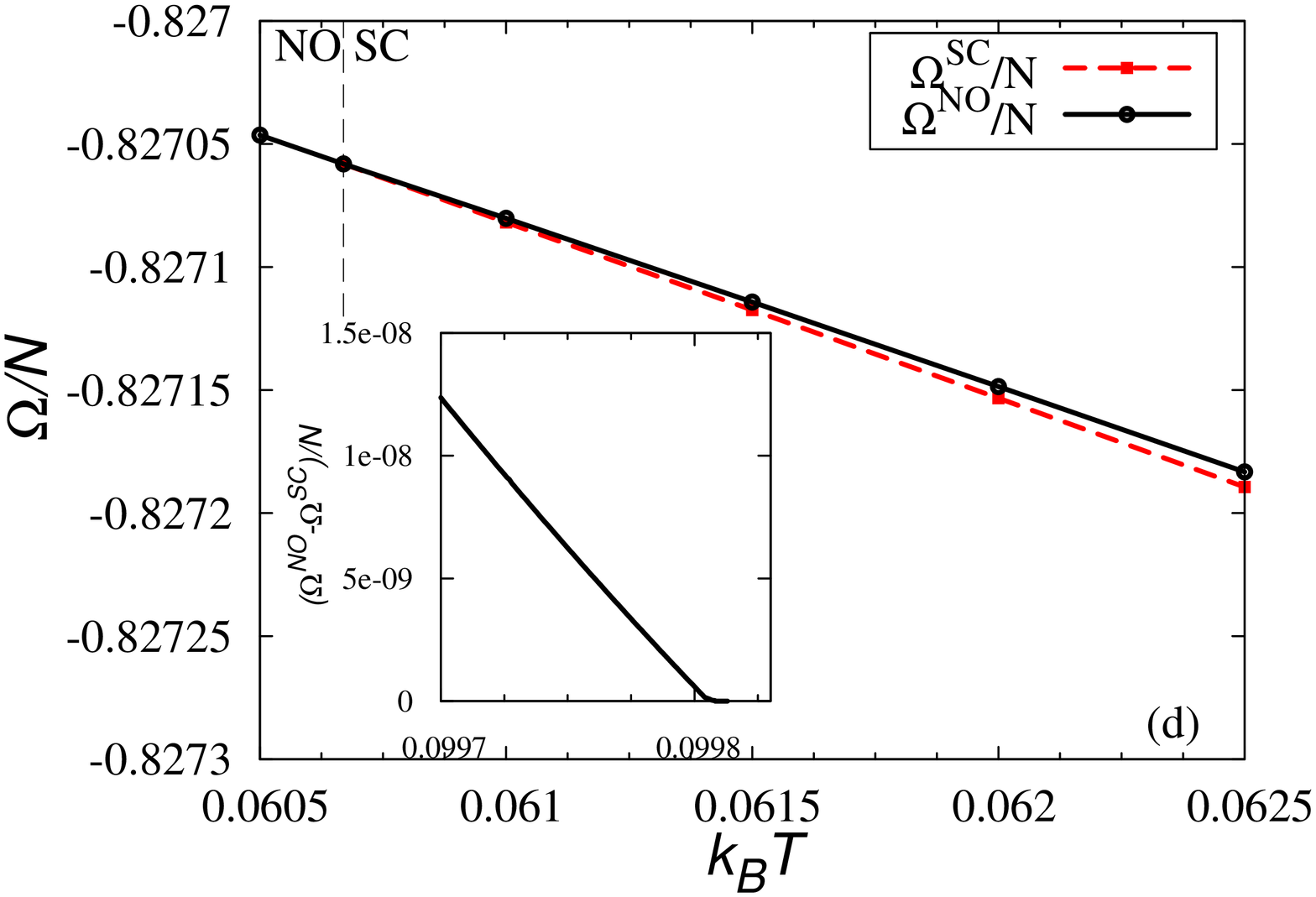}\\
\hspace*{-0.8cm}
\includegraphics[width=0.35\textwidth,angle=270]{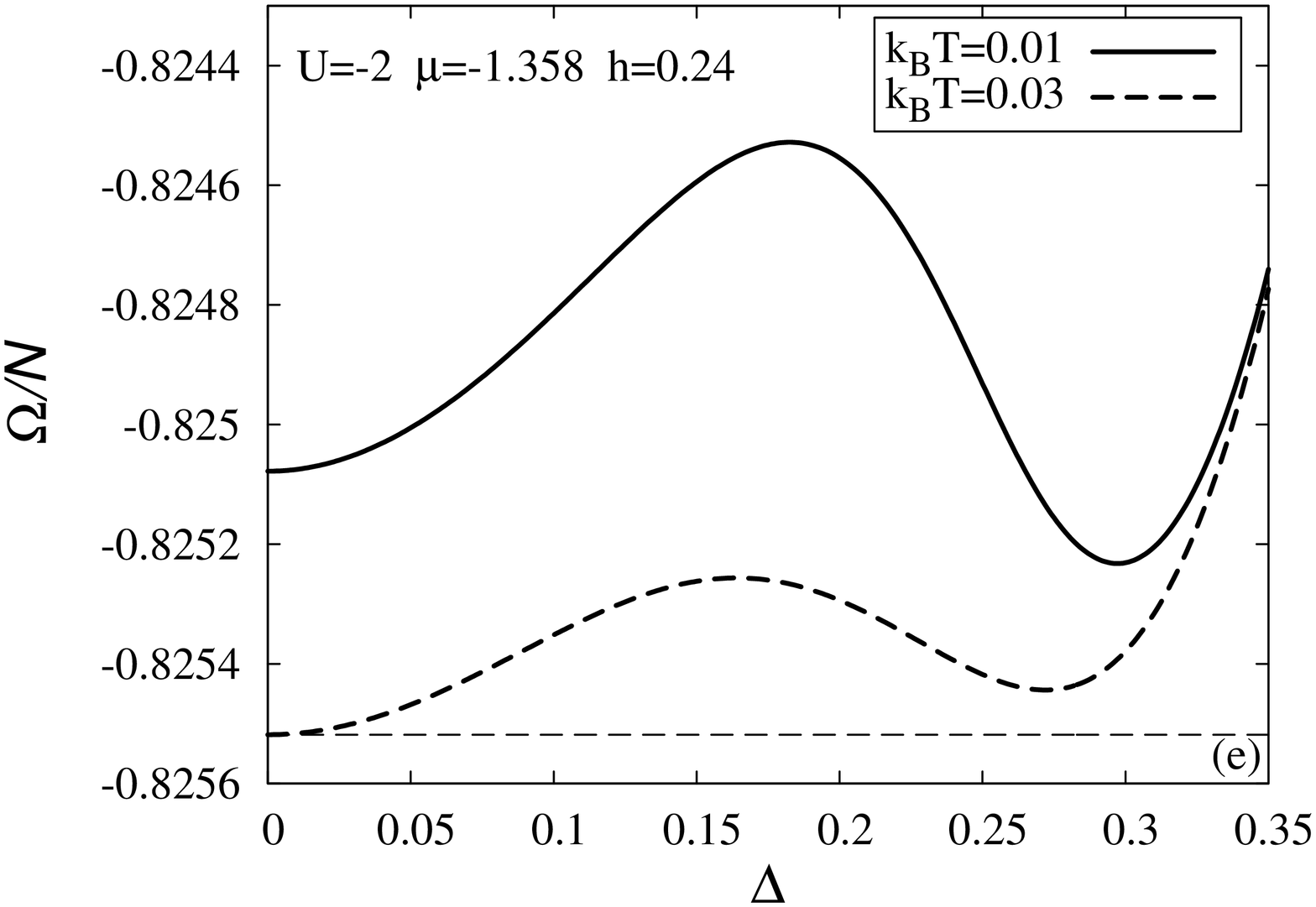}
\hspace*{-0.6cm}
\includegraphics[width=0.35\textwidth,angle=270]{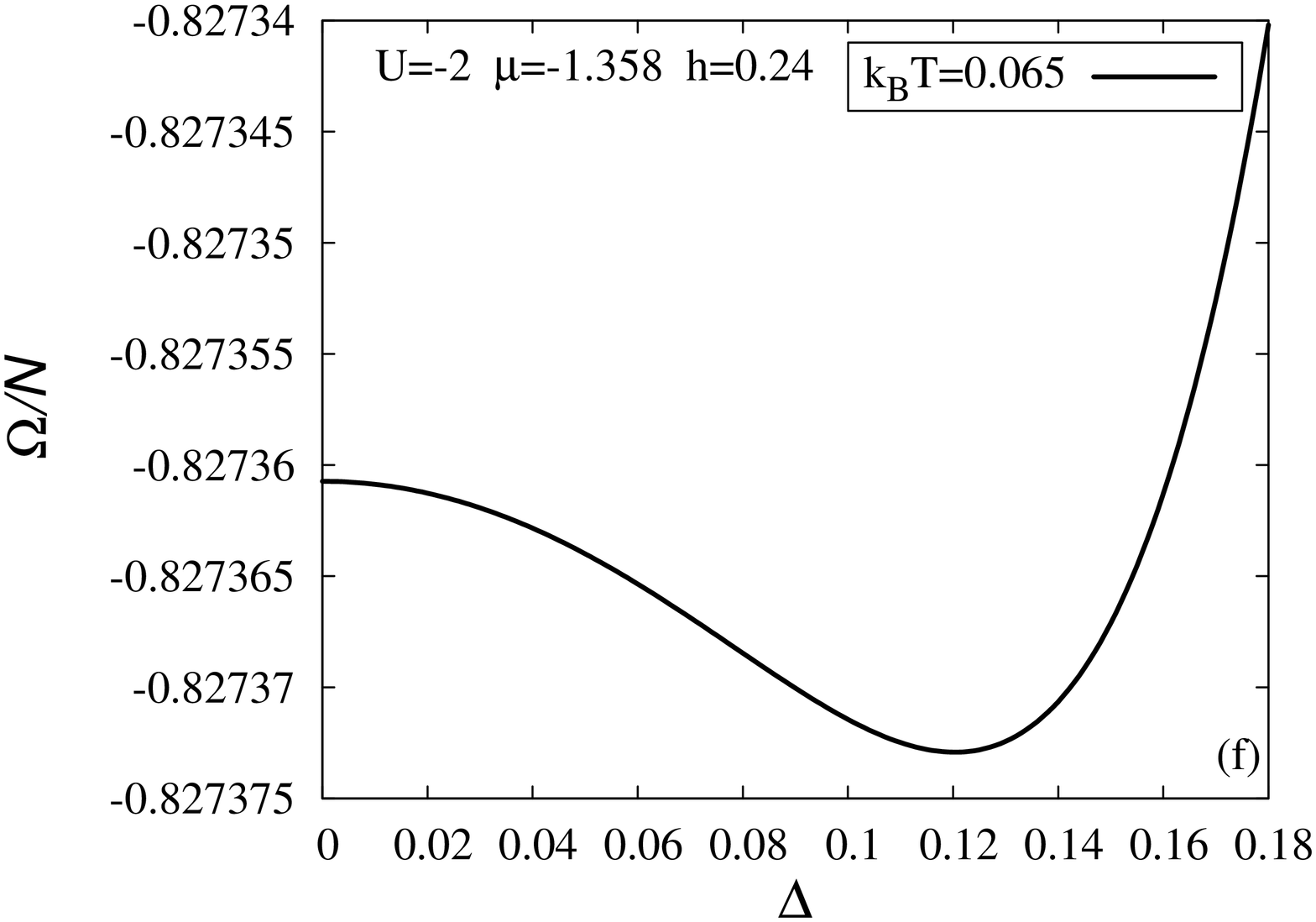}\\
\caption[Dependence of the order parameter on the magnetic field (a) and temperature (b), $d=2$, $U=-2$, for a fixed $\mu \approx -1.358$. The grand canonical potential vs. temperature for $h=0.24$: (c) the first order phase transition to the normal state, (d) the details of the reentrant transition: second order transition from NO to SC and from SC to NO (inset).]{\label{delT} Dependence of the order parameter on the magnetic field (a) and temperature (b), $d=2$, $U=-2$, for a fixed $\mu \approx -1.358$. In Fig (b), for $h=0.23$ the lower branch is unstable. For $h=0.24$ the vertical dashed line denotes the first order phase transition to the normal state and there are two second order transitions in the reentrant case (at $T=0.06067$ and $T=0.09891$). The grand canonical potential vs. temperature for $h=0.24$: (c) the first order phase transition to the normal state, (d) the details of the reentrant transition: second order transition from NO to SC and from SC to NO (inset). The vertical dashed lines mark 
the H-F phase transition temperatures. The dependence of the grand canonical potential on the order parameter for $h=0.24$, $T=0.01$ (solid line) and $T=0.03$ (dashed line) (e), and for $T=0.065$ (f).}
\end{figure}

Fig. \ref{delT} shows the dependence of the order parameter on magnetic field, for a few fixed values of temperature (a) and the order parameter vs. temperature, for a few fixed values of magnetic field (b), for a fixed $\mu$, in $d=2$. As mentioned above, at $T=0$ only the solutions with $\Delta \neq 0$ and $P=0$ are energetically favorable against NO, in the presence of the magnetic field. Hence, the BCS state is stable. The situation is more interesting for $T>0$. For low magnetic fields, $\Delta$ vanishes continuously with increasing temperature. However, there arise two non-zero solutions for the order parameter, for $h=0.23$ (Fig. \ref{delT}(b)). The lower branch is unstable. 
We have also found very interesting behavior of the order parameter for fixed
$h=0.24$. In this case, $\Delta$ vanishes discontinuously at $T=0.0225$ and
there is a first order phase transition. The solutions with $\Delta \neq 0$
become energetically unfavorable, i.e. the upper branch is metastable and the
lower branch is unstable for $T>T_c$. Moreover, the lower branch is also
unstable for $T<T_c$, which is clearly visible in Fig. \ref{delT}(c) and
\ref{delT}(e). If we investigate the behavior of the grand canonical potential
vs. the order parameter, for fixed $h$ and $T$, we find that at $h=0.24$ and
$T=0.01$ (on the SC side) there is a maximum for the solution from the lower
branch (i.e. it is unstable) and a global minimum for the solution from the
upper branch (i.e. it is stable). For the same value of the magnetic field but
for a higher value of temperature, $T=0.03$ (which is on NO side), we still find
a maximum for the solution from the lower branch, but for the solution from the
upper branch there is a local minimum, indicating a metastable solution. 
Moreover, for $T\geq 0.06067$ the superconducting solution becomes favorable
again (second order transition to SC state) and $\Delta$ vanishes continuously
(second order phase transition to NO at $T=0.0998$), which is shown in Fig.
\ref{delT}(d) and \ref{delT}(f). Such behavior points out that for
sufficiently high fields a reentrant transition takes place. Hence, an
increase in temperature can induce superconductivity. 

\begin{figure}
\hspace*{-0.8cm}
\includegraphics[width=0.38\textwidth,angle=270]{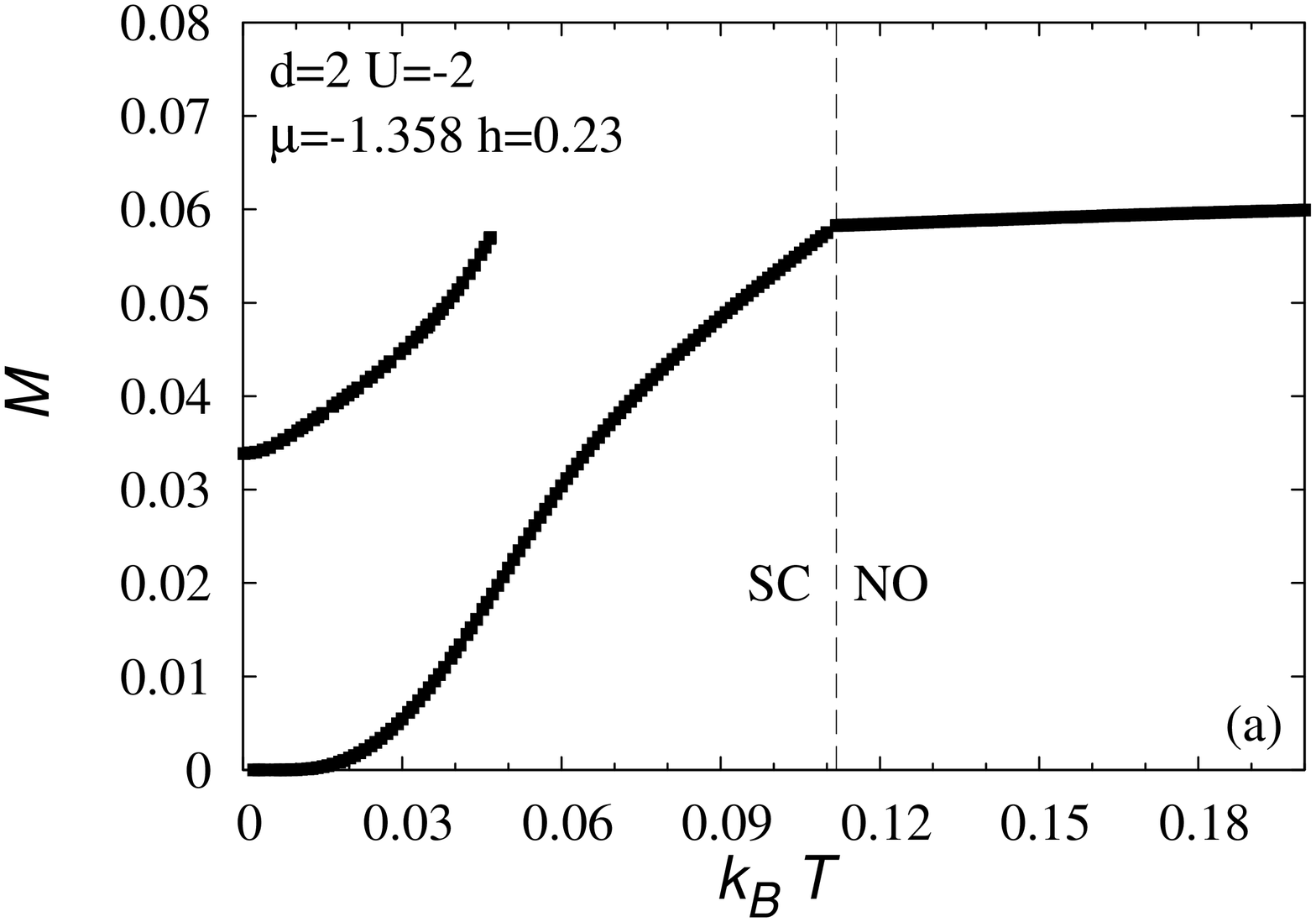}
\hspace*{-0.6cm}
\includegraphics[width=0.38\textwidth,angle=270]{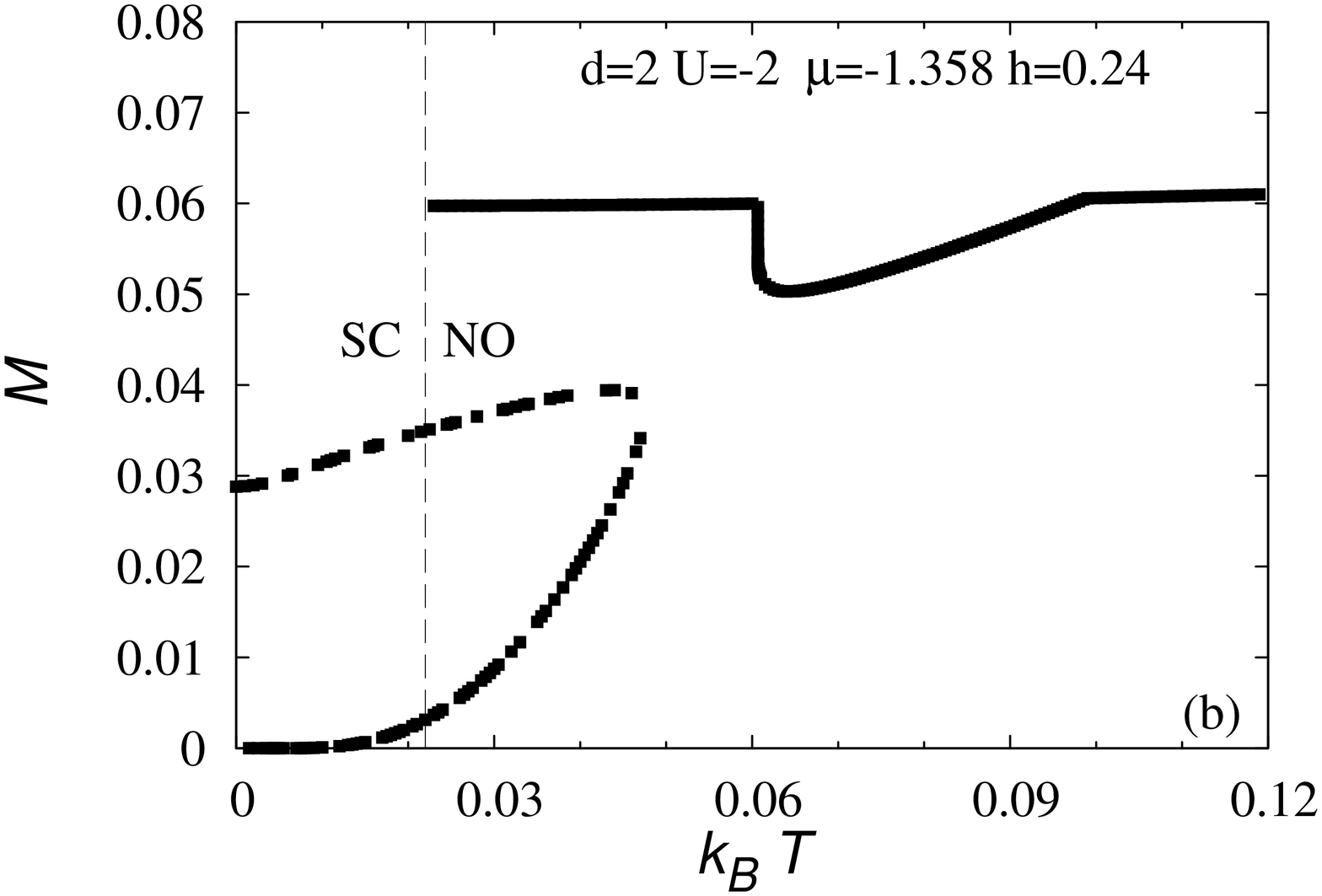}\\
\caption{\label{MagT} Magnetization vs. temperature, $d=2$, $U=-2$, for a fixed $\mu \approx -1.358$. (a) $h=0.23$, (b) $h=0.24$.}
\end{figure}

\begin{figure}[t!]
\hspace*{-0.8cm}
\includegraphics[width=0.38\textwidth,angle=270]{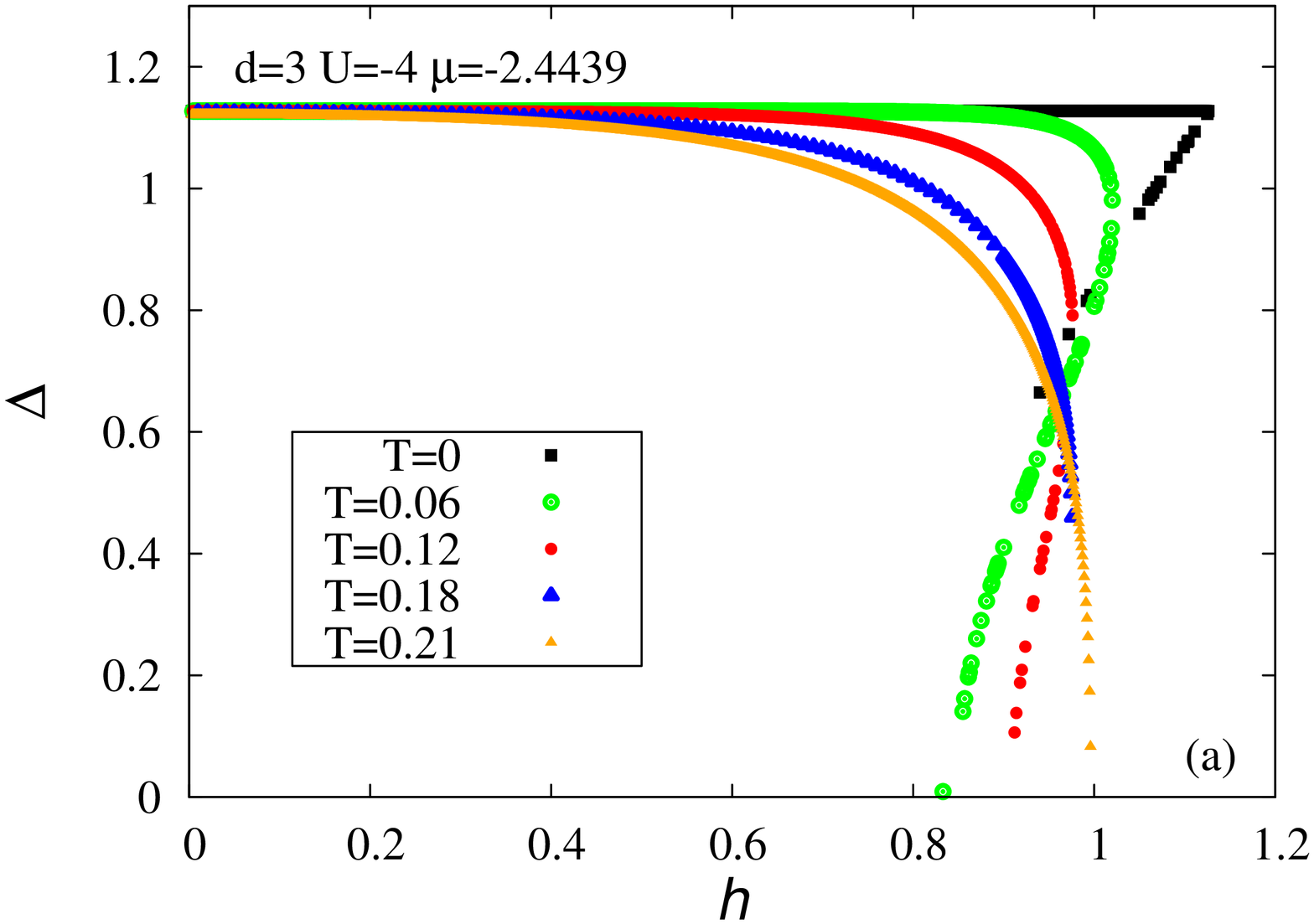}
\hspace*{-0.6cm}
\includegraphics[width=0.38\textwidth,angle=270]{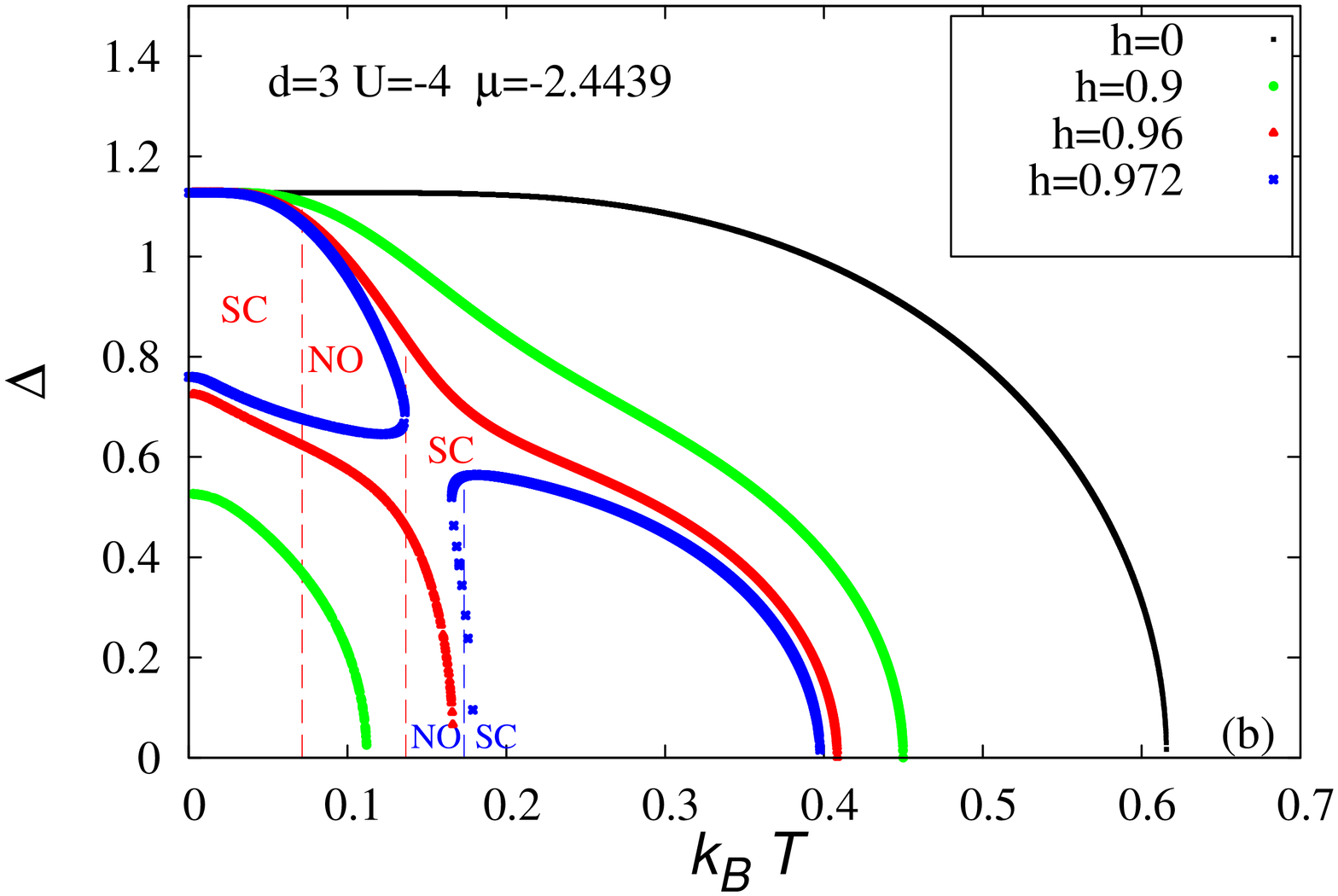}
\caption[Dependence of the order parameter on the magnetic field (a) and temperature (b), $d=3$, $U=-4$, for a fixed $\mu \approx -2.4439$.]{\label{DelT3d} Dependence of the order parameter on the magnetic field (a) and temperature (b), $d=3$, $U=-4$, for a fixed $\mu \approx -2.4439$. In Fig (b), for $h=0.96$ the lower branch is unstable. For $h=0.972$ the solutions from $T=0$ to $T=0.1362$ are unstable. The first order phase transition from the normal to the superconducting state (at $T=  0.1734$) and the second order transition from the superconducting to the normal state (at $T=0.3975$) take place.}
\end{figure}

This interesting behavior at $T\neq 0$, for higher magnetic fields, is
reflected by the temperature dependence of magnetization shown in Figs.
\ref{MagT}(a) and \ref{MagT}(b). At $h=0.23$, there is a second order
transition to NO (lower branch). The upper branch is unstable. For a higher
magnetic field
($h=0.24$), the first order phase transition is revealed through a jump in
magnetization. In the NO state, magnetization is continuous. After the second
order transition to SC (reentrant transition), magnetization decreases abruptly.
With increasing temperature, $M$ increases and there is a second order
transition to NO at $T=0.0998$.

We also perform an analysis, similar to the above, for \textcolor{czerwony}{a} simple cubic lattice, at fixed $\mu \approx -2.4439$ and $U=-4$. 

At $T=0$, in the weak coupling limit, only the solutions with $\Delta\neq 0$ and
$P=0$ are favorable against NO, as in the 2D case. At $T>0$, the dependencies of
the order parameter on the magnetic field or temperature are also similar to
those in the square lattice case. However, the sequences of the SC-NO or NO-SC
transitions are more complex.

\begin{figure}
\hspace*{-0.8cm}
\includegraphics[width=0.38\textwidth,angle=270]{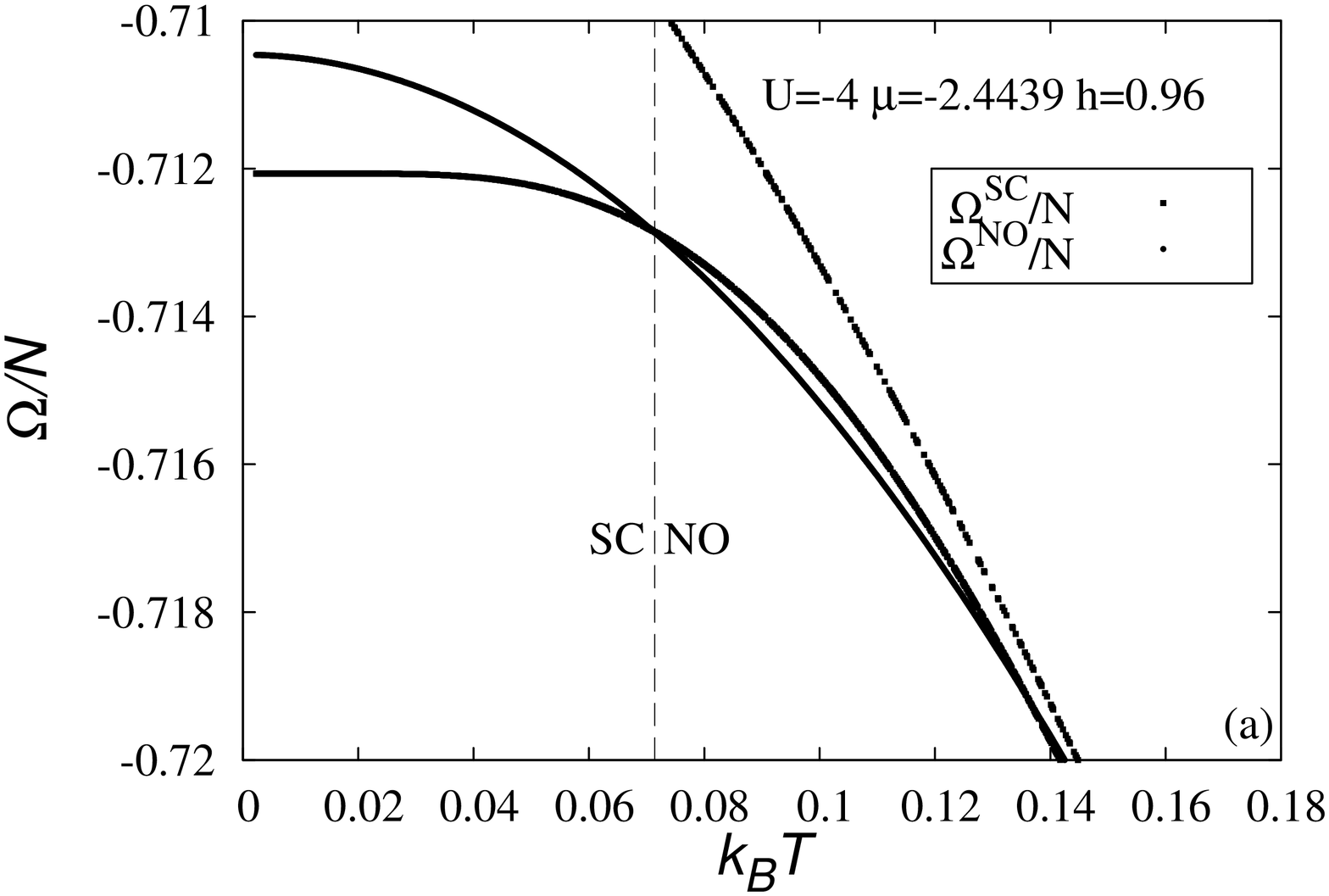}
\hspace*{-0.6cm}
\includegraphics[width=0.38\textwidth,angle=270]{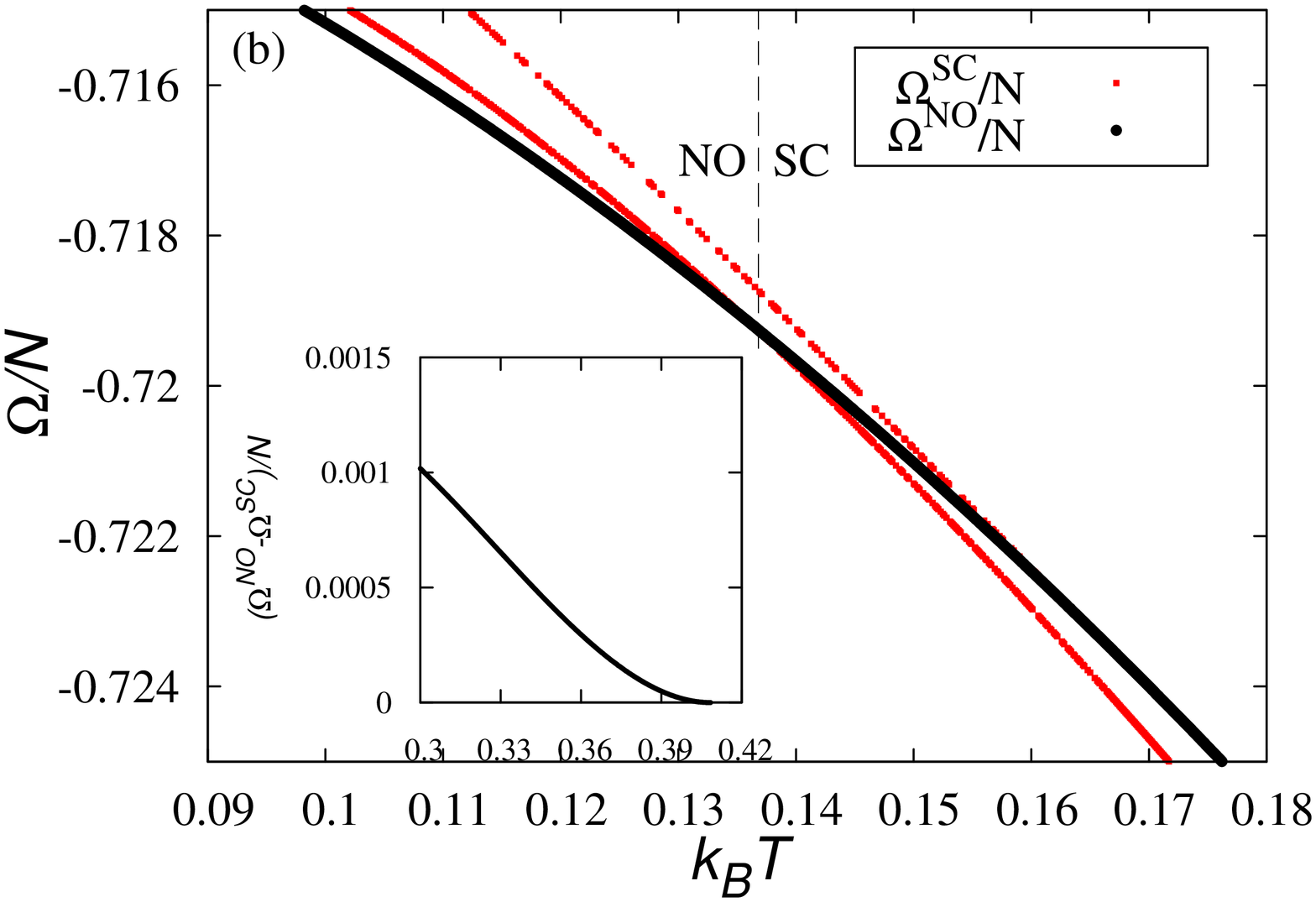}\\
\hspace*{-0.8cm}
\includegraphics[width=0.38\textwidth,angle=270]{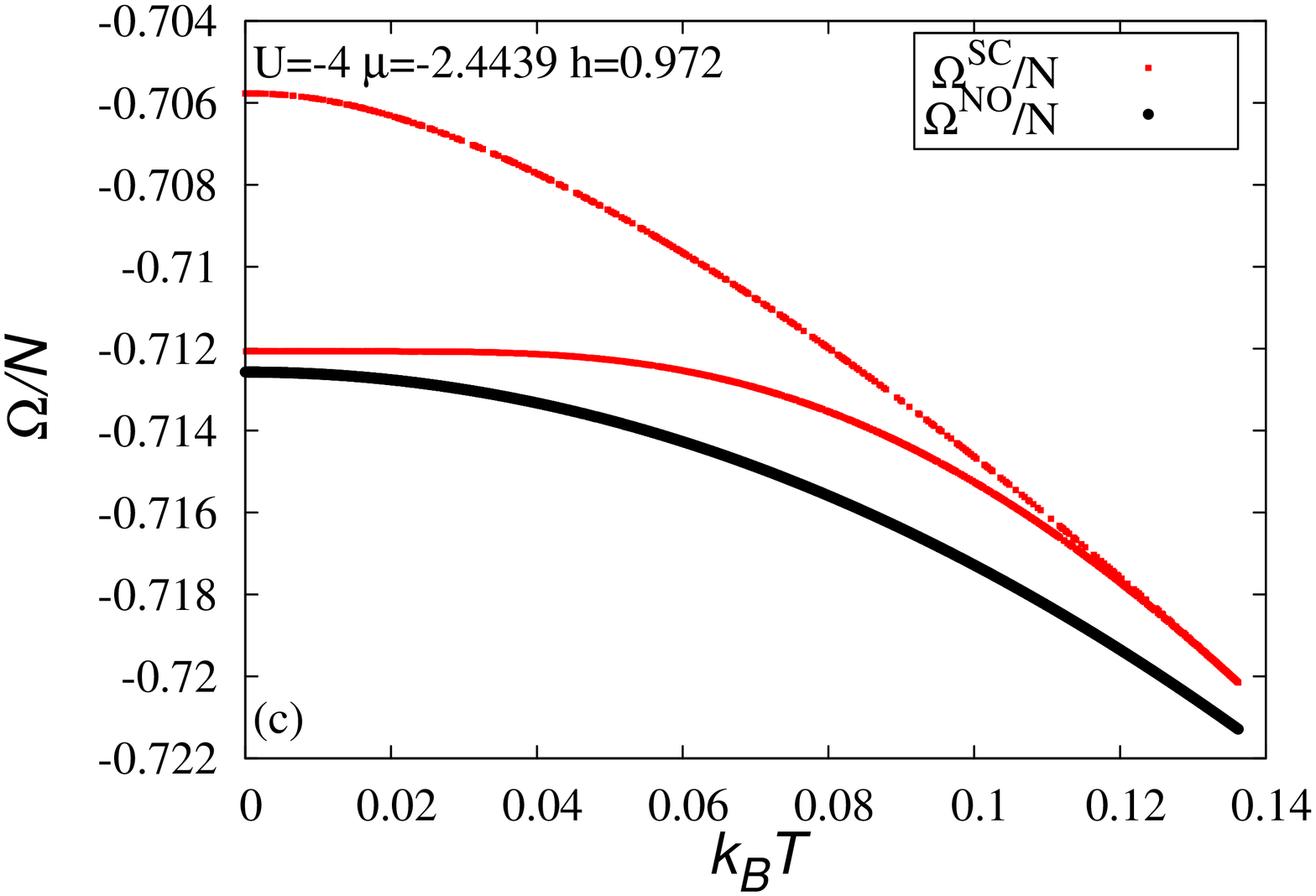}
\hspace*{-0.6cm}
\includegraphics[width=0.38\textwidth,angle=270]{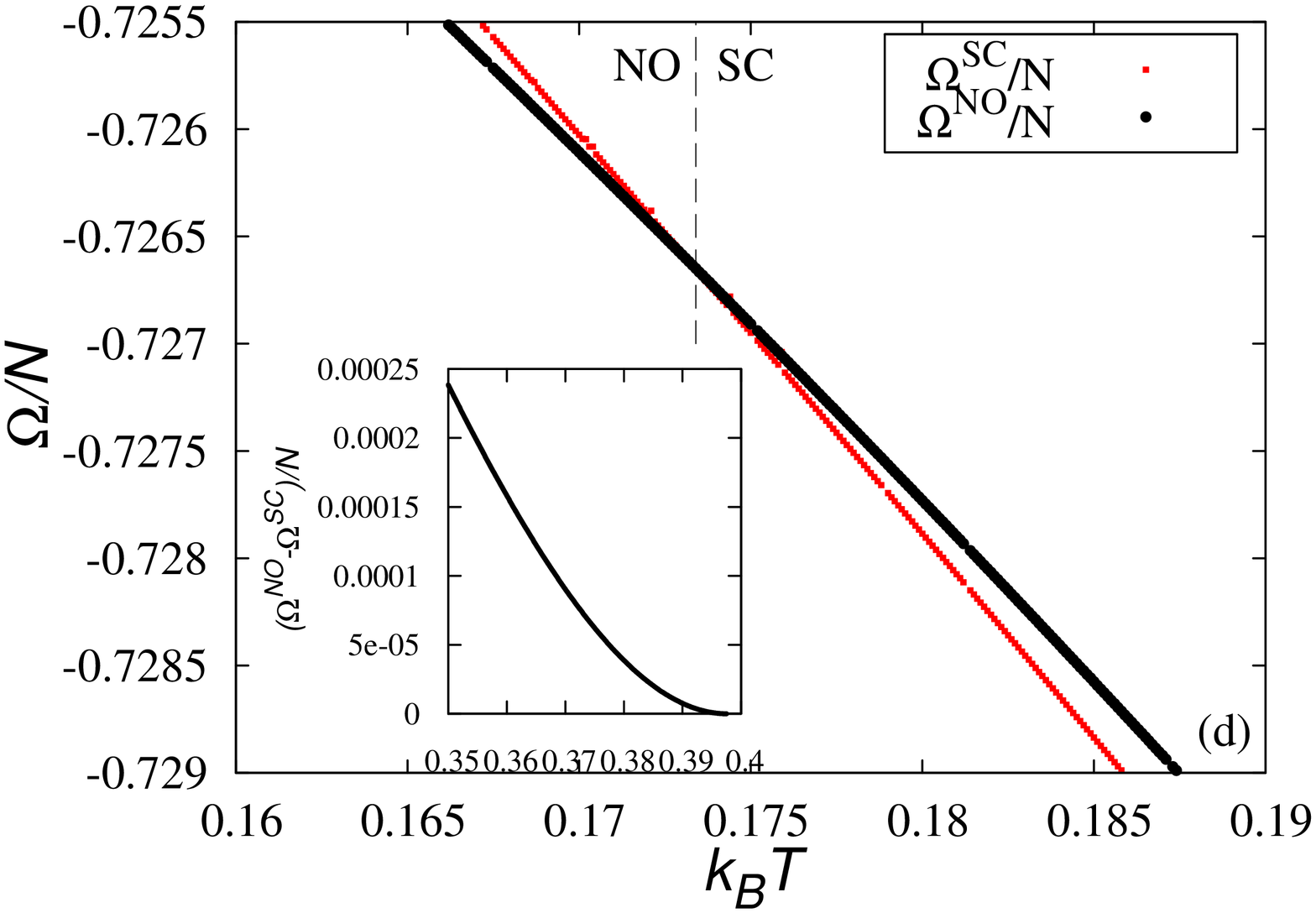}\\
\caption[Grand canonical potential vs. temperature, $d=3$, $U=-4$, for a
fixed $\mu \approx -2.4439$, at $h=0.96$ and $h=0.972$.]{\label{Omega-T} Grand canonical potential vs. temperature, $d=3$, $U=-4$, for a fixed $\mu
\approx -2.4439$, at $h=0.96$: \textcolor{green}{(a)} the first order phase transition to the
normal state, \textcolor{green}{(b)} the details of the reentrant transition: first order
transition from NO to SC and second order transition from SC to NO (inset). The
grand canonical potential vs. temperature for $h=0.972$; \textcolor{green}{(c)} normal state is
stable, \textcolor{green}{(d)} first order transition from NO to SC and second order transition
from SC to NO (inset). The vertical dashed lines mark the phase transition
temperatures.}
\end{figure}

\begin{figure}
\hspace*{-0.8cm}
\includegraphics[width=0.38\textwidth,angle=270]{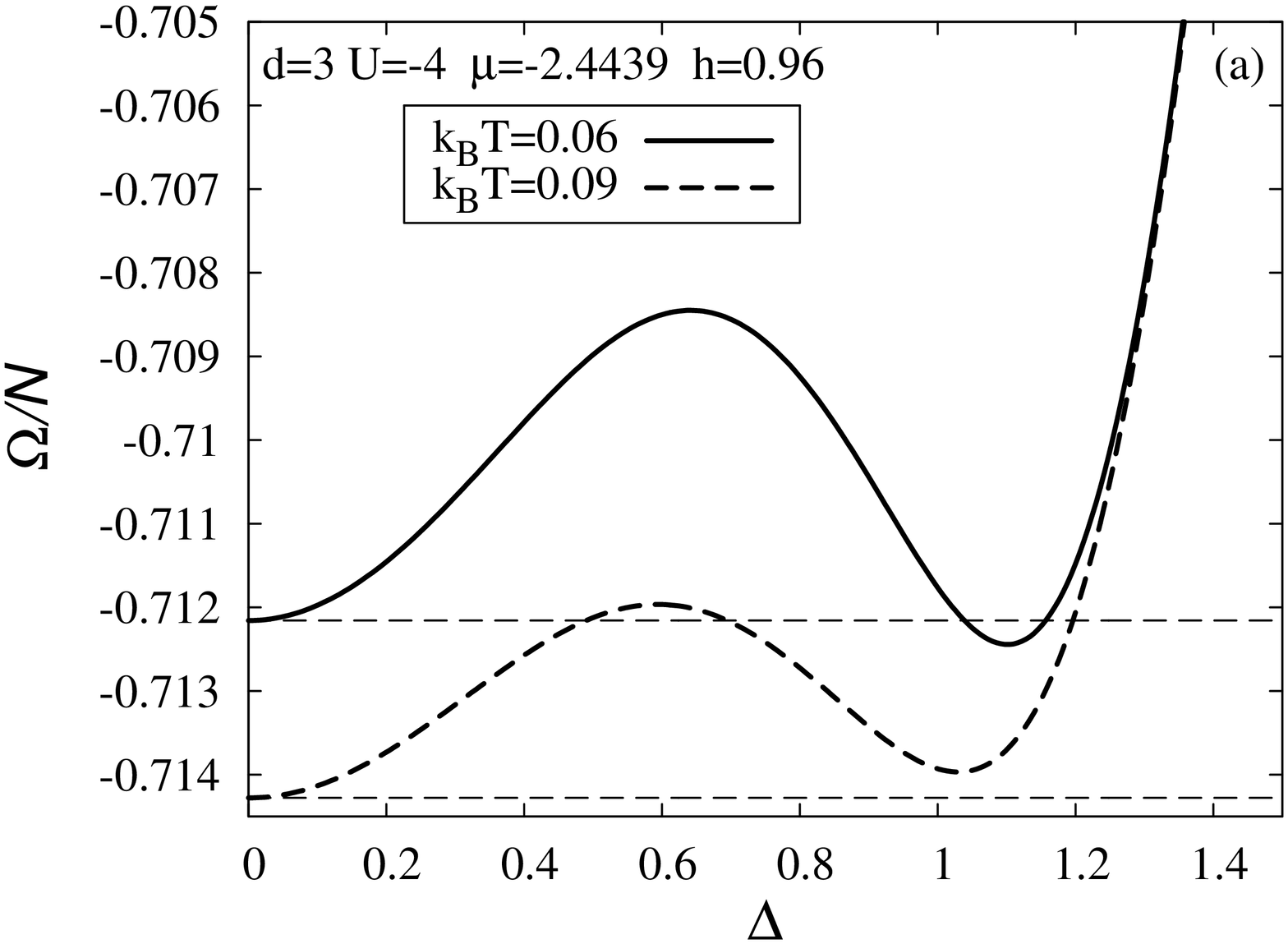}
\hspace*{-0.6cm}
\includegraphics[width=0.38\textwidth,angle=270]{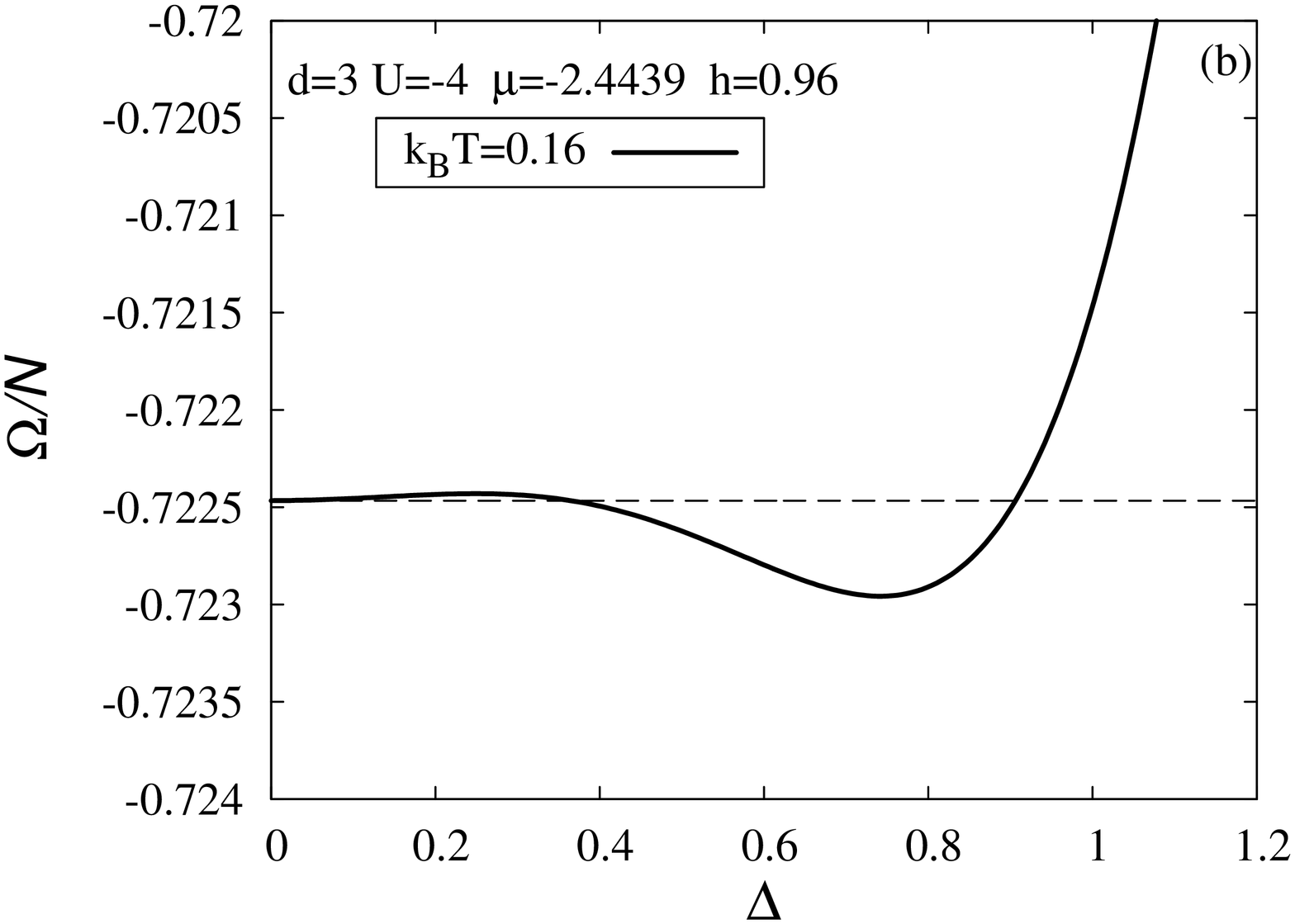}\\
\hspace*{-0.8cm}
\includegraphics[width=0.38\textwidth,angle=270]{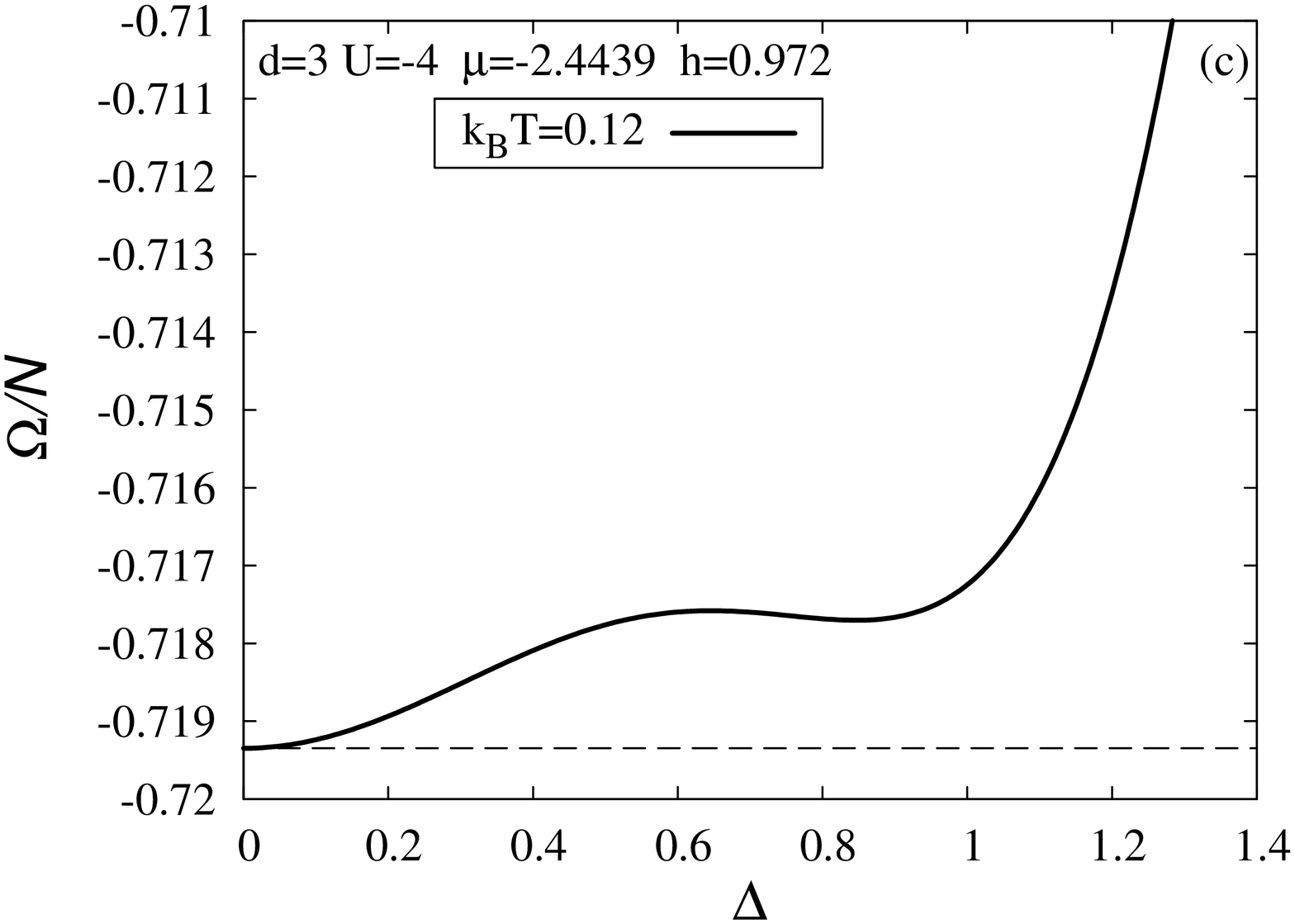}
\hspace*{-0.6cm}
\includegraphics[width=0.38\textwidth,angle=270]{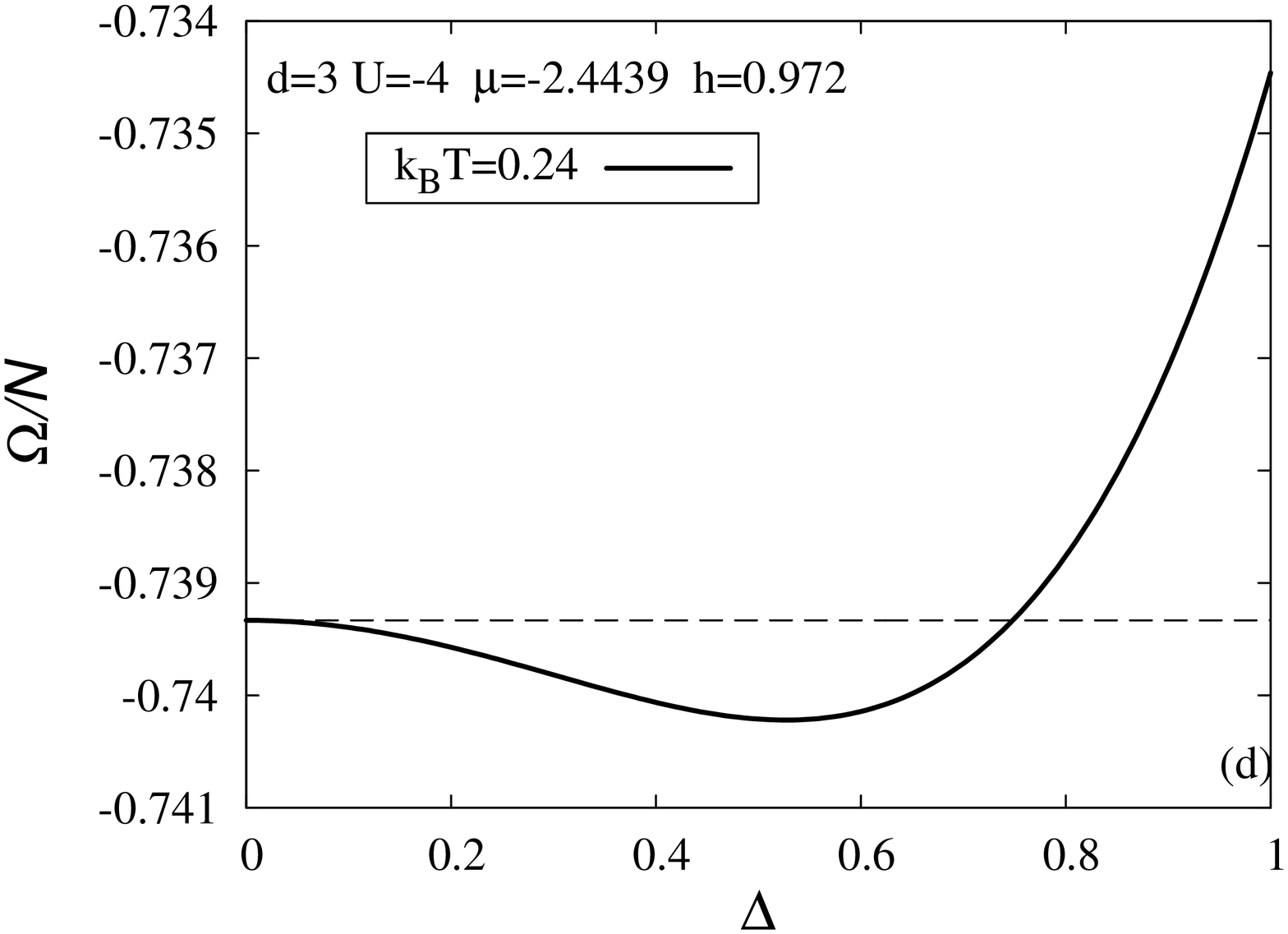}\\
\caption{\label{Omega-delta} Dependence of the grand canonical potential on the order parameter, $d=3$, $U=-4$, for a fixed $\mu \approx -2.4439$, at $h=0.96$, $T=0.06$ (solid line) $T=0.09$ (dashed line) (a), $T=0.16$ (b) and for $h=0.972$, $T=0.12$ (c), $T=0.24$ (d).}
\end{figure}

\begin{figure}
\hspace*{-0.8cm}
\includegraphics[width=0.38\textwidth,angle=270]{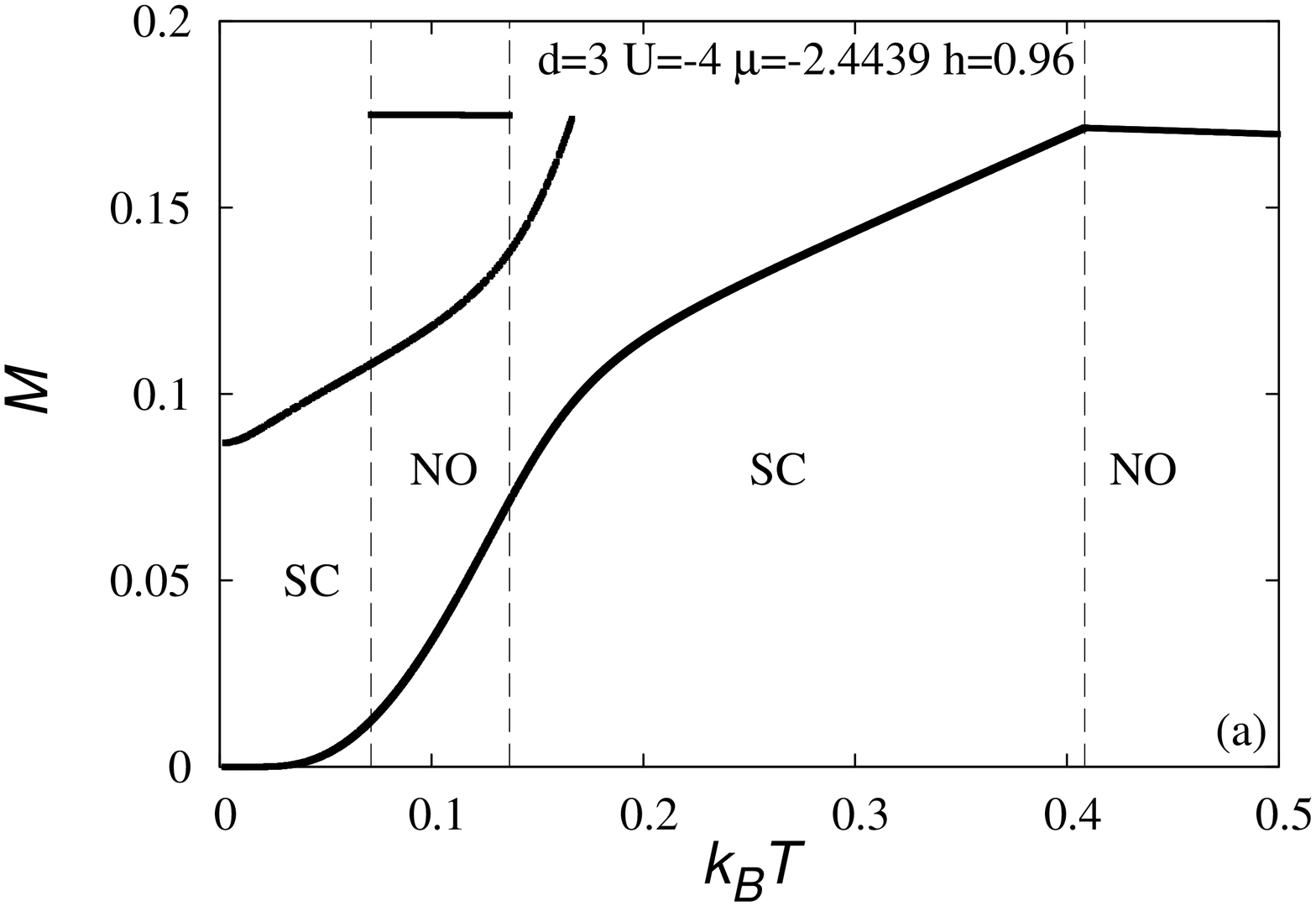}
\hspace*{-0.6cm}
\includegraphics[width=0.38\textwidth,angle=270]{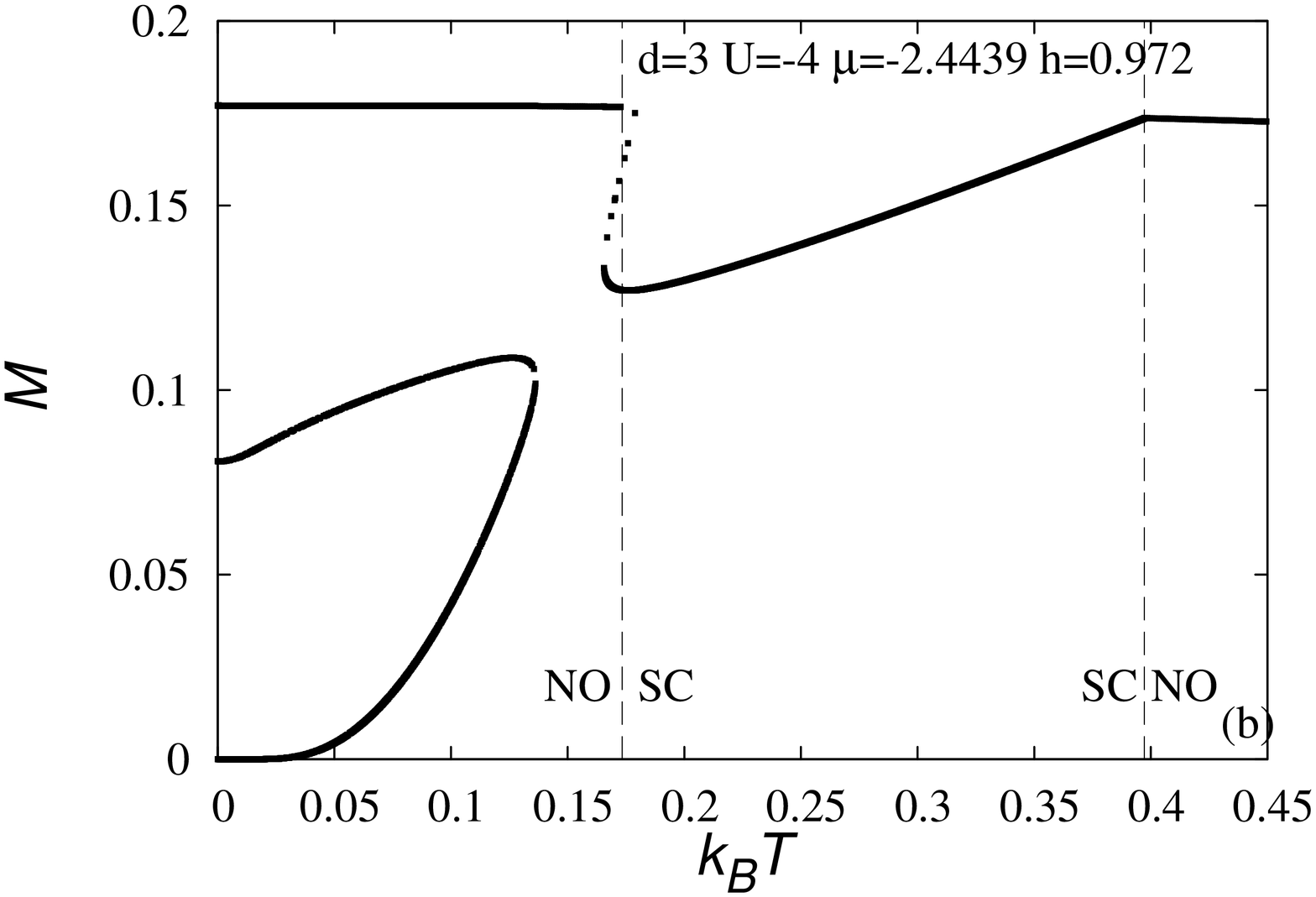}\\
\caption{\label{M-3D} Magnetization vs. temperature for fixed $h=0.96$ (a) and
$h=0.972$ (b); $d=3$, $U=-4$, fixed $\mu \approx -2.4439$.}
\end{figure}

For low $h$, the order parameter tends to zero continuously, with increasing
temperature. For higher magnetic fields, there is a change in the phase
transition: from the second to the first order. There can arise two non-zero
solutions for the order parameter (Fig. \ref{DelT3d}(b)).

At $h=0.9$, the upper branch is stable while the lower branch is
thermodynamically unstable. At $h=0.96$, the situation is more
interesting. The lower branch is also unstable which is clearly visible in Figs.
\ref{Omega-T}(a)-(b) and \ref{Omega-delta}(a). There is a maximum \textcolor{czerwony}{of $\Omega$} for the
solution from this branch. However, there is a global minimum for the solution
from the upper branch. After the first order phase transition from SC to NO, at
$T\approx 0.0714$, the solution from the upper branch becomes metastable (a
local minimum at $T=0.09$). At $T\approx 0.1368$, there is yet another
first order transition from the NO to the SC state and up to $T\approx 0.42$
(the second order phase transition from SC to NO) the solutions from the upper
branch are stable. Therefore, there is the following sequence of transitions
in the system, for fixed $h=0.96$: SC$\rightarrow$NO (1$^{st}$ order),
NO$\rightarrow$SC (1$^{st}$ order) and SC$\rightarrow$NO (2$^{nd}$ order).

At $h=0.972$, the solutions from $T=0$ to $T=0.1362$ are metastable (the upper
branch -- a local minimum) and unstable (the lower branch -- a maximum) (\textcolor{green}{Fig. \ref{Omega-delta}(c)}). The
system is in the normal state. At $T\approx 0.1734$, the first order phase
transition from the normal to the superconducting state takes place. Then, at
\textcolor{green}{$T\approx T_c=0.3975$}, the system goes to the normal state (2$^{nd}$ order
transition). 

Therefore, for sufficiently high magnetic fields a reentrant transition takes
place, as in the 2D case. However, this transition is more complex in $d=3$,
which is also reflected in the behavior of spin magnetization (Fig.
\ref{M-3D}(a)-(b)).

\subsection{The superfluid density}
\begin{figure}[t!]
\begin{center}
\includegraphics[width=0.55\textwidth,angle=270]{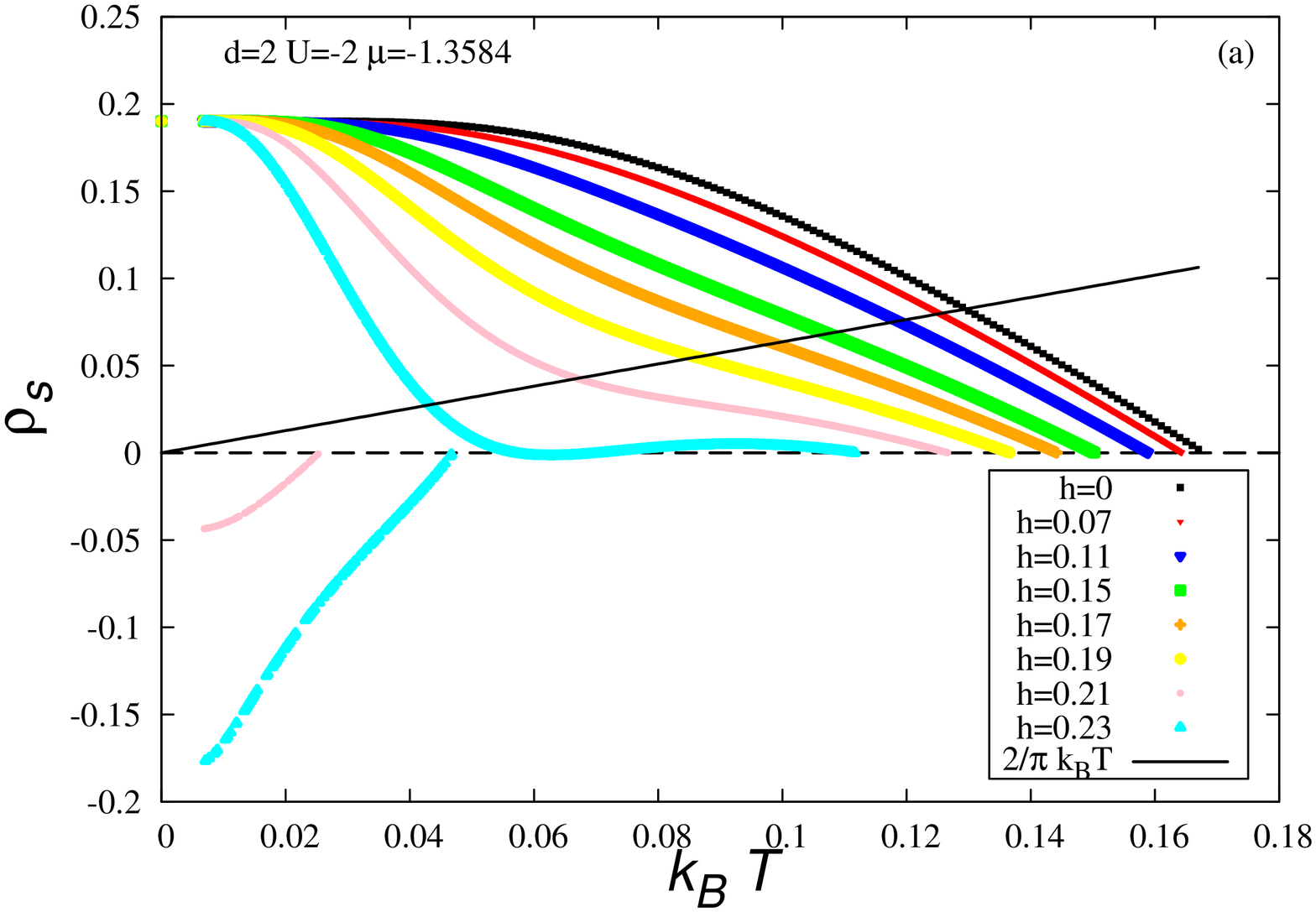}\hspace{-0.2cm}
\includegraphics[width=0.55\textwidth,angle=270]{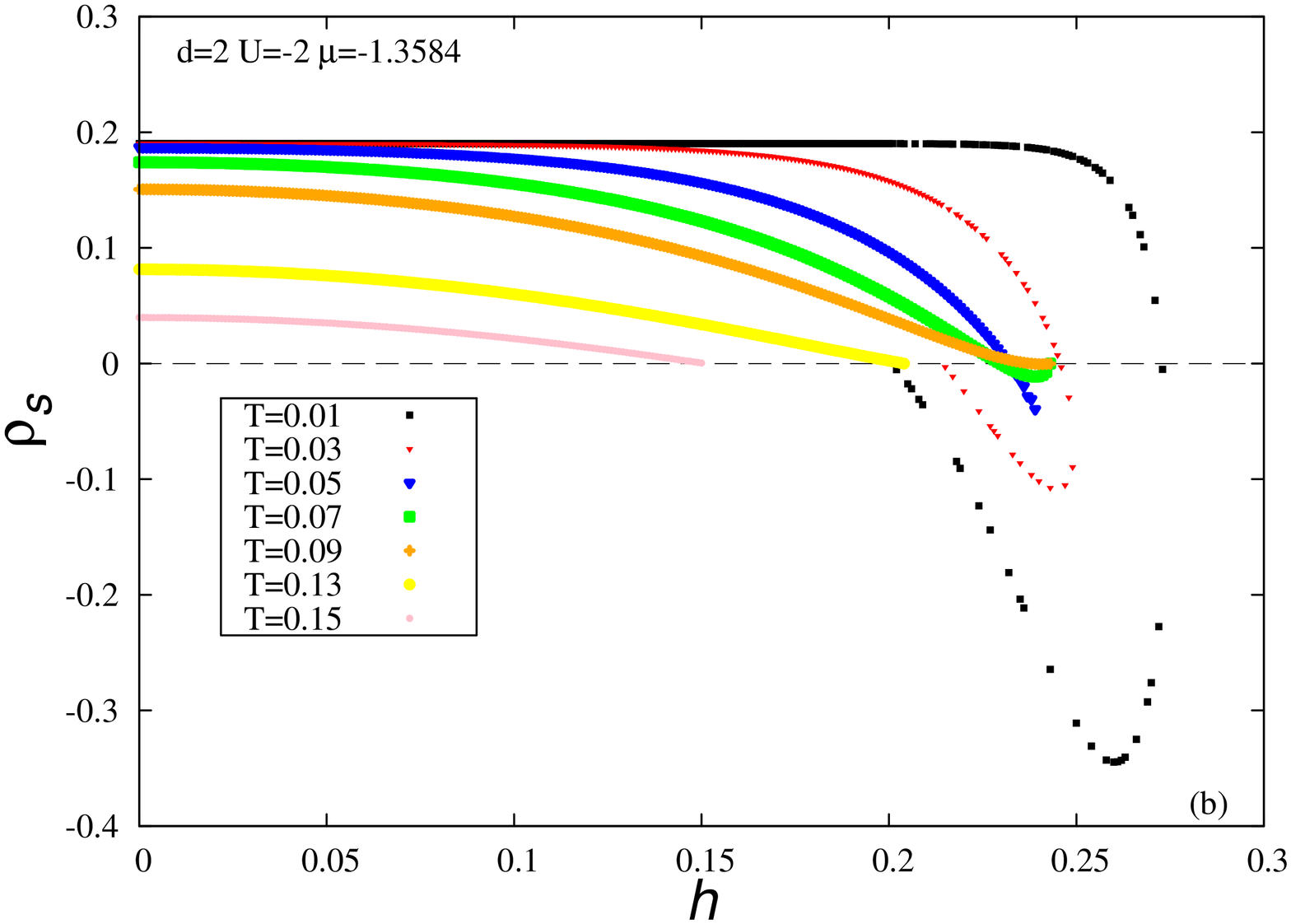}\\
\caption{\label{ro_s-2D} Superfluid density vs. temperature (a) and magnetic field (b) for the square lattice, $\mu \approx -1.358$, $U=-2$. The Kosterlitz-Thouless temperatures in 2D are found from the intersection point of the straight line $\frac{2}{\pi} k_B T$ with the curve $\rho_s(T)$.}
\end{center}
\end{figure}

\begin{figure}[t!]
\begin{center}
\includegraphics[width=0.55\textwidth,angle=270]{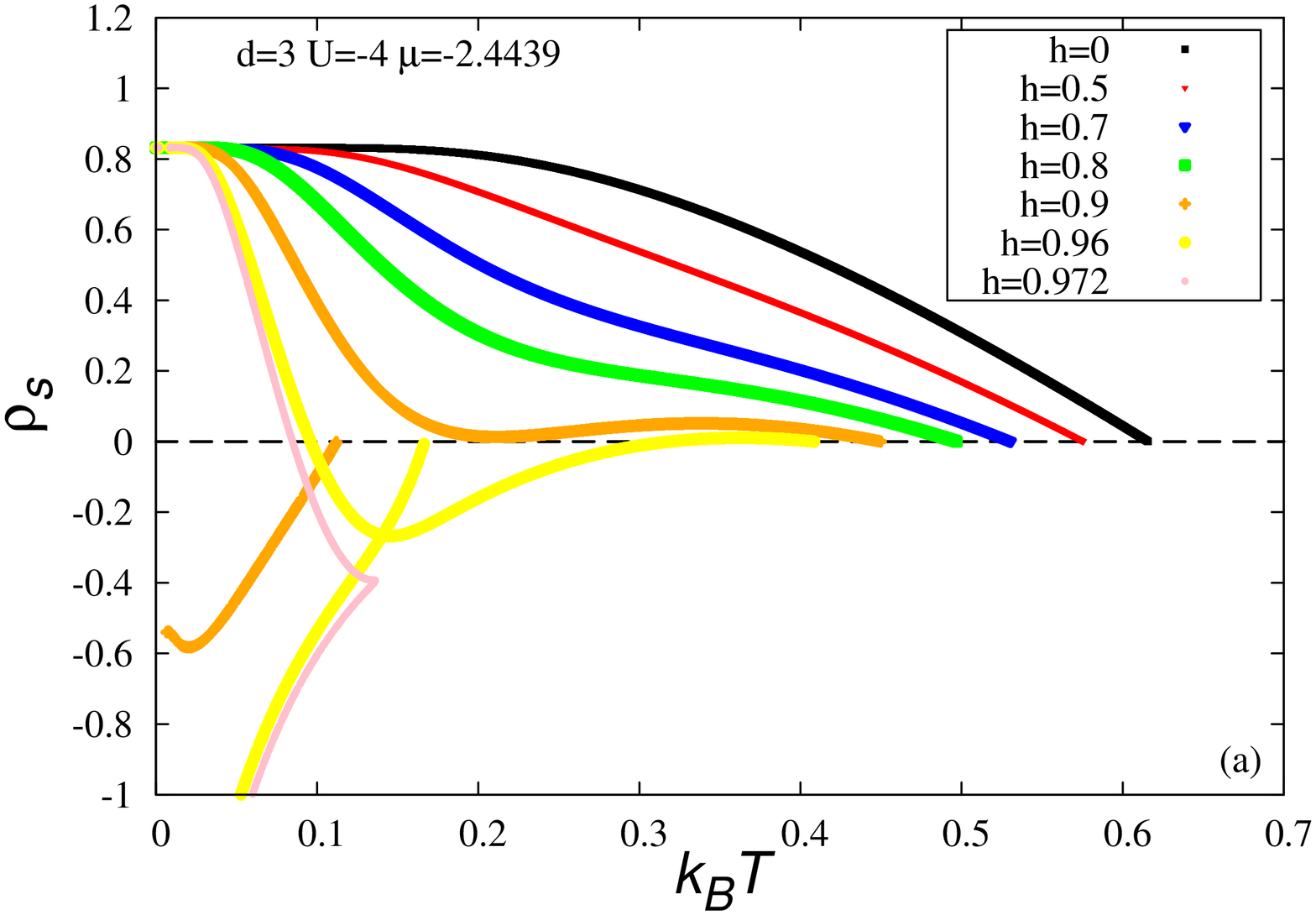}\hspace{-0.2cm}
\includegraphics[width=0.55\textwidth,angle=270]{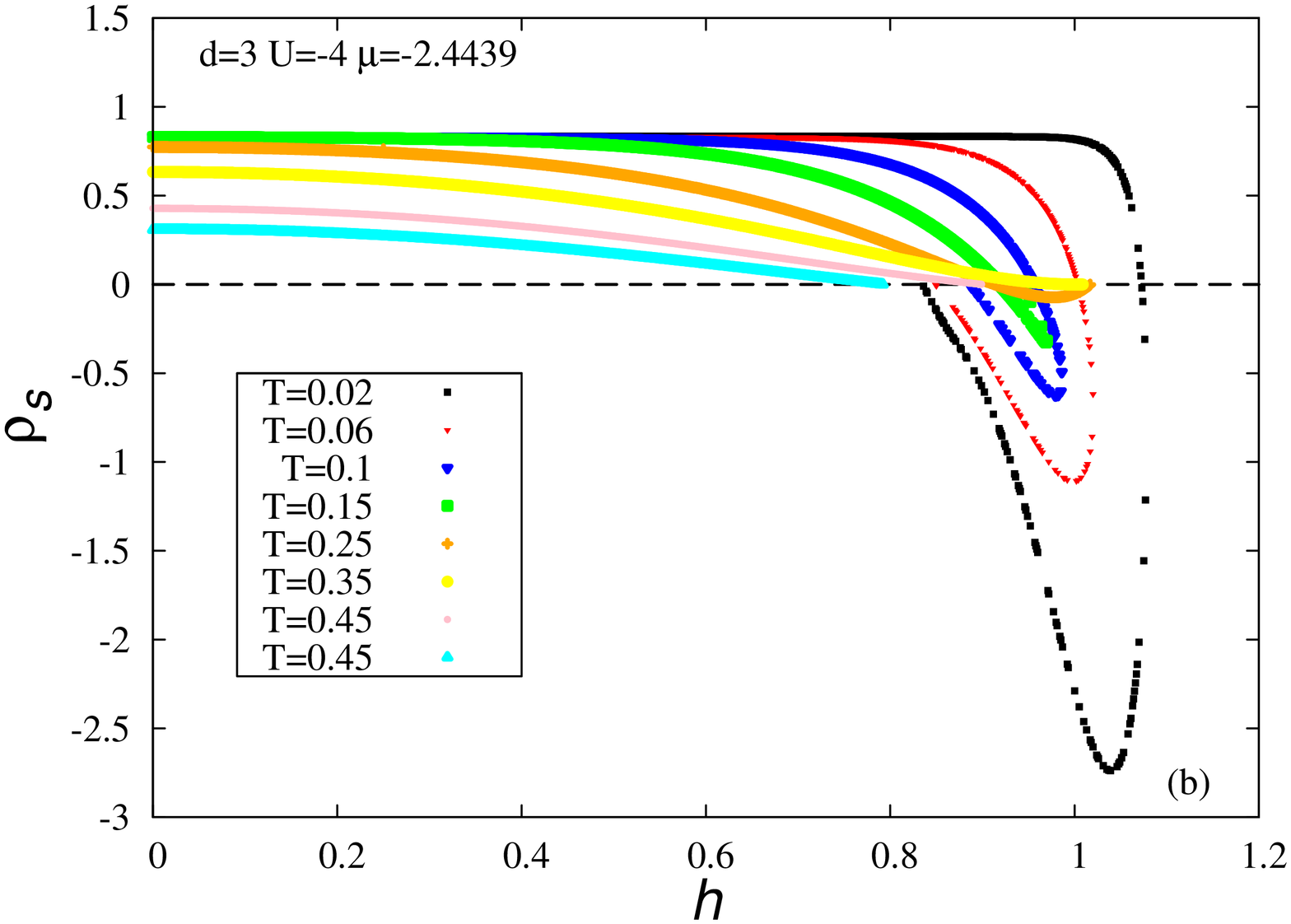}\\
\caption{\label{ro_s-3D} Superfluid density vs. temperature (a) and magnetic field (b) for the simple cubic lattice, $\mu \approx -2.4439$, $U=-4$.}
\end{center}
\end{figure}

In this section, the properties of the superfluid stiffness -- $\rho_s$, in the Zeeman magnetic field, will be discussed.

This physical quantity is important both because of a relation to the London
penetration depth $\lambda$ ($\rho_s \sim \lambda^{-2}$) \cite{Tinkham} and
its relevance in the Kosterlitz-Thouless temperature determination procedure.

First, we solve the system of the self-consistent equations for the order
parameter, chemical potential and magnetization, to determine $\rho_s$. For the
s-wave pairing symmetry case, at $t^{\uparrow}=t^{\downarrow}$ and $h\neq 0$, we
solve Eqs.~\eqref{del}, \eqref{particle_eq} and \eqref{Magn_s}. We determine
from these equations the values of $\Delta$, $\bar{\mu}$, $M$ and then we
calculate the value of the superfluid stiffness, from Eq.~\eqref{ro_s-2}.

Let us consider the temperature dependencies of $\rho_s$. Figs. \ref{ro_s-2D}-\ref{ro_s-3D} show the superfluid density vs. magnetic field (a) and temperature (b), \textcolor{czerwony}{at fixed attractive interaction and a chemical potential}, for $d=2$ and $d=3$, respectively. 

The superfluid density is the highest in the ground state and falls to zero
at the Hartree-Fock critical temperature. The decrease in $\rho_s$ with the
increasing temperature results from the paramagnetic part of the superfluid
stiffness:
\begin{equation}
\label{ro_s_para}
\rho_s^{para}(T)= \frac{1}{4N}\sum_{\vec{k}}\Bigg\{\Bigg(\frac{\partial \epsilon_{\vec{k}}}{\partial k_x} \Bigg)^2\Bigg[\frac{\partial f(E_{\vec{k}\uparrow})}{\partial E_{\vec{k}\uparrow}} +\frac{\partial f(E_{\vec{k}\downarrow})}{\partial E_{\vec{k}\downarrow}} \Bigg] \Bigg\},
\end{equation}
which increases with temperature and compensates the diamagnetic part of $\rho_s$:
\begin{equation}
\rho_s^{dia}(T)= \frac{1}{4N} \sum_{k} \Bigg\{ \frac{\partial^2\epsilon_{\vec{k}}}{\partial k_x^2}\Bigg[ 1-\frac{\bar{\epsilon}_{\vec{k}}}{2\omega_{\vec{k}}} \Bigg(\tanh\Bigg(\frac{\beta E_{\vec{k}\uparrow}}{2}\Bigg)+\tanh\Bigg(\frac{\beta E_{\vec{k}\downarrow}}{2}\Bigg) \Bigg)\Bigg]\Bigg\}.
\end{equation}
If the excitations in the system have non-zero
minimum energy at $h=0$ (e.g. for the s-wave pairing symmetry), there is no
significant decrease in $\rho_s (T)$ in low temperatures. 

As mentioned above, there is a change in the character of the phase transition
from the second to the first order with increasing magnetic field, which is
also reflected in the $\rho_s$ dependences. If the transition is of second
order, the superfluid stiffness tends to zero continuously, at the mean-field
transition point (in which the order parameter drops to zero). However, the
situation is more complex for the first order phase transition. Then, there
arise two non-zero solutions for the order parameter and also for the superfluid
density. The lower branches are always energetically unstable. Therefore,
$\rho_s<0$ which means that these solutions are also dynamically unstable (see:
Fig. \ref{ro_s-2D}(a), $h=0.21$, $h=0.23$ and Fig. \ref{ro_s-3D}(a), $h=0.9$,
$h=0.96$, $h=0.972$). However, one can find that $\rho_s$ drops below zero for
$h<h_c^{HF}$ ($h<h_c^{HF}$ -- Hartree-Fock critical magnetic field) (see: Fig.
\ref{ro_s-2D}(b), $T=0.05$, $T=0.07$, $T=0.09$ and Fig. \ref{ro_s-3D}(b),
$T=0.25$, $T=0.35$), afterwards increases to zero in the HF critical magnetic
field. Therefore, there exists a region for which $\rho_s$ is negative, despite
the energetic stability. Then, the whole range of the reentrant transition,
which is discussed above in details, turns out to be unstable, which is
particularly important in the $d=3$ case. This aspect of the reentrant
transition will be analyzed in next section, which concerns the determination of
the Hartree-Fock and Kosterlitz-Thouless critical temperature and also the phase
diagrams construction.

\subsection{Phase diagrams}

In this subsection, we focus on the temperature phase diagrams for fixed
chemical potential and electron concentration. We analyze the weak attraction
case for $d=2$ and $d=3$.

One of the basic quantities characterizing the superconductor is the critical
temperature ($T_c$). The theory of the second-order phase transitions was
intensively studied by Landau in the 1930s. He found that a change in the
symmetry of the system accompanies each second-order phase transition.
Therefore, Landau introduced a quantity called the order parameter to the
theory. The order parameter is nonzero below $T_c$, but becomes zero above
$T_c$, in the disordered state \cite{landau3}. 

According to the Ginzburg-Landau theory \cite{GL}, the order parameter $\psi$ is
zero above $T_c$ of the superconductor, in the normal state and below $T_c$ it
is nonzero, in the superconducting phase. Then, one can write:
\begin{equation}
\psi = \left\{ \begin{array}{ll} 
 0 & \textrm{if} \,\, T>T_c \\ 
\psi(T)\neq 0 & \textrm{if} \,\, T<T_c 
\end{array} \right.
\end{equation}
where $\psi$ is a complex function.

Gor'kov showed that from the microscopic point of view, \textcolor{czerwony}{the order parameter
$\psi$ is related to the local
value of the superconducting gap $\Delta$ \cite{Gorkov}}.

The free energy $F$ of the superconducting state shows a continuous dependence
on $\psi$. As $F$ is real and $\psi$ is
complex, the free energy can be dependent on $|\psi|$. Now, we can Taylor expand
$F$ in powers of $|\psi|$, because $\psi \rightarrow 0$ at the critical
temperature. Therefore, the free energy density ($f\equiv F/V$) takes the form:
\begin{equation}
 f_s (T)=f_n (T) + a(T)|\psi|^2+\frac{1}{2} b(T) |\psi|^4+...,
\end{equation}
where $f_s(T)$ and $f_n(T)$ are the superconducting and normal free energy
densities, respectively. $a(T)$, $b(T)$ -- phenomenological parameters of the
theory. Parameter $b(T)$ must be positive, so that the free energy density has a
minimum. Above $T_c$, $a(T)$ is positive and the free energy density has one
minimum at $\psi=0$ and the system is in the normal state.
However, if $a(T)$ decreases with decreasing temperature, the state of the
system changes at $a(T)=0$, i.e. for $T<T_c$ the free energy has a minimum at
$\psi \neq 0$. Therefore, the temperature at $a(T)=0$ can be identified as the
critical temperature $T_c$.

As mentioned above, according to the BCS theory, it is the temperature at
which the energy gap vanishes. We have to solve Eqs. \eqref{del}-\eqref{Magn_s}
to determine $T_c^{HF}$. If $\Delta =0$, the set of equations
\eqref{del}-\eqref{Magn_s} for the s-wave pairing symmetry takes the form:
\begin{eqnarray}
1=-\frac{U}{N} \sum_{\vec{k}}\frac{1}{2(\epsilon_{\vec{k}}-\bar{\mu})}\Bigg(\frac{1}{2}\tanh \Bigg[\frac{\beta_c}{2}\Big(\frac{UM}{2}+h+\epsilon_{\vec{k}}-\bar{\mu}\Big)\Bigg]\nonumber \\
+\frac{1}{2}\tanh \Bigg[\frac{\beta_c}{2}\Big(-\frac{UM}{2}-h-\epsilon_{\vec{k}}+\bar{\mu}\Big)\Bigg]\Bigg),
\end{eqnarray}
\begin	{eqnarray}
 n=1&-&\frac{1}{N} \sum_{\vec{k}}\frac{1}{4} \Bigg(\tanh \Bigg[\frac{\beta_c}{2}\Big(\frac{UM}{2}+h+\epsilon_{\vec{k}}-\bar{\mu}\Big)\Bigg]\nonumber \\
&-&\tanh \Bigg[\frac{\beta_c}{2}\Big(-\frac{UM}{2}-h-\epsilon_{\vec{k}}+\bar{\mu}\Big)\Bigg]\Bigg),
\end{eqnarray}
\begin{eqnarray}
 M=\frac{1}{2N}\sum_{\vec{k}} \Bigg(\tanh \Bigg[\frac{\beta_c}{2}\Big(\frac{UM}{2}+h+\epsilon_{\vec{k}}-\bar{\mu}\Big)\Bigg]\nonumber\\
-\tanh \Bigg[\frac{\beta_c}{2}\Big(-\frac{UM}{2}-h-\epsilon_{\vec{k}}+\bar{\mu}\Big)\Bigg]\Bigg).
\end{eqnarray}
It should be emphasized that the temperature in which the gap vanishes is
closely related to the value of $\Delta$ in the ground state, i.e. the higher
value of $\Delta$ at $T=0$, the higher critical temperature can be expected.

The above method of the critical temperatures determination for the 2D
superconductors can lead to an overestimate of $T_c$, because
we neglect the fluctuation effects. However, the fluctuations play a crucial
role in these systems. As mentioned above, the influence of the
fluctuations
can be included by the assumption that the phase transition from the
superconducting to the normal state is the Kosterlitz-Thouless transition, with
a characteristic jump in the superfluid stiffness $\frac{2}{\pi}$ at $T_c$.
According to Eq.~\eqref{KT}, the KT transition temperature is found from the
intersection point of the straight line $\frac{2}{\pi} k_B T$ with the curve
$\rho_s(T)$ in 2D (Fig. \ref{ro_s-2D}(a)).

The properties of the superfluid density which determines $T_c^{KT}$ are
described in the previous section. Now, we focus on the temperature phase
diagrams for fixed chemical potential and electron concentration.

\begin{figure}
\hspace*{-0.8cm}
\includegraphics[width=0.38\textwidth,angle=270]{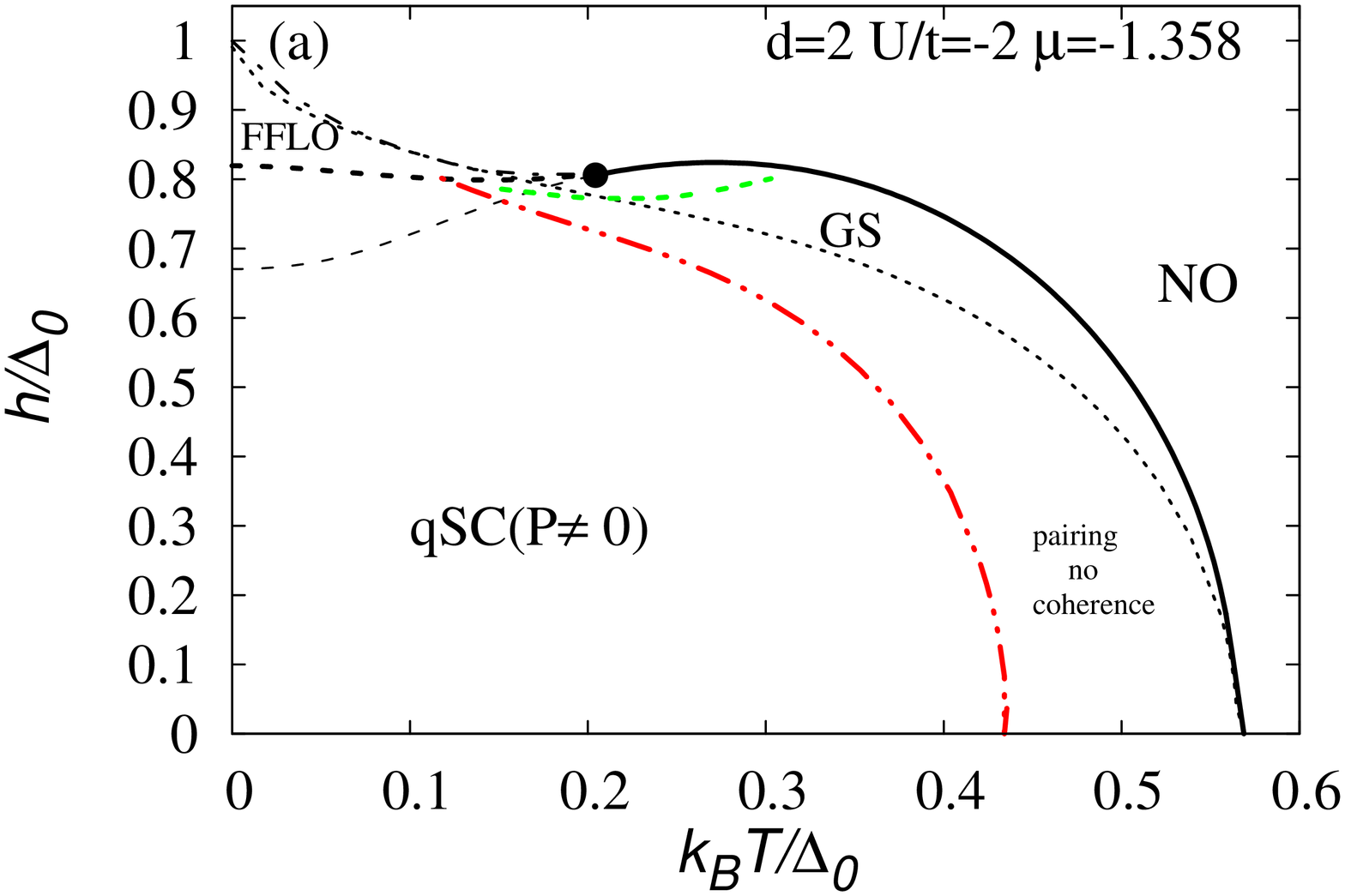}
\hspace*{-0.6cm}
\includegraphics[width=0.38\textwidth,angle=270]{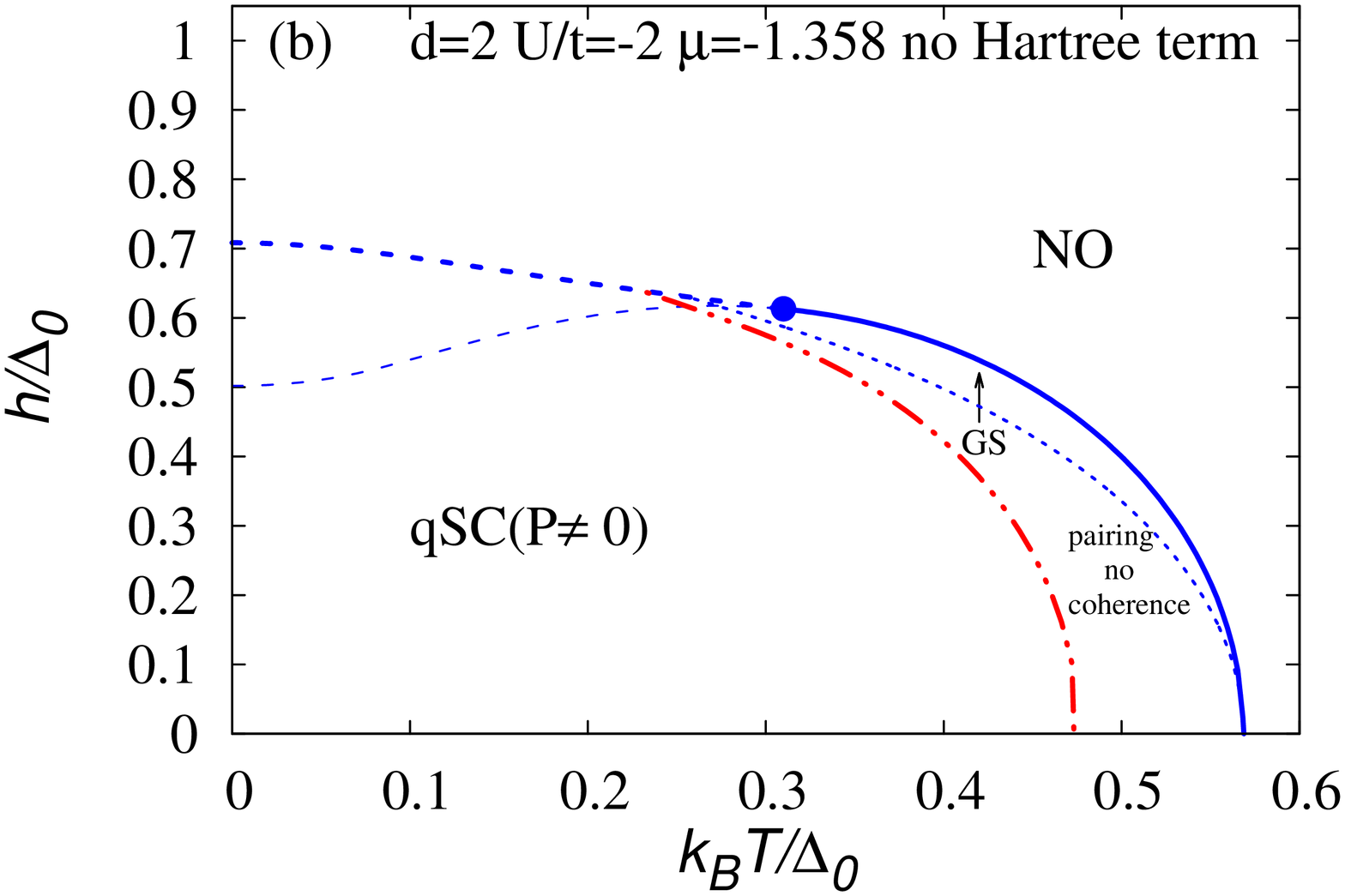}\\
\hspace*{-0.8cm}
\includegraphics[width=0.38\textwidth,angle=270]{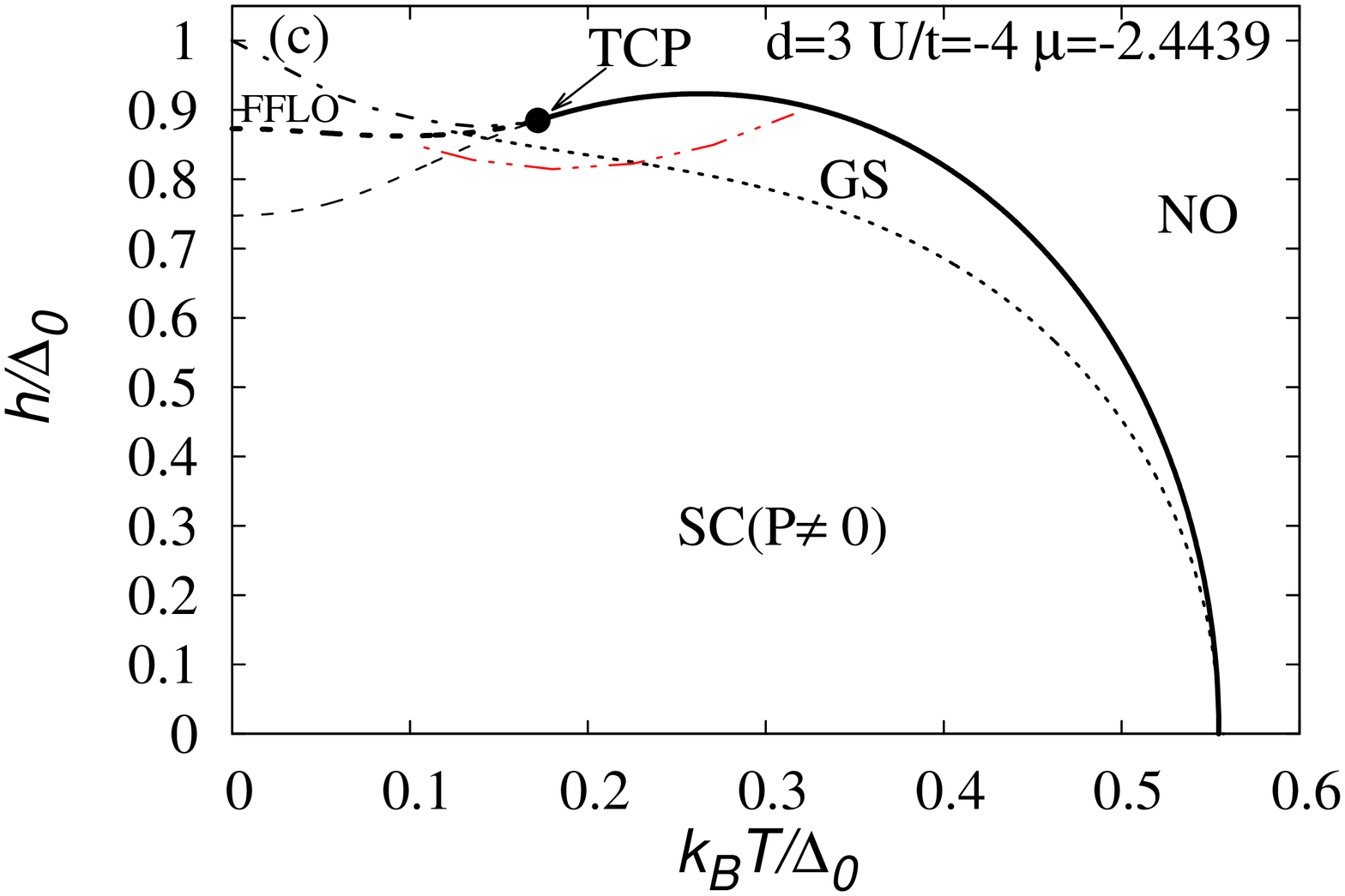}
\hspace*{-0.6cm}
\includegraphics[width=0.38\textwidth,angle=270]{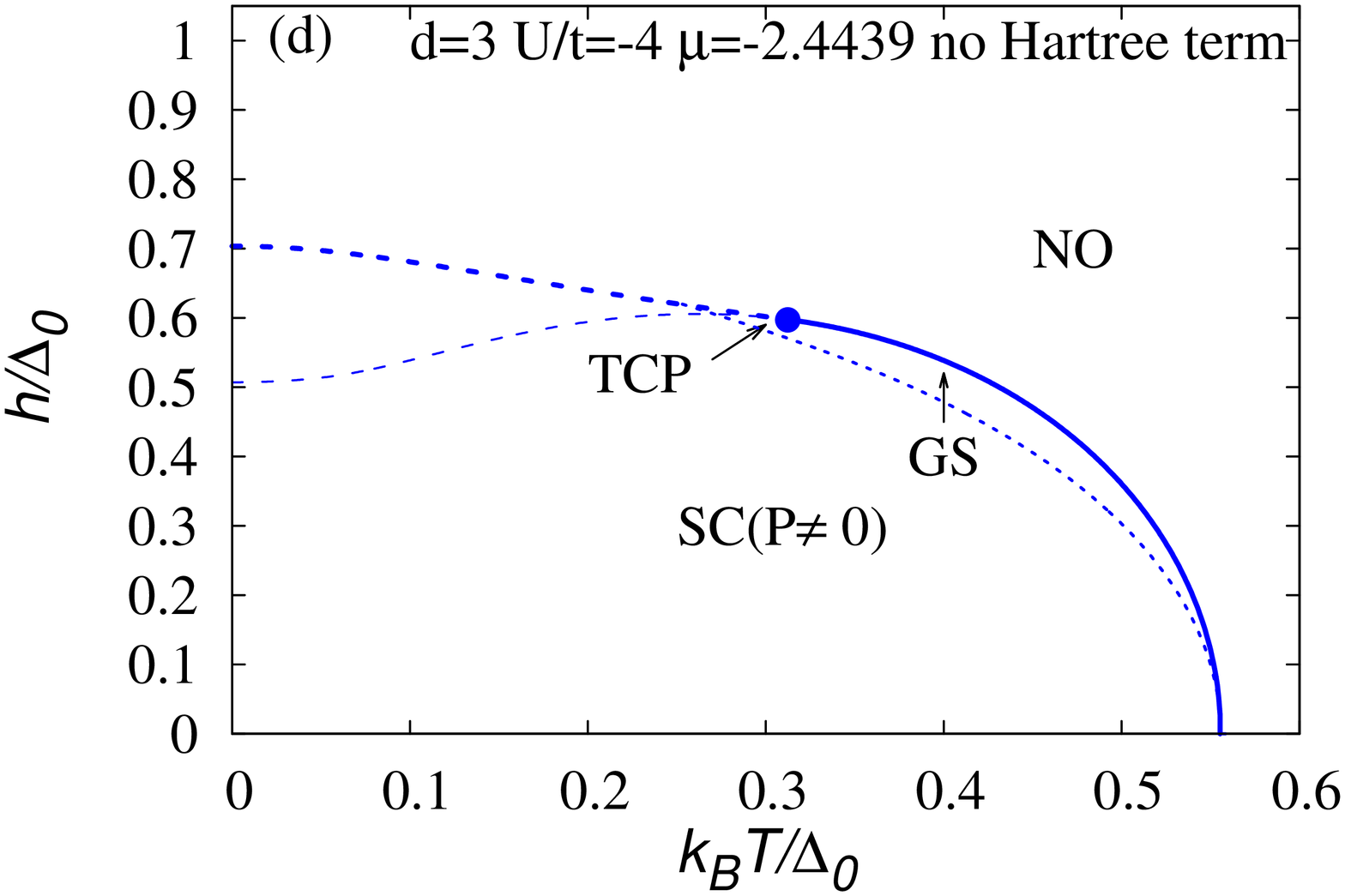}
\caption[Temperature vs. magnetic field phase diagrams for $d=2$, $U=-2$ and
fixed chemical potential $\mu \approx -1.358$. (a) diagram with Hartree term,
(b) without Hartree term (blue color). (c) phase diagrams for $d=3$,
$U=4$, $\mu \approx -2.4439$.]{\label{diag_T_fixed_mu} Temperature vs. magnetic
field phase diagrams for $U=-2$ and fixed chemical potential $\mu \approx
-1.358$. (a) diagram with Hartree term, (b) without Hartree term (blue
color). (c) phase diagrams for $d=3$, $U=4$, $\mu \approx -2.4439$. The
thick solid line is the second order phase transition line from pairing without
coherence region to NO (Hartree approximation). The thin dashed line is an
extension of the $2^{nd}$ order transition line (metastable solutions). The
dashed-dotted line indicates the upper limit for the occurrence of the FFLO
state. The thick dashed-double dotted line (red color) is the
Kosterlitz-Thouless transition line. The dashed green line limits the region
with $\rho_s<0$. The thick dotted line denotes the first order phase transition
to NO, GS -- gapless region. The chemical potential \textcolor{czerwony}{is} chosen to yield $n
\approx0.75$ in the case of the Hartree term included, at $T=0$ and $h=0$. $\Delta_{0}$ denotes the gap at $T=0$ and
$h=0$.}
\end{figure}
\begin{figure}
\begin{center}
\hspace*{-1.0cm}
\includegraphics[width=0.38\textwidth,angle=270]{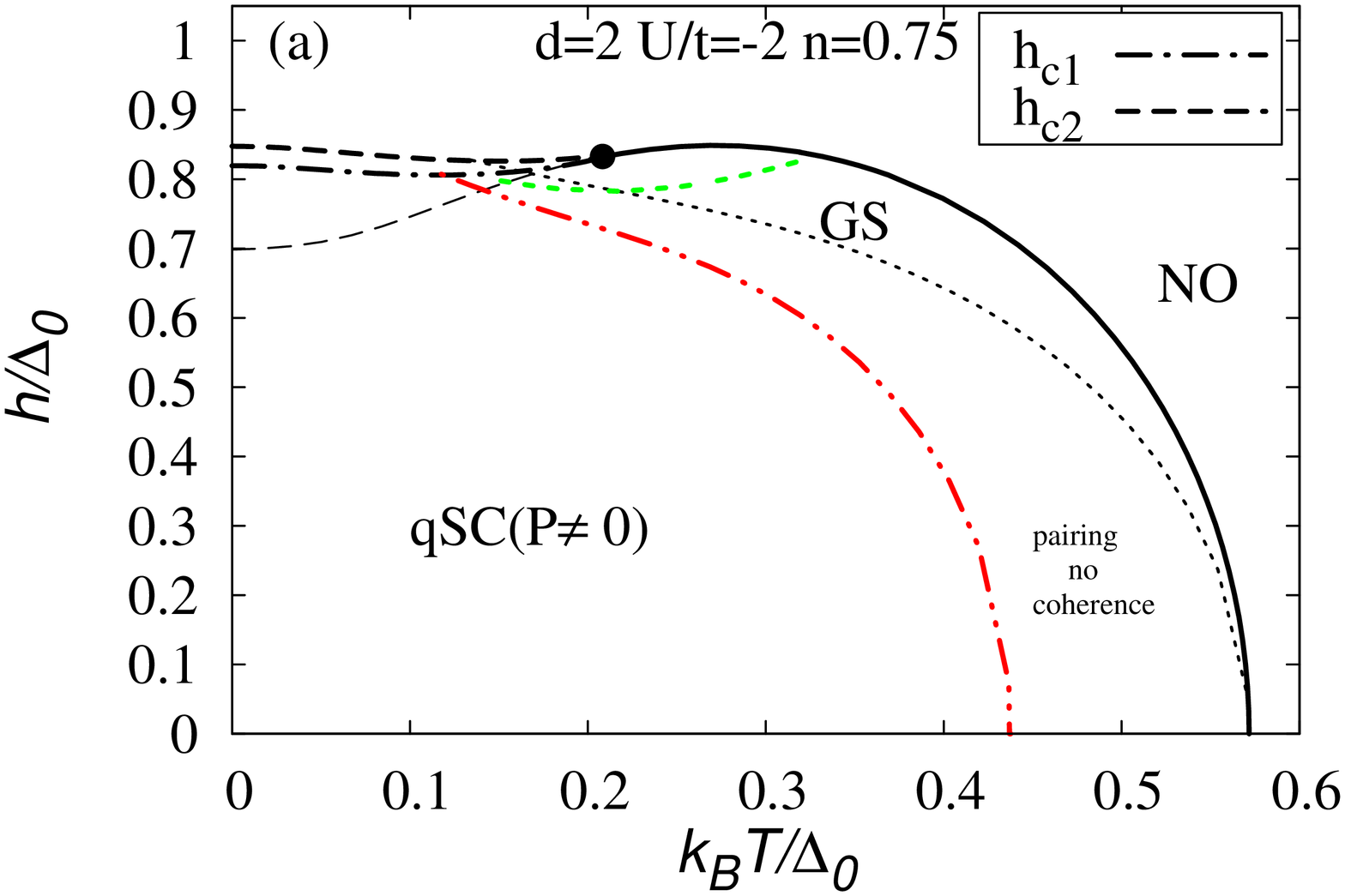}
\hspace*{-0.8cm}
\includegraphics[width=0.38\textwidth,angle=270]{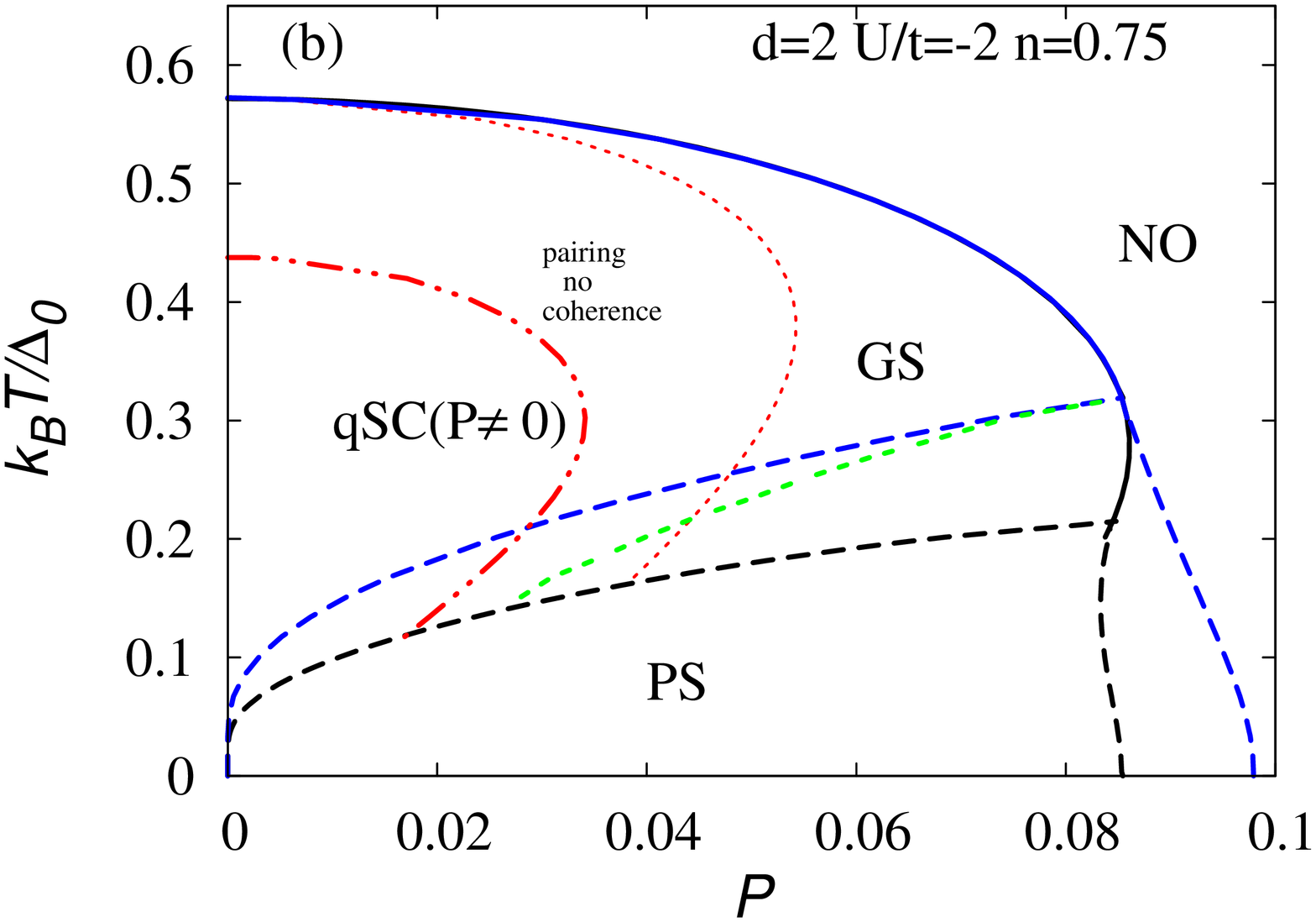}
\includegraphics[width=0.38\textwidth,angle=270]
{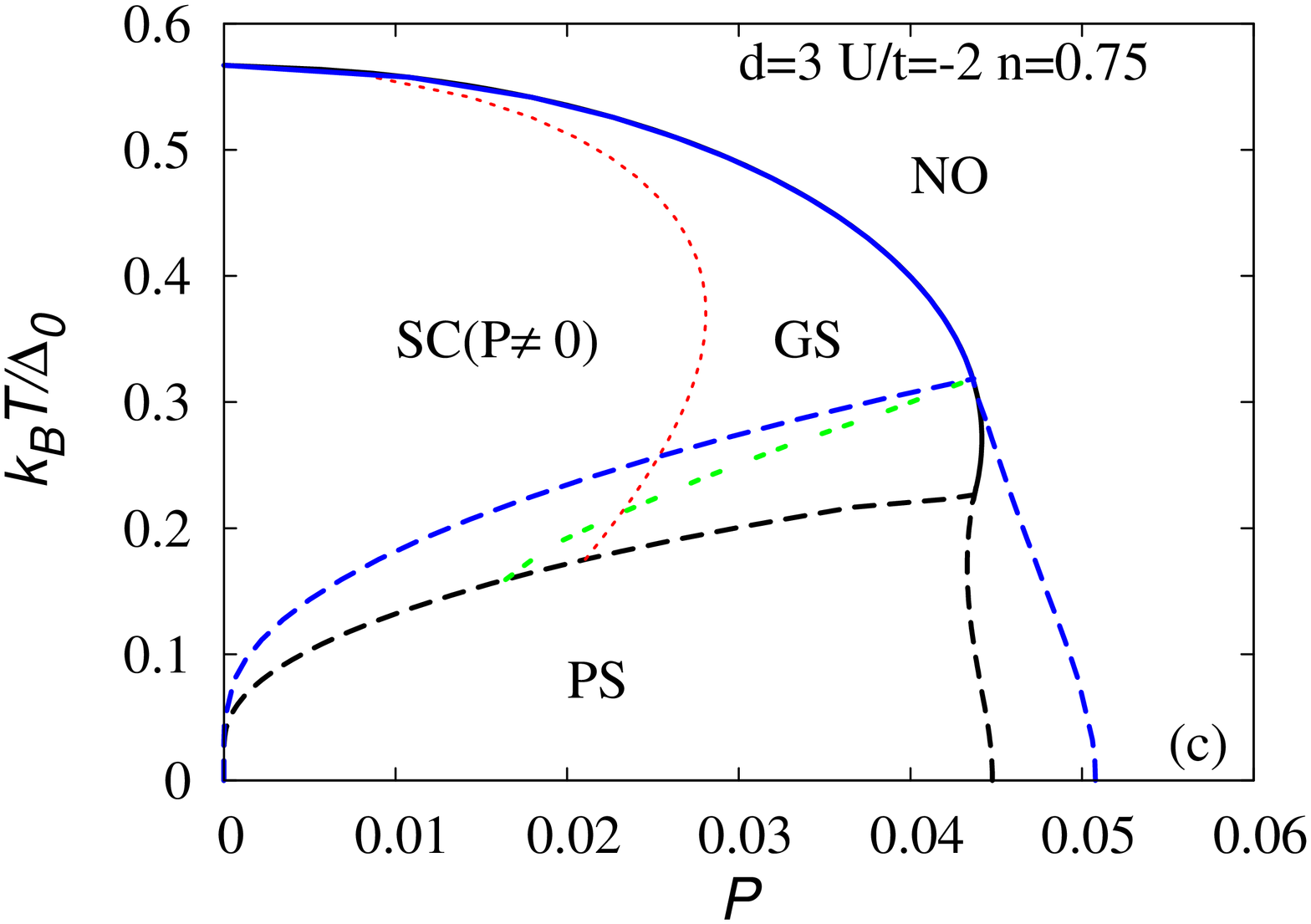}
\caption[Temperature vs. magnetic field (a) and polarization (b) phase diagrams
for square lattice, (c) $T$ vs. $P$ phase diagrams for simple cubic lattice;
fixed $U=-2$ and $n=0.75$.]{\label{diag_T_fixed_n} Temperature vs. magnetic
field (a) and polarization (b) phase diagrams for square lattice, (c) $T$ vs.
$P$ phase diagrams for simple cubic lattice; fixed $U=-2$ and $n=0.75$. The
thick dashed lines are the first order phase transition lines. The upper set of
curves denotes the diagram without the Hartree term (blue color), red line
limits the GS region. The dashed green line limits the region
with $\rho_s<0$. $\Delta_{0}$ denotes the gap at $T=0$ and $P=0$.}
\end{center}
\end{figure}

\begin{figure}
\hspace*{-0.8cm}
\includegraphics[width=0.38\textwidth,angle=270]{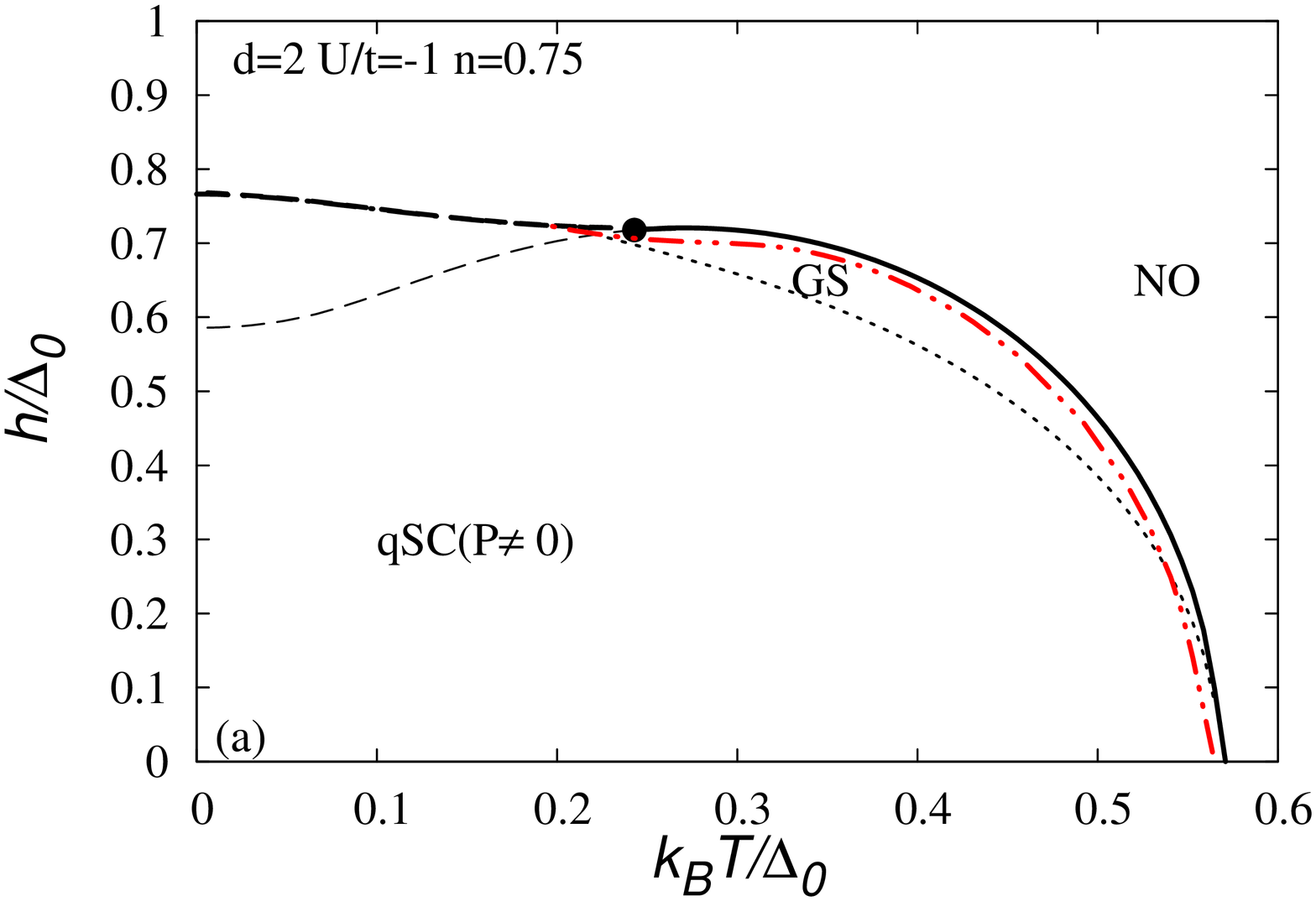}
\hspace*{-0.6cm}
\includegraphics[width=0.38\textwidth,angle=270]{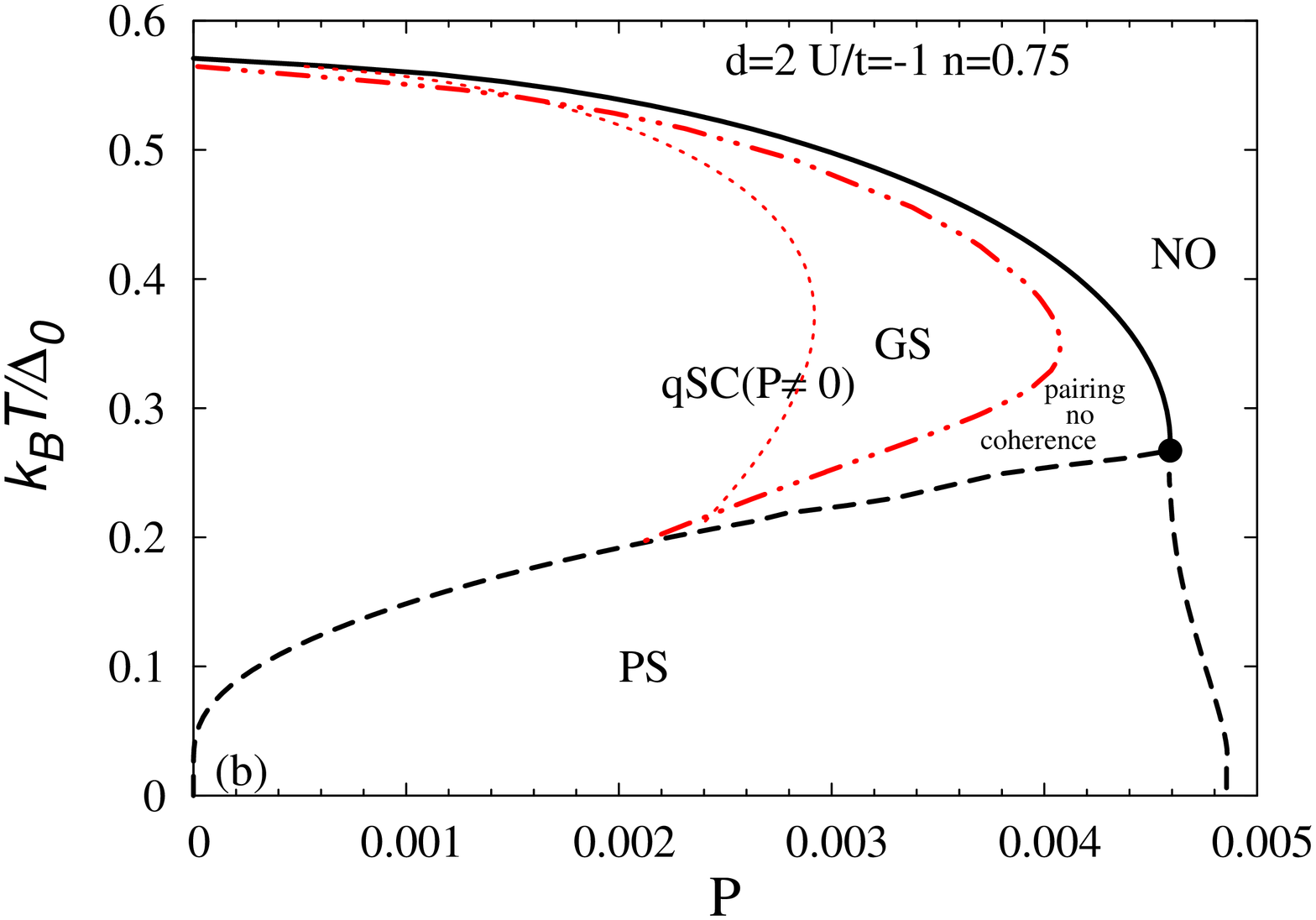}\\
\caption[Temperature vs. magnetic field (a) and polarization (b) phase diagrams
with Hartree term, for square lattice, fixed $U=-1$ and
$n=0.75$.]{\label{diag_U-1} Temperature vs. magnetic field (a) and polarization
(b) phase diagrams with the Hartree term, for the square lattice, fixed $U=-1$
and $n=0.75$. The thick dashed-double dotted line (red color) is the
Kosterlitz-Thouless transition line, thick dashed lines are the first order
phase transition lines, GS -- gapless region. $\Delta_{0}$ denotes the gap at
$T=0$ and $P=0$.}
\end{figure}

Fig. \ref{diag_T_fixed_mu} shows $h-T$ phase diagrams, for the square lattice.
Both the case of a fixed chemical potential with the Hartree term (a) and the
case without the Hartree term (b) are presented. The chemical potential is
chosen to yield $n \approx0.75$ at $T=0$ and $h=0$. Similar results for a simple
cubic lattice were obtained using the Hartree approximation (Fig.
\ref{diag_T_fixed_mu}(c)). Fig. \ref{diag_T_fixed_n} shows $h-T$ phase diagrams
for fixed $n=0.75$ and $d=2$. Thus, one can compare the fixed chemical potential
and the fixed electron concentration cases in finite temperatures.

The temperature at which the amplitude of the order parameter reaches zero is
that at which the Cooper pairs no longer exist. According to the
BCS theory, this is the temperature of the pairs breaking and also the
temperature
of the second order transition from the superconducting to the normal state. 

For a sufficiently low magnetic field, there is a KT transition from qSC$(P\neq
0)$ to NO through the region of incoherent pairs (so-called \emph{pseudogap}
region), below the HF critical temperature. With increasing magnetic field, the
character of the qSC-NO transition changes from the second order to the first
order. The curve below the first order transition line on the phase diagrams
(the thin dashed line) is the extension of the HF second order transition line
(metastable solutions). The dash-dotted line (above the first order transition
line) constrains the range of occurrence of the FFLO state (the upper limit).
Above this line, solutions with $\Delta \neq 0$ no longer exist. Moreover, one
obtains the phase separation region in the $h-T$ diagram for the fixed $n\neq
1$, as opposed to the fixed $\mu$ case.

$T_c^{KT}$ are much lower than $T_c^{HF}$. The range in which differences
between the $T_c^{KT}$ and $T_c^{HF}$ temperatures occur can be read from
the order parameter and $\rho_s$ temperature dependences. As shown in
the previous section, $\rho_s$ reaches a maximum value in the ground state
and drops to zero at $T_c^{HF}$. The Kosterlitz-Thouless temperature is also
limited
by the value of the superfluid stiffness at $T=0$ and, simultaneously, it must
be lower than $T_c^{HF}$, which is proportional to the value of the order
parameter
in the ground state. Thus, comparing the values of $\Delta$ and $\rho_s$ at
$T=0$, we can expect that whenever the order parameter is much smaller than
the superfluid density, the quantity which limits $T_c^{KT}$ is $T_c^{HF}$. The
Kosterlitz-Thouless temperatures approach the Hartree-Fock temperatures in
this region. If $\rho_s$ is smaller than the order parameter, the value
of $T_c^{KT}$ can be estimated from the value of $\rho_s$ at $T=0$:
\begin{equation}
 k_B T_c^{KT} \approx \frac{\pi}{2} \rho_s (T=0).
\end{equation}
The above dependence gives the upper limit for the temperatures in which the
phase ordering appears. The smaller the value of $\rho_s$ than the value of the
order parameter, the larger the difference between the Kosterlitz-Thouless and
Hartree-Fock temperatures. The order parameter increases with increasing
attractive interaction, while the value of the superfluid stiffness decreases
with increasing $|U|$ \cite{Denteneer-2, Schneider}. Therefore, one can expect
an increase in the differences between $T_c^{KT}$ and $T_c^{HF}$. In turn, if
the attraction decreases, the differences between $T_c^{KT}$ and $T_c^{HF}$ are
smaller, which is clearly visible in Fig. \ref{diag_U-1} at fixed $n=0.75$ and
$U=-1$, with respect to Fig. \ref{diag_T_fixed_n} at $U=-2$. 

An important aspect of the analysis is the influence of the Hartree term on the
phase diagrams. Firstly, the presence of the Hartree term leads to the reentrant
transition (RT), which is not observed in the phase diagrams without the Hartree
term (blue color in Figs. \ref{diag_T_fixed_mu}(b), (d)). We also find
the region in which $\rho_s<0$ although $\Omega^{SC}< \Omega^{NO}$ (green
double-dashed lines), in the phase diagrams with the Hartree term. If RT exists,
it becomes unstable because $\rho_s<0$. However, the existence of RT does not
influence the existence of the region with  $\rho_s<0$, i.e. even if there is no
RT in the phase diagrams with the Hartree term, there is a region with
$\rho_s<0$ in the weak coupling limit. It is highly relevant especially for
$d=3$ (Fig. \ref{diag_T_fixed_n}(c)), because it means that there exists an
instability of the homogeneous superconducting state. This would suggest
the existence of a stable FFLO state in the weak coupling regime. The Hartree
term induces an increase in the Chandrasekhar-Clogston limit. It is worth
mentioning that the Hartree term narrows the range of PS occurrence, which
is clearly seen in $P-T$ diagrams (Fig. \ref{diag_T_fixed_n}(b)-(c)). 

Analyzing the spectrum of quasiparticle excitations, we also
find a gapless region (GS), for $h> \Delta$ and at $T\neq 0$, i.e. the region,
which has a gapless spectrum for the majority spin species (see: Appendix
\ref{appendix3}).

Therefore, for fixed $n$ one can distinguish the following states in $P-T$
diagrams. At $T\geq 0$ and $P=0$ there is the unpolarized SC phase. At $T\neq
0$, $\Delta \neq 0$ and $P\neq 0$ (i.e. homogeneous superconductivity in the
presence of the spin polarization), the system is also in SC (or in qSC in
$d=2$), but this region is narrowed (in diagrams with the Hartree term) through
the area with $\rho_s<0$. One can observe GS at sufficiently high values of the
magnetic field. In the PS region, not only the polarizations, but also the
particle densities in SC and NO are different. From a comparison of the two- and
the three- dimensional cases, we find that the critical values of polarizations
for the simple cubic lattice are lower than for the square lattice. Taking into
account the phase fluctuations in $d=2$, the region of incoherent pairs can be
distinguished. 

\section{The influence of the pure d-wave pairing symmetry on the polarized superconducting phase stability}

In this section, we briefly discuss the superfluid properties of the Extended
Hubbard Model \eqref{extham'} with spin independent hopping integrals, in a
magnetic field. We take into account only the pure d-wave pairing symmetry case.
Then $W<0$ and the equation for the order parameter takes the form:
\begin{equation}
 \frac{4}{|W|}=\frac{1}{N} \sum_{\vec{k}} \frac{\eta_{\vec{k}}^2}{2\omega_{\vec{k}}}(1-f(E_{\vec{k}\uparrow})-f(E_{\vec{k}\downarrow})),
\end{equation}
where: $\eta_{\vec{k}}=2(\cos \textcolor{czerwony}{k_x}-\cos \textcolor{czerwony}{k_y})$, \textcolor{czerwony}{$E_{\vec{k}\downarrow, \uparrow}=
\pm h+\omega_{\vec{k}}$}, \textcolor{czerwony}{$\omega_{\vec k}=\sqrt{\bar{\epsilon}_{\vec k}+(\eta_{\vec k}\Delta_{\vec k})^2}$}
 
Our motivation to study this kind of pairing symmetry is not only the interest
in high-temperature superconductivity, but also the possibility of existence of
new phases with non-trivial Cooper pairing mechanism in imbalanced Fermi
gases \cite{Acta}.

\begin{figure}[t!]
\hspace*{-0.8cm}
\includegraphics*[width=0.38\textwidth,angle=270]{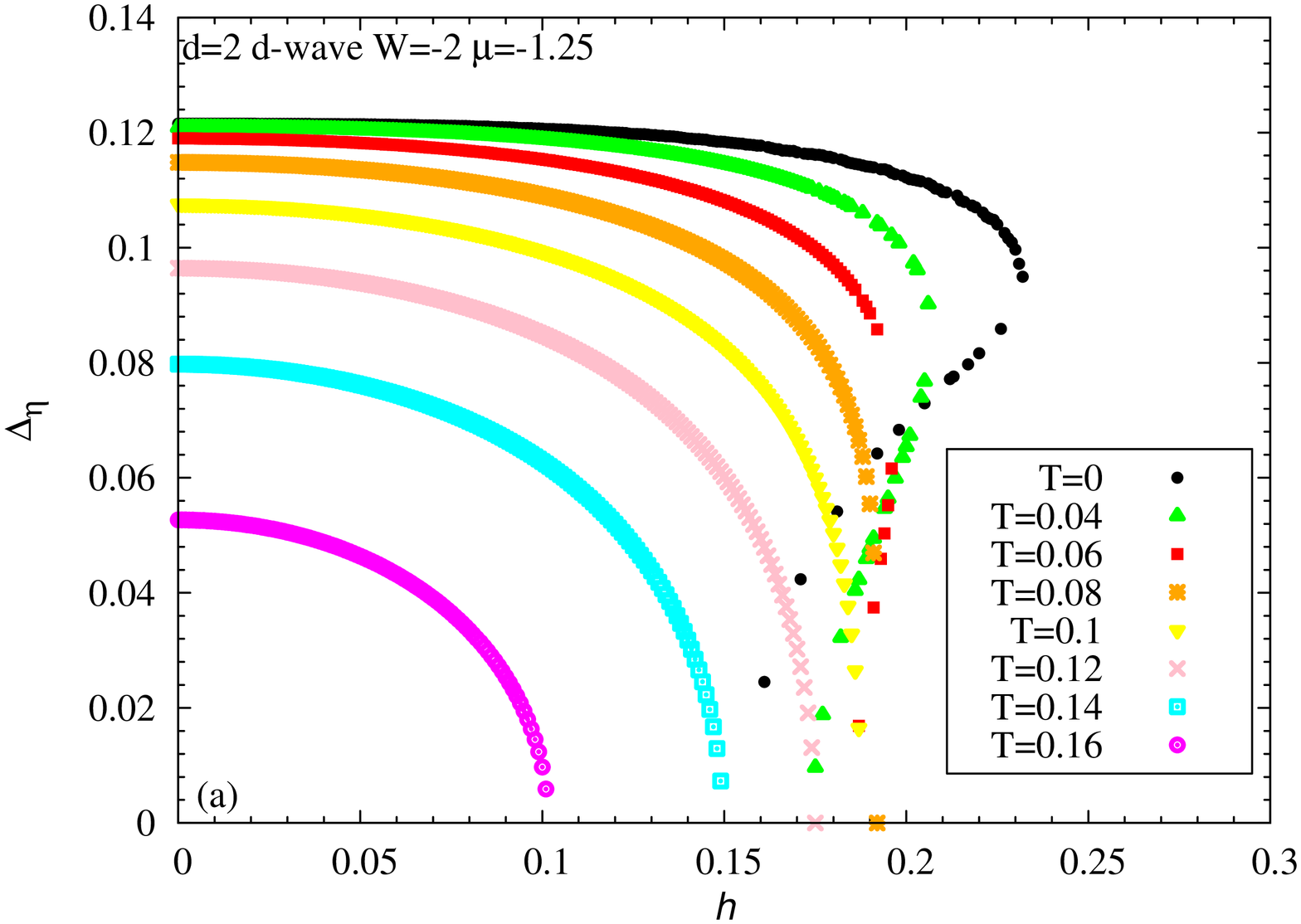}
\hspace*{-0.6cm}
\includegraphics*[width=0.38\textwidth,angle=270]{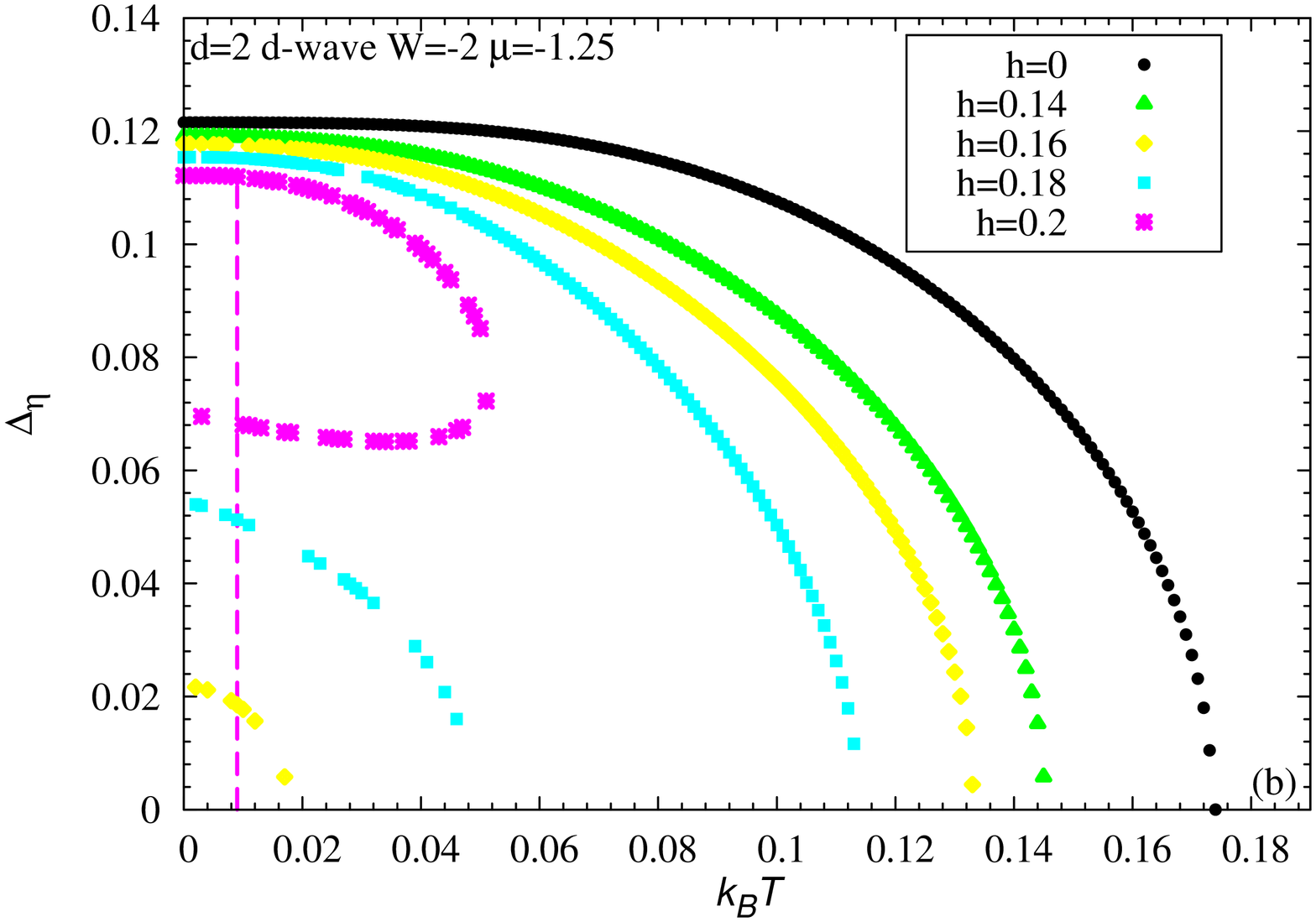}
\hspace*{-0.8cm}
\includegraphics*[width=0.38\textwidth,angle=270]{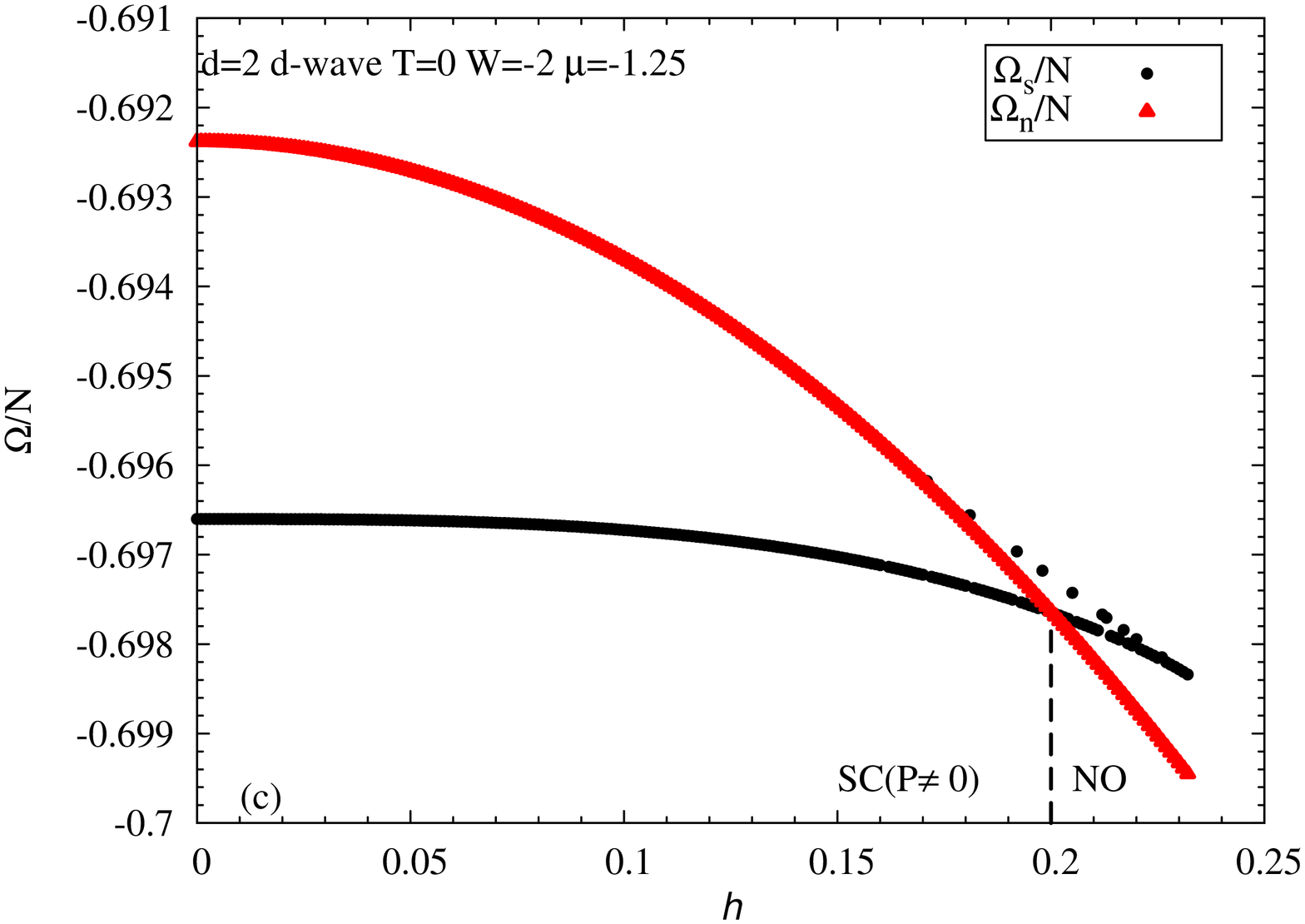}
\hspace*{-0.6cm}
\includegraphics*[width=0.38\textwidth,angle=270]{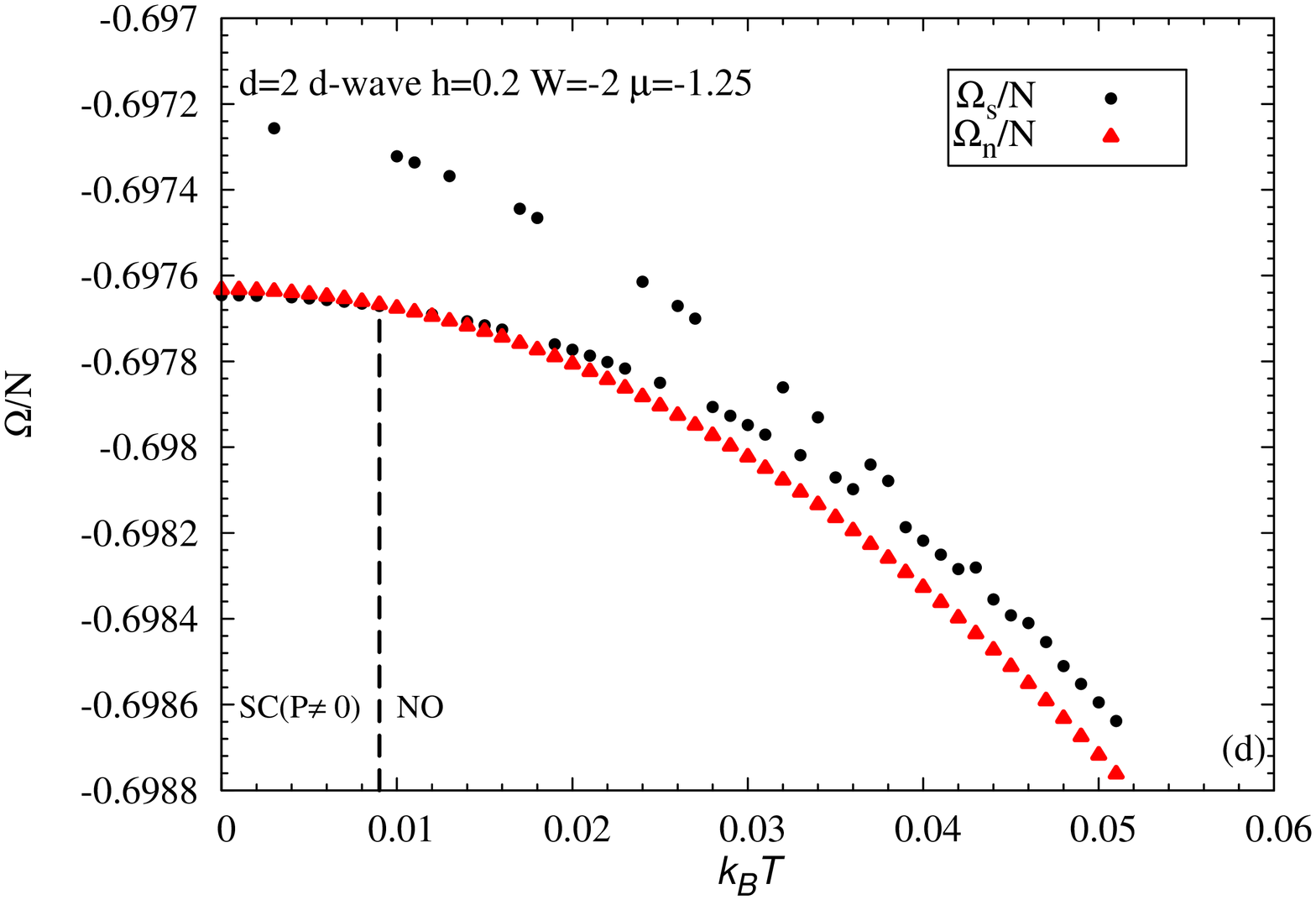}
\caption[Dependence of $\Delta_{\eta}$ on the magnetic field (a) and temperature
(b), \textcolor{green}{$W/t=-2$}, $U=0$, for a fixed $\mu=-1.25$. (c) The grand canonical potential
($\Omega$) vs. $h$ at $T=0$ (first order phase transition); (d) $\Omega$ vs. $T$
at $h=0.2$.]{\label{del_eta} Dependence of $\Delta_{\eta}$ on the magnetic field
(a) and temperature (b), \textcolor{green}{$W/t=-2$}, $U=0$, for a fixed $\mu=-1.25$. In Fig. (b),
for $h=0.16$ and $h=0.18$ the lower branches are unstable. For $h=0.2$ the
vertical dashed line denotes the first order phase transition from the
magnetized superconducting state (SC$(P\neq0)$) to the normal state (NO). (c)
The grand canonical potential ($\Omega$) vs. $h$ at $T=0$ (first order phase
transition); (d) $\Omega$ vs. $T$ at $h=0.2$.  
}
\end{figure}
One of the most important quantities related to superconductivity is the gap
parameter. As well known, the BCS theory predicts the existence of an
isotropic order parameter, which vanishes at the temperature of the
superconductor-normal phase transition. However, intensive studies of the gap
parameter for high-T$_c$ superconductors indicate significant
deviations from the predictions of the BCS theory. Accurate determination of
the gap parameter encounters many difficulties. Most of the measurements show
that its value in the ground state is much higher than the value of $\Delta$
in conventional superconductors \cite{Maple}. The width of the gap is usually
expressed in $k_B T$ units. For the superconductors to which the
BCS theory applies, this ratio is around 3.5, while for the high-T$_c$
superconductors it varies from 5 to 8, which corresponds to the energy gap
value of around 50 meV. The symmetry of the energy gap can be determined
from measurements of changes in its magnitude for different momentum
directions $|\Delta_{\vec{k}}|$. Most studies indicate the $d_{x^2-y^2}$ pairing
symmetry (with the energy gap $\Delta_{\vec{k}}=2\Delta_{\eta} (\cos (k_x)-\cos
(k_y))$) \cite{Shen, Tsuei, Harlingen, Annett, Bulut, Bulut-2}. 

The d-wave pairing symmetry is also very interesting from the point of view of
the breached pair (BP) state (or Sarma phase) in imbalanced ultracold Fermi
gases.  

As in section \ref{ground_state}, we start from the analysis of the influence of magnetic field on the order parameter characteristics. 

If $h=0$, in the case of the d-wave pairing symmetry, the energy gap vanishes
for some values of the wave vector $\vec{k}$, i.e. along the lines
$|k_x|=|k_y|$. Disappearance of the gap on the Fermi surface leads to the
existence of zero energy quasiparticles. Therefore, it is justified to believe
that the Sarma-type phases ($SC(P\neq 0)$) will be stable at $h\neq 0$, as opposed to the s-wave
pairing symmetry case in 2D.

Fig. \ref{del_eta} shows the dependence of the order parameter amplitude ($\Delta_{\eta}$) on magnetic field (a) and temperature (b), for $W=-2$, $\mu =-1.25$. 
As shown in Fig. \ref{del_eta}(a), there are two different solutions
for $\Delta_{\eta} \neq 0$, at $T=0$, as in the s-wave pairing symmetry case. 
However, as opposed to the isotropic order parameter case (Fig. \ref{fig1}), the
upper branch of the solutions of $\Delta_{\eta}$ is dependent on the magnetic
field in the ground state. Therefore\textcolor{green}{,} a finite polarization occurs in the system,
for an arbitrarily small value of the magnetic field, even at $T=0$. This is
explained by the creation of polarized excitations in the nodal points of the
gap \cite{Tempere'', Yang}. 
Moreover, this branch is stable up to $h\approx 0.2$. At this point, the first
order phase transition from the polarized superconducting to the normal state
occurs. On the other hand, the lower branch, which also depends on $h$, is
unstable at $T=0$.

Obviously, the polarization increases with increasing temperature. Thus, the
range of occurrence of the Sarma-type phase (the superconducting state with
$P\neq 0$) increases.

\begin{figure}
\begin{center}
\includegraphics*[width=0.55\textwidth,angle=270]{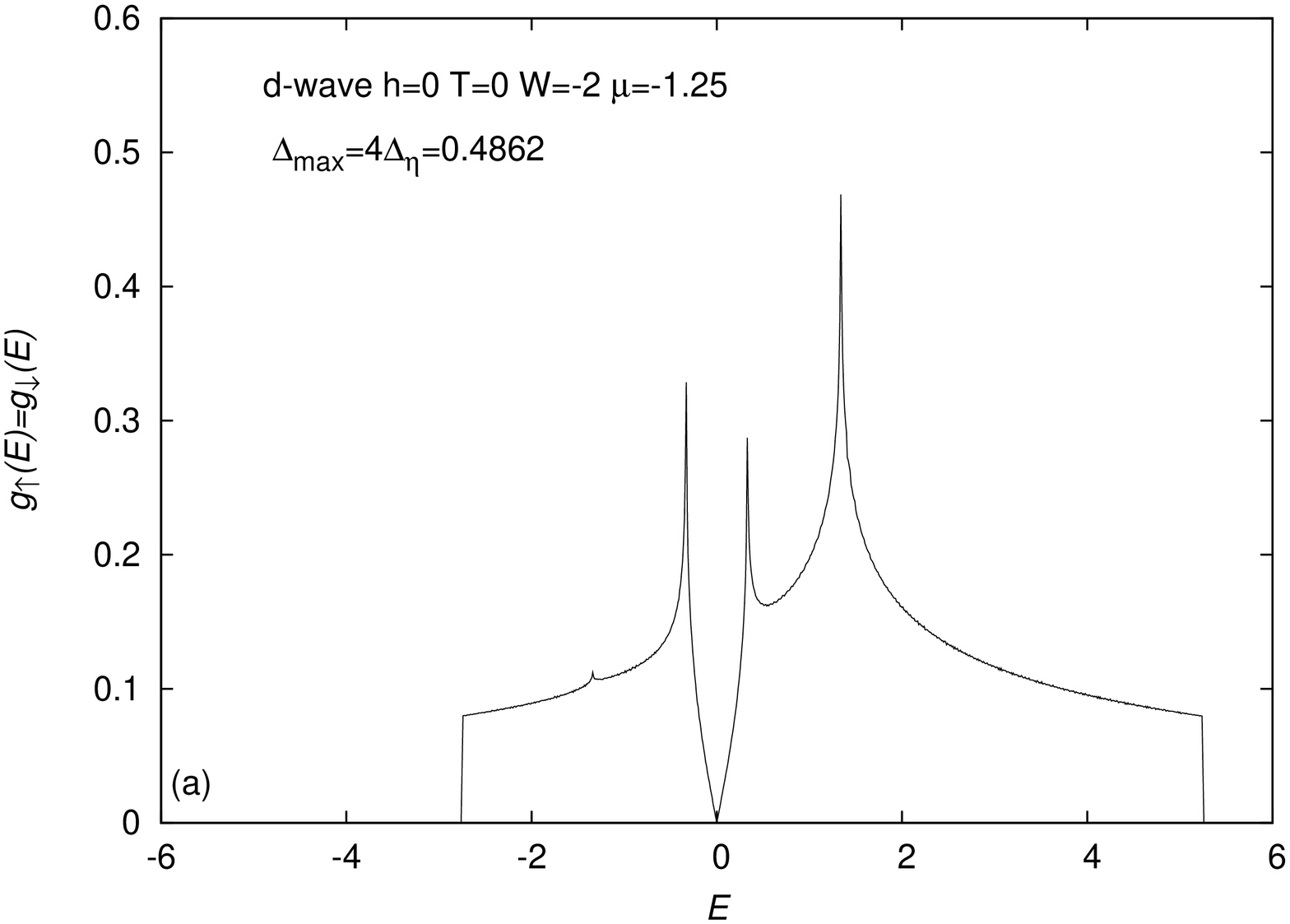}\hspace{-0.2cm}
\includegraphics*[width=0.55\textwidth,angle=270]{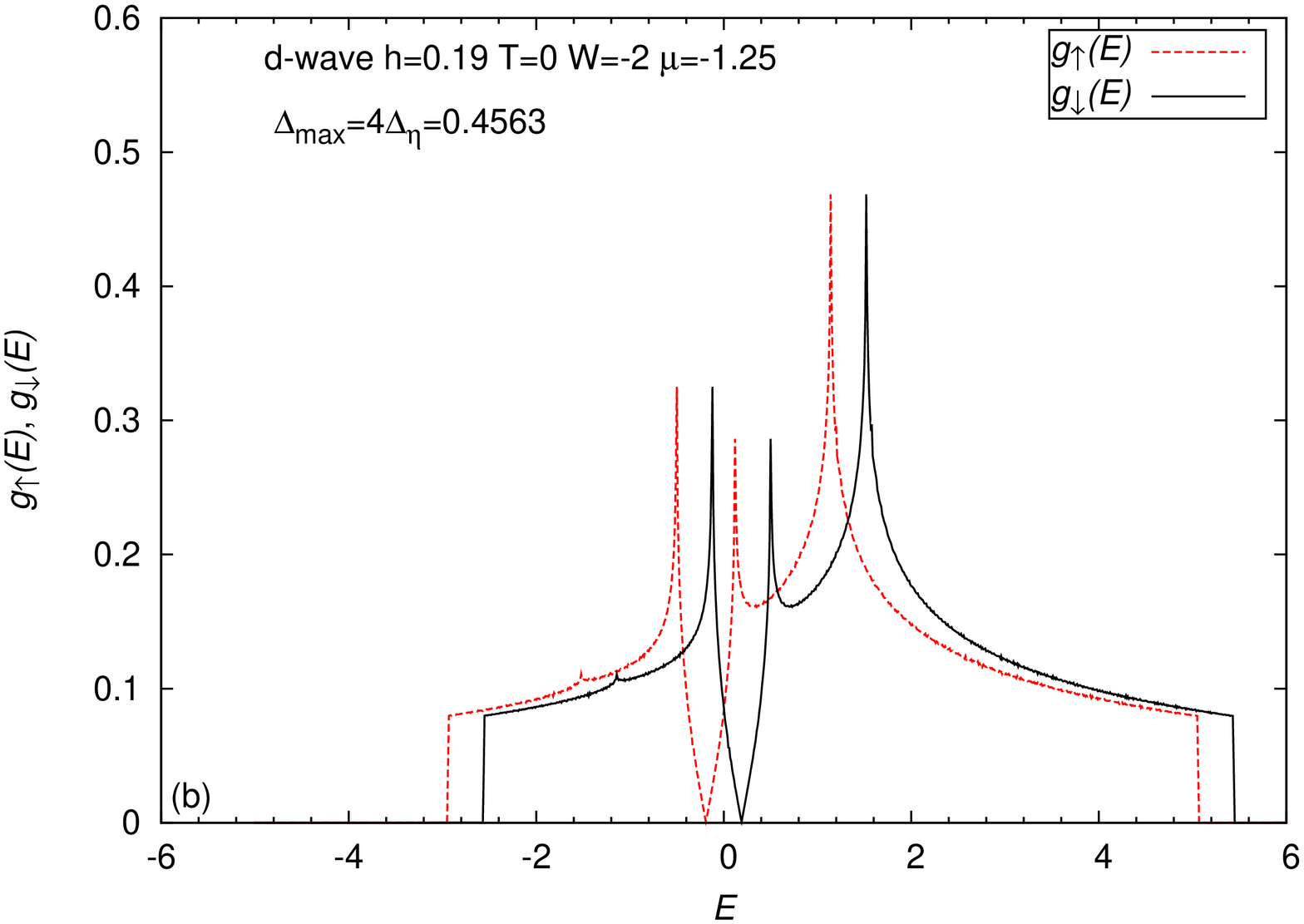}
\caption{\label{Density_d} Density of states for $W=-2$, $U=0$, $\mu=-1.25$, (a) $h=0$ and (b) $h=0.19$ at $T=0$.}
\end{center}
\end{figure}

As mentioned above, the d-wave pairing symmetry at $h=0$ is gapless
($\Delta_{\vec{k}}=\Delta_{\eta} \eta_{\vec{k}}=0$ at $|k_x|=|k_y|$) in four
nodal points on the Fermi surface \cite{zhou}. In the weak coupling limit, at
$h=0$, the nodal points are located at $\approx$ ($\pm \frac{\pi}{2}$, $\pm
\frac{\pi}{2}$). However, the gap ($E_g$) in the density of states is fixed by
the location of the logarithmic singularities. The value of $E_g$ is determined
by the maximum value of the energy gap: $E_g=2\Delta_{max}$, where
$\Delta_{max}=4\Delta_{\eta}$. An example of the density of states for the
d-wave pairing symmetry, at $h=0$ is shown in Fig. \ref{Density_d}(a).
At $h=0$, the quasiparticle energies (Fig. \ref{n_k_dwave}(a)) for spin-up and
spin-down are equal ($E_{\vec{k}\uparrow}=E_{\vec{k}\downarrow}
\equiv E_{\vec{k}}$) and the quasiparticle spectrum function has one zero
at the Fermi level. There is one (equal for both directions of spin) Fermi
surface in the system. The momentum distribution function is a step function, as
for the Fermi distribution in the ground
state. The step change of $n_{\vec{k}}$  occurs in the nodal points on the
diagonal of the Fermi surface. 


The influence of the Zeeman magnetic field on the density of states, shown in
Fig. \ref{Density_d}(b), is
analogous to that in the s-wave pairing symmetry case (see Fig.
\ref{fig3}),
i.e. $g_\uparrow(E)$ is shifted to the left and $g_\downarrow(E)$ to the right
by the value of the magnetic field $h$.
The corresponding quasiparticle energies are plotted in Fig. \ref{n_k_dwave}(b).
The gap appears in the spectrum for the minority spin
species and equals $h$, while the spectrum for the majority spin species is
gapless. The occurrence of the gap in $E_{\vec{k}\downarrow}$ is
caused by the existence of some minimum non-zero quasiparticle energy. 
The quasiparticle branch for the
majority spin species ($E_{\vec{k}\uparrow}$) has two zeros, which indicates
the occurrence of two Fermi surfaces in the system.
This is confirmed in the plots of the momentum distribution along the
diagonal $|k_x|=|k_y|$. If $h>0$, there appears a region where
$n_{\vec{k}\downarrow}$ has already vanished, while still
$n_{\vec{k}\uparrow}=1$, i.e. the Fermi momentum for the majority spin
species is higher than the one for the minority spin species.

\begin{figure}[h!]
\begin{center}
\includegraphics*[width=0.55\textwidth,angle=270]{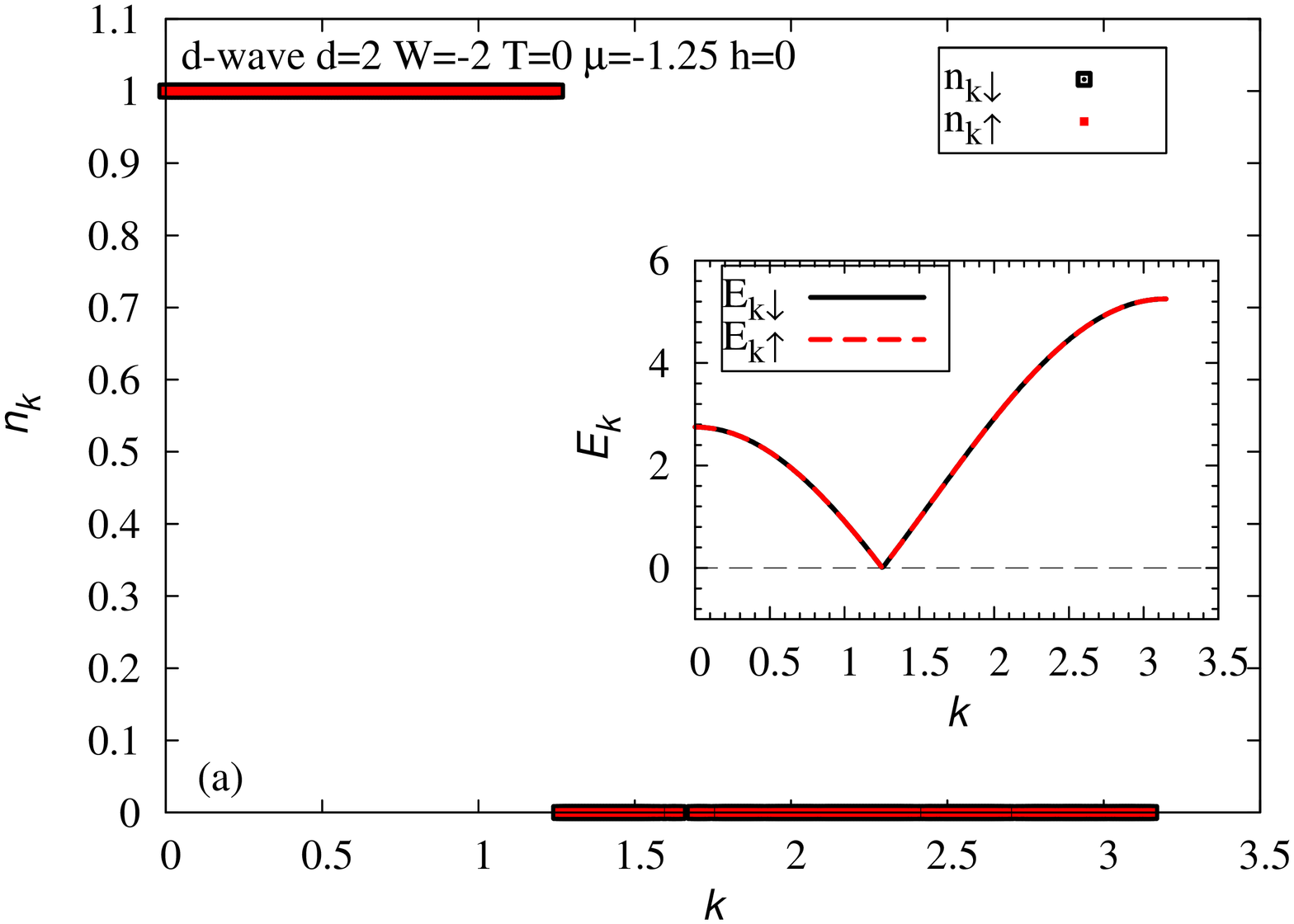}\hspace{-0.2cm}
\includegraphics*[width=0.55\textwidth,angle=270]{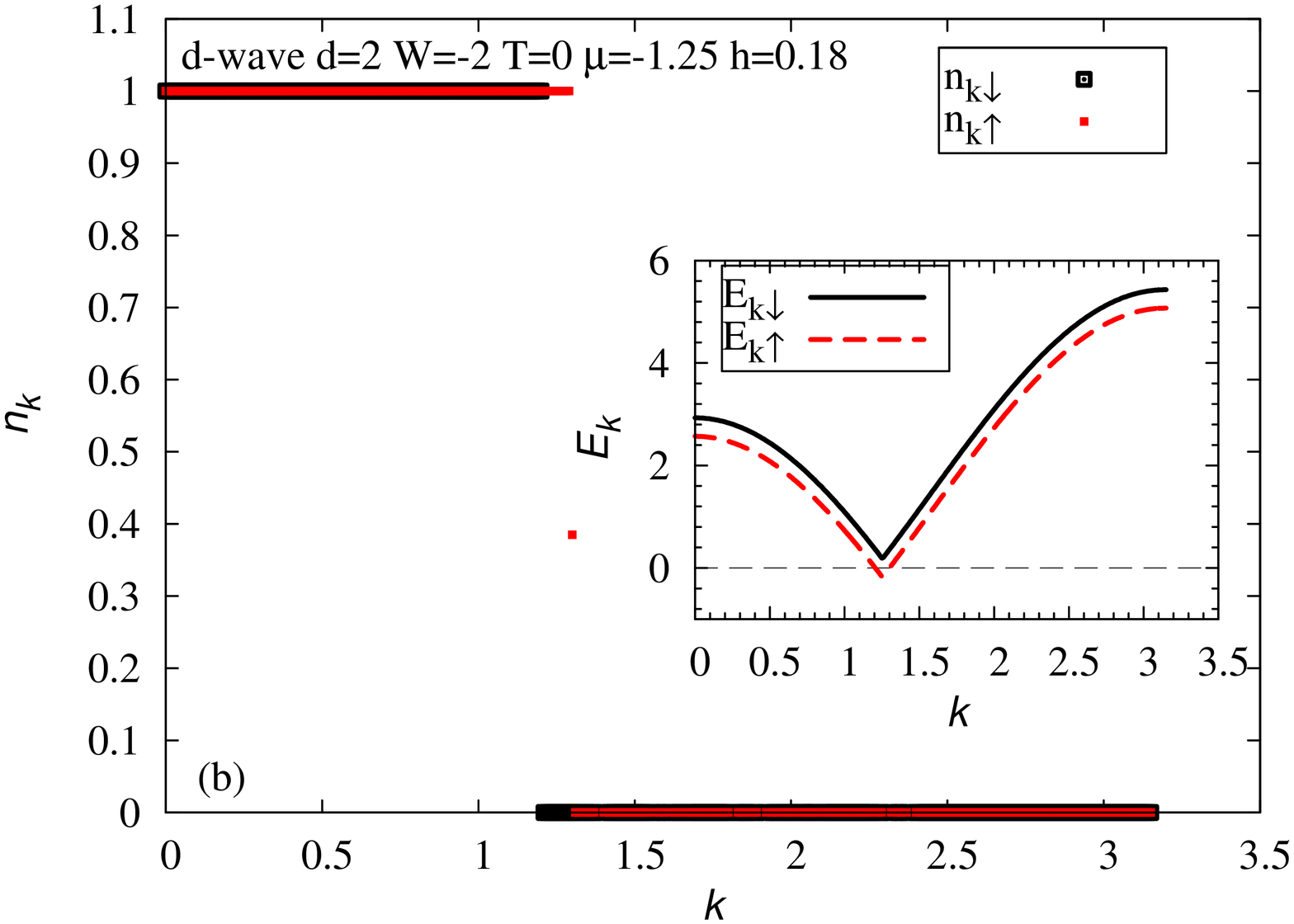}
\caption{\label{n_k_dwave} Plots of momentum occupation numbers
$n_{\vec{k}\uparrow}$ (red points), $n_{\vec{k}\downarrow}$ (black points) vs.
$k\equiv |k_x|=|k_y|$ and the corresponding quasiparticle spectra
$E_{\vec{k}\uparrow}$, $E_{\vec{k}\downarrow}$ (inset) for $W=-2$, $\mu=-1.25$,
$h=0$ (a) and $h=0.18$ (b).}
\end{center}
\end{figure}

Now, let us consider the ground state phase diagrams for the d-wave pairing
symmetry case, in the weak coupling regime. Both the fixed chemical potential
and the fixed electron concentration case are analyzed. The influence
of the Hartree term on the stability of the magnetized superfluid phases is not
considered.

In the s-wave case and in the weak coupling regime, the SC$_0$ superconducting
phase (with $P=0$) is stable at $T=0$. However, the analysis of the
d-wave order parameter behavior, the density of states and the momentum
distributions characteristics indicate the possibility of the occurrence of stable $SC(P\neq 0)$ phase at $T=0$, even in the weak coupling regime, as opposed to the s-wave
pairing symmetry case in 2D (Figs. \ref{mu_diagram}(a), \ref{n_diagram}(a)). As
shown above, for infinitesimally low value of the magnetic field, the $SC(P\neq 0)$
state is stable. Due to the vanishing of the gap at four nodal points on the
Fermi surface, there is a spin-polarized normal state, at $h\neq 0$. Therefore,
the $SC(P\neq 0)$ phase is the superfluid state of coexisting Cooper
pairs and excess fermions, with the latter responsible for finite
polarization (magnetization) and the gapless excitations characteristic for this
state. 

At higher values of the Zeeman magnetic field, $SC(P\neq 0)$ is destroyed by the
paramagnetic effect or by population imbalance. Then, there is the first order
phase transition from the polarized superconducting phase to the polarized
normal state \textcolor{green}{(Fig. \ref{diag_d-wave_T0}(a)-(b))}. The first order transition is manifested by the presence of the
phase separation (PS) region in the phase diagrams at fixed $n$ (see: Fig.
\ref{diag_d-wave_T0}(b)). The phase separation occurs between $SC (P\neq 0)$ with the
number of particles $n_s$ and NO with the number of particles $n_n$. It is worth
mentioning that the d-wave superfluidity is stable around the half-filled
band in the weak coupling limit and its range of occurrence widens with
increasing attractive interaction.  

\begin{figure}
\begin{center}
\includegraphics*[width=0.55\textwidth,angle=270]{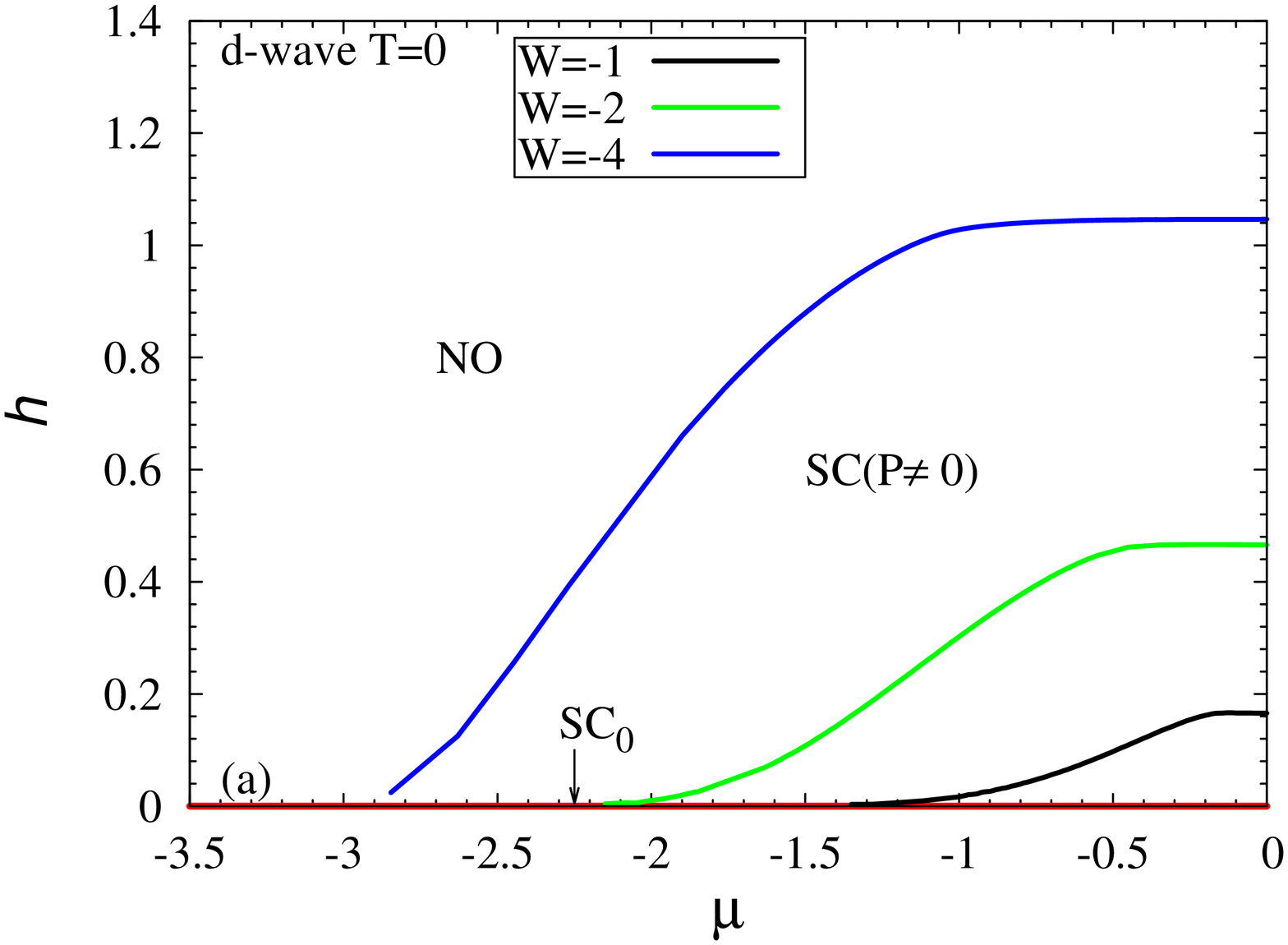}\hspace{-0.2cm}
\includegraphics*[width=0.55\textwidth,angle=270]{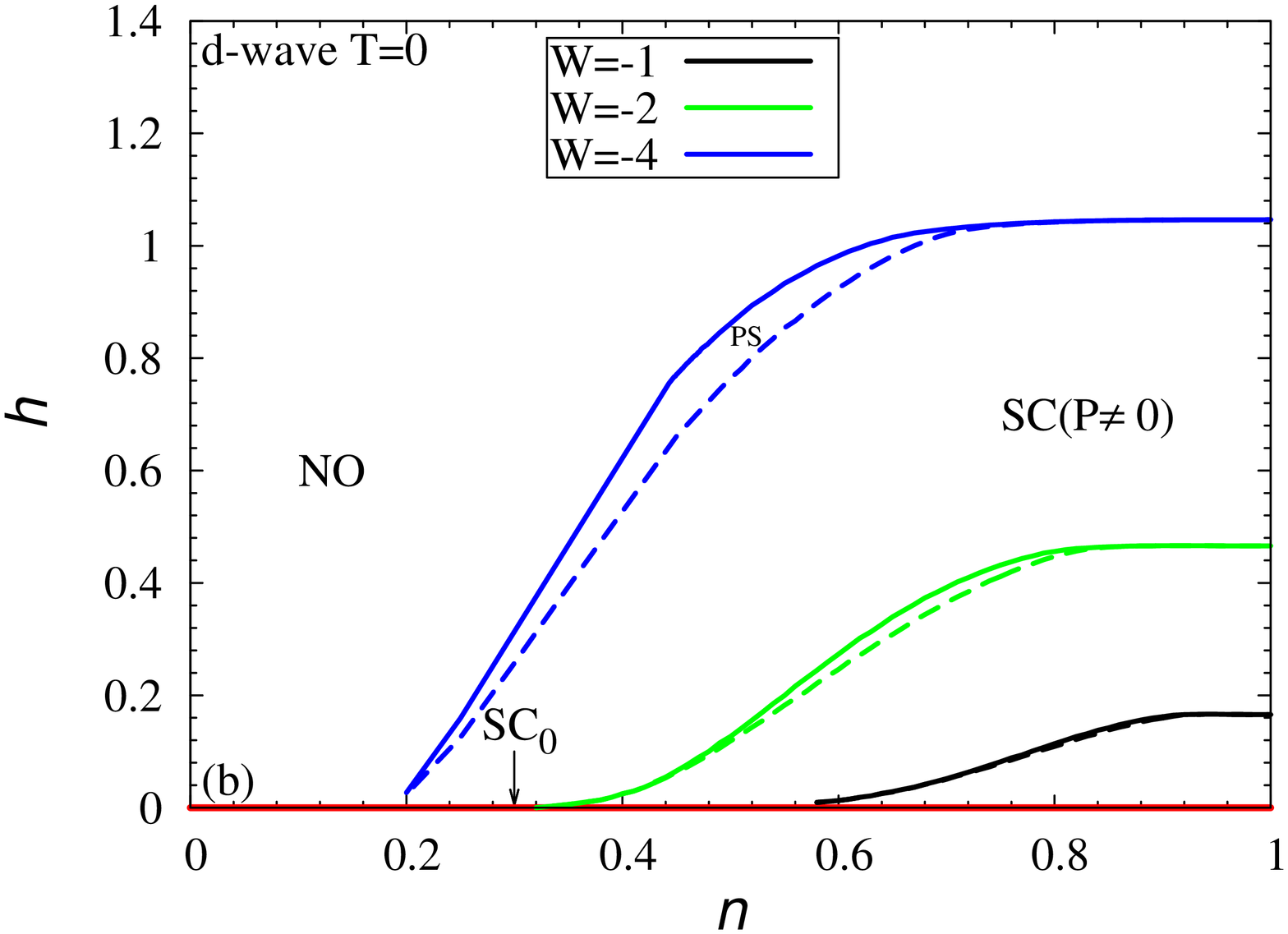}
\caption{\label{diag_d-wave_T0} Critical magnetic field vs. chemical
potential (a) and electron concentration (b) for the first order SC$(P\neq
0)$-NO transition, at $T=0$; three different values of the attractive
interaction.}
\end{center}
\end{figure}

Let us discuss the finite temperature phase diagrams. However, first let us
analyze briefly the influence of the Zeeman magnetic field on the superfluid
density characteristics. Fig. \ref{ro_s-dwave} shows the dependence of $\rho_s$
on temperature (a) and magnetic field (b), for the d-wave pairing symmetry,
$W=-2$ and $\mu=-1.25$. As mentioned before, the superfluid stiffness is the
largest in the ground state and drops to zero at $T_c^{HF}$. The pairing
symmetry plays the crucial role in the temperature dependences. For the d-wave
pairing symmetry, at $h=0$, $\rho_s$ decreases linearly. This dependence is
significantly different for the s-wave pairing symmetry case (Fig.
\ref{ro_s-2D}(a)).

\begin{figure}[t!]
\hspace*{-0.8cm}
\includegraphics*[width=0.38\textwidth,angle=270]{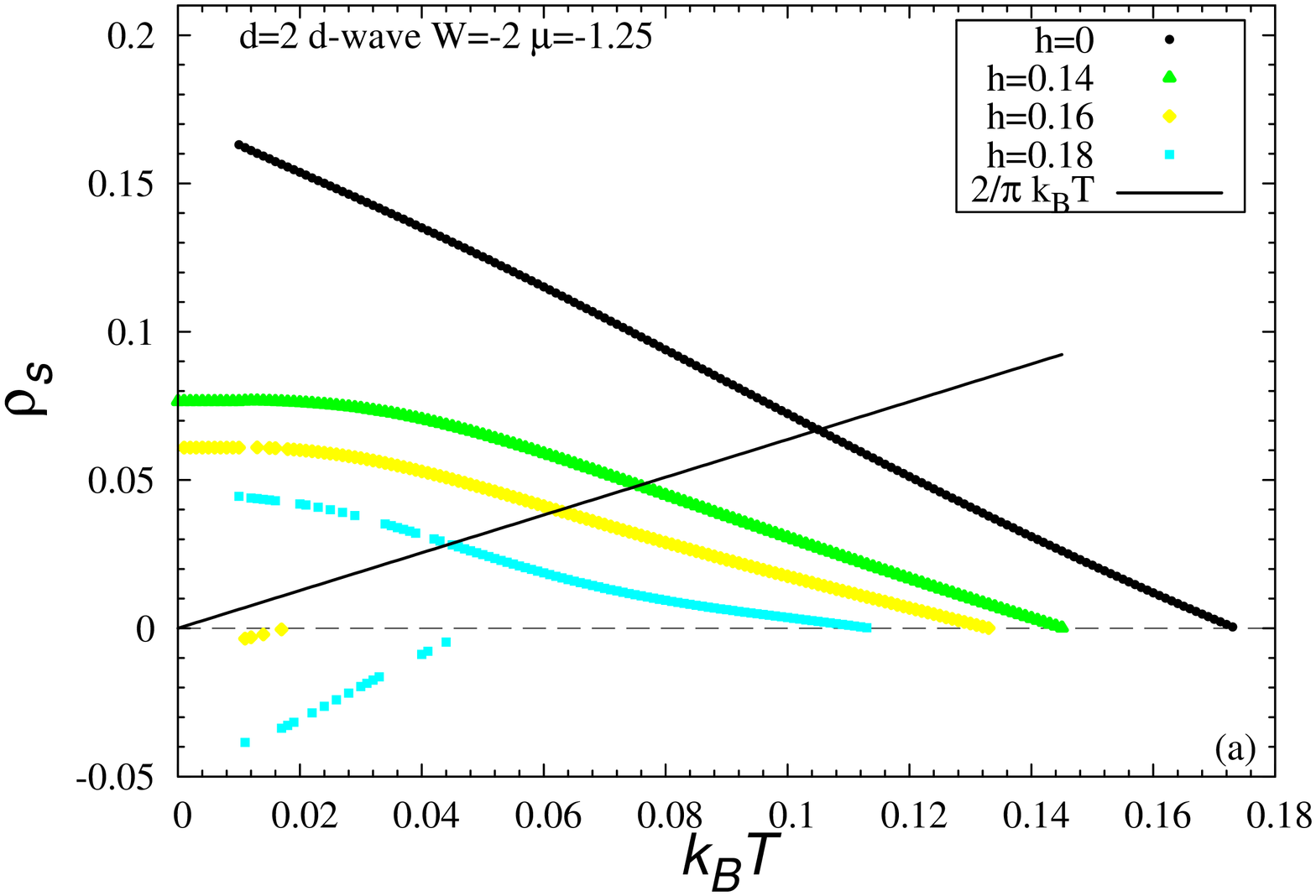}
\hspace*{-0.6cm}
\includegraphics*[width=0.38\textwidth,angle=270]{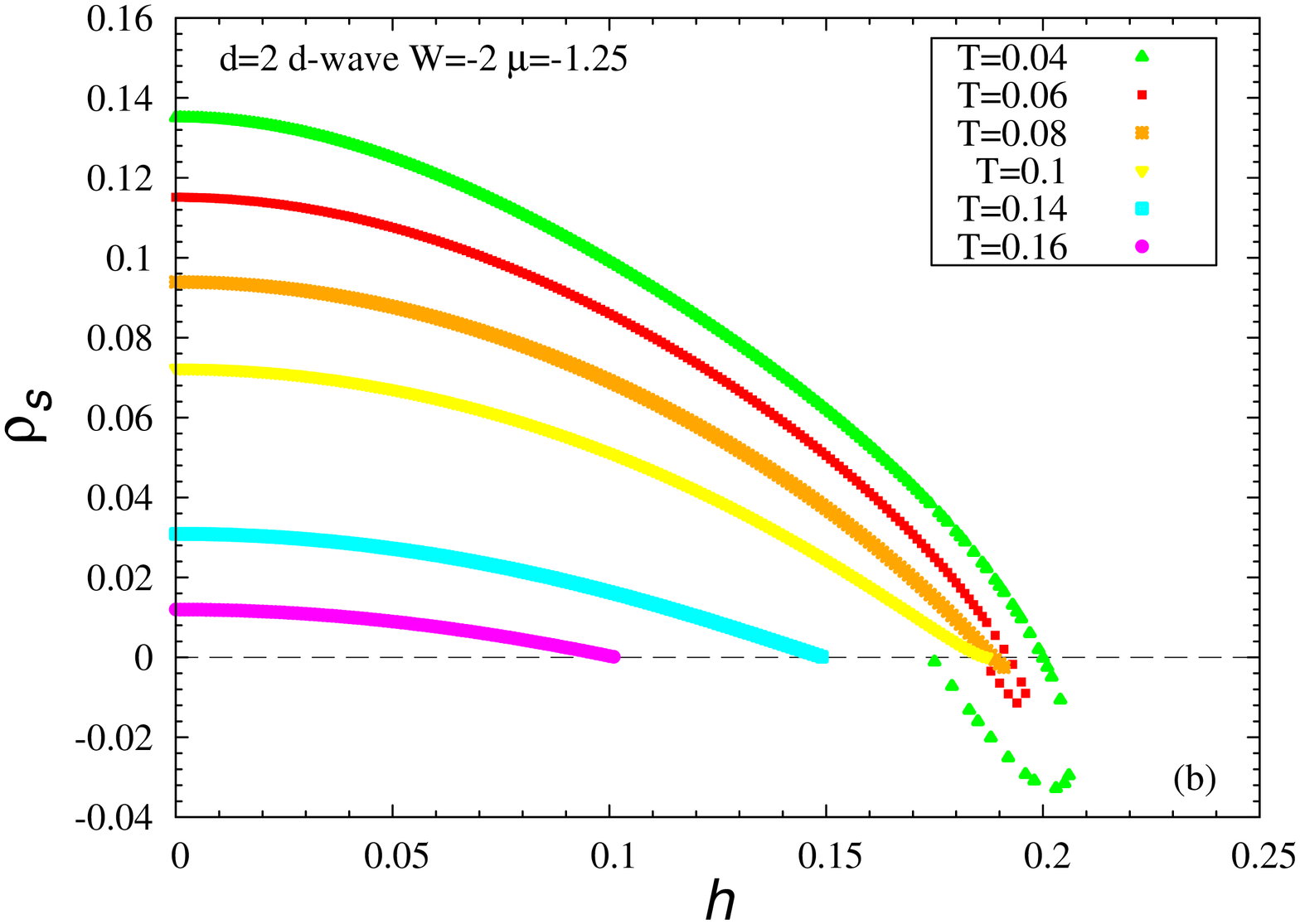}
\caption[Superfluid density vs. temperature (a) and magnetic field (b) for 2D d-wave pairing symmetry, $\mu =-1.25$, $W=-2$, $U=0$.]{\label{ro_s-dwave} Superfluid density vs. temperature (a) and magnetic field (b) for 2D d-wave pairing symmetry, $\mu =-1.25$, $W=-2$, $U=0$. The Kosterlitz-Thouless temperatures in 2D are found from the intersection point of the straight line $\frac{2}{\pi} k_B T$ with the curve $\rho_s(T)$.}
\end{figure}

The magnetic field has a clear influence on the temperature characteristics of
the superfluid stiffness. The example dependences, for some chosen
values of $h$, are presented in Fig. \ref{ro_s-dwave}(a). One can see that the
differences in the shapes of the plots for s-wave and d-wave pairing symmetries
decrease at $h\neq 0$. The linearity of the plot vanishes for the d-wave pairing
symmetry, in the presence of the magnetic field.

As mentioned above, the quasiparticle energy can be zero, for the d-wave
pairing symmetry. It takes place when $\Delta_{\vec{k}}=0$ on the Fermi surface
(nodal points). The existence of the zero-energy excitations can increase the
value of $\rho_s^{para}$ significantly, because of the presence of the
derivatives in Eq.~\eqref{ro_s_para}. In this case, one observes a sudden
decrease of $\rho_s$ and the temperature dependence has a linear character.
However, if $h\neq 0$, the excitations in the system have some minimum non-zero
energy for the minority spin species ($E_{\vec{k}\downarrow}$) and there is no
significant decrease of $\rho_s (T)$ at low temperatures.

As mentioned in the introduction to this thesis, the impact of the
Zeeman coupling between the spins of the electrons and the applied magnetic
field on
superconductivity has been being analyzed for many years \cite{clogston, Fulde,
Larkin, Sarma, gruenberg}. One of the well-known results concerning this
influence is the existence of the \emph{so-called} Chandrasekhar-Clogston limit. In the weak
coupling regime, for the s-wave pairing symmetry case, at $T=0$, the
superconductivity is destroyed through the paramagnetic effect and the first
order phase transition to the normal state at a universal value of the critical
magnetic field $h_{c}=\Delta_0/\sqrt{2}\approx 0.707\Delta_0$, where $\Delta_0$
is, as usual, the gap at $T=0$ and $h=0$. In turn, this universal value of the
magnetic field in which the superconducting state is destroyed in the ground
state, for the d-wave pairing symmetry case is:
$h_{c}^{d-wave}=0.56\Delta_{max}$ \cite{Yang}, where
$\Delta_{max}=4\Delta_{\eta}$ at $T=0$ and $h=0$. 

Fig. \ref{diag_T_d-wave} shows the temperature vs. magnetic field ($T-h$) and
polarization ($T-P$) phase diagrams for $W=-2$ and three values of the chemical
potential. These fixed values of $\mu$ correspond to lower values of $n$ than
$n=1$, therefore the d-wave Chandrasekhar-Clogston limit is not reached. However, our results
for $\mu=0$ ($n=1$) agree with the ones from the paper \cite{Yang}, i.e. indeed
$h_{c}^{d-wave}=0.56\Delta_{max}$ for this case.

We take into account the phase fluctuations in $d=2$ within the KT scenario, as
in the s-wave pairing symmetry case (Figs.: \ref{diag_T_fixed_mu}(a)-(b),
\ref{diag_T_fixed_n}(a)-(b), \ref{diag_U-1}). In such way, we can estimate the
phase coherence temperatures, in addition to the MF temperatures. The solid
lines (2$^{nd}$ order transition lines) and the PS region are obtained within
the MF approximation. The curves below the first order phase transition lines on
the phase diagrams (the thin dotted lines) are merely extensions of the
2$^{nd}$ order transition lines below tricritical points. The thick dash-double
dotted lines denote the KT transition. The system is a quasi superconductor
(qSC) below $T_c^{KT}$. Between $T_c^{KT}$ and $T_c^{HF}$ pairs still exist, but
without a long-range phase coherence (the pseudogap behavior). The KT
temperatures are much smaller than $T_c^{HF}$. It can be seen particularly
clearly for fixed $\mu=-0.5$ or $n=0.75$ (Fig. \ref{diag_T_d-wave-n}) and $W=-2$
cases -- the difference between $T_c^{KT}$ and $T_c^{HF}$ amounts to nearly
50\%. However, this difference decreases with decreasing $\mu$ (decreasing $n$)
and decreasing attraction (in the weak coupling limit). Both for fixed $\mu$
and fixed $n$, a finite temperature second order phase transition takes place
from the pairing without coherence region to the normal state at sufficiently
low values of the magnetic field. With increasing $h$, the character of the
transition between the pairing without coherence region and the normal state
changes from the second to the first order, which is manifested by the existence
of the MF tricritical point on the phase diagrams. In this MF TCP, the state of
incoherent pairs, the normal phase and the phase separation region (in the case
of fixed $n$) coexist. Therefore, the topology of the ($T-h$) diagrams is the
same as in the s-wave case.

\begin{figure}
\begin{center}
\includegraphics[width=0.55\textwidth,angle=270]{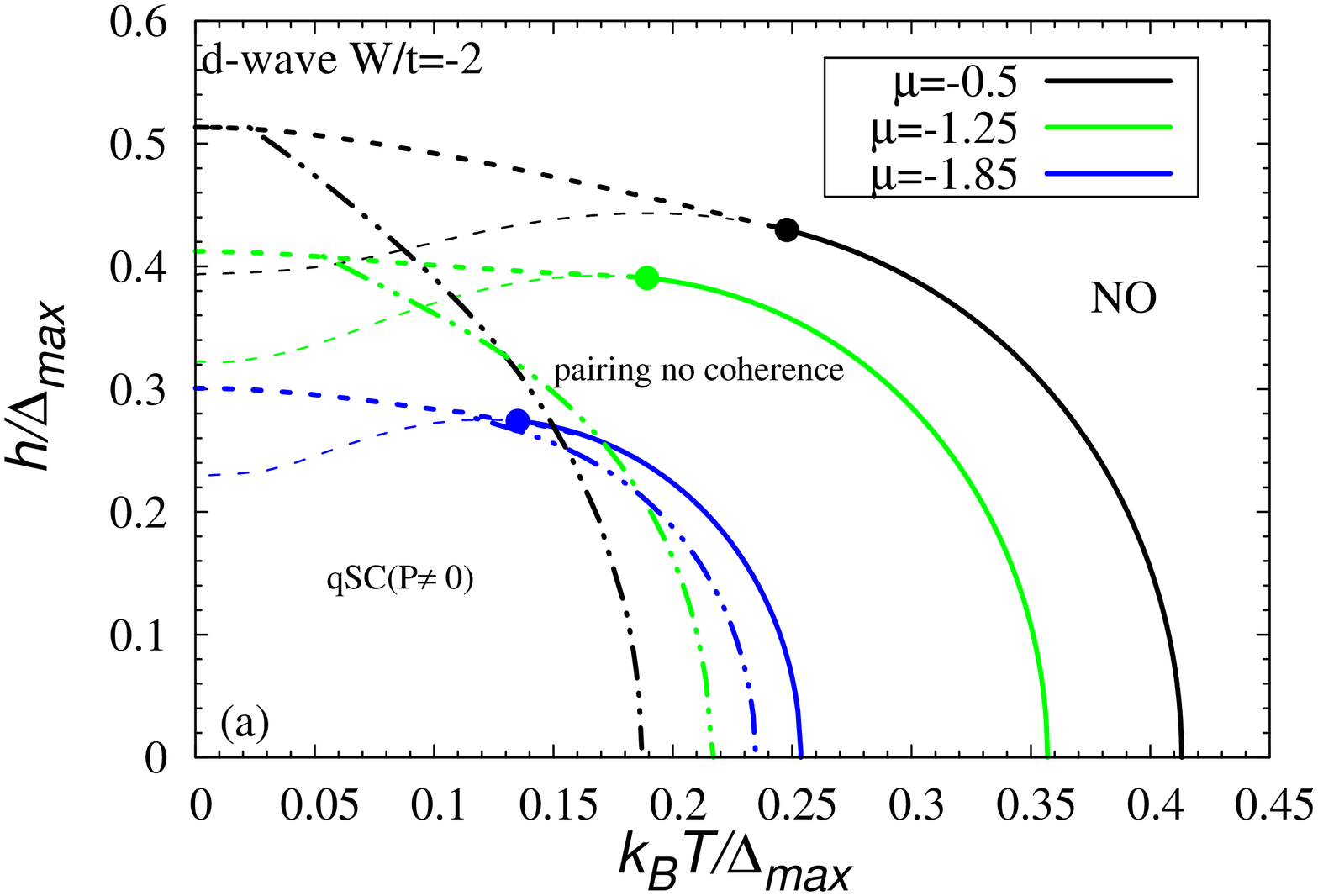}\hspace{-0.2cm}
\includegraphics[width=0.55\textwidth,angle=270]{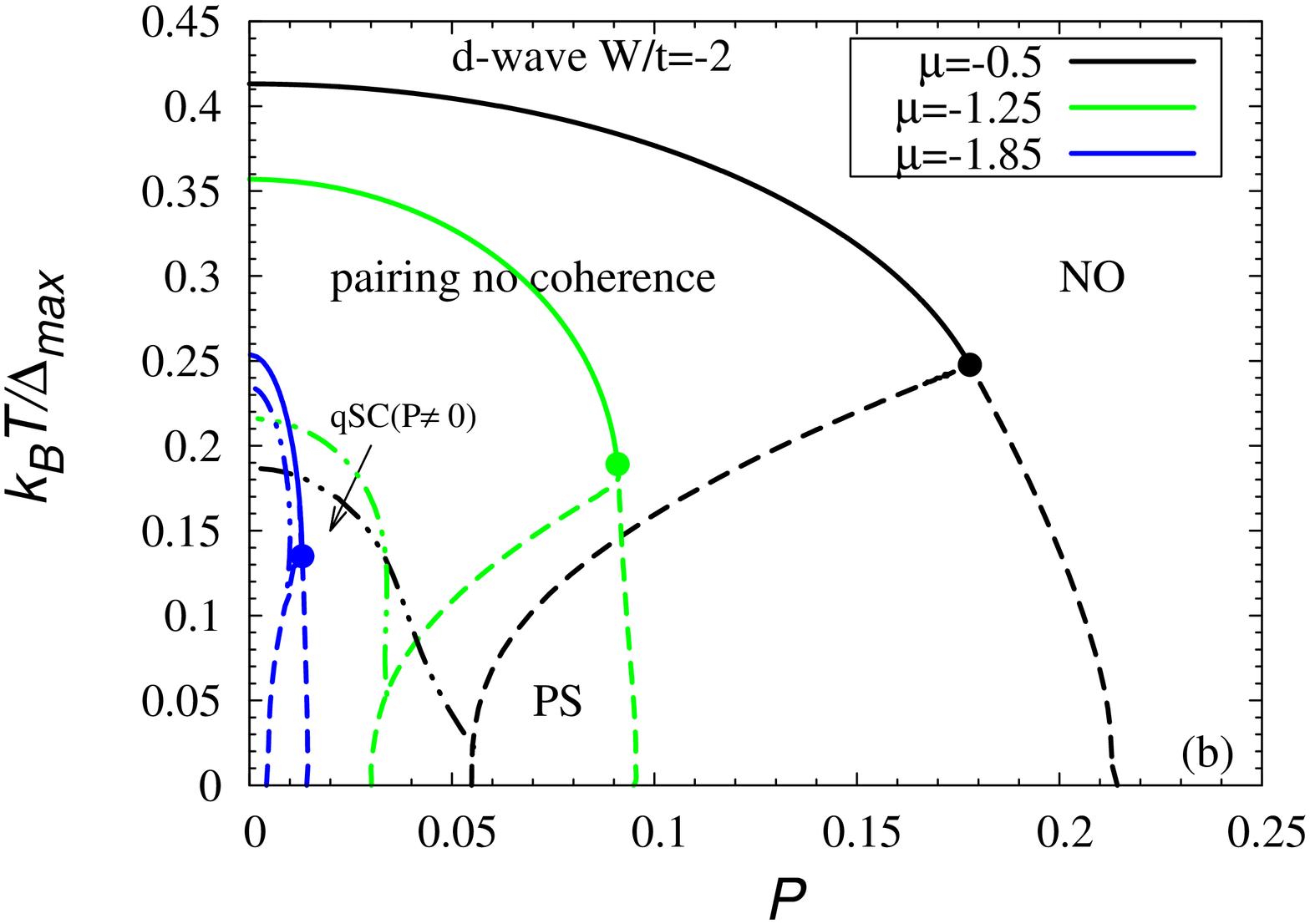}
\caption[Temperature vs. magnetic field (a) and polarization (b) phase diagrams
for $W=-2$, $U=0$, three values of $\mu$.]{\label{diag_T_d-wave} Temperature vs.
magnetic field (a) and polarization (b) phase diagrams for $W=-2$, $U=0$, three
values of $\mu$; SC$_{0}$ -- non-polarized superconducting state ($P=0$),
SC$(P\neq 0)$ -- 2D superconductor in the presence of polarization, NO --
partially-polarized normal state. The thick solid line is the second order phase
transition line from the pairing without coherence region to NO. The thin dashed
line is an extension of the $2^{nd}$ order transition line (metastable
solutions). The thick dashed-double dotted line is the Kosterlitz-Thouless
transition line. The thick dotted line denotes the first order phase transition
to NO. $\Delta_{max}=4\Delta_{\eta}$ denotes the gap at $T=0$ and $h=0$. }
\end{center}
\end{figure}

\begin{figure}
\begin{center}
\includegraphics[width=0.55\textwidth,angle=270]{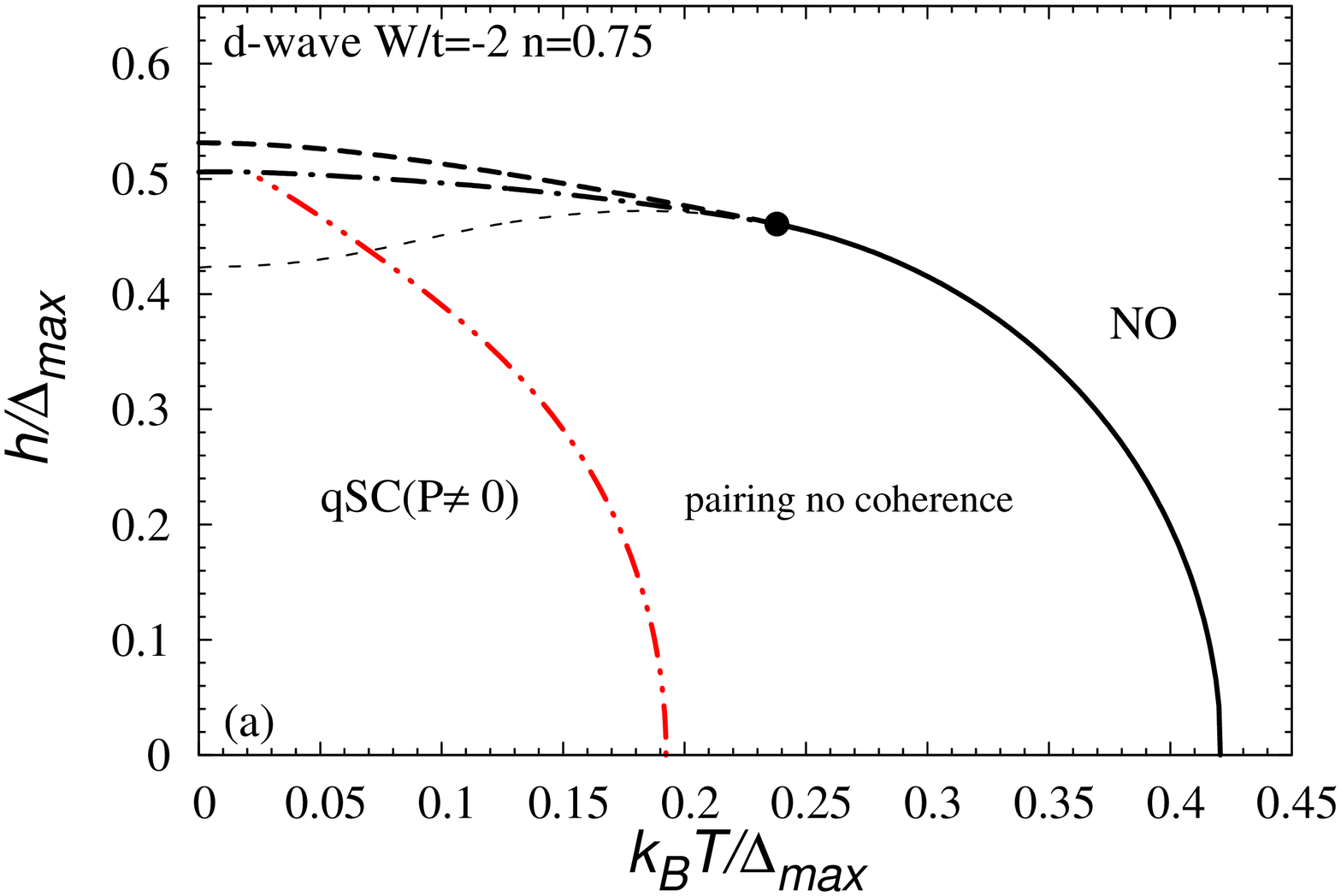}\hspace{-0.2cm}
\includegraphics[width=0.55\textwidth,angle=270]{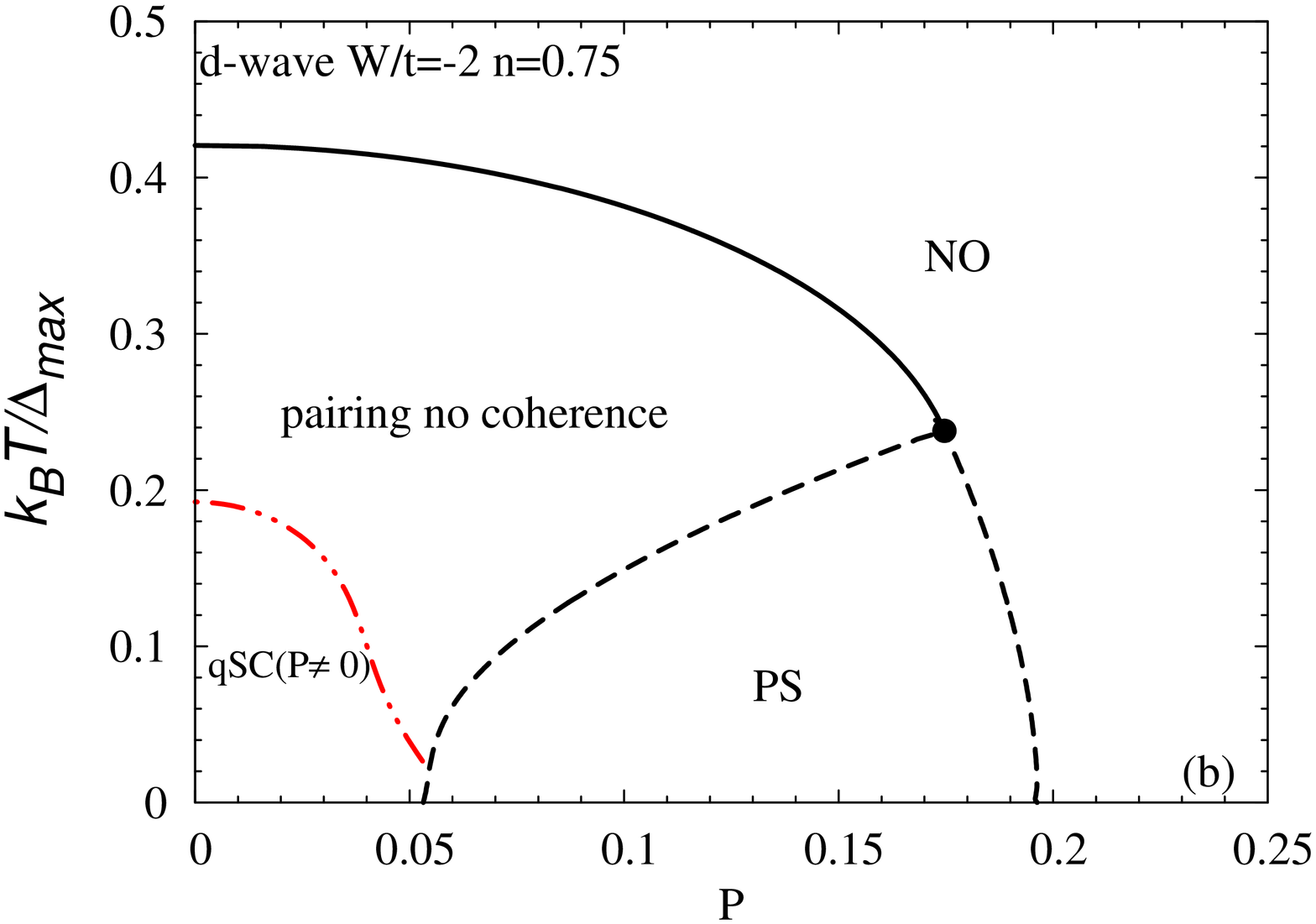}
\caption[Temperature vs. magnetic field (a) and polarization (b) phase diagrams
for $W=-2$, $U=0$, $n=0.75$.]{\label{diag_T_d-wave-n} Temperature vs. magnetic
field (a) and polarization (b) phase diagrams for $W=-2$, $U=0$, $n=0.75$;
SC$_{0}$ -- non-polarized superconducting state ($P=0$), SC$(P\neq 0)$ -- 2D
superconductor in the presence of polarization, NO -- partially-polarized
normal state. The thick solid line is the second order phase transition line
from the pairing without coherence region to NO. The thin dashed line is an
extension of the $2^{nd}$ order transition line (metastable solutions). The
thick dashed-double dotted line is the Kosterlitz-Thouless transition line. The
thick dotted line denotes the first order phase transition to NO.
$\Delta_{max}=4\Delta_{\eta}$ denotes the gap at $T=0$ and $h=0$.}
\end{center}
\end{figure}

However, qualitative differences between the s-wave and d-wave pairing
symmetries are clearly visible in $(P-T)$ phase diagrams. At $T\geq 0$ and
$P=0$, there is the unpolarized superconducting phase, both for the s-wave and
the d-wave pairing symmetry case. At $T=0$, there is only the PS region, for the
whole range of polarizations, i.e. $P>0$, for the isotropic order parameter
case. In turn, in the d-wave pairing symmetry case, there is the spin-polarized
superconducting phase ($SC(P\neq 0)$) at $T=0$, up to some critical value of the
polarization, for which the first order phase transition to the normal state
takes place. In the PS region, not only the polarizations, but also the particle
densities in SC and NO are different. At $T=0$ and for the s-wave pairing
symmetry, this separation region is between the non-polarized superconducting
phase and the normal state, while in the d-wave pairing symmetry it is between
$SC (P\neq 0)$ and NO. At $T \neq 0$, $\Delta \neq 0$ and $P\neq 0$, the system is also
in the polarized qSC phase (i.e. homogeneous superconductivity in the presence
of the spin polarization) in the s-wave pairing symmetry case up to $T_c^{KT}$.
The KT phase is restricted to the weak coupling region and low values of $P$, as
in the d-wave pairing symmetry case. Increasing polarization favors the phase of
incoherent pairs. As shown in Fig. \ref{diag_T_d-wave}(b), the range of
occurrence of qSC in the presence of $P$ widens in the weak coupling regime with
increasing $n$ (increasing $\mu$). In the s-wave pairing symmetry case, one can
distinguish the gapless region at sufficiently high values of the magnetic
field and temperature. As mentioned before, the d-wave pairing symmetry at
$h=0$ is gapless ($\Delta_{\vec{k}}=\Delta_{\eta} \eta_{\vec{k}}=0$ at
$|k_x|=|k_y|$) in four nodal points on the Fermi surface. Therefore, a natural
consequence of this is the occurrence of the gapless region also for
infinitesimally low values of $h$, even at $T=0$.

\newpage
\thispagestyle{empty}
\mbox{}

\chapter{The BCS-BEC crossover at $T=0$ in the \textcolor{czerwony}{s}pin-\textcolor{czerwony}{p}olarized Attractive Hubbard Model with spin independent hopping integrals}
\label{chapter5}
In this chapter, we analyze the influence of magnetic field on the BCS-LP
(BEC) crossover at $T=0$, for $d=2$ (square lattice) and $d=3$ (simple cubic
lattice), within the spin-polarized \textcolor{czerwony}{AHM}.
Development of experimental techniques in cold atomic Fermi gases with tunable
attractive interactions (through Feshbach resonance) has allowed the study of
the BCS-BEC crossover and the properties of exotic states in these systems. 

In the strong coupling limit of AHM the tightly bound local pairs of fermions behave
as hard-core bosons (see also Chapter \ref{6.0} \textcolor{czerwony}{and 7}) and can exhibit a superfluid state similar to that of
$^{4}\textrm{He}$ II \cite{MicnasModern}. The evolution from the weak attraction
(BCS-like) limit to that of the strong attraction (LP) takes place when
the
interaction is increased or the electron concentration is decreased at moderate
fixed attraction. According to the Leggett criterion \cite{leggett}, the Bose
regime begins when the modified chemical potential $\bar{\mu}$ drops below the
lower band edge (the limiting value is $\bar{\mu}/t=-4$ and $\bar{\mu}/t=-6$, at
$h=0$, for $d=2$ and $d=3$, respectively). 

For spin independent hopping integrals ($t^{\uparrow}=t^{\downarrow}$, $r=1$),
at $T=0$ we find no magnetized superconducting phase in the strong attraction
limit in the two dimensional case -- PS is energetically favorable. However, for
strong attraction and in the dilute limit we show that the existence of the
homogeneous magnetized superconducting phase (SC$_M$) is possible for 3D
case. The SC$_M$ phase  is a specific
superfluid state, being a coherent mixture of LP's (hard-core bosons) and excess
of spin-up fermions. We also briefly discuss the influence of different lattice
geometries (or densities of states) on the stability of the SC$_M$ phase. Some
of our results have been published in Refs. \cite{Kujawa3, Kujawa4}.

\section{2D square lattice}
\begin{figure}[t!]
\hspace*{-0.8cm}
\includegraphics[width=0.38\textwidth,angle=270]{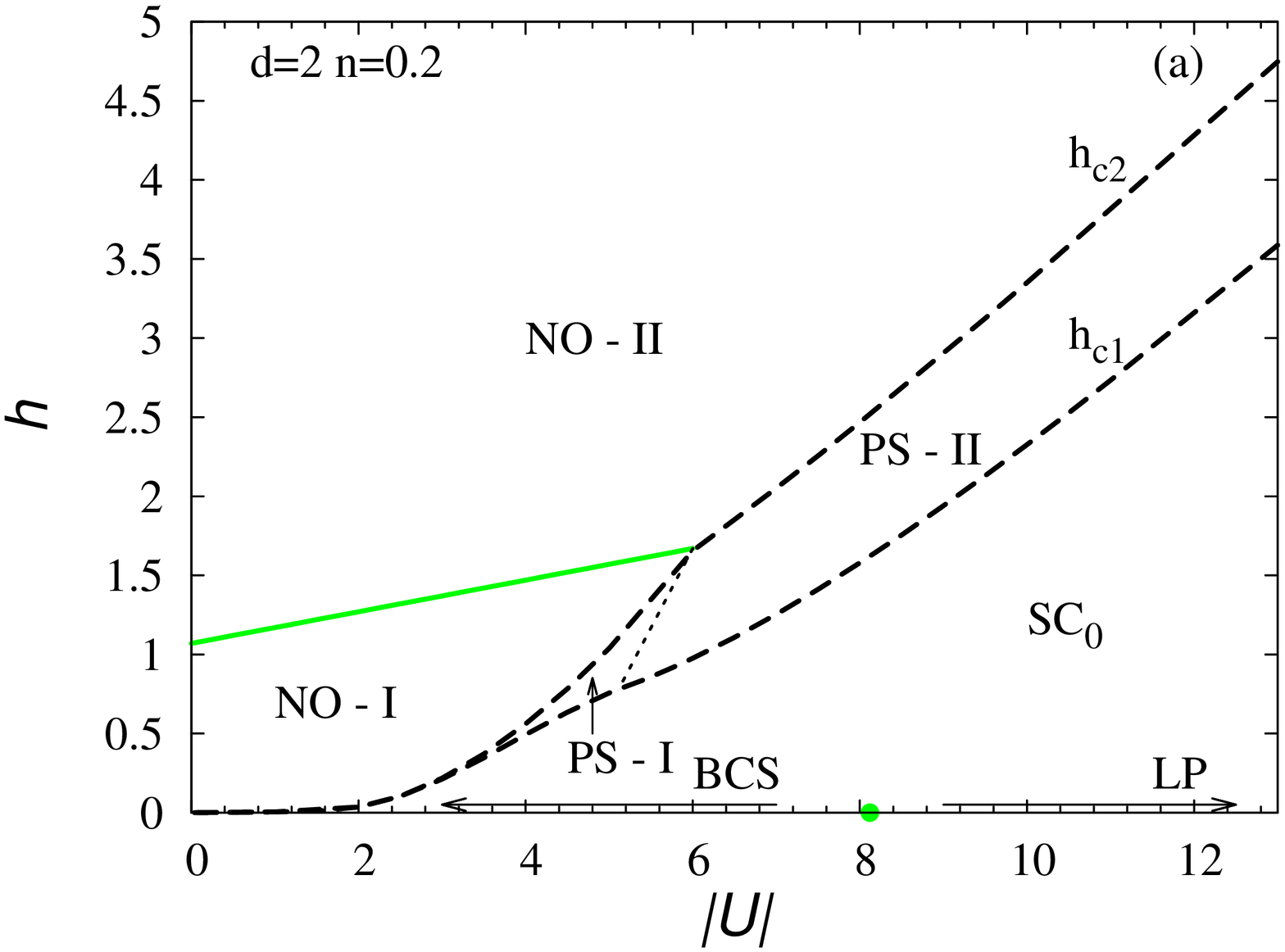}
\hspace*{-0.6cm}
\includegraphics[width=0.38\textwidth,angle=270]{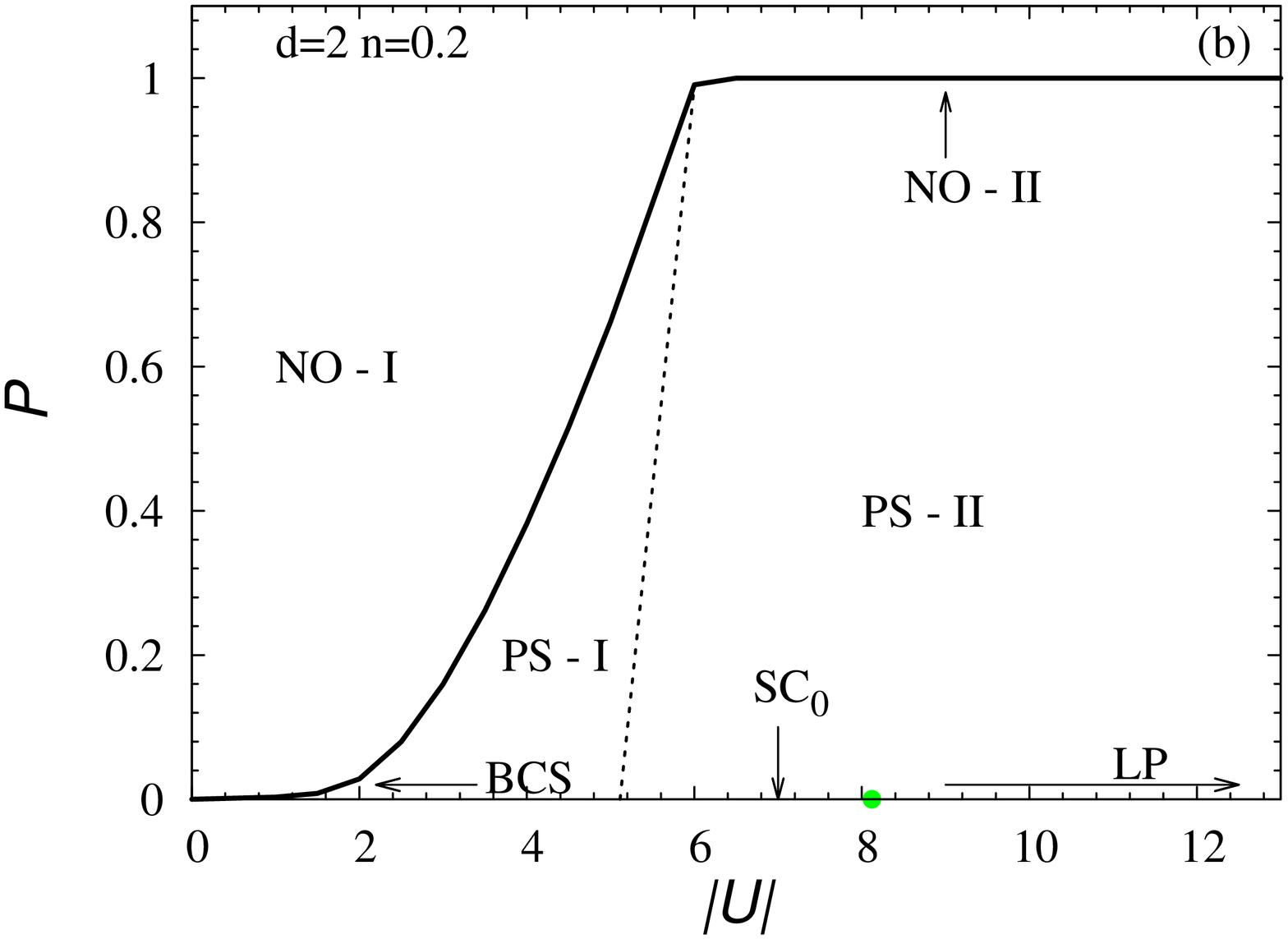}
\caption[Magnetic field vs. on-site attraction (a) and polarization (b) phase
diagrams, at $T=0$ and fixed $n=0.2$, for the square
lattice.]{\label{2D_n02_crossover} Magnetic field vs. on-site attraction (a) and
polarization (b) phase diagrams, at $T=0$ and fixed $n=0.2$, for the square
lattice. SC$_0$ -- unpolarized superconducting state with
$n_{\uparrow}=n_{\downarrow}$, LP -- tightly bound local pairs. Green solid line
separates partially polarized (NO-I) and fully polarized (NO-II) normal states.
PS-I (SC$_0$+NO-I) -- partially polarized phase separation, PS-II (SC$_0$+NO-II)
-- fully polarized phase separation, $h_{c1}$, $h_{c2}$ -- critical fields
defining the PS region. \textcolor{czerwony}{The green point in (a)-(b) shows} the BCS-LP crossover point
($U/t\approx -8.12$).}
\end{figure}

\begin{figure}
\begin{center}
\includegraphics[width=0.55\textwidth,angle=270]{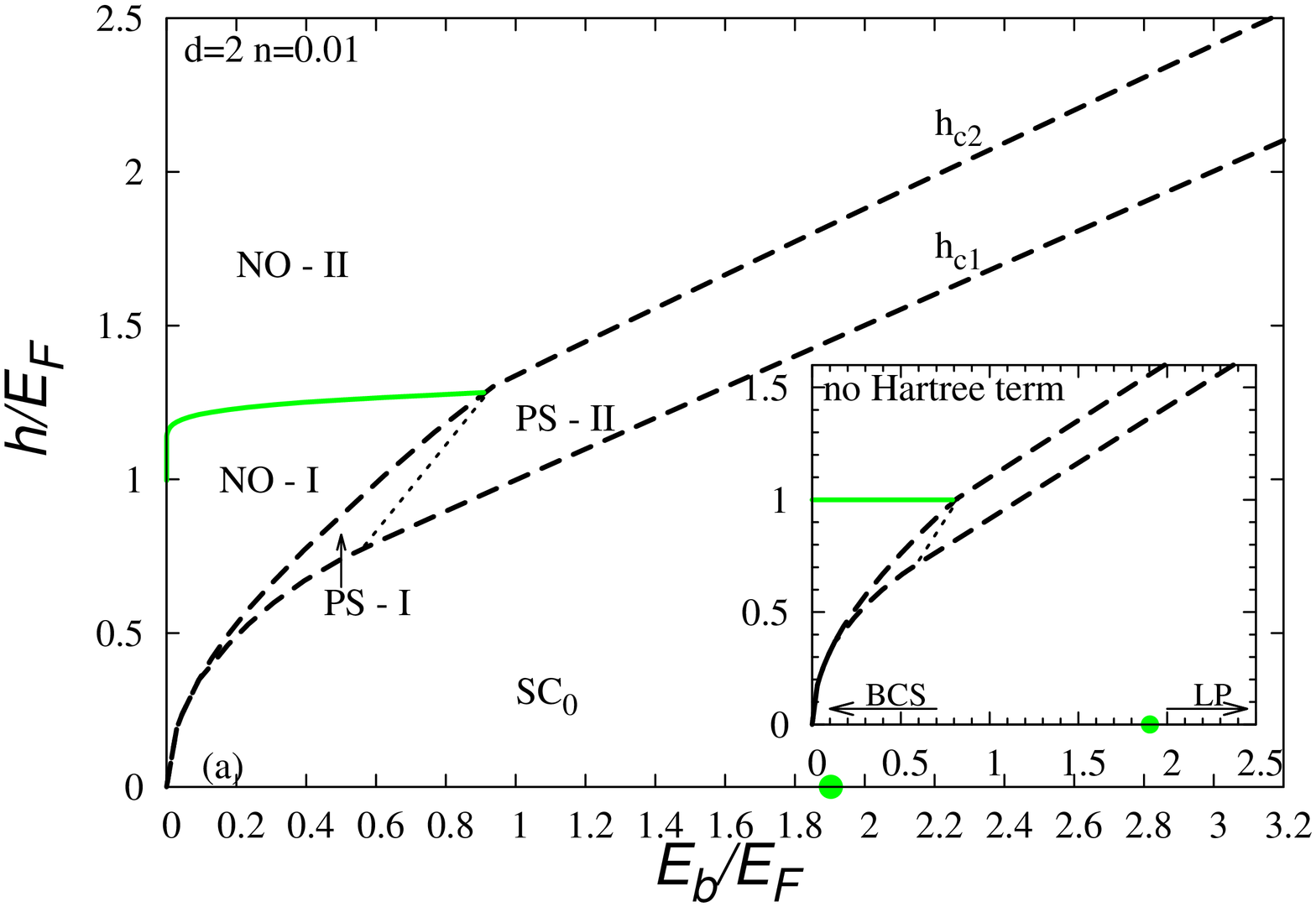}\hspace{-0.2cm}
\includegraphics[width=0.55\textwidth,angle=270]{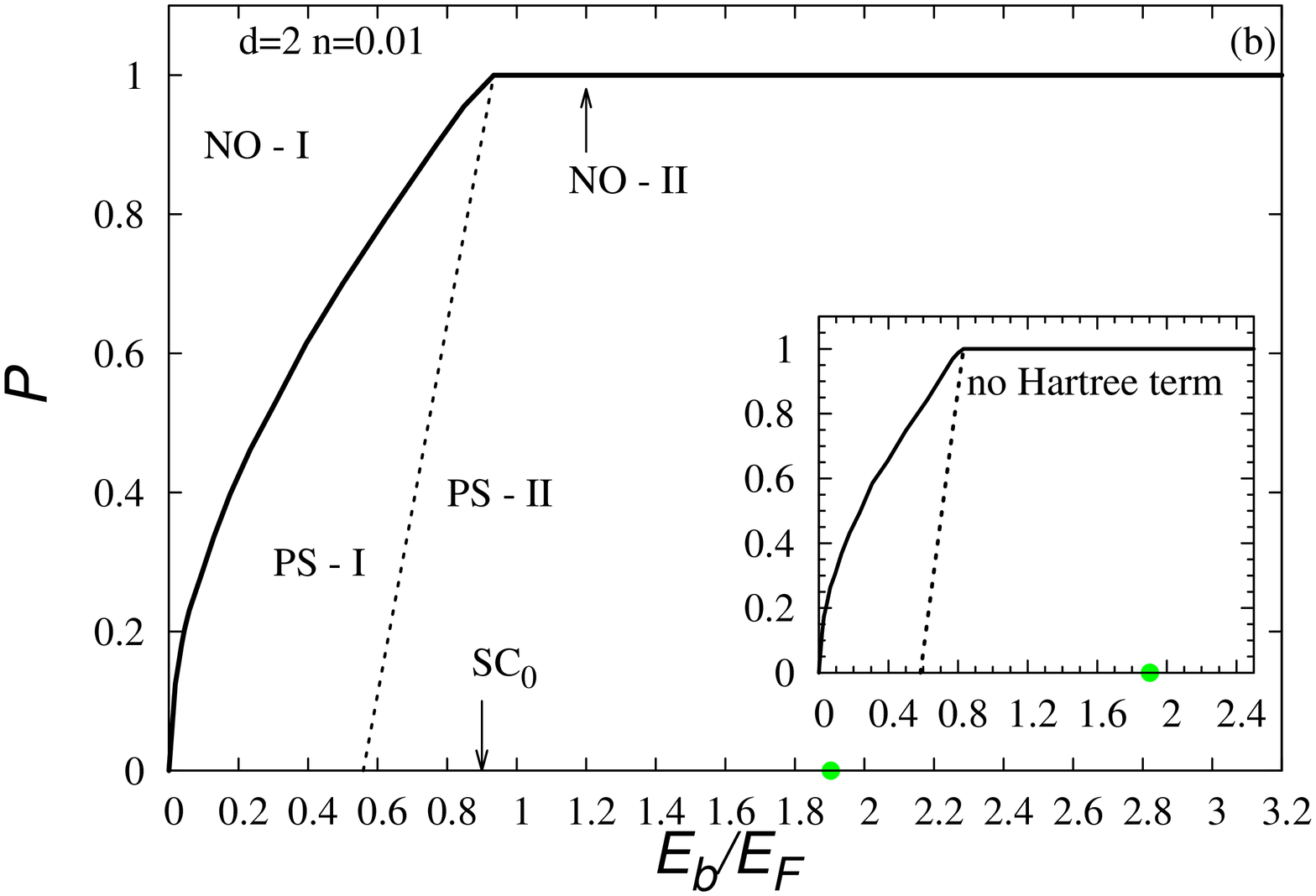}\\
\caption[Magnetic field vs. binding energy (a) and polarization (b) phase
diagrams both with and without the Hartree term (inset), in units of the lattice
Fermi energy, at $T=0$ and fixed $n=0.01$ for $d=2$.]{\label{2D_n001_crossover}
\textcolor{czerwony}{$h/E_F$} vs. binding energy (a) and polarization (b) phase diagrams both
with and without the Hartree term (inset),
at $T=0$ and fixed $n=0.01$ for $d=2$. $SC_0$ -- unpolarized superconducting
state with $n_{\uparrow}=n_{\downarrow}$, LP -- tightly bound local pairs. Green
solid line separates partially polarized (NO-I) and fully polarized (NO-II)
normal states. PS-I ($SC_0$+NO-I) -- partially polarized phase separation, PS-II
($SC_0$+NO-II) -- fully polarized phase separation, $h_{c1}$, $h_{c2}$ --
critical fields defining the PS region. \textcolor{czerwony}{The green point in (a)-(b) shows} the BCS-LP
crossover point ($U/t=-4.01959$). Labels of states in the insets are the same as
in the main figures.}
\end{center}
\end{figure}

From the point of view of the BCS-LP crossover, construction of the phase
diagrams which describe the evolution of the system with increasing
attraction is very important.  

Here, we present the results of calculations for rather low electron
concentrations and arbitrary values of on-site attraction $|U|$, for the
case of $d=2$. The particle concentration is fixed at $n=0.2$ (Fig.
\ref{2D_n02_crossover}). For fixed $n$, the 1$^{st}$ order SC-NO transition line
in the ($\mu -h$) plane is replaced by the PS region bounded by two critical
Zeeman fields $h_{c1}$ and $h_{c2}$. We find an unmagnetized SC$_0$ phase in the
strong attraction limit. With increasing magnetic field, PS is energetically
favored i.e. even in the strong attraction limit, the Sarma or breached (BP-1)
phase is unstable. Superconductivity is destroyed by pair breaking in the weak
coupling regime. On the other hand, in the strong coupling regime, the
transition from the superconducting to the normal state goes in addition through
phase separation (SC$_0$ $\rightarrow$ PS-II $\rightarrow$ NO-II).  

In the limiting case of $n=0$, one can perform an exact analysis. In this case,
the BCS equation can be reduced to the Schr\"odinger equation for a single pair,
where the chemical potential plays the role of the pair binding energy $E_b$. In
the lattice fermion model, for the case of $s$-wave pairing, the
two-particle binding energy is given by \cite{MicnasModern}:
\begin{equation}
\label{benergy}
\frac{2D}{U}=-\frac{1}{N} \sum_{\vec{k}} \frac{1}{(\frac{E_b}{2D}+1)-\frac{\gamma_{\vec{k}}}{z}},
\end{equation}
where: $D=zt$, $z=2d$ is the coordination number, $\gamma_{\vec{k}} = 2\Theta_{\vec{k}}$.
\\
Since in $d=2$ the two-body bound state is formed for any attraction (if $h=0$),
one can
replace the pairing potential $|U|$ by $E_b$.
This is of interest as far as a comparison with the continuum model of a dilute
gas of fermions is concerned.

A mapping of the phase diagrams (Fig. \ref{2D_n02_crossover}(a) and
\ref{2D_n02_crossover}(b)) was also performed, using $E_b$ in units of the \textcolor{czerwony}{lattice}
Fermi energy $E_F$ (\textcolor{czerwony}{calculated from: $n=\frac{2}{N}\sum_{\vec k}\Theta (\mu - \epsilon_{\vec k}+\epsilon_0)$}), instead of $|U|$ and the results for lower $n$ are shown in
Fig. \ref{2D_n001_crossover}(a)-(b).  In the dilute limit, the lattice effects
are smaller and the lattice model gives results similar to those of the model in
continuum. To show this, we fix the electron concentration to $n=0.01$.
In this case, one finds the same phases as in Fig.
\ref{2D_n02_crossover}(a) and \ref{2D_n02_crossover}(b). The phase diagrams
without the Hartree term have also been constructed and shown in the inset of
Fig. \ref{2D_n001_crossover}(a)-(b). Let us point out that these diagrams are in
good agreement with the results for the continuum fermion model in $d=2$
\cite{He}. However, if we consider the continuum model, the
chemical potential changes its sign exactly at $E_b=2E_F$, which indicates the
point of the BCS-BEC crossover. Our analysis gives a smaller ratio $E_b/E_F$.
The differences are attributed to lattice effects, because the electron
concentration is relatively small, but still non-zero. On the other hand,
our results do not agree with the analysis in Ref. \cite{Du}, according to which
the BP-2 state is stable at $r=1$ in $d=2$. However, we have examined the
stability of all phases very thoroughly.

If one takes corrections beyond mean-field (MF) into account, the existence of the spin imbalanced superfluid 
mixture of bosonic molecules and Fermi atoms can not be excluded in the BEC limit \cite{Tempere2}. 
We should also add that the deep BEC side is better described by a Boson-Fermion mixture of 
hard-core bosons and spin-up fermions. 

\section{3D simple cubic lattice}
\label{3Dcase}

\begin{figure}[t!]
\hspace*{-0.8cm}
\includegraphics[width=0.38\textwidth,angle=270]{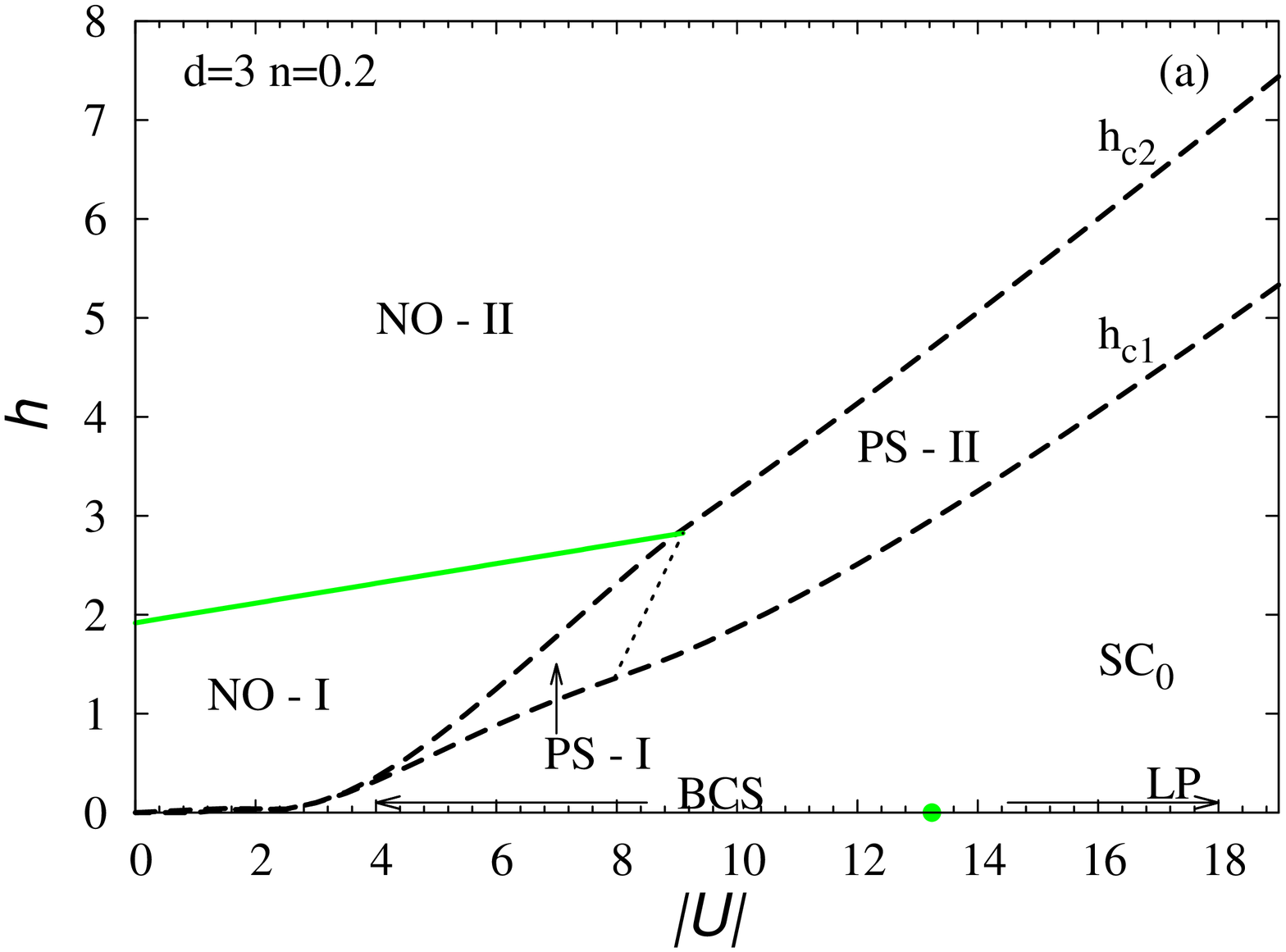}\hspace{-0.2cm}
\hspace*{-0.6cm}
\includegraphics[width=0.38\textwidth,angle=270]{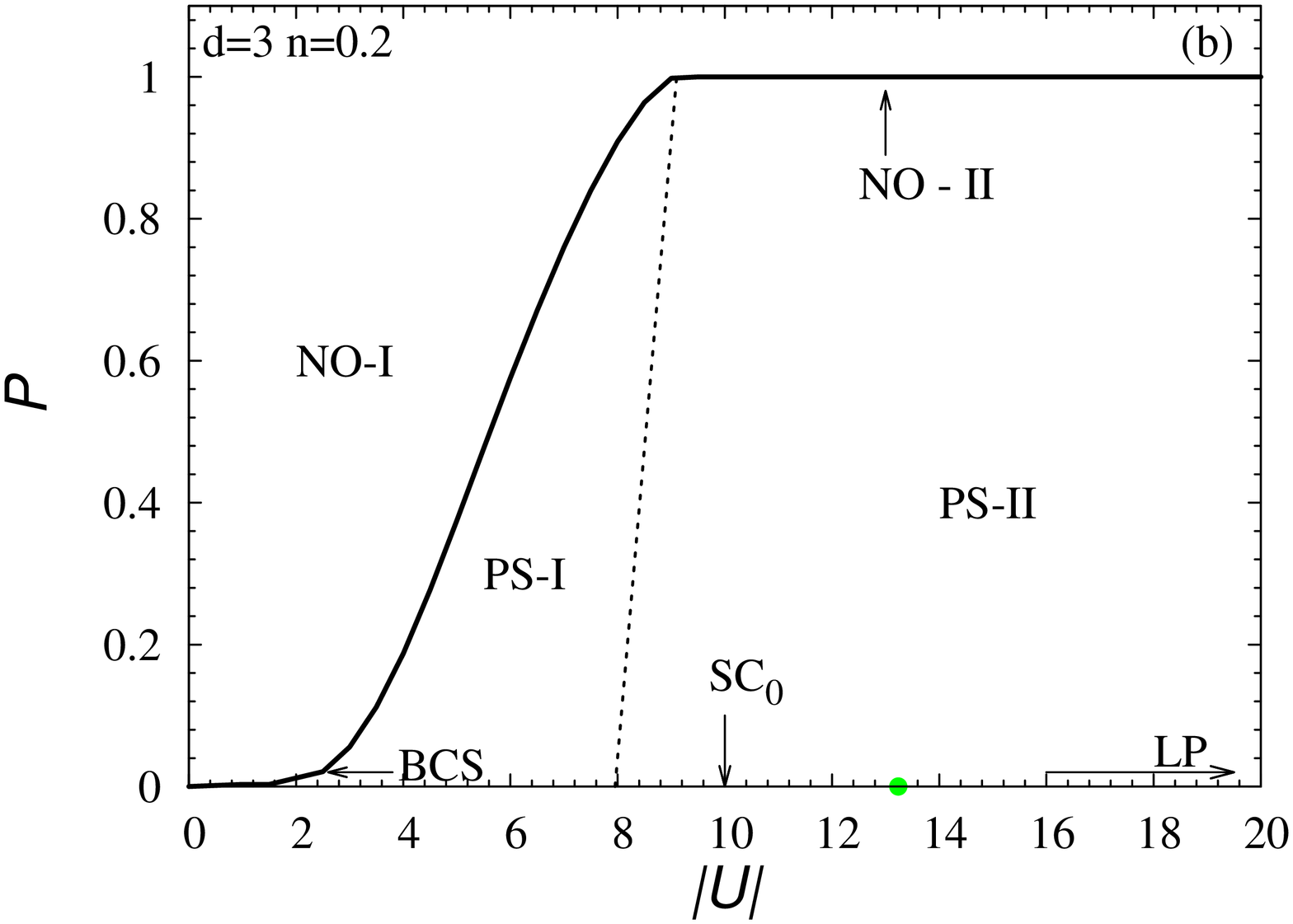}
\caption[Magnetic field vs. on-site attraction (a) and polarization (b) phase
diagrams, at $T=0$ and fixed $n=0.2$, for the simple cubic
lattice.]{\label{3D_n02_crossover} Magnetic field vs. on-site attraction (a) and
polarization (b) phase diagrams, at $T=0$ and fixed $n=0.2$, for the simple
cubic lattice. SC$_0$ -- unpolarized superconducting state with
$n_{\uparrow}=n_{\downarrow}$, LP -- tightly bound local pairs. Green solid line
separates partially polarized (NO-I) and fully polarized (NO-II) normal states.
PS-I (SC$_0$+NO-I) -- partially polarized phase separation, PS-II (SC$_0$+NO-II)
-- fully polarized phase separation, $h_{c1}$, $h_{c2}$ -- critical fields
defining the PS region. The green point in (a)-(b) shows the BCS-BEC crossover point.}
\end{figure}

We also investigate the BCS-BEC crossover diagrams in the presence of a Zeeman
magnetic field  in 3D for a simple cubic lattice. The diagrams in ($h$, $|U|$)
and ($P$, $|U|$) planes are displayed in Figs. \ref{3D_n02_crossover}(a) and
\ref{3D_n02_crossover}(b) for $n=0.2$. These diagrams are typical for $d=3$
and a relatively low $n$, with possible FFLO state on the BCS side. A
similar behavior persists for higher $n$. 

However, for strong attraction and in the dilute limit, the homogeneous
magnetized superconducting phase occurs. In general, the solutions of this
type (Sarma-type with $\Delta (h)$) appear when $h>\Delta$ (on the BCS side) or
when 
$h>E_{g}/2$, where $E_g=2\sqrt{(\bar{\mu}-\epsilon_0)^2+|\Delta |^2}$ (on the
BEC side) (see: Appendix \ref{appendix3}). As shown above, the SC$_M$
phase is
unstable in the weak coupling regime, but can be stable in the strong coupling
LP limit ($E_b >> \Delta$). Consequently, the quasiparticle branches in that
regime are given by:
\begin{equation}
 E_{\vec{k}\downarrow}=\epsilon_{\vec{k}}-\mu_{\downarrow}+g(\vec{k})\approx \epsilon_{\vec{k}} -\epsilon_{0}+\frac{1}{2} E_b+g(\vec{k}),
\end{equation}
\begin{equation}
 E_{\vec{k}\uparrow}=\epsilon_{\vec{k}}-\mu_{\uparrow}+g(\vec{k}),
\end{equation}
where: $g(\vec{k})=\frac{|\Delta|^2}{2(\epsilon_{\vec{k}}-\epsilon_0)+E_b}$ is a
self-energy correction for fermions due to the interaction with the Bose
condensate of LP, $\mu_{\sigma}=\bar{\mu}+\sigma (\frac{UM}{2}+h)$. $E_b$ is the
LP binding energy calculated from Eq.~\eqref{benergy}.

Deep in the BEC limit, unpaired spin down fermions do not exist. Hence, the
SC$_M$ phase is a specific superfluid state,which is a coherent mixture of LP's
and an excess of spin-up fermions.

\begin{figure}[t!]
\hspace*{-0.8cm}
\includegraphics[width=0.38\textwidth,angle=270]{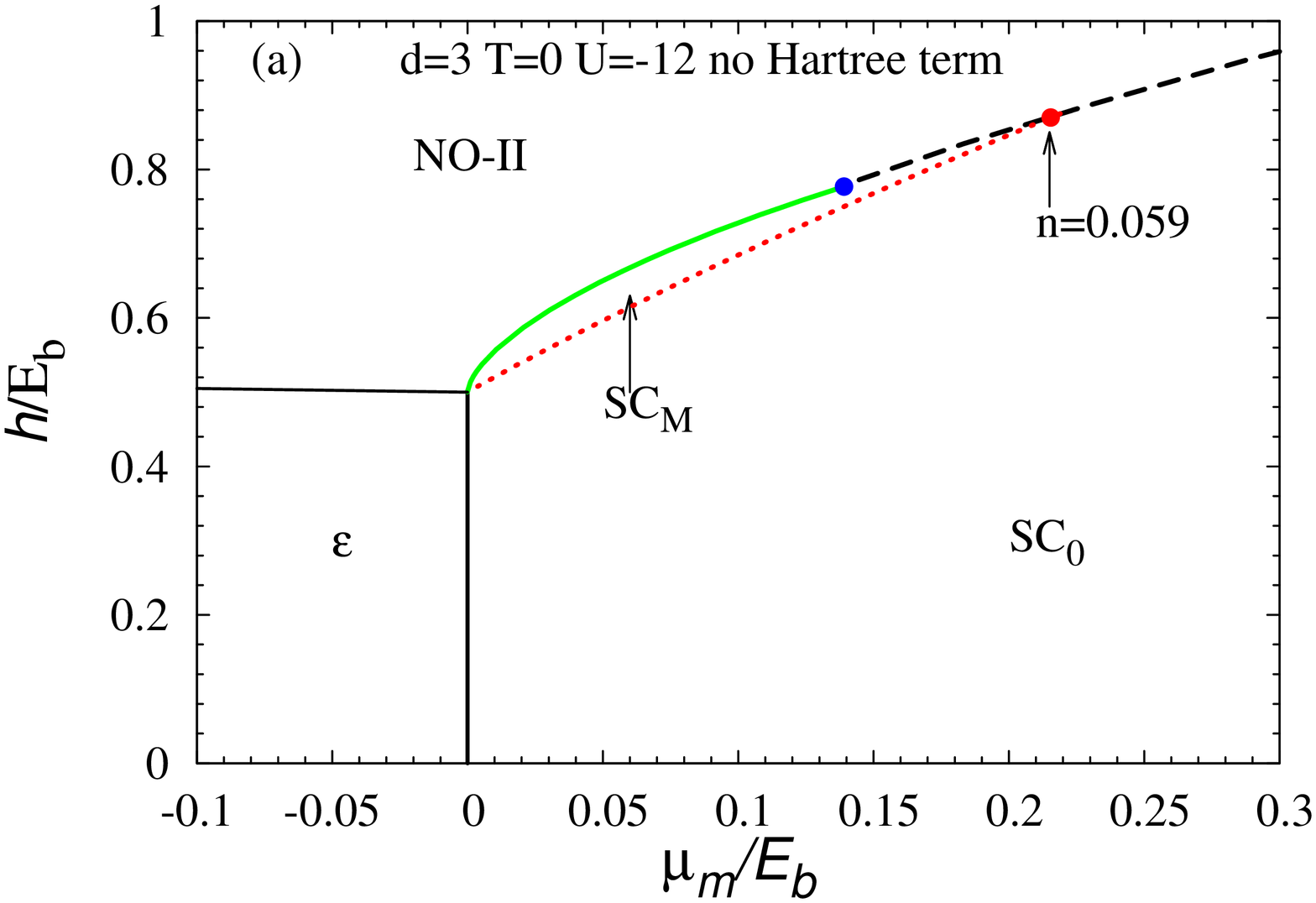}\hspace{-0.2cm}
\hspace*{-0.6cm}
\includegraphics[width=0.38\textwidth,angle=270]{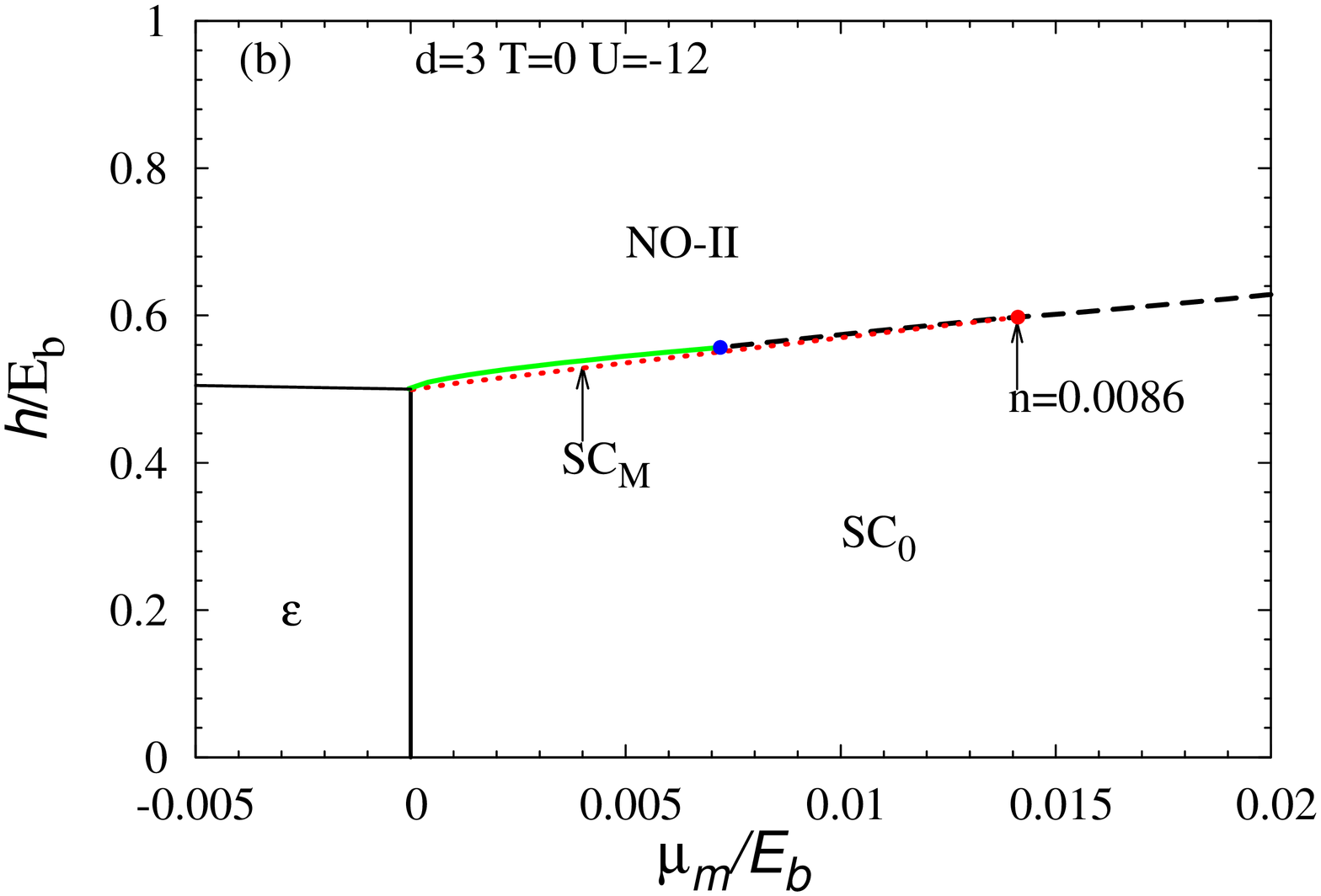}
\caption[Critical magnetic field vs. chemical potential for $d=3$, $U=-12$ on LP
side.]{\label{U-12_r1_mi} Critical magnetic field vs. chemical potential for
$d=3$, $U=-12$ on LP side. Diagram without (a) and with (b) Hartree term. SC$_0$
-- unpolarized SC state with $n_{\uparrow}=n_{\downarrow}$, SC$_M$ -- magnetized
SC state, NO-II -- fully polarized norma state, $\varepsilon$ -- empty state,
$\mu_{m}$ -- half of the pair chemical potential defined as:
$\mu-\epsilon_0+\frac{1}{2} E_b$, where $\epsilon_0=-6t$, $E_b$ is the binding
energy for two fermions in an empty lattice. Red point -- $h_{c}^{SC_M}$
(quantum critical point (QCP)), blue point -- tricritical point. The dotted red
and the solid green lines are continuous transition lines.}
\end{figure}

Let us start our analysis from the phase diagrams at fixed $\mu_{m}$ and $h$
(Fig. \ref{U-12_r1_mi}). We define $\mu_{m}=\mu-\epsilon_0+\frac{1}{2} E_b$ as
one half of the pair chemical potential (molecular potential). The structure of
these diagrams is different from the ones illustrated in Fig.
\ref{mu_diagram}(b), where one can distinguish only first order phase transition
from pure SC$_0$ to the NO phase. Here, we observe also the continuous phase
transition from SC$_0$ to SC$_M$ with decreasing chemical potential and
increasing magnetic field. The transition from SC$_M$ to NO can be
either of the first or second order. The character of this transition
changes
with decreasing $\mu$. Hence, we also find the tricritical point in these
diagrams (blue point). Therefore, we can distinguish the following sequences of
transitions: SC$_0$ $\rightarrow$ NO or SC$_0$ $\rightarrow$ SC$_M$
$\rightarrow$ NO. In fact SC$_0$ $\rightarrow$ SC$_M$ is a topological quantum
phase transition (Lifshitz type) (see: Appendix \ref{appendix3}). There is a cusp in the order parameter and polarization vs. magnetic field plots (for fixed $n$, for $\mu$ vs. $h$ as well), which is clearly visible in Fig.
\ref{3D_del_n001_h_U_multiplot}. After transition to the SC$_M$ phase, the
polarization increases up to its maximum value and there is the 2$^{nd}$ order
transition to the fully polarized normal state. There is also a change in the
electronic structure. In the SC$_0$ phase there is no Fermi surface (FS), but
in the SC$_M$ state there is one FS for excess fermions. One can notice that
the presence of the Hartree term restricts the range of occurrence of the SC$_M$
phase, except for a very dilute limit (it will be discussed later).

\begin{figure}
\hspace*{-0.8cm}
\includegraphics[width=0.38\textwidth,angle=270]
{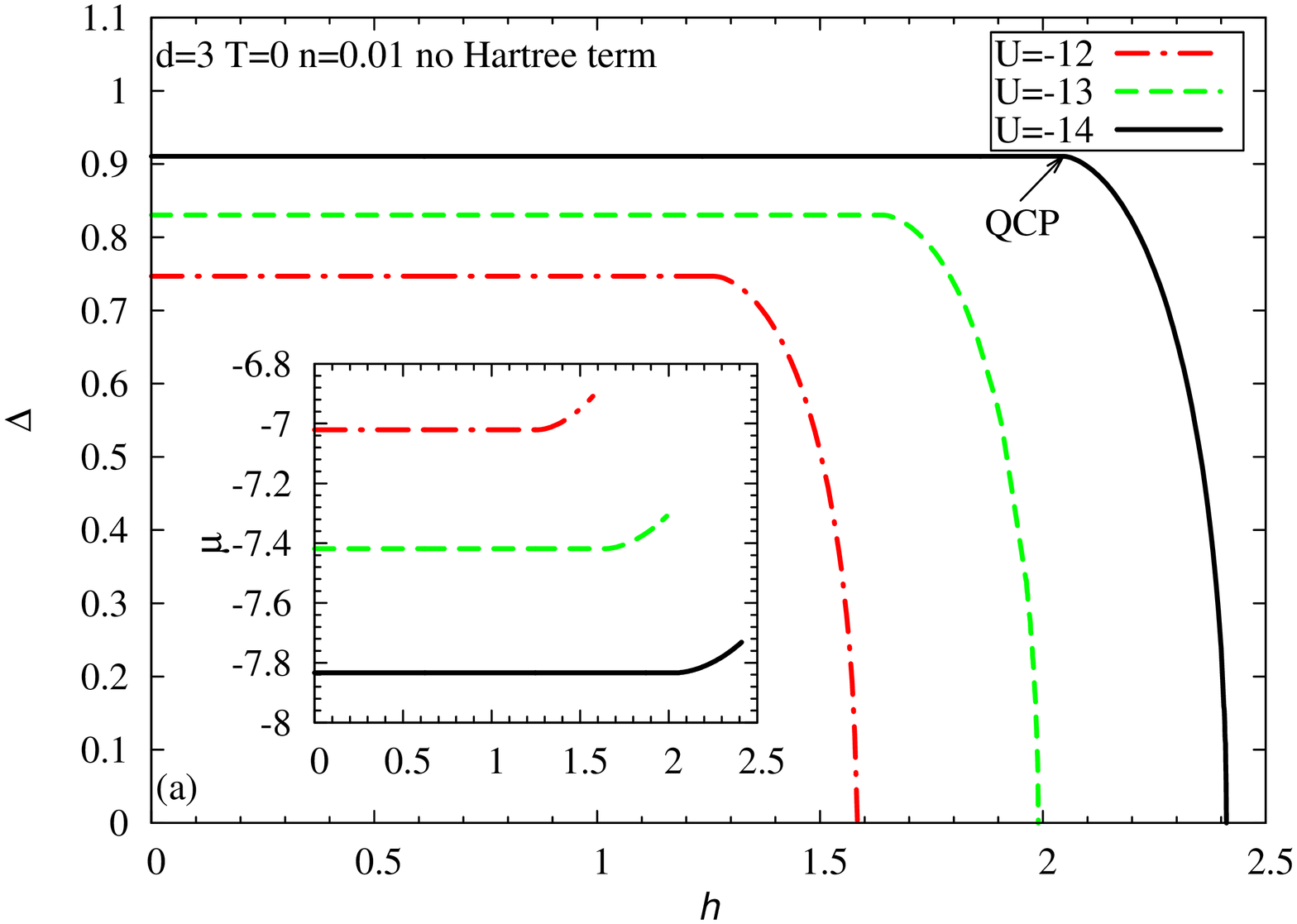}
\hspace*{-0.6cm}
\includegraphics[width=0.38\textwidth,angle=270]
{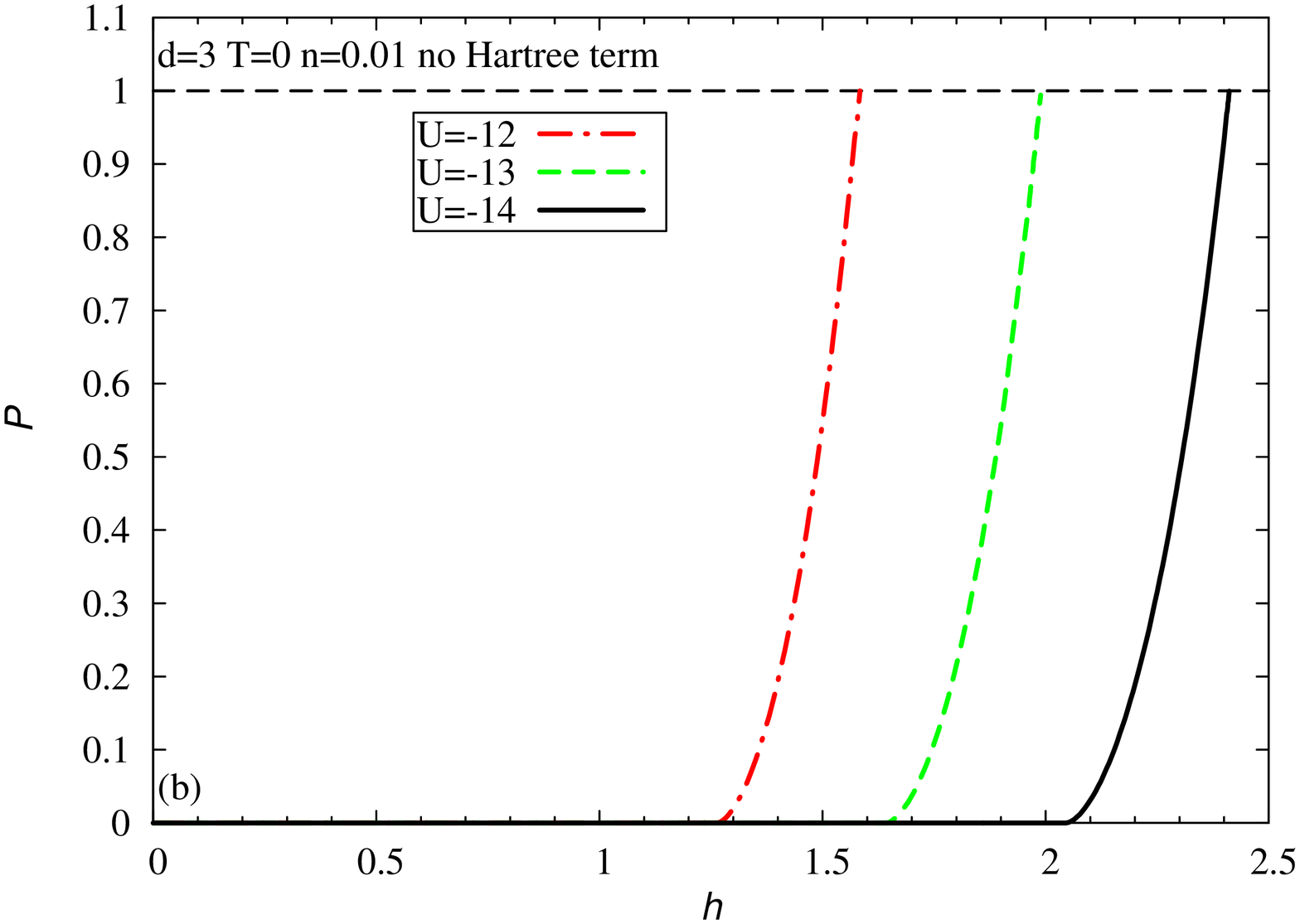}
\caption[Dependence of the order parameter (a), chemical potential (inset) and the polarization (b) on the magnetic field for $d=3$, at $T=0$, fixed $n=0.01$ and three values of the attraction.]{\label{3D_del_n001_h_U_multiplot} Dependence of the order parameter (a), chemical potential (inset) and the polarization (b) on the magnetic field for $d=3$, at $T=0$, fixed $n=0.01$ and three values of the attraction. The arrow points the quantum critical point (QCP).}
\end{figure}

Fig. \ref{3D_diag_n001vsU_U-12_vsn} shows the critical magnetic fields vs.
attractive interaction, for fixed $n=0.01$ (a) and $h$ vs. $n$ for fixed $U=-12$
(b), without the Hartree term. In a very weak coupling limit (Fig.
\ref{3D_diag_n001vsU_U-12_vsn}(a)) superconductivity is destroyed by the pair
breaking. However, there exists a critical value of $|U_c|^{SC_M}$ (red point in
the diagrams), above which the SC$_M$ state becomes stable. As shown in Fig.
\ref{3D_diag_n001vsU_U-12_vsn}(a), for $|U|<|U_c|^{S_CM}$ the transition from
SC$_0$ to NO takes place, as previously, through PS-I or PS-II. However, the
transition from SC$_M$ to NO can be accomplished in two ways: (i) through PS-III
(SC$_M$+NO-II) or (ii) through the second order phase transition (for higher
$|U|$). Hence, TCPs are found in the $(h-|U|)$ and $(h-n)$ diagrams. As
mentioned above, magnetized superconducting state is stable in the deep LP
limit (for high $|U|$ and very low $n$ (see  Fig.
\ref{3D_diag_n001vsU_U-12_vsn}(b))). 

\begin{figure}[t!]
\hspace*{-0.8cm}
\includegraphics[width=0.38\textwidth,angle=270]
{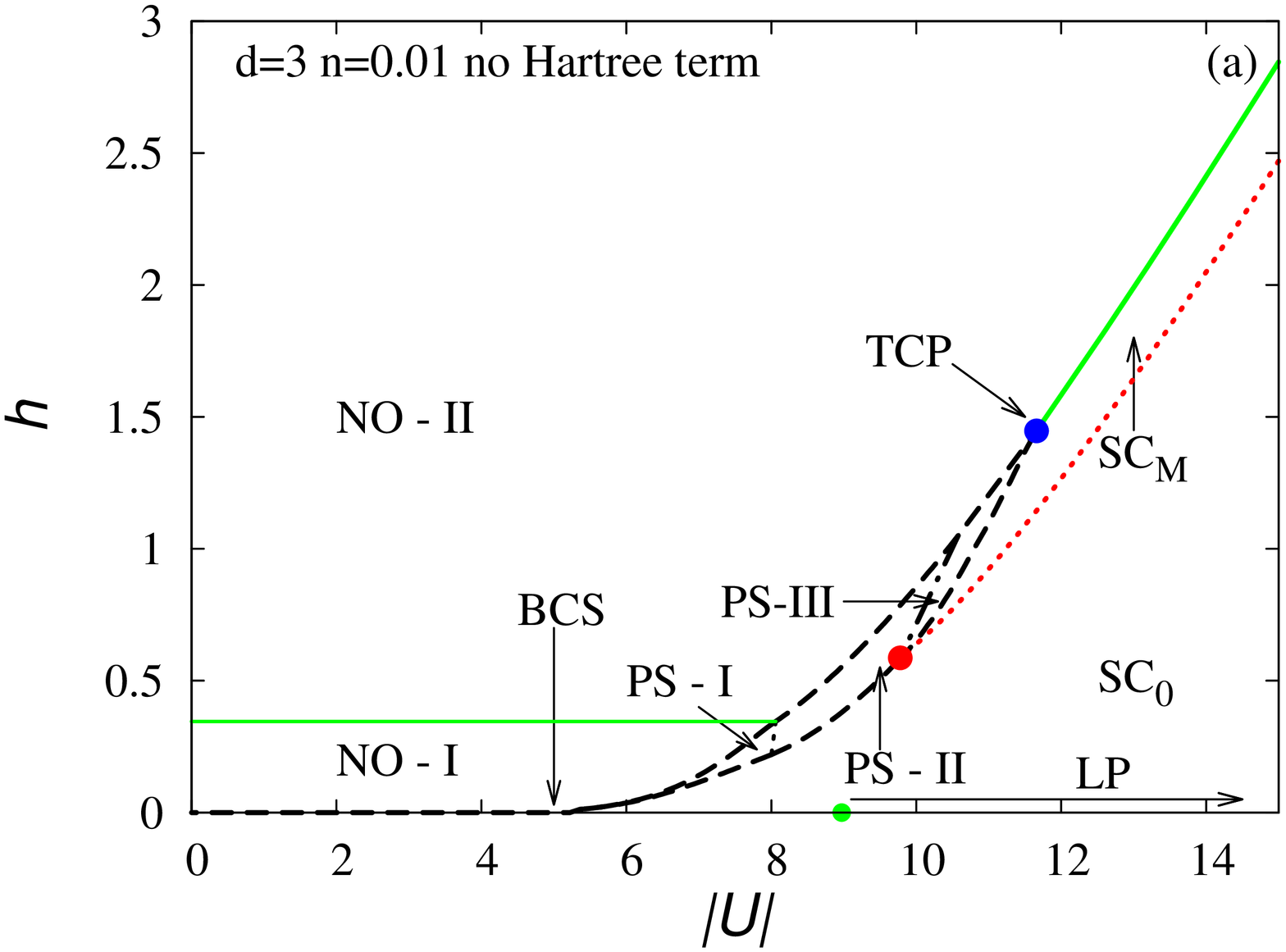}\hspace{-0.2cm}
\hspace*{-0.6cm}
\includegraphics[width=0.38\textwidth,angle=270]
{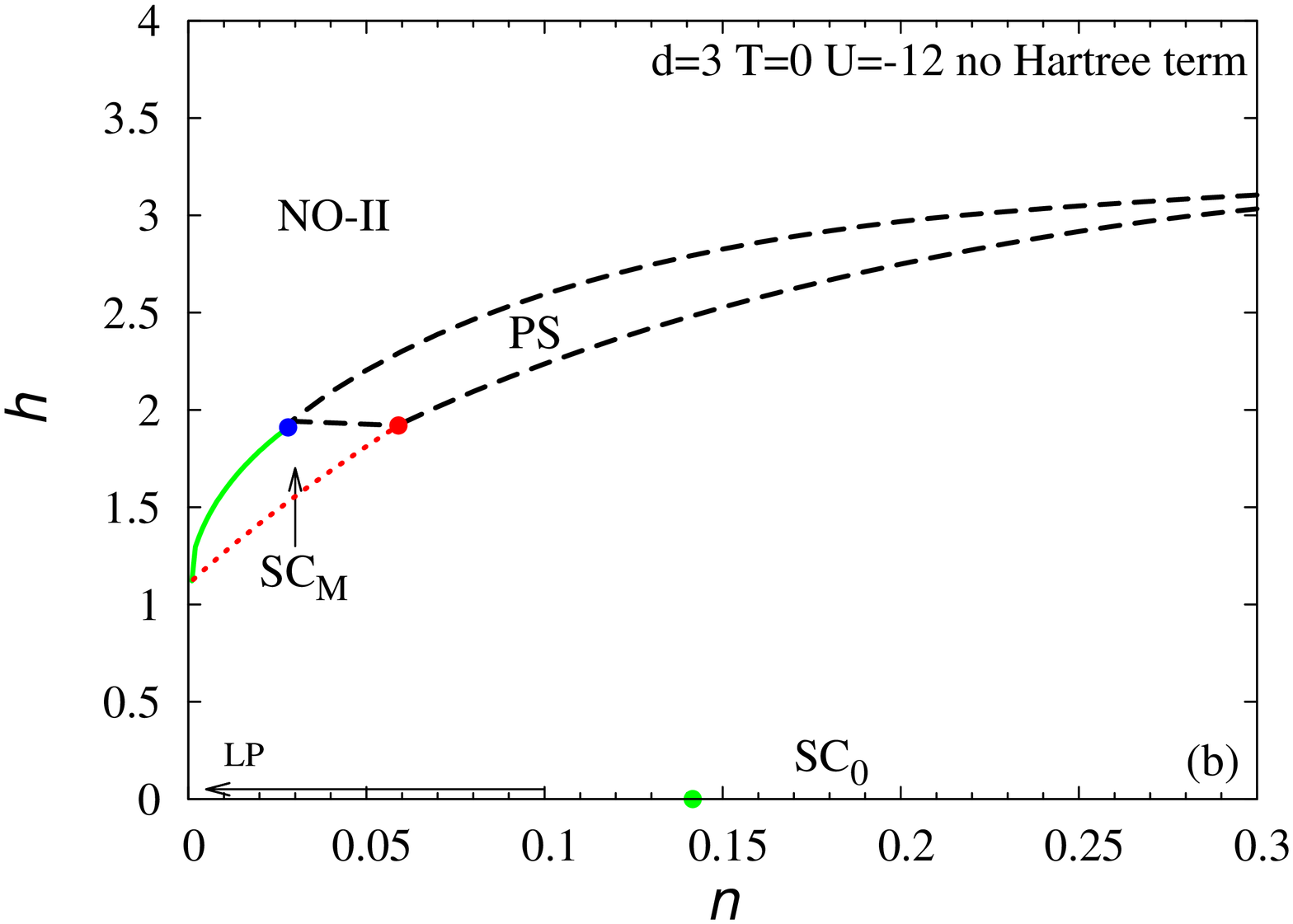}
\caption[Critical magnetic field vs. $|U|$ for fixed $n=0.01$ (a) and electron
concentration for fixed $U=-12$ (b) phase diagrams, $T=0$,
$d=3$.]{\label{3D_diag_n001vsU_U-12_vsn} Critical magnetic field vs. $|U|$ for
fixed $n=0.01$ (a) and electron concentration for fixed $U=-12$ (b) phase
diagrams, $T=0$, $d=3$. SC$_0$ -- unpolarized superconducting state, SC$_M$ --
magnetized superfluid phase, LP -- tightly bound local pairs. Green solid line
separates partially polarized (NO-I) and fully polarized (NO-II) normal states.
PS-I (SC$_0$+NO-I) -- partially polarized phase separation, PS-II (SC$_0$+NO-II)
-- fully polarized phase separation, PS-III -- SC$_M$+NO-II. Red point --
quantum critical point, blue point -- tricritical point.\textcolor{czerwony}{The green point in (a)-(b) shows}
the BCS-BEC crossover point.}
\end{figure}

The diagrams with and without the Hartree term for \textcolor{czerwony}{a} sc lattice in a very
dilute limit are in good agreement with the ones for the continuum model of a
dilute
gas of fermions, plotted in terms of 
\textcolor{czerwony}{lattice} $E_F$ 
(\textcolor{czerwony}{calculated from: $n=\frac{2}{N}\sum_{\vec k}\Theta (\mu - \epsilon_{\vec k}+\epsilon_0)$}) 
and $k_F a_s$ \cite{Sheehy, Parish, Parish2, Pilati} (where $a_s$ is the scattering length between fermions for a two-body
problem \textcolor{czerwony}{in vacuum}). \textcolor{czerwony}{The scattering length is related to the contact
interaction potential $g$ via the Lippmann-Schwinger equation: $m/(4\pi a_s)=1/g +1/V \sum_{\vec k} 1/2\epsilon_{\vec k}$. 
On the lattice: $m=1/2t$. We take $g=U$. Then, $E_F=k_F^2t$ ($a=1$, $\hbar=1$), from Eq. \eqref{benergy} $1/|U_c| =\frac{1}{N}\sum_{\vec k}\frac{1}{2(\epsilon_{\vec k}-\epsilon_0)}$
and
after short calculations we have: $1/k_F
a_s=\Big(\frac{1}{|U_c|^{d=3}}-\frac{1}{|U|}\Big)\frac{8\pi}{\sqrt{E_F}}$.}

Fig. \ref{3D_diag_n0001vsU_U_Hartree} shows the critical magnetic field (a) and
polarization (b) vs. $1/k_F
a_s=\Big(\frac{1}{|U_c|^{d=3}}-\frac{1}{|U|}\Big)\frac{8\pi}{\sqrt{E_F}}$, where
$|U_c|^{d=3}/12=0.659$, for $n=0.001$, with the Hartree term. In the strong
coupling limit at $P=0$, the system is in the $SC_0$ phase. For arbitrarily low
values of polarization, the SC$_M$ state becomes stable (as opposed to the
$n=0.2$ case (Fig. \ref{3D_n02_crossover}(b))). When polarization increases,
local pairs are broken and at $P=1$ there is the second order phase transition
to the fully-polarized normal state. With decreasing attractive interaction, the
system goes through phase separation to the NO state. One can perform comparison with the case of the model in continuum within mean-field analysis \cite{Parish} and also quantum Monte Carlo (QMC) method \cite{Pilati}. The tricritical point (blue point) on the ground state phase diagram ($h/E_F$ - $1/k_F a_s$) (Fig. \ref{3D_diag_n0001vsU_U_Hartree}) is located at $h/E_F \approx 6.854$ and $1/k_F a_s \approx 2.41$. Similar to the case of the model in continuum -- $h/E_F \approx 6.876$, $1/k_F a_s \approx 2.368$. In turn, \textcolor{czerwony}{the} values in the QMC method \textcolor{czerwony}{are}: $1/k_F a_s \approx 2.142$ \textcolor{czerwony}{with $P=1$}. Moreover, there is also phase separation between magnetized superfluid and partially-polarized phases as opposed to the mean-field case.             

\begin{figure}[h!]
\hspace*{-0.8cm}
\includegraphics[width=0.38\textwidth,angle=270]
{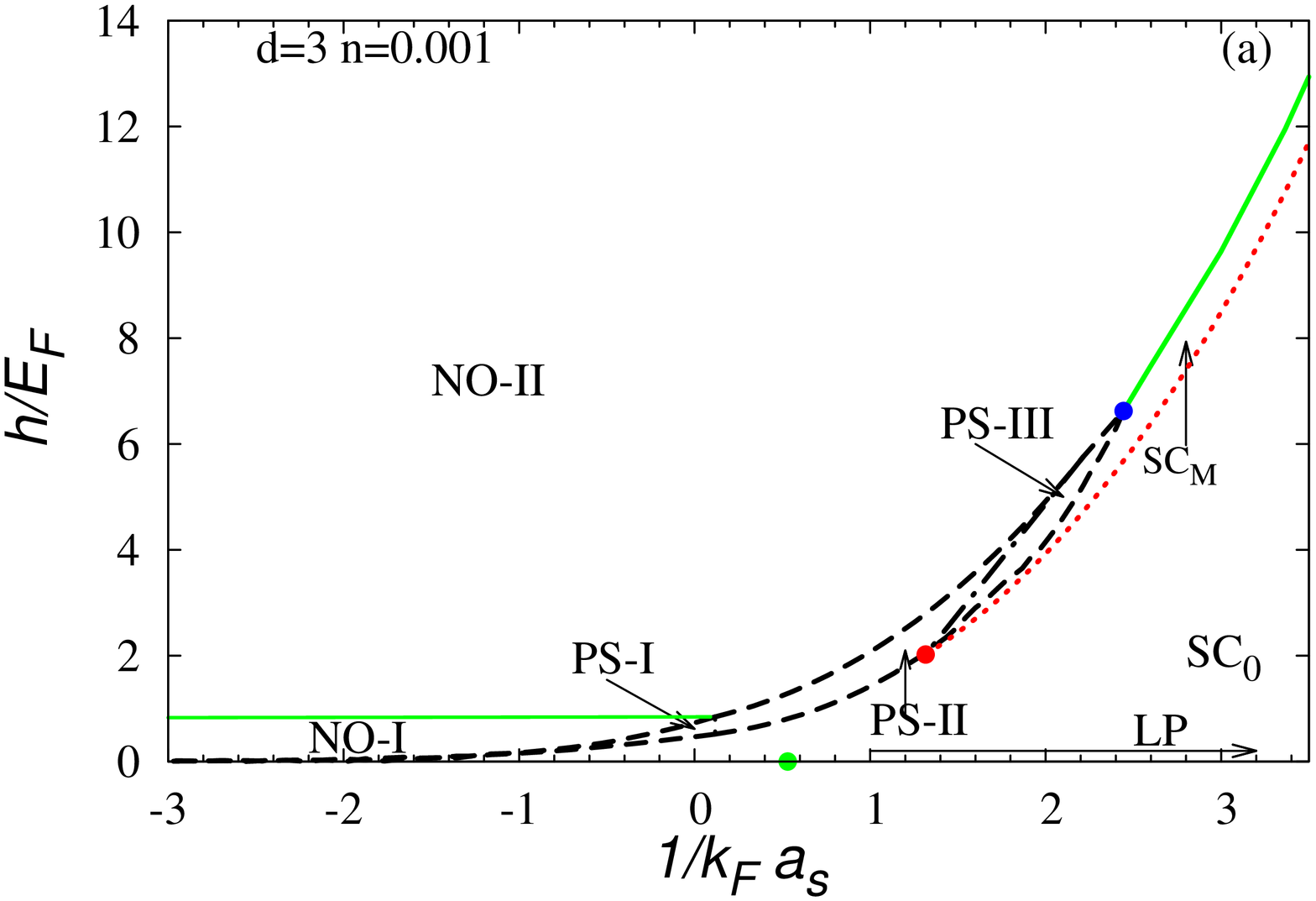}\hspace{-0.2cm}
\hspace*{-0.6cm}
\includegraphics[width=0.38\textwidth,angle=270]
{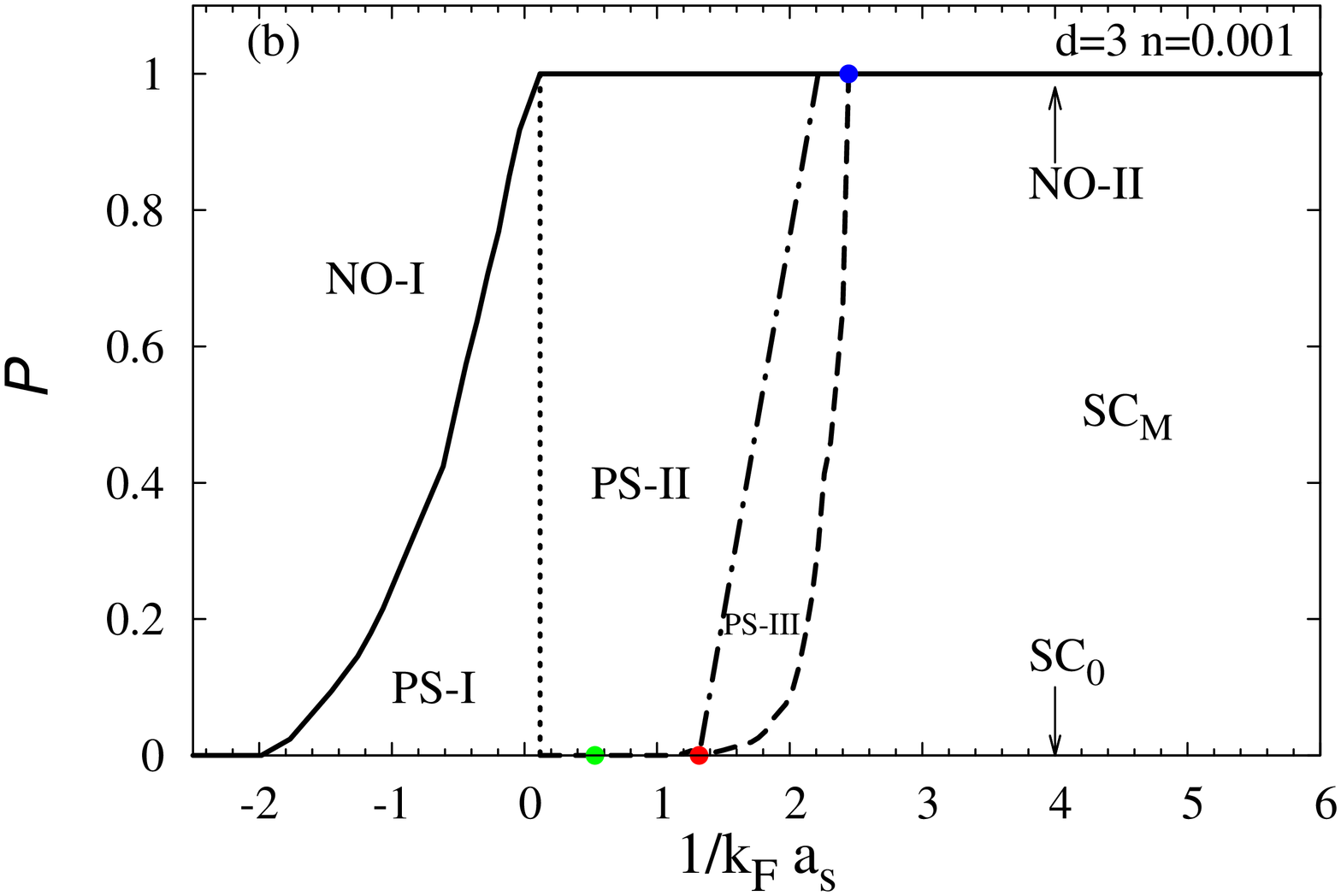}
\caption[$h/E_F$ (a) and polarization (b) vs. $1/k_F
a_s=\Big(\frac{1}{|U_c|^{d=3}}-\frac{1}{|U|}\Big)\frac{8\pi}{\sqrt{E_F}}$, where
$|U_c|^{d=3}/12t=0.659$, for the sc lattice.]{\label{3D_diag_n0001vsU_U_Hartree} \textcolor{czerwony}{$h/E_F$} (a) and polarization (b) vs. $1/k_F
a_s=\Big(\frac{1}{|U_c|^{d=3}}-\frac{1}{|U|}\Big)\frac{8\pi}{\sqrt{E_F}}$, where
$|U_c|^{d=3}/12t=0.659$, for the sc lattice. $SC_M$ -- magnetized
superconducting state, PS-III -- ($SC_M$+ NO-II). Red point -- $h_{c}^{SC_M}$,
blue point -- tricritical point. The dotted red and the solid green lines are
continuous transition lines. The diagrams are with the Hartree term.}
\end{figure}

\begin{figure}[t!]
\hspace*{-0.8cm}
\includegraphics[width=0.38\textwidth,angle=270]
{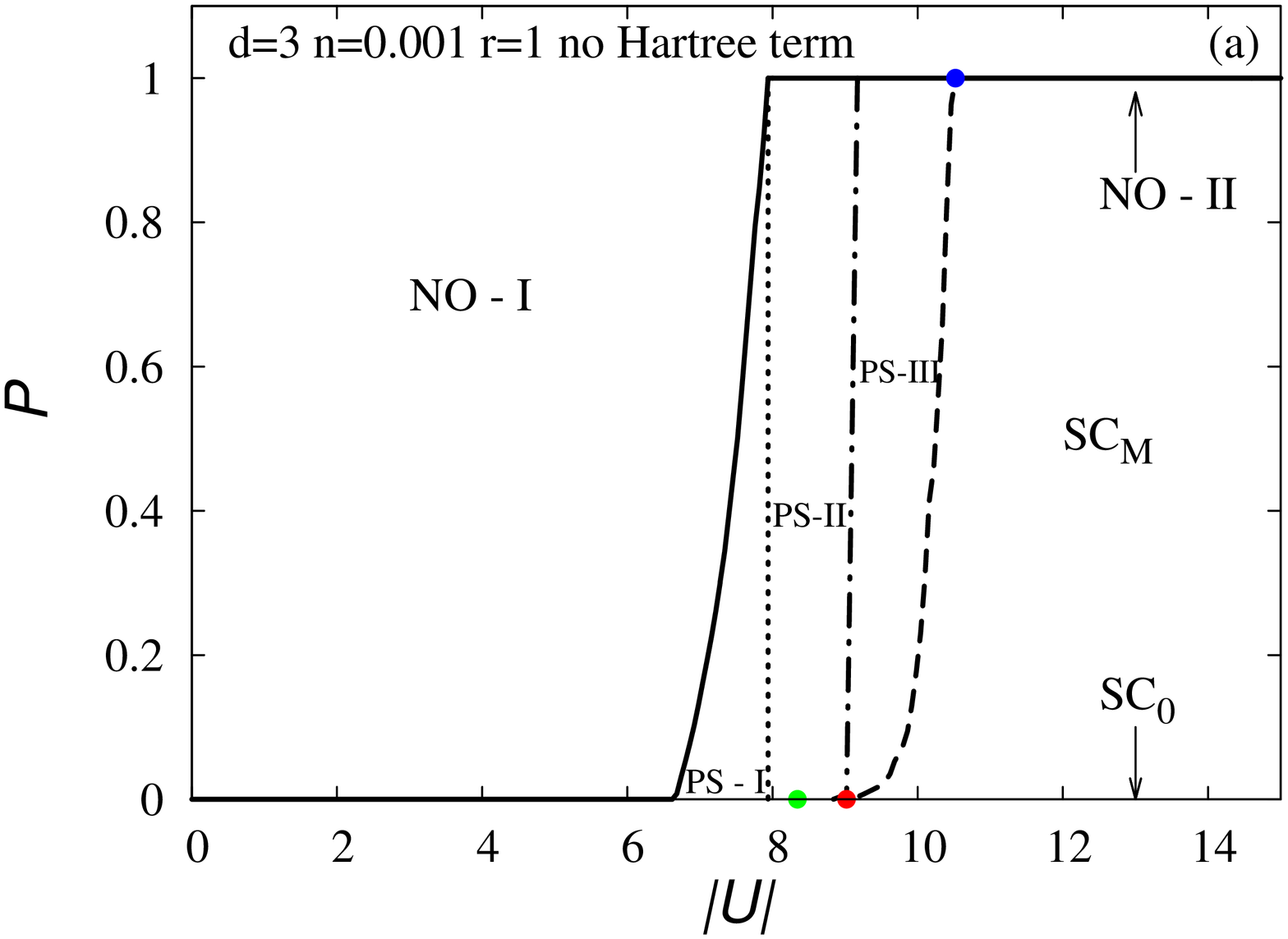}
\hspace*{-0.6cm}
\includegraphics[width=0.38\textwidth,angle=270]
{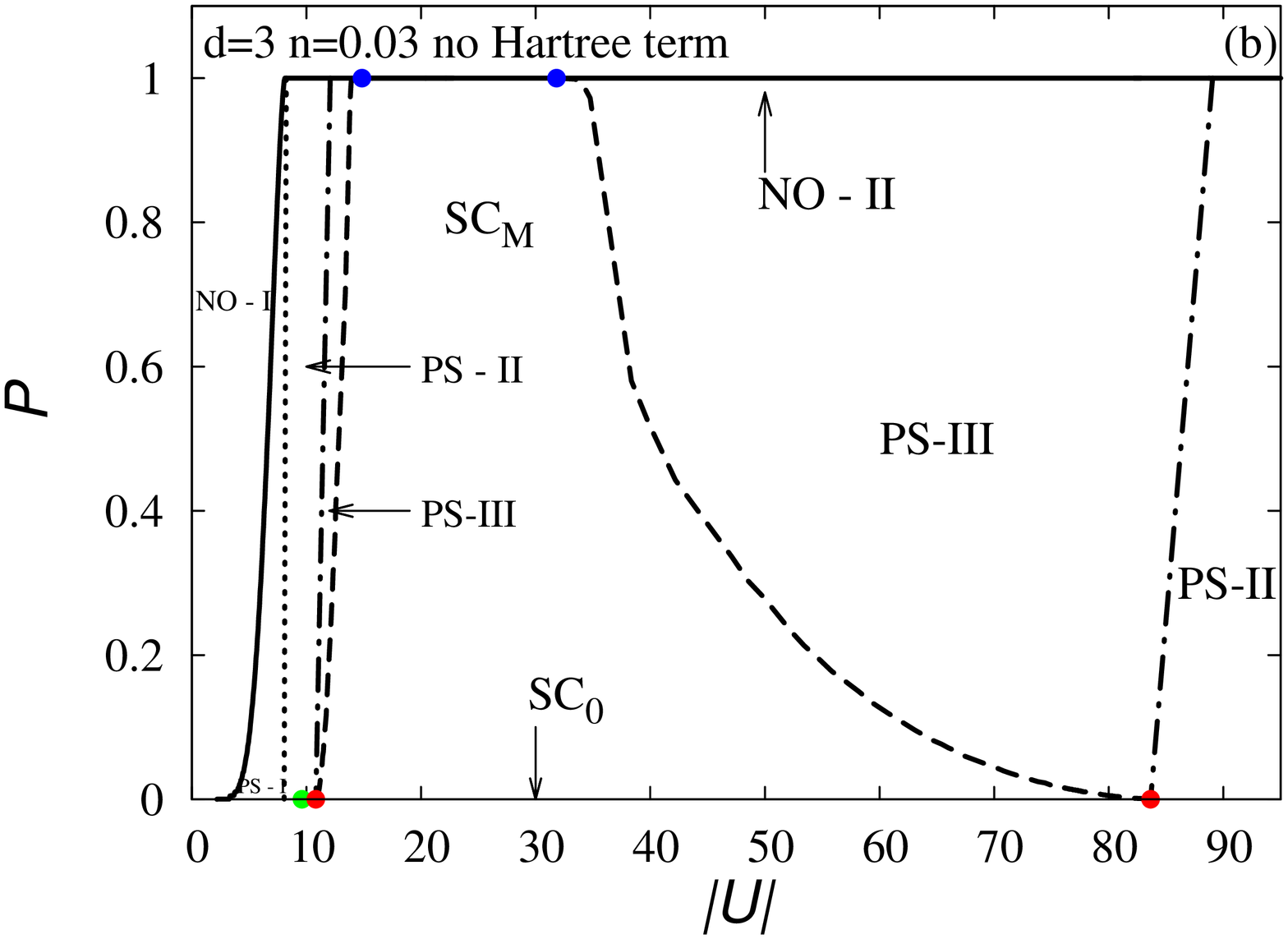}
\hspace*{-0.8cm}
\includegraphics[width=0.38\textwidth,angle=270]
{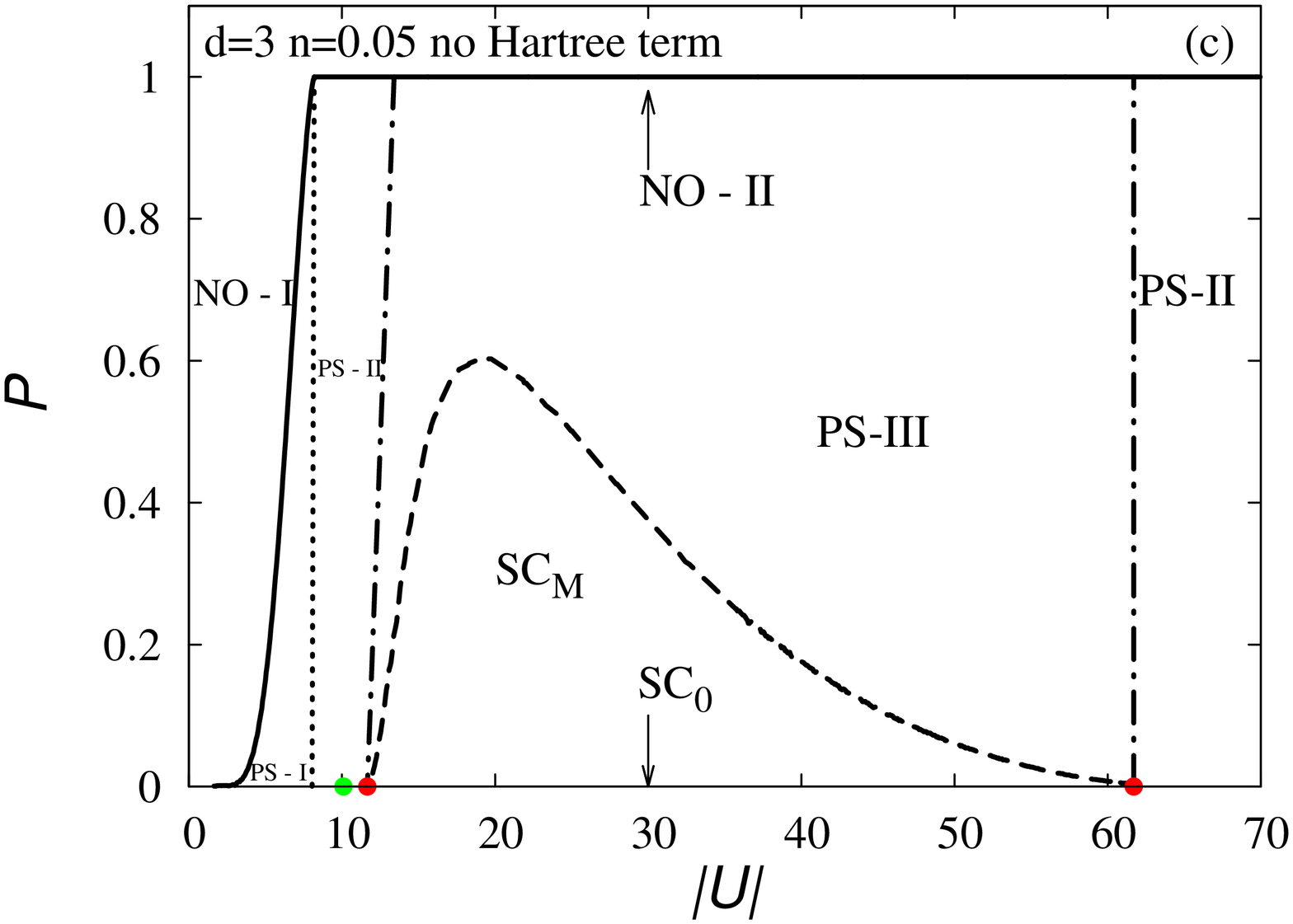}
\hspace*{-0.6cm}
\includegraphics[width=0.38\textwidth,angle=270]
{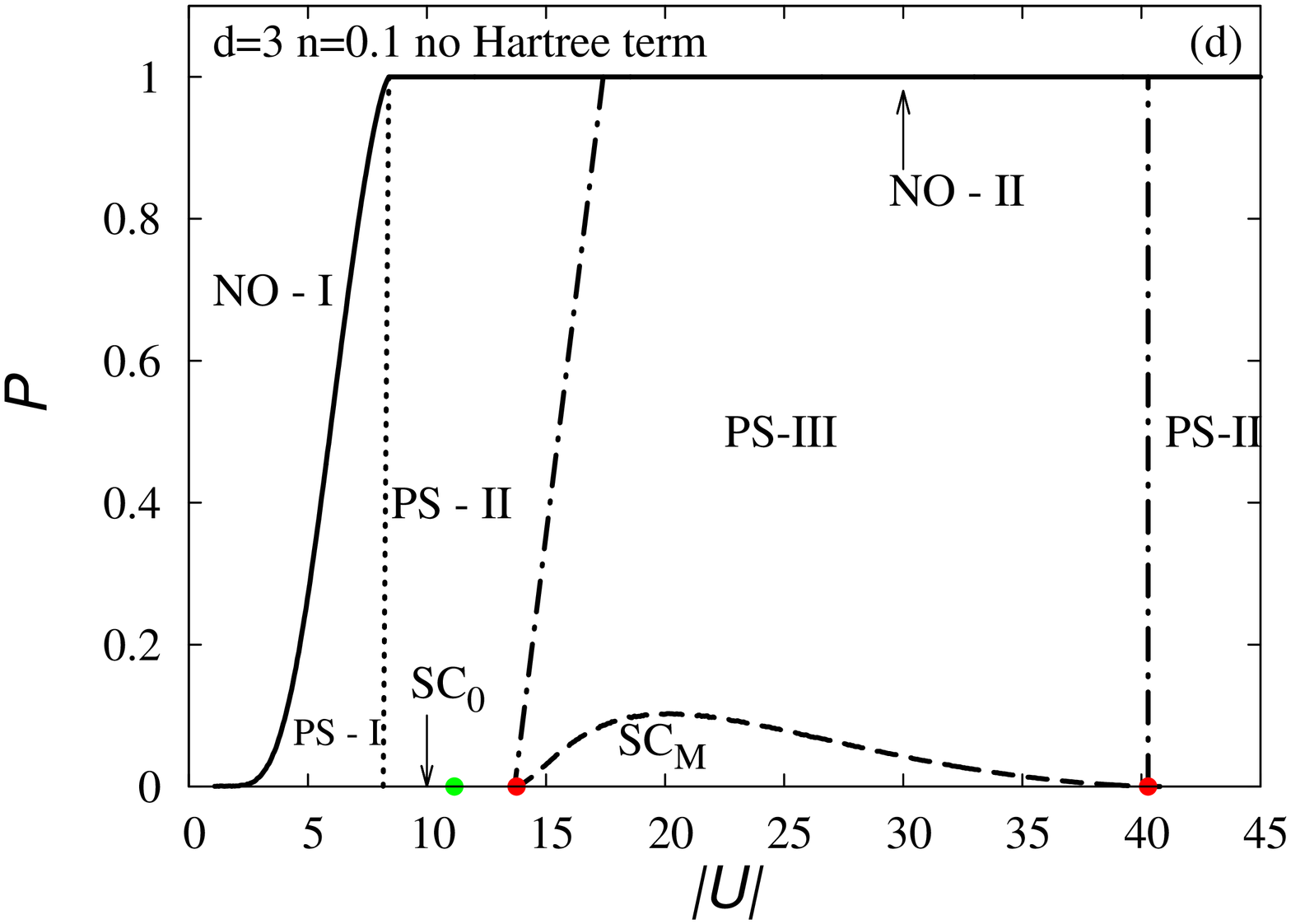}
\caption[Polarization vs. on-site attraction phase diagrams, at $T=0$ and fixed
$n=0.001$ (a), $n=0.03$ (b), $n=0.05$ (c), $n=0.1$ (d), for the simple cubic
lattice.]{\label{czerwony} Polarization vs. on-site attraction phase diagrams, at
$T=0$ and fixed $n=0.001$ (a), $n=0.03$ (b), $n=0.05$ (c), $n=0.1$ (d), for the
simple cubic lattice. SC$_0$ -- unpolarized superconducting state with
$n_{\uparrow}=n_{\downarrow}$, SC$_M$ -- magnetized superconducting state, PS-I
(SC$_0$+NO-I) -- partially polarized phase separation, PS-II (SC$_0$+NO-II) --
fully polarized phase separation, PS-III (SC$_M$+NO-II). Green point shows the
BCS-BEC crossover point, blue point -- tricritical point, red point --
$|U|_{c}^{SC_M}$. The diagrams are without the Hartree term.}
\end{figure}

Here we discuss in detail the BCS-BEC crossover for the simple cubic lattice
without the Hartree term. Fig. \ref{czerwony} shows (P-$|U|$) phase diagrams for
four fixed values of the particle concentration. The diagram for $n=0.001$ with
the Hartree term (Fig. \ref{3D_diag_n0001vsU_U_Hartree}(b)) has \textcolor{green}{already} been 
discussed. When we plot  Fig. \ref{czerwony}(a) in terms of $E_F$ and $k_F a_s$, we
can compare the case with and without the Hartree term in a very dilute limit.
The results are the same, both qualitatively and quantitatively. However, if the
number of particles increases, the range of occurrence of SC$_M$ narrows. At
$n=0.03$, for $|U|<30$, the diagram looks very similar to the diagram for
$n=0.001$. However, if the polarization and the attractive interaction increase,
the character of the transition changes from second to first order and a
second tricritical point appears. In the very strong coupling limit, at $P\neq
0$, phase separation is energetically more stable than the SC$_M$ phase. For
higher values of $n$ (see Fig. \ref{czerwony}(c) and \ref{czerwony}(d)) the range of
SC$_M$ is still decreasing and the system goes through phase separation to
the NO state for the whole range of parameter values. The tricritical point does
not exist. Thus, there are qualitative differences between the results for
the spin polarized attractive Hubbard model and for the continuum model of a
dilute gas of fermions. 

\begin{figure}[t!]
\hspace*{-0.8cm}
\includegraphics[width=0.38\textwidth,angle=270]
{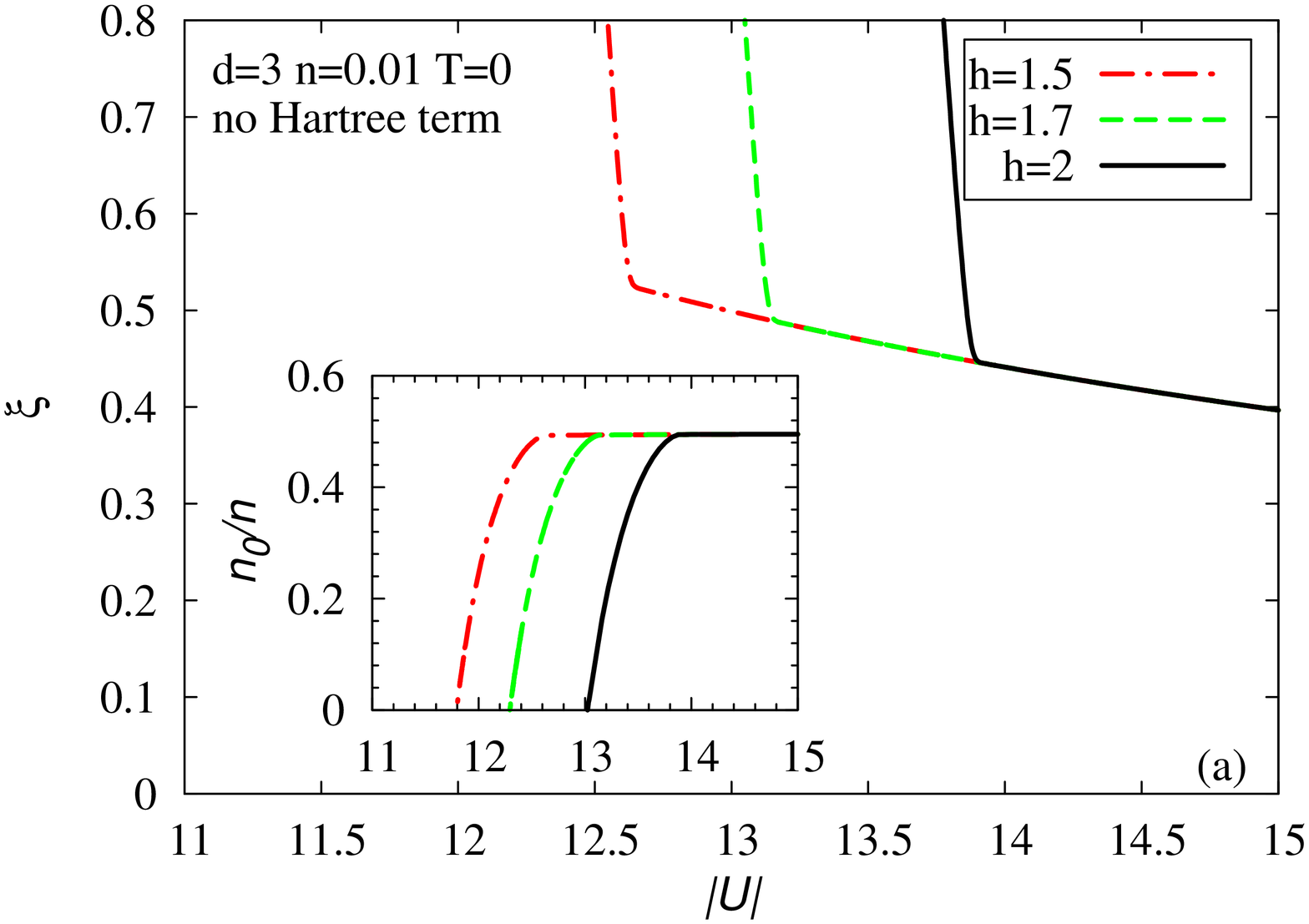}
\hspace*{-0.6cm}
\includegraphics[width=0.38\textwidth,angle=270]
{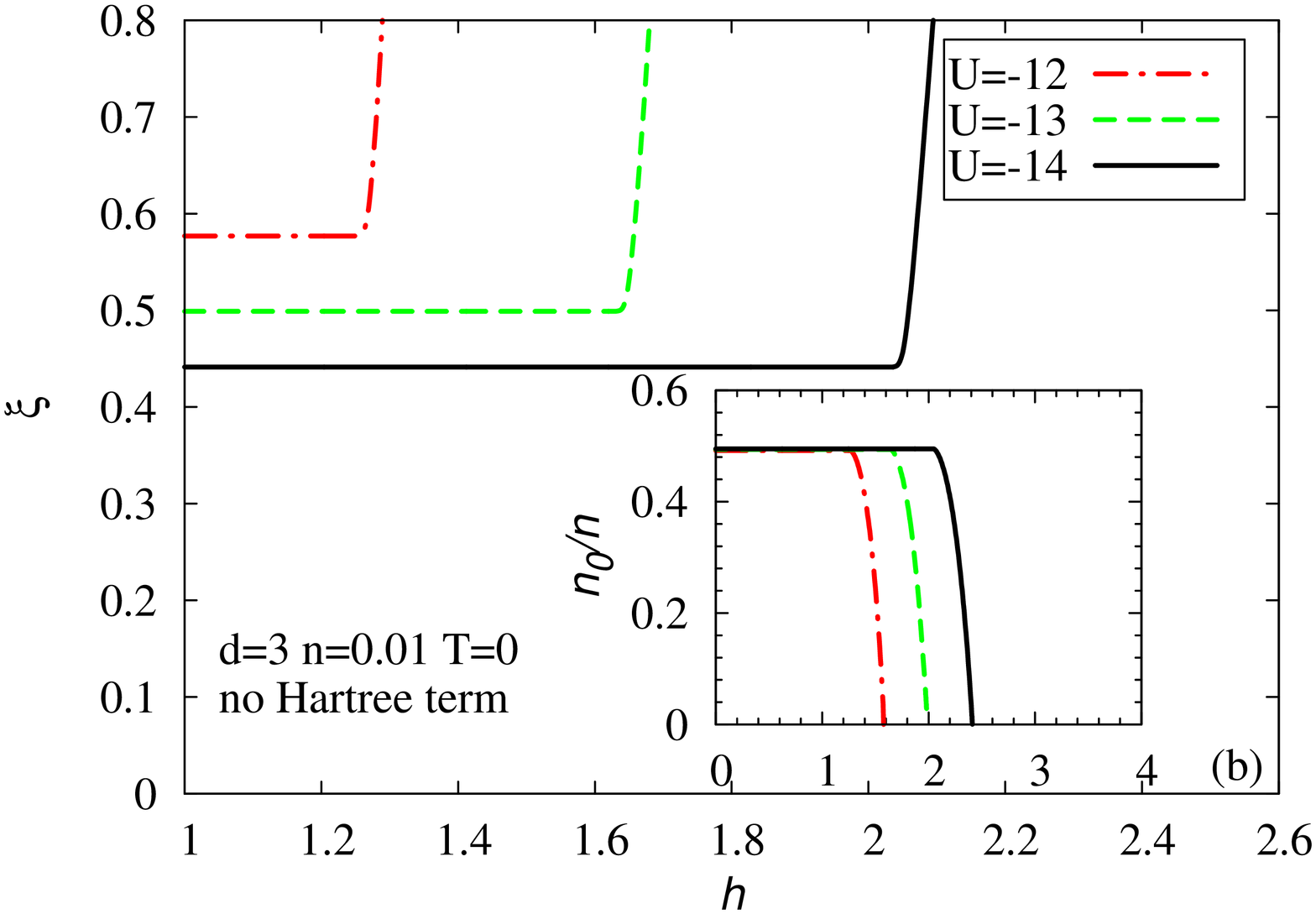}
\caption{\label{3D_xi_n001_r1} Dependence of the coherence length $\xi$ (in units of the lattice constant) on $h$ (a) and $|U|$ (b). \textcolor{czerwony}{$\xi$ is computed for finite but small $T$ ($T=5\times 10^{-3}$).} Fraction of pairs in condensate $n_0/n$ vs. $h$ (inset (a)) and $|U|$ (inset (b)); $d=3$, $n=0.01$.}
\end{figure}

A very important quantity from the point of view of the BCS-BEC crossover is
the coherence length $\xi$. The suggestions to study this quantity in this
context follow from the observations of the properties of exotic
superconductors. In agreement with the Uemura's plot \cite{uemura}, the
unconventional superconductors have a considerably shorter coherence length
($\sim 20-50$ \AA) than the conventional superconductors ($\xi \sim 10^3-10^4$
\AA). The coherence length $\xi$ can be defined \textcolor{czerwony}{through} the pair correlation function \textcolor{czerwony}{with opposite spins}
\cite{tobi, Pistolesi, Misawa}:
\begin{equation}
g(\vec{r} )=\frac{1}{n^2} \vert \langle \Phi \vert \Psi_{\uparrow}^{\dag}(\vec{r}) \Psi_{\downarrow}^{\dag} (0) \vert \Phi \rangle \vert^2,     
\end{equation}
(where: $\Psi_{\sigma} (\vec{r})$ -- fermionic field operator) in such a way that:
\begin{equation}
\label{xi}
\xi^2=\frac{\int d\vec{r}g(\vec{r}) \vec{r}^2}{\int d\vec{r} g(\vec{r})}=\frac{\sum_k \vert \nabla_k \phi_k \vert^2}{\sum_k \vert \phi_k \vert^2},
\end{equation}
\textcolor{czerwony}{where $\phi_{\vec{k}}$ is the pair wave function in the BCS ground state.}
\textcolor{czerwony}{For finite temperatures we use Eq. \eqref{xi} with $\phi_{\vec k} \rightarrow \mathcal{F}_{\vec k}$ $=u^{*}_{\vec{k}}\nu_{\vec{k}}[1-f(E_{\vec{k}\uparrow})-f(E_{\vec{k}\downarrow})]=\Delta \slash (E_{\vec{k}\uparrow}+E_{\vec{k}\downarrow})[1-f(E_{\vec{k}\uparrow})-f(E_{\vec{k}\downarrow})]$ which is the pair wave function at finite $T$ and $E_{\vec{k}\downarrow, \uparrow}$ are given by Eq.~\eqref{quasiparticles}. 
In the ground state, the pair wave function is: $\phi_{\vec{k}}=u^{*}_{\vec k}\nu_{\vec k}=\Delta \slash (E_{\vec{k}\uparrow}+E_{\vec{k}\downarrow})$ if $E_{\vec k \uparrow}>0$.}  

The condensate density is given:
\begin{equation}
 n_0=\frac{1}{N}\sum_{\vec k}|\mathcal{F}_{\vec k}|^2=\frac{1}{N}\sum_{\vec k} \frac{|\Delta|^2}{(E_{\vec{k}\uparrow}+E_{\vec{k}\downarrow})^2}[1-f(E_{\vec{k}\uparrow})-f(E_{\vec{k}\downarrow})]^2
\end{equation}
and is in general temperature dependent.
\textcolor{czerwony}{The fraction of pairs in the condensate} can be expressed by the formula:
\begin{equation}
\frac{n_0}{n}=\frac{1}{N} \sum_{\vec{k}} \frac{|\mathcal{F}_{\vec k}|^2}{n},
\end{equation}
where $n$ is the particle concentration.

Fig. \ref{3D_xi_n001_r1} shows the dependence of the coherence length $\xi$ (in
units of the lattice constant) and the fraction of condensed particles on
the magnetic field and attractive interaction, for $d=3$ and $n=0.01$. In
the strong coupling limit, the pair size is smaller than the lattice constant.
Only \textcolor{czerwony}{tightly bound} local pairs exist in the system (SC$_0$ phase) and the coherence
length does not depend on the magnetic field. The transition from SC$_0$ to
SC$_M$ (see Fig. \ref{3D_diag_n001vsU_U-12_vsn}(a)) is manifested through
\textcolor{czerwony}{sharp} increase in $\xi$, because the Pauli Exclusion Principle prohibits the
bound pairs and the excess of spin-up fermions to occupy the same lattice site. 
\textcolor{czerwony}{This behaviour of $\xi$ seems to be characteristic for a topological transition to gapless superfluid state.}
\textcolor{czerwony}{The $SC_M$ phase can be considered as a miscible mixture of tightly bound pairs (composite bosons) and excess of spin-up fermions. The interaction between bosons and unpaired fermions is repulsive.
If $n_p=n_{\downarrow}$ is number of pairs then $n_F=n_{\uparrow}-n_{\downarrow}$ is number of unpaired fermions such that $n=2n_{p}+n_{F}$ is constant. This interpretation is consistent with the fact that
the chemical potential (see Fig. \ref{3D_del_n001_h_U_multiplot}) is below the band bottom indicating the bosonic regime \cite{Pilati}.} 

The character of the magnetized superfluid phase is also reflected in the
features of the fraction of condensed particles. The maximum value
of the
ratio $n_0/n$ is 0.5, because if all particles in the system form pairs and
condense, their concentration is twice lower than the initial electron
concentration. If the system is in the SC$_0$ state (in the strong coupling
regime), $n_0/n$ does not depend on the magnetic field and takes the largest
possible value. The fraction of condensed pairs decreases rapidly after the
transition to SC$_M$ and equals zero in the NO state. There is a cusp in the
$n_0/n$ vs. magnetic field and $|U|$ plots, at the point of SC$_0$ $\rightarrow$
SC$_M$ transition.

\begin{figure}[p!]
\begin{center}
\includegraphics[width=0.44\textwidth,angle=270]
{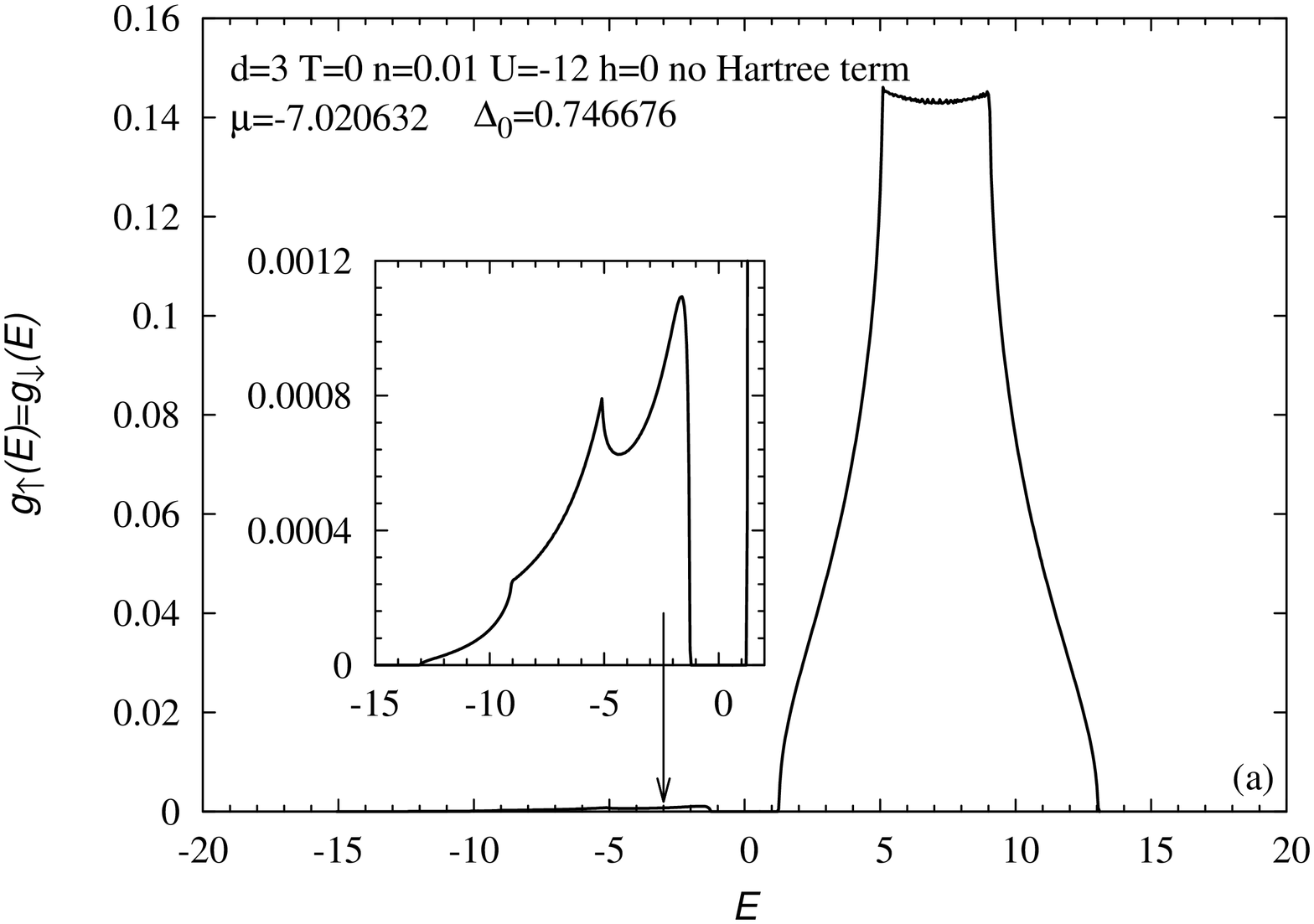}
\includegraphics[width=0.44\textwidth,angle=270]
{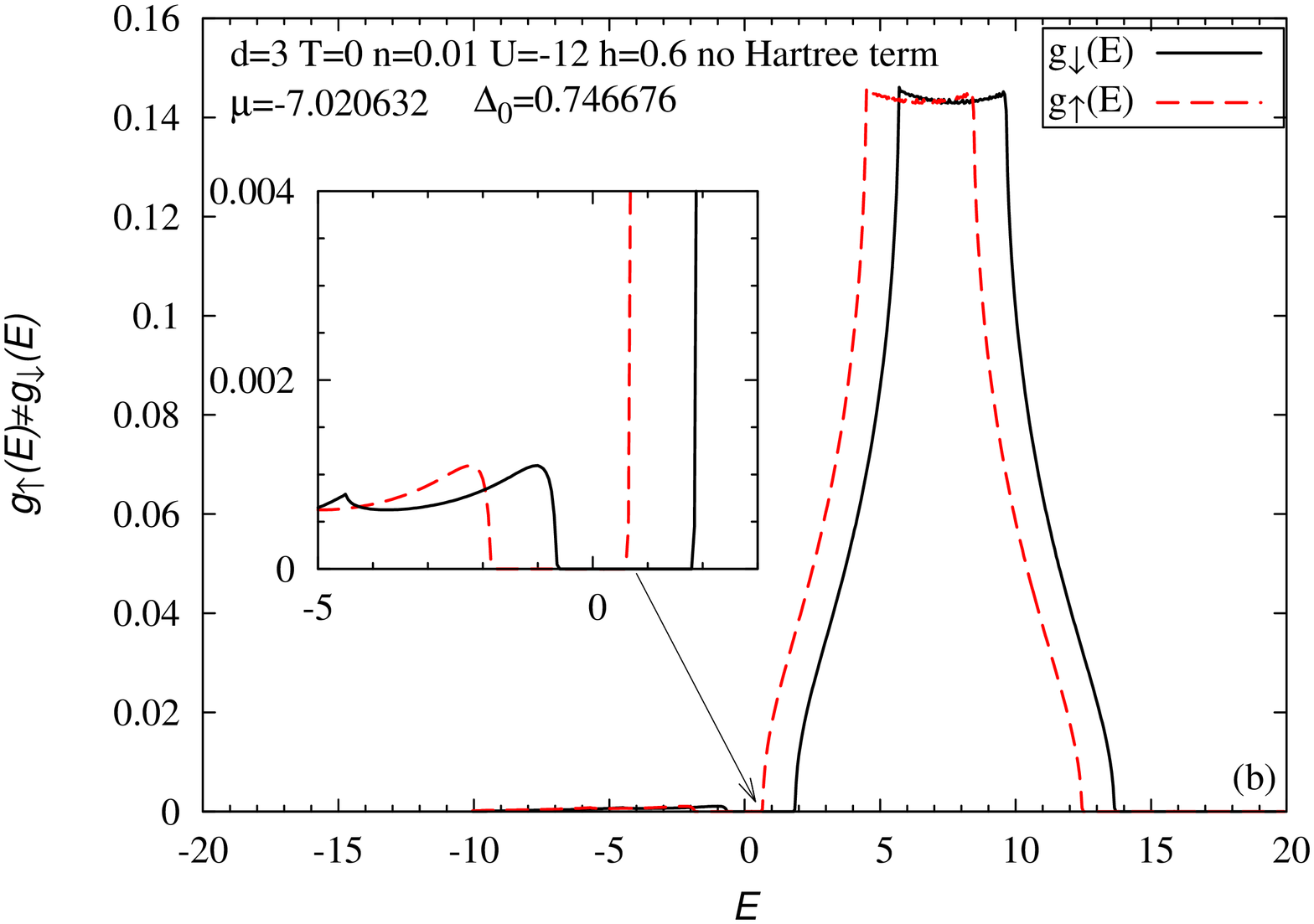}
\includegraphics[width=0.44\textwidth,angle=270]
{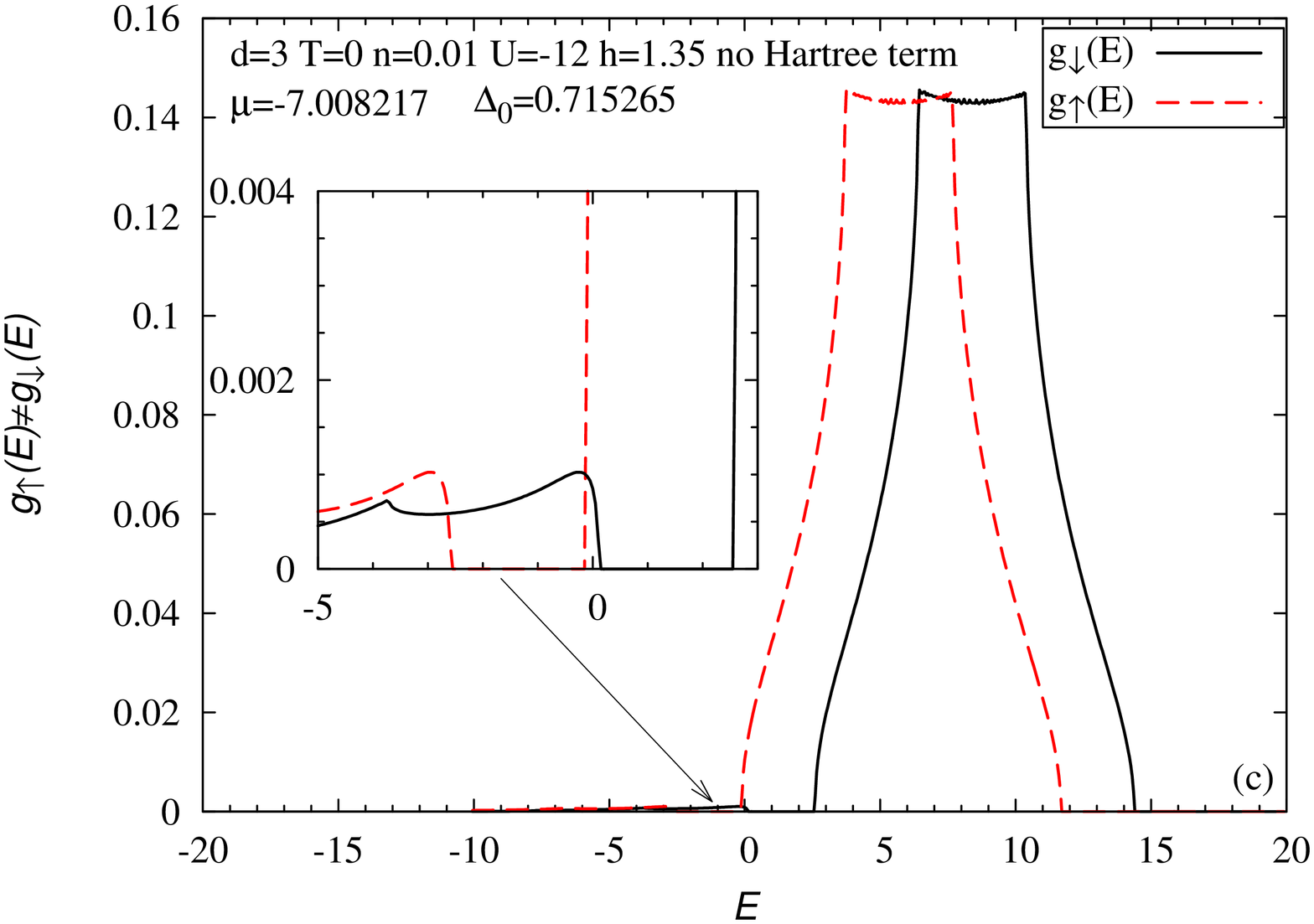}
\caption[\label{figDensity} Density of states for $d=3$, $n=0.01$, $U=-12$,
$h=0$ and two different values of the magnetic field.]{\label{figDensity}
Density of states for $d=3$, $n=0.01$, $U=-12$, $h=0$ and two different values
of the magnetic field. 
}
\end{center}
\end{figure}

Now, we also discuss the density of states features in the strong
coupling regime, paying special attention to changes in the quasiparticle
spectrum in the SC$_M$ phase.

The density of states in the superconducting ground state is determined from
Eq.~\eqref{DOS}.

Fig. \ref{figDensity} shows the density of states plot for the simple cubic
lattice, $n=0.01$, $U=-12$ (LP region). There exists an energy gap at $h=0$,
both in the weak and strong coupling regime. The gap in the density of states is
fixed by the location of the logarithmic singularities. However, in the BCS and
LP limits the gap width is determined in a different way. As mentioned
above,
for the s-wave pairing symmetry case, the energy gap width in the density of
states equals $E_g=2\Delta$, in the BCS regime. However, since $\bar{\mu}$
lies below the bottom of the band in the LP regime, the behavior of the
quasiparticle energies changes and the energy gap width in the density of
states equals $E_g=2\sqrt{(\epsilon_0-\bar{\mu})^2+|\Delta|^2}$.

In the presence of a magnetic field, the densities of states are
shifted by $h$ to the left ($g_\uparrow(E)$) and to the right
($g_\downarrow(E)$).
For $h=0.6$ (Fig. \ref{figDensity}(b)), the gapped regions $g_\uparrow(E)=0$
and $g_\downarrow(E)=0$ still overlap and the total density of states $g(E)$
has a gap.
However, for magnetic fields higher than
$E_{g}/2=\sqrt{(\epsilon_0-\bar{\mu})^2+|\Delta|^2}$ (Fig.
\ref{figDensity}(c), for this case $h=1.35$, while $E_g/2\approx1.23$),
i.e. after the
transition to the SC$_M$ state (see Fig. \ref{3D_diag_n001vsU_U-12_vsn}(a)), the
energy gap does not exist in the total density of states $g(E)$.
This also implies that the quasiparticle energy $E_{\vec{k}\uparrow}$ becomes
gapless.
Hence, the SC$_M$ state is characterized by
a gapless spectrum for the majority spin species when $h>E_{g}/2$.

\section{The influence of different lattice geometries (densities of states) on the stability of the magnetized superfluid phase}

\begin{figure}[h!]
\begin{center}
\hspace*{-1.0cm}
\includegraphics[width=0.38\textwidth,angle=270]{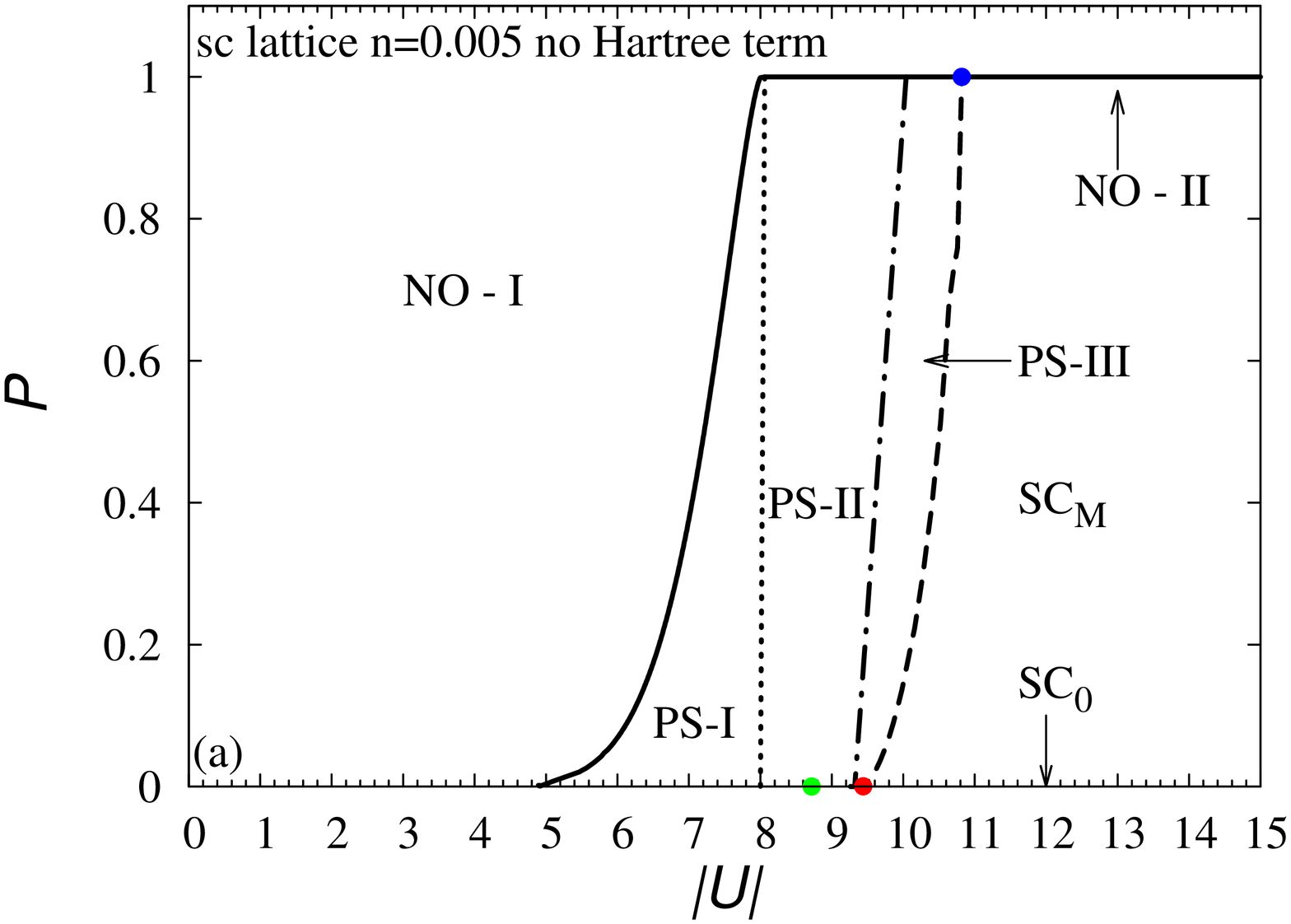}
\hspace*{-0.8cm}
\includegraphics[width=0.38\textwidth,angle=270]{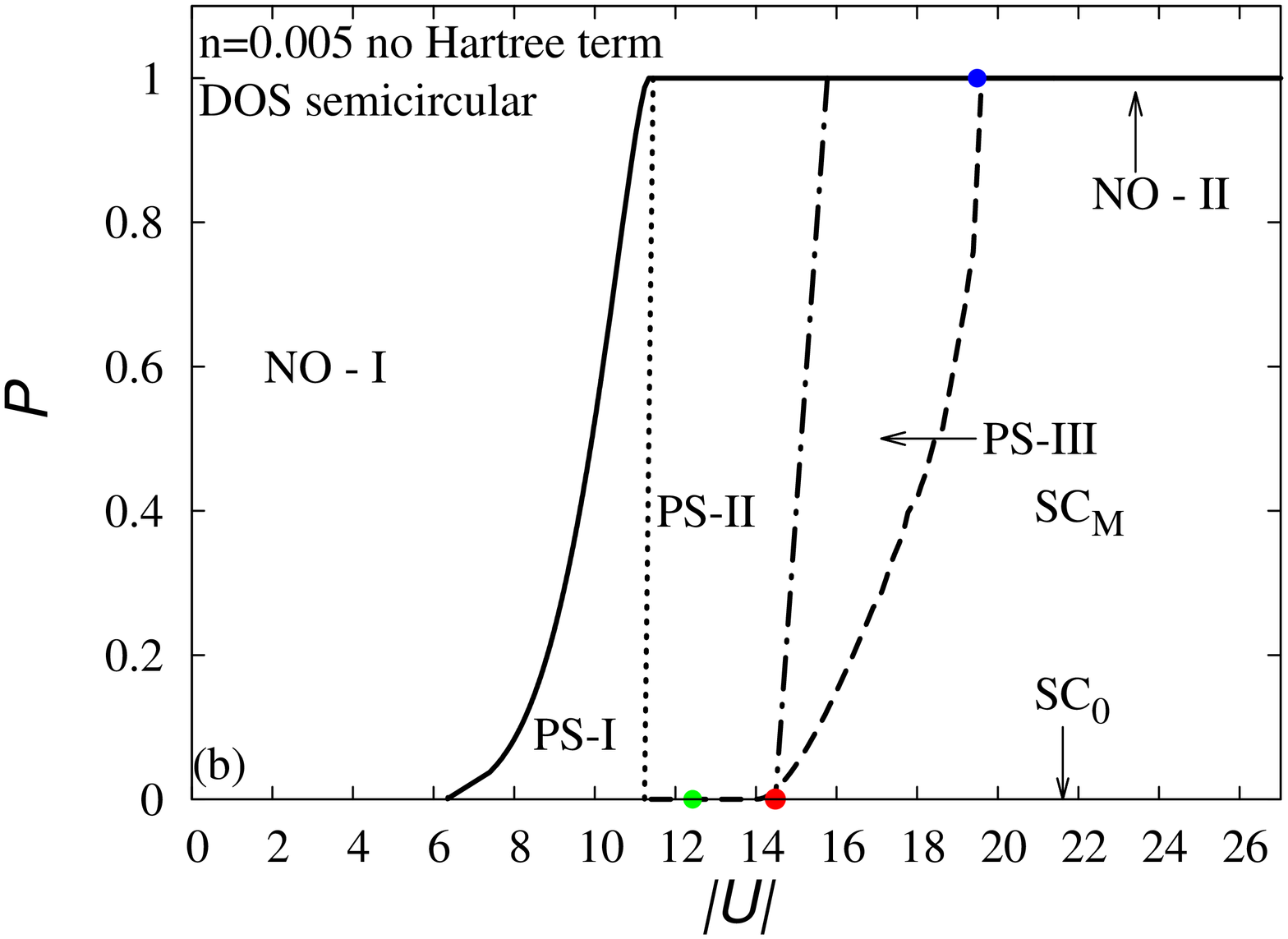}
\includegraphics[width=0.38\textwidth,angle=270]{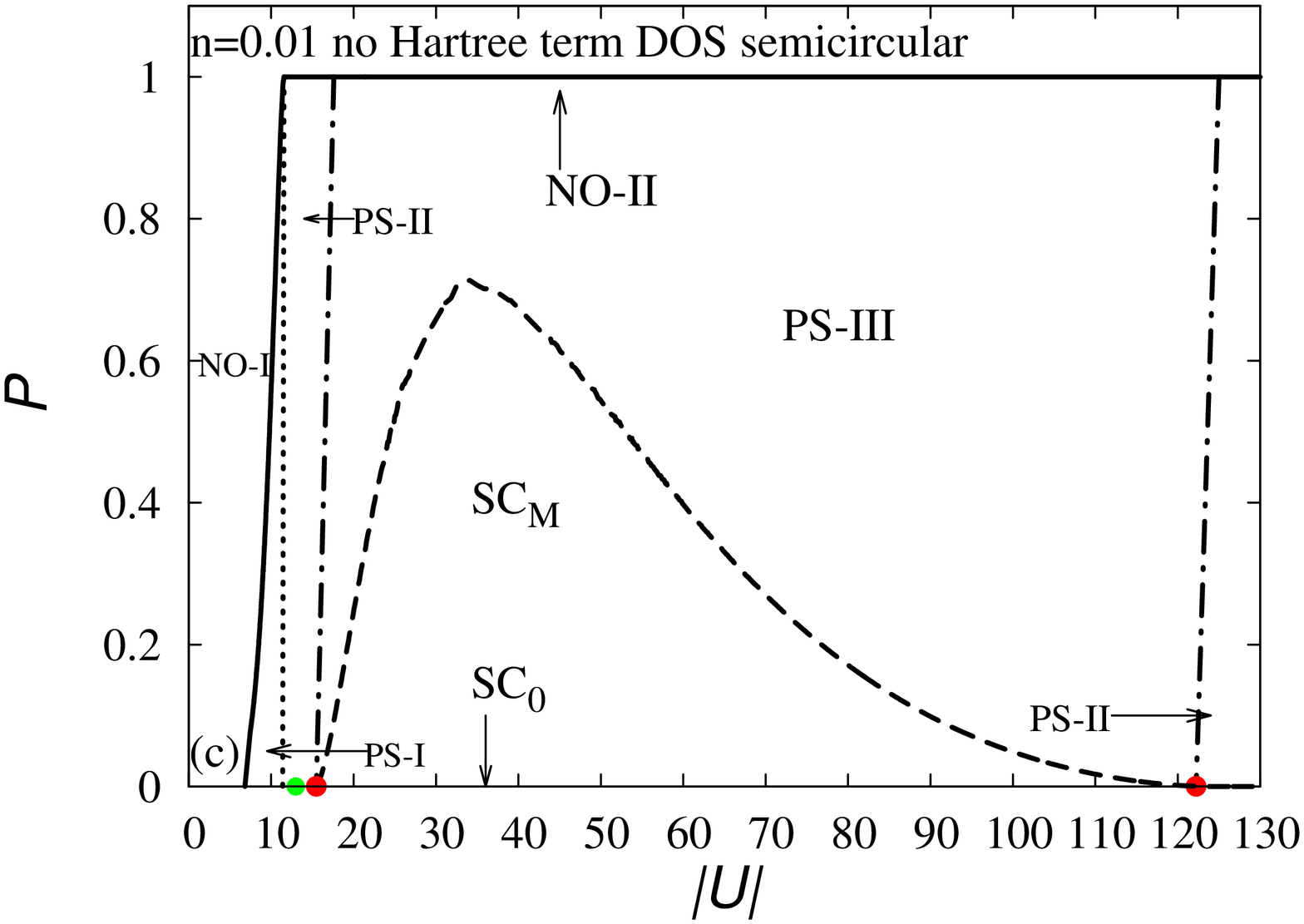}
\caption[Polarization vs. $|U|$ ground state phase diagrams for (a) sc lattice $n=0.005$, (b) semicircular DOS $n=0.005$, (c) semicircular DOS $n=0.01$. Diagrams without the Hartree term.]{\label{diagrams_3D_semicircular_n0005} Polarization vs. $|U|$ ground state phase diagrams for (a) sc lattice $n=0.005$, (b) semicircular DOS $n=0.005$, (c) semicircular DOS $n=0.01$. SC$_0$ -- unpolarized superconducting state, SC$_M$ -- magnetized superconducting state, PS-I (SC$_0$+NO-I) -- partially polarized phase separation, PS-II (SC$_0$+NO-II) -- fully polarized phase separation, PS-III (SC$_M$+NO-II). Green point -- BCS-BEC crossover point, blue point -- tricritical point, red point -- $|U|_{c}^{SC_M}$. \textcolor{green}{The diagrams are constracted in units of $t$.}}
\end{center}
\end{figure}

\begin{figure}[t!]
\begin{center}
\hspace*{-1.0cm}
\includegraphics[width=0.38\textwidth,angle=270]{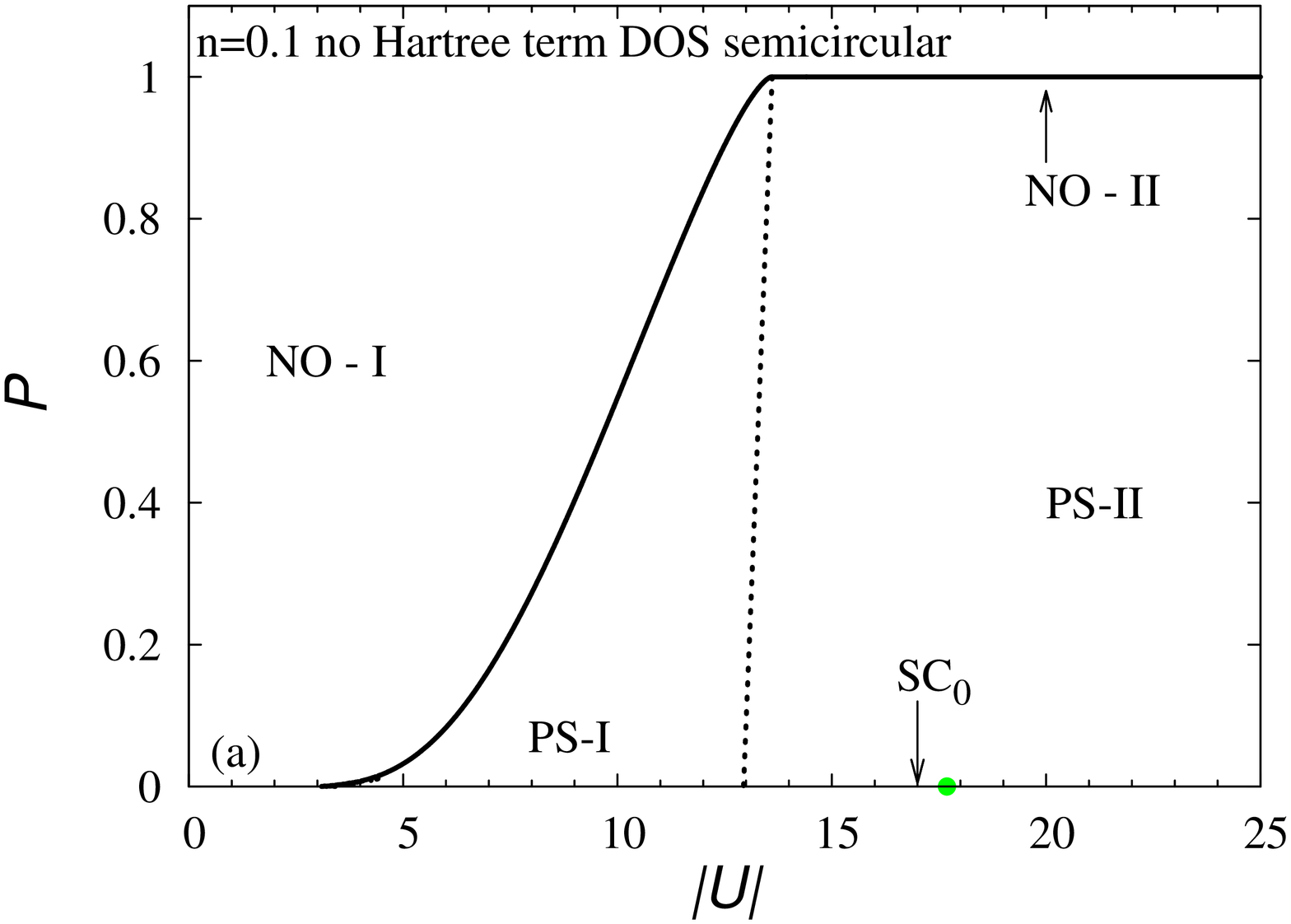}
\hspace*{-0.8cm}
\includegraphics[width=0.38\textwidth,angle=270]{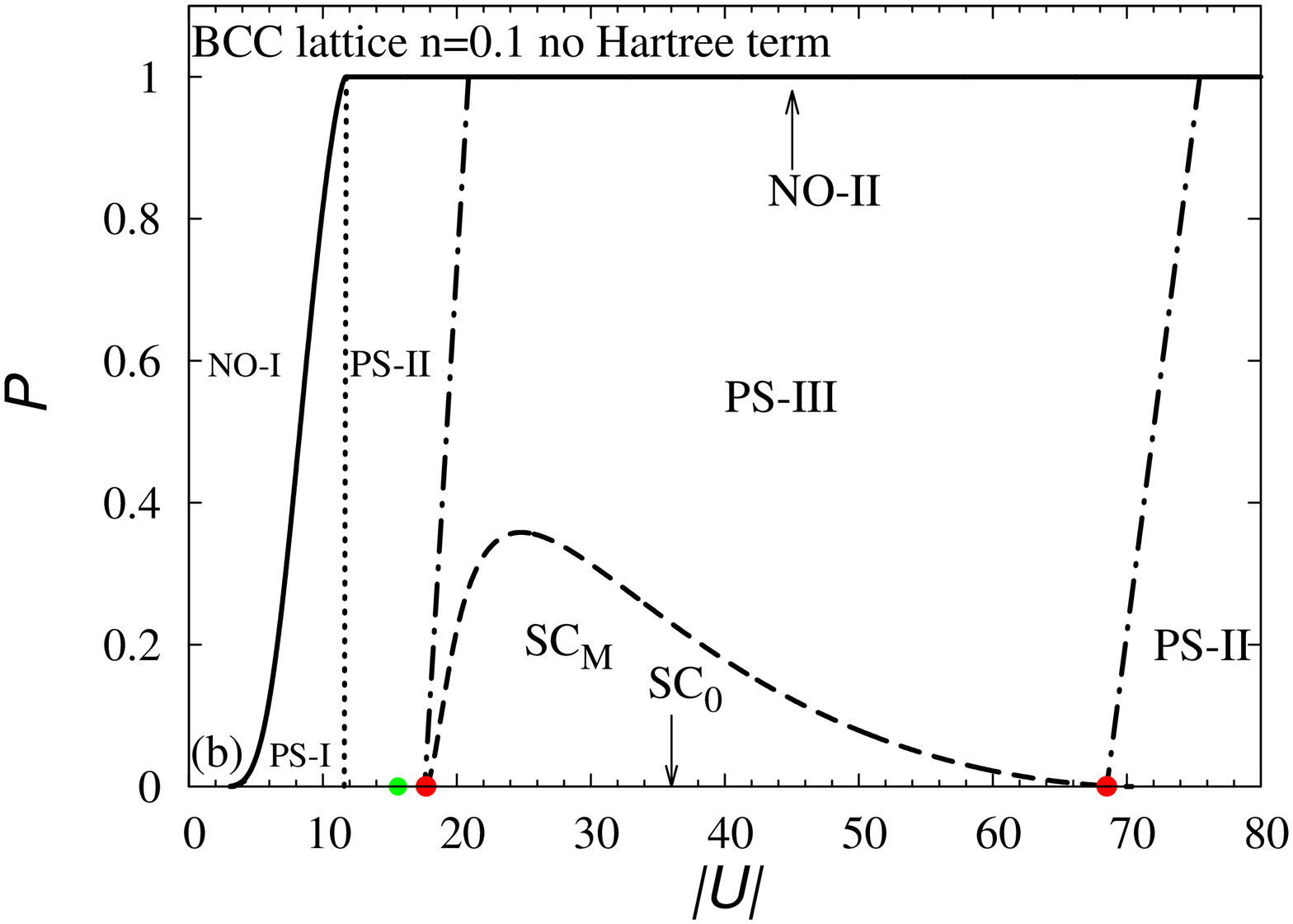}
\includegraphics[width=0.38\textwidth,angle=270]{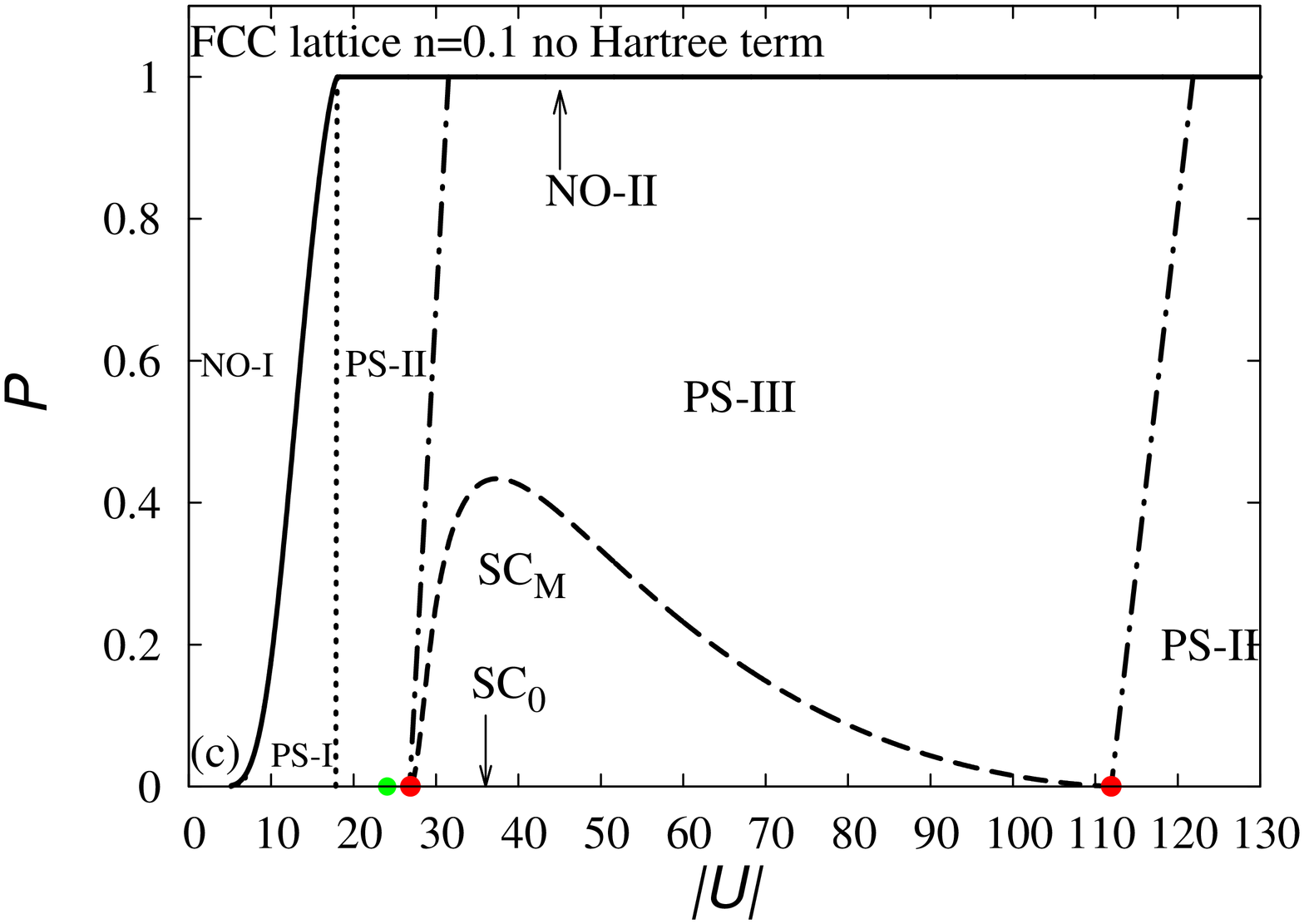}
\caption[Polarization vs. $|U|$ ground state phase diagrams, at fixed $n=0.1$, for (a) semicircular DOS, (b) BCC lattice, (c) FCC lattice. Diagrams without the Hartree term.]{\label{diagrams_3D_densities_n01} Polarization vs. $|U|$ ground state phase diagrams, at fixed $n=0.1$, for (a) semicircular DOS, (b) BCC lattice, (c) FCC lattice. SC$_0$ -- unpolarized superconducting state, SC$_M$ -- magnetized superconducting state, PS-I (SC$_0$+NO-I) -- partially polarized phase separation, PS-II (SC$_0$+NO-II) -- fully polarized phase separation, PS-III (SC$_M$+NO-II). Green point -- BCS-BEC crossover point, blue point -- tricritical point, red point -- $|U|_{c}^{SC_M}$.}
\end{center}
\end{figure}

\begin{figure}[t!]
\hspace*{-0.8cm}
\includegraphics[width=0.38\textwidth,angle=270]
{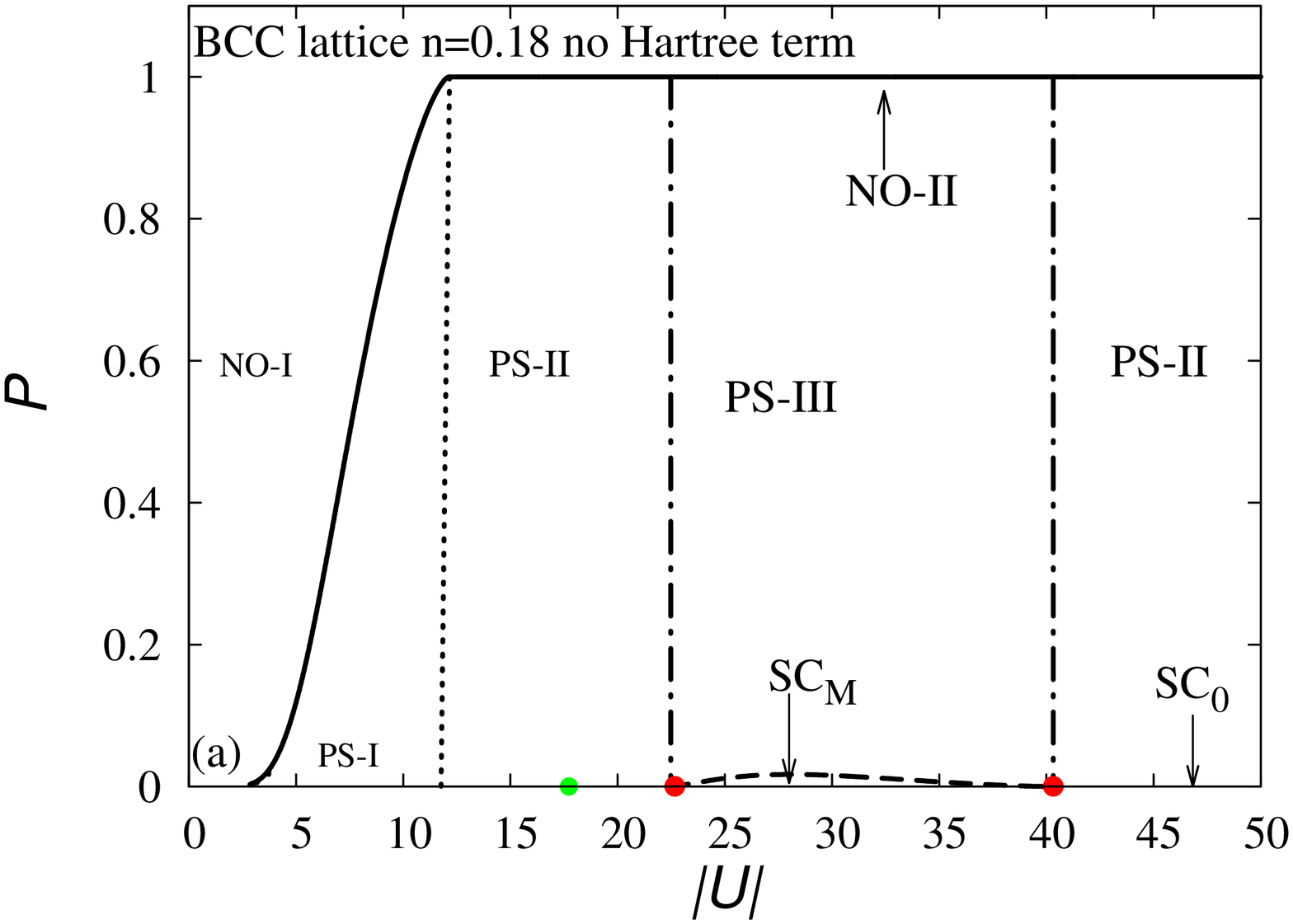}
\hspace*{-0.6cm}
\includegraphics[width=0.38\textwidth,angle=270]
{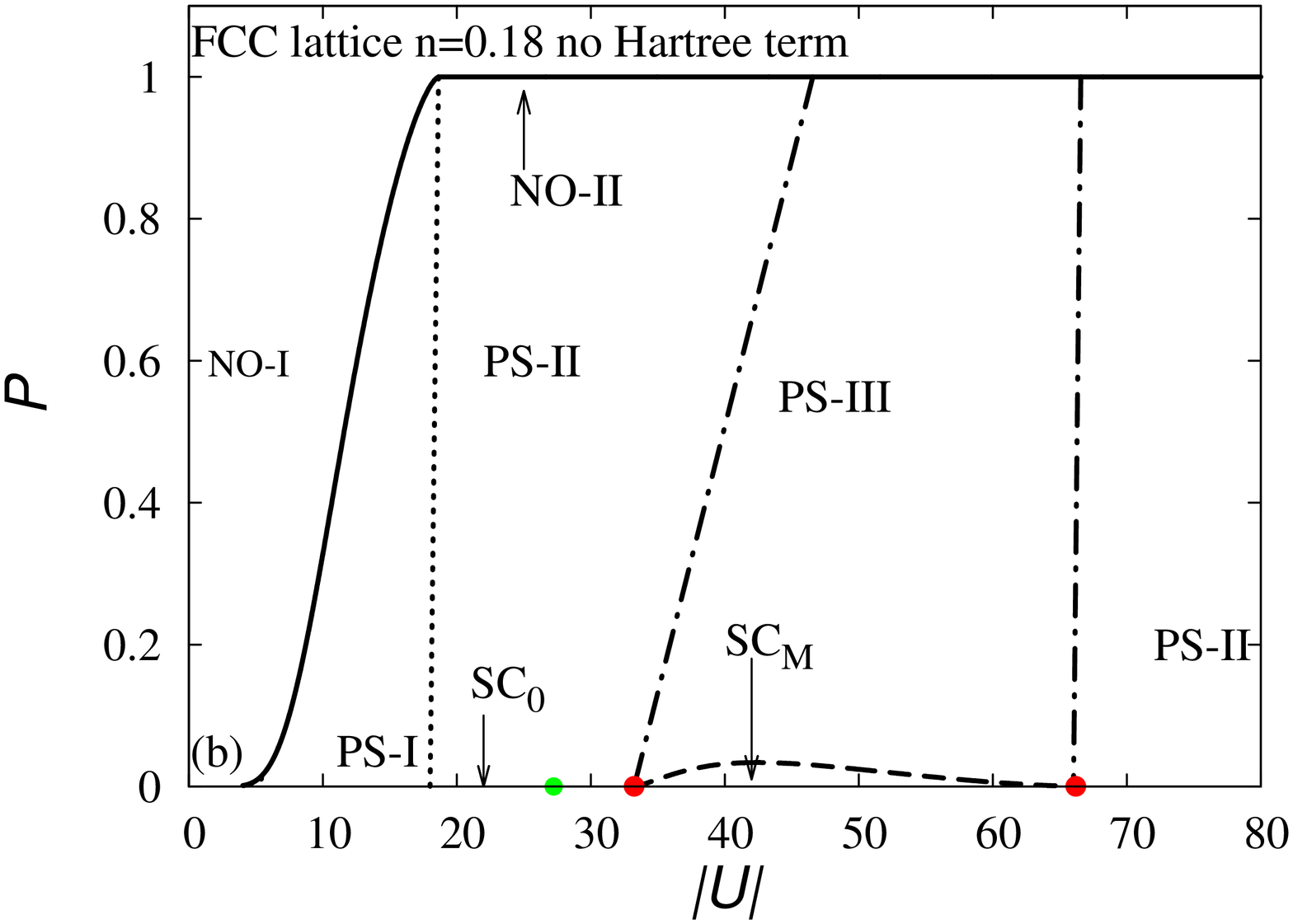}
\caption[Polarization vs. $|U|$ ground state phase diagrams, at fixed $n=0.18$, for (a) BCC lattice (b) FCC lattice. Diagrams without the Hartree term.]{\label{diagrams_3D_densities_n018} Polarization vs. $|U|$ ground state phase diagrams, at fixed $n=0.18$, for (a) BCC lattice, (b) FCC lattice. SC$_0$ -- unpolarized superconducting state, SC$_M$ -- magnetized superconducting state, PS-I (SC$_0$+NO-I) -- partially polarized phase separation, PS-II (SC$_0$+NO-II) -- fully polarized phase separation, PS-III (SC$_M$+NO-II). Green point -- the BCS-BEC crossover point, blue point -- tricritical point, red point -- $|U|_{c}^{SC_M}$.}
\end{figure}

\begin{figure}[t!]
\begin{center}
\includegraphics[width=0.55\textwidth,angle=270]{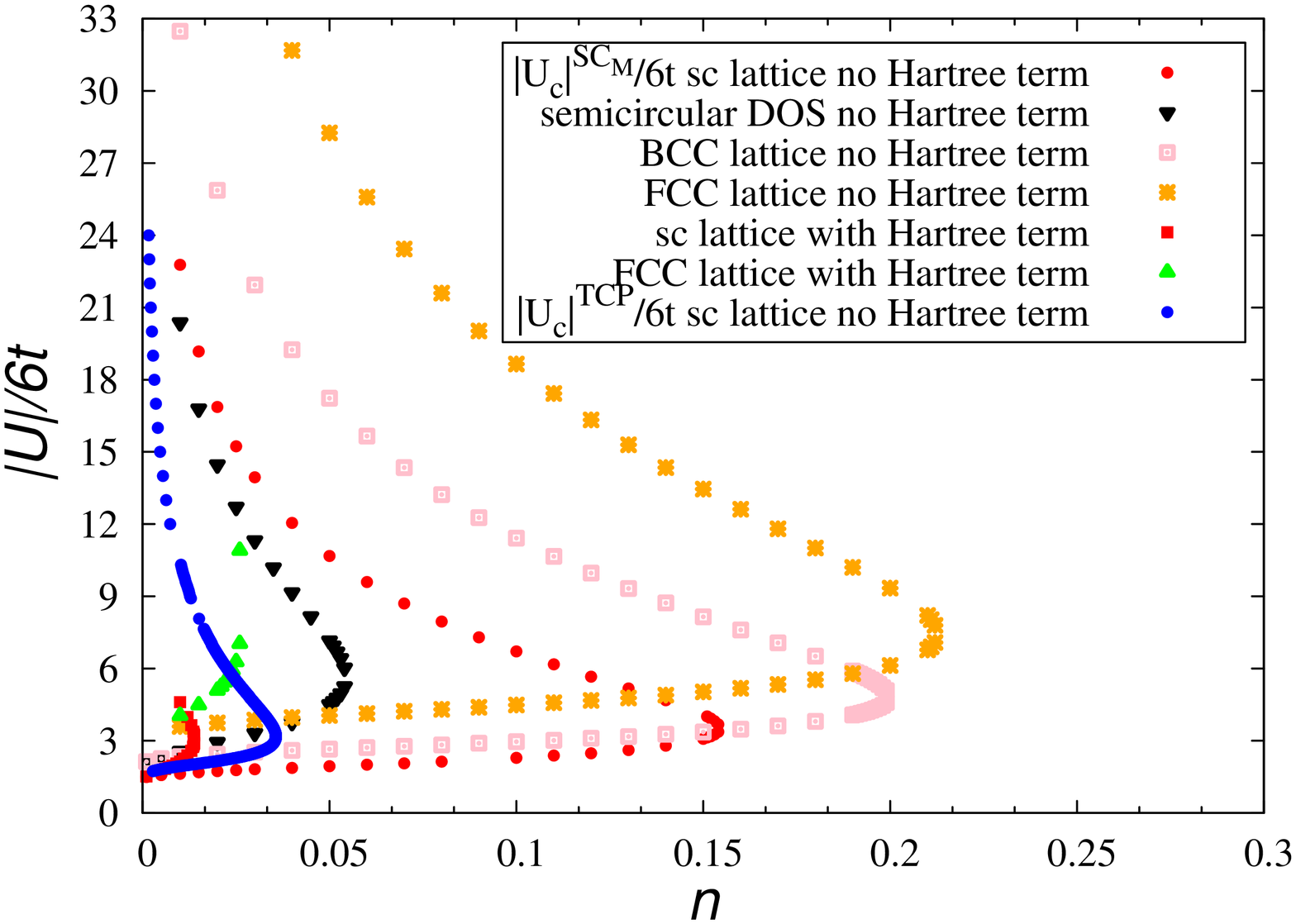}
\caption[Critical values of the attraction, for which the SC$_M$ state arises stable and tricritical points at $T=0$ vs. electron concentration.]{\label{magnetized_r1} Critical values of the attraction, for which the SC$_M$ state arises stable and tricritical points at $T=0$ vs. electron concentration. $h_c^{SC_M}=\sqrt{(\bar{\mu}-\epsilon_0)^2+|\Delta|^2}$, where $\Delta=\Delta(h=0)$.}
\end{center}
\end{figure}

In this section, we briefly discuss the influence of different lattice
geometries on the stability of the SC$_M$ phase. In the previous section, we
have analyzed the impact of the Zeeman magnetic field on the BCS-BEC crossover
at $T=0$ for $d=3$ simple cubic lattice (sc) and shown that the existence of
the SC$_M$ phase is possible for the sc lattice case and for spin independent
hopping integrals. 

Here, we construct the BCS-BEC crossover phase diagrams at $T=0$, for the
body-centered cubic (BCC) and face-centered cubic (FCC) lattices. As mentioned
above, the results contained in this thesis have been obtained by
numerical
calculations. However, the sums over the first Brillouin zone were
performed with the use of the density of states, whenever possible. The formulas
for the density of states are presented in Appendix \ref{appendix4}. We also use
the semicircular (s-circ) DOS  \cite{Privitera} in our calculations, which can
be realized as the Bethe lattice (tree lattice in infinite dimensions) or as an approximate DOS in $d=3$ but without Van Hove's singularities. 

Obviously, the Hubbard model is a lattice model and is expected to give similar
results as its continuum counterpart (universal behavior) only in the dilute
limit, when both $n\rightarrow 0$ and $a\rightarrow 0$ ($a$ -- lattice
constant). Therefore, the specifics of the lattice are relevant, at every finite
$n$. Hence, the range of occurrence of the SC$_M$ phase can be essentially
different for different lattice geometries.

Now, we briefly discuss the ground state phase diagrams for different types of lattices. 

As mentioned above, the critical value of attraction for bound state
formation in the empty lattice equals $|U_c|^{sc}/12t=0.659$, for sc. In turn,
for the semicircular DOS one finds $|U_c|^{s-circ}\approx 10.80732$
\cite{Privitera}.
Therefore, one can expect that there are some quantitative differences in the
phase diagrams between the sc and semicircular DOS, even at low electron
concentration. 

Fig. \ref{diagrams_3D_semicircular_n0005}(a)-(b) shows the (P-$|U|$) diagrams at
fixed $n=0.005$, for the sc and semicircular DOS. The topology of these
diagrams is the same. There is the SC$_0$ state for low values of attraction.
With increasing polarization, there is the 1$^{st}$ order transition to the
normal phase, through the PS-I or PS-II regions. If $|U|$ increases, the SC$_M$
state appears in the phase diagrams, both for the sc and semicircular DOS.
However, the SC$_M$ phase moves towards higher $|U|$ values, in the
semicircular DOS case. Therefore, the location of TCP is also different in this
case. Moreover, the PS-III region is larger for the s-circ DOS than for the sc
lattice. Obviously, the range of occurrence of the SC$_M$ state decreases with
increasing electron concentration (Fig.
\ref{diagrams_3D_semicircular_n0005}(c)).

We have performed a similar comparative analysis for different types of
lattices, at fixed $n=0.1$ (Fig. \ref{diagrams_3D_densities_n01}). In the s-circ
DOS case, with increasing polarization, PS is energetically favored, i.e. even
on the LP side, SC$_M$ is unstable. The situation is different in the sc
case. As shown in section \ref{3Dcase} (Fig. \ref{czerwony}(d)), SC$_M$ is
stable for fixed $n=0.1$, but the range of SC$_M$ is decreasing and the system
goes through phase separation to the NO state for the whole range of
parameter values. The tricritical point does not exist. One can observe a
similar behavior for the BCC and FCC lattices. However, the range
of
occurrence of SC$_M$ is much larger than in the other cases (s-circ and sc DOS).
While, for the sc lattice, a critical value of $n$ above which the SC$_M$ state
becomes unstable equals $n\approx 0.154$ (for semicircular DOS this value is
even smaller), one can find the SC$_M$ phase in the diagrams for FCC and BCC
lattices case, even at fixed $n=0.18$ (Fig. \ref{diagrams_3D_densities_n018}). 

We have also performed an analysis of the evolution of the critical values of
the attraction, for which the SC$_M$ state becomes stable for different types of
lattices and of the tricritical points (blue circle points, without the Hartree
term in the sc lattice case) with increasing $n$ (Fig. \ref{magnetized_r1}). The
occurrence of the SC$_M$ phase depends on lattice structure. As shown in
this chapter, SC$_M$ is unstable for $d=2$, but it can be realized for $d=3$
lattices.

We have also investigated the influence of the Hartree term on the SC$_M$ phase
stability. 

In the very dilute limit, there is only one value of $|U_c|^{SC_M}$ and it is
the same for the case with and without the Hartree term, for all types of
lattices. \textcolor{czerwony}{For each value of $n<n_c$ ($n_c$ defined below), there are two critical values of the
attraction for which the $SC_M$ state becomes stable
(except for the very dilute limit, where there is only the lower critical value,
 i.e. the upper critical value becomes infinite).
The system is in the $SC_M$ phase between
the lower and upper critical points in this plot.} 
\textcolor{czerwony}{However}, with increasing particle concentration, the range of stability of
SC$_M$ is smaller and there exists a critical value of $n$ ($n_c$) above which
the SC$_M$ state becomes unstable. For the semicircular DOS case, without the
Hartree term, this value is the lowest and equals $n_c\approx 0.054$, while for
the sc lattice $n_c=0.154$, for the BCC lattice $n_c\approx 0.199$ and the
highest for the FCC lattice $n_c\approx 0.212$. Additionally, as opposed to the
results for the continuum model of a dilute gas of fermions, there are
always two values of $|U_c|^{SC_M}$, for higher values of $n$. We have also
found the
range of $n$, for which the transition from SC$_M$ to NO is of the first order,
even for a very strong attraction. 

The presence of the Hartree term restricts the range of occurrence of the SC$_M$
phase, which is clearly visible in Fig. \ref{magnetized_r1} (red square points
for the sc lattice and green triangular points for the FCC lattice). A critical
value of $n$ above which the SC$_M$ state becomes unstable equals $n\approx
0.0145$ (the sc lattice) and $n\approx 0.026$ (the FCC lattice). The Hartree
term, usually promoting ferromagnetism in the Stoner model ($U>0$), here
($U<0$), strongly competes with superconductivity. Thus, such a term restricts
the SC$_M$ state to lower densities. However, as we will show in the following
chapters, the mass imbalance can change this behavior even for $d=2$ due to
spin polarization stemming from the kinetic energy term.

\newpage
\thispagestyle{empty}
\mbox{}
\chapter{The BCS-BEC crossover at finite temperatures in the \textcolor{czerwony}{s}pin-\textcolor{czerwony}{p}olarized AHM with spin independent hopping integrals}
\label{6.0}
In this chapter, we extend the BCS-BEC crossover analysis to finite
temperatures by taking into account phase fluctuations in $d=2$ within the
Kosterlitz-Thouless scenario and pairing fluctuations in $d=3$ (self-consistent
T-matrix scheme). The T-matrix scheme goes beyond the standard mean-field,
since it includes the effects of non-condensed pairs with $\vec{q}\neq 0$ and
allows a description of the BEC regime of the crossover. In general, the
method works in the low density regime. The crossover diagrams include
the pseudogap state. Some of our results have been published in Ref.
\cite{Kujawa4}.

\section{2D square lattice. The Kosterlitz-Thouless scenario}

\begin{figure}[t!]
\hspace*{-0.8cm}
\includegraphics[width=0.38\textwidth,angle=270]{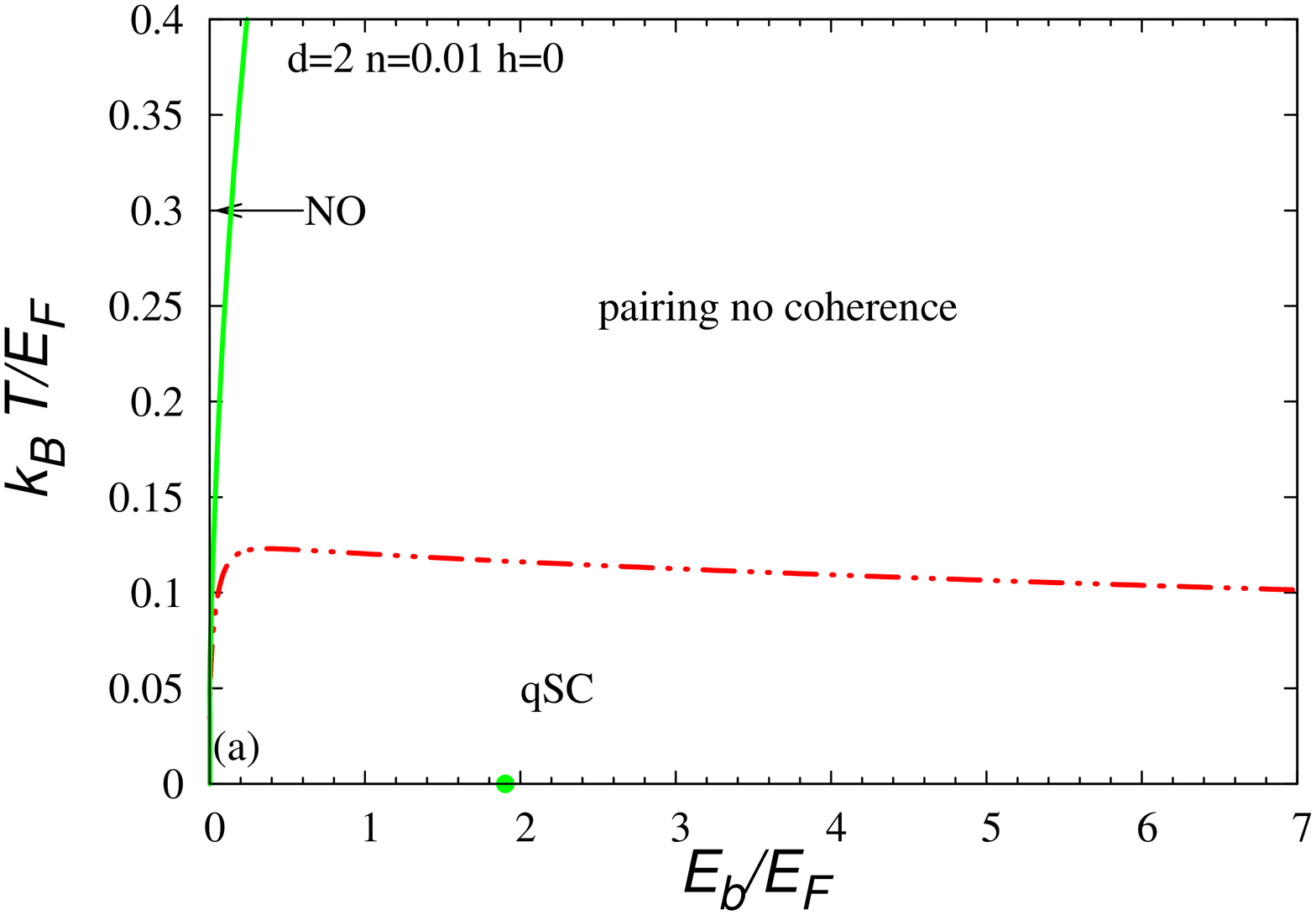}
\hspace*{-0.6cm}
\includegraphics[width=0.38\textwidth,angle=270]{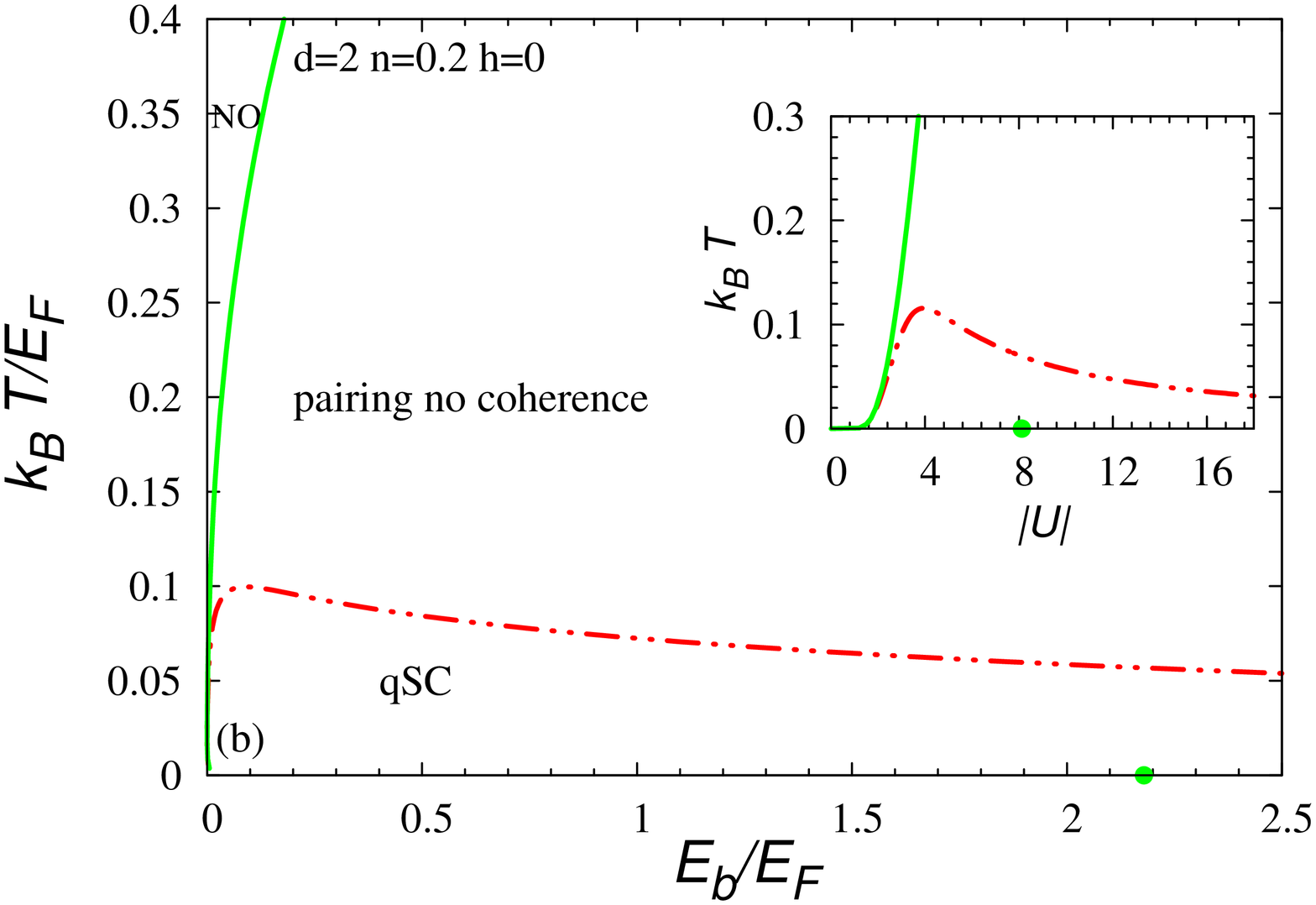}
\caption[Temperature vs. binding energy phase diagrams in units of the lattice Fermi energy at $h=0$, (a) $n=0.01$, (b) $n=0.2$ and T vs. $|U|$ phase diagram (inset of (b)).]{\label{kTvsE_b_h0} Temperature vs. binding energy phase diagrams in units of the lattice Fermi energy at $h=0$, (a) $n=0.01$, (b) $n=0.2$ and T vs. $|U|$ phase diagram (inset of (b)). Thick dashed-double dotted line (red color) is the KT transition line, thick solid line denotes transition from pairing without coherence region to NO within the Hartree approximation. $qSC$ -- superconductor (algebraic order). \textcolor{czerwony}{The green point in (a)-(b) shows} the BCS-LP crossover point at $T=0$.}
\end{figure}

Here, the results concerning the influence of \textcolor{czerwony}{a} magnetic field on superfluidity at 
finite temperatures are presented.

One should emphasize that phase transitions in quantum systems are
dependent on the dimensionality of these systems. Two-dimensional Fermi systems
exhibit features which are not observed in the three-dimensional case.
According to the Mermin-Wagner theorem, the long-range order does not exist in
one- and two-dimensional lattice spin systems, at non-zero temperatures, if
the system has a continuous symmetry group (e.g. the Heisenberg model or the XY
model).

As mentioned above, for $d=2$ AHM at $h=0$, the SC-NO transition is of the
Kosterlitz-Thouless type, i.e. below $T_c^{KT}$ the system has a
quasi-long-range (algebraic) order which is characterized by a power law decay
of the order parameter correlation function and non-zero superfluid stiffness.
According to Eq.~\eqref{KT}, the KT transition temperature is found from the
intersection point of the straight line $\frac{2}{\pi} k_B T$ with the curve
$\rho_s(T)$. In such a way, we can estimate the phase coherence temperatures and
extend the analysis of the crossover from the weak to strong coupling to
finite $T$ \cite{tobi}. The results obtained for the Kosterlitz-Thouless
temperatures give upper bounds on actual transition temperatures.

First, we analyze the influence of increasing attractive interaction on the
critical temperatures at $h=0$.
Fig. \ref{kTvsE_b_h0} shows the temperature vs. binding energy phase diagrams in
units of the lattice Fermi energy $E_F$ at $h=0$ and low values of the electron
concentration. The temperatures $T_c^{KT}$ (dashed-double dotted line (red
color) in the diagram) are generally much smaller than $T_c^{HF}$ (solid line
(green color) in the diagram) but on reducing the attraction, in the absence of
magnetic field, the difference between $T_c^{KT}$ and $T_c^{HF}$ decreases.

In the weak coupling regime, in which the value of the order parameter is low
in comparison with the value of the superfluid stiffness, $T_c^{KT}$ is
limited by the temperature determined in the Hartree-Fock approximation.
However, with increasing attraction, the order parameter increases, the value of
the superfluid density decreases and the quantity which limits the KT
temperatures is $\rho_s (T=0)$. In the strong coupling regime, $T_c^{KT}$ are
very well approximated by the value of $\frac{\pi}{2} \rho (0)$. This means that
despite the unlimited increase in $T_c^{HF}$ with increasing $|U|$ ($E_b$),
$T_c^{KT}$, after the increase to the maximum value for the optimal value of the
interaction, begin to decrease and are limited by the finite value of $\rho_s
(0)$ \cite{Denteneer-2, Denteneer, Schneider, chorowska, bak-2}. Similar
behavior of the Kosterlitz-Thouless critical temperatures (i.e. an increase
followed by a decrease) was also observed for the s-extended- and
d-wave pairing
symmetry cases, within the Extended Hubbard Model \cite{tobi}. 

\begin{figure}[h!]
\hspace*{-0.8cm}
\includegraphics[width=0.38\textwidth,angle=270]
{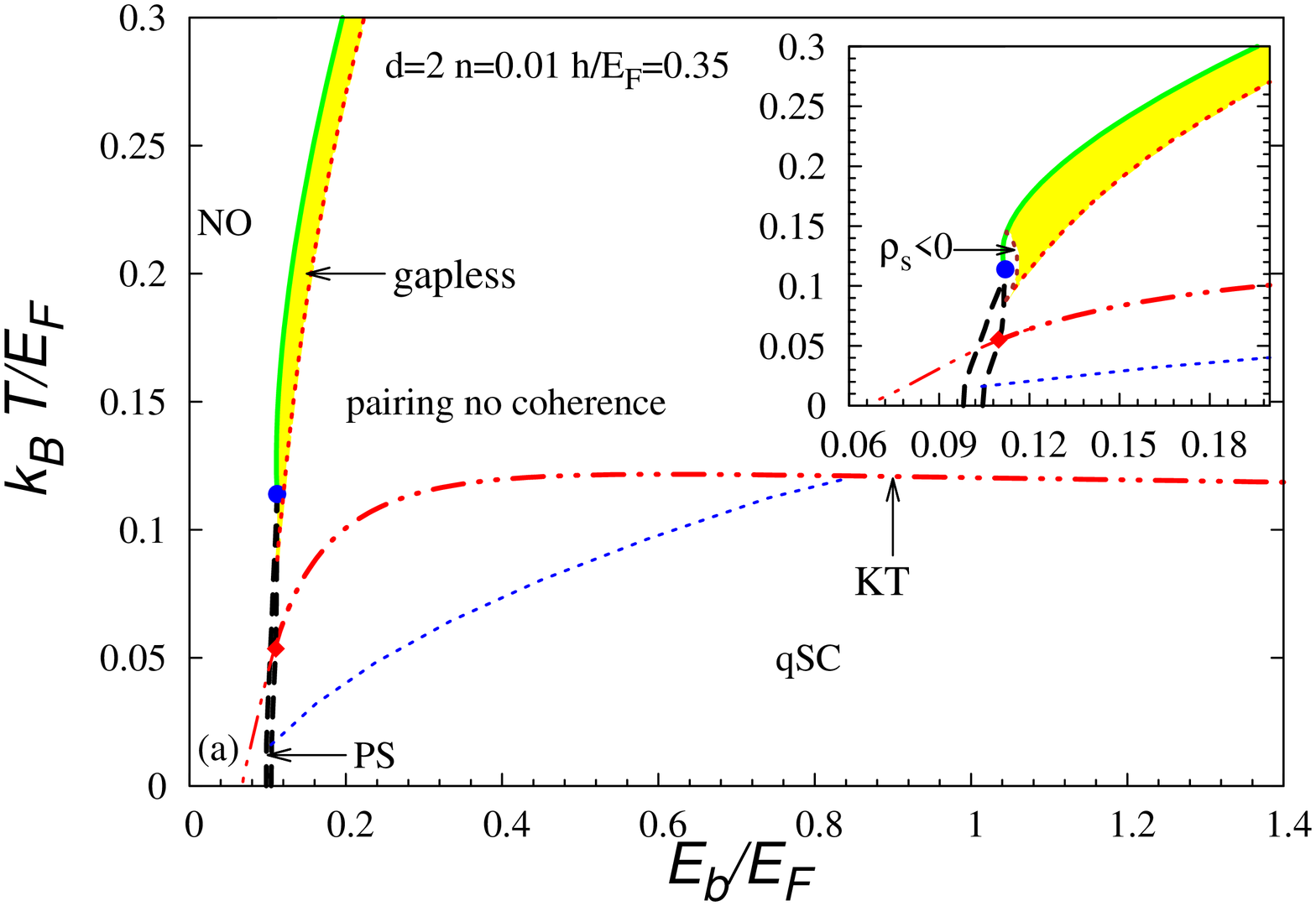}
\hspace*{-0.6cm}
\includegraphics[width=0.38\textwidth,angle=270]
{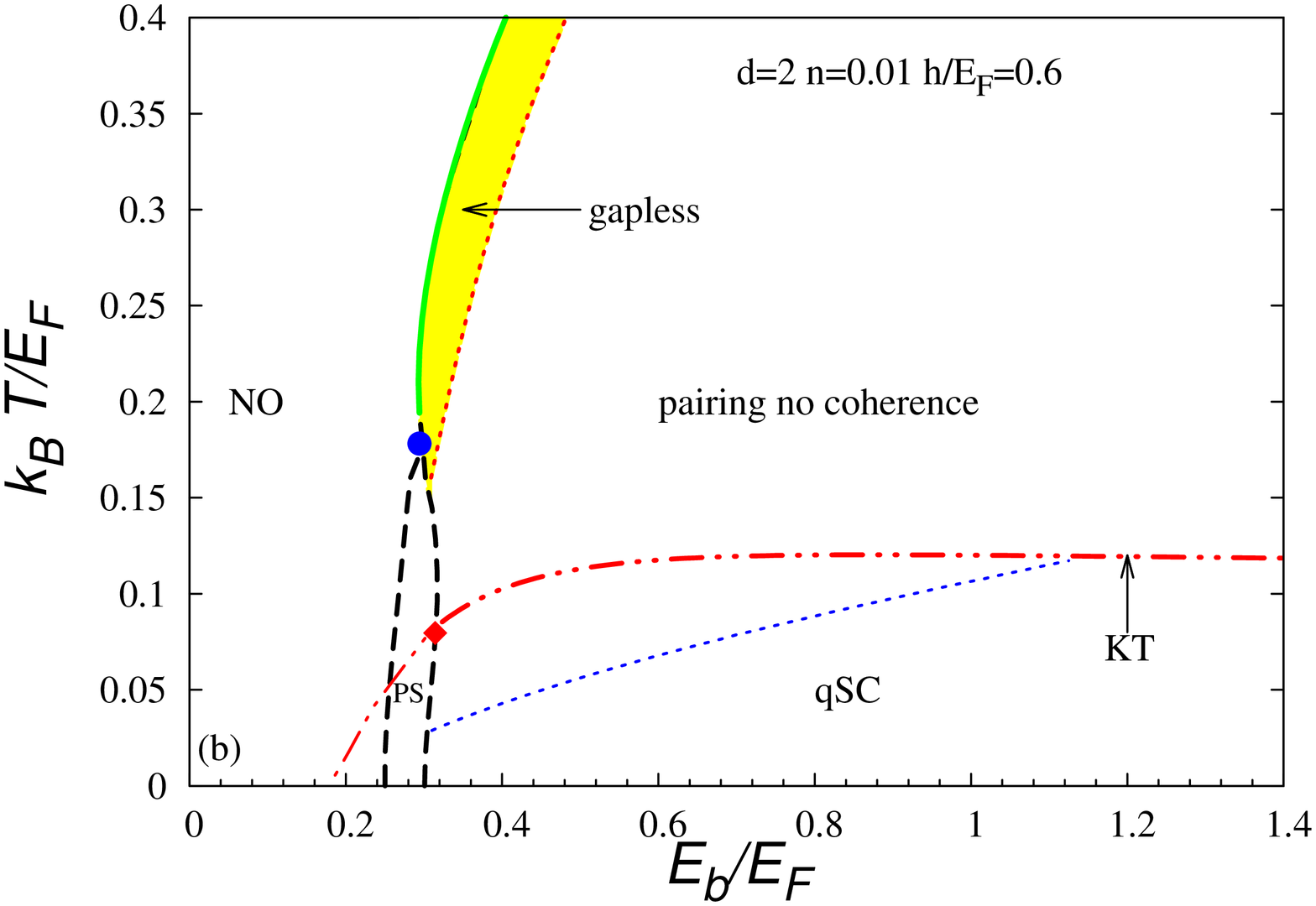}
\hspace*{-0.8cm}
\includegraphics[width=0.38\textwidth,angle=270]
{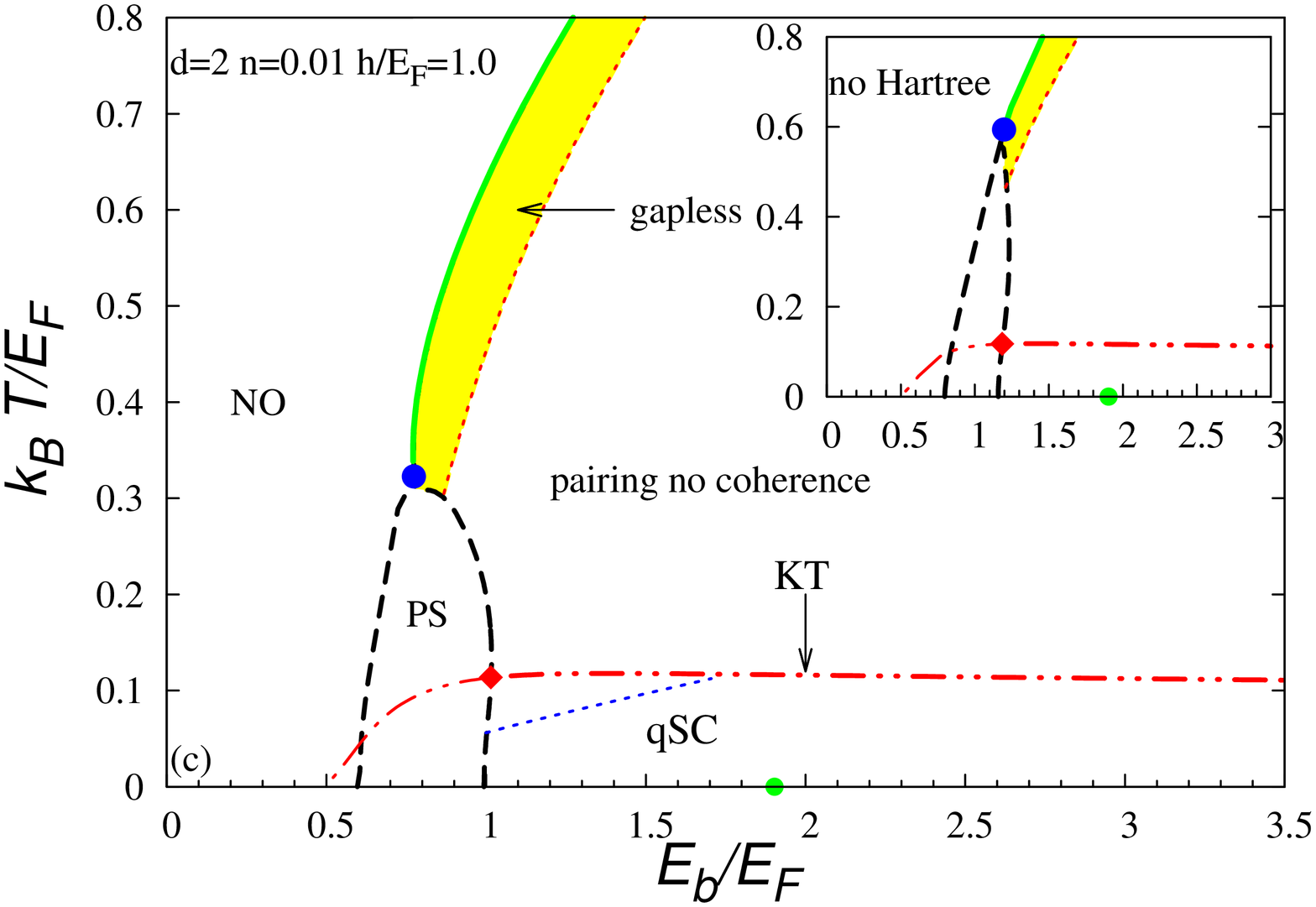}
\hspace*{-0.6cm}
\includegraphics[width=0.38\textwidth,angle=270]
{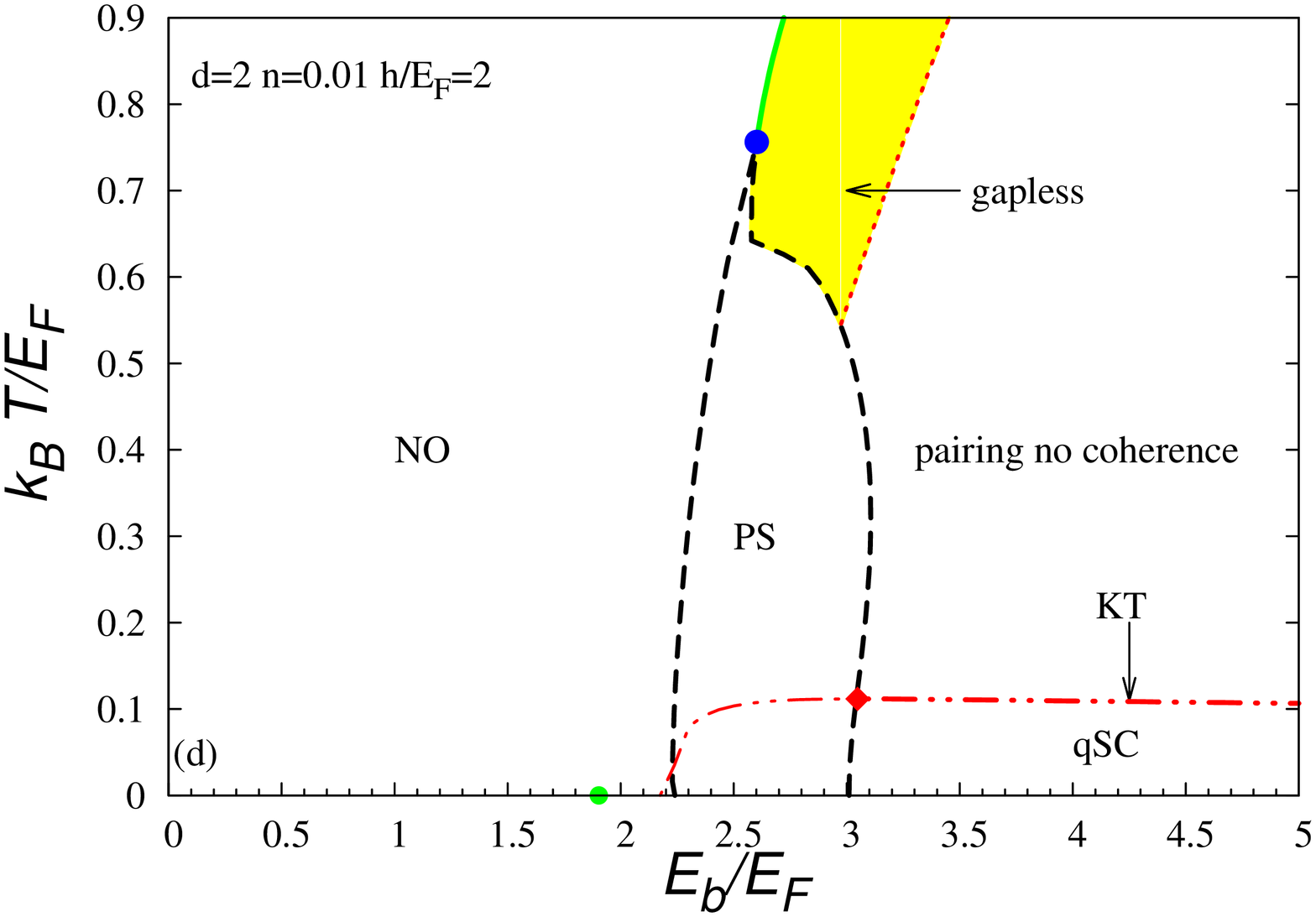}
\caption[Temperature vs. binding energy phase diagrams in units of the lattice
Fermi energy at (a) $h/E_F=0.35$ (inset -- details of a region around TCP), (b)
$h/E_F=0.6$, (c) $h/E_F=1$ (inset -- diagram without the Hartree term) and (d)
$h/E_F=2$.]{\label{kTvsE_b_h_finite} Temperature vs. binding energy phase
diagrams in units of the lattice Fermi energy at (a) $h/E_F=0.35$ (inset --
details of the region around TCP), (b) $h/E_F=0.6$, (c) $h/E_F=1$ (inset -- diagram without
the Hartree term) and (d) $h/E_F=2$. Thick dashed-double dotted
line (red color) is the KT transition line. Thin dash-double dotted line is the
KT transition line to the metastable superfluid state. Thick solid line denotes
transition from pairing without coherence region to NO within the Hartree
approximation, with distinguished gapless region (yellow color). $qSC$ -- 2D
quasi superconductor, PS -- phase separation. Below thin dotted line (blue
color) -- $P<10^{-4}$ (i.e. zero polarization within numerical precision). The
green point in (c)-(d) shows the BCS-LP crossover point at $T=0$; blue point is the MF TCP
point.}
\end{figure}

\begin{figure}[h!]
\hspace*{-0.8cm}
\includegraphics[width=0.38\textwidth,angle=270]
{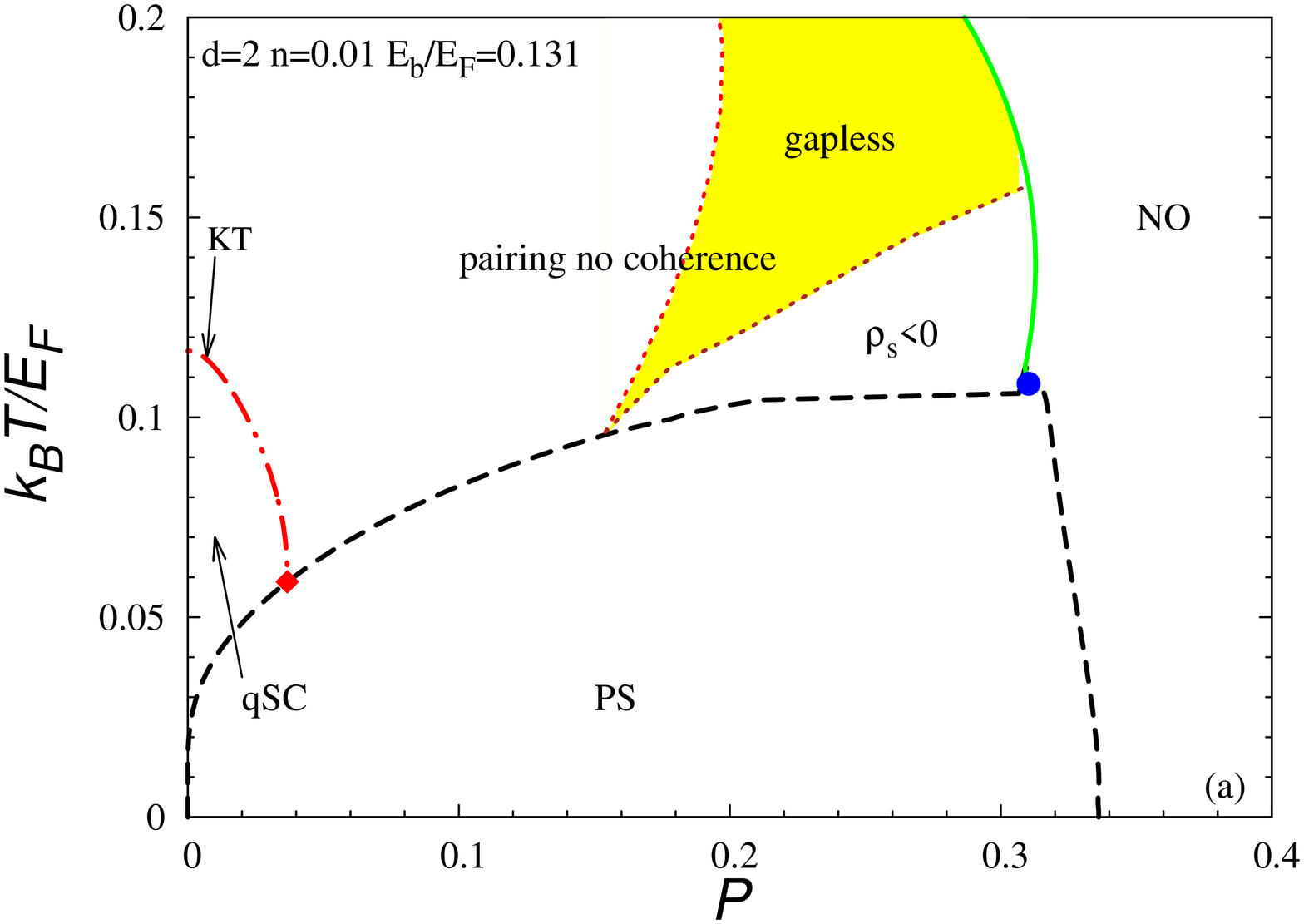}
\hspace*{-0.6cm}
\includegraphics[width=0.38\textwidth,angle=270]
{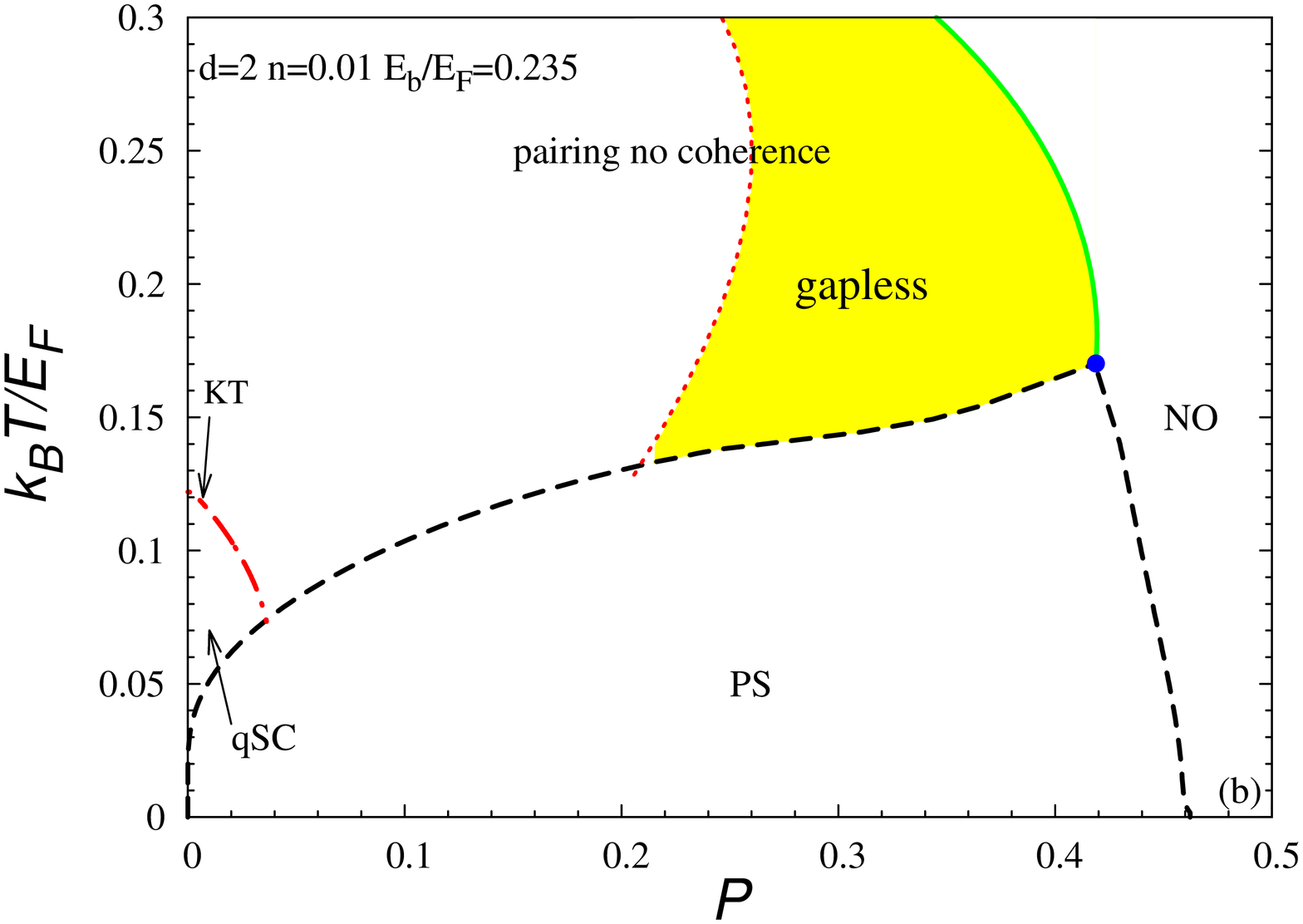}
\hspace*{-0.8cm}
\includegraphics[width=0.38\textwidth,angle=270]
{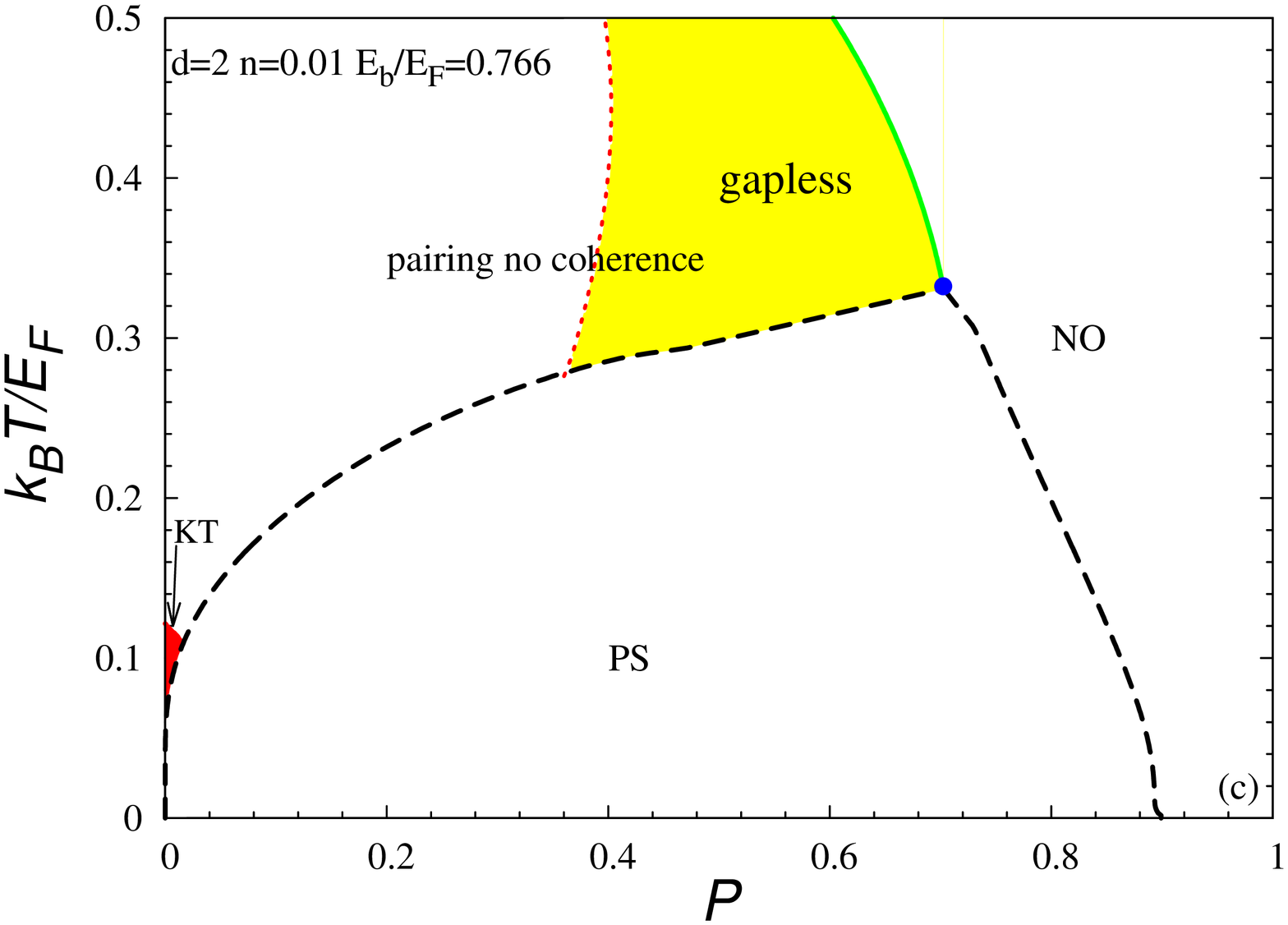}
\hspace*{-0.6cm}
\includegraphics[width=0.38\textwidth,angle=270]
{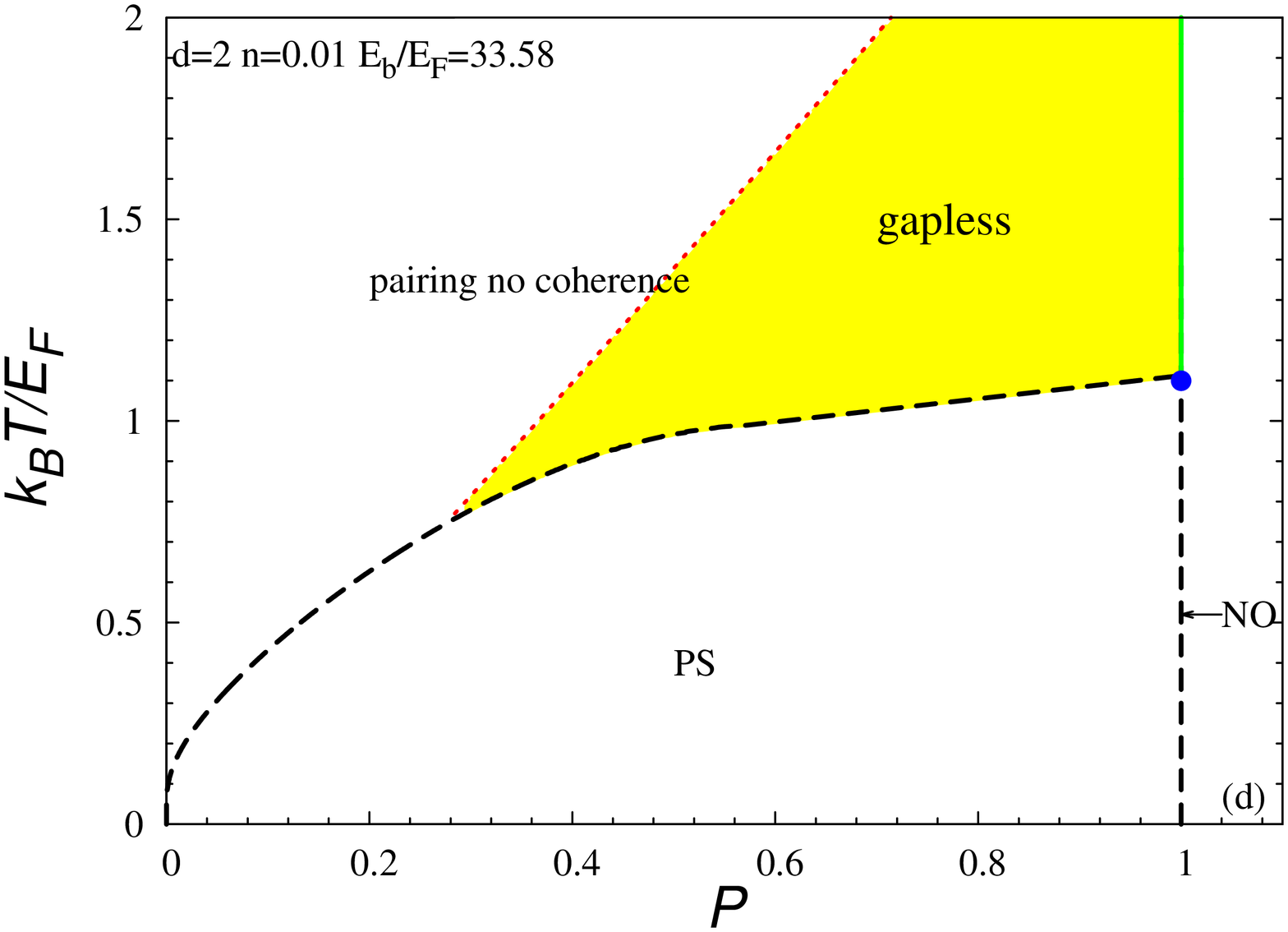}
\caption[Temperature vs. polarization at (a) $E_b/E_F=0.131$ ($U=-2.8$), (b)
$E_b/E_F=0.235$ ($U=-3$), (c) $E_b/E_F=0.766$ ($U=-3.5$) and (d)
$E_b/E_F=33.58$ ($U=-8$).]{\label{kTvsP_Ebfinite} Temperature vs. polarization at (a)
$E_b/E_F=0.131$ ($U=-2.8$), (b) $E_b/E_F=0.235$ ($U=-3$), (c) $E_b/E_F=0.766$ ($U=-3.5$) and (d)
$E_b/E_F=33.58$ ($U=-8$). Thick dashed-double dotted line (red color) is the KT
transition line. Thick solid line denotes transition from pairing without
coherence region to NO within the Hartree approximation, with distinguished
gapless region (yellow color). $qSC$ -- 2D quasi superconductor, PS -- phase
separation, blue point is the MF TCP point.}
\end{figure}

Here, we perform an analysis of the influence of \textcolor{czerwony}{a} Zeeman magnetic field on the BCS-BEC crossover phase diagrams.  
Fig. \ref{kTvsE_b_h_finite} shows ($T-E_b$) phase diagrams in units of the Fermi
energy, for fixed values of the magnetic field: (a) $h/E_F=0.35$ (on the BCS
side), (b) $h/E_F=0.6$, (c) $h/E_F=1$ (in the intermediate couplings) and (d)
$h/E_F=2$ (on the LP side). The solid lines (2$^{nd}$ order transition lines)
and PS regions are obtained within the Hartree approximation. The thick
dash-double dotted line (red color) denotes the KT transition determined from
Eqs. (\ref{ro_s-2}), (\ref{KT}). 

The system is a quasi superconductor (qSC) below $T_c^{KT}$. Polarization
can be induced by thermal excitations of quasiparticles. Above $T_c^{KT}$, but
below $T_c^{HF}$ (pair breaking temperature), pairs exist but without a
long-range phase coherence. In this region a pseudogap behavior is observed.
$T_{pair}\sim T_c^{HF}$ can be interpreted as the temperature at which the
amplitude of the order parameter takes a finite value, but short-range phase
correlations are negligible (superconducting correlation length is of the order
of the lattice spacing). It corresponds to the occurrence of incoherent
pairs, either Cooper pairs in the weak coupling or LP's in the strong coupling
regime. With lowering the temperature, these fluctuating pairs develop
short-range phase correlations. Near but above $T_c^{KT}$, the phase correlation
length increases and becomes much longer than the lattice spacing. At
$T_c^{KT}$, the phase transition occurs and the system becomes superconducting.
Superconducting long-range order sets in at $T=0$.
 
As mentioned above, the temperatures $T_c^{KT}$ are generally lower than
$T_c^{HF}$, but on reducing the attraction, in the absence of magnetic field,
the difference between $T_c^{KT}$ and $T_c^{HF}$ decreases in the weak coupling
limit. At $h=0$ and $E_b \ll 1$, $T_c^{HF}\neq 0$ and $T_c^{KT}\neq 0$. When the
magnetic field increases, $T_c^{HF}=0$, below some value of the
binding energy. This critical $E_b$ increases with $h$. In the strict BCS-MFA
diagram the tricritical point (TCP) exists at finite magnetic fields. In this
mean-field TCP, the SC MF phase, the NO state and PS coexist. There is also
``TCP'' on the KT curve in which three states meet: the qSC phase, the state of
incoherent pairs and PS. The PS range widens with increasing $h$ and the
distance between the TCP MF and KT ``TCP'' is longer. As shown in Fig.
\ref{kTvsP_Ebfinite}, the effect of finite P on the KT superfluid state is
strong. If $t^{\uparrow}=t^{\downarrow}$, the KT phase is restricted to the weak
coupling region and low values of P, which is clearly visible in Fig.
\ref{kTvsE_b_h_Pfinite}. The phase diagrams are plotted in ($T-E_b$) variables
in units of the Fermi energy, for fixed values of population imbalance $P=0.02$
(a) and $P=0.1$ (b). One can observe that the region of the KT phase existence
is rather narrow, even at very low $P=0.02$ and is essentially restricted to
weak couplings. Moreover, the $qSC$ state is not gapless. For higher value of
$P=0.1$, even in the weak coupling limit, the $qSC$ state disappears. One can
distinguish the gapless region in the weak coupling limit, but only within the
state of incoherent pairs. 

With increasing attractive interaction, the range of occurrence of the KT phase
becomes narrower in the intermediate couplings region (Fig.
\ref{kTvsP_Ebfinite}(b)-(c)) and disappears on the LP side (Fig.
\ref{kTvsP_Ebfinite} (d)), in favor of the phase separation region. Therefore,
the $qSC$ state is highly reduced with increasing attractive interaction even
for low population imbalance. Increasing polarization favors the phase of
incoherent pairs. The range of occurrence of $qSC$ in the presence of $P$ widens
in the weak coupling regime with increasing $n$.  

\begin{figure}[t!]
\hspace*{-0.8cm}
\includegraphics[width=0.38\textwidth,angle=270]{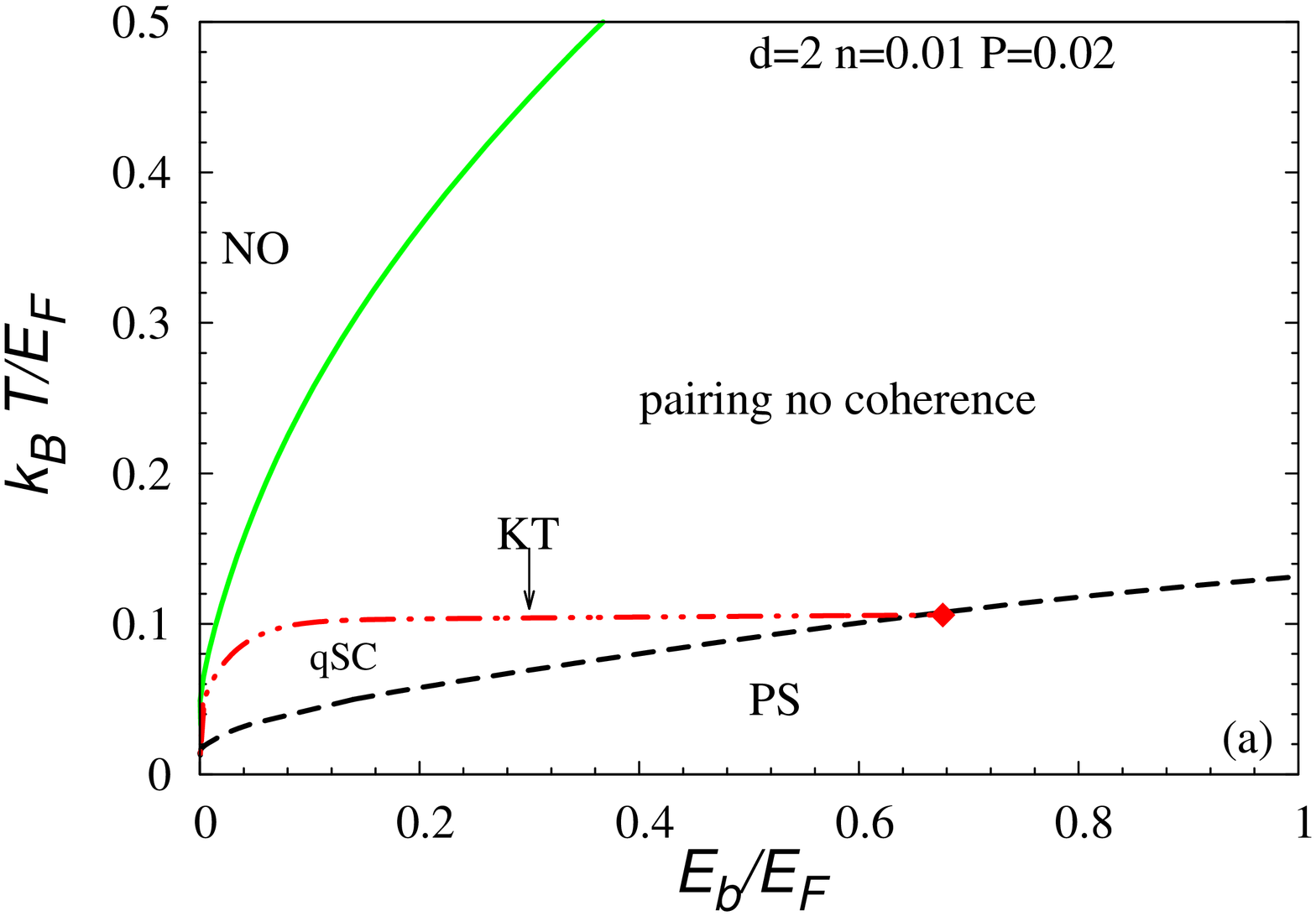}
\hspace*{-0.6cm}
\includegraphics[width=0.38\textwidth,angle=270]{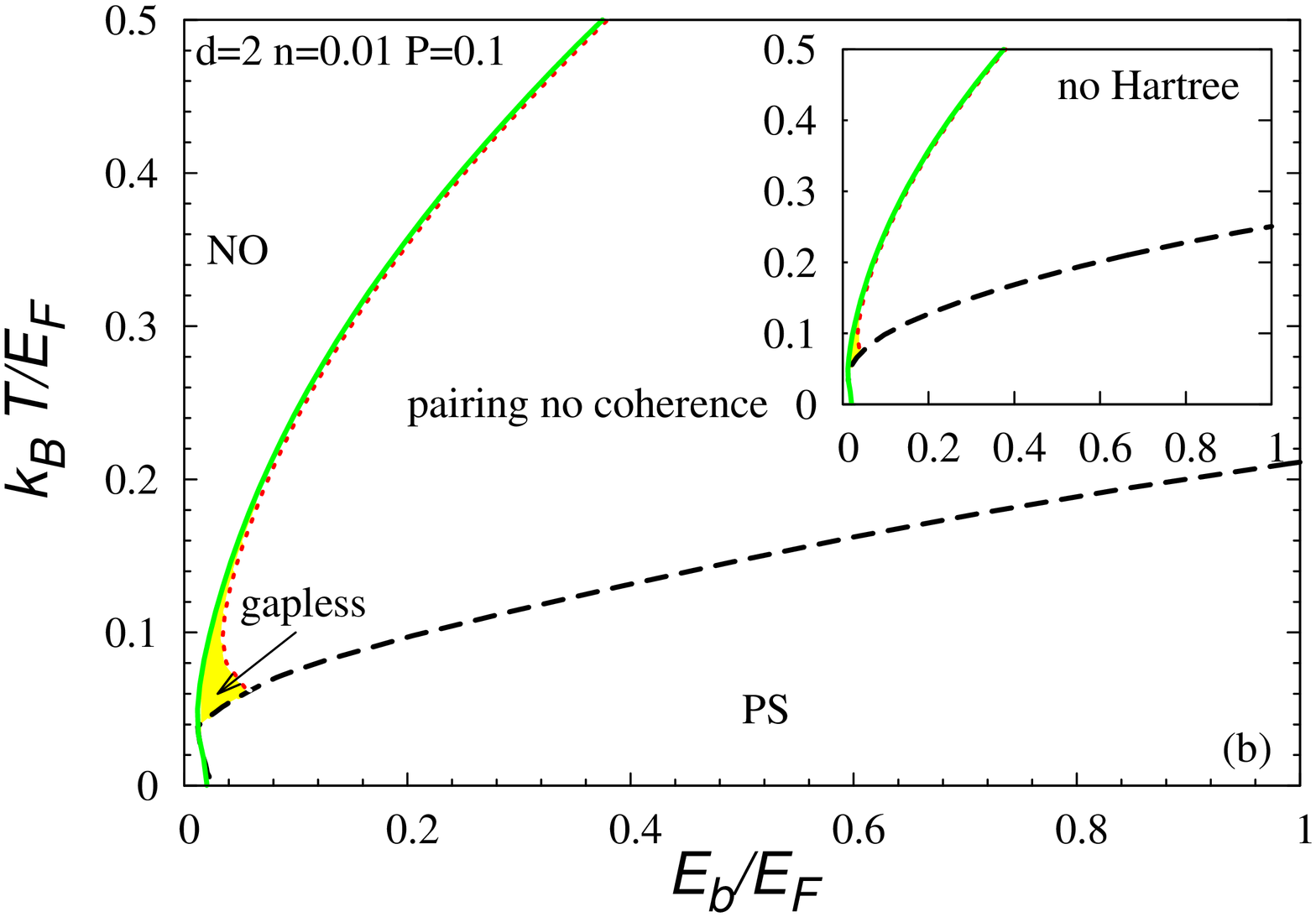}
\caption[Temperature vs. binding energy phase diagrams in units of the lattice
Fermi energy at (a) $P=0.02$, (b) $P=0.1$ (inset -- diagram without the Hartree
term).]{\label{kTvsE_b_h_Pfinite} Temperature vs. binding energy phase diagrams
in units of the lattice Fermi energy at (a) $P=0.02$, (b) $P=0.1$ (inset --
diagram without the Hartree term). Thick dashed-double dotted line (red color)
is the KT transition line. Thick solid line denotes transition from pairing
without coherence region to NO within the Hartree approximation, with
distinguished gapless region (yellow color) in Fig. (b). $qSC$ -- \textcolor{czerwony}{2D quasi 
superconductor}, PS -- phase separation.}
\end{figure}

As shown before, in the analysis of the quasiparticle excitation spectrum, we
also find a gapless region (yellow color in diagrams), for $h> \Delta$ (the BCS
side) and for $h>E_{g}/2$, where $E_g=2\sqrt{(\bar{\mu}-\epsilon_0)^2+|\Delta
|^2}$ (on the LP side). If $t^{\uparrow}=t^{\downarrow}$, this gapless region
can only be realized at $T>0$ and has excess fermions with two FS in the weak
coupling limit. In Figs. \ref{kTvsE_b_h_finite} and \ref{kTvsP_Ebfinite}, the
gapless region is distinguished within the state of incoherent pairs, i.e. is
non-superfluid. In the strong coupling regime, the temperature can induce the
spin-polarized gapless region (in the state of incoherent pairs) with one FS.
This is in contrast to the 3D case, where the BP-1 phase can be stable even
without mass imbalance at $T=0$ and low $n$ \cite{Kujawa3}. In the strong
coupling limit, $T_c^{KT}$ does not depend on magnetic field and approximately
approaches $k_BT_c^{KT}/E_F \approx \frac{t}{2|U|} (1-\frac{n}{2})$ for $|U|\gg
t$, $E_F =2\pi tn$, in contrast to the continuum model which yields in that
limit $k_B T_{c}^{KT}/E_F=\frac{1}{8}$ \cite{tempere}. For $k_B T<<|U|$, there
exist only LPs not broken by the magnetic field and the system is equivalent to
that of a hard-core Bose gas on a lattice. The thin dash-double dotted line in
Fig. \ref{kTvsE_b_h_finite} inside the NO state marks the region where $\Omega$
has two minima (below the curve): lower at $\Delta =0$ and higher at $\Delta
\neq 0$. It means that there can exist a metastable superconducting state.

As discussed in chapter \ref{chapter4}, an \textcolor{czerwony}{important} aspect of the analysis
is the influence of the Hartree term on the phase diagrams. First, the presence
of the Hartree term leads to the reentrant transition (RT) in the weak coupling
limit (Figs. \ref{kTvsE_b_h_finite}(a) and \ref{kTvsP_Ebfinite}(a)), which is
not observed in the phase diagrams without the Hartree term. We also find a
region around MF TCP in which formally $\rho_s<0$, although $\Omega^{SC}<
\Omega^{NO}$ (Figs. \ref{kTvsE_b_h_finite}(a) inset and
\ref{kTvsP_Ebfinite}(a)), in the phase diagram on the BCS side with the Hartree
term. If RT exists, it becomes dynamically unstable because $\rho_s<0$. In
addition, the Hartree term causes an increase in the Chandrasekhar-Clogston
limit \cite{chandrasekhar, Kujawa, Kujawa2}.

\section{3D simple cubic lattice. T-matrix approach}
\subsection{Formalism}
In this section, we discuss the T-matrix approach. As mentioned earlier,
the most interesting and promising ideas assume that the properties of
High-T$_c$ superconductors (and also other unconventional superconductors) place
them between two regimes: BCS and BEC. To extend the analysis of the crossover
from weak to strong coupling to finite temperatures in 3D, it is
necessary to take into account the pairing fluctuations effects. For
intermediate couplings, the normal state can have a pseudo-gap (PG) in the
single particle energy spectrum. It is connected with the presence of preformed
fermionic pairs in the normal state. These pairs are formed at the temperature
$T_p$, which is much higher than $T_c$ at which the long-range phase coherence
occurs and a phase transition to the superconducting state takes place. $T_c$
corresponds to the condensation of pairs with $\vec q=0$ and the pairs with finite
center-of-mass momentum ($\vec q \neq 0$) remain as excitations. In the BCS
limit, pairs form and condense at the same temperature. For many years, many
different approaches have been developed to extend the BCS-BEC crossover to
finite temperatures -- most of them concern different T-matrix schemes in
which one considers coupled equations between particles (with the propagator
$G$) and pairs (with the propagator $T(q)$, hence the name T-matrix). 

As mentioned before, Eagles \cite{Eagles} and Leggett \cite{leggett} discussed 
the BCS-BEC crossover concept in the ground state; Eagles for superconductors with low carrier concentration while Leggett in the context of p-wave pairing symmetry in 
$^{3}$He. Leggett suggested that the system evolves
continuously from BCS to BEC with increasing attractive interaction. Furthermore, Robaszkiewicz, Micnas and Chao analyzed the evolution from BCS to LP limit within the extended attractive Hubbard model \cite{Robaszkiewicz, Robaszkiewicz2,MicnasModern}. In 1985
Nozieres and Schmitt-Rink (NSR) \cite{nozieres} extended Leggett's analysis to
calculate the critical temperature. A very important point of their work was the introduction of the
self-energy in the scattering T-matrix. They used the so-called $(G_0G_0)G_0$
scheme, where, in general, the symbols in parentheses denote the type of Green
functions that enter the equation for the pairing susceptibility
($\chi(q)=\sum_k G_0(k)G_0(q-k)$, see below for notation and further details)
and the latter symbol stands for the type of Green function in the self-energy
equation ($\Sigma(k)=\sum_q T(q) G_0(q-k)$). Here, $G_0$ is the bare
(non-interacting) fermionic Green function. As a consequence, in this approach,
the self-energy is not included in the gap equation, hence it is not completely
self-consistent.

The BCS-BEC crossover concept has been adapted to explain some different
features of HTC superconductors by Micnas \cite{MicnasModern, Micnas-BCS-BEC} Uemura \cite{uemura} and Randeria
\cite{randeria}. Many works of the Levin's
group \cite{Levin2, Kosztin, Chen2} on the BCS-BEC crossover are based on the
s-wave pairing symmetry in the 3D jelium. 

Many theoretical approaches to derivation of the formula for
the transition temperature in the presence of the pseudo-gap are based on
Kadanoff and Martin (KM) works \cite{Kadanoff}. This scheme was also extended by
Patton \cite{Patton} to analyze superconducting fluctuations in a dirty low dimensional superconductor near the
critical temperature. The KM scheme relies on the reformulation of the BCS
theory within Green functions formalism (Green function equations of motion or
the correlation functions) and it is very similar to the Thouless formalism
\cite{Thouless2}. This is the so-called $(GG_0)G_0$ scheme, where $G$ denotes
the dressed (interacting or full) Green function. One can also determine the
equations for $T_c$ within the $(GG)G_0$ scheme, which is discussed in detail
below. 

We start from the consideration of the single component model of fermions:
\begin{equation}
{\cal {H}} = \sum_{\vec{k}\sigma} {\bar \epsilon}_{\vec{k}}c^{\dagger}_{\vec{k}\sigma}
c_{\vec{k}\sigma} 
+\frac{1}{N}\sum_{\vec{k},\vec{k'},\vec{q}}U_{\vec{k},\vec{k'}}
c^{\dagger}_{\vec{k}+\vec{q}/2,\uparrow}c^{\dagger}_{-\vec{k}+\vec{q}/2,\downarrow}
c_{-\vec{k'}+\vec{q}/2,\downarrow}
c_{\vec{k'}+\vec{q}/2,\uparrow},
\end{equation}
where: $U_{\vec{k},\vec{k'}}=g\phi_{\vec{k}}\phi_{\vec{k'}}$ -- separable
attractive potential, $\phi_{\vec{k}}$ -- specifies the pairing symmetry,
${\bar \epsilon_{\vec{k}}}=\epsilon_{\vec{k}}-\mu$. In the Attractive Hubbard
Model on the simple cubic lattice: $g=U$, $\phi_{\vec{k}}=1$ and
$\epsilon_{\vec{k}}=-2t(\cos (k_x)+\cos (k_y)+\cos (k_z))+6t$, the momentum
summation is restricted to the first Brillouin zone.

The formalism that we will employ makes use of the one- and two-particle Green
functions, i.e. expectation values of time-ordered products of creation and
annihilation operators. The equation for the one-particle Green function is used
in the exact form, while the formulas for the two-particle Green functions are
approximate, following KM. The corresponding two-particle correlation functions
can be conveniently expressed in terms of a T-matrix (which can be shown to play
the role of a pair propagator, see below) and one-particle Green functions.
For a detailed derivation of the relevant equations satisfied by the Green
functions, self-energies, pairing susceptibilities and the T-matrix in real and reciprocal
space, we refer the Reader to Appendix \ref{appendix-Tmatrix}. After Fourier
transforming
to momentum space, one obtains the self-consistent equations in the normal
state:
\begin{equation}
\label{t-matrix1}
T^{-1}(q)=g^{-1}+\sum_{k}G(k)G_0(q-k)\phi^{2}_{{\vec{k}-\vec{q}/2}}
\end{equation}
is the T-matrix in separable channels,
\begin{equation}
\Sigma(k)=\sum_{q}T(q)G_0(q-k)\phi^{2}_{{\vec{k}-\vec{q}/2}}
\end{equation}
is the self-energy equation. The full Green function satisfies the Dyson
equation:
\begin{equation}
\label{green_f_dressed}
G^{-1}(k) = G_{0}^{-1}(k)-\Sigma(k), 
\end{equation}
where: $G_{0}^{-1}(k)= i\omega_{n}-{\bar \epsilon_{\vec{k}}}$ -- free fermionic Green function, $\beta=1/k_{B}T$.

Here, we introduce the following notation: 
$k=({\vec{k}}, i\omega_{n}), q=({\vec{q}}, i\nu_{n})$, 
$ \sum_{k}= \frac{1}{\beta N}\sum_{{\vec{k}}, \omega_{n}}$
$ \sum_{q}= \frac{1}{\beta N}\sum_{{\vec{q}}, \nu_{n}}$, $\omega_n=\frac{2\pi}{\beta}(n+\frac{1}{2})$, $\nu_n=\frac{2\pi}{\beta}n$ -- odd, even Matsubara frequencies, respectively, $n$ -- \textcolor{czerwony}{integer}.

The pairing susceptibility equation takes the form:
\begin{equation}
\label{Pair_suss}
\chi(q)=\sum_{k}G(k)G_0(q-k)\phi^{2}_{{\vec{k}-\vec{q}/2}}.
\end{equation}

Eq. \eqref{t-matrix1} can be written as:
\begin{equation}
T^{-1}(q)=g^{-1}+\chi(q).
\end{equation}
For high temperatures, the T-matrix is finite for all values of $q$. 
When the temperature is lowered, metastable or long-lived pairs appear. 
At the superconducting (superfluid) transition temperature, 
the T-matrix becomes divergent for $q=0$ (the Thouless criterion), i.e.:
\begin{equation}
\label{thouless_crit}
g^{-1}+\chi(\vec{0},0,T_c)=0.
\end{equation}

If we assume slow fluctuations of the pairing field close to $T_c$, i.e. the term with $q\approx 0$ is dominant, then the fermionic self-energy can be approximated in the following way:
\begin{equation}
\Sigma(k) =\sum_{q}T(q)G_0(q-k)\phi^{2}_{{\vec{k}-\vec{q}/2}}
\approx G_{0}(-k)\phi^{2}_{{\vec{k}}}
\sum_{q}T(q)
= \frac{\Delta_{pg}^2\phi^{2}_{{\vec{k}}}}{i\omega_{n}+{\bar \epsilon_{-\vec{k}}}}. 
\end{equation}
where: $\Delta_{pg}^2$ is the pseudogap parameter:
\begin{eqnarray}
\label{Delta_pg}
\Delta_{pg}^2\equiv-\sum_{q}T(q)=-\frac{1}{\pi N}\sum_{{\vec q}\neq 0}\int_{-\infty}^{\infty}
\textrm{Im}\,T({\vec q},\Omega)b(\Omega)d\Omega,
\end{eqnarray}
where $b(\Omega)=1/(\exp{(\beta\Omega)}-1)$ is the Bose function.

From the Dyson equation \ref{green_f_dressed}: 
\begin{equation}
G^{-1}(k)=i\omega_n-\bar{\epsilon}_{\vec k}-\frac{\Delta_{pg}^{2}\phi_{\vec k}^{2}}{i\omega_n+\bar{\epsilon}_{\vec k}}.
\end{equation}
Therefore, the full fermionic Green function reads:
\begin{equation}
\label{Green_full}
 G(k)=\frac{i\omega_{n}+{\bar \epsilon_{\vec{k}}}}
{(i\omega_{n})^2-({\bar \epsilon_{\vec{k}}}^2+\Delta_{pg}^2\phi^{2}_{{\vec{k}}})}
\end{equation}

and it has poles in $\pm E_{\vec{k}}=\sqrt{{\bar \epsilon_{\vec{k}}}^2+\Delta_{pg}^2\phi^{2}_{{\vec k}}}$. This form of the Green function is analogous to the one in the standard BCS theory.

Using \eqref{Green_full} we can evaluate the pairing susceptibility \eqref{Pair_suss}. Summing over Matsubara frequencies, one obtains:
\begin{equation}
\label{pair_suss_prim}
\chi(q)=-\frac{1}{N}\sum_{\vec k}\left[
\frac{f(E_{\vec k}) +f({\bar \epsilon}_{\vec{q}-\vec{k}})-1}
{{\bar \epsilon}_{\vec{q}-\vec{k}}+E_{\vec k}-i\nu_{n}}u_{\vec k}^2
+\frac{f({\bar \epsilon_{\vec{q}-\vec{k}})-f(E_{\vec k})}}
{\bar \epsilon_{\vec{q}-\vec{k}}-E_{\vec k}-i\nu_{n}}v_{\vec k}^2\right]\phi^{2}_{{\vec{k}-\vec{q}/2}}~,
\end{equation}
where, as usual: $u_{\vec k}^2+v_{\vec k}^2=1$, $u_{\vec k}^2=\frac{1}{2}(1+\bar \epsilon_{\vec k}/E_{\vec k})$,
$f(\omega)=1/(\exp{(\beta\omega)}+1)$ is the Fermi function.

The equation for the number of particles is:
\begin{equation}
n=\frac{2}{\beta N}\sum_{{\vec k}, \omega_{n}}e^{i\eta\omega_{n}}G(k).
\end{equation}
where: $\eta=0^+$.
\\ 
Performing summation over Matsubara frequencies, the above equation takes the form:
\begin{equation}
n=\frac{1}{N}\sum_{\vec k}\left[1-\frac{{\bar \epsilon_{\vec k}}}{E_{\vec k}}\tanh{(\beta_{c}E_{\vec k}/2)}\right].
\end{equation} 

From the Thouless criterion (Eq.~\eqref{thouless_crit}) with the use of Eq.~\eqref{pair_suss_prim} we get:
\begin{equation}
\label{thouless_5}
1=|g|\frac{1}{N}\sum_{\vec k}\phi_{\vec k}^2\frac{\tanh(\beta_{c}E_{\vec k}/2)}
{2E_{\vec k}}.
\end{equation}

To calculate the pseudogap parameter in the neighborhood of the transition
region, it is sufficient to approximate the imaginary part of the T-matrix by
the value at small $\Omega$ and $|\vec{q}|$. Performing analitical continuation \textcolor{czerwony}{$(i\nu_n \rightarrow \Omega + i\eta)$},
the real and imaginary parts of the inverse of the T-matrix can be written as:
\begin{equation}
\label{A0prim}
 \textrm{Re}\,T^{-1}({\vec q},\Omega)\approx A_0'(\Omega-\Omega_{\vec q}),
\end{equation}
\begin{equation}
\label{A0bis}
 \textrm{Im}\,T^{-1}({\vec q},\Omega)\approx A_0''\Omega,
\end{equation}
where $\Omega_{\vec q}$ -- fermion pairs energy dispersion and the parameters
$A_0'$, $A_0''$ are approximately constant and close to $T_c$ their ratio
$\varepsilon\equiv A_0''/A_0'\ll1$. 

Writing:
\begin{equation}
 T^{-1}({\vec q},\Omega)=\textrm{Re}\,T^{-1}({\vec q},\Omega)+i\,\textrm{Im}\,T^{-1}({\vec q},\Omega),
\end{equation}
one has:
\begin{equation}
 T({\vec q},\Omega)=\frac{1}{\textrm{Re}\,T^{-1}({\vec q},\Omega)+i\,\textrm{Im}\,T^{-1}({\vec q},\Omega)}=\frac{\textrm{Re}\,T^{-1}({\vec q},\Omega)-i\,\textrm{Im}\,T^{-1}({\vec q},\Omega)}{(\textrm{Re}\,T^{-1}({\vec q},\Omega))^2+(\textrm{Im}\,T^{-1}({\vec q},\Omega))^2}.
\end{equation}
Hence:
\begin{equation}
 \textrm{Im}\,T({\vec q},\Omega)=-\frac{\textrm{Im}\,T^{-1}({\vec q},\Omega)}{(\textrm{Re}\,T^{-1}({\vec q},\Omega))^2+(\textrm{Im}\,T^{-1}({\vec q},\Omega))^2}.
\end{equation}
Using now Eqs. \eqref{A0prim}-\eqref{A0bis}, one obtains:
\begin{equation}
 \textrm{Im}\,T({\vec q},\Omega)\approx-\frac{A_0''\Omega}{A_0'^2(\Omega-\Omega_{\vec q})^2+A_0''^2\Omega^2}=-\frac{1}{A_0'\Omega}\frac{\varepsilon}{(1-\Omega_{\vec q}/\Omega)^2+\varepsilon^2}.
\end{equation}
In this formula, one can recognize the Poisson kernel $\eta_\varepsilon(x)=\frac{1}{\pi}\frac{\varepsilon}{x^2+\varepsilon^2}$, which in the limit of $\varepsilon\rightarrow0$ becomes the Dirac delta function. Finally:
\begin{equation}
 \textrm{Im}\,T({\vec q},\Omega)\approx-\frac{\pi}{A_0'}\delta(\Omega-\Omega_{\vec q}).
\end{equation}
This can be rewritten, using the Weierstrass theorem, as:
\begin{equation}
 T({\vec q},\Omega)\approx\frac{A_0'^{-1}}{\Omega-\Omega_{\vec q}+i\varepsilon\Omega},
\end{equation}
which has the natural interpretation of the pair propagator.

The pair dispersion is determined from Eqs.~\eqref{A0prim} and \eqref{pair_suss_prim}, and is given by:
\begin{eqnarray}
\label{omega_q}
\Omega_{\vec q}&=&\frac{1}{A_{0}'}\Bigg\{
\frac{1}{N}\sum_{\vec k}\Big[ 
\frac{f(E_{\vec k})+f({\bar \epsilon}_{\vec{q}-\vec{k}})-1}
{{\bar \epsilon}_{\vec{q}-\vec{k}} +E_{\vec k}}u_{\vec k}^2 
+\frac{f({\bar \epsilon}_{\vec{q}-\vec{k}})-f(E_{\vec k})}
{{\bar \epsilon}_{\vec{q}-\vec{k}} -E_{\vec k}}v_{\vec k}^2\Big]\phi_{\vec{k}-\vec{q}/2}^2
\nonumber\\ 
&+&\frac{1-2f(E_{\vec k})}{2E_{\vec k}}\phi_{\vec k}^2\Bigg\},
\end{eqnarray}
where:
\begin{eqnarray}
\label{A0prime}
A_{0}'=\frac{1}{2\Delta_{pg}^2}
\left[n -2\frac{1}{N}\sum_{\vec k}f({\bar \varepsilon_{\vec k}})\right]\nonumber.
\end{eqnarray}
For small $|\vec{q}|$, $\Omega_{\vec q}$ is quadratic in $\vec{q}$:
\begin{equation}
 \Omega_{\vec q}\approx\frac{\vec{q}^2}{2M^*}, 
\end{equation}
where $M^*$ is the effective pair mass, which can be determined from the expansion of $\Omega_{\vec q}$.

Finally, the pseudogap parameter equation \eqref{Delta_pg} can be written in terms of the Bose function:
\begin{equation}
\Delta_{pg}^2=\frac{1}{A_{0}'}{\frac{1}{N}}
\sum_{{\vec q}\neq 0}b\left(\Omega_{{\vec q}}\right).
\end{equation}

To summarize, we gather here all equations for the superconducting transition temperature:
\begin{equation}
\label{Thouless}
1=|g|\frac{1}{N}\sum_{\vec k}\phi_{\vec k}^2\frac{\tanh(\beta_{c}E_{\vec k}/2)}
{2E_{\vec k}},
\end{equation}
\begin{equation}
\label{Fermion_number}
n=\frac{1}{N}\sum_{\vec k}\left[1-\frac{{\bar \epsilon_{\vec k}}}{E_{\vec k}}\tanh{(\beta_{c}E_{\vec k}/2)}\right],
\end{equation} 
\begin{equation}
\label{Delta_pg_prim}
\Delta_{pg}^2=\frac{1}{A_{0}'}{\frac{1}{N}}
\sum_{{\vec q}\neq 0}b\left(\Omega_{{\vec q}}\right),
\end{equation}
where $\Omega_{\vec q}$ and $A_0'$ are given by Eqs. \eqref{omega_q} and \eqref{A0prime}, respectively. 

The form of the first two Eqs. \eqref{Thouless}-\eqref{Fermion_number} is
analogous to the standard BCS theory (but without the Hartree term), with
the gap parameter replaced by the pseudogap parameter. The third Eq.
\eqref{Delta_pg_prim} gives 
the pseudogap parameter in terms of the Bose function. These equations are solved self-consistently for a given
band structure and filling and together they determine $T_{c}$.

The above scheme is the so-called $(GG_{0})G_{0}$ scheme, i.e. the pair
susceptibility is expressed by
$GG_{0}$ and the self-energy is expressed by $G_{0}$.

\subsubsection{\boldmath{$(GG)G_{0}$} scheme}

Now, we briefly discuss the $(GG)G_{0}$ scheme \cite{RMUnpublished}.

The starting point is to use the self-consistent T-matrix equations for the normal state, i.e. for $T(q)$, $\Sigma(k)$ and $G(k)$, as given by Eqs.~\eqref{t-matrix1}-\eqref{green_f_dressed}, where all the Green functions are fully dressed. 
We introduce the pseudogap parameter in the fermionic Green function as above,
in the calculations of $T_{c}$ from the T-matrix equations. However, the Green
functions in the pairing susceptibility are fully dressed ($G$).
Therefore, one gets: 
\begin{eqnarray}
&\chi(q)&=\sum_{k} G(k)G(q-k) \phi^2_{\vec{k}-\vec{q/2}}=\frac{1}{N}\sum_{\vec k} \frac{1}{\beta} \sum_{\omega_n}G(\vec{k},i\omega_n)G(\vec{q}-\vec{k},i\nu_m-i\omega_n) \phi^2_{\vec{k}-\vec{q/2}}=\nonumber\\
&=&\frac{1}{N}\sum_{\vec k}\left(\frac{v_{\vec k}^2 v_{\vec{q}-\vec{k}}^2}{i\nu_{n}+E_{\vec k}+E_{\vec{q}-\vec{k}}} -
\frac{ u_{\vec k}^2 u_{\vec{q}-\vec{k}}^2}{i\nu_{n}-E_{\vec k}-E_{\vec{q}-\vec{k}}}\right)\left[1-f(E_{\vec k}) -f(E_{\vec{q}-\vec{k}})   \right]\phi_{\vec k-\vec{q}/2}^{2}\nonumber\\
&+&\frac{1}{N}\sum_{\vec k}\left(\frac{u_{\vec k}^2 v_{\vec{q}-\vec{k}}^2}{i\nu_{n}+E_{\vec{q}-\vec{k}}-E_{\vec k}} -
\frac{u_{\vec{q}-\vec{k}}^2v_{\vec k}^2}{i\nu_{n}-E_{\vec{q}-\vec{k}}+E_{\vec k}} \right)\left[f(E_{\vec k}) -f(E_{\vec{q}-\vec{k}})\right]\phi_{\vec k-\vec{q}/2}^{2}.
\end{eqnarray}
 
Formally, the pseudogap and the number equations keep the same form as in the
$(GG_0)G_0$ scheme. We use again the small $\Omega$, $|\vec{q}|$ expansion of
the T-matrix in the evaluation of the pseudogap, i.e. we take at $T_{c}$:
$T({\vec q},\Omega)=\frac{Z^{-1}}{\Omega-\Omega_{\vec q}}$, where $Z$ is
specified below. This scheme has also been recently considered by T. Ozawa and  G. Baym
\cite{Ozawa}.
The Thouless criterion is:
\begin{equation}
1+g\chi (\vec{0},0,T_c)=0. 
\end{equation}
Therefore, the equations for $T_c$ in the $(GG)G_{0}$ scheme take the final form:
\begin{equation}
\label{thoulles_GG}
\frac{1}{|g|}=\frac{1}{N}\sum_{\vec k}\phi_{\vec k}^{2}
\left[\frac{1}{2}\left(1+\frac{{\bar \epsilon}_{\vec k}^2}{E_{\vec k}^2}\right)\frac{\tanh(\beta_{c}E_{\vec k}/2)}{2E_{\vec k}}
 - \frac{\Delta_{pg}^2}{2E_{\vec k}^2} f'(E_{\vec k}) \right],                   
\end{equation}
\begin{equation}
n=\frac{1}{N}\sum_{\vec k}\left[1-\frac{{\bar \epsilon_{\vec k}}}{E_{\vec k}}\tanh{(\beta_{c}E_{\vec k}/2)}\right],
\end{equation}
\begin{equation}
\Delta_{pg}^2=\frac{1}{Z}{\frac{1}{N}}
\sum_{{\vec q}\neq 0}b\left(\Omega_{{\vec q}}\right),
\end{equation}
where:
\begin{eqnarray}
\Omega_{\vec q}&=&\frac{1}{Z}\Bigg\{\frac{1}{N}\sum_{\vec k}\Bigg[\left(\frac{u_{\vec k}^2 u_{\vec{q}-\vec{k}}^2+v_{\vec k}^2 v_{\vec{q}-\vec{k}}^2}{E_{\vec k}+E_{\vec{q}-\vec{k}}} 
\right)\left[f(E_{\vec k}) +f(E_{\vec{q}-\vec{k}})-1\right]\phi_{\vec k-\vec{q}/2}^{2}  \nonumber \\
&+& \left(\frac{u_{\vec k}^2 v_{\vec{q}-\vec{k}}^2+u_{\vec{q}-\vec{k}}^2v_{\vec k}^2}
{E_{\vec{q}-\vec{k}}-E_{\vec k}} \right)\left[f(E_{\vec{q}-\vec{k}})-f(E_{\vec k})\right] \Bigg]\phi_{\vec k-\vec{q}/2}^{2}
+\chi({\vec 0},0)\Bigg\},
\end{eqnarray}
\begin{equation}
Z=\frac{1}{N}\sum_{\vec k}\left[\frac{{\bar \epsilon_{\vec k}}}{2E_{\vec k}^2}
\left(\frac{1-2f(E_{\vec k})}{2E_{\vec k}}+f'(E_{\vec k})\right)\right],
\end{equation}
where: $ f'=\partial f(\omega)/\partial\omega$ -- derivative of the Fermi function.

The pair dispersion for small $|\vec{q}|$ is taken as previously: $\Omega_{{\vec
q}}=q^2/2M^*$, where $M^*$ is the effective pair mass.

Comparing $(GG_0)G_0$ and $(GG)G_0$ T-matrix schemes we see the differences in
the first equation for $T_c$ and also in the pair dispersion. For a small
pseudogap parameter, Eq.~\eqref{thoulles_GG} goes over to
Eq.~\eqref{thouless_5}.
The numerical solutions for $T_c$ in both schemes are discussed later.

\subsubsection{Hubbard model \textcolor{czerwony}{with U<0} in a magnetic field -- \boldmath{$(GG_0)G_{0}$} scheme}
Now, we discuss the generalization of the $T_c$ equations in non-zero Zeeman
magnetic field case, for the 3D simple cubic lattice and the s-wave pairing
symmetry case. For more details, we refer the Reader to Appendix \ref{appD3}. In
particular, we discuss there the nature of the approximation involved. 

Let us introduce
the symmetrized pairing susceptibility \cite{Levin2007}:
\begin{equation}
\label{chi_magn}
\chi(q)=\frac{1}{2}\left[\chi_{\uparrow\downarrow}(q)+\chi_{\downarrow\uparrow}(q)\right],
\end{equation}
where:
\begin{equation}
\chi_{\uparrow\downarrow}(q)= \sum_{k}G_{0\uparrow}(q-k)G_{\downarrow}(k),
\end{equation}
\begin{equation}
\label{chi_compl}
\chi_{\downarrow\uparrow}(q)= \sum_{k}G_{0\downarrow}(q-k)G_{\uparrow}(k),
\end{equation}
\begin{equation}
G_{0\sigma}^{-1}(k)= i\omega_{n}-{\bar{\epsilon}}_{\vec{k}} -\sigma h. 
\end{equation}
The full Green functions for fermions with spin $\sigma$ are given by the Dyson equation:
\begin{equation}
G^{-1}_{\sigma}(k) = 
G_{0\sigma}^{-1}(k)-\Sigma_{\sigma}(k),
\end{equation}
In the pairing approximation: 
\begin{equation}
\Sigma_{\sigma}(k)=\sum_{q}T(q)G_{o\bar{\sigma}}(q-k) \approx -\Delta_{pg}^2G_{0{\bar{\sigma}}}(k),
\end{equation}
where: ${\bar{\sigma}}=-\sigma$. Therefore:
\begin{equation}
G_{\uparrow,\downarrow}(k)=\frac{u_{\vec k}^2}{i\omega_{n}\pm h -E_{\vec k}}
+\frac{v_{\vec k}^2}{i\omega_{n}\pm h +E_{\vec k}},
\end{equation}
where, as usual:
$E_{\vec{k}}=\sqrt{{\bar \epsilon_{\vec k}}^2+\Delta_{pg}^2}$,  
$u_{\vec k}^2=\frac{1}{2}(1+\frac{{\bar \epsilon_{\vec k}}}{E_{\vec k}})$, 
$u_{\vec k}^2+v_{\vec k}^2=1$. 

The pseudogap parameter is determined as previously: 
\begin{equation}
\Delta_{pg}^2=-\sum_{q}T(q)=-\frac{1}{\pi N}\sum_{\vec q\neq 0}\int_{-\infty}^{\infty}
\textrm{Im} T({\vec q},\Omega)b(\Omega)d\Omega
\end{equation}
and
\begin{equation} 
 T^{-1}(q)=1/g+\chi(q).
\end{equation}
The pairing susceptibility is obtained after performing summation over the Matsubara frequencies in Eqs.~\eqref{chi_magn}-\eqref{chi_compl}:
\begin{eqnarray}
\chi(q)=-\frac{1}{N}\sum_{\vec k}\left[
\frac{{\bar f}(E_{\vec k}) +{\bar f}({\bar \epsilon}_{\vec{q}-\vec{k}})-1}
{{\bar \epsilon}_{\vec{q}-\vec{k}}+E_{\vec k}-i\nu_{n}}u_{\vec k}^2
+\frac{{\bar f}({\bar \epsilon_{\vec{q}-\vec{k}}})-{\bar f}(E_{\vec k})}
{{\bar \epsilon_{\vec{q}-\vec{k}}}-E_{\vec k}-i\nu_{n}}v_{\vec k}^2\right],
\end{eqnarray}
where:
\begin{equation}
{\bar f}(x)=\frac{1}{2}\left[f(x+h)+f(x-h)\right]
\end{equation} 
is the symmetrized Fermi function.

Finally, the equations for $T_c$ in the presence of the Zeeman magnetic field take the form:
\begin{equation}
\label{thouless_crit2}
0=1+U\chi(0)=1+U\frac{1}{N}\sum_{\vec k}\frac{1-2{\bar f}(E_{\vec k})}{2E_{\vec
k}}, 
\end{equation}
\begin{equation}
n=2\frac{1}{N}\sum_{\vec k}\left[v_{\vec k}^2 +
\frac{\bar\epsilon_{\vec k}}{E_{\vec k}}{\bar f}(E_{\vec
k})\right],
\end{equation}
\begin{equation}
\label{T-matrix-pol}
 Pn=\frac{1}{N}\sum_{\vec k}\left[{ f}(E_{\vec k}-h)-{ f}(E_{\vec
k}+h)\right],
\end{equation}
\begin{equation}
\label{Del_pg-Tmatrix}
\Delta_{pg}^2=\frac{1}{A_{0}'}{\frac{1}{N}}
\sum_{{\vec q}\neq 0}b\left(\Omega_{{\vec q}}\right),
\end{equation}
where: $P=(n_{\uparrow}-n_{\downarrow})/n$ -- spin polarization, $f(x)=1/[\exp(\beta x)+1]$ and $b(x)=1/[\exp(\beta x)-1]$ 
are Fermi and Bose functions, respectively. Eqs.~\eqref{thouless_crit2}-\eqref{T-matrix-pol} have the form of the BCS equations (without the Hartree term), with the gap parameter replaced by the pseudogap parameter. 

The pairing susceptibility $\chi({\vec {q}},\Omega)$ and the
pair dispersion $\Omega_{\vec q}$ are given as previously, but the Fermi function 
$f(x)$ is replaced by the symmetrized one $\bar f(x)$:
\begin{eqnarray}
\Omega_{\vec q}&=&\frac{1}{A_{0}'}\Bigg\{
\frac{1}{N}\sum_{\vec k}\left[ 
\frac{{\bar f}(E_{\vec k})+{\bar f}({\bar \epsilon}_{\vec{q}-\vec{k}})-1}
{{\bar \epsilon}_{\vec{q}-\vec{k}} +E_{\vec k}}u_{\vec k}^2 
+\frac{{\bar f}({\bar \epsilon}_{\vec{q}-\vec{k}})-{\bar f}(E_{\vec k})}
{{\bar \epsilon}_{\vec{q}-\vec{k}} -E_{\vec k}}v_{\vec k}^2\right]
\nonumber\\ 
&+&\frac{1-2\bar{f}(E_{\vec k})}{2E_{\vec k}}\Bigg\},
\end{eqnarray}
\begin{equation}
A_{0}'=\frac{1}{2\Delta_{pg}^2}
\left[n -2\frac{1}{N}\sum_{\vec k} {\bar f}({\bar \epsilon_{\vec
k}})\right].
\end{equation}

For small $|\vec{q}|$, $\Omega_{\vec q}$ is quadratic in $\vec{q}$:
\begin{equation}
\Omega_{\vec q}\approx\frac{\vec{q}^2}{2M^*}, 
\end{equation}
where $M^*$ is the effective pair mass, which can be determined from the expansion of $\Omega_{\vec q}$:
\begin{eqnarray}
\frac{1}{M^*}=\frac{1}{A'_{0}}\frac{1}{\Delta_{pg}^2 N}\sum_{\vec k}\left[
\frac{1}{2}\left[(1-2{\bar f}({\bar \epsilon}_{\vec k}))-
\frac{{\bar \epsilon}_{\vec k}}{E_{\vec k}}(1-2{\bar f}(E_{\vec k}))\right]
(2t \cos{k_{x}}) -\nonumber \right. \\
\left. \left\{2 {\bar f}'({\bar \epsilon}_{\vec k})+
\frac{E_{\vec k}}{\Delta_{pg}^2}\left[(1+\frac{{\bar \epsilon}_{\vec k}^2}
{E_{\vec k}^2})\left(1-2{\bar f}(E_{\vec k})\right)-2\frac{{\bar \epsilon}_{\vec k}} {E_{\vec k}}
(1-2{\bar f}({\bar \epsilon}_{\vec k}))\right]\right\}(2t
\sin{k_{x}})^2\right],
\end{eqnarray}
where ${\bar f}'(x)$ is the derivative of ${\bar f}(x)$.

Eq.~\eqref{Del_pg-Tmatrix} gives the pseudogap parameter in terms of the Bose function. 

Next, we explicitly sum over the vectors $\vec{q}\neq 0$ in Eq.~\eqref{Del_pg-Tmatrix}. We replace the sum over $\vec{q}$ with an integral over spherical coordinates. Therefore:
\begin{equation}
\frac{1}{N} \sum_{\vec{q}\neq 0} b(\Omega_{\vec{q}})=\frac{1}{(2\pi)^3}\int_{0}^{2\pi}\int_{0}^{\pi}\int_{0}^{\Lambda} q^2 dq \frac{1}{e^{\beta_c \Omega_{\vec{q}}}-1}, 
\end{equation}
where: $\Lambda$ -- the sphere radius. To evaluate this integral, we extend the integration range to infinity. We also write the expression $(e^x-1)^{-1}$ as the geometric series.
Then:
\begin{equation}
 \frac{1}{N} \sum_{\vec{q}\neq 0} b(\Omega_{\vec{q}})=\frac{1}{2\pi^2}\int_0^{\infty} dq\,q^2\sum_{l=1}^{\infty}e^{-l\beta_c \Omega_{\vec q}}=\frac{1}{2\pi^2} \sum_{l=1}^{\infty} \int_0^{\infty}dq\,q^2 e^{\frac{-l\beta_c q^2}{2M^*}}.
\end{equation}
 After changing the variables $l\beta_c q^2=x^2$, we get:
\begin{equation}
\frac{1}{N} \sum_{\vec{q}\neq 0} b(\Omega_{\vec{q}})=\frac{1}{2 \pi^2} \sum_{l=1}^{\infty}\Big(\frac{\beta_c l}{2M^*}\Big)^{-3/2} \int_{0}^{\infty} dx \,x^2 e^{-x^2}=\frac{1}{2\pi^2} \sum_{l=1}^{\infty} \frac{1}{l^{3/2}} \int_{0}^{\infty} dx\, x^2 e^{-x^2}. 
\end{equation}
Using: $\sum_{l=1}^{\infty} \frac{1}{l^{3/2}} \equiv \zeta (3/2)$ -- the Riemann zeta function and $\int_{0}^{\infty} dx\, x^2 e^{-x^2}=\sqrt{\pi}/4$ (a Gaussian integral), we get:
\begin{equation}
\frac{1}{N} \sum_{\vec{q}\neq 0} b(\Omega_{\vec{q}})=\frac{1}{8}\zeta (3/2) \Big(\frac{2 M^*}{\pi \beta_c} \Big)^{3/2}. 
\end{equation}
Thus, finally the equation for the pseudogap parameter (Eq.~\eqref{Del_pg-Tmatrix}) takes the form:
\begin{equation}
\label{Delta_pg_num}
\Delta_{pg}^2 =\frac{1}{A_0'} \frac{1}{8}\zeta (3/2) \Big(\frac{2 M^*}{\pi \beta_c} \Big)^{3/2}.
\end{equation}
To determine $T_c$, this equation together with Eqs.~\eqref{thouless_crit2}-\eqref{T-matrix-pol} are solved numerically.

\subsection{Numerical results}
\label{6.2.2}

In this section we present the numerical results concerning the BCS-BEC
crossover in non-zero temperatures. In the following analysis, we
consider the 3D case for AHM in the Zeeman magnetic field, extending the results of
section \ref{3Dcase} on the BCS-BEC crossover in the ground state. The
transition temperature from the superconducting state to the pseudogap region
(PG) is determined from Eqs.~\eqref{thouless_crit2}-\eqref{T-matrix-pol},
together with Eq.~\eqref{Delta_pg_num}, within the $(GG_0)G_0$ scheme for $h\neq
0$.
\subsubsection{\boldmath{$h=0$} case} 

\begin{figure}[p!]
\begin{center}
\includegraphics[width=0.55\textwidth,angle=270]{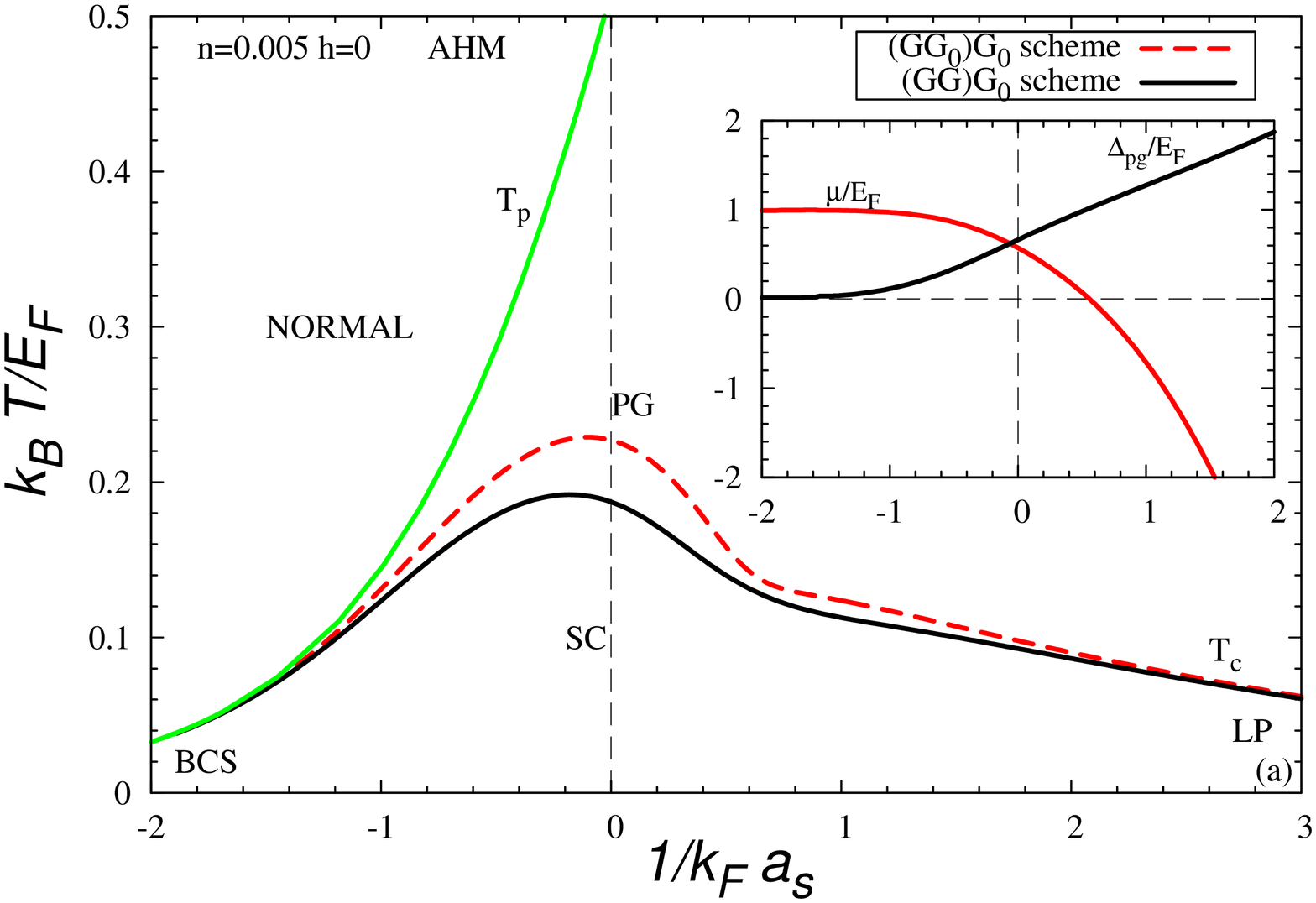}\hspace{-0.2cm}
\includegraphics[width=0.55\textwidth,angle=270]{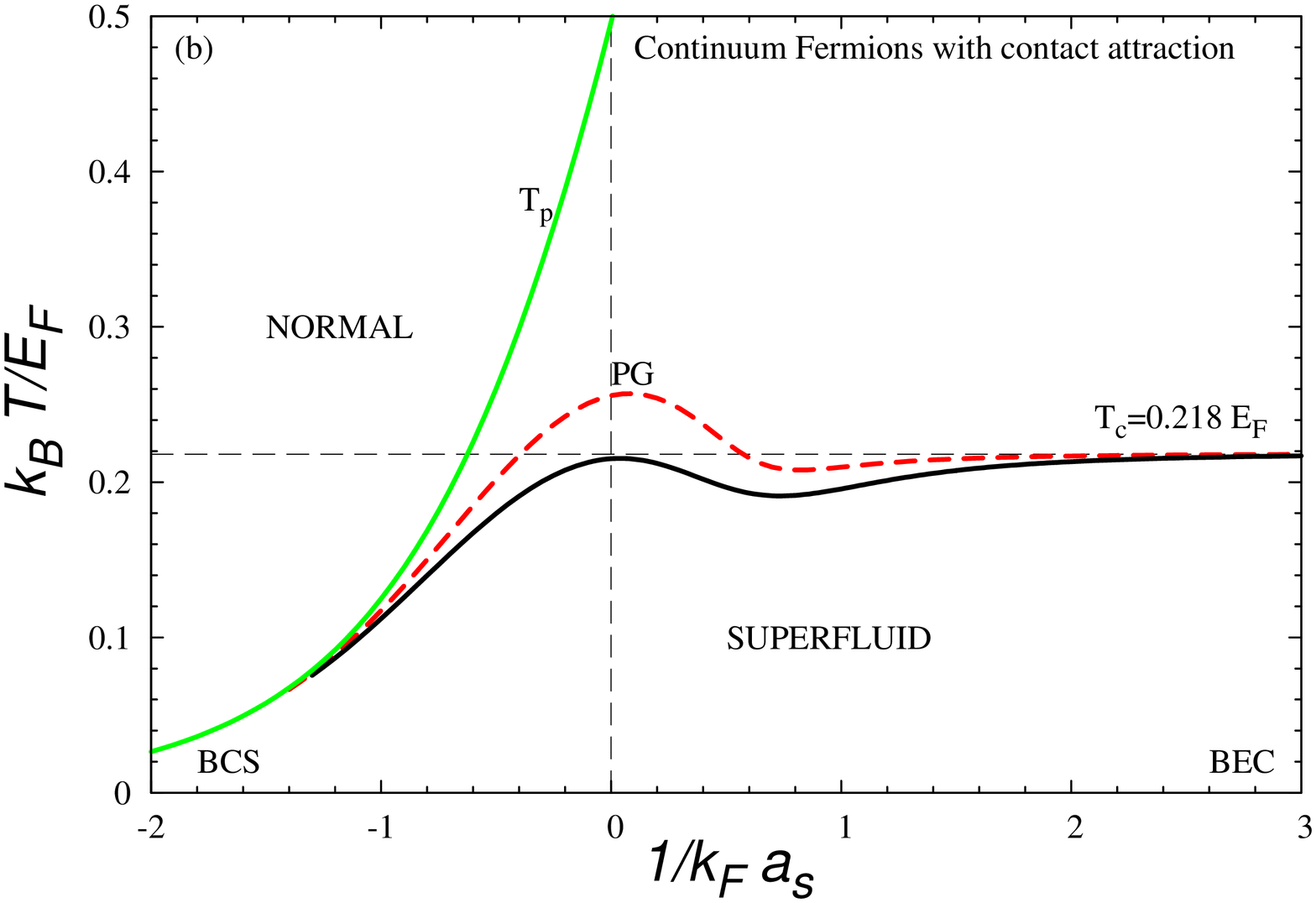}
\caption[\textcolor{czerwony}{(a)} $k_BT/E_F$ ($E_F$ -- \textcolor{czerwony}{lattice} Fermi energy) 
vs. $1/k_Fa_s$ phase
diagram of the AHM at fixed $n=0.005$ for the sc lattice. \textcolor{czerwony}{(b)} the \textcolor{czerwony}{analogous} diagram for the 3D continuum
model with contact attraction; $h=0$.]{\label{T-matrixn0005-h0} \textcolor{czerwony}{(a)} $k_BT/E_F$ ($E_F$ -- \textcolor{czerwony}{lattice} Fermi energy) 
vs. $1/k_Fa_s$ phase
diagram of the AHM at fixed $n=0.005$ for the sc lattice. \textcolor{czerwony}{(b)} the \textcolor{czerwony}{analogous} diagram for the 3D continuum
model with contact attraction; $h=0$. The red dashed line -- $T_c$
determined within $(GG_0)G_0$ scheme, the black solid line -- $T_c$ determined
within $(GG)G_0$ scheme, the green solid line -- $T_c$ determined within MF
approximation (denoted as $T_p$). Inset shows $\mu/E_F$ and $\Delta_{pg}/E_F$ vs. $1/k_F
a_s$ at $T_c$.} 
\end{center}
\end{figure}

Here, we start from a brief discussion of the BCS-BEC crossover at $h=0$.
Within this analysis, we perform a comparison of the results obtained from the
$(GG_0)G_0$ and $(GG)G_0$ scheme.

Fig. \ref{T-matrixn0005-h0} shows typical temperature phase diagrams,
illustrating the changes in the 3D system which evolves from the weak to strong
coupling limit \cite{RMUnpublished}. We analyze the BCS-BEC crossover diagrams at non-zero
temperatures within both AHM (Fig.
\ref{T-matrixn0005-h0}(a)) and the continuum model of a dilute gas of
fermions (Fig. \ref{T-matrixn0005-h0}(b)). In the first case, the fixed particle
concentration is very low ($n=0.005$), i.e. the system is very diluted. These
two diagrams are plotted in terms of $E_F$ and $k_Fa_s$ (where $a_s$ is the
scattering length between fermions for a two-body problem).
In the continuum model with the Hamiltonian:
\begin{equation}
 H=\sum_{\vec k \sigma}\xi_{\vec k} c_{\vec k \sigma}^{\dag}c_{\vec k \sigma}
+\frac{g}{V} \sum_{\vec k \vec{k'} \vec q} c_{\vec{k}\uparrow}^{\dag}
c_{\vec{k}+\vec{q}\uparrow} c_{\vec{k'}\downarrow}^{\dag}
c_{\vec{k'}-\vec{q}\downarrow},
\end{equation}
where: $\xi_{\vec k}=\epsilon_{\vec k}-\mu=\frac{k^2}{2m}-\mu$,
the scattering length is related to the contact
interaction potential $g$ via the Lippmann-Schwinger equation: $m/(4\pi
a_s)=1/g +1/V \sum_{\vec k} 1/2\epsilon_{\vec k}$.
For comparison, we remind
here also the definition of $1/k_F a_s$ in the lattice model case: $1/k_F
a_s=\Big(\frac{1}{|U_c|^{d=3}}-\frac{1}{|U|}\Big)\frac{8\pi}{\sqrt{E_F}}$ ($|U|$
and $E_F$ are in units of $t$), where
$|U_c|^{d=3}/12=0.659$ and $E_F$ is the lattice Fermi energy. 
Maximum $T_c$ is
around the unitarity point ($1/k_Fa_s=0$). The minimum in $T_c$ plots
corresponds to the change of sign of the chemical potential (beginning of the
bosonic regime).

As discussed above, in the weak coupling limit, the fermionic pairs form
and condense at the same temperature $T_c$. In the BCS theory, the
gap ($\Delta_{sc}$) decreases with increasing temperature and drops to zero at
$T_c$. When the attractive interaction increases, the pseudogap ($\Delta_{pg}$)
appears in the excitation spectrum, which is non-zero also above $T_c$. The
pseudogap parameter behavior in the critical temperature is shown in Fig.
\ref{T-matrixn0005-h0}(a) (inset). In the BCS limit, $\Delta_{pg}\rightarrow 0$, while it
becomes non-zero for the intermediate and strong couplings. In the pseudogap
region (below $T_p$ and above $T_c$) there are long-lived, incoherent pair
excitations (i.e. non-condensed pairs with $\vec{q}\neq 0$). The range of
occurrence of PG widens with increasing attraction. Therefore, the temperature
of the pair condensation is much lower than the temperature of the pairs
creation -- $T_p$ in the strong coupling limit. 

One can also observe changes in the chemical potential behavior when the
system evolves from the weakly interacting fermionic pairs to the strong
coupling local pairs \ref{T-matrixn0005-h0}(a) (inset). In the BCS limit,
$\mu=E_F$ and there is a Fermi surface in the system. With increasing attractive
interaction, $\mu$ drops below zero, the Fermi surface disappears and the
system becomes bosonic in the strong coupling limit. 

\begin{figure}[t!]
\hspace*{-0.8cm}
\includegraphics[width=0.36\textwidth,angle=270]
{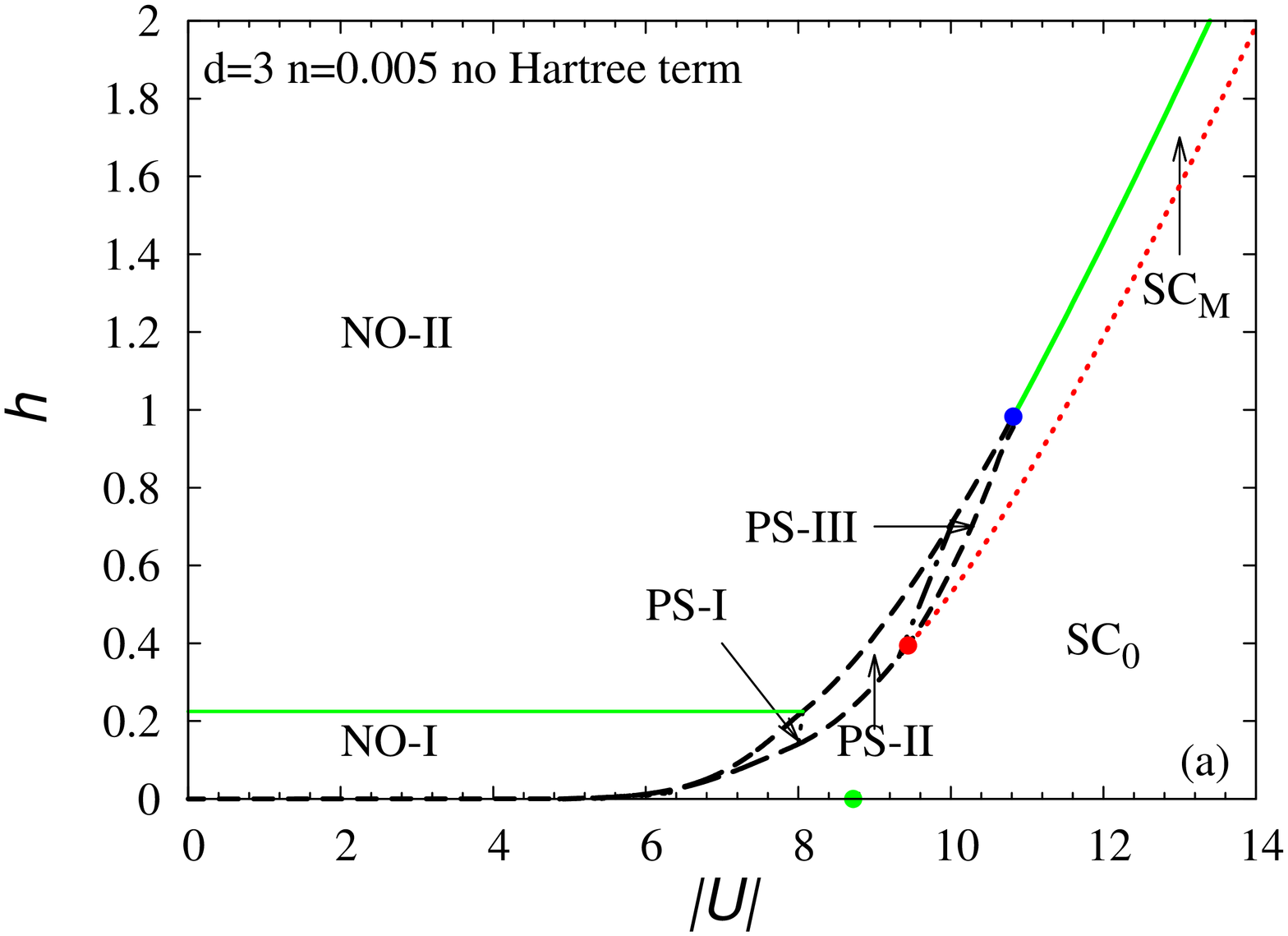}
\includegraphics[width=0.38\textwidth,angle=270]
{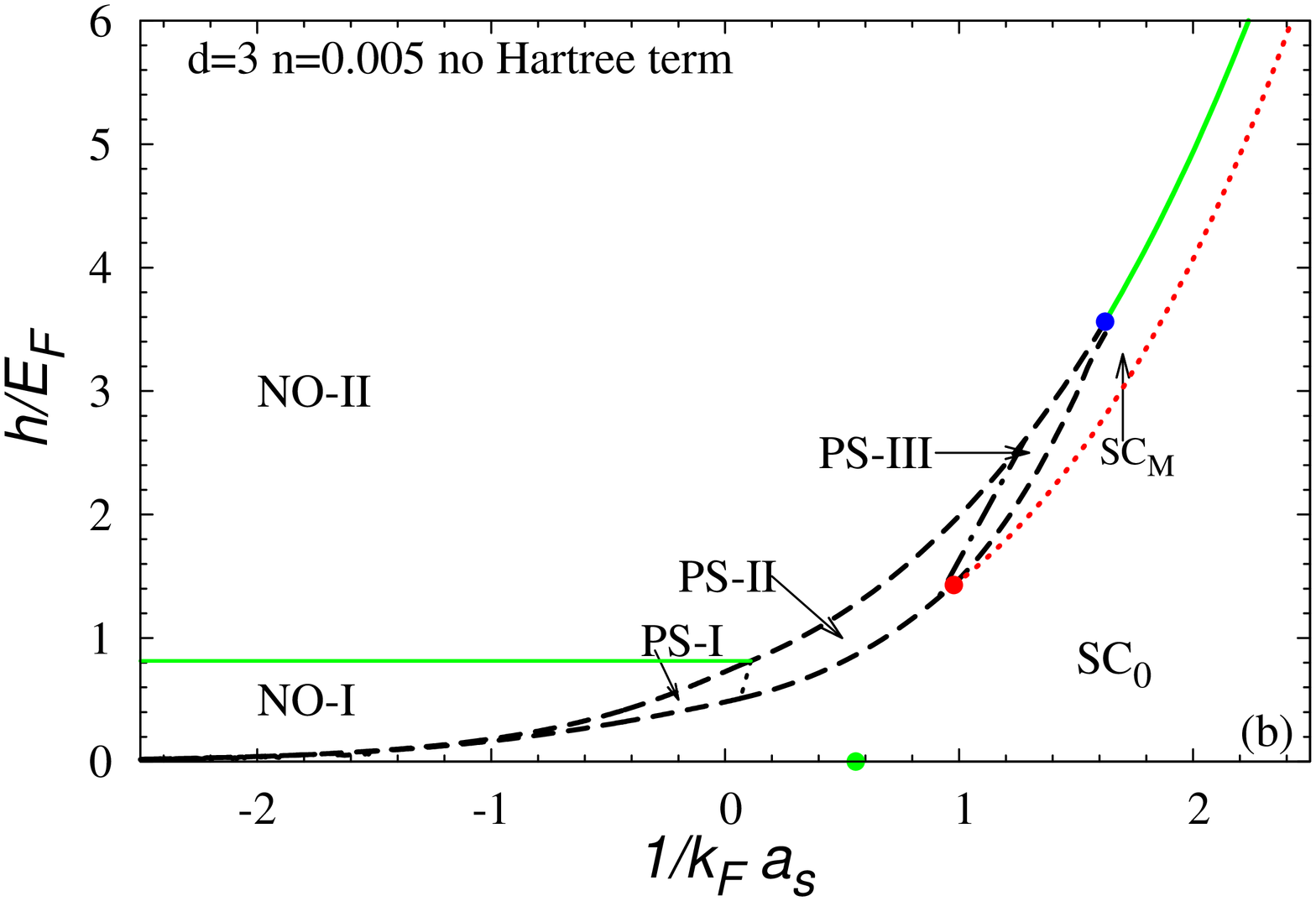}
\caption[$h$ vs. $|U|$ ground state phase diagram \textcolor{czerwony}{of spin polarized AHM} (a) and critical magnetic field in units of the lattice Fermi energy $E_F$
vs. $1/k_F
a_s$ (b).]{\label{3D_diag_n0005vsU_noHartree} $h$ vs. $|U|$ ground state phase
diagram \textcolor{czerwony}{of spin polarized AHM} (a) and critical magnetic field in units of
the lattice Fermi energy $E_F$ vs. $1/k_F
a_s$ (b), sc lattice, $n=0.005$. SC$_0$ --
unpolarized superconducting state, $SC_M$ -- magnetized
superconducting state, PS-I (SC$_0$+NO-I) -- partially polarized phase separation, PS-II (SC$_0$+NO-II) --
fully polarized phase separation, PS-III -- ($SC_M$+ NO-II). Red point -- $h_{c}^{SC_M}$,
blue point -- tricritical point. The dotted red and the solid green lines are
the second order transition lines.}
\end{figure}

\begin{figure}[p!]
\hspace*{0.8cm}
\includegraphics[width=0.32\textwidth,angle=270]{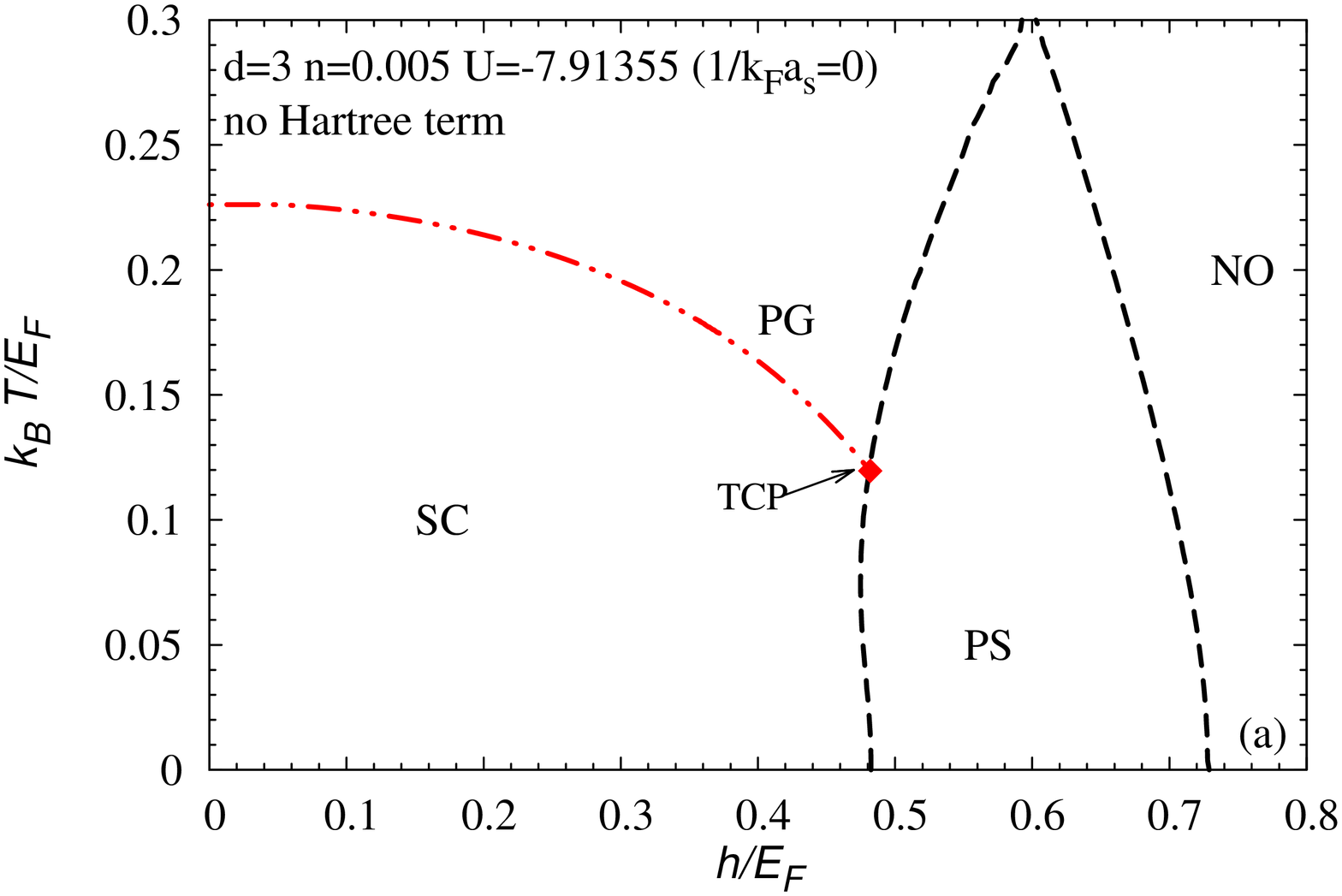}
\includegraphics[width=0.3\textwidth,angle=270]{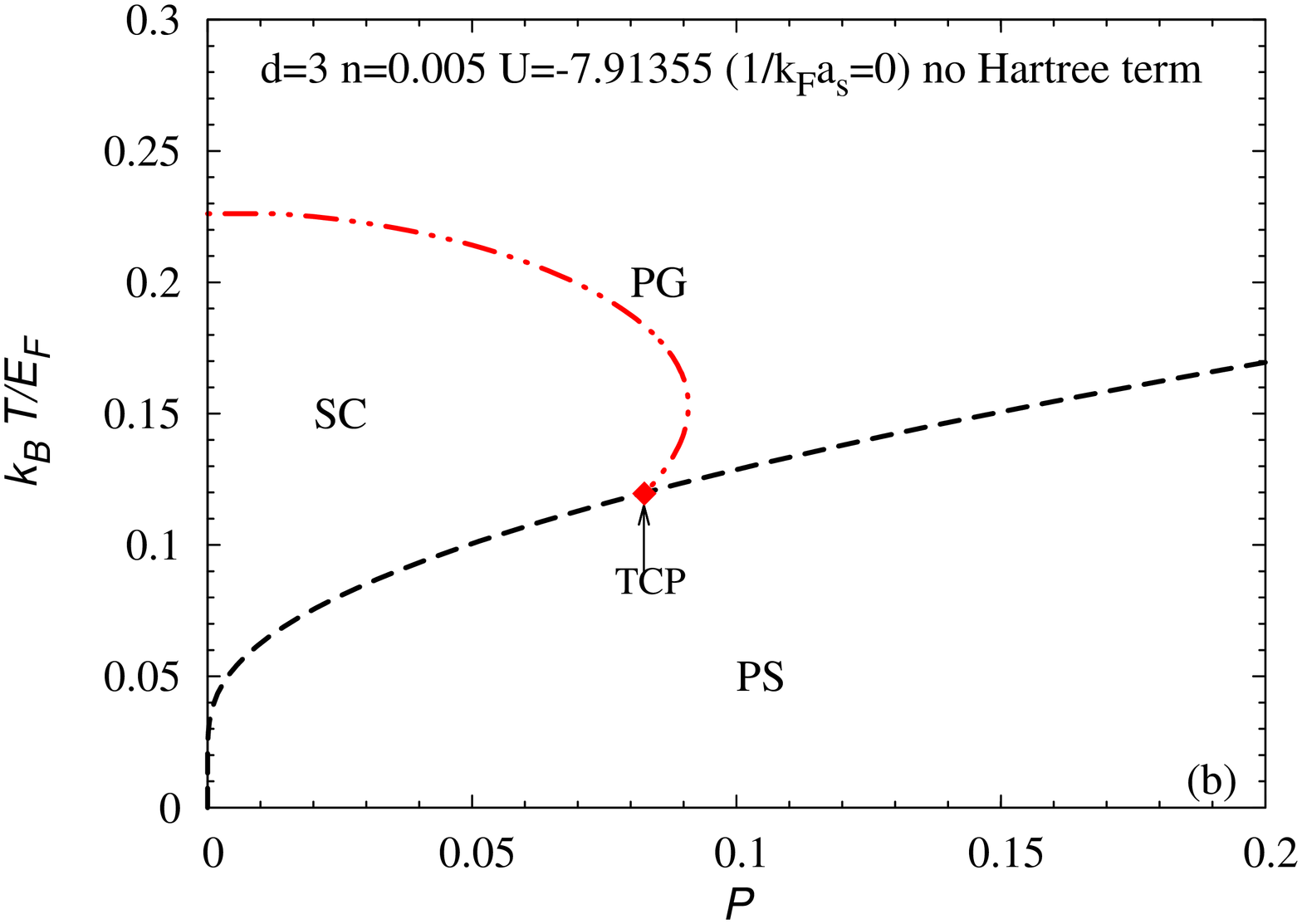}\\
\vspace*{-4mm}
\hspace*{0.8cm}
\includegraphics[width=0.32\textwidth,angle=270]{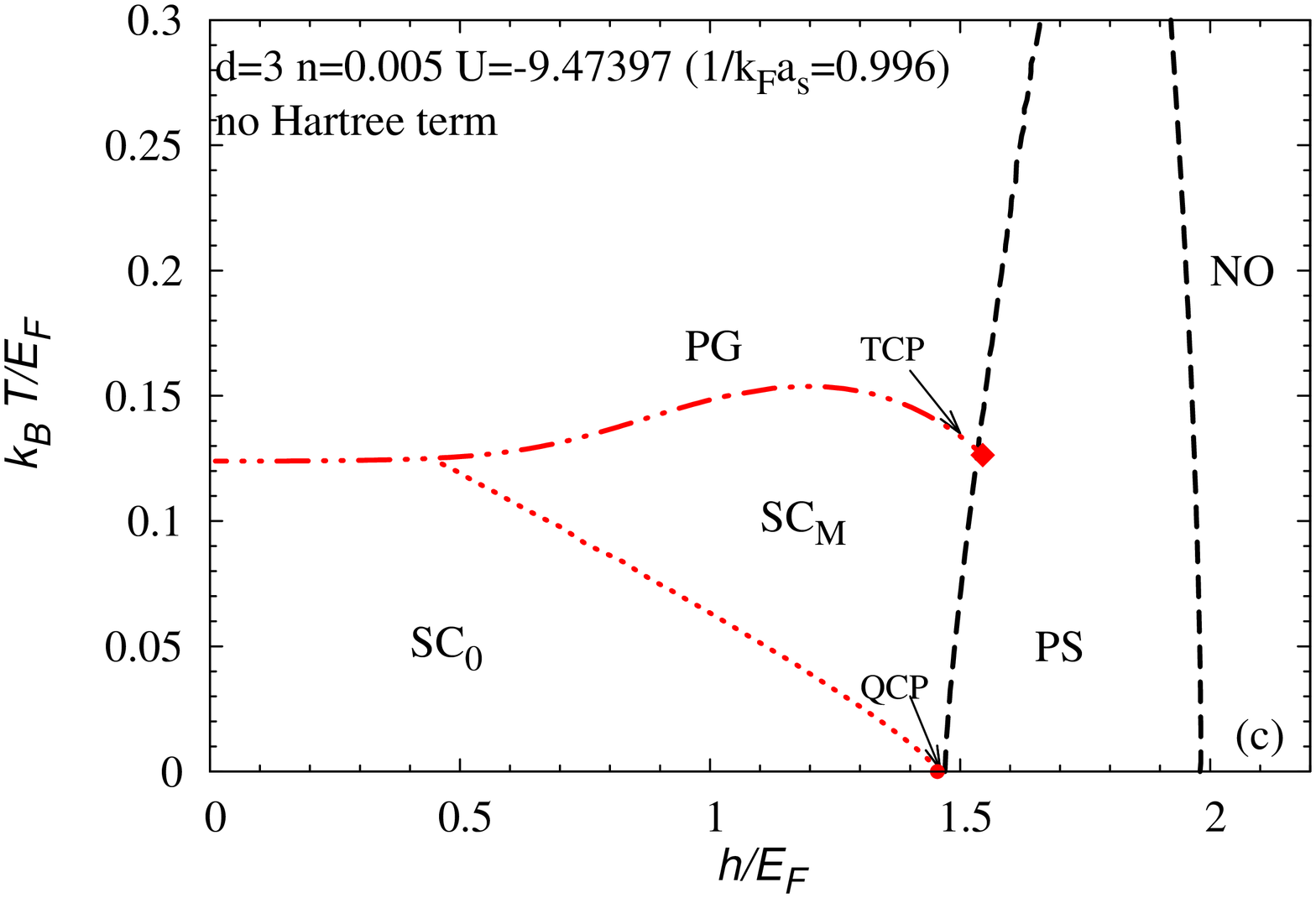}
\includegraphics[width=0.3\textwidth,angle=270]{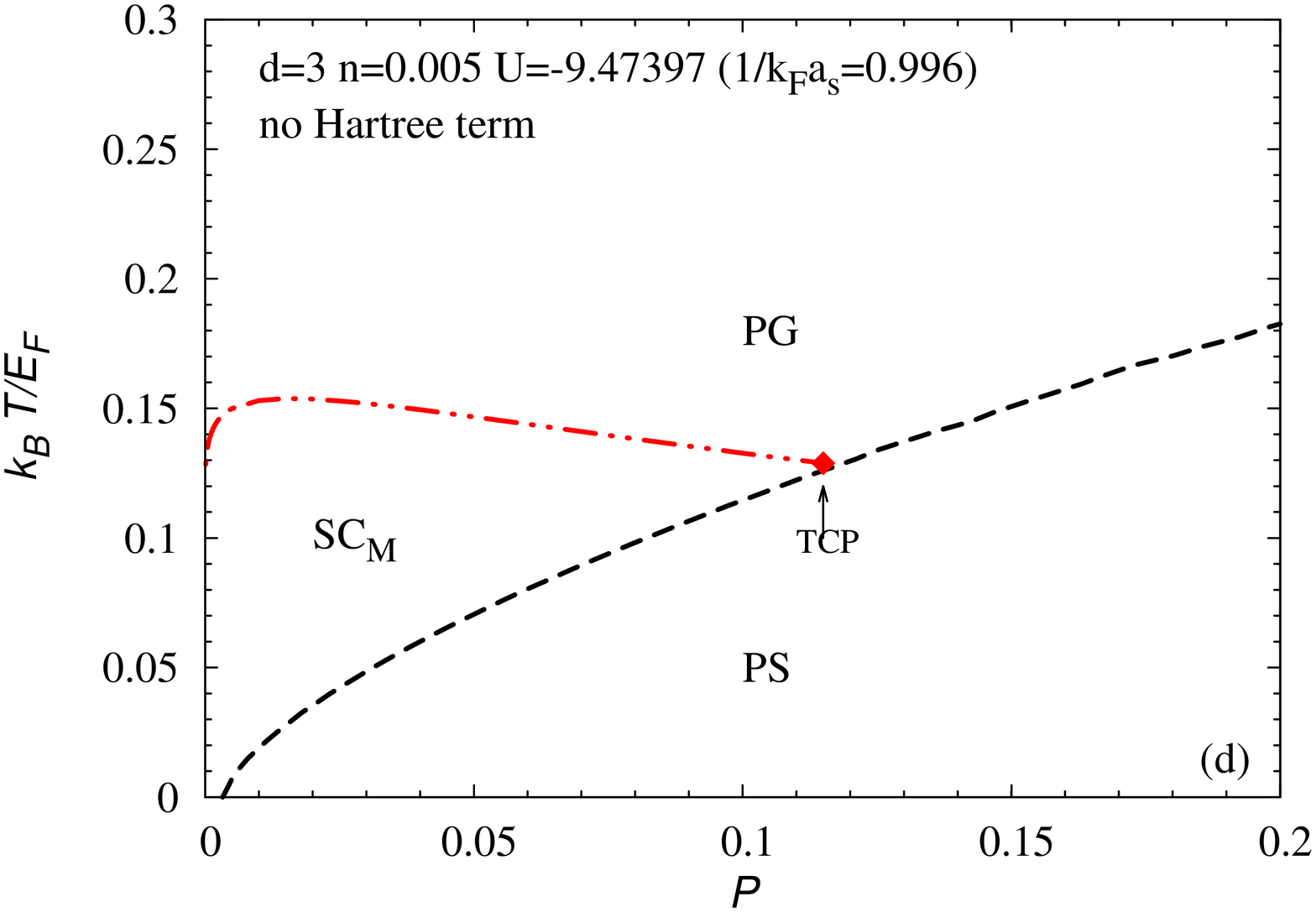}\\
\vspace*{-4mm}
\hspace*{0.8cm}
\includegraphics[width=0.32\textwidth,angle=270]{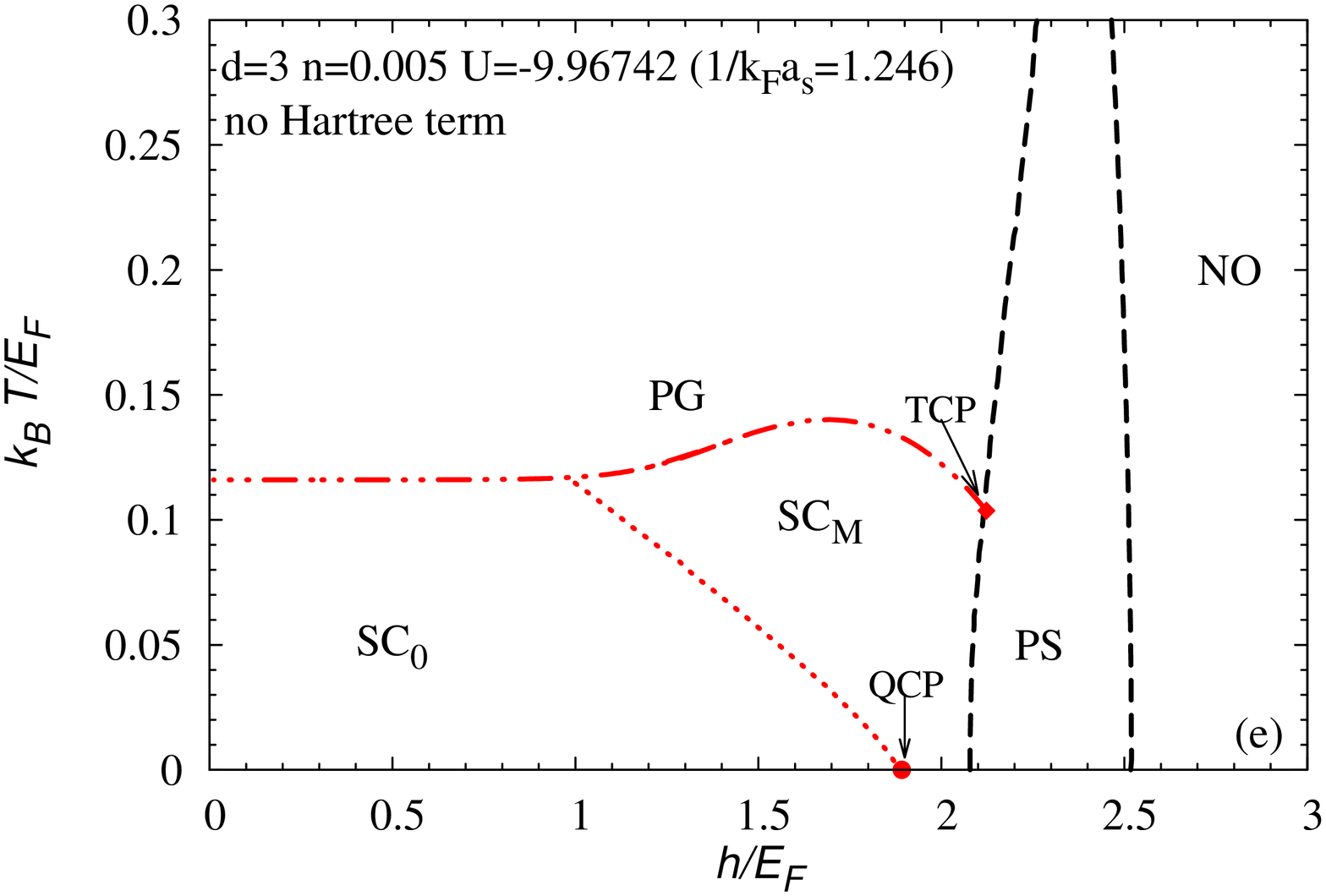}
\includegraphics[width=0.3\textwidth,angle=270]{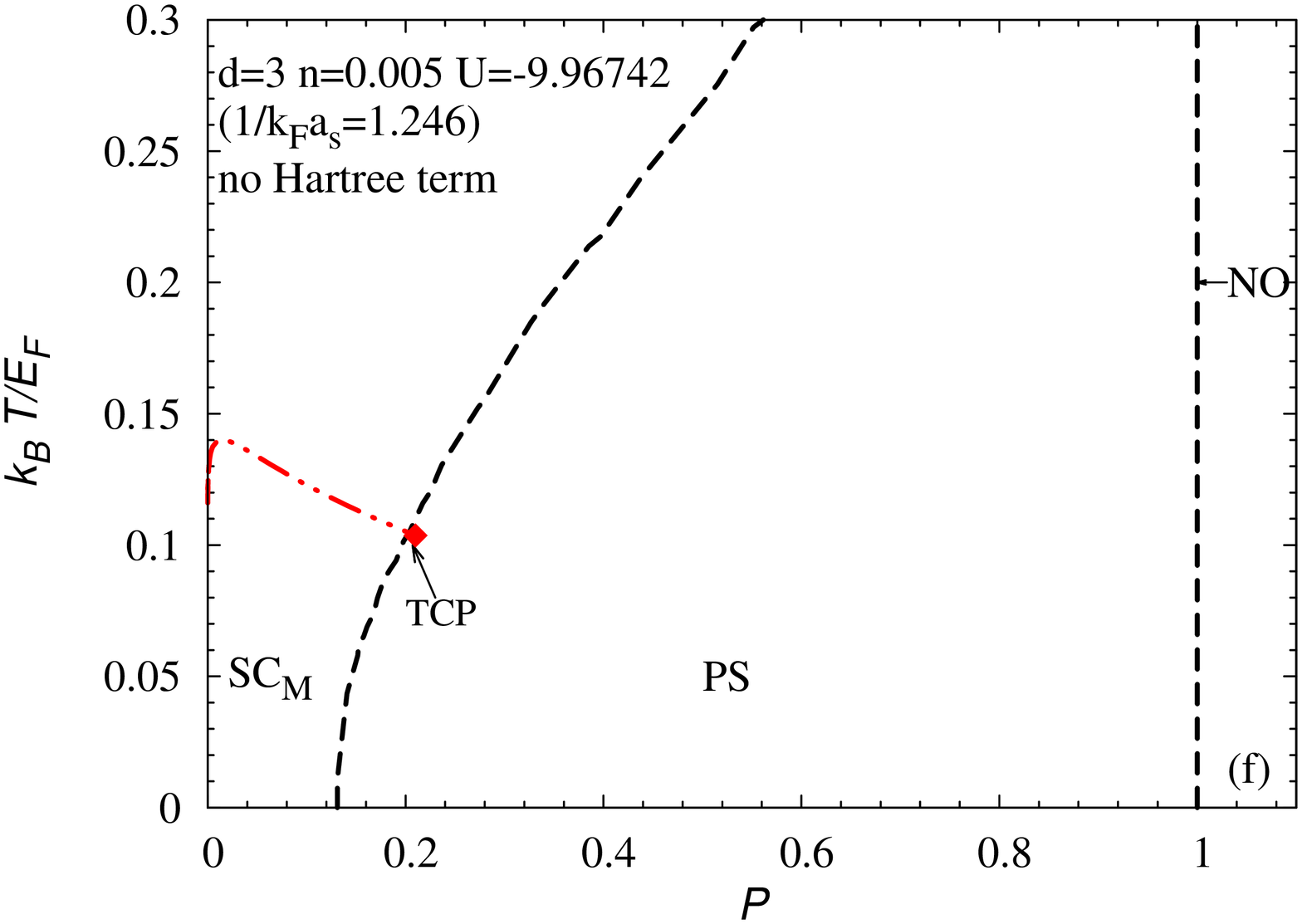}\\
\vspace*{-4mm}
\hspace*{0.8cm}
\includegraphics[width=0.32\textwidth,angle=270]{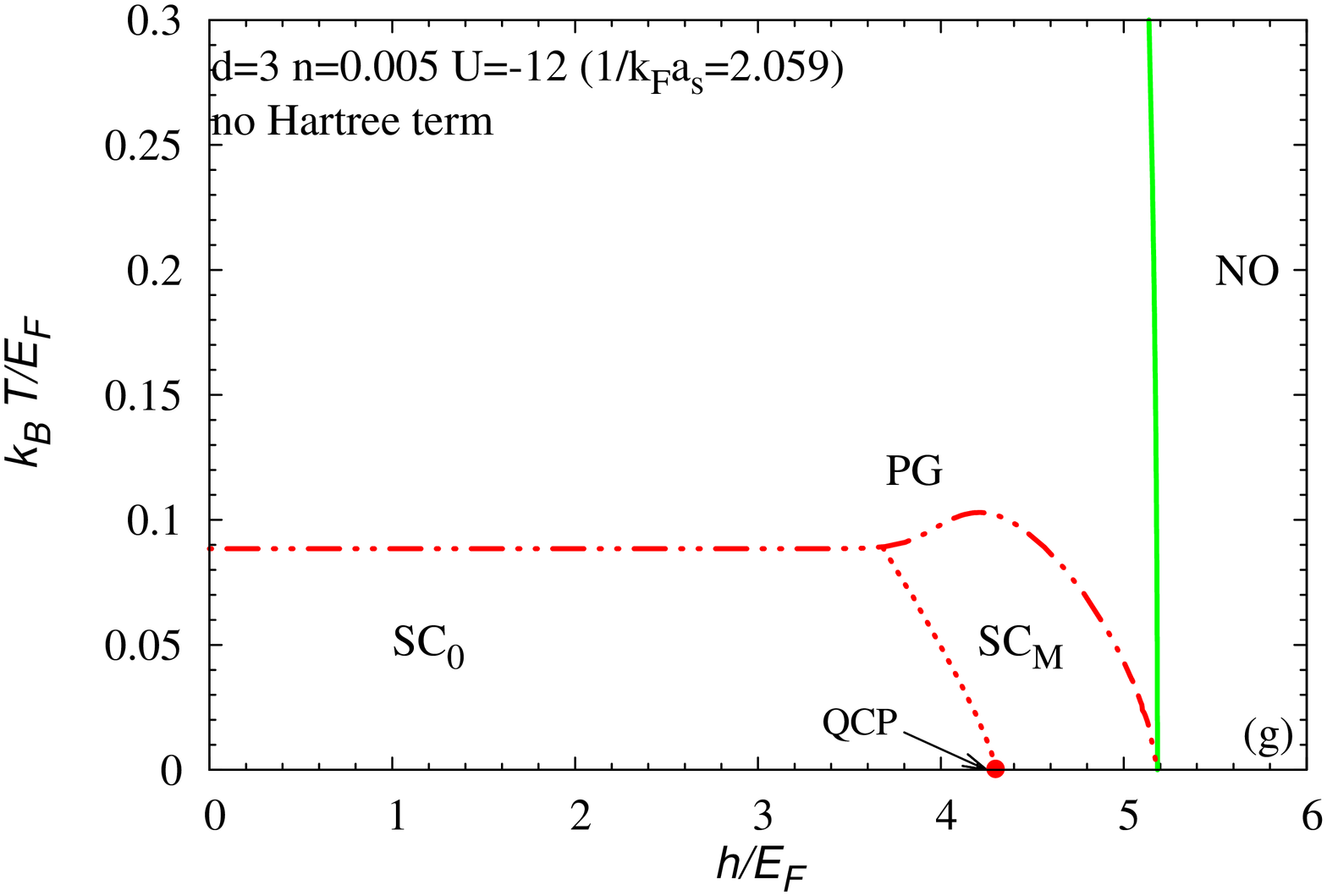}
\includegraphics[width=0.3\textwidth,angle=270]{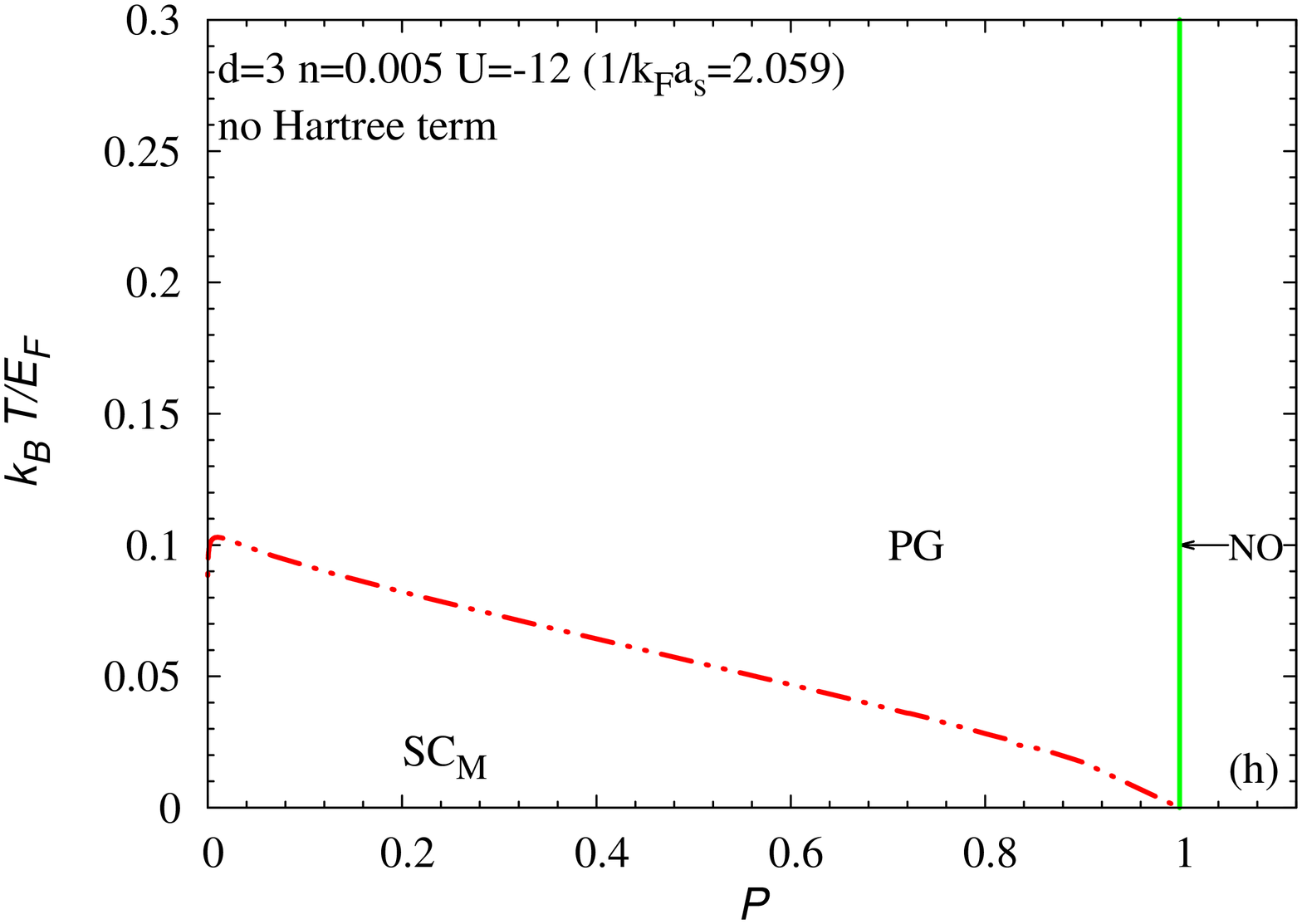}\\%
\caption[$T$ vs. $h$ and $T$ vs. $P$ phase diagrams in units of the lattice Fermi
energy, at fixed $n=0.005$ for the sc lattice.]{\label{Tvsh_n0005_Tmatrix} $T$
vs. $h$ and $T$ vs. $P$ phase diagrams \textcolor{czerwony}{of spin polarized AHM} in units of the lattice Fermi energy, at fixed
$n=0.005$ for the sc lattice. (a), (b) -- unitarity limit, (c), (d), (e), (f),
(g), (h) -- on the LP side. Dashed-double dotted red line -- $T_c$ determined
from the $(GG_0)G_0$ scheme, SC$_M$ -- magnetized
superconducting state, PG -- pseudogap, TCP -- tricritical point, QCP -- quantum critical point (Lifshitz-type).}
\end{figure}
In Fig. \ref{T-matrixn0005-h0}, $T_c$ is determined within the $(GG_0)G_0$ and
the $(GG)G_0$ scheme. First, let us consider a continuum model of a gas of
fermions in 3D. In this case, $T_c=\frac{8e^{-2}\gamma}{\pi}E_F
e^{\frac{-\pi}{2k_F a_s}}$, where $\gamma =1.78$ in the BCS limit, while in the
BEC limit $T_c/E_F=0.218$. The latter comes from the formula for the temperature
of BEC for a gas of molecules with mass $2m_F$. The calculations within
the $(GG_0)G_0$ and
the $(GG)G_0$ schemes give very similar results in these two extreme limits. However, there
are differences in the unitarity regime ($1/k_F a_s=0$), i.e. the critical
temperature determined within the $(GG_0)G_0$ scheme equals: $T_c/E_F=0.256$,
while $T_c$ calculations within the $(GG)G_0$ scheme give: $T_c/E_F=0.215$. 
For comparison, the value of $T_c$ obtained from the QMC method for the model
of continuum fermions with contact attraction in
the unitarity regime varies from $T_c/E_F=0.15$ \cite{Burovski} to
$T_c/E_F=0.18$ \cite{Bulgac}. Therefore, the results \textcolor{czerwony}{obtained} from the $(GG)G_0$
scheme are more consistent with the QMC \textcolor{czerwony}{studies} (than the ones from the
$(GG_0)G_0$ scheme).

A comparison of the above results (Fig. \ref{T-matrixn0005-h0}(b)) with those
obtained within the Attractive Hubbard model (Fig.
\ref{T-matrixn0005-h0}(a)) is quite interesting. As shown in Fig.
\ref{T-matrixn0005-h0}(a), the value of $T_c$ in the BCS limit does not change
significantly in comparison to the continuum model case.
However, there
are large differences between the results obtained within AHM and those
obtained within the 3D continuum model in the BEC limit. The value of $T_c$
which is determined within the model on the lattice is much lower then
$T_c/E_F=0.218$.
As we know, for the AHM, in the LP limit, the effective mass of the hard-core
bosons is $m_B=U/4t^2$ and increases with $|U|$  and it is reflected by the
results of the T-matrix calculations.
At the unitarity, the results for AHM ($n=0.005$) are as follows:
$T_c/E_F=0.227$ in the $(GG_0)G_0$ scheme and $T_c/E_F=0.187$ in the $(GG)G_0$
scheme. 

\subsubsection{\boldmath{$h\neq0$} case}
Fig. \ref{3D_diag_n0005vsU_noHartree} shows the critical magnetic field vs.
$|U|$ (a) and
 vs. $-1/k_Fa_s$ (b), for $n=0.005$, at $T=0$. These diagrams are only shown
so that we could refer to them in our further considerations, when the
zero-temperature results
are extended to finite temperatures. The topology of the
diagrams \ref{3D_diag_n0005vsU_noHartree} is the same as that in Fig.
\ref{3D_diag_n0001vsU_U_Hartree}. Therefore, we do not discuss Fig.
\ref{3D_diag_n0005vsU_noHartree} in detail.

Now, the results concerning the influence of the magnetic field on
superconductivity in 3D at finite temperatures are presented. As mentioned
above, $T_c$ is determined within the $(GG_0)G_0$ scheme \cite{KujawaMicnas}.

\begin{figure}[t!]
\hspace*{-0.8cm}
\includegraphics[width=0.38\textwidth,angle=270]{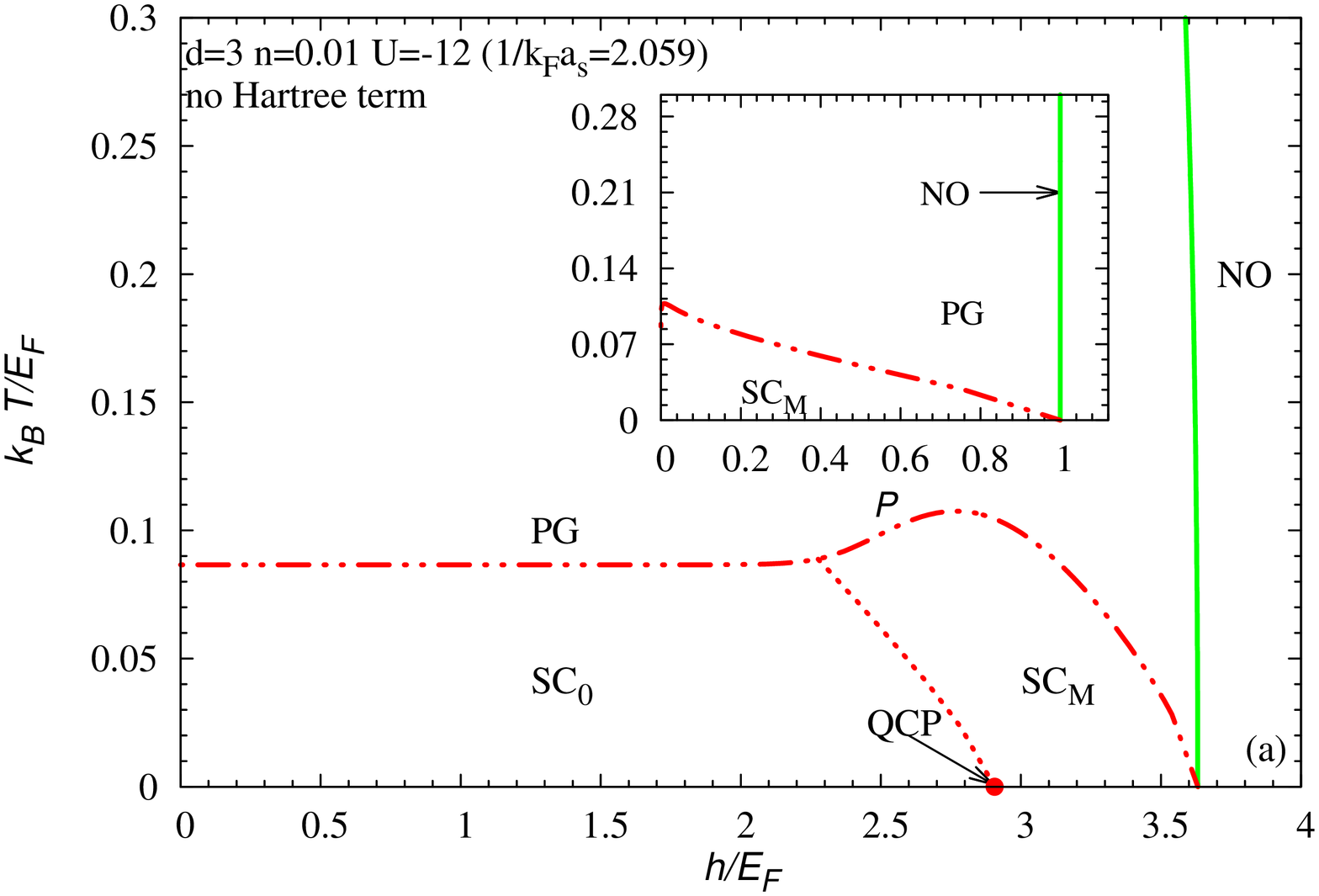}
\hspace*{-0.6cm}
\includegraphics[width=0.38\textwidth,angle=270]{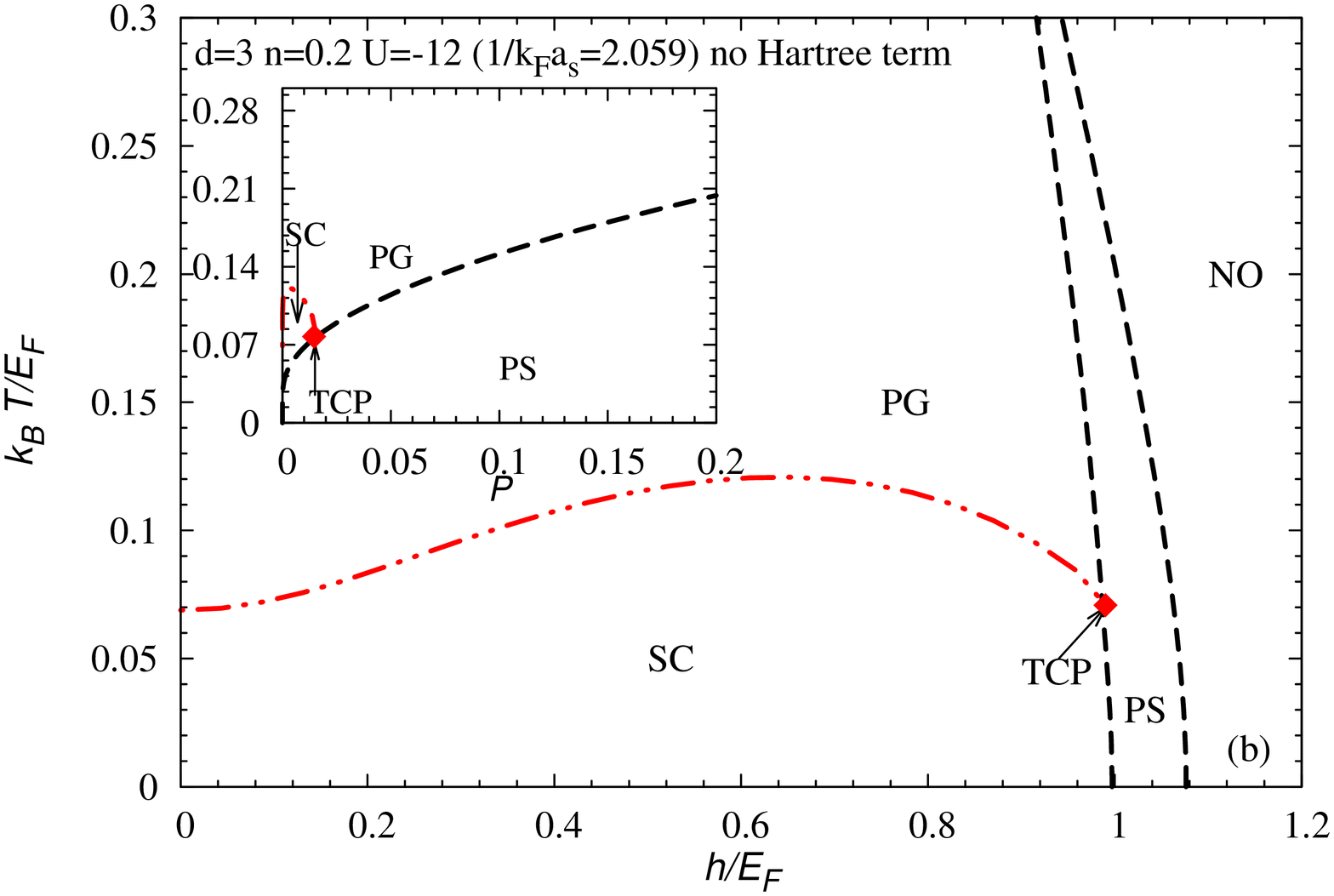}
\caption[$T$ vs. $h$ and $T$ vs. $P$ (insets) phase diagrams in units of the lattice
Fermi energy, at fixed $n=0.01$ (a) and $n=0.2$ (b) for the sc lattice,
$U=-12$.]{\label{Tvsh_n001_n02_U-12_Tmatrix}  $T$ vs. $h$ and $T$ vs. $P$
(insets) phase diagrams \textcolor{czerwony}{of spin polarized AHM} in units of the lattice Fermi energy, at fixed $n=0.01$ (a) and
$n=0.2$ (b) for the sc lattice, $U=-12$. Dashed-double dotted red line -- $T_c$
determined from the $(GG_0)G_0$ scheme, SC$_M$ --
magnetized
superconducting state, PG -- pseudogap, TCP -- tricritical point, QCP -- quantum critical point (Lifshitz-type).}
\end{figure}

Fig. \ref{Tvsh_n0005_Tmatrix} shows temperature vs. magnetic field and $T$ vs.
$P$ phase diagrams\footnote{Note that in the context of imbalanced
two-component Fermi gases on optical lattices, magnetic field translates to
chemical potential difference, while polarization to population imbalance.}
in units of the Fermi energy, at fixed $n=0.005$ and four
values of $|U|$ ($1/k_F a_s$). The PS regions are obtained within the Hartree
approximation. The thick dash-double dotted line (red color) denotes the $T_c$
curve. In other words, above this curve (above $T_c$ but below $T_p$ ($T_c^{MF}$) -- not shown in the figures) there are long lived incoherent pairs
with $\vec{q}\neq 0$ (PG region) in the system. At $T_c$ these pairs
condense and below the $T_c$ curve the system is superconducting with
zero-momentum coherent pairs. It should be noted that the pair
breaking temperature is much higher than the superconducting critical temperature.
This difference between $T_p$ and $T_c$ increases with increasing attractive
interaction. 
The evolution of TCP which is presented in Fig. \ref{Tvsh_n0005_Tmatrix} shows
that the tricritical point at $T\neq 0$ tends to TCP at $T=0$.  

In the unitarity regime (Fig. \ref{Tvsh_n0005_Tmatrix}(a)-(b)), one can
distinguish four states in the phase diagrams. There is the SC state at
sufficiently low temperatures. The system goes from SC to the normal state
across the PS region with increasing magnetic field. When temperature
increases, one observes a transition to the PG region. As shown in Fig.
\ref{Tvsh_n0005_Tmatrix}(b), the effect of finite polarization on the
superconducting state is not so strong as in the 2D case in which the KT
superfluid phase is restricted only to the weak coupling region and low values
of $P$ (see: Fig. \ref{kTvsP_Ebfinite}). 

One should mention that our results are different from those obtained by the Levin's group
results concerning the unitarity regime, for the continuum model of
two component Fermi gas with population imbalance \cite{Levin2006}.
Considering the phase diagram $T$ vs. $P$, according to their predictions, the
superfluid phase is stable at low temperatures and low polarizations, whereas
from our calculations we find the phase separation region (e.g. around $P=0.01$,
according to \cite{Levin2006} (Fig. 7), the SF phase is stable down to
$T/T_F\approx0.03$, while as follows from our calculations the SC phase is
stable only down to $T/T_F\approx0.065$). This difference can stem from two
possible reasons. First, our model is the
lattice model with a very low ($n=0.005$), but still finite electron
concentration. Second, the methods of analysis of the superfluid phase
stability are different. The Levin's group has applied the stability condition
which involves the calculation of $\partial^2 \Omega /\partial
\Delta^2$, where $\Omega$ is the thermodynamic potential and $\Delta$ --
fermionic excitation gap. Then, they impose the requirement that the
quantity $\partial^2 \Omega
/\partial \Delta^2$ has to be non-negative, i.e. $\partial^2 \Omega
/\partial \Delta^2\geq 0$. However, this stability condition does not take into
account the possibility of a first order transition and the existence of phase
separation regions in the system. Therefore, our approach seems to be more
appropriate and the condition that $\partial^2 \Omega/\partial \Delta^2\geq 0$
can be considered to be the lower bound for the stability of the superfluid
phase.

The situation is different in the strong coupling case (Fig.
\ref{Tvsh_n0005_Tmatrix}(c)-(h)). The most important feature is the occurrence
of the stable spin polarized superfluid phase. Because of the occurrence of the
magnetized
superconducting state at $T=0$ for higher attractive interactions in 3D (see:
Fig.\ref{3D_diag_n0005vsU_noHartree}), this phase can persist to non-zero
temperatures. For sufficiently high value of $|U|$, below $T_{c}$, the spin
polarized superfluid state with gapless spectrum and one FS can be stable. The
appearance of the SC$_M$ phase in the diagrams is manifested by the
existance of QCP at $T=0$. The transition from the non-polarized
superconducting state SC$_0$ to the SC$_M$ phase is of the Lifshitz type. With
increasing attractive interaction, the range of occurrence of SC$_M$ widens,
even up to $P=1$ (Fig. \ref{Tvsh_n0005_Tmatrix}(h)). An increase in the
temperature causes an increase in the polarization and, as a consequence,
the widening of the range of the SC$_M$ phase. The character of the
transition from the SC$_M$ to the NO state changes with increasing $|U|$. For
$|U|=12$ this transition is of the second order. Therefore, it is possible
to observe the polarized superfluid state even at finite temperatures in
the strong coupling regime in 3D, as opposed to the 2D case in which this phase
is unstable. 

\begin{figure}[t!]
\hspace*{-0.8cm}
\includegraphics[width=0.38\textwidth,angle=270]
{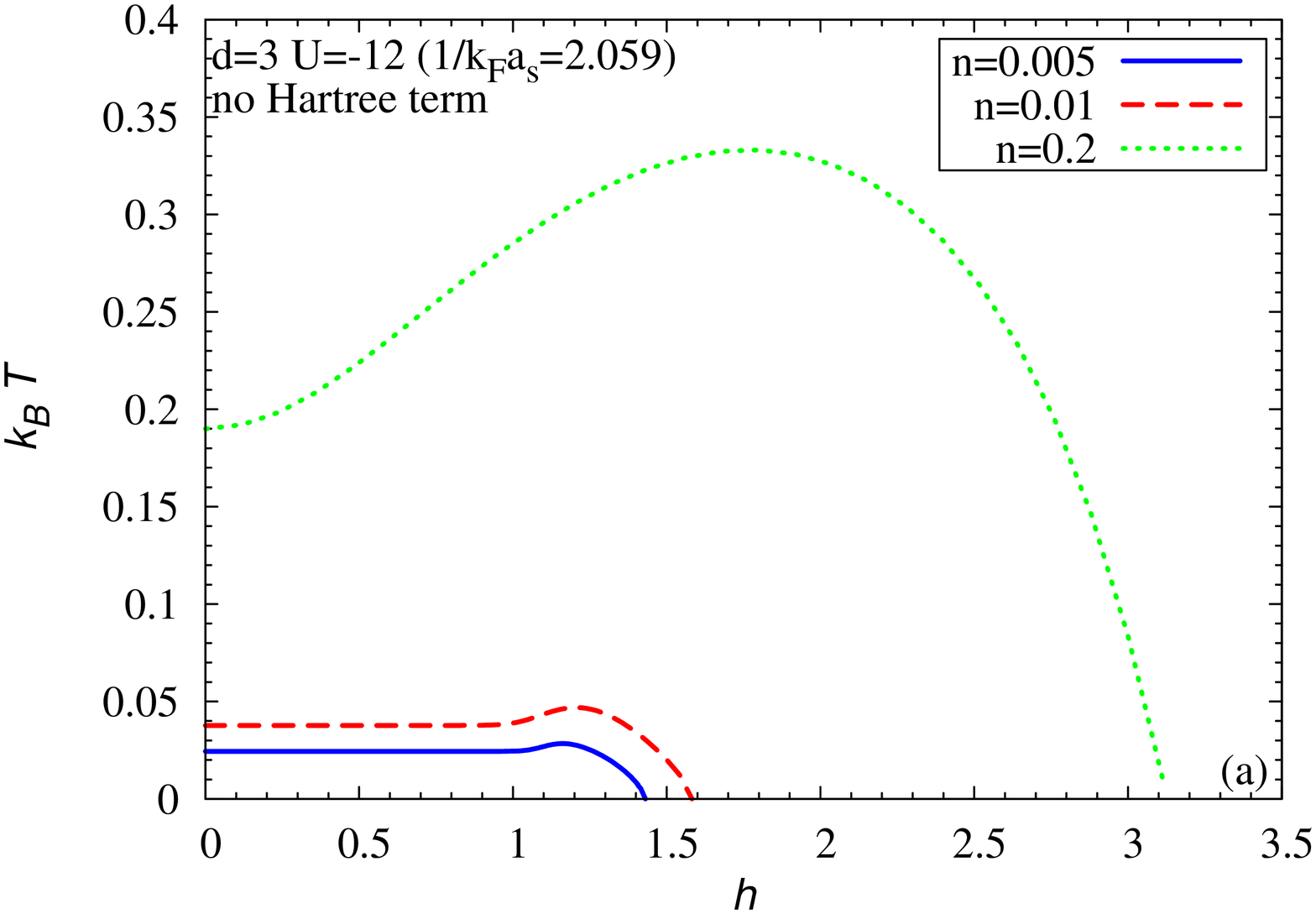}
\hspace*{-0.6cm}
\includegraphics[width=0.38\textwidth,angle=270]
{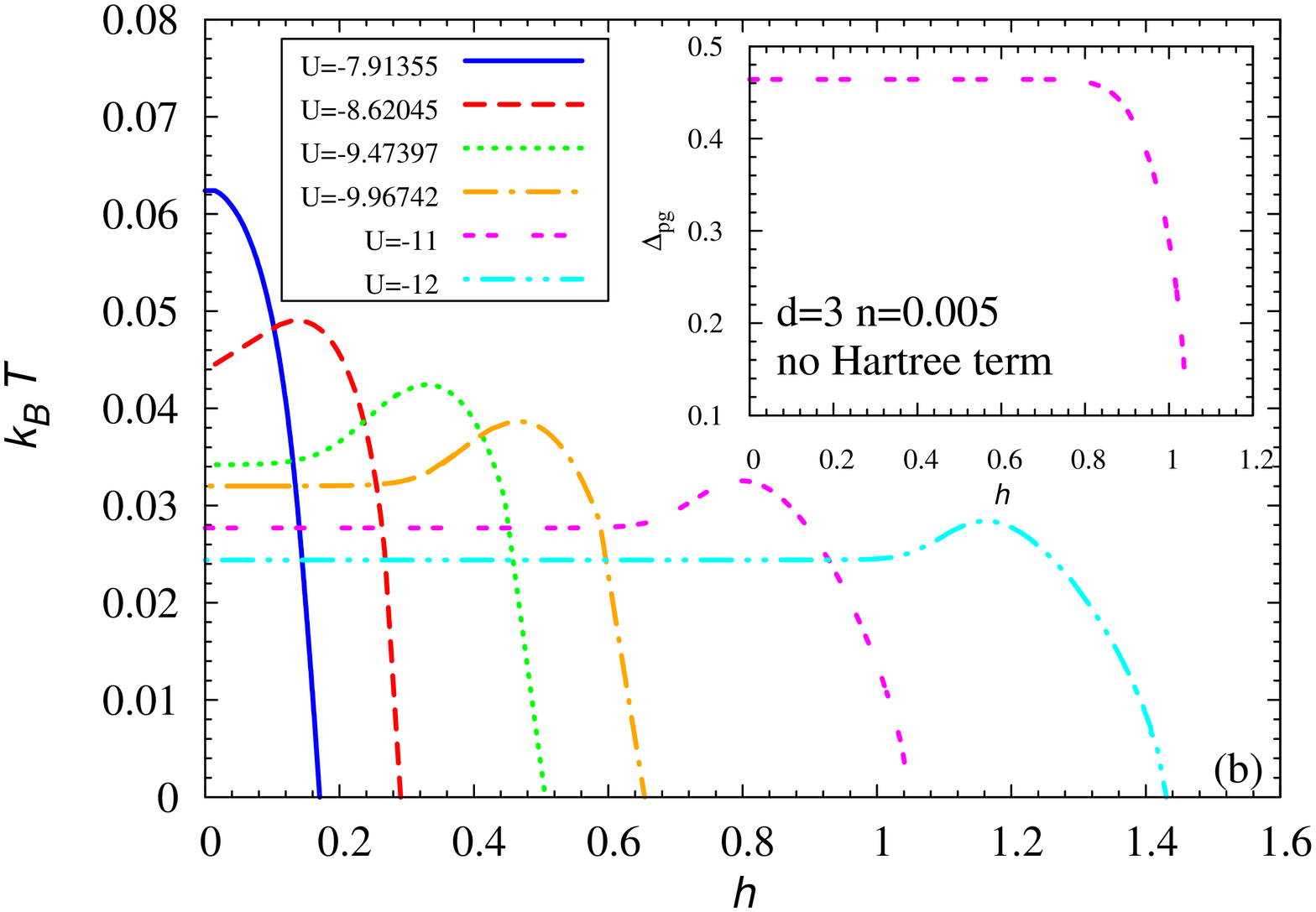}
\caption{\label{T_matrix_scheme} $T_c$ determined from the
$(GG_0)G_0$ scheme vs. $h$ for three values
of $n$ at $U=-12$ (a) and at fixed $n=0.005$ for six values of $|U|$ (b). Inset of (b) shows the behavior of $\Delta_{pg}$ vs. $h$ for fixed $n=0.005$ and $U=-11$.}
\end{figure}

However, the SC$_M$ state is stable for rather low electron concentrations,
at $T\neq 0$. It is clearly visible in Fig. \ref{Tvsh_n001_n02_U-12_Tmatrix}.
For $|U|=12$ and $n=0.01$ the SC$_M$ phase is stable. However, for higher value
of the electron concentration ($n=0.2$) the SC phase is restricted to low
polarizations. Moreover, this phase is not gapless for higher $n$. Here, one
should emphasize, that the SC phase and the SC$_M$ state are formally different
phases.

The evolution of the $T_c$ curve with increasing attractive interaction or
electron concentration is also very interesting. Fig. \ref{T_matrix_scheme}
shows $T_c$ determined from the $(GG_0)G_0$ scheme vs.
$h$ for three values of $n$ at $U=-12$ (a) and at fixed $n=0.005$ for six
values of $|U|$.
For low values of the electron concentration and high values of attraction
(Fig. \ref{T_matrix_scheme}(b)), the critical temperature and the pseudogap parameter do not depend on the
magnetic field for a wide range of values of $h$. The system is in the $SC_0$
phase. When non-zero polarization appears, the critical temperature increases
and after reaching some maximum, it starts to decrease. 
Moreover, in this non-zero polarization region, an increase in $h$ leads to a decrease in the pseudogap parameter, which tends to zero (as does the critical temperature) for some critical value of the magnetic field.

When the electron concentration is increased (Fig.
\ref{T_matrix_scheme}(a)), the range of values of $h$ for which the critical
temperature does not depend on $h$ shrinks and disappears for large enough $n$,
similarly as for lower values of attraction (Fig. \ref{T_matrix_scheme}(b)).

\chapter{The influence of spin dependent hopping integrals (mass imbalance) on
the BCS-BEC crossover}

Two-component Fermi systems with different fermion masses in optical
lattices \cite{Wille} are characterized by a new parameter, i.e. the ratio of
the spin dependent hopping integrals ($t^{\uparrow}/t^{\downarrow}\equiv r$,
where $t^{\uparrow}\neq t^{\downarrow}$), which enables the
appearance of various exotic states in these systems. \textcolor{czerwony}{The case of spin dependent masses is also 
studied in papers of Spałek group in the context of unconventional superconducting materials including the heavy fermion systems   
\cite{spalek}.}

In this chapter, we examine the influence of spin dependent hopping integrals on
the stability of the  SC$_M$ phase. We study the evolution from the weak
coupling (BCS-like limit) to the strong coupling limit of tightly bound local
pairs with increasing attraction, within the attractive Hubbard model in a
magnetic field with spin-dependent hopping integrals, for square and simple
cubic lattices. The broken symmetry Hartree approximation, as well as the strong
coupling expansion are used. In the 2D case, we also apply the KT scenario to
determine the phase coherence temperatures.

Some theoretical studies of Fermi condensates in systems with spin and mass
imbalances have shown that the BP state can have excess fermions with two FS's
(BP-2 or interior gap state) \cite{Wilczek,Wilczek3,Iskin,Iskin-2}. According to
some investigations, the interior gap state \cite{Wilczek} is always unstable
even for large mass ratio $r$ and PS is favored \cite{Parish,Parish2}.
Therefore, the problem of stability of the BP-2 state is still open. 

At strong attraction, the SC$_M$ phase occurs in three-dimensional imbalanced
Fermi gases \cite{Sheehy, Parish, Parish2}, as well as in the spin-polarized
attractive Hubbard model in the dilute limit (for $h\neq 0$, $r=1$
\cite{Kujawa3} and $r\neq 1$ \cite{Kujawa5}). As noted in the previous chapter,
this homogeneous magnetized superfluid state consisting of a coherent mixture of
LP's (hard-core bosons) and excess spin-up fermions (Bose-Fermi mixture) can
only have one Fermi surface (BP-1).

We also show that if $r\neq 1$, the SC$_M$ phase can be realized in $d=2$ for
the intermediate and strong coupling regimes, as opposed to the $r=1$ case. In
other words, the simultanious presence of mass and population imbalance can
stabilize
the BP-1 phase in 2D, on the BEC side of the crossover. We determine the
critical value of $n$ above which SC and the charge density wave ordered (CO)
state can form the PS state at $h=0$ and $r\neq 1$. We find that the BP-2 state
is unstable in the whole range of parameters, in the $d=2$ one-band
spin-polarized AHM. Nevertheless, one can suppose that the Liu-Wilczek (BP-2)
phase can be realized within the two-band model.

Some of our results have been published in Refs. \cite{Kujawa4,
Kujawa6}.

In this chapter, we study the superfluid phases in the AHM in a magnetic
field with spin-dependent hopping. The Hamiltonian \eqref{extham'} is reduced
to
the form \cite{Kujawa4,Kujawa6}:
\begin{equation}
\label{ham-rneq1}
H=\sum_{ij\sigma} (t_{ij}^{\sigma}-\mu
\delta_{ij})c_{i\sigma}^{\dag}c_{j\sigma}+U\sum_{i}
n_{i\uparrow}n_{i\downarrow}-h\sum_{i}(n_{i\uparrow}-n_{i\downarrow}),
\end{equation}
where: $t_{ij}^{\sigma}$ -- hopping integrals, $\sigma=\uparrow,\downarrow$.

\section{The large-$U$ \textcolor{czerwony}{limits}}
In this section, we use the canonical transformation method of the
Schrieffer-Wolff type \cite{Schrieffer} to map the Hamiltonian
\eqref{ham-rneq1} ($h=0$, $r\neq 1$) onto the pseudo-spin model in the
large-$|U|$ \textcolor{czerwony}{limits} \cite{Robaszkiewicz,MicnasModern,Cazalilla}.

The canonical transformation method allows us to isolate the interactions
that dominate the dynamics of the system. Strongly correlated Hubbard
models have been investigated for many years. The large-positive-$U$ (LPU) limit
was used to explain some aspects of antiferromagnetism and superconductivity
\cite{Baskaran}. Harris and Lange \cite{Harris} have carried out the
calculations for
the Hubbard model with the use this transformation. Then, Chao, Spałek and Oleś
have proposed an iterative method for the derivation of an appropriate
transformation for
higher orders of perturbation theory \cite{Chao}. This scheme has been improved
by
McDonald, Girvin and Yoshioka \cite{McDonald}. The canonical transformation
method has also been used in Refs.: \cite{Gros, Oles, Kolley}. The
large-negative-$U$ limit has also been widely discussed in literature
\cite{Robaszkiewicz, MicnasModern, Kulik, Kulik-2, Oppermann}. 

Here, we briefly introduce the concept of the canonical transformation. Let us
consider the Hamiltonian:
\begin{equation}
\label{ham_pert}
 H=H_0+H_{I},
\end{equation}
where: $H_0$ -- unperturbed part, $H_{I}$ -- perturbation. 

Now, we define the canonical transformation as: 
\begin{equation}
\label{transHam}
 \tilde{H}=e^{S}He^{-S},
\end{equation}
where: $S^{\dag}=-S$ -- anti-Hermitian operator. 

The Hamiltonian \eqref{transHam} can be expanded:
\begin{equation}
\tilde{H}=H+[S,H]+\frac{1}{2} [S,[S,H]]+\ldots, 
\end{equation}
We decompose the Hamiltonian \eqref{ham-rneq1} ($h=0$) into:
\begin{equation}
\label{H1}
H_1=\sum_{ij\sigma} t_{ij}^{\sigma} (c_{i\sigma}^{\dag}
(1-n_{i\bar{\sigma}})c_{j\sigma}
(1-n_{j\bar{\sigma}})+c_{i\sigma}^{\dag}n_{i\bar{\sigma}}c_{j\sigma}n_{j\bar{
\sigma}}), 
\end{equation}
\begin{equation}
 H_{mix}=\sum_{ij\sigma}
t_{ij}^{\sigma}(c_{i\sigma}^{\dag}(1-n_{i\bar{\sigma}})n_{j\bar{\sigma}}+c_{
i\sigma}^{\dag}n_{i\bar{\sigma}}c_{j\sigma}(1-n_{j\bar{\sigma}})),
\end{equation}
\begin{equation}
\label{Hu}
 H_U=U\sum_{i} n_{i\uparrow}n_{i\downarrow}.
\end{equation}
where: $\bar{\sigma}=-\sigma$.
Therefore, the Hamiltonian \eqref{ham_pert} can be rewritten as:
$H=H_U+H_1+H_{mix}$, where $H_0=H_U+H_1$ and $H_I=H_{mix}$. In this way:
\begin{equation}
\label{ham_trans'}
\tilde{H}=H_U+H_1+H_{mix}+[S,H_U]+[S,H_1]+[S,H_{mix}]+\frac{1}{2}[S,[S,H]]
+\ldots  
\end{equation}
By means of the canonical transformation, the perturbation ($H_I=H_{mix}$) can
be eliminated to first order. Therefore, we demand that the generator $S$
satisfies the condition: $H_{mix}+[S,H_U]=0$. Then, the Hamiltonian
\eqref{ham_trans'} takes the form:
\begin{equation}
\label{ham_trans''}
 \tilde{H}=H_U+H_1+[S,H_1]+\frac{1}{2}[S,H_{mix}]+\ldots
\end{equation}
In the lowest order, the Hamiltonian \eqref{ham_trans''} is well approximated
by:
\begin{equation}
 \tilde{H}'=H_U+H_1+\frac{1}{2}[S,H_{mix}]+\ldots
\end{equation}
Let us introduce the Hubbard operators. We have the following four local basis
states in the Hubbard model \cite{Fazekas}:
\begin{equation}
\begin{array}{ll}
|0\rangle_j	&	\textrm{site j is empty}\\
|+\rangle_j=c_{j\uparrow}^{\dag}|0\rangle_j	&	\textrm{site j is
occupied by an} \uparrow \textrm{-electron}\\
|-\rangle_j=c_{j\downarrow}^{\dag}|0\rangle_j	&	\textrm{site j is
occupied by an} \downarrow \textrm{-electron}\\
|2\rangle_j=c_{j\uparrow}^{\dag}c_{j\downarrow}^{\dag}|0\rangle_j	&
\textrm{site j is doubly occupied}.\\
\end{array}
\end{equation}
The corresponding local projectors take the form:
\begin{equation}
\label{P0}
\hat{P}_{j0}=|0\rangle_{j}\,_{j}\!\langle0|=
(1-n_{j\uparrow})(1-n_{j\downarrow}),
\end{equation}
\begin{equation}
\hat{P}_{j+}=|+\rangle_{j}\,_{j}\!\langle+|= n_{j\uparrow}(1-n_{j\downarrow}),
\end{equation}
\begin{equation}
\hat{P}_{j-}=|-\rangle_{j}\,_{j}\!\langle-|= n_{j\downarrow}(1-n_{j\uparrow}),
\end{equation}
\begin{equation}
\label{P2}
\hat{P}_{j2}=|2\rangle_{j}\,_{j}\!\langle2|=n_{j\uparrow}n_{j\downarrow},
\end{equation}
and satisfy:
\begin{equation}
\hat{P}_{j0}+\hat{P}_{j+}+\hat{P}_{j-}+\hat{P}_{j2}=\hat{1}, 
\end{equation}
where: $\hat{1}$ -- the unit operator.

Now, we can define the Hubbard operator as:
\begin{equation}
 X_{j}^{ba}=|b\rangle_{j}\,_{j}\!\langle a|.
\end{equation}
One can notice that the projection operators \eqref{P0}-\eqref{P2} are the
diagonal Hubbard operators: $\hat{P}_{ja}=|a \rangle \langle a |=X_{j}^{a a}$,
e.g. $X_{j}^{22}=n_{j\uparrow}n_{j\downarrow}$. In turn, an example of the
off-diagonal Hubbard operator takes the form:
\begin{equation}
 X_{j}^{\sigma 0}=c_{j\sigma}^{\dag}(1-n_{j\bar{\sigma}}),
\end{equation}
or:
\begin{equation}
 X_j^{20}=\eta (\sigma) c_{j\sigma}^{\dag} c_{j\bar{\sigma}}^{\dag},
\end{equation}
where the sign factor:
\begin{equation}
 \eta (\sigma)= \left\{ \begin{array}{ll}
+1 & \textrm{if} \,\, \sigma = \uparrow (+)  \\    
-1 & \textrm{if} \,\, \sigma = \downarrow (-)
\end{array} \right.
\end{equation}

The products of the Hubbard operator are given by:
\begin{equation}
 X_{j}^{cf}X_{j}^{ba}=\delta_{bf}X_{j}^{cb}X_{j}^{ba}=\delta_{bf}X_{j}^{ca}.
\end{equation}
The Hubbard operators description is an alternative approach to the standard
creation and anihilation operators picture. 

Let us consider the $c_{j\uparrow}^{\dag}$ operator to understand the
correspondences between these two descriptions. As is well known,
$c_{j\uparrow}^{\dag}$ operator acting on an empty state gives $|+\rangle$ or,
acting on an $|-\rangle$, gives a doubly occupied state $|2\rangle$. Then, one
can write:
\begin{equation}
\label{c_up}
c_{j\uparrow}^{\dag}=X_{j}^{+0}+X_{j}^{2-} 
\end{equation}
and similarly $c_{j\downarrow}^{\dag}$:
\begin{equation}
\label{c_down}
c_{j\downarrow}^{\dag}=X_{j}^{-0}-X_{j}^{2+}.
\end{equation}
Now, one can write \eqref{c_up} and \eqref{c_down} in a compact form:
\begin{equation}
 c_{j\sigma}^{\dag}=X_{j}^{\sigma 0}+\eta (\sigma) X_{j}^{2\bar{\sigma}},
\end{equation}

It is worth mentioning that the Hubbard operators obey the following commutation
relations:
\begin{equation}
 [X_{i}^{ba},X_{j}^{dc}]_{\pm}=\delta_{i,j}(\delta_{ad}X_{i}^{bc}\pm
\delta_{bc}X_{i}^{da}),
\end{equation}
where:
\begin{equation}
 \begin{array}{ll}
+ & \textrm{if both operators are fermionic type,}  \\    
- & \textrm{if at least one of them is bosonic type.} 
\end{array}
\end{equation}
Now, parts of the Hamiltonian \eqref{ham-rneq1} (\eqref{H1}-\eqref{Hu}) can
be rewritten in terms of the Hubbard operators:
\begin{equation}
H_{1}=\sum_{ij\sigma} t_{ij}^{\sigma} \Big(X_{i}^{\sigma
0}X_{j}^{0\sigma}+X_{i}^{2\bar{\sigma}}X_{j}^{\bar{\sigma}2}\Big), 
\end{equation}
\begin{equation}
H_{mix}=\sum_{ij\sigma}\eta (\sigma) t_{ij}^{\sigma} \Big(X_{i}^{\sigma
0}X_{j}^{\bar{\sigma}2}+X_{i}^{2\bar{\sigma}}X_{j}^{0\sigma}\Big), 
\end{equation}
\begin{equation}
H_U=U\sum_{i} X_{i}^{22}. 
\end{equation}
The generator $S$ of the canonical transformation is given by:
\begin{equation}
\label{generator}
 S=\frac{1}{2i}\Big(\int_0^{\infty} (H_{mix}(t)-H_{mix}(-t))e^{-\epsilon
t}dt\Big) \Big|_{\epsilon \rightarrow 0},
\end{equation}
where:
\begin{equation}
\label{Hmix'}
 H_{mix}(t)=Ae^{-iUt}+Be^{iUt},
\end{equation}
 $A=\sum_{ij\sigma} \eta (\sigma) t_{ij}^{\sigma} X_{i}^{\sigma
0}X_{j}^{\bar{\sigma} 2}$, $B=\sum_{ij\sigma} \eta (\sigma) t_{ij}^{\sigma}
X_{i}^{2\bar{\sigma}}X_{j}^{0\sigma}$.
\\
\\
Substituting Eq.~\eqref{Hmix'} into Eq.~\eqref{generator} we get:
\begin{equation}
S=-\frac{1}{U} (A-B). 
\end{equation}
Now, let us calculate the expression:
\begin{equation}
\label{Hdoublebar}
 \bar{\bar{H}}\equiv
\frac{1}{2}[S,H_{mix}]=-\frac{1}{2U}[A-B,A+B]=-\frac{1}{2U}\Big([A,B]-[B,A]
\Big)=-\frac{1}{U}[A,B],
\end{equation}
where:
\begin{equation}
 [A,B]=\sum_{ij\sigma}\sum_{kl\sigma'} \eta (\sigma) \eta (\sigma')
t_{ij}^{\sigma} t_{kl}^{\sigma'}[X_{i}^{\sigma
0}X_{j}^{\bar{\sigma}2},X_{k}^{2\bar{\sigma'}}X_{l}^{0\sigma}],
\end{equation}
$t_{ij}^{\sigma}=t_{ji}^{\sigma}$. We take into account the terms for which
$i=k$, $j=l$ or $i=l$, $k=j$ and omit the three-site terms. In this way, we get
two commutators:
\begin{equation}
 \begin{array}{ll}
[X_{i}^{\sigma 0}X_{j}^{\bar{\sigma}2}, X_{i}^{2\bar{\sigma'}}X_{j}^{0\sigma}]
\,\,\,\, \textrm{or} & [X_{i}^{\sigma 0}X_{j}^{\bar{\sigma}2},
X_{j}^{2\bar{\sigma'}}X_{i}^{0\sigma'}].\\    
\end{array}
\end{equation}
After the simple calculations, first of them equals:
\begin{equation}
\label{commutator1}
 [X_{i}^{\sigma 0}X_{j}^{\bar{\sigma}2},
X_{i}^{2\bar{\sigma'}}X_{j}^{0\sigma}]=\delta_{\sigma'\bar{\sigma}}X_{i}^{20}X_{
j}^{02},
\end{equation}
and the second:
\begin{equation}
\label{commutator2}
[X_{i}^{\sigma 0}X_{j}^{\bar{\sigma}2}, X_{j}^{2\bar{\sigma'}}X_{i}^{0\sigma'}]=
X_{i}^{\sigma \sigma'} X_{j}^{\bar{\sigma}\bar{\sigma'}}-\delta_{\sigma
\sigma'}X_{i}^{00}X_{j}^{22}.
\end{equation}
Substituting Eqs.~\eqref{commutator1}, \eqref{commutator2}
into Eq.~\eqref{Hdoublebar} we get:
\begin{equation}
\bar{\bar{H}}=\frac{1}{U}\sum_{ij\sigma} t_{ij}^{\sigma} t_{ij}^{\bar{\sigma}}
X_{i}^{\sigma\bar{\sigma}}X_{j}^{\bar{\sigma}\sigma}-\frac{1}{U}\sum_{ij\sigma}
(t_{ij}^{\sigma})^2 (X_{i}^{\sigma
\sigma}X_{j}^{\bar{\sigma}\bar{\sigma}}-X_{i}^{00}X_{j}^{22})
+\frac{1}{U}\sum_{ij\sigma}
t_{ij}^{\sigma}t_{ij}^{\bar{\sigma}}X_{i}^{20}X_{j}^{02}.
\end{equation}
The effective Hamiltonian can be written as:
\begin{equation}
\label{Heff}
H_{eff}=H_U+H_1+\bar{\bar{H}}.
\end{equation}
Let us introduce the spin operators:
\begin{equation}
\begin{array}{ll}
S_{i}^{+}=c_{i\uparrow}^{\dag}c_{i\downarrow}=X_{i}^{+-},\\    
S_{i}^{-}=c_{i\downarrow}^{\dag}c_{i\uparrow}=X_{i}^{-+},\\
S_{i}^{z} = \frac{1}{2} (n_{i\uparrow}-n_{i\downarrow})=X_{i}^{++}-X_{i}^{--}.
\end{array}
\end{equation}
The effective Hamiltonian \eqref{Heff} for large $U$ takes the form:
\begin{eqnarray}
\label{Heff_full}
H_{eff}&=&(U+K_0)\sum_{i}X_{i}^{22}+\sum_{ij\sigma} t_{ij}^{\sigma}
(X_{i}^{\sigma
0}X_{j}^{0\sigma}+X_{i}^{2\bar{\sigma}}X_{j}^{\bar{\sigma}2})+\nonumber \\
&+&\frac{1}{2} \sum_{ij}J_{ij} (S_{i}^{+}S_{j}^{-}+S_{i}^{-}S_{j}^{+})
+ \sum_{ij} K_{ij} (S_{i}^{z}S_{j}^{z}-\frac{1}{4}n_{i}n_{j})+\nonumber\\
&+&\sum_{ij}J_{ij}X_{i}^{20}X_{j}^{02}.  
\end{eqnarray}
where: $K_0=\sum_{j} K_{ij}$,
$K_{ij}=\frac{(t_{ij}^\uparrow)^2+(t_{ij}^{\downarrow})^2}{U}$,
$J_{ij}=2\frac{t_{ij}^{\uparrow} t_{ij}^{\downarrow}}{U}$.  

If we take into account large positive $U$ ($U>0$), the double occupancies have
to be projected out. Therefore, in a limited space:
$X_{i}^{22}=X_{i}^{2\sigma}=X_{i}^{20}=0$. In
this way, the effective Hamiltonian for the large positive $U$ case and $n=1$,
$r\neq 1$ takes the
form:
\begin{equation}
\label{Heff_Heis}
H_{eff}^{Heisenberg}=\sum_{ij}J_{ij} (S_{i}^{x}S_j^x+S_i^yS_j^y)
+\sum_{ij} K_{ij} (S_{i}^{z}S_{j}^{z}-\frac{1}{4}),
\end{equation}
where: $S_{i}^{x}=\frac{1}{2}(S_i^{+}+S_{i}^{-})$,
$S_{i}^{y}=\frac{1}{2i}(S_i^{+}-S_{i}^{-})$. The above Hamiltonian
\eqref{Heff_Heis} is the anisotropic spin-1/2 Heisenberg model. 

In turn, if $r=1$ and $n\leq 1$ the Hamiltonian \eqref{Heff_full} is reduced to
the t-J model:
\begin{eqnarray}
\label{Heff_t-J}
 H_{eff}^{t-J}&=&\sum_{ij\sigma}t_{ij}(1-n_{i\bar{\sigma}})c_{i\sigma}^{\dag}c_{
j\sigma}(1-n_{j\bar{\sigma}})+\sum_{ij}J_{ij}
(\vec{S}_{i}\cdot\vec{S}_j-\frac{1}{4}n_{i}n_{j}),
\end{eqnarray}
operating in the lower Hubbard subband. It is worth mentioning that we
neglected 3-site terms.

Finally, in the most general case ($n\leq 1$) the Hamiltonian \eqref{Heff_full}
for the large positive $U$ case takes the form:
\begin{eqnarray}
\label{Heff_pos}
 H_{eff}^{U>0}&=&\sum_{ij\sigma}t_{ij}^{\sigma}(1-n_{i\bar{\sigma}})c_{i\sigma}^
{\dag}c_{j\sigma}(1-n_{j\bar{\sigma}})+\sum_{ij}J_{ij}
(S_{i}^{x}S_j^x+S_i^yS_j^y)\nonumber\\
&+& \sum_{ij} K_{ij} (S_{i}^{z}S_{j}^{z}-\frac{1}{4}n_{i}n_{j}),
\end{eqnarray}
which is the generalized (anisotropic) t-J model.

 
In the large-negative-$U$ limit ($|U| \gg t^{\uparrow}, t^{\downarrow}$) and
wihout spin imbalance, the
effective Hamiltonian operates in the subspace of states without single
occupancies. Hence, one should impose an appropriate constraint:
\begin{equation}
n_{i\uparrow}-n_{i\downarrow}=0. 
\end{equation}
Therefore, $X_{i}^{\sigma 0}=X_{i}^{2\sigma}=X_{i}^{+-}=S_{i}^{z}=0$. Then, the
effective Hamiltonian \eqref{Heff_full} takes the form:
\begin{equation}
\label{pseudospin}
H_{eff}^{U<0}=-\frac{1}{2}\sideset{}{'}{\sum}_{i, j} J_{ij} (\rho_i^+ \rho_j^-
+h.c.) +\sideset{}{'}{\sum}_{i, j} K_{ij} \rho_i^z \rho_j^z-\tilde{\mu} \sum_i
(2\rho_i^z+1)-\frac{N}{4}K_0,
\end{equation}
and: $n=\frac{1}{N} \sum_{i} \langle 2\rho_i^z+1\rangle$,
$\tilde\mu=\mu+\frac{|U|}{2}$, $K_0=\sum_{j} K_{ij}$,
$J_{ij}=2\frac{t_{ij}^{\uparrow}
t_{ij}^{\downarrow}}{|U|}=4\frac{t^2}{|U|}\frac{2r}{(r+1)^2}$,
$K_{ij}=2\frac{(t_{ij}^\uparrow)^2+(t_{ij}^{\downarrow})^2}{2|U|}=4\frac{t^2}{
|U|}\frac{1+r^2}{(1+r)^2}$. The pseudo-spin operators are:
$\rho_i^+=X_{i}^{20}=c_{i\uparrow}^{\dag}c_{i\downarrow}^{\dag}$,
$\rho_i^-=X_{i}^{02}=c_{i\downarrow}c_{i\uparrow}$,
$\rho_i^z=X_{i}^{22}-\frac{1}{2}=\frac{1}{2}(n_{\uparrow}+n_{\downarrow}-1)$,
primed sum excludes terms with $i=j$. The Hamiltonian \eqref{pseudospin} is of
the form of the anisotropic pseudospin Heisenberg model in a magnetic field
(chemical potential). The pseudo-spin operators satisfy the
commutation rules of $s=\frac{1}{2}$ operators, i.e.: 
\begin{equation}
[\rho_{i}^{+},\rho_{j}^{-}]=2\rho_i^z\delta_{ij},
\end{equation}
\begin{equation}
 [\rho_{i}^{\pm},\rho_{j}^{z}]=\mp \rho_{i}^{\pm}\delta_{ij},
\end{equation}
\begin{equation}
(\rho_{i}^{+})^2=(\rho_{i}^{-})^2=0. 
\end{equation}

After the transformation to the bosonic operators: $\rho_i^{+}=b_i^{\dag}$, 
$\rho_i^{-}=b_i$, $\rho_i^{z}=-\frac{1}{2}+b_i^{\dag}b_i$, the Hamiltonian
\eqref{pseudospin} takes the form: 
\begin{equation}
\label{pseudospin_boson}
H=-\frac{1}{2}\sideset{}{'}{\sum}_{i, j}J_{ij} (b_i^{\dag}b_j +h.c.)
+\sideset{}{'}{\sum}_{i, j} K_{ij}n_i n_j
-\tilde{\mu} \sum_i n_i.
\end{equation}
The above Hamiltonian describes a system of hard-core bosons on a lattice
with the commutation relations \cite{MicnasModern,Micnas4,Micnas5}:
$[b_i,b_j^{\dag}]=(1-2n_i)\delta_{ij}$, $b_i^{\dag}b_i + b_ib_i^{\dag}=1$, where
$n_i=b_i^{\dag}b_i$, $\tilde\mu=2\mu+|U|+K_0$ -- chemical potential for bosons.

\begin{figure}[t!]
\begin{center}
\includegraphics[width=0.55\textwidth,angle=270]{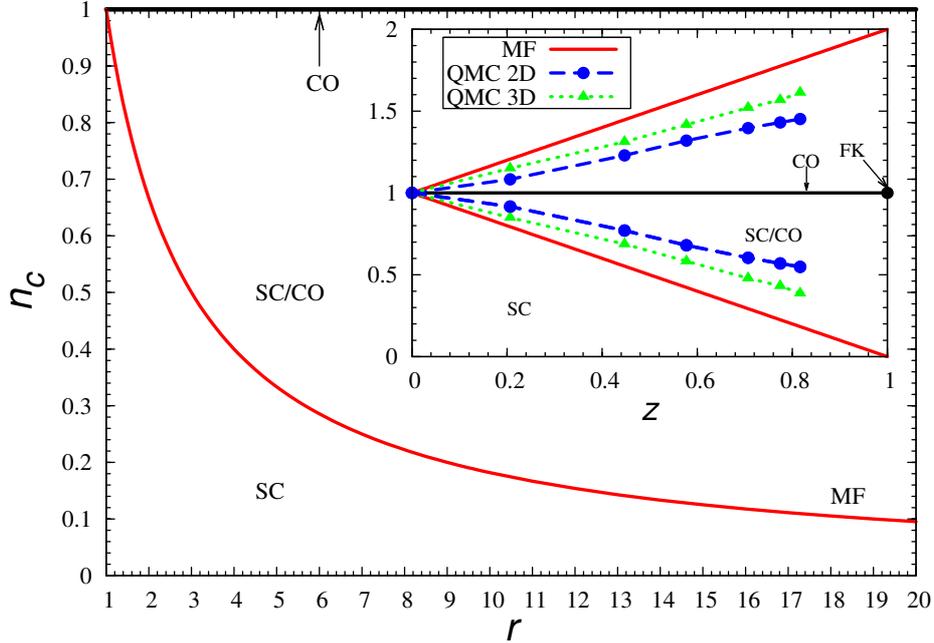}
\caption[The critical $n$ ($n_c$) (within the mean field (MF)
approximation) above which superconductivity can coexist with
commensurate CO for different values of $r$. Inset -- $n_c$ vs.
$z=\Big|\frac{r-1}{r+1}\Big|$ above which superconductivity can coexist with
commensurate CO. The results from the Quantum Monte Carlo (QMC) method for the
anisotropic Heisenberg model (Yunoki \cite{Yunoki}) are also shown -- for 2D and 3D.]{\label{SS-CO} The critical $n$ ($n_c$) (within the mean field (MF)
approximation -- red solid line) above which superconductivity can coexist with
commensurate CO for different values of $r$. Inset -- $n_c$ vs.
$z=\Big|\frac{r-1}{r+1}\Big|$ above which superconductivity can coexist with
commensurate CO. The results from the Quantum Monte Carlo (QMC) method for the
anisotropic Heisenberg model (Yunoki \cite{Yunoki}) are also shown -- blue
circles (2D) and green triangles (3D). The blue dashed and the green dotted
lines are a guide to the eye. FK -- Falicov-Kimball model limit.} 
\end{center}
\end{figure}

With the mass imbalance, it is possible that the charge (density wave) ordered
(CO) state can develop for any particle concentration. The SC to CO transition
is of first order at $h=0$, $t^{\uparrow}\neq t^{\downarrow}$ and $n
\neq 1$. The critical $n$ ($n_c$) (within the mean field approximation) above
which superconductivity can coexist with commensurate CO is given by
\cite{MicnasModern,Dao}:
\begin{equation}
\label{nc}
 |n_c-1|=\sqrt{\frac{K_{0}-J_{0}}{K_{0}+J_{0}}}.
\end{equation}
Substituting expressions for $J$ and $K$, one obtains:
\begin{equation}
 n_c=1\pm \Big| \frac{r-1}{r+1} \Big|.
\end{equation}
SC/CO is in fact the region of phase separation of the SC domain and CO
domain (with $n=1$). 

Fig. \ref{SS-CO} shows $n$ above which SC can coexist with commensurate CO. If
$r=1$, the SC and CO phases are degenerate at $n=1$. When the hopping imbalance
increases, the CO phase becomes energetically favorable and in the limit of
infinite $r$ (the Falicov-Kimball model \cite{Falicov}) only the CO state is
possible at half filling.
However, away from half filling the quantum fluctuations can extend the region
of stability of the SC phase at $T=0$ and enhance $n_c$
\cite{Micnas4,Micnas5,Yunoki}. This is clearly visible in Fig. \ref{SS-CO}
(inset), showing the results of QMC simulations (blue circles and green
triangles), in comparison to MF (red solid line).

\section{AAHM on 2D square lattice}
In this section, we investigate the influence of hopping imbalance on the
evolution from the weak to strong coupling limit of tightly bound
local pairs with increasing attraction for $d=2$.
We focus on the analysis of stability of the SC$_M$ phase in the crossover
diagrams in the ground state. Most of the phase diagrams \textcolor{czerwony}{of the hopping asymmetric AHM (AAHM)} of  are constructed without
the Hartree term. We do not study CO stability, because we fix $n<n_c$ and
consider mostly low electron concentrations in our analysis. We also
extend our analysis to finite temperatures. We apply the KT scenario to
determine the phase coherence temperatures.

\begin{figure}[t!]
\hspace*{-0.8cm}
\includegraphics[width=0.38\textwidth,angle=270]
{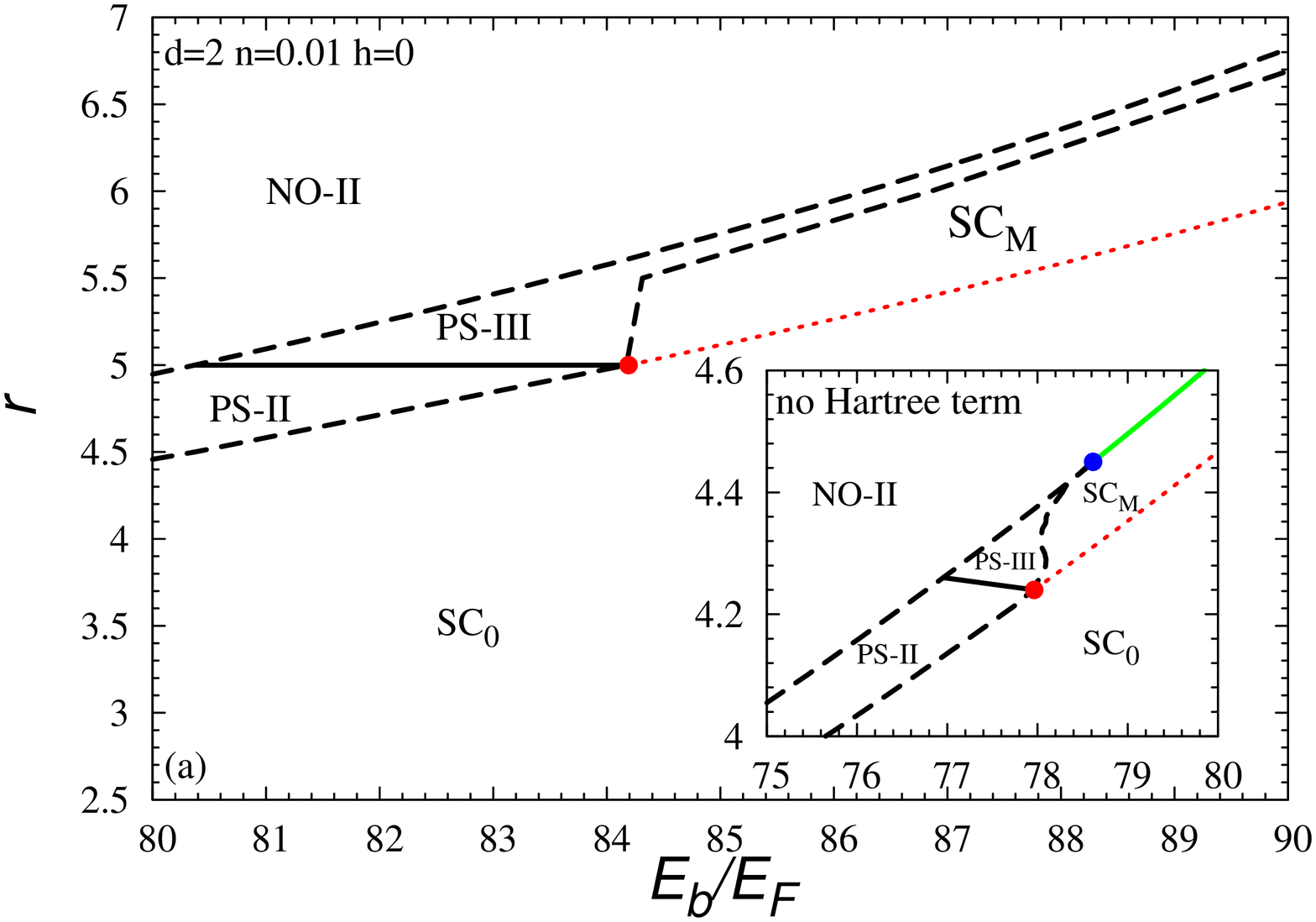}
\hspace*{-0.6cm}
\includegraphics[width=0.38\textwidth,angle=270]
{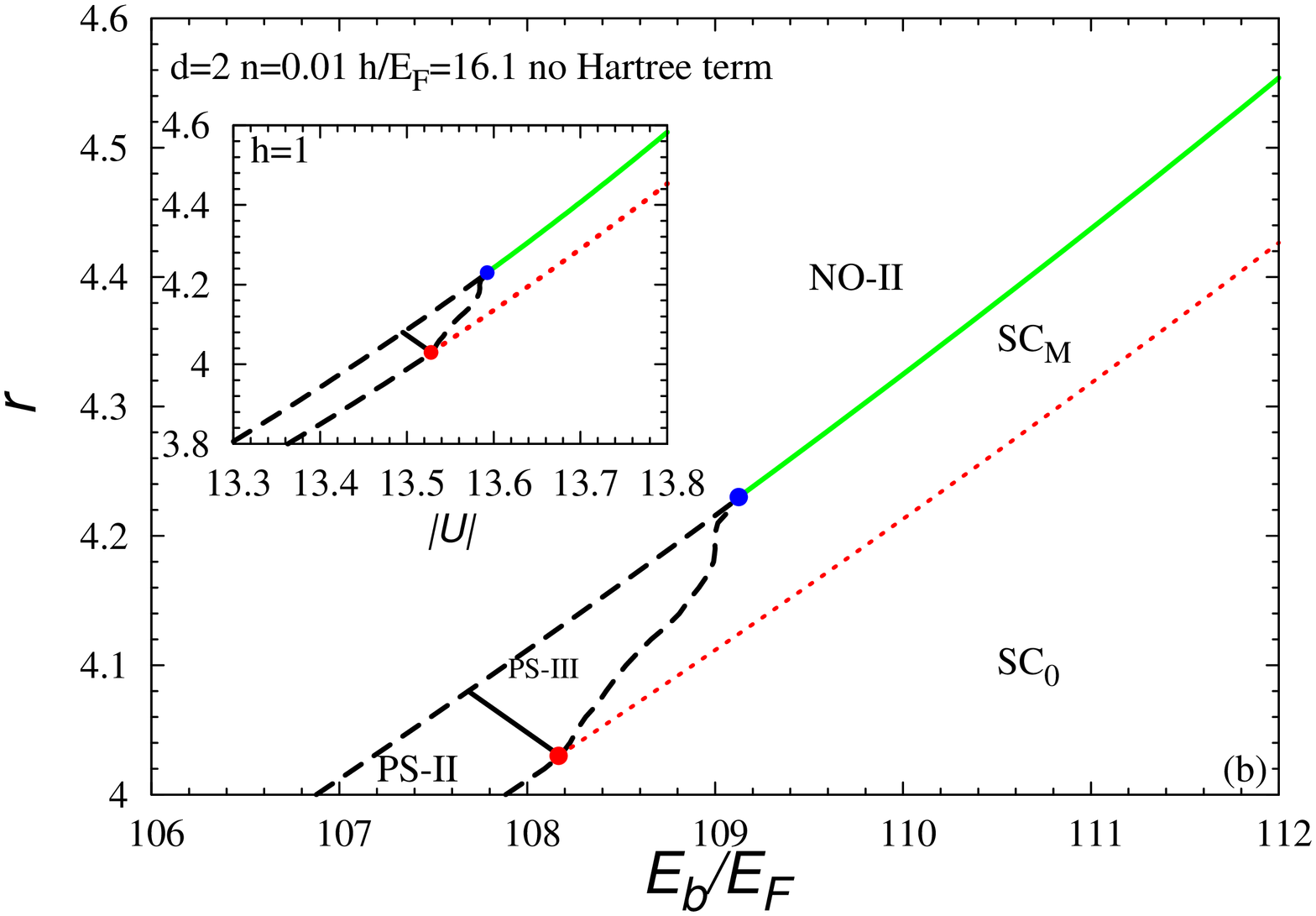}
\caption[Ground state phase diagrams for the $d=2$ square lattice: mass
imbalance vs. $E_b/E_F$ (where $E_F$ is the lattice Fermi energy of
unpolarized, non-interacting fermions with hopping $t$) for $n=0.01$, (a)
$h=0$ with and without Hartree term (inset), (b) $h/E_F=16.1$ (inset -- $r$ vs.
$|U|$, $h=1$).]{\label{rvsEb_n001} Ground state phase diagrams \textcolor{czerwony}{of AAHM} 
for the $d=2$
square lattice: mass imbalance vs. $E_b/E_F$ (where $E_F$ is the lattice Fermi
energy of unpolarized, non-interacting fermions with hopping $t$, $E_b$ is the
binding energy for two fermions in an empty lattice with hopping
$t$ ($t=(t^\uparrow+t^\downarrow)/2$)) for $n=0.01$,
(a) $h=0$ with and without Hartree term (inset), (b) $h/E_F=16.1$ (inset -- $r$
vs. $|U|$, $h=1$). SC$_M$ -- magnetized superconducting state, NO-I -- partially
polarized normal state, NO-II -- fully polarized normal state, PS-I --
(SC$_0$+NO-I), PS-II -- (SC$_0$+NO-II), PS-III -- (SC$_M$+NO-II). Red points --
$h_{c}^{SC_M}$, blue points  -- tricritical points, green points -- the BCS-LP
crossover points in the SC$_0$ phase ($r=1$). The dotted red and the solid green
lines are continuous transition lines.}
\end{figure}

\subsection{Ground state phase diagrams}
\label{T0r2D}
\begin{figure}[t!]
\hspace*{-0.8cm}
\includegraphics[width=0.38\textwidth,angle=270]
{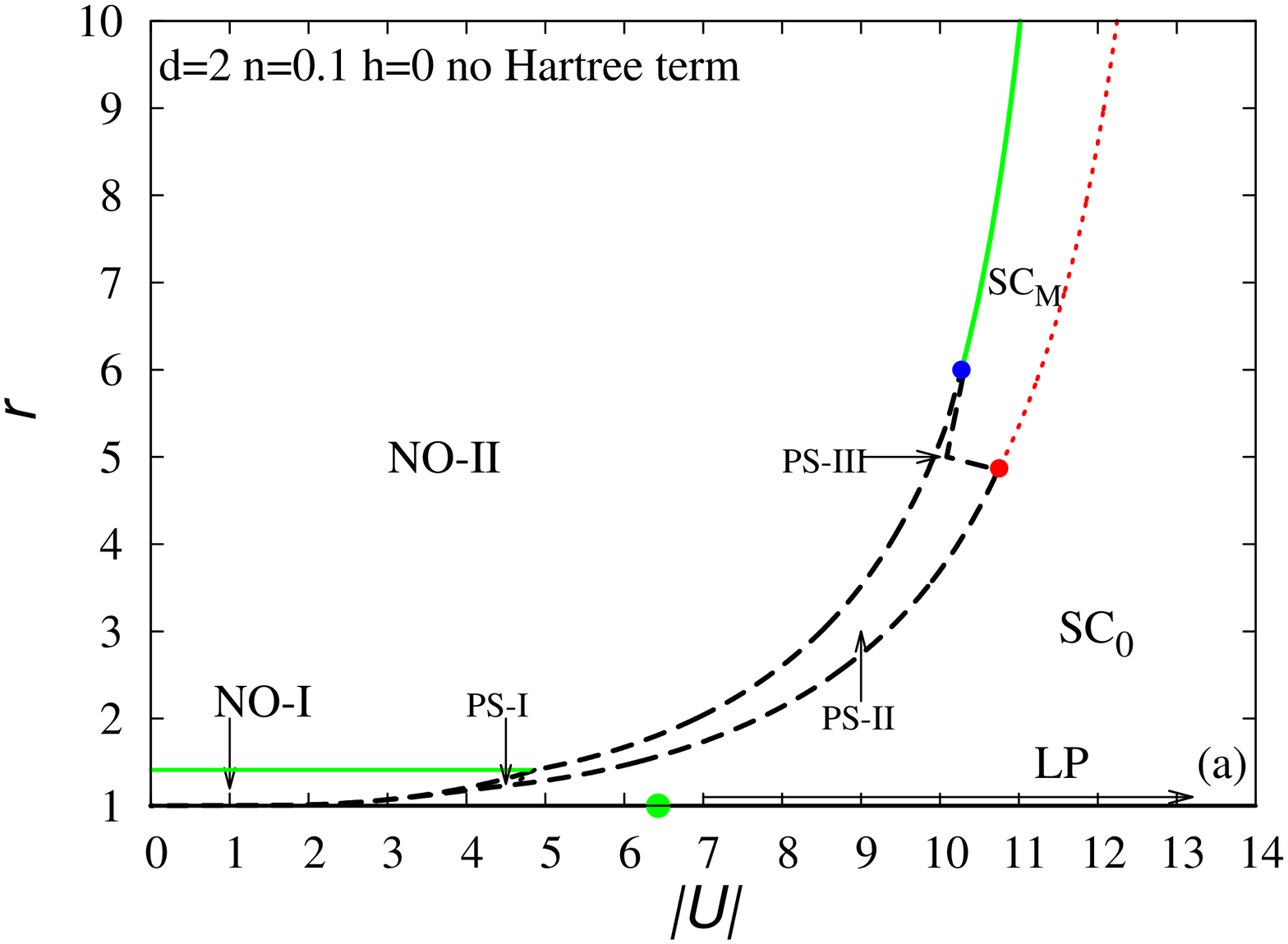}
\hspace*{-0.6cm}
\includegraphics[width=0.38\textwidth,angle=270]
{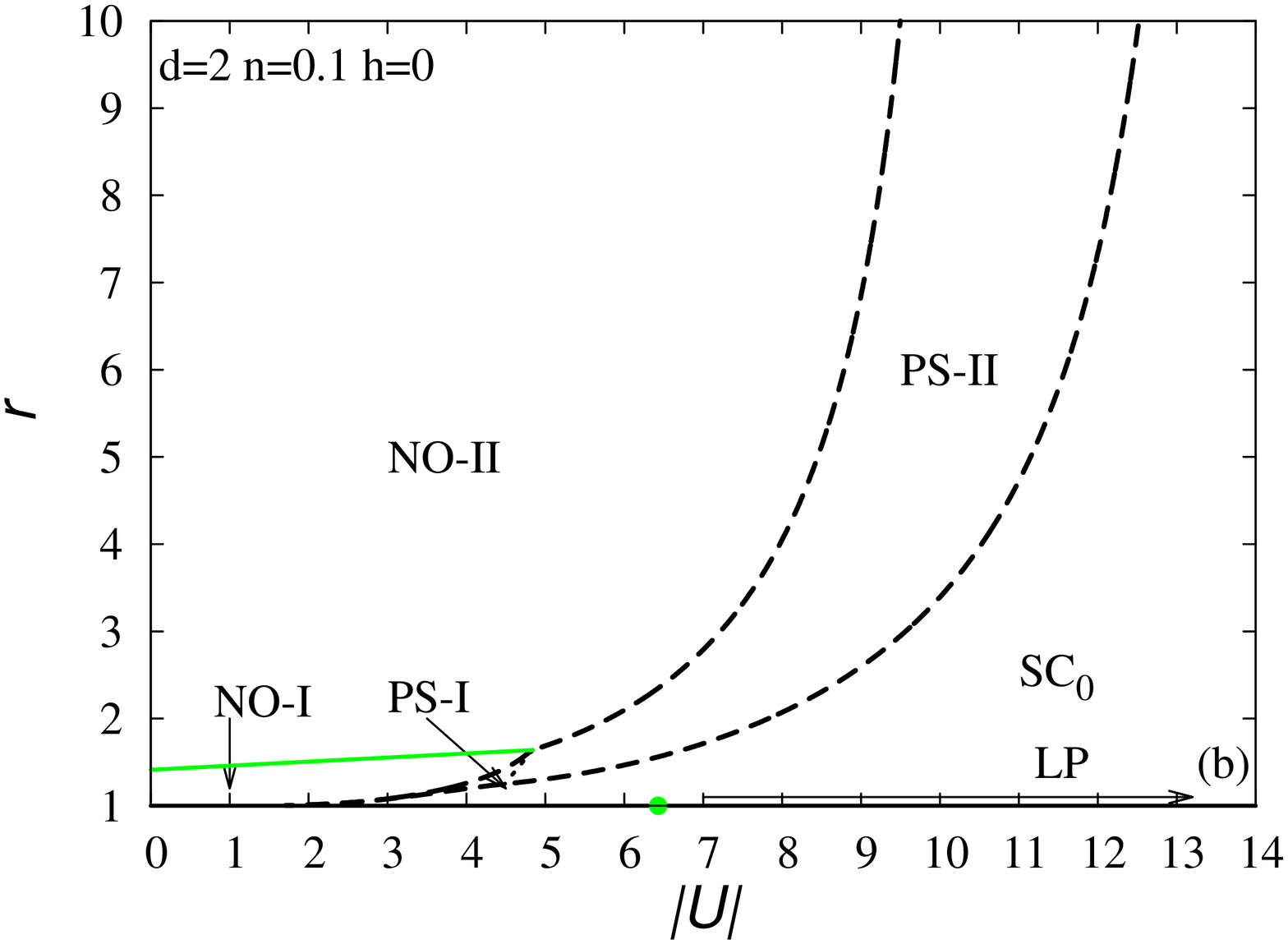}
\caption[Ground state phase diagrams $r$ vs. $|U|$ for $n=0.1$, $h=0$ (a)
without the Hartree term, (b) with Hartree term.]{\label{rvsU_n01} Ground state
phase diagrams \textcolor{czerwony}{of AAHM} $r$ vs. $|U|$ for $n=0.1$, $h=0$ 
(a) without the Hartree term,
(b) with Hartree term. SC$_0$ -- unpolarized SC state, SC$_M$ -- magnetized SC
state, NO-I (NO-II) -- partially (fully) polarized normal states. PS-I
($SC_0$+NO-I) -- partially polarized phase separation, PS-II ($SC_0$+NO-II) --
fully polarized phase separation, PS-III -- ($SC_M$+NO-II). Red point --
$|U|_{c}^{SC_M}$ (quantum critical point), blue point  -- tricritical point,
green point -- the BCS-BEC crossover point in the SC$_0$ phase ($t=1$).}
\end{figure}

\begin{figure}[t!]
\begin{center}
\hspace*{-1.0cm}
\includegraphics[width=0.38\textwidth,angle=270]
{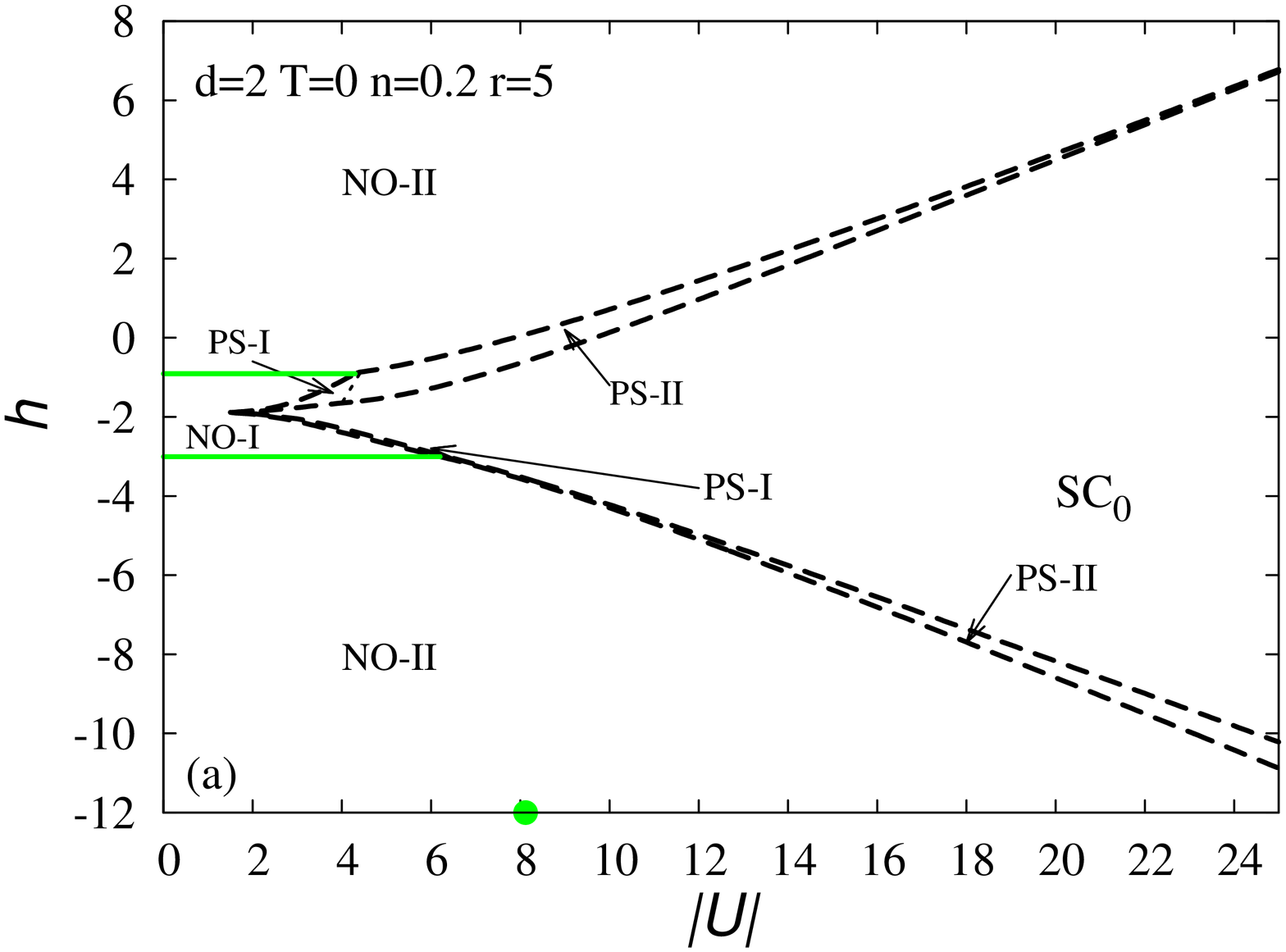}
\hspace*{-0.8cm}
\includegraphics[width=0.38\textwidth,angle=270]
{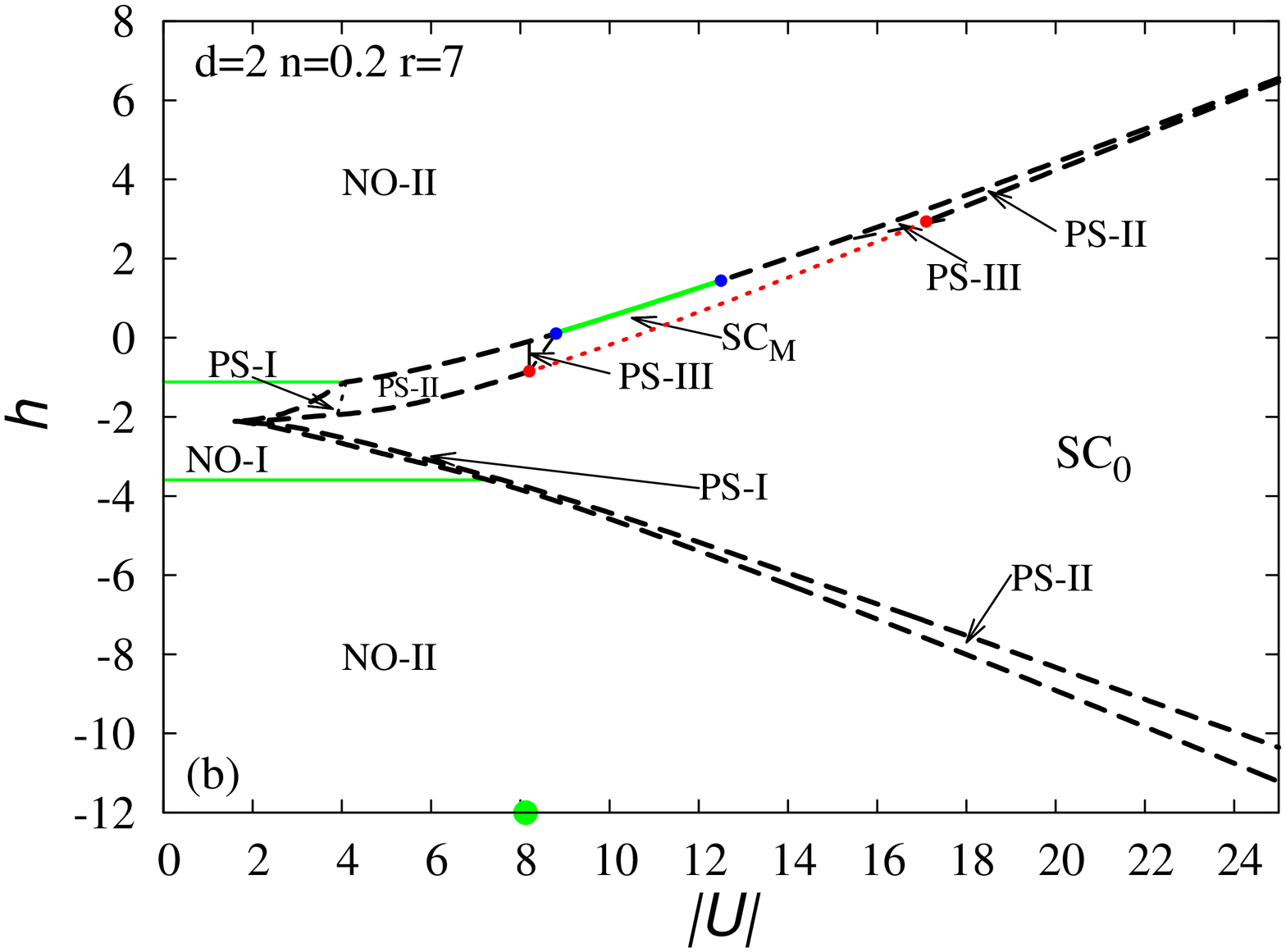}
\includegraphics[width=0.38\textwidth,angle=270]
{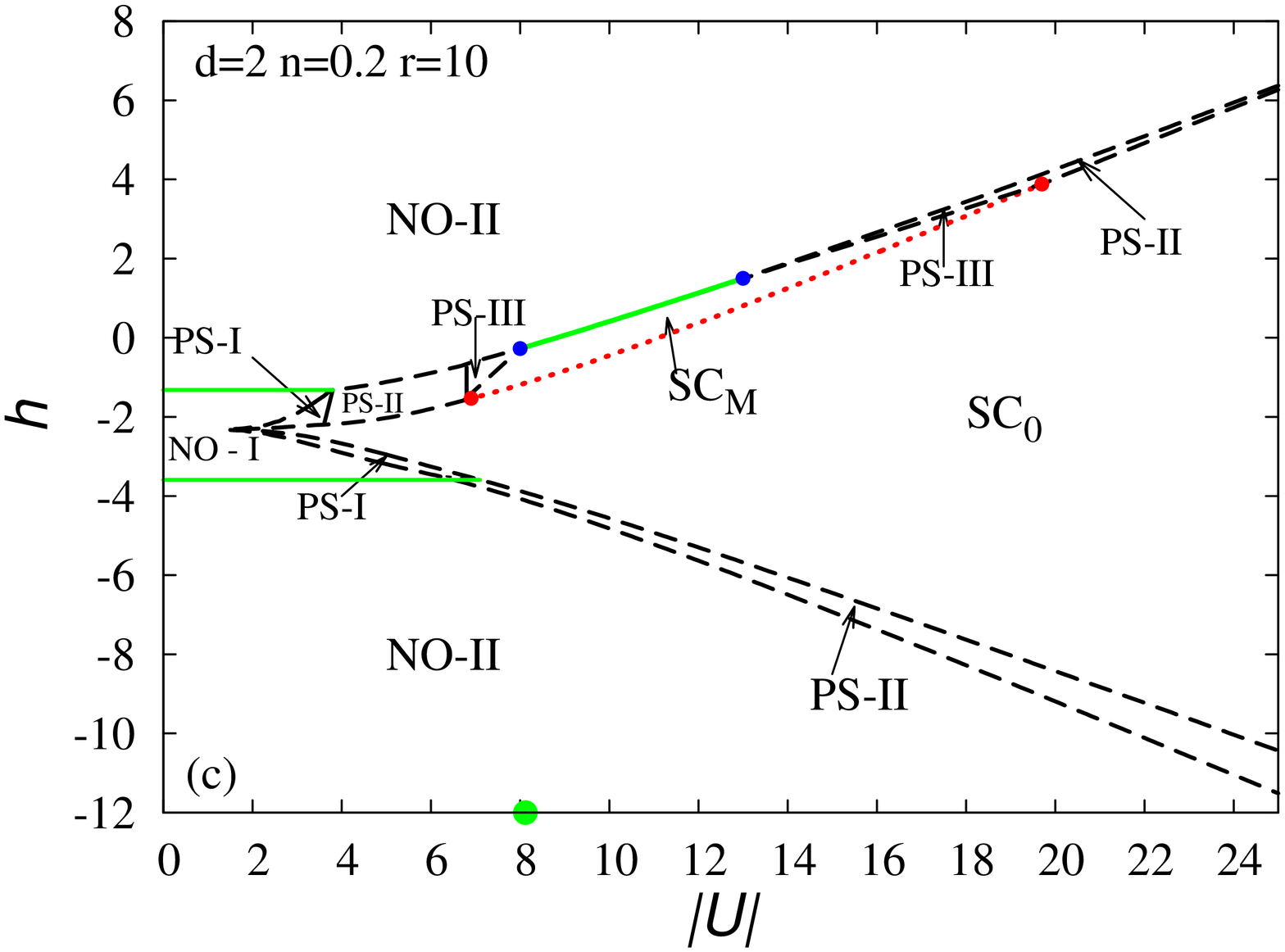}
\caption[Magnetic field vs. attractive interaction phase diagrams for three
values of $r$: (a) $r=5$, (b) $r=7$, $r=10$ and $n=0.2$, without Hartree
term.]{\label{hvsU_n02_r} Magnetic field vs. attractive interaction phase
diagrams \textcolor{czerwony}{of AAHM} for three values of $r$: 
(a) $r=5$, (b) $r=7$, $r=10$ and $n=0.2$,
without Hartree term. SC$_M$ -- magnetized superconducting state, NO-I --
partially polarized normal state, NO-II -- fully polarized normal state, PS-I --
(SC$_0$+NO-I), PS-II -- (SC$_0$+NO-II), PS-III -- ($SC_M$+NO-II). Red points --
$h_{c}^{SC_M}$, blue points  -- tricritical points, green points -- the BCS-LP
crossover points in the SC$_0$ phase ($r=1$). The dotted red and the solid green
lines are continuous transition lines ($t=1$).}
\end{center}
\end{figure}

The BCS-LP crossover diagrams in the presence of a Zeeman magnetic field for
$r\neq 1$ exhibit a novel behavior. As opposed to the $r=1$ case, for strong
attraction, SC$_M$ occurs at $T=0$. For $r>1$, the solutions of this type
(Sarma-type with $\Delta (h)$) appear when $h>(\frac{r-1}{r+1})\bar{\mu}+2\Delta
\frac{\sqrt{r}}{r+1}$ (on the BCS side) or when 
$h>\sqrt{(\bar{\mu}-\epsilon_0)^2+|\Delta |^2}-D\frac{r-1}{r+1}$ (on the LP
side). The $SC_M$ phase is unstable in the weak coupling regime at $T=0$, but
can be stable in the intermediate and strong coupling LP limit. As mentioned
before, deep in the LP limit, unpaired spin down fermions do not exist. Hence,
the SC$_M$ phase is the superfluid state of coexisting LP's (hard-core bosons)
and single-species fermions, with the latter responsible for finite polarization
(magnetization) and the gapless excitations characteristic of this state of
Bose-Fermi mixture. Note also that the $SC_0$ solutions are such as for
particles with the effective hopping integral $t=(t^\uparrow+t^\downarrow)/2$
or,
equivalently, for the effective particle mass
of \textcolor{czerwony}{$2/m=1/m^\uparrow+1/m^\downarrow$}.

The structure of the ground state diagrams in Fig. \ref{rvsEb_n001} is
different from that shown in Fig. \ref{2D_n001_crossover} (Chapter
\ref{chapter5}), where one has only a first order phase transition from pure
SC$_0$ to the NO phase in the ($\mu -h$) plane. In addition, there are
critical values of $|U|$ ($|U_c|^{SC_M}$ -- QCP, red points in the diagrams),
for which the SC$_M$ state becomes stable, instead of PS. However, one should
mention that there is a critical value of $r$, for which SC$_M$ is stable, as
opposed to the 3D case, in which the magnetized superconducting phase can be
stable even at $r=1$. 

Fig. \ref{rvsEb_n001} shows the ground state ($r-E_b/E_F$) phase diagrams for
fixed $n=0.01$, (a) $h=0$, (b) $h/E_F \approx 16.1$. If $h=0$, the SC$_M$ state
does not appear (is not stable) up to $r_c\approx 5$ in the diagram with the
Hartree term and also up to $r_c\approx 4.2$ in the diagram without the Hartree
term. In turn, if $h\neq 0$ (Fig. \ref{rvsEb_n001}(b), diagram without the
Hartree term), the QCP point moves towards smaller values of $r$ (for
$h/E_F\approx 16.1$, $r_c\approx 4$). However, in this case, the SC$_M$ state
is stable for higher values of attraction.

The influence of the Hartree term on the stability of the SC$_M$ state is
reflected in Fig. \ref{rvsU_n01}, which shows the $(r-|U|)$ ground state phase
diagrams for fixed $n=0.1$, both without (a) and with (b) the Hartree term. If
the Hartree term is not included in the construction of the diagram (Fig.
\ref{rvsU_n01}(a)), we have the following sequences of transitions with
increasing $|U|$: NO-II$\rightarrow$ SC$_M$ $\rightarrow$ SC$_0$ (for higher
values of $r$, up from $r_c\approx 4.87$) or NO-II (NO-I)$\rightarrow$ PS
$\rightarrow$ SC$_0$ (for lower values of $r$). In the diagram with the Hartree
term  (Fig. \ref{rvsU_n01}(b)), the SC$_M$ state is not stable even
at $r=10$. Therefore, the presence of this term restricts the range of
occurrence of SC$_M$, except for a dilute limit. Moreover, this term widens the
range of occurrence of the PS region, compared to the case without the
Hartree term. It is worth mentioning that the value of \textcolor{czerwony}{$|U|/t$}
for which
$\bar{\mu}$ reaches the lower band edge does not depend on the mass imbalance in
the SC$_0$ state.   

The diagrams ($h-|U|$) for higher filling ($n=0.2$) and fixed $r$ are presented
in Fig. \ref{hvsU_n02_r}. For higher $n$, the region of SC$_M$ is narrowing. The
SC$_M$ state is unstable even at $r=5$, in the diagram without the Hartree term
(Fig. \ref{hvsU_n02_r}(a)). SC$_M$ occurs in the phase diagrams for higher
values of $r$ (Fig. \ref{hvsU_n02_r}(b)-(c)).   
The transition from SC$_M$ to NO-II can be accomplished in two ways: through
PS-III (SC$_M$+NO-II) or through a 2$^{nd}$ order phase transition. The
character of this transition changes with increasing $|U|$. In the very strong
coupling limit, PS is more stable than the SC$_M$ phase. Hence, we also find two
TCP in these diagrams. The distance between these two critical points changes
with increasing $r$. For fixed $r=7$, TCPs are located at $|U|\approx 8.8$ and
$|U|\approx 12.5$, while for $r=10$ at $|U|\approx 8$ and $|U|\approx 13$. 
However, in the very dilute limit ($n\rightarrow 0$) there is only one TCP in
the ($h-|U|$) diagram. 

The BP-2 phase in $d=2$ is unstable, even for large mass ratio. If $r\neq 1$,
the  symmetry with respect to $h=0$ is broken. However, this symmetry is
restored upon replacement $r\to r^{-1}$. 

\begin{figure}[t!]
\hspace*{-0.8cm}
\includegraphics[width=0.38\textwidth,angle=270]
{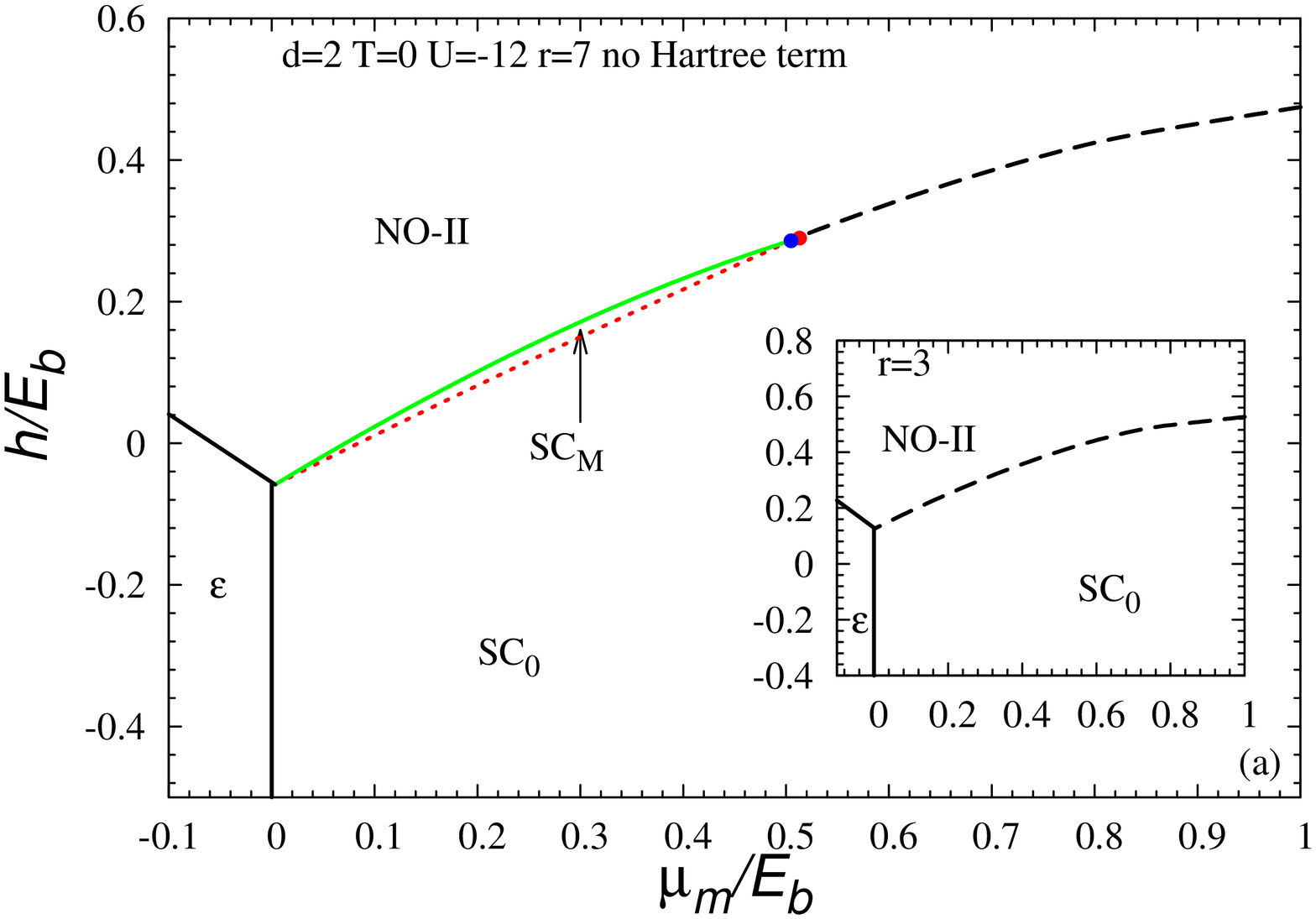}
\hspace*{-0.6cm}
\includegraphics[width=0.38\textwidth,angle=270]
{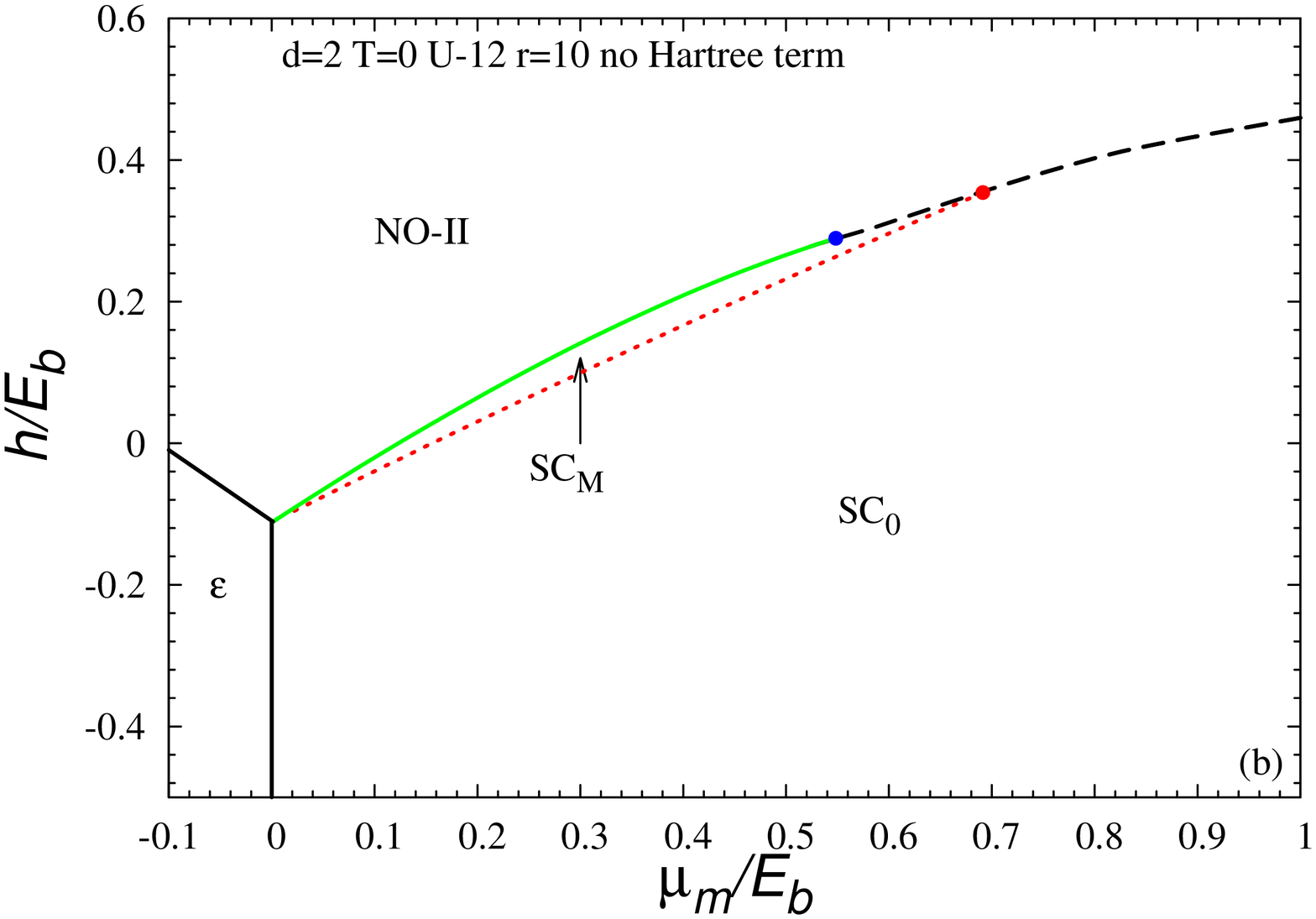}
\hspace*{-0.8cm}
\includegraphics[width=0.38\textwidth,angle=270]
{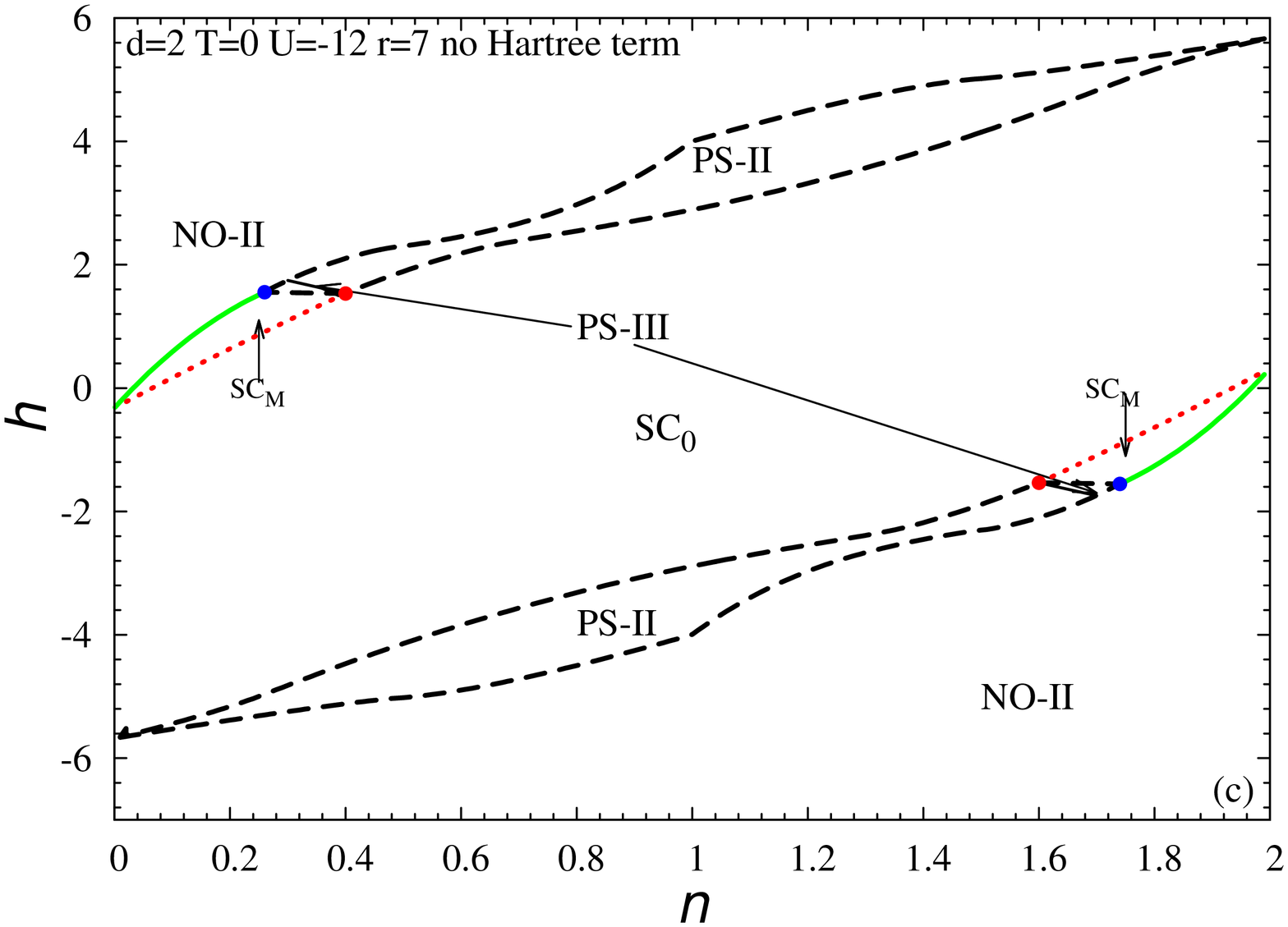}
\hspace*{-0.6cm}
\includegraphics[width=0.38\textwidth,angle=270]
{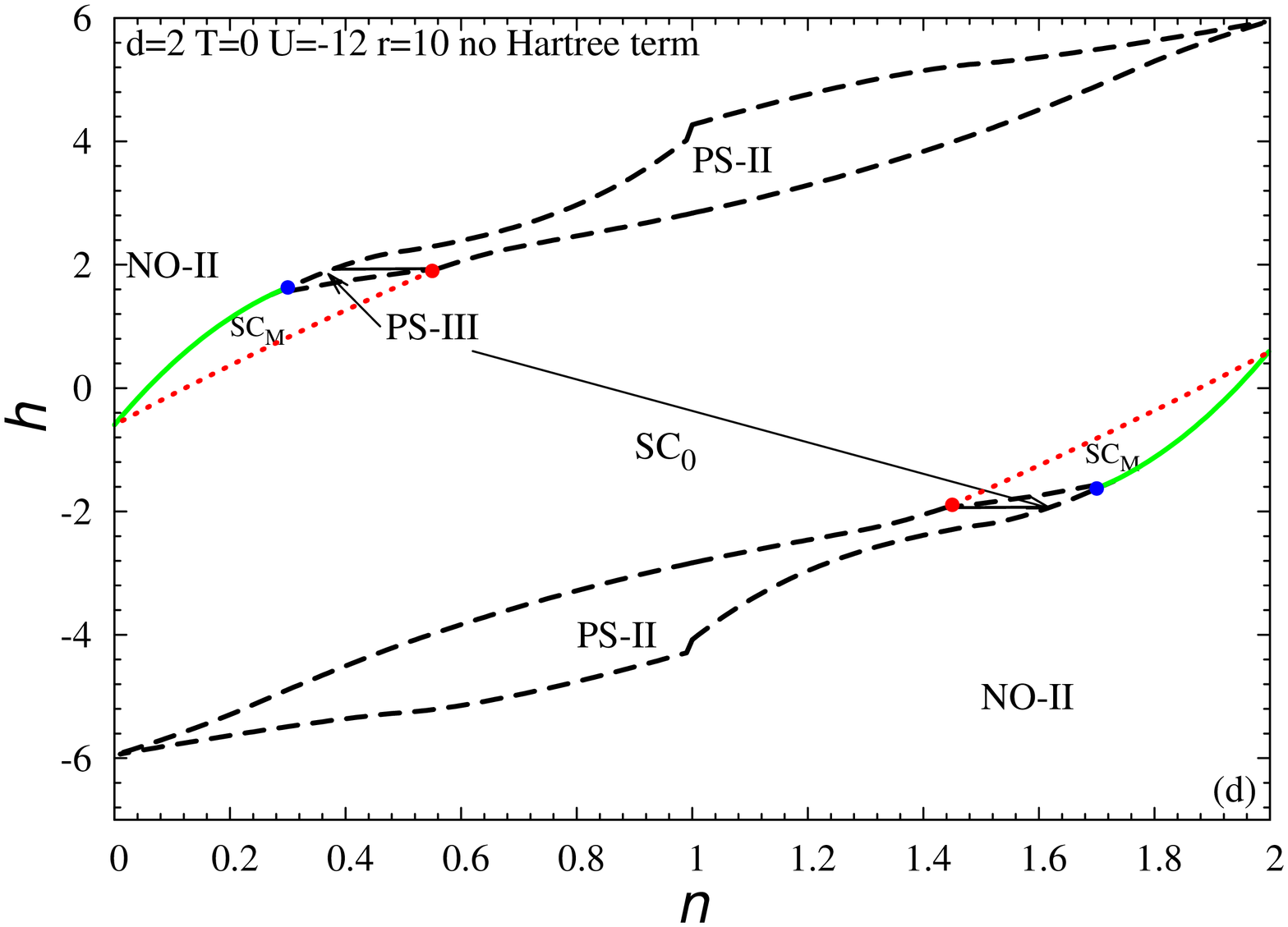}
\caption[Critical magnetic field vs. molecular chemical potential diagrams for
$d=2$ at fixed $U=-12$ (a) $r=7$  ($r=3$ -- inset), (b) $r=10$. (c), (d) $h$ vs.
$n$ phase diagrams.]{\label{h_mu_r7_r10} Critical magnetic field vs. chemical
potential diagrams \textcolor{czerwony}{of AAHM} for $d=2$ at fixed $U=-12$ 
(a) $r=7$  ($r=3$ -- inset), (b)
$r=10$. (c), (d) $h$ vs. $n$ phase diagrams. SC$_0$ -- unpolarized SC state,
SC$_M$ -- magnetized SC state, NO-II -- fully polarized normal state,
$\varepsilon$ -- empty state, $\mu_m$ -- half of the pair chemical potential
defined as: $\mu_m=\mu -\epsilon_0+\frac{1}{2}E_b$, where $\epsilon_0=-4t$,
$E_b$ is the binding energy for two fermions in an empty lattice with hopping
$t$. Red point --
$h_{c}^{SC_M}$, blue point -- tricritical point. These points are close to each
other for $r=7$. The dotted red and the solid green lines are continuous
transition lines. The dashed black line is the 1$^{st}$ order transition
line. The possible region of existence of the commensurate CO phase is not
shown in these diagrams -- for the relevant values of parameters, see Fig.
\ref{SS-CO}.}
\end{figure}

\begin{figure}[t!]
\begin{center}
\includegraphics[width=0.55\textwidth,angle=270]
{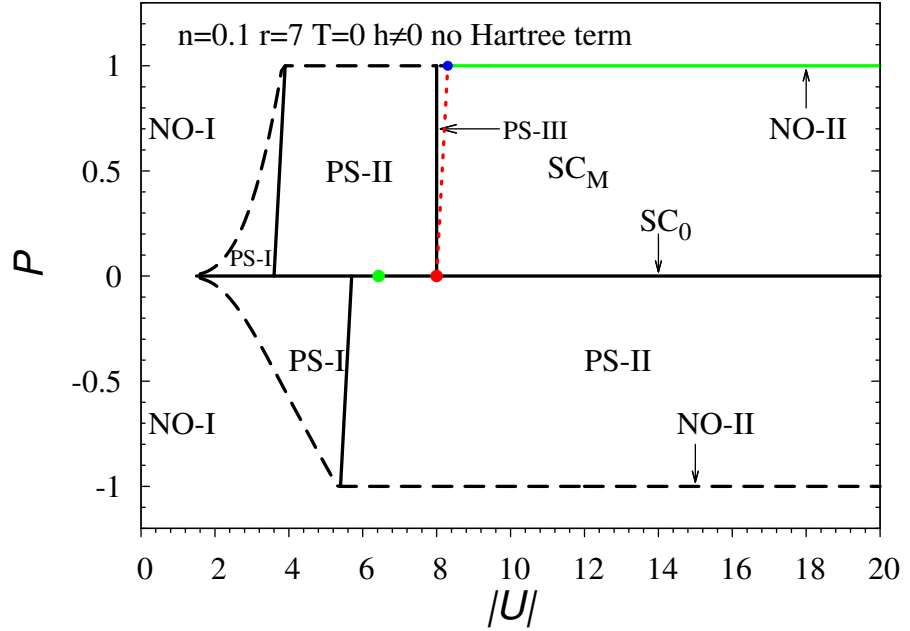}
\caption[Polarization vs. on-site attraction ground state phase diagram at fixed
$n=0.1$ and $r=7$, for the square lattice.]{\label{PvsU_n01_r7_T0} Polarization
vs. on-site attraction ground state phase diagram \textcolor{czerwony}{of AAHM} 
at fixed $n=0.1$ and $r=7$,
for the square lattice. $SC_0$ -- unpolarized SC state with
$n_{\uparrow}=n_{\downarrow}$, $SC_M$ -- magnetized SC state, NO-I (NO-II) --
partially (fully) polarized normal states. PS-I ($SC_0$+NO-I) -- partially
polarized phase separation, PS-II ($SC_0$+NO-II) -- fully polarized phase
separation, PS-III -- ($SC_M$+NO-II). Red point -- $|U|_{c}^{SC_M}$ (quantum
critical point), blue point  -- tricritical point, green point -- the BCS-BEC
crossover point in the SC$_0$ phase.
} 
\end{center}
\end{figure}
Fig. \ref{h_mu_r7_r10} shows the phase diagrams at a fixed molecular potential
$\mu_{m}$ and $h$ (a)-(b), on the LP side and $h$ vs. n (c)-(d). As usual, we
define $\mu_{m}=\mu-\epsilon_0+\frac{1}{2} E_b$ as one half of the pair chemical
potential. In Fig. \ref{h_mu_r7_r10}(a) (inset) there is
only the first order phase transition from pure SC$_0$ to the NO phase (with
increasing $h$) at fixed $r=3$. However, for higher values of the hopping ratio
($r=7$, $r=10$), we observe also a continuous phase transition from SC$_0$ to
SC$_M$, with decreasing chemical potential and increasing magnetic field.
This fact confirms the earlier observations that there exists a critical value
of $r$ above which the SC$_M$ state is stable in the 2D case. The character of
the transition from the superconducting to the normal phase changes with
decreasing $\mu$. Hence, we also find TCP in the diagrams (blue point), at which
the second order transition from SC$_M$ to NO-II terminates. The tricritical
and quantum critical points are close to each other for $r=7$. However, for
higher values of $r$ (Fig. \ref{h_mu_r7_r10}(b)), the distance between these
points increases. This increase is due to the fact that the SC$_M$ phase becomes
stable for higher values of $\mu_{m}$, which corresponds to higher values of
$n$ (Fig. \ref{h_mu_r7_r10}(d)). As shown in Figs. \ref{h_mu_r7_r10}(c)-(d), the
results for the fixed chemical potential have been mapped onto the case of fixed
$n$. This implies the PS region occurrence in the $h-n$ plane. These diagrams
are good examples of the $(h,n)\rightarrow(-h,2-n)$ symmetry, i.e. a combined
particle-hole symmetry and the symmetry with respect to a change in the
direction of the magnetic field. We emphasize again that we have not considered
the commensurate CO phase in these diagrams. This phase can occur above some
critical value $n_c$ which depends on the hopping imbalance -- for more details
about the relevant region of parameters, see Fig. \ref{SS-CO}.

\subsection{Finite temperature phase diagrams}

Because of the occurrence of the SC$_M$ state at $T=0$ for higher $r$, one can
expect that this phase persists to non-zero temperatures.

\begin{figure}[t!]
\hspace*{-0.8cm}
\includegraphics[width=0.38\textwidth,angle=270]
{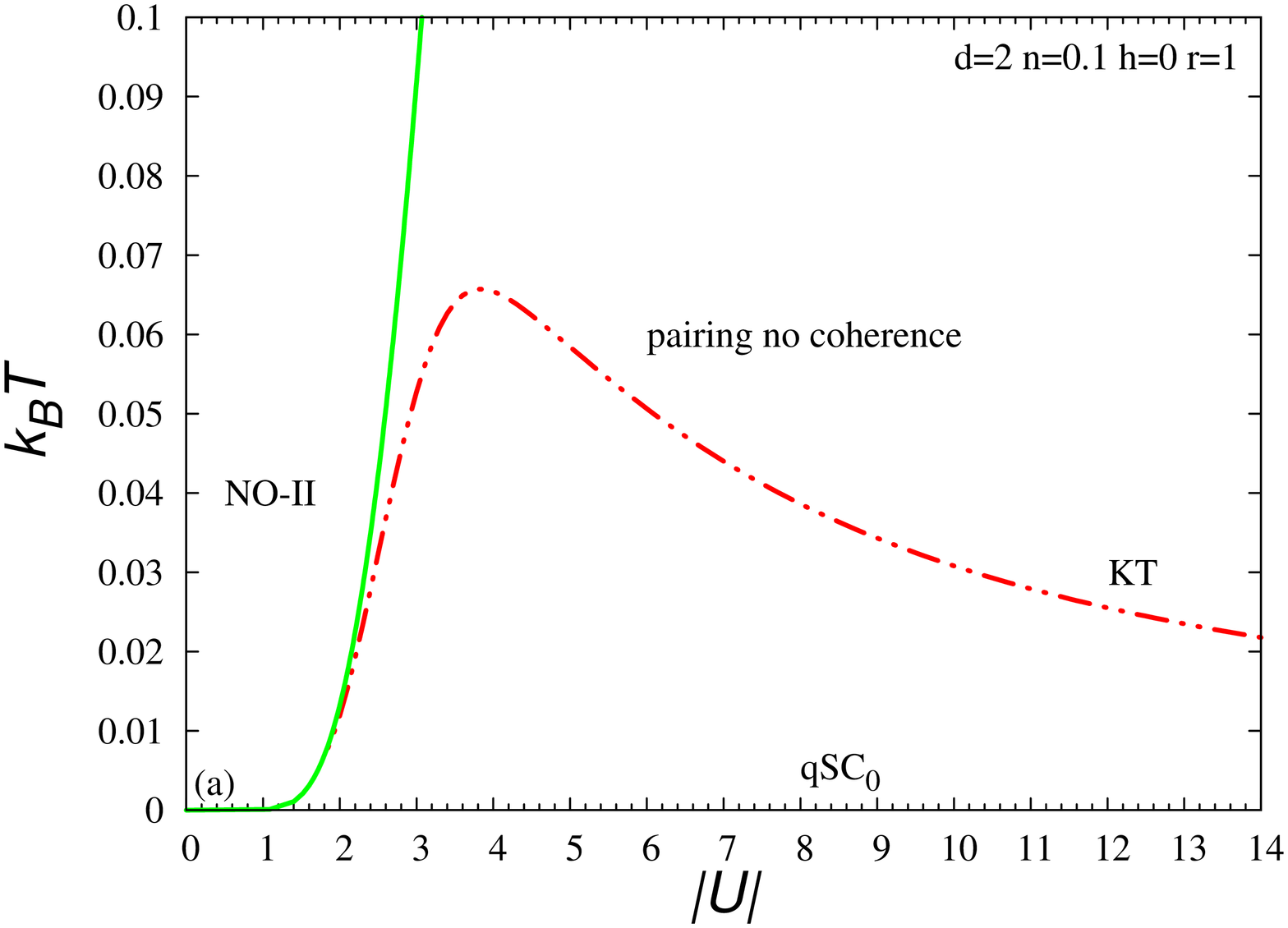}
\hspace*{-0.6cm}
\includegraphics[width=0.38\textwidth,angle=270]
{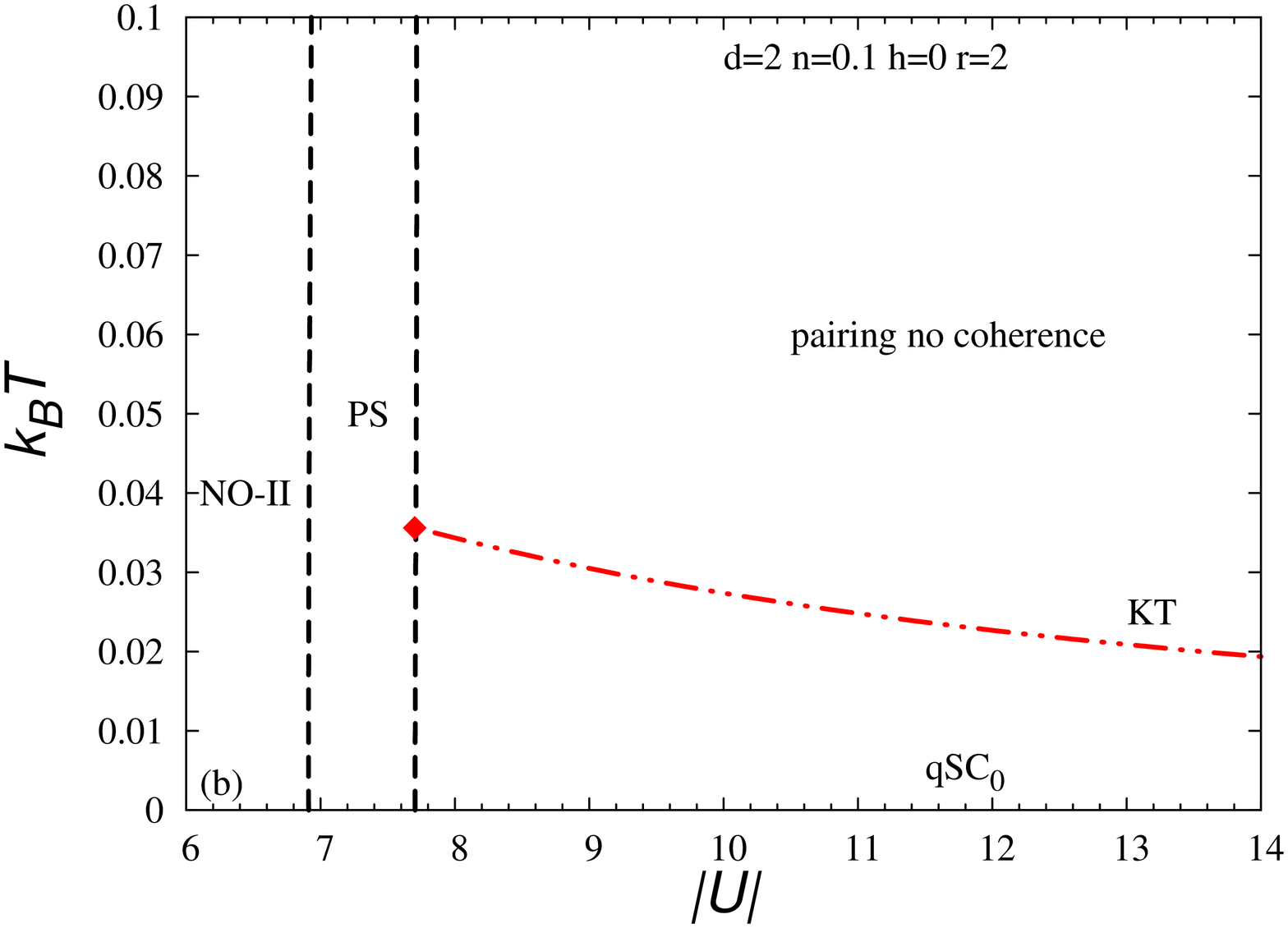}
\hspace*{-0.8cm}
\includegraphics[width=0.38\textwidth,angle=270]
{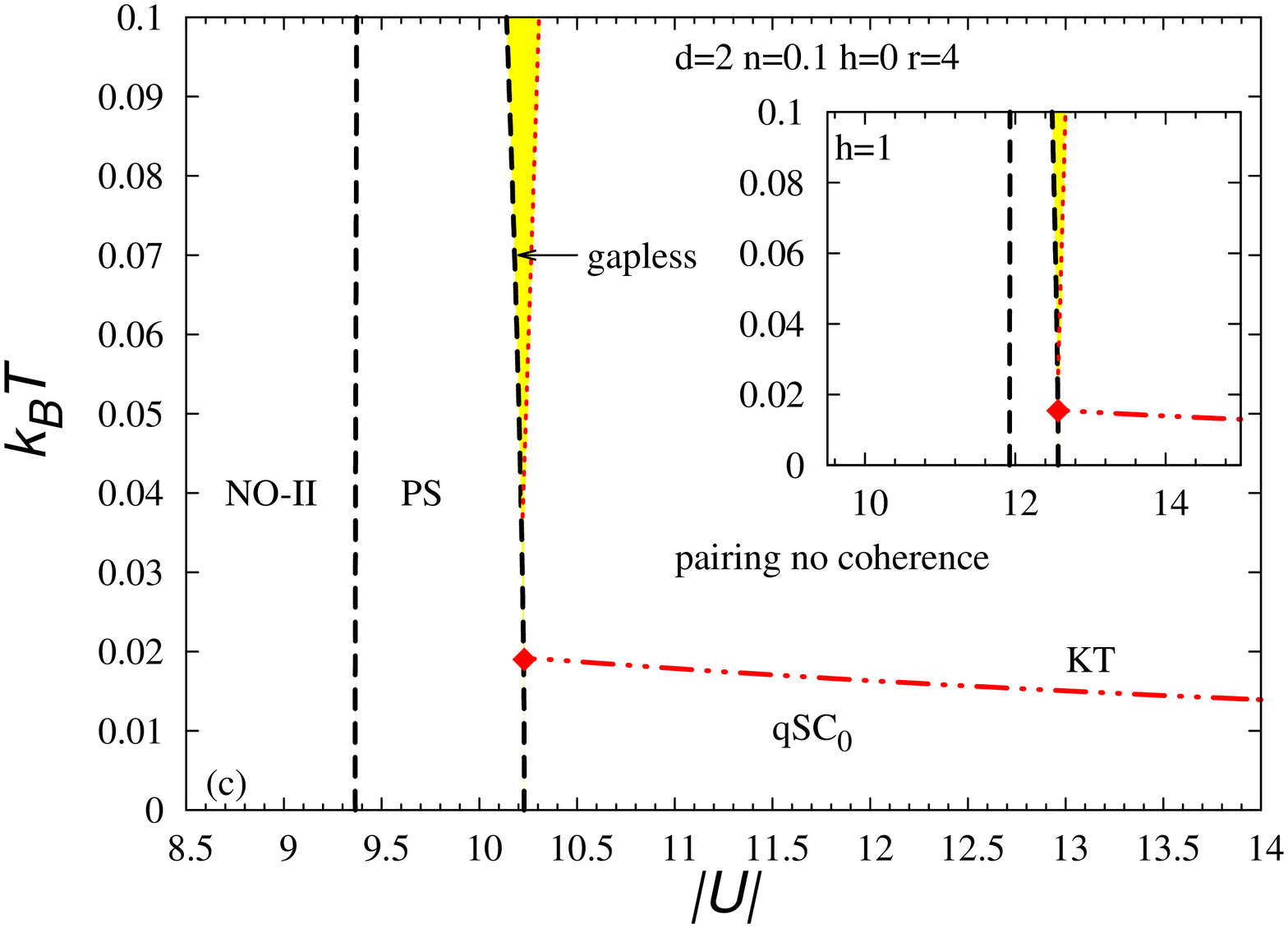}
\hspace*{-0.6cm}
\includegraphics[width=0.38\textwidth,angle=270]
{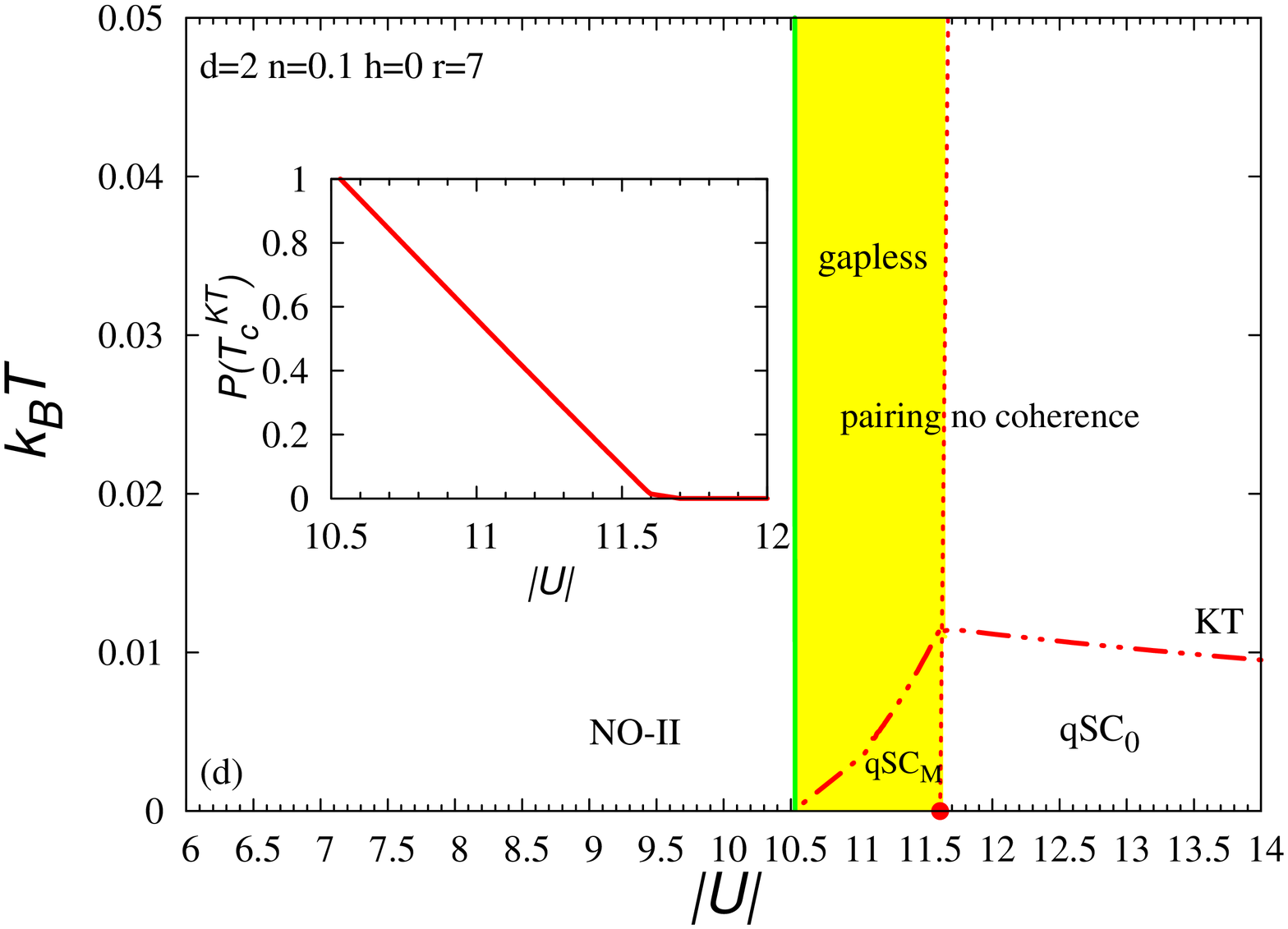}
\caption[Temperature vs. $|U|$ phase diagrams, at fixed $n=0.1$, $h=0$ for the
square lattice. (a) $r=1$, (b) $r=2$, (c) $r=4$ (inset -- $h=1$) (d) $r=7$
(inset -- $P\big(T_{c}^{KT}\big)$ vs. $|U|$).]{\label{2d_TvsU_h0_r} Temperature
vs. $|U|$ phase diagrams \textcolor{czerwony}{of AAHM}, 
at fixed $n=0.1$, $h=0$ for the square lattice. (a)
$r=1$, (b) $r=2$, (c) $r=4$ (inset -- $h=1$) (d) $r=7$ (inset --
$P\big(T_{c}^{KT}\big)$ vs. $|U|$). The thick dashed-double dotted line (red
color) is the KT transition line.  Thick solid line denotes transition from
pairing without coherence region to NO-II within the BCS approximation, dotted
line (red color) is continuous transition line from qSC$_0$ to qSC$_M$ state at
$T=0$ or from qSC$_0$ to the gapless region (yellow color) at $T\neq 0$. PS --
phase separation, qSC$_0$ -- 2D qSC without polarization, qSC$_M$ -- 2D qSC in
the presence of polarization (a spin polarized KT superfluid). Red point at
$T=0$ (see Fig. (d)) -- QCP for Lifshitz transition.}
\end{figure}

In Fig. \ref{PvsU_n01_r7_T0}, we present the $P-|U|$ ground state diagram for
low electron concentration $n=0.1$ and fixed $r$, at $h\neq 0$. This diagram is
only shown so that we could refer to it in our further considerations, when
the zero-temperature results are extended to finite temperatures. As
mentioned above, in the 2D system at $r=1$, for $h\neq 0$, the SC$_M$ phase
is unstable even in the strong coupling limit and phase separation is
favorable. This is in opposition to the 3D case in a Zeeman magnetic field in
which for $r=1$ the SC$_M$ phase occurs for strong attraction and in the dilute
limit. However, as shown in Sec. \ref{T0r2D}, for $r\neq 1$ SC$_M$ in $d=2$ can
be stable. There is a critical value of $|U_c|^{SC_M}$ (red point in the
diagram), for which the SC$_M$ state becomes stable, instead of PS. The
transition from SC$_M$ to NO can be accomplished in two ways for fixed $n$:
through PS-III (SC$_M$+NO-II) or through the second order phase transition for
higher $|U|$. The change in the character of this transition is manifested
through TCP. For so chosen
parameters ($n=0.1$, $r=7$), the magnetized superconducting state is stable only
on the BEC side (see Fig. \ref{PvsU_n01_r7_T0}, $P>0$), but as shown in Sec.
\ref{T0r2D}, SC$_M$ can also be stable in the intermediate couplings (for
higher values of $r$, e.g. $r=10$). If $r\neq 1$, the symmetry with respect to
$h=0$ is broken. Hence, the diagram is not symmetric with respect to $P=0$ and
for $P<0$ the PS is favorable instead of SC$_M$ in the LP limit (Fig.
\ref{PvsU_n01_r7_T0}). The presented phase diagram has been constructed without
the Hartree term.

Here, we perform a detailed analysis of the BCS-LP crossover phase diagrams for
finite temperatures and show that if $r\neq 1$, the appearance of a new phase is
possible at $T\neq 0$.

\begin{figure}[t!]
\hspace*{-0.8cm}
\includegraphics[width=0.38\textwidth,angle=270]
{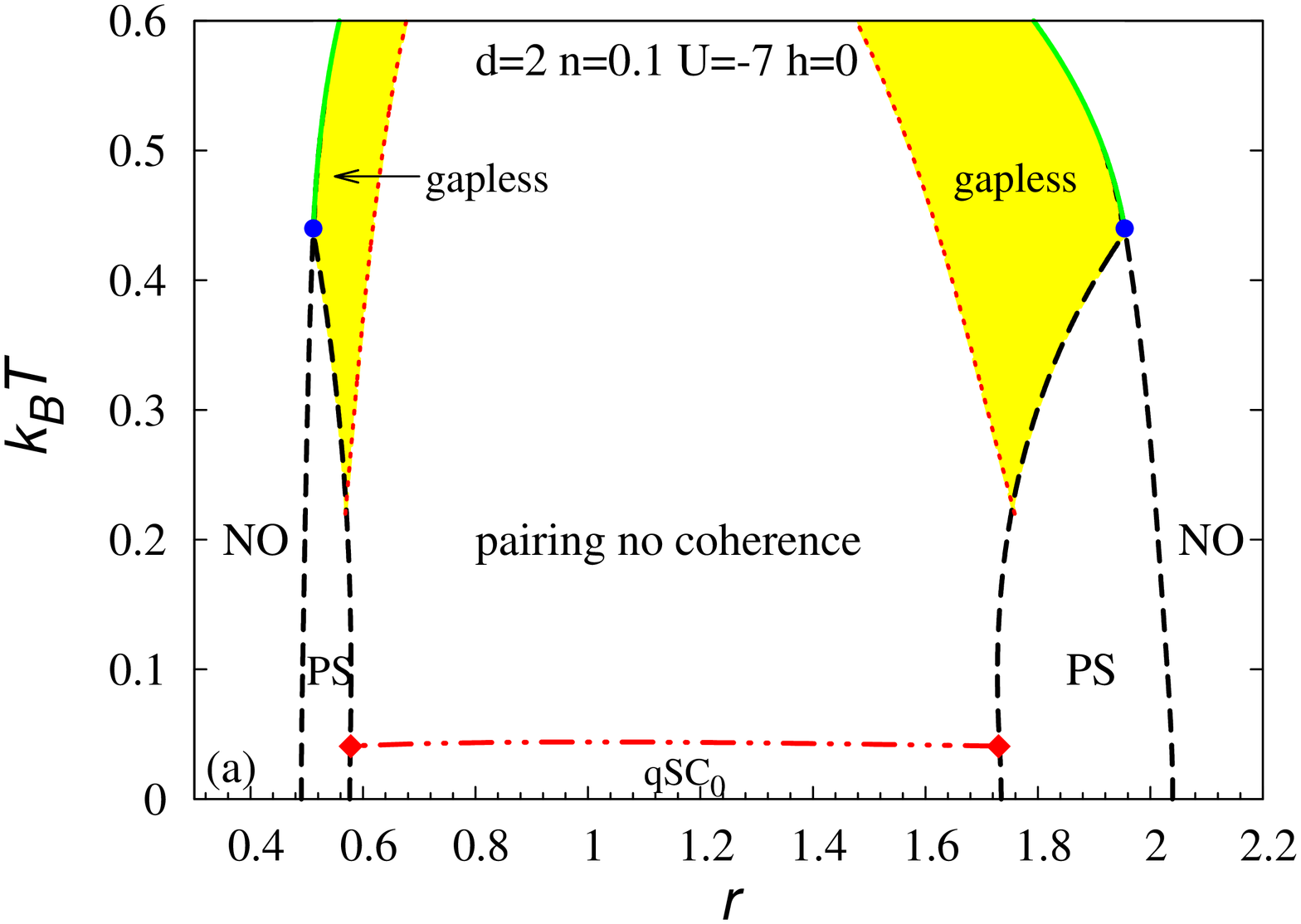}
\hspace*{-0.6cm}
\includegraphics[width=0.38\textwidth,angle=270]
{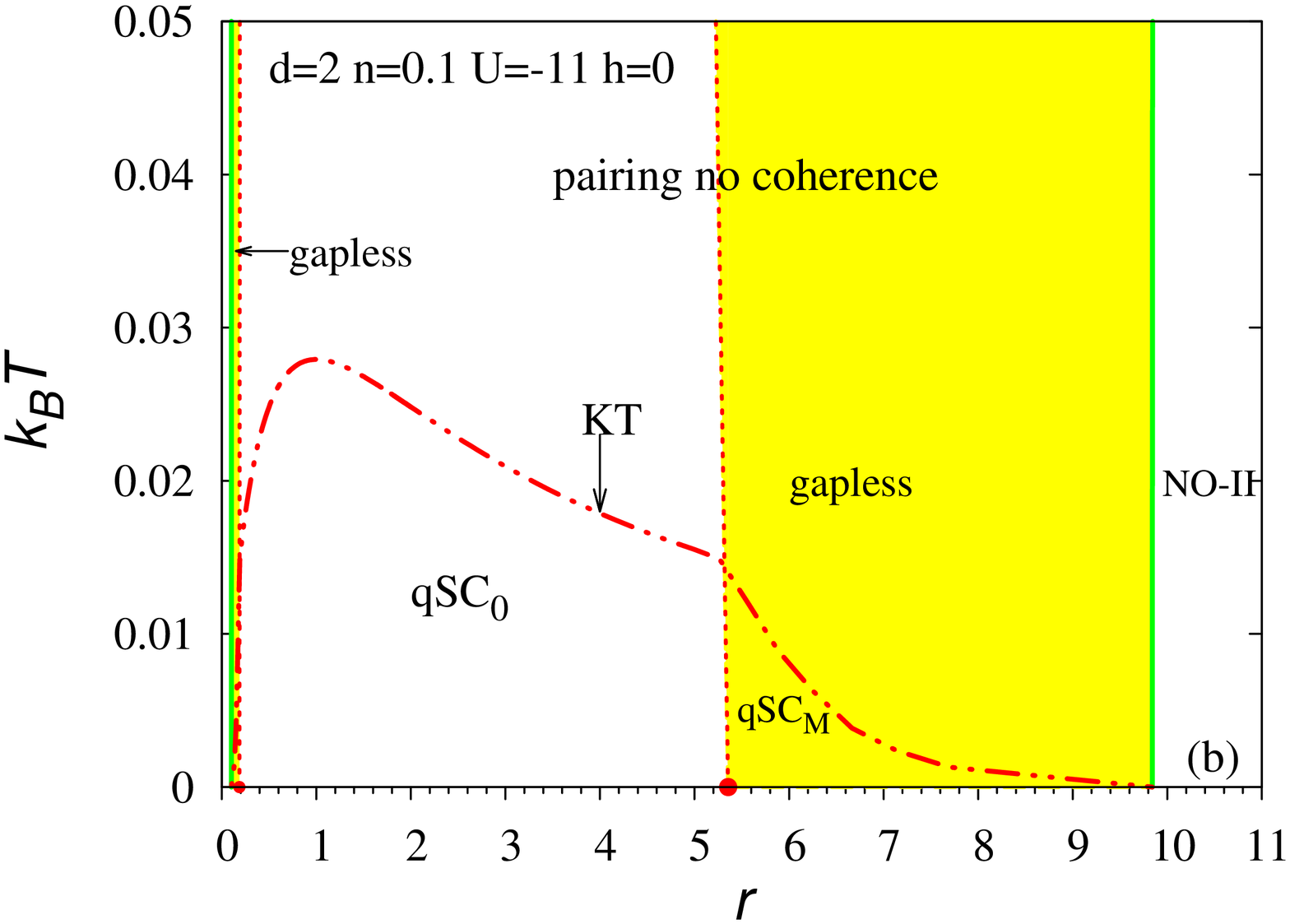}
\caption[Temperature vs. $r$ phase diagrams, at fixed $n=0.1$, $h=0$ for the
square lattice. (a) $U=-7$, (b) $U=-11$.]{\label{Tvsr_h0_n01} Temperature vs.
$r$ phase diagrams \textcolor{czerwony}{of AAHM}, 
at fixed $n=0.1$, $h=0$ for the square lattice. (a) $U=-7$,
(b) $U=-11$. The thick dashed-double dotted line (red color) is the KT
transition line. Dotted line (red color) is continuous transition line from
qSC$_0$ to qSC$_M$ state at $T=0$ or from qSC$_0$ to the gapless region (yellow
color) at $T\neq 0$. PS -- phase separation, qSC$_0$ -- 2D qSC without
polarization, qSC$_M$ -- 2D qSC in the presence of polarization (a spin
polarized KT superfluid). Red point at $T=0$ (see Fig. (b)) -- QCP for Lifshitz
transition.}
\end{figure}

Fig. \ref{2d_TvsU_h0_r} shows ($T-|U|$) phase diagrams for $h=0$, $n=0.1$ and
four fixed values of $r$: (a) $r=1$, (b) $r=2$, (c) $r=4$ and (d) $r=7$. As
usual, the solid lines (2$^{nd}$ order transition lines) and the PS regions are
obtained within the BCS approximation, while the thick dash-double dotted
line (red color) denotes the KT transition determined from Eqs.~\eqref{KT} and
\eqref{ro_s}. As mentioned before, the temperatures $T_c^{KT}$ are much smaller
than $T_c^{HF}$. However, as shown in Fig. \ref{2d_TvsU_h0_r}(a) ($h=0$, $r=1$),
the difference between $T_c^{KT}$ and $T_c^{HF}$ decreases with decreasing
attractive interaction and in the weak coupling limit these temperatures are
comparable. The KT temperature increases with increasing $|U|$, reaches its
maximum for some value of $|U|$ and then decreases. At $h=0$ and $r=1$, both
$T_c^{HF}$ as $T_c^{KT}$ are nonzero, for very low values of $|U|$. However,
$r\neq 1$ gives a similar effect to $h\neq 0$, i.e. $T_c^{HF}=0$, below a
definite value of attractive interaction. This critical $|U|$ increases with
$r$, which is clearly visible in Fig. \ref{2d_TvsU_h0_r}(b)-(d). Moreover, if
additionally $h\neq 0$, one can see (Fig. \ref{2d_TvsU_h0_r}(c) -- inset) that
$T_c^{HF}=0$ moves towards larger $|U|$ (e.g. if $h=0$, $r=4$ -- $|U_c| \approx
9.4$ and if $h=1$, $r=4$ -- $|U_c| \approx -11.9$).

Because of the occurrence of the magnetized superconducting state at $T=0$ for
higher $r$, this phase persists to non-zero temperatures (as shown in Fig.
\ref{2d_TvsU_h0_r}(d), $r=7$). However, if SC$_M$ is unstable at $T=0$ (compare
with Fig. \ref{rvsU_n01}(a)) for lower $r$ (Fig. \ref{2d_TvsU_h0_r}(b)-(c)), the
gapless region can still occur at some temperatures (with one FS in the strong
coupling). Moreover, the combination of $r\neq 1$ and $h\neq 0$ causes a
decrease in temperature at which this gapless region occurs (compare: Fig.
\ref{2d_TvsU_h0_r}(c) and its inset). It is worth mentioning that there is a
gapless region in the diagram for fixed $r=2$, but at definitely higher
temperatures. 

The system is a quasi superconductor below $T_c^{KT}$. Apart from the
unpolarized qSC$_0$ state, qSC$_M$ occurs, which can be termed \emph{a spin polarized
KT superfluid} (Fig. \ref{2d_TvsU_h0_r}(d)).  Above $T_c^{KT}$, we have an
extended region of incoherent pairs which is bounded from above by the pair
breaking temperature. In the strong coupling limit, $T_c^{KT}$ does not depend
on magnetic field, but it depends on mass imbalance and its upper bound takes
the form:
\begin{equation}
k_{B}T_c^{KT}=2\pi\frac{r}{(1+r)^2}\frac{t^2}{|U|}n(2-n), 
\end{equation}
for $r>0$.  
In that limit only LP's exist and the system is equivalent to that of hard-core
Bose gas on a lattice, described by the Hamiltonian (\ref{pseudospin}).

In Fig. \ref{Tvsr_h0_n01} we present the ($T-|U|$) phase diagrams for $n=0.1$
and fixed $|U|$, at $h=0$ to illustrate, among others, the symmetry ($h$, $r$)
$\rightarrow$ ($-h$, $1/r$). At $T=0$, for $U=-7$ (Fig. \ref{Tvsr_h0_n01}(a))
the transition from the superconducting to the normal state goes through PS both
for $r>1$ and $r<1$. The phase separation region is limited by two critical
values of $r$: for $r>1$ -- $r_{c1}^{r>1}\approx 1.735$, $r_{c2}^{r>1}\approx
2.0392$,while for $r<1$ -- $r_{c1}^{r<1}\approx 0.576$, $r_{c2}^{r<1}\approx
0.49$. As can be easily seen, $r_{c1}^{r>1}\rightarrow 1/r_{c1}^{r<1}$ and
$r_{c2}^{r>1}\rightarrow 1/r_{c2}^{r<1}$. At low temperatures, the system is a
quasi superconductor without polarization up to $T_c^{KT}$. There exist only
local pairs. For higher temperatures (above $T_c^{KT}$ and below $T_c^{HF}$)
there is a region of pairs without the phase coherence. Obviously, we can
distinguish two gapless regions both for $r>1$ and $r<1$. Two MF TCP, which
terminate the second order transition are located at the same value of
temperature because of the $r\rightarrow 1/r$ symmetry. 

For higher attractive interaction ($U=-11$), because of the existence of the
SC$_M$ phase at $T=0$ (see: Fig. \ref{PvsU_n01_r7_T0}), this state is continued
to finite temperatures. The transition from the KT superfluid to the normal
state is of the second order for $r>1$ and $r<1$. 
The critical values $r_c^{r>1}$ (for $t^\uparrow>t^\downarrow$) and $r_c^{r<1}$
(for $t^\uparrow<t^\downarrow$) for which SC$_M$ becomes stable (gapless region
-- yellow color in the diagrams) and for which the 2$^{nd}$ order
transition to the NO phase takes place are related by $r_c^{r<1}=1/r_c^{r<1}$,
which is another confirmation of the $r\rightarrow 1/r$ symmetry at $h=0$.
As in the $U=-7$ case, for lower temperatures, below
$T_c^{KT}$ there are only local pairs and the system is equivalent to that of
hard-core Bose gas on a lattice. The maximum value of $T_c^{KT}$ is located at
$r=1$ and decreases with increasing and decreasing $r$. The transition from
qSC$_0$ to qSC$_M$ is manifested by a cusp on the KT critical temperature
line, both for $r>1$ and $r<1$. The region of the SC$_M$ occurrence at $T=0$ is
between $r\approx 5.356$ (red point in Fig. \ref{Tvsr_h0_n01}(b)) and $r\approx
9.842$ for $r>1$. Because of the $r\rightarrow 1/r$ symmetry this region is
visually much narrower for $r<1$.

\begin{figure}[t!]
\hspace*{-0.8cm}
\includegraphics[width=0.38\textwidth,angle=270]
{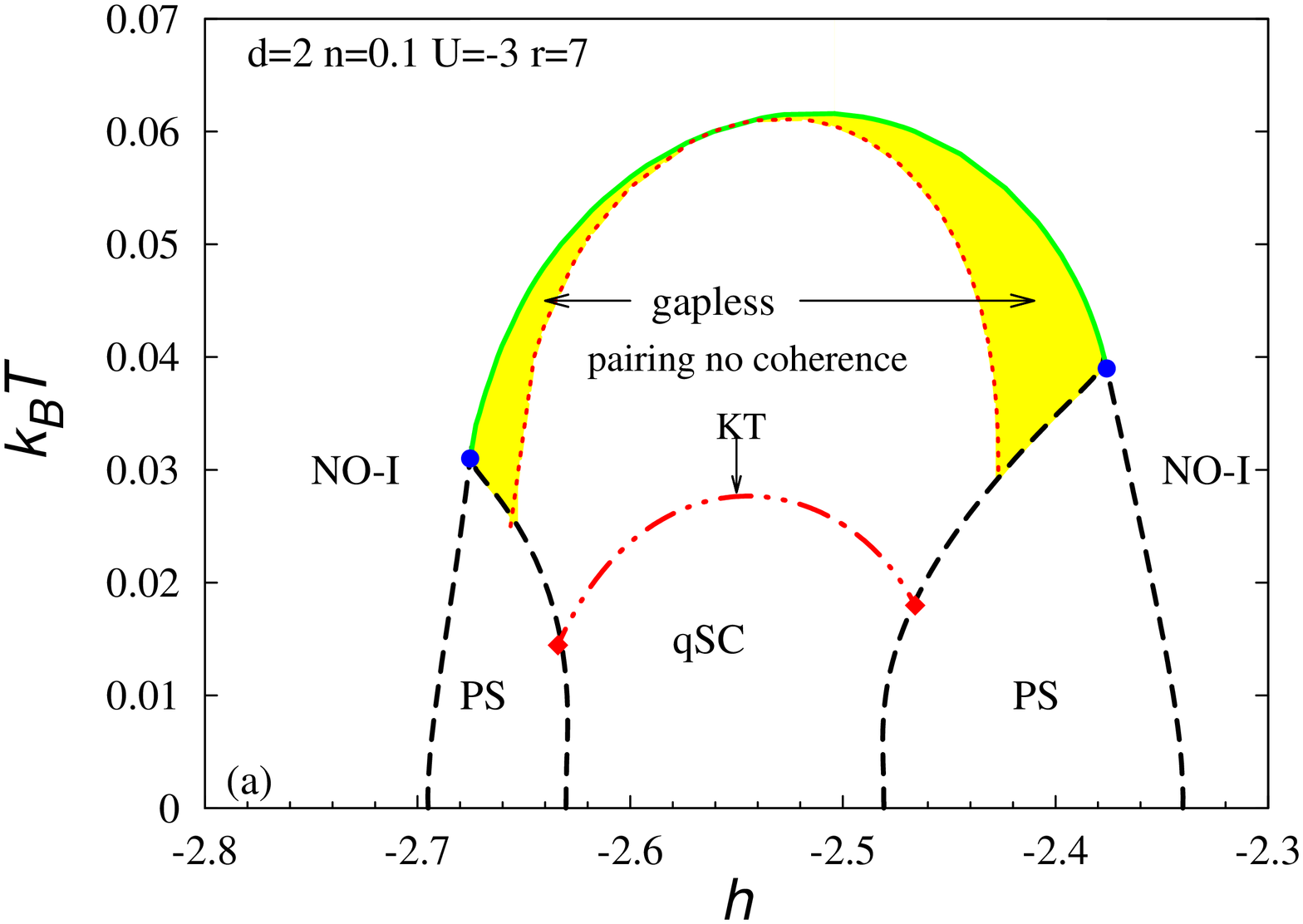}
\hspace*{-0.6cm}
\includegraphics[width=0.38\textwidth,angle=270]
{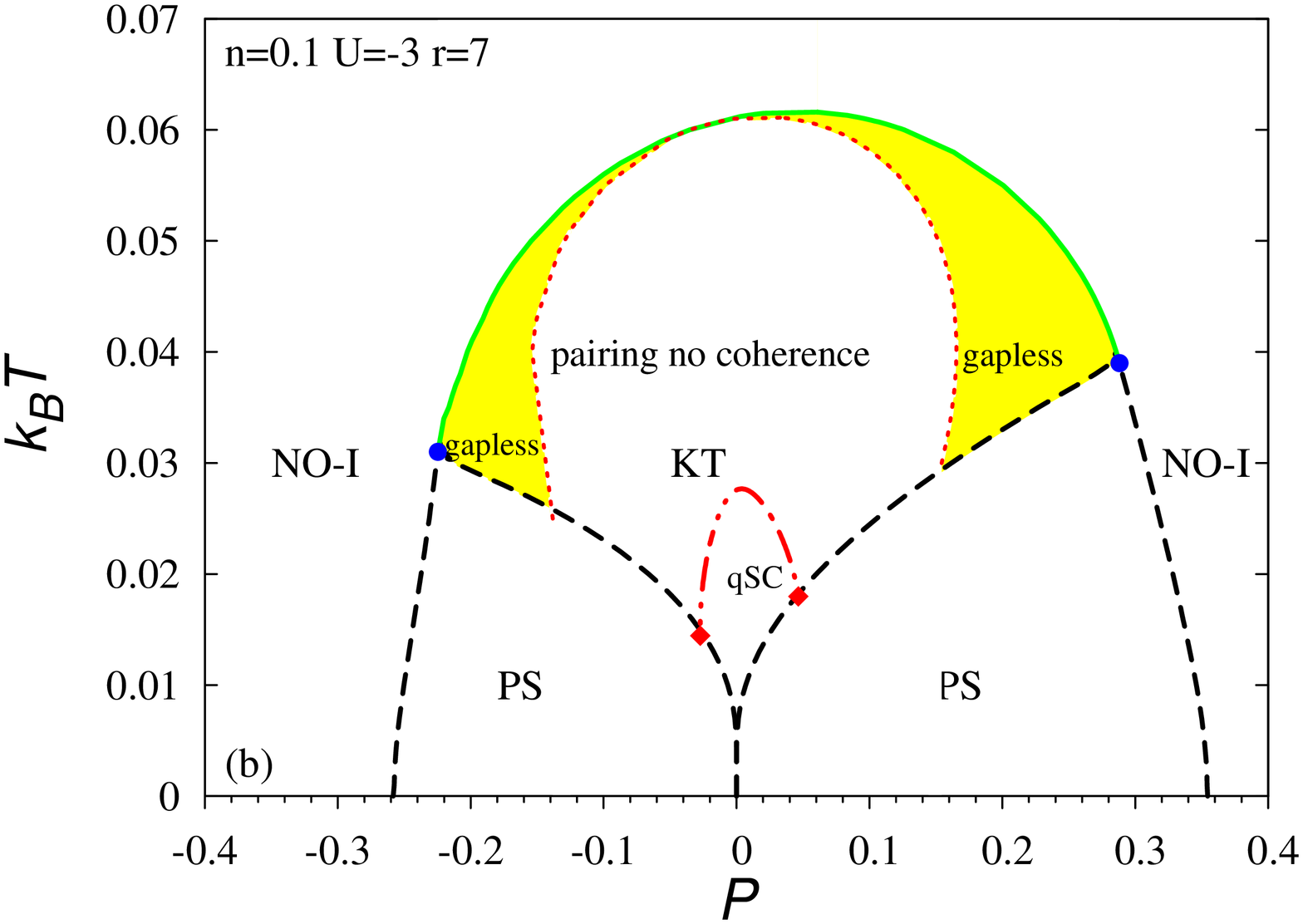}
\hspace*{-0.8cm}
\includegraphics[width=0.38\textwidth,angle=270]
{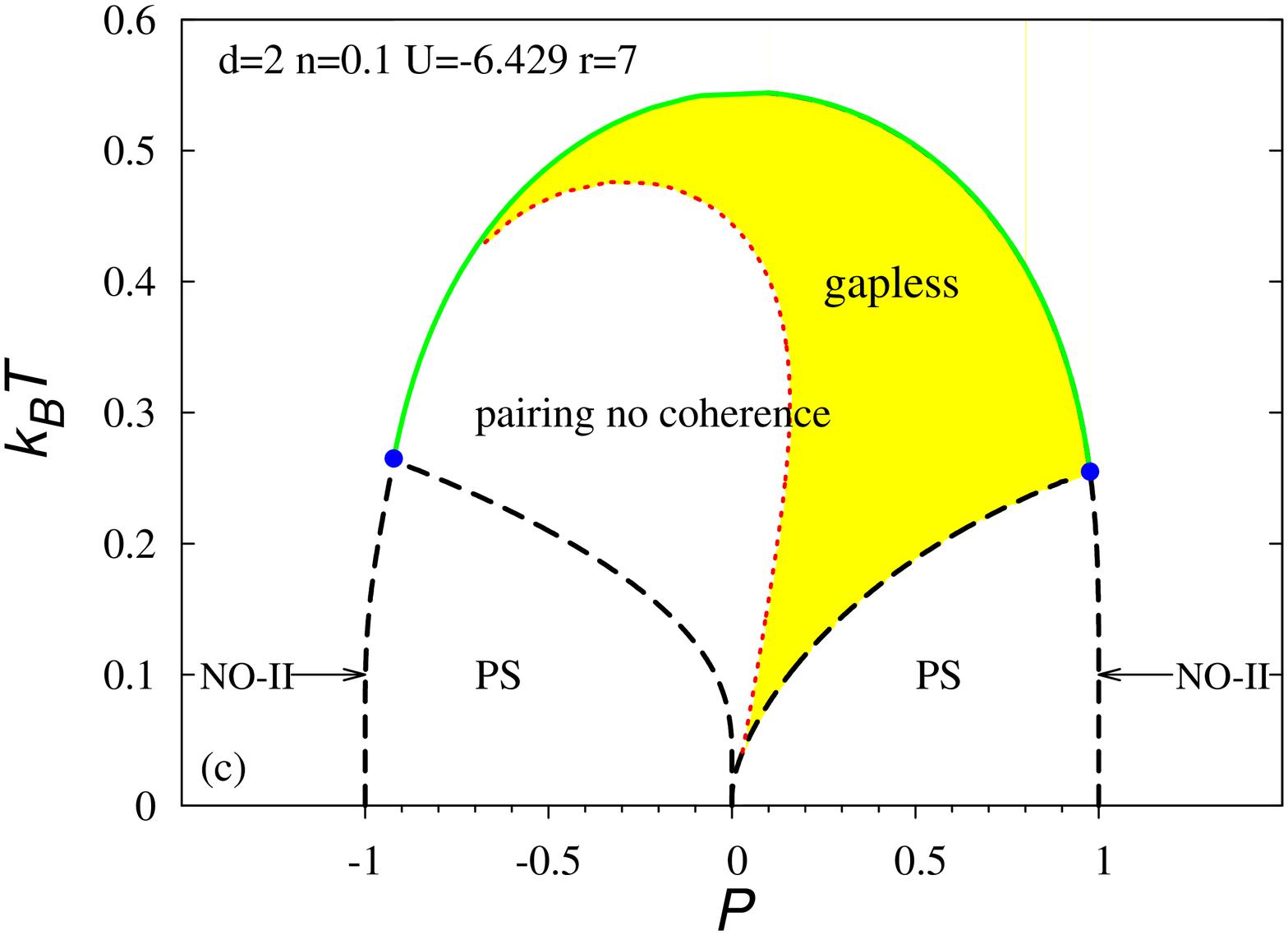}
\hspace*{-0.6cm}
\includegraphics[width=0.38\textwidth,angle=270]
{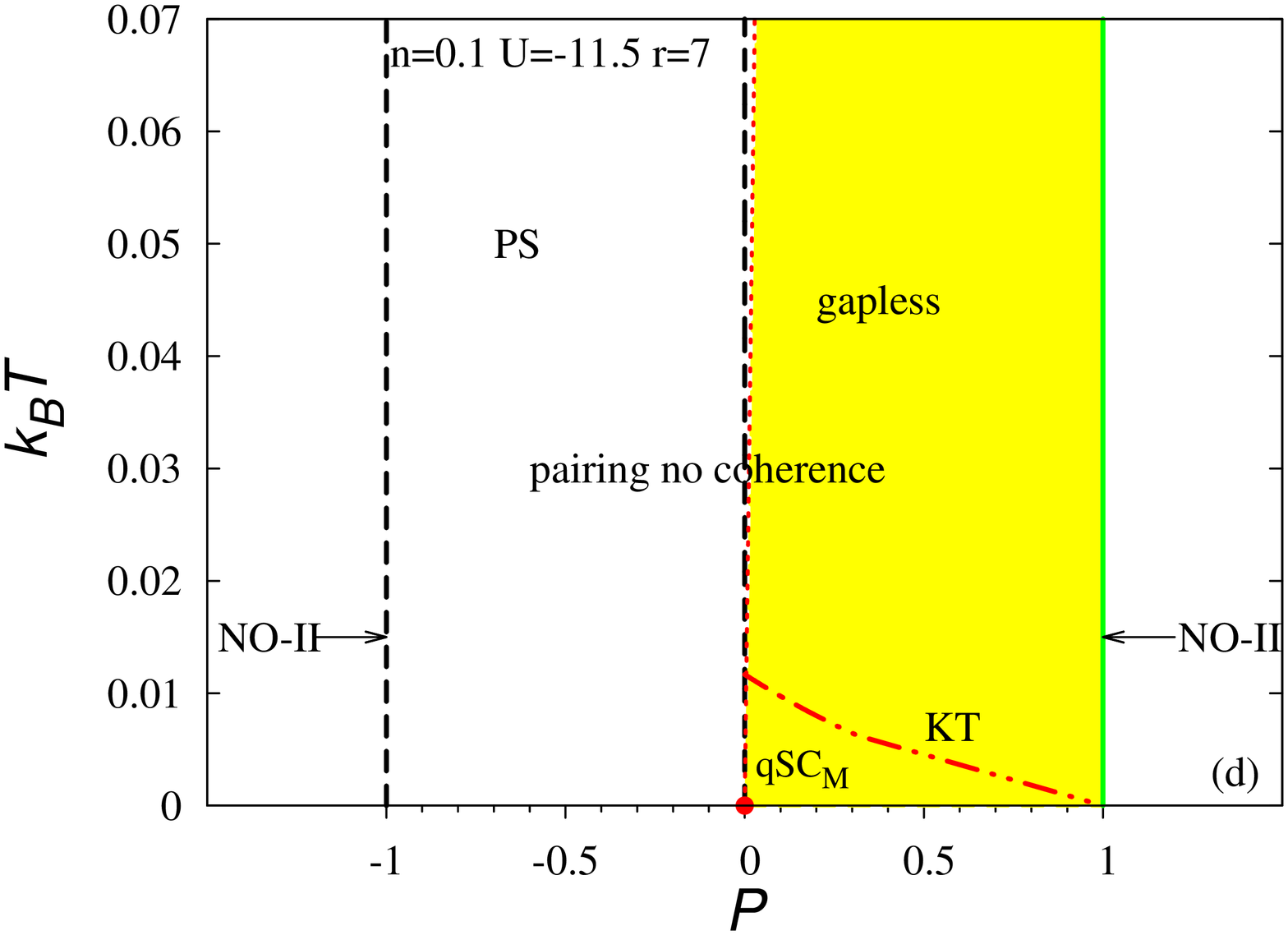}
\caption[Temperature vs. magnetic field (a) and polarization (b)-(d) phase
diagrams without the Hartree term, at fixed $n=0.1$, $r=7$, for the square
lattice. (a)-(b) $U=-3$, (c) $U=-6.429$, (d) $U=-11.5$.]{\label{2d_TvsP_h_r7}
Temperature vs. magnetic field (a) and polarization (b)-(d) phase diagrams
\textcolor{czerwony}{of AAHM}
without the Hartree term, at fixed $n=0.1$, $r=7$, for the square lattice.
(a)-(b) $U=-3$, (c) $U=-6.429$, (d) $U=-11.5$. The thick dashed-double dotted
line in (red color) is the KT transition line. Thick solid line denotes
transition from pairing without coherence region to NO within the BCS
approximation. Above the dotted line (red color) -- gapless (yellow color) --
the region which has a gapless spectrum for the majority spin species. $qSC$ --
2D KT superconductor, SC$_M$ -- gapless KT SC with one FS in the presence of
polarization (a spin polarized KT superfluid). Red point at $T=0$ (see Fig. (d))
-- QCP for Lifshitz transition.}
\end{figure}

Fig. \ref{2d_TvsP_h_r7} shows $T-h$ and $T-P$ phase diagrams for $n=0.1$, $r=7$,
at $h\neq 0$ and three values of attraction -- moderately weak ($U=-3$,
$E_b/E_F=0.024$), intermediate ($U=-6.429$ -- critical value of $U$ at which the
chemical potential drops below the lower band edge) and strong ($U=-11.5$)
coupling. These diagrams have also been constructed within the mean field
approximation (the solid lines (2$^{nd}$ order transition lines), PS and gapless
regions), but the phase coherence temperatures have been obtained as usual
within the KT scenario (thick dash-double dotted line (red color)). In our
approach the qSC phase is characterized by a non-zero gap ($\Delta \neq 0$)
and non-zero superfluid stiffness ($\rho_s \neq 0$). In the weak coupling
regime, the KT superconductor exists at low $|P|$ and low $T$ (Fig.
\ref{2d_TvsP_h_r7}(a)-(b)). Despite the fact that $r\neq 1$, the SC phase is
restricted to low $|P|$, while for larger $|P|$ the PS region is favored. There
is also the nonsuperfluid region (pairs without coherence), formally defined by
$\Delta \neq 0$, $\rho_s=0$. As mentioned above, in this region one observes
a
pseudogap behavior. Therefore, the region of incoherent pairs is different from
the normal phase. In the $T-P$ diagrams (see: Fig. \ref{2d_TvsP_h_r7}(b)-(c)),
one finds MF TCPs at which the thermal transition changes from the second to the
first order. We also show the gapless area within the state of pairing without
coherence. This gapless region is above the KT coherence temperatures in the
weak coupling limit.  

In the intermediate coupling (Fig. \ref{2d_TvsP_h_r7}), below $T^{KT}_c$ the SC
state is strongly reduced to very low $|P|$. At the BCS-LP crossover point a
polarized SC does not exist even for $r\neq 1$. 

\begin{figure}[t!]
\hspace*{-0.8cm}
\includegraphics[width=0.38\textwidth,angle=270]
{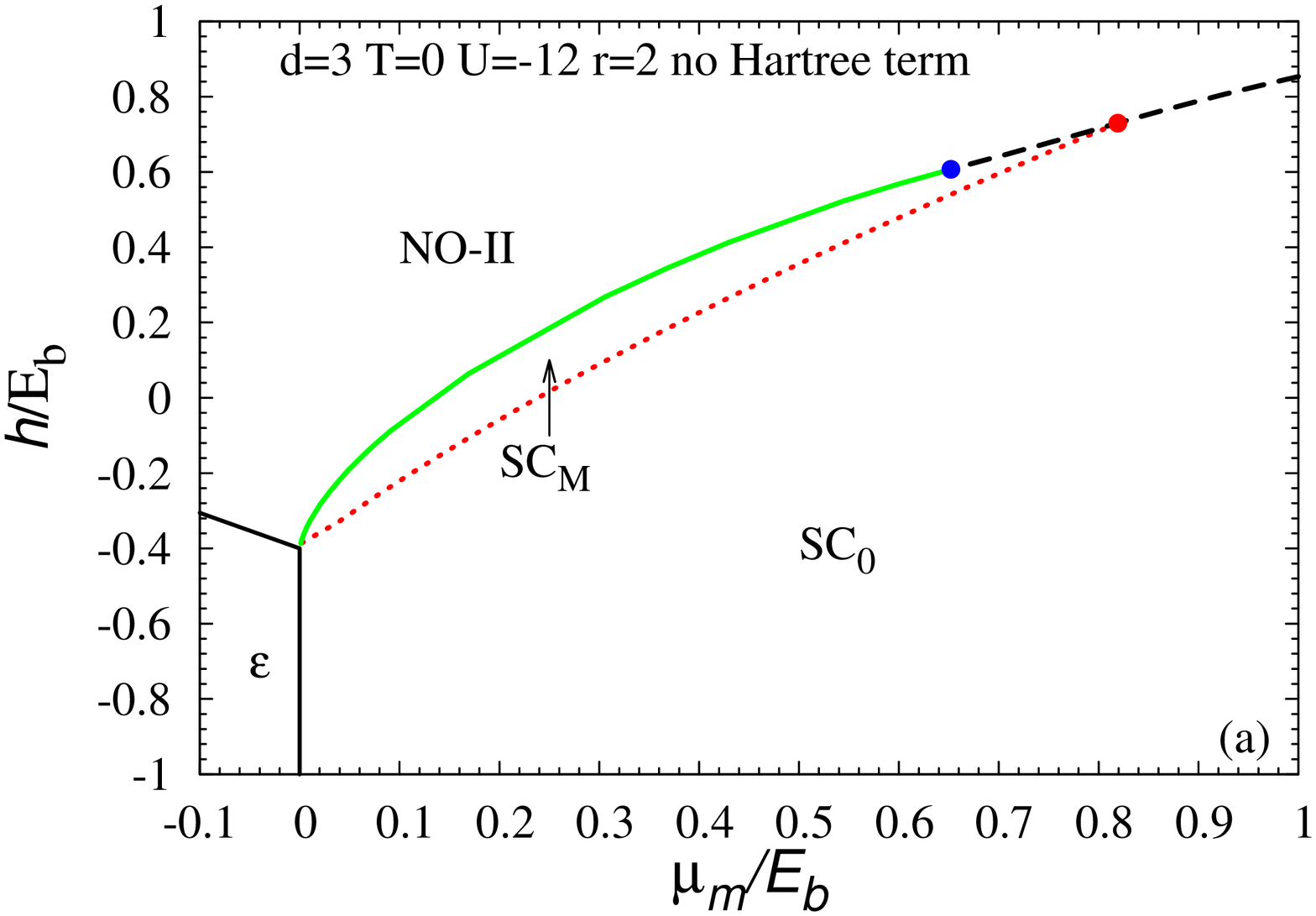}\hspace{-0.2cm}
\hspace*{-0.6cm}
\includegraphics[width=0.38\textwidth,angle=270]
{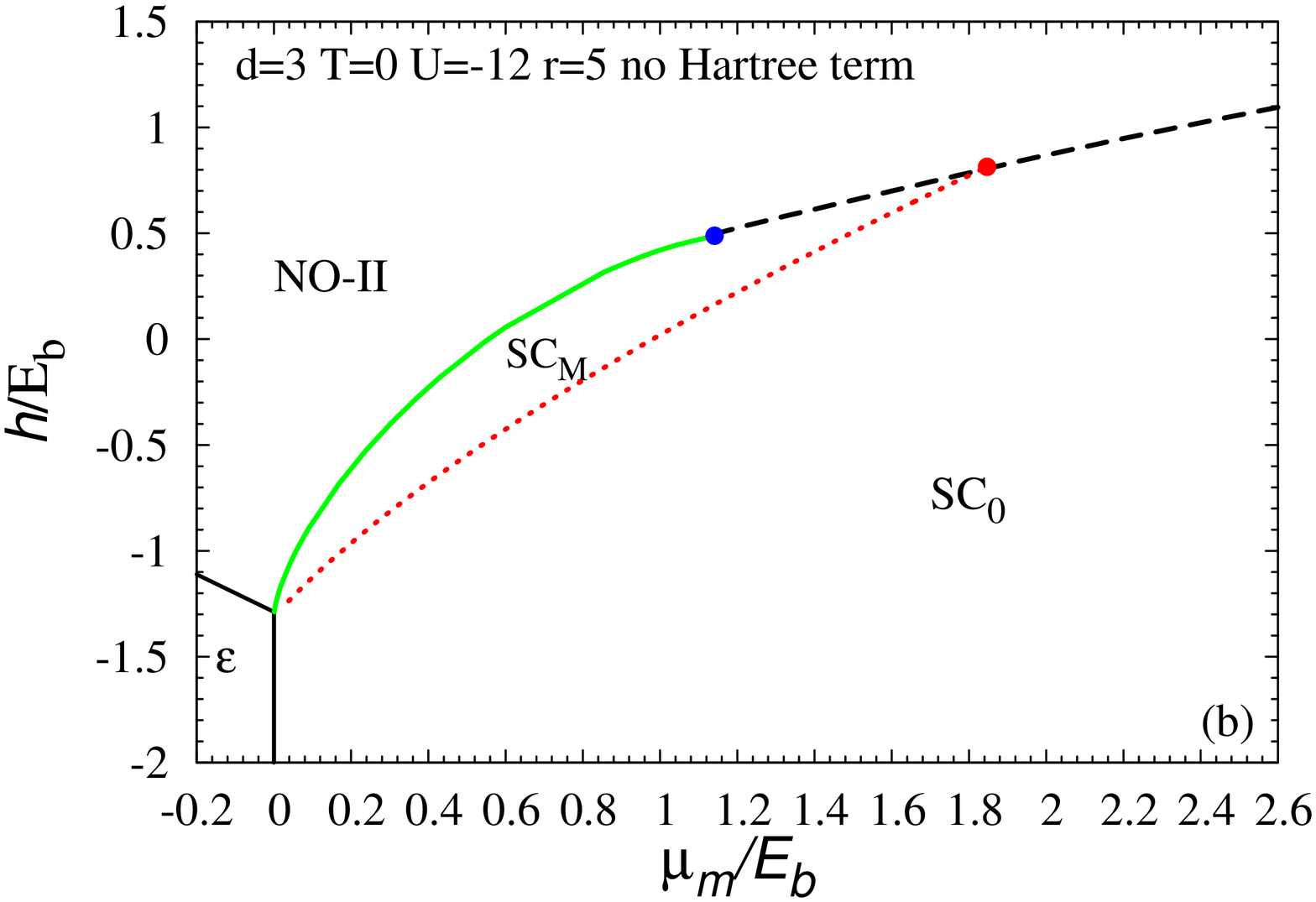}
\caption[Critical magnetic field vs. chemical potential for $d=3$, $U=-12$ on LP
side, (a) $r=2$ and (b) $r=5$. Diagram without the Hartree
term.]{\label{U-12_r2_r5_mi} Critical magnetic field vs. chemical potential for
$d=3$, $U=-12$ on LP side, (a) $r=2$ and (b) $r=5$. SC$_0$
-- unpolarized SC state with $n_{\uparrow}=n_{\downarrow}$, SC$_M$ -- magnetized
SC state, NO-II -- fully polarized normal state, $\varepsilon$ -- empty state,
$\mu_{m}$ -- half of the pair chemical potential defined as:
$\mu-\epsilon_0+\frac{1}{2} E_b$, where $\epsilon_0=-6t$, $E_b$ is the binding
energy for two fermions in an empty lattice with hopping $t$. Red point --
$h_{c}^{SC_M}$
(quantum critical point (QCP)), blue point -- tricritical point. The dotted red
and the solid green lines are continuous transition lines.}
\end{figure}

The situation is radically different in the spin asymmetric hopping and strong
coupling case (Fig. \ref{2d_TvsP_h_r7}(d)). As mentioned above, for
sufficiently high value of $r$, below $T_{c}^{KT}$, the spin polarized KT
superfluid state with gapless spectrum and one FS can be stable for all $P>0$.
If $P<0$, there is the PS region at low $T$. 

We notice however that the polarized KT superfluid is restricted to low
temperatures
$T_c^{KT}/E_F\sim 0.02$ for this value of n. For lower n $T_c^{KT}/E_F$ will be
higher and upper limit is set by the value for continuum case. 
  

\section{AAHM on 3D simple cubic lattice}
In chapter \ref{chapter5}, among other things, we analyzed the
BCS-BEC crossover ground state phase diagrams in the presence of the Zeeman
magnetic field in 3D for a simple cubic lattice. For strong attraction and in
the dilute limit, the homogeneous magnetized superconducting phase and the
tricritical point were found in the $(h-\mu)$ and $(h-n)$ diagrams. In this
section, we briefly study the hopping asymmetric AHM (i.e.
$t^{\uparrow}\neq t^{\downarrow}$) and show that the introduction of a mismatch
between the hopping integrals causes an extension of the range of occurrence
of
the  SC$_M$ phase in the 3D case. The importance of the Hartree term in the
broken symmetry Hartree-Fock approximation is also indicated.

\begin{figure}[t!]
\hspace*{-0.8cm}
\includegraphics[width=0.38\textwidth,angle=270]
{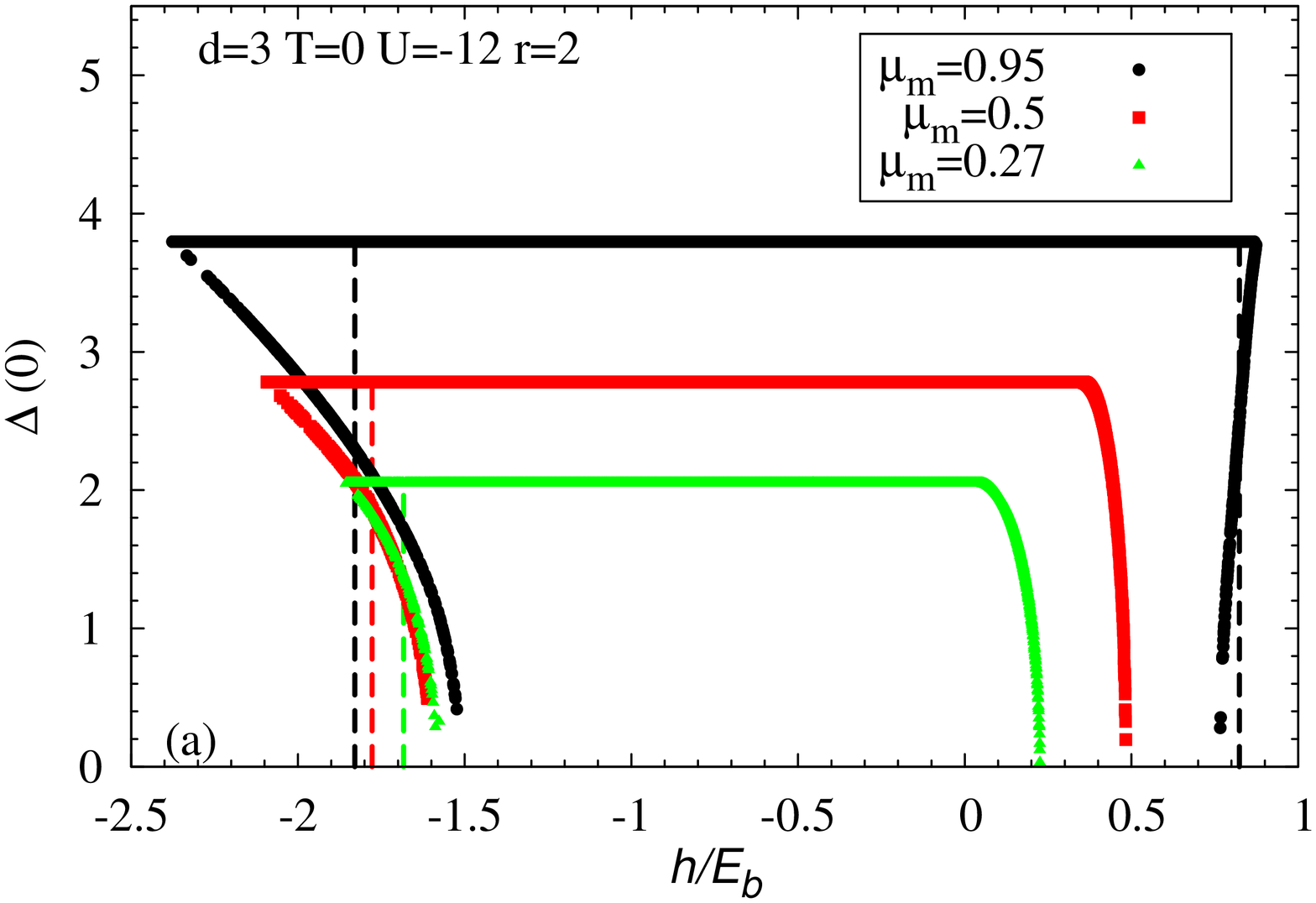}\hspace{-0.2cm}
\hspace*{-0.6cm}
\includegraphics[width=0.38\textwidth,angle=270]
{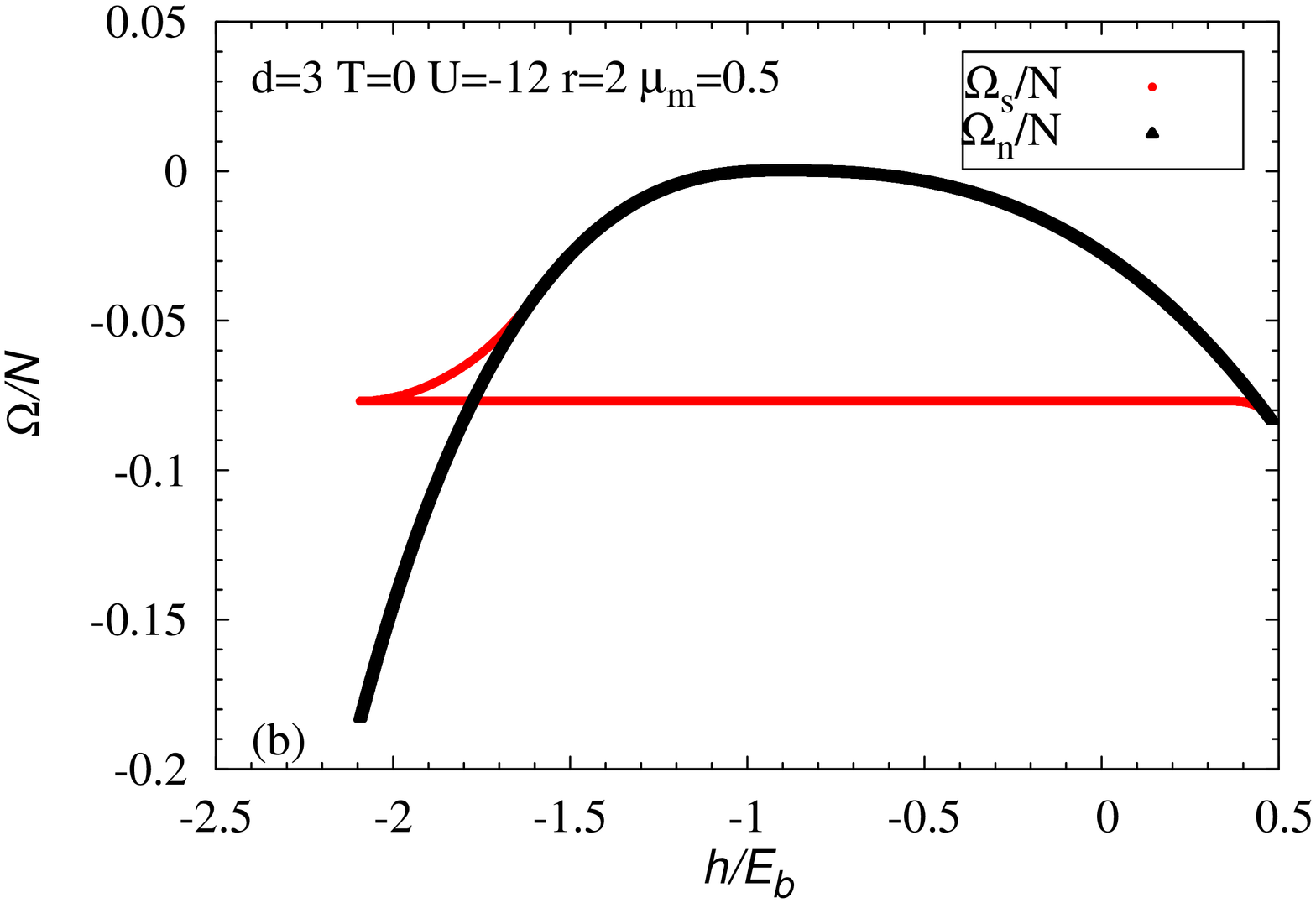}
\caption[Dependence of the order parameter (a) and the grand canonical potential
(b) on the magnetic field in the units of the binding
energy $E_b$  at $T=0$, $d=3$, $U=-12$, $r=2$, for three values of
$\mu_{m}=\mu-\epsilon_0+\frac{1}{2} E_b$.]{\label{U-12_r2_mi_del} Dependence of
the order parameter (a) and the grand canonical potential
(b) on the magnetic field in the units of the binding
energy $E_b$  at $T=0$, $d=3$, $U=-12$, $r=2$, for three values of
$\mu_{m}=\mu-\epsilon_0+\frac{1}{2} E_b$. The vertical dashed lines mark the
Hartee-Fock phase
transition magnetic field -- the first order phase transition to the normal
state at $T=0$.}
\end{figure}

We start our analysis from a comparison of three ground state phase diagrams of
AHM
at fixed $\mu_{m}$ and $h$, for $r=1$ (Fig. \ref{U-12_r1_mi}), $r=2$ (Fig.
\ref{U-12_r2_r5_mi}(a)) and $r=5$ (Fig. \ref{U-12_r2_r5_mi}(b)). As usual, we
define $\mu_{m}=\mu-\epsilon_0+\frac{1}{2} E_b$ as one half of the pair chemical
potential (molecular potential). As mentioned above, many experiments
have indicated
that in the density profiles of trapped Fermi mixtures with population (mass)
imbalance, there is an unpolarized superfluid core in the center of the trap and
a polarized normal state surrounding this core, in the BCS and unitarity
regimes. These experimental observations are in good agreement with our earlier
analysis (see: chapter \ref{chapter4}) according to which there is non-polarized
($n_{\uparrow}=n_{\downarrow}$) superfluid phase for higher $\mu$ and polarized
($n_{\uparrow}\neq n_{\downarrow}$) normal state for lower $\mu$ at fixed $h$
and attractive interaction (see: Fig. \ref{mu_diagram}(b)). On the BEC
side, in contrast to the BCS limit, there can appear a shell of the
SC$_M$
phase (with a finite polarization and a gapless spectrum) between the
SC$_0$ and the NO phases, as shown in Figs. \ref{U-12_r1_mi},
\ref{U-12_r2_r5_mi}(a) and \ref{U-12_r2_r5_mi}(b). The structure of the above
diagrams has been discussed in detail in chapter \ref{chapter5} for the $r=1$
case. The topology of the diagrams in Fig. \ref{U-12_r2_r5_mi} for $r\neq 1$
and \ref{U-12_r1_mi} for $r=1$ is the same. We still observe the first order
phase transition from the SC$_0$ to the NO phase or the continuous transition
from SC$_0$ to SC$_M$ and then first or second order to the normal state.
However, there are quantitative differences. While for the $r=1$ case, the
SC$_M$ phase appears only in the dilute limit (in particular in the diagrams
with the Hartree term), e.g. at $U=-12$ a critical value of $n$ ($n_c$) below
which
the SC$_M$ state becomes stable equals around 0.059 (without the Hartree term
case) or $n_c \approx 0.0086$ (with the Hartree term included), for $r=2$ $n_c
\approx 0.238$ and for $r=5$ $n_c\approx 0.56$. Hence, TCP moves towards higher
values of $\mu$ ($n$). There are also differences in the phase diagrams around
the empty state, due to the $h\rightarrow -h$ symmetry breaking for $r\neq 1$.
For $r=1$, $\mu_{m}/E_b=0$ for $h/E_b=0.5$, while for $r\neq 1$ the ratio
$h/E_b$ which gives $\mu_{m}/E_b=0$ moves toward negative values with increasing
$r$. Moreover, as one can see in Fig. \ref{U-12_r2_mi_del} which shows the order
parameter vs. magnetic field in the units of the binding energy at $T=0$, the
transition from the SC$_0$ to the NO phase is always of the first order for
$h<0$ and negative values of $\mu=\mu_m+\epsilon_0-\frac{1}{2}E_b$ (left part of
Fig. \ref{U-12_r2_mi_del}). In turn, as mentioned above, the transition
from SC$_0$ to NO can take place in two ways in the $h-n$ plane for
$h>0$: through phase separation (for higher $\mu_m$, e.g. $\mu_m=0.95$ in
Fig. \ref{U-12_r2_mi_del}, which corresponds to $n\approx0.279$) or through the
SC$_M$ state (for lower $\mu_m$, e.g. $\mu_m=0.5$, $\mu_m=0.27$  in Fig.
\ref{U-12_r2_mi_del}, which corresponds to $n\approx 0.142$ and $n\approx
0.0756$, respectively.).

\begin{figure}[t!]
\hspace*{-0.8cm}
\includegraphics[width=0.38\textwidth,angle=270]
{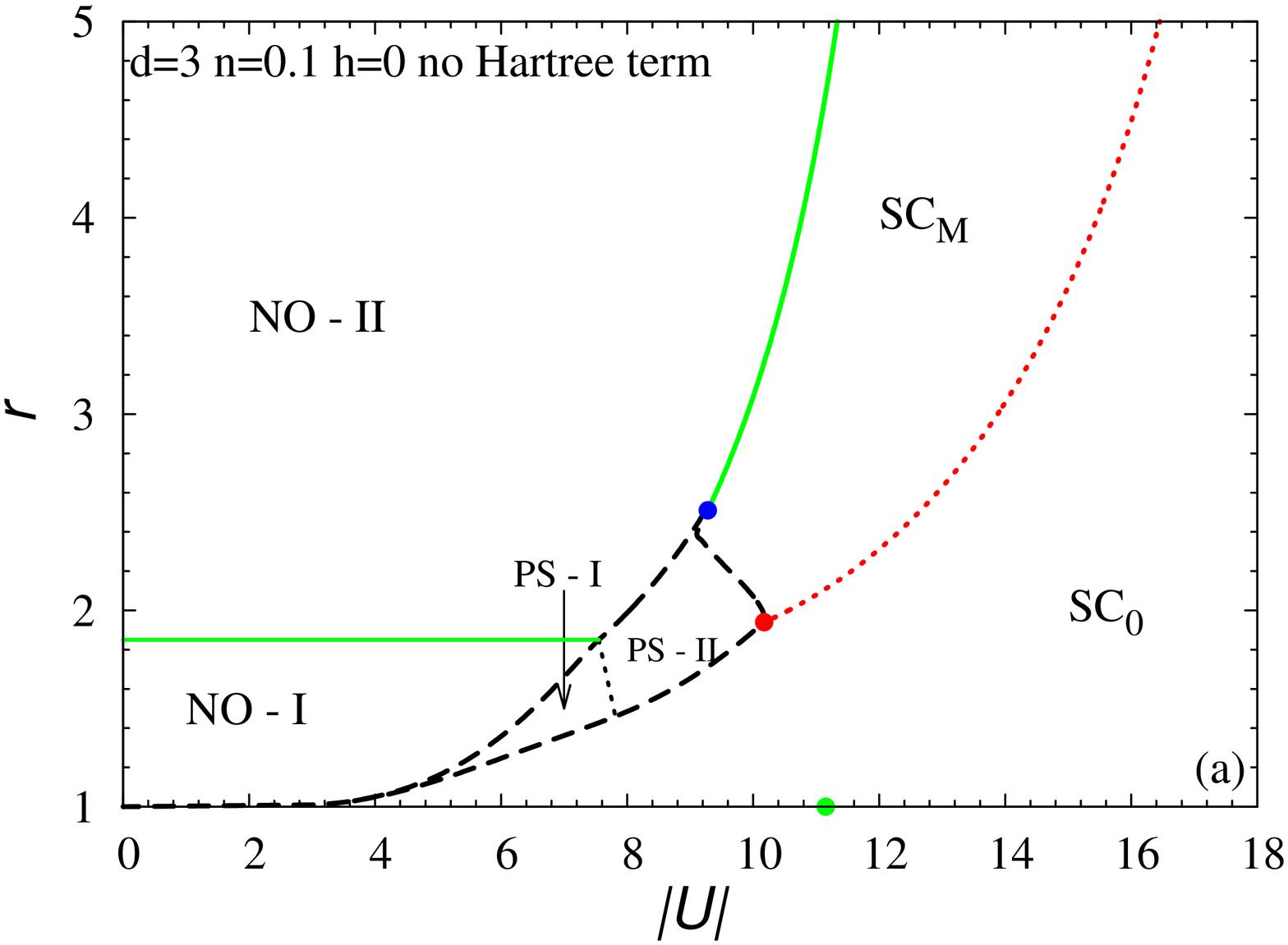}\hspace{-0.2cm}
\hspace*{-0.6cm}
\includegraphics[width=0.38\textwidth,angle=270]
{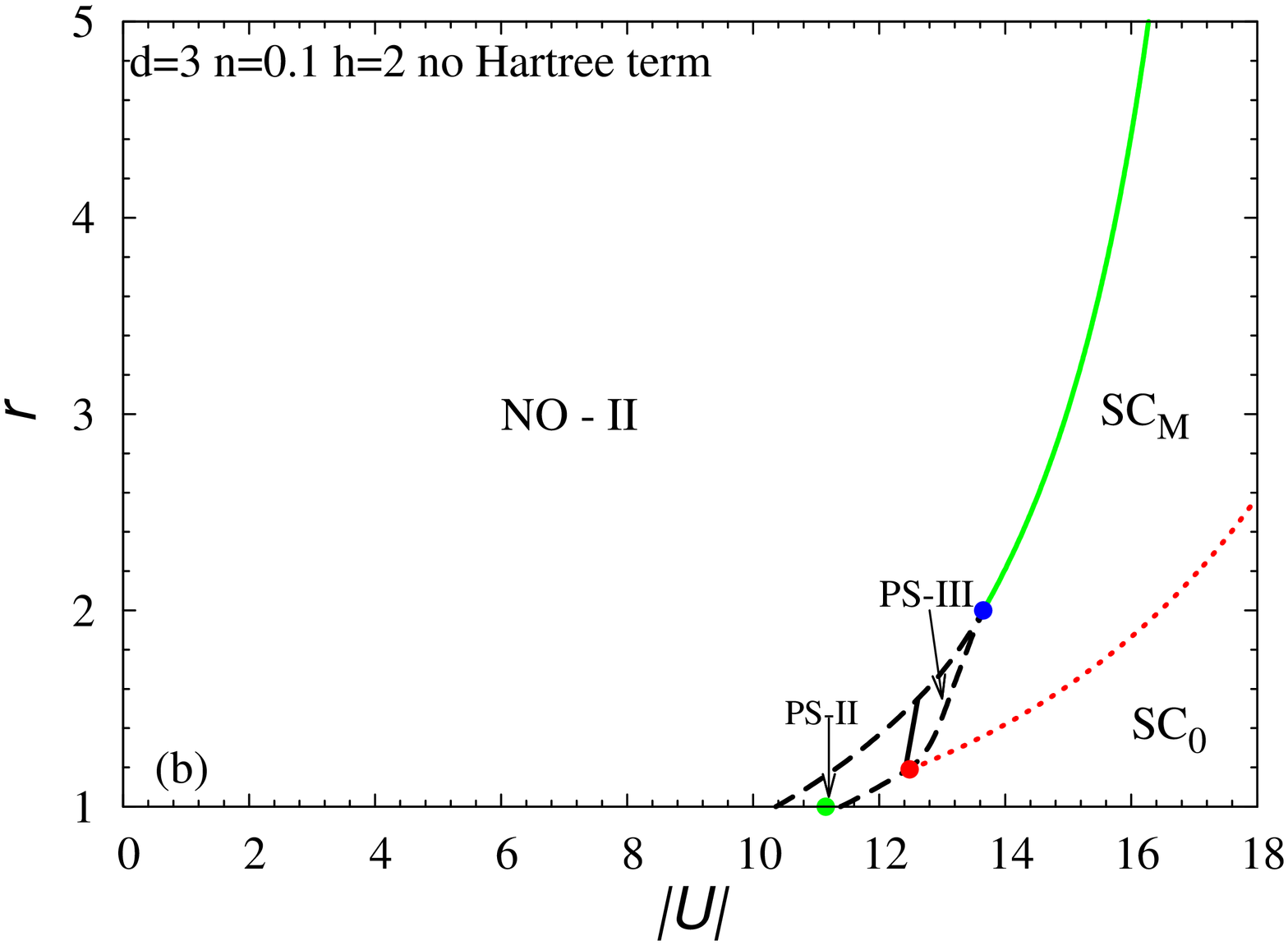}
\caption[Ground state phase diagrams of AHM $r$ vs. $|U|$ for $d=3$, $n=0.1$,
(a) $h=0$ (b) $h=2$.]{\label{U_r_n01_h0_h2} Ground state phase diagrams $r$ vs.
$|U|$ for $d=3$, $n=0.1$, (a) $h=0$ (b) $h=2$. SC$_0$ -- unpolarized SC state,
SC$_M$ -- magnetized SC state, NO-I (NO-II) -- partially (fully) polarized
normal states. PS-I (SC$_0$+NO-I) -- partially polarized phase separation, PS-II
(SC$_0$+NO-II) -- fully polarized phase separation, PS-III -- (SC$_M$+NO-II).
Red point -- $|U|_{c}^{SC_M}$ (quantum critical point), blue point  --
tricritical point, green point -- the BCS-BEC crossover point in the SC$_0$
phase.}
\end{figure}

Fig. \ref{U_r_n01_h0_h2} shows ($r-|U|$) phase diagrams for fixed $n=0.1$,
$h=0$ and $h=2$. Here, we consider a rather low value of $n$, but still not
the very dilute case. Hence, there is the critical value of $r\neq 1$ for which
the SC$_M$ is stable both for $h=0$ as for $h=2$, while the magnetized
superfluid state can be stable even for $r=1$ in the 3D case. The structure of
the diagram \ref{U_r_n01_h0_h2}(a) is very similar to that from section
\ref{T0r2D} (Fig. \ref{rvsU_n01}(a)), for the 2D case. However, $r_c$ is much
lower in the simple cubic lattice case than in the square lattice case.
Moreover, the SC$_M$ state is stable in the intermediate region for 3D as
opposed to the 2D case for which this phase is stable only in the strong
coupling limit, for so chosen parameters. Next, if $h\neq 0$ (Fig.
\ref{U_r_n01_h0_h2}(b)), the SC$_M$ phase moves towards higher values
of attractive interactions. However, the critical value $r_c$ is lower than in
the
case without the Zeeman magnetic field. The presence of $h$ breaks the
$r\rightarrow 1/r$ symmetry. Moreover, there is neither PS-I, nor the
partially polarized normal state in the phase diagram at finite $h$, even for
$r=1$.

\begin{figure}[t!]
\hspace*{-0.8cm}
\includegraphics[width=0.38\textwidth,angle=270]
{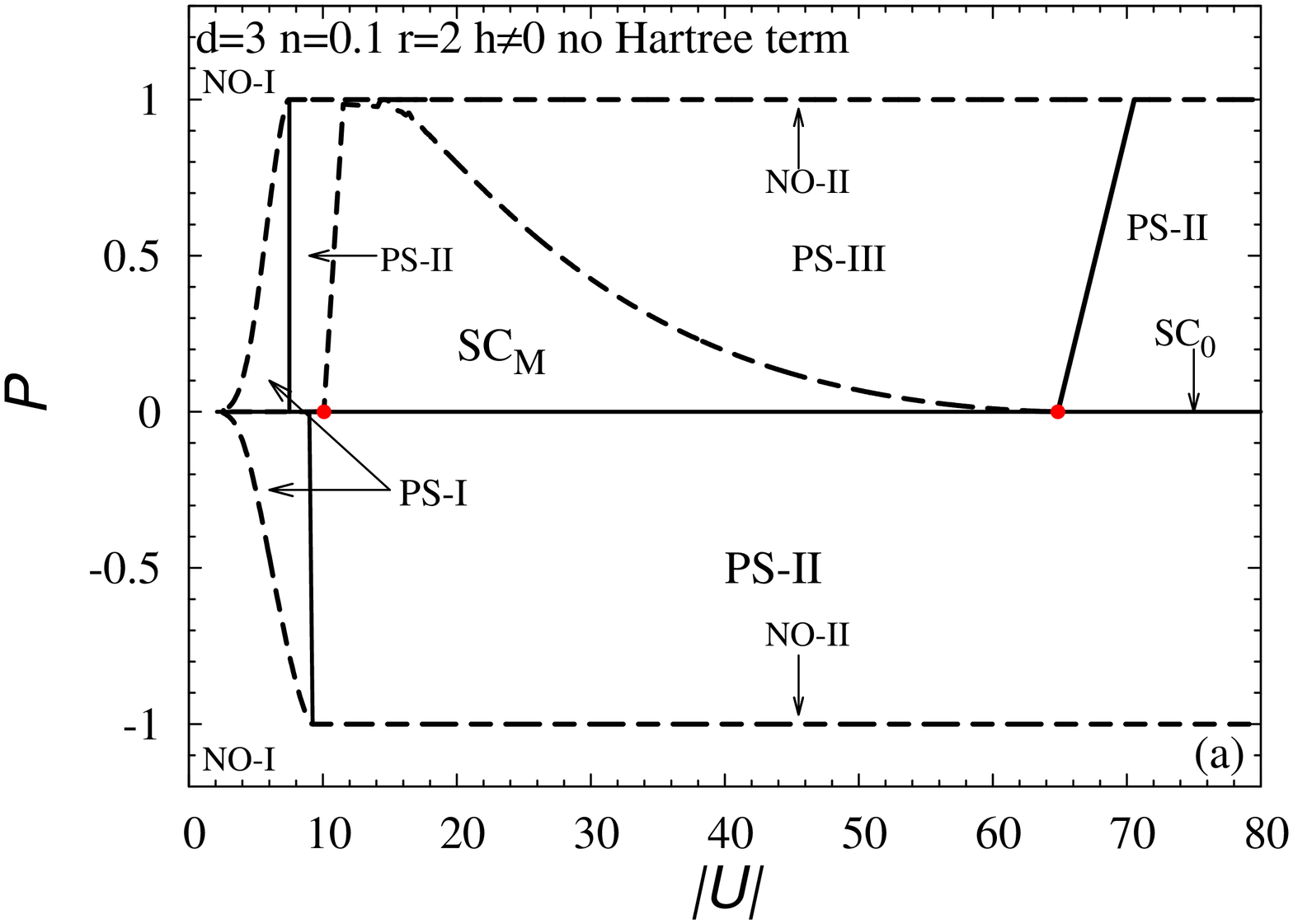}
\hspace*{-0.6cm}
\includegraphics[width=0.38\textwidth,angle=270]
{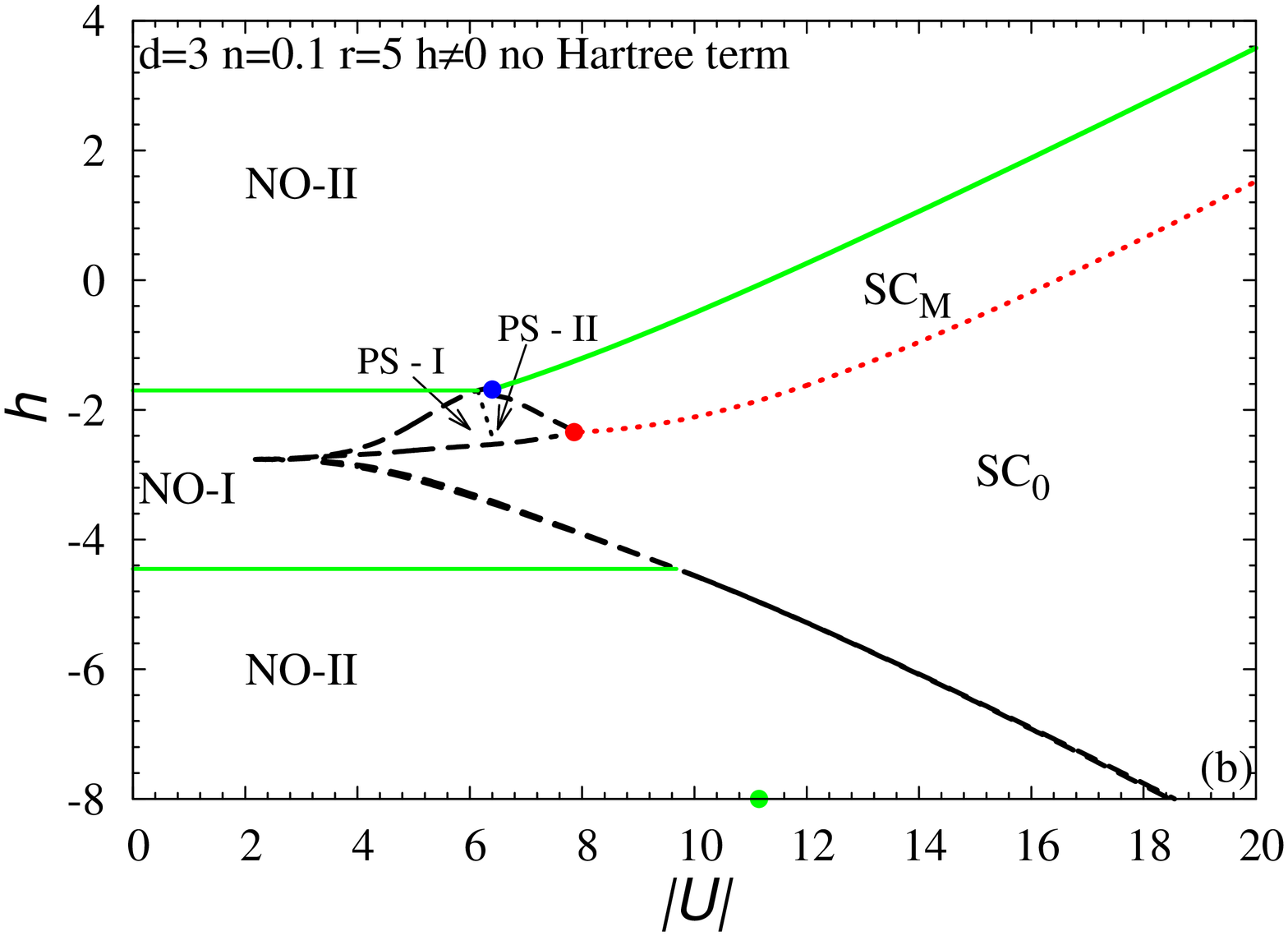}
\hspace*{-0.8cm}
\includegraphics[width=0.38\textwidth,angle=270]
{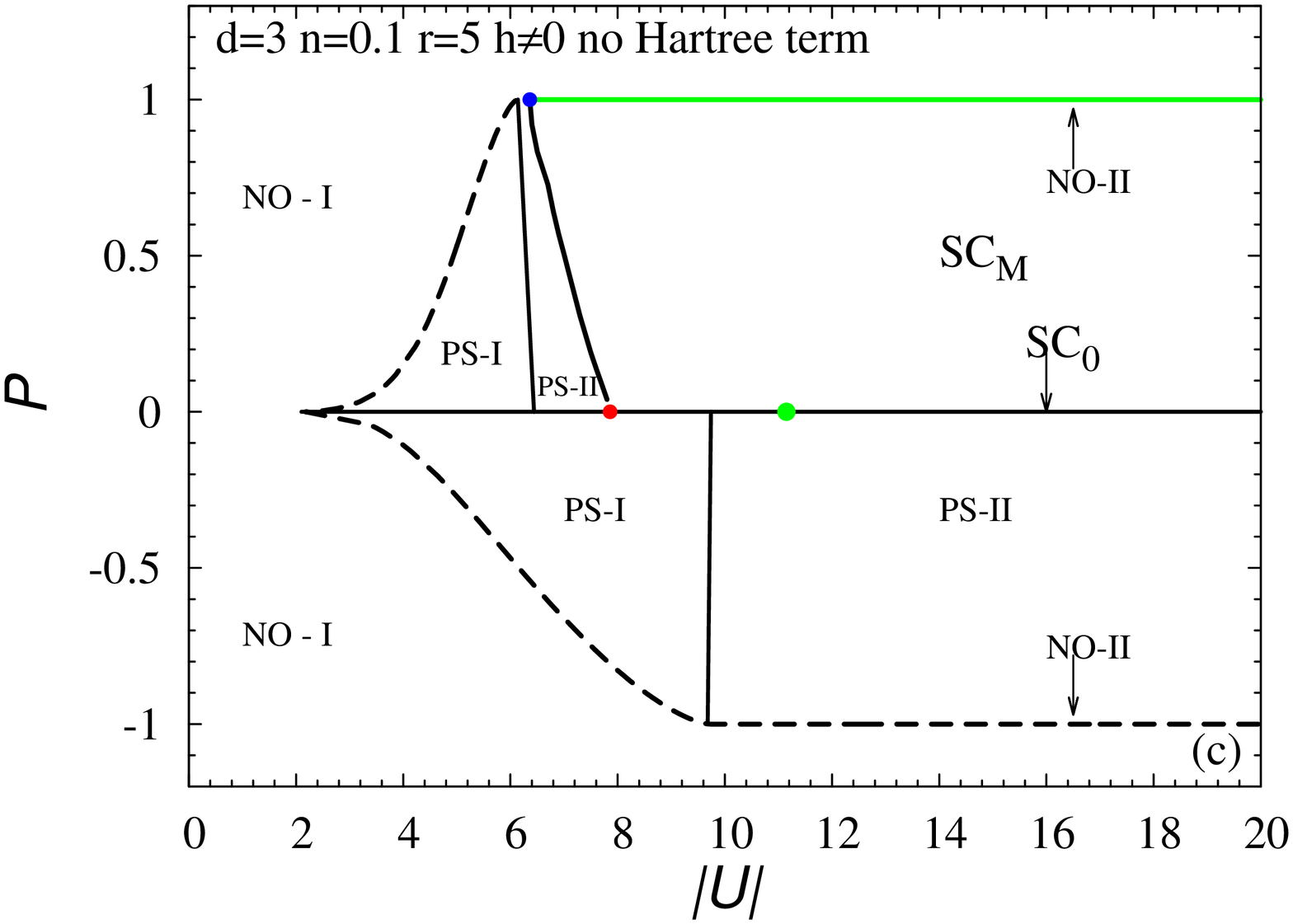}
\hspace*{-0.6cm}
\includegraphics[width=0.38\textwidth,angle=270]
{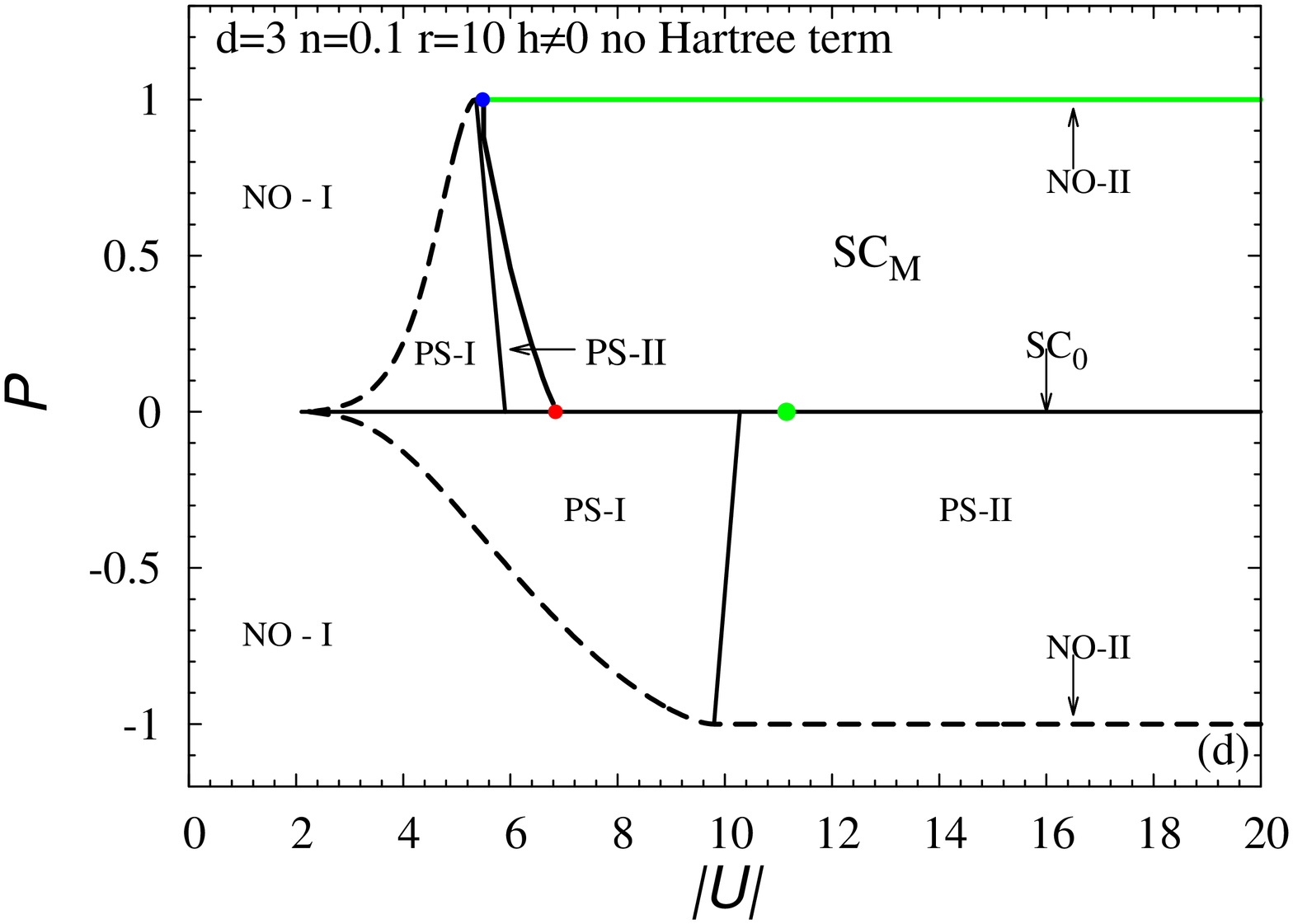}
\caption[Polarization vs. $|U|$ ground state phase diagrams, at fixed $n=0.1$,
for (a) $r=2$, (b), (c) $r=5$, (d) $r=10$.]{\label{n_01_r_P} Polarization vs.
$|U|$ ground state phase diagrams \textcolor{czerwony}{of spin polarized AHM}, 
at fixed $n=0.1$, for (a) $r=2$, (b), (c)
$r=5$, (d) $r=10$. SC$_0$ -- unpolarized superconducting state, SC$_M$ --
magnetized superconducting state, PS-I (SC$_0$+NO-I) -- partially polarized
phase separation, PS-II (SC$_0$+NO-II) -- fully polarized phase separation,
PS-III (SC$_M$+NO-II). Green point -- the BCS-BEC crossover point, blue point --
tricritical point, red point -- $|U|_{c}^{SC_M}$.}
\end{figure}

\begin{figure}[t!]               
\hspace*{-0.8cm}
\includegraphics[width=0.38\textwidth,angle=270]
{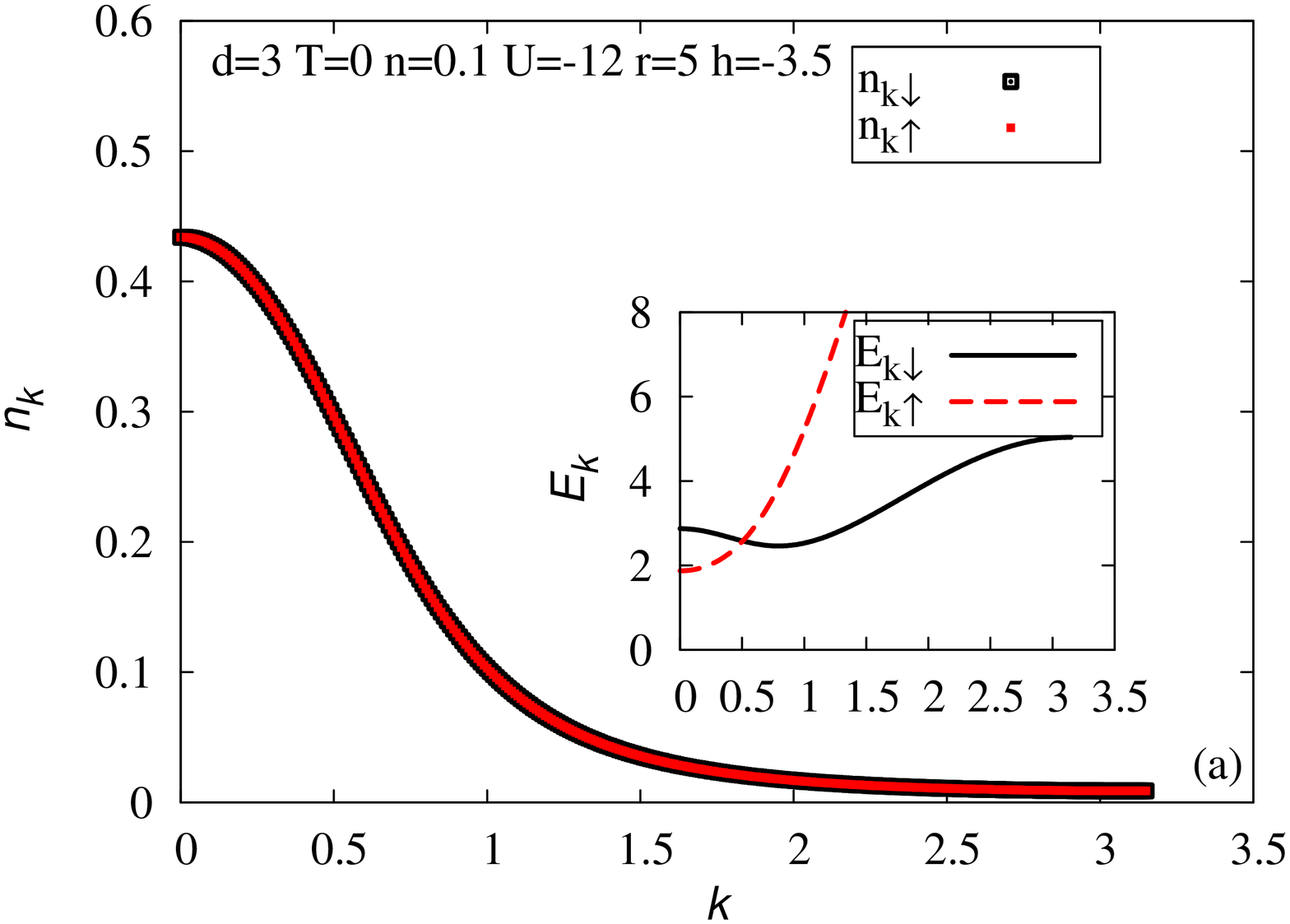}
\hspace*{-0.6cm}
\includegraphics[width=0.38\textwidth,angle=270]
{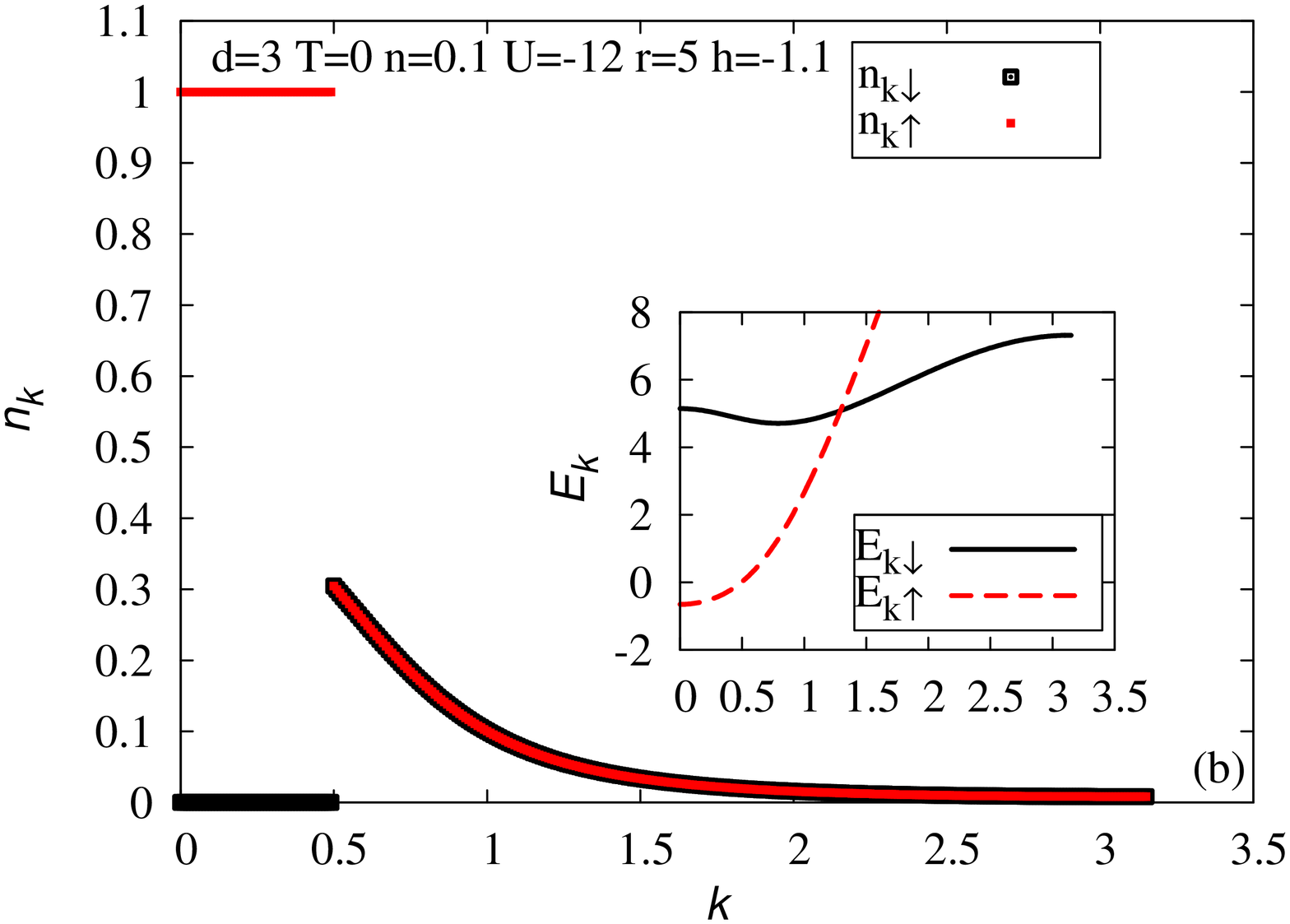}
\hspace*{-0.8cm}
\includegraphics[width=0.38\textwidth,angle=270]
{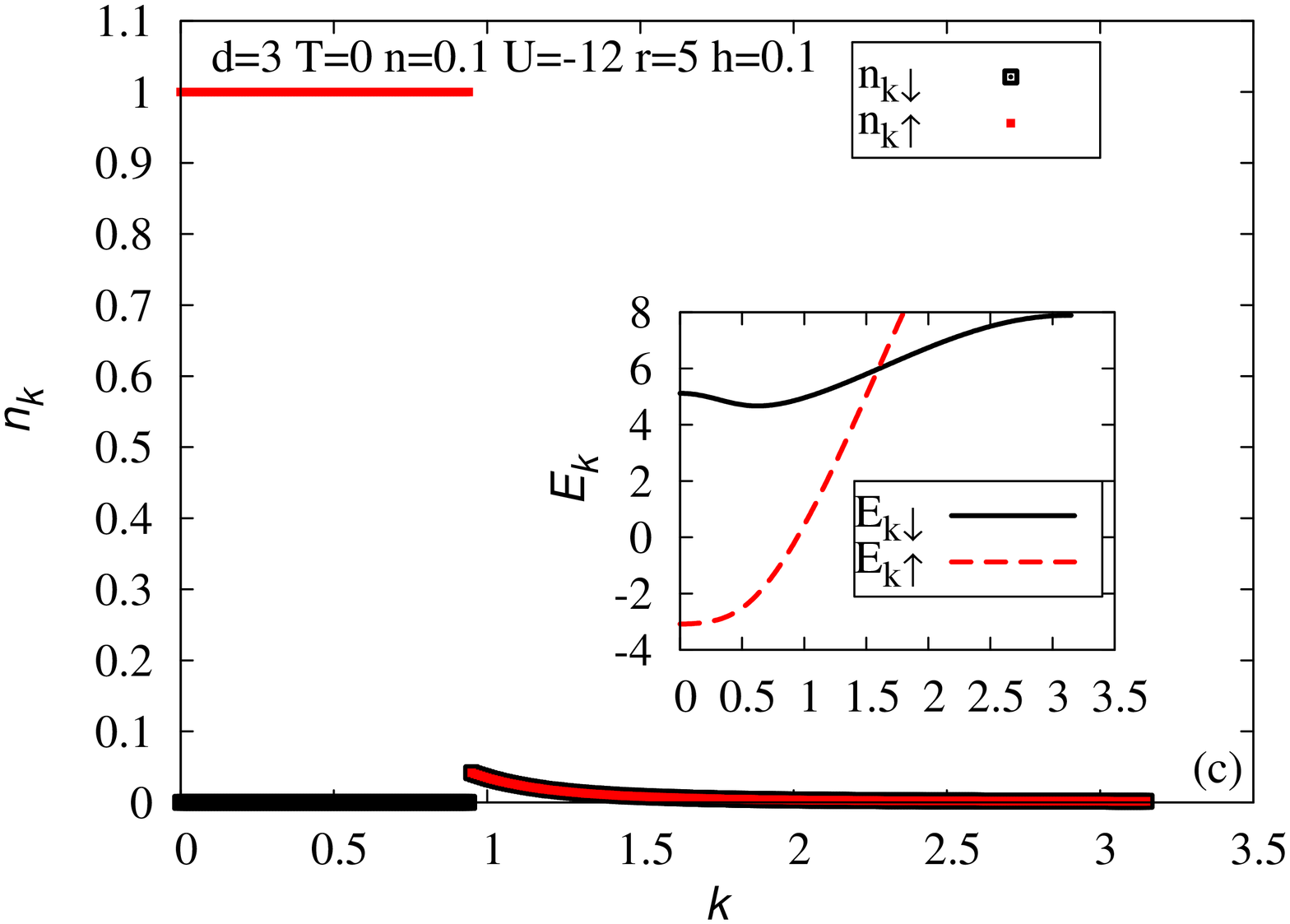}
\hspace*{-0.6cm}
\includegraphics[width=0.38\textwidth,angle=270]
{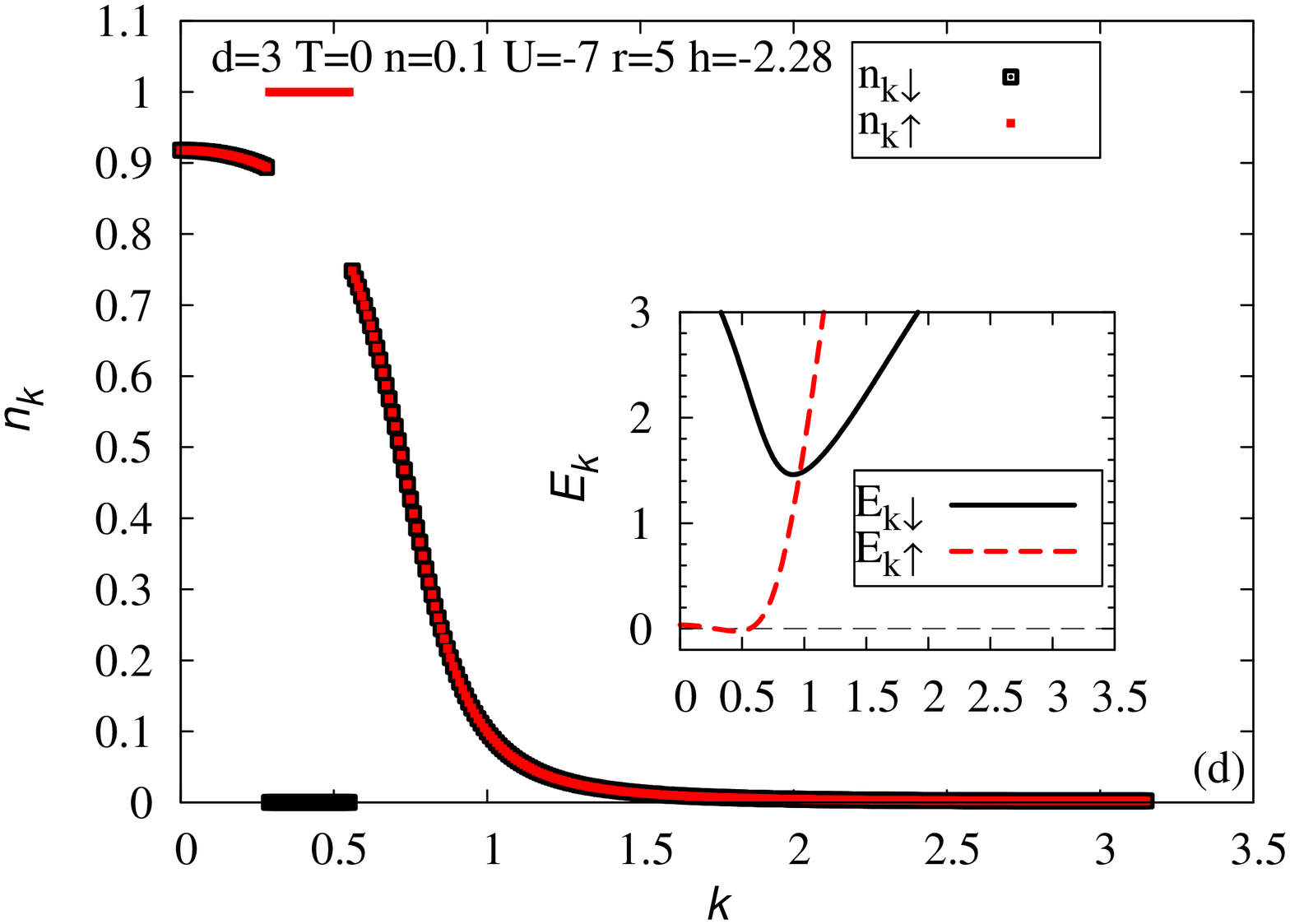}
\caption[Plots of momentum occupation numbers $n_{\vec{k}\uparrow}$ (red
points), $n_{\vec{k}\downarrow}$ (black points) vs. $k\equiv
k_x=k_y=k_z$ and the corresponding quasiparticle spectra $E_{\vec{k}\uparrow}$,
$E_{\vec{k}\downarrow}$ (inset) for $d=3$, $n=0.1$, $r=5$
$U=-12$.]{\label{n_k_E_k_r5_n01} Plots of momentum occupation numbers
$n_{\vec{k}\uparrow}$ (red points), $n_{\vec{k}\downarrow}$ (black points) vs.
$k\equiv k_x=k_y=k_z$ and the corresponding quasiparticle spectra
$E_{\vec{k}\uparrow}$, $E_{\vec{k}\downarrow}$ (inset) for $d=3$, $n=0.1$, $r=5$
$U=-12$. (a) $h=-3.5$ -- on the LP side, SC$_0$ phase; (b) $h=-1.1$ -- on the LP
side, SC$_M$ phase; (c) $h=0.1$ -- strongly correlated superconductor, SC$_M$
phase; (d) $U=-7$, $h=-2.28$  -- unstable SC$_M$ phase (PS region).}
\end{figure}

It is interesting that the combination of $h\neq 0$ and higher values of $r$
widens the range of occurrence of the SC$_M$ phase and the PS-III region
disappears in the crossover phase diagrams, which is clearly visible in Fig.
\ref{n_01_r_P}. If $r=2$, the SC$_M$ is stable but limited by the two critical
values of the attractive interaction: $|U|_{c}^{SC_M}\approx 10.08$ and 
$|U|_{c}^{SC_M}\approx 64.86$. There is only the first order phase transition
from SC$_M$ to the fully polarized normal state, through PS-III. For higher
hopping asymmetry of $r$ (Fig. \ref{n_01_r_P}(b)-(d)), the region of SC$_M$
stability is
much larger. As mentioned above, the region in which the SC$_M$ phase and
the
NO-II state separate spatially (PS-III) is unstable with regard to the pure
spatially homogeneous magnetized superfluid. Therefore, there is the second
order phase transition from the SC$_M$ to the NO-II phase in the whole range of
SC$_M$ occurrence. The Lifshitz type quantum critical point (red color) moves
towards lower $|U|$, with increasing $r$ and even for relatively
low hopping imbalance ($r=5$), the SC$_M$ phase is stable in the
intermediate couplings region. However, the position of the tricritical point
does not move with increasing $r$, starting from some definite value of $r
\approx 5$. Moreover, it is worth mentioning that the BP-II phase is unstable in
the whole range of parameters in 3D, as well as in the 2D case. To confirm this
hypothesis, let us analyze in detail the behavior of the superconducting
solutions for different fixed parameters. For this purpose, we show in Fig.
\ref{n_k_E_k_r5_n01} the plots of the momentum occupation numbers vs. $k\equiv
k_x=k_y=k_z$ and the corresponding quasiparticle spectra $E_{\vec{k}\uparrow}$,
$E_{\vec{k}\downarrow}$ for fixed $r=5$, $n=0.1$ and different values of $h$ and
$|U|$. For $U=-12$ and $h=-3.5$, the system is in the unpolarized
superconducting state on the LP side. Therefore, there is no Fermi surface and
both of the quasiparticle branches are gapped, as shown in Fig.
\ref{n_k_E_k_r5_n01}(a). For higher values of magnetic field, on
the LP side, there is a continuous transition from the SC$_0$ to the SC$_M$
state. There is one Fermi surface which comes from the excess of spin-up
fermions in the SC$_M$ phase with the gapless spectrum (Fig.
\ref{n_k_E_k_r5_n01}(b)-(c)). These unpaired fermions occupy the region around
momentum $k=0$.
\begin{figure}[t!]
\begin{center}
\includegraphics[width=0.55\textwidth,angle=270]{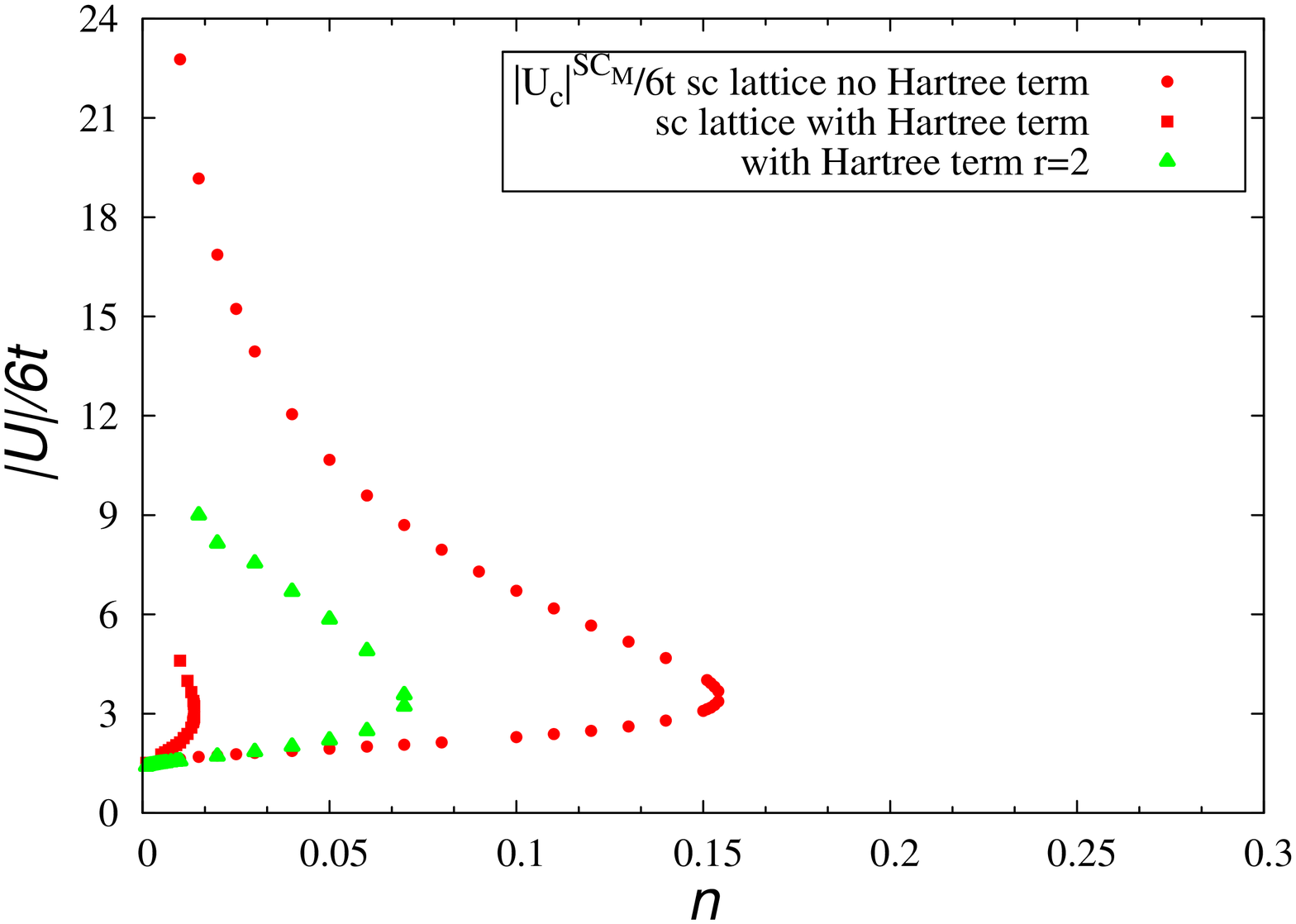}
\caption[Critical values of the attraction for which the SC$_M$ state becomes
stable at $T=0$ vs. electron concentration. A comparison of two cases: $r=1$
vs. $r=2$ with and without the Hartree term.]{\label{magnetized_r_neq_1}
Critical values of the attraction for which the SC$_M$ state becomes stable at
$T=0$ vs. electron concentration. A comparison of two cases: $r=1$ vs. $r=2$
with and without the Hartree term.
$h_c^{SC_M}=\sqrt{(\bar{\mu}-\epsilon_0)^2+|\Delta|^2}-D\frac{r-1}{r+1}$, where
$\Delta=\Delta(h=0)$.}
\end{center}
\end{figure}

We also analyze a superconducting solution from
the phase separation region, for an intermediate value of coupling (Fig.
\ref{n_k_E_k_r5_n01}(d)). 
A solution of this type (interior gap superfluidity or the BP-II phase) was
analyzed in \cite{Wilczek}. We show that the occurrence of such a solution is
possible, but it is energetically unstable. As mentioned above, the BP-II
state
can have excess of fermions with two FS’s at $T=0$ and a gapless spectrum for
the majority spin species. For $r=t^{\uparrow}/t^{\downarrow}=5$, the heavier
fermions are the ones with spin-down, but the majority spin species are the
spin-up fermions. As shown in Fig. \ref{n_k_E_k_r5_n01}(d), the unpaired
fermions occupy the state around the Fermi momentum, in contrast to
the BP-I
case. The energy spectrum for the lighter fermions (the majority spin species)
is gapless (Fig. \ref{n_k_E_k_r5_n01}(d) -- inset). There are two zeros in the
plot of $E_{\vec{k}\uparrow}$.

Another important aspect of this section is the influence of hopping
imbalance on the SC$_M$ phase stability in the presence of the Hartree term. As
discussed in Chapter \ref{chapter5}, the occurrence of the BP-I phase depends on
the lattice
structure, i.e. if $r=1$, SC$_M$ is unstable for $d=2$ but it can be realized
for $d=3$ lattices. However, in the AHM, the very existence of the BP-I phase is
restricted to low fillings. The Hartree term, usually promoting ferromagnetism
in the Stoner model ($U>0$), here ($U<0$) strongly competes with
superconductivity. Thus, such term restricts the SC$_M$ state to lower
densities as shown in Figs. \ref{magnetized_r1} and also
\ref{magnetized_r_neq_1}. If $r=1$, the critical value of $n$ above which the
SC$_M$ state becomes unstable equals $n\approx 0.0145$ for the simple cubic
lattice case. However, the mass imbalance can change this behavior because
of spin
polarization stemming from the kinetic energy term. In this way, SC$_M$ can be
realized for the intermediate and strong coupling regimes for higher values of
$n$, even in the presence of the Hartree term. Then, the hopping imbalance
increases a critical value of $n$ above which the SC$_M$ state becomes unstable
and for $r=2$ $n\approx 0.07$, as shown in Fig. \ref{magnetized_r_neq_1}. 
\textcolor{czerwony}{For each value of $n$, there are two critical values of the
attraction for which the $SC_M$ state becomes stable
(except for the very dilute limit, where there is only the lower critical value,
 i.e. the upper critical value becomes infinite).
The system is in the $SC_M$ phase between
the lower and upper critical points in this plot.}


\chapter*{Conclusions and prospects}
\addcontentsline{toc}{chapter}{Conclusions and prospects}

The main aim of this thesis was to investigate the influence of the Zeeman
magnetic field on superconducting properties and the BCS-BEC crossover in the
systems with local fermion pairing. An extensive analysis of the
superconducting properties of the spin polarized Hubbard model, with on-site
attractive interaction $|U|$ (s-wave pairing symmetry), both for the square (2D)
and the simple cubic (3D) lattices, has been performed. We have also studied
the stability of the superfluid phases with the pure d-wave pairing symmetry in
2D, within the extended Hubbard model with \textcolor{czerwony}{nearest neighbor}
attractive interaction $|W|$ in the presence of \textcolor{czerwony}{a} Zeeman
magnetic field. In
addition, we have analyzed in detail the phase diagrams and BCS-BEC crossover
\textcolor{czerwony}{for} the case of spin dependent hopping integrals ($t^{\uparrow}\neq
t^{\downarrow}$)
within the spin polarized attractive Hubbard model, both for the 2D and 3D.

\section*{The main results}             

\begin{itemize}
 \item We have investigated the influence of the Zeeman magnetic field on the
superfluid characteristics of the attractive Hubbard model, both in the weak
coupling ($|U|<2zt$, where $z$ -- the number of nearest neighbors) and in the
strong coupling limit ($|U|\gg t$). The ground state phase diagrams have been
obtained in the cases of a fixed chemical potential ($\mu$) and a fixed electron
concentration, within the Hartree-Fock approximation. This method includes
the spin-dependent Hartree term and can be called a BCS-Stoner approach. If the
number of particles is fixed and $n\neq 1$, one obtains two critical Zeeman
magnetic fields, which limit the phase separation of the superconducting (SC)
and the normal (NO) states.  
 Superconductivity is destroyed by the pair breaking in a very weak coupling
regime. On the other hand, in the intermediate or strong coupling regimes and
$d=2$, the transition from the unpolarized
($P=0$) superconducting (SC$_0$) to the spin
polarized normal state (NO-I/NO-II -- partially ($P\neq 0$)/fully ($P=1$)
polarized NO states) goes in addition through phase separation. Therefore,
for the 2D square lattice and spin independent hopping integrals
($t^{\uparrow}=t^{\downarrow}$), we find no stable homogeneous polarized
superfluid state for the strong attraction and obtain that for two-component
Fermi system on a 2D lattice with population imbalance, the phase separation is
energetically favorable for a fixed particle concentration, even on the LP (BEC)
side.
The ground state phase diagrams in the ($h-|U|$) and ($P-|U|$) planes which have
been constructed without the Hartree term in the $d=2$ and $d=3$ case for
low electron
concentration, agree well with those obtained in the framework of the continuum
model of a dilute gas of fermions in $d=2$ and $d=3$.

\item We have also investigated the ground state BCS-BEC crossover diagrams in
the presence of the Zeeman magnetic field in 3D, for the simple cubic lattice.
As opposed to the $d=2$ case, for strong attraction and in the dilute limit,
a homogeneous magnetized superconducting phase (SC$_M$ (or breached pairing
(BP)))
has been found in the phase diagrams. The SC$_M$ phase is a specific superfluid
state consisting of a coherent mixture of LP's (hard-core bosons)  and excess
spin-up fermions (Bose-Fermi mixture). This state can only have one Fermi
surface (FS), hence can be called BP-1. Therefore, the occurrence of the BP-1
phase depends on the lattice structure, i.e. if $t^{\uparrow}=t^{\downarrow}$,
SC$_M$ is unstable for $d=2$ but it can be realized for $d=3$ lattices.  
We have also found a topological quantum phase transition (Lifshitz type) from
the unpolarized superfluid phase to SC$_M$ and tricritical points in the ground
state phase diagrams. Analysis of the influence of the Hartree term on
the BCS-BEC crossover diagrams in a magnetic field shows that the presence
of such a term restricts the range of occurrence of the SC$_M$ phase. The
critical electron concentration ($n_c$) above which the SC$_M$
state becomes unstable against the phase separation region has been estimated
(with the Hartree term $n_c=0.0145$, without the Hartree term $n_c=0.154$).  The
Hartree term, usually promoting ferromagnetism in the Stoner model ($U>0$), here
($U<0$) strongly competes with superconductivity. We have analyzed the influence
of different lattice
geometries on the stability of the SC$_M$ phase. The
body-centered cubic (BCC) and face-centered cubic (FCC) lattices have been
taken into account. We have also constructed the BCS-BEC crossover phase
diagrams at $T=0$ with the use of the semicircular density of states, which
can be realized as the Bethe lattice. 
For the latter, in the case without the
Hartree term, $n_c$ is the lowest and equals $n_c\approx 0.054$, while for the
BCC lattice $n_c\approx 0.199$ and the
highest for the FCC lattice $n_c\approx 0.212$. 

\item The finite temperature phase diagrams of \textcolor{czerwony}{AHM} with
magnetic field 
have been obtained (both at fixed
$\mu$ and fixed $n$) for 2D and 3D, including an analysis of the role
of the Hartree term. In $d=2$ ($h=0$), the transition from the superconducting
to the normal phase is of the Kosterlitz-Thouless (KT) type from the
topologically ordered to the non-ordered state. Below the critical temperature
$T_c^{KT}$, such topological ordering is manifested by the existence of
tightly bound vortex-antivortex pairs, which are broken by thermal fluctuations
when temperature increases above $T_c^{KT}$. The phase transition is
characterized by a universal jump of the superfluid density ($\rho_s$) at
$T_c^{KT}$. In this way, we have partly included the effects of phase
fluctuations at $h\neq 0$ in our calculations and we have estimated the KT
critical temperatures.      

\item The ($T-h$) and ($T-P$) phase diagrams have been obtained for the
2D and 3D Hubbard model with $U<0$ in the Zeeman magnetic field, in the weak
coupling regime.
 In the weak coupling regime and for fixed $n$, the following states have been
found in 2D: the quasi-superconducting state qSC (below $T_c^{KT}$), region of
pairs without the phase coherence (bounded by the Hartree temperature), the
spin
polarized region with a gapless spectrum for the majority spin species, the PS
region and the normal state. We have also found an energetically stable region
with $\rho_s<0$, both for the $d=2$ and the $d=3$ case, in the diagrams with the
Hartree term. It suggests the existence of a stable FFLO state in $d=3$, in
the
weak coupling limit. As far as the role of the Hartree term is concerned, we
have found that the Hartree term leads to an increase in the
Chandrasekhar-Clogston limit. 

\item We have investigated the influence of the Zeeman magnetic field on the
superfluid characteristics of the extended Hubbard model with $W<0$ and
$t^{\uparrow}=t^{\downarrow}$. We have analyzed the pure d-wave pairing symmetry
case. 
At $T=0$, in the presence of the magnetic field, the ground state is the
spatially homogeneous spin-polarized superfluid state, which has a gapless
spectrum for the majority spin species and two FS (BP-2), for weak attraction,
as opposed to the s-wave pairing symmetry case in 2D.
We have also extended our analysis to finite temperatures in $d=2$ by invoking
the KT scenario. 
At finite temperatures, in the weak coupling regime and for fixed $\mu$, the
following states have been found in the 2D system: at $h=0$ -- the SC$_0$ phase;
at $T=0$, $h\neq 0$ -- polarized superfluid state with a gapless spectrum for
the majority spin species; at $T>0$ -- qSC$(P\neq 0)$ (below $T_c^{KT}$); region
of pairs without coherence (below the Hartree temperature); the PS region and
NO. PS terminates at the mean-field tricritical point, in ($T-P$) phase
diagrams. 

\item We have also studied in detail the mass imbalance case in the
spin
polarized attractive Hubbard model with spin dependent hopping integrals
($t^{\uparrow}\neq t^{\downarrow}$, $t^{\uparrow}/t^{\downarrow}\equiv r$).
The BCS-LP crossover diagrams in the presence of the Zeeman magnetic field for
$d=2$ and $t^{\uparrow}\neq t^{\downarrow}$ exhibit a novel behavior. As opposed
to the $t^{\uparrow}= t^{\downarrow}$ case, for strong attraction, SC$_M$ occurs
at $T=0$. In general, the solutions of these type (Sarma-type with
$\Delta (h)$)
appear (for $r> 1$) when $h>(\frac{r-1}{r+1})\bar{\mu}+2\Delta
\frac{\sqrt{r}}{r+1}$ (on the BCS side) or when 
$h>\sqrt{(\bar{\mu}-\epsilon_0)^2+|\Delta |^2}-D\frac{r-1}{r+1}$ (on the LP
side). If $r\neq 1$, the SC$_M$ phase (or BP-2 phase) is unstable in the weak
coupling regime at $T=0$, but BP-1 can be stable in the strong coupling LP
limit,
both for $d=2$ and $d=3$.
The influence of the Hartree term on the SC$_M$ state stability has been
investigated in the hopping asymmetric case. If $t^{\uparrow}\neq
t^{\downarrow}$
(see above) such a term restricts the SC$_M$ state to lower densities. However,
the mass imbalance can change this behavior even for $d=2$ due to spin
polarization stemming from the kinetic energy term. 


\item We have used the strong coupling expansion \textcolor{czerwony}{to map AAHM onto the pseudospin model} and determine the critical
value of particle concentration above which the SC phase can coexist with
charge ordering (CO) at $h=0$ and $r\neq 1$. SC/CO is a region of the phase
separation with \textcolor{czerwony}{domain} of SC and \textcolor{czerwony}{domain} of CO (with $n=1$).

\item  We have also extended the analysis of the crossover to finite
temperatures in $d=2$ by invoking the KT scenario. The KT transition
temperatures are much lower than those determined in the BCS
scheme. Moreover, spin polarization has a strong destroying influence on the KT
superfluid state at $r=1$ and allows this phase in the weak coupling regime, in
agreement with the results for the continuum case. In the strong coupling limit,
 $T_c^{KT}$ does not depend on magnetic field (below $h_{c1}$), but it depends
on mass imbalance and its upper bound takes the form:
$k_{B}T_c^{KT}=2\pi\frac{r}{(1+r)^2}\frac{t^2}{|U|}n(2-n)$ ($r>0$). For
$k_BT<<|U|$, only unbroken LP's exist, which can form unpolarized qSC below
$T_c^{KT}$ or remain phase disordered. The system is equivalent to that of
a hard-core Bose gas on a lattice.
If $r\neq 1$ in $d=2$, a spin polarized KT superfluid state can be stable even
in the intermediate and strong coupling region.

\item The BCS-BEC crossover has also been studied at finite temperatures in
the spin polarized Hubbard model with $U<0$ ($t^{\uparrow}=t^{\downarrow}$), for
$d=3$, going beyond the standard mean field approximation. The critical
temperatures of the superconducting transition have been determined within the
self-consistent T-matrix approach. We have performed a comparison of the
results
obtained from the $(GG_0)G_0$ and $(GG)G_0$ schemes, both for the 3D continuum
model with contact attraction and the AHM at fixed low electron concentration.
At finite temperatures, at $h=0$, the following states have been found in the
3D system: the SC/SF phase (below $T_c$), the pseudogap region (above the
temperature of the pair condensation $T_c$ and below the temperature of the pair
formation $T_p$) and normal state (above $T_p$). In the pseudogap region there
are long-lived, incoherent pair excitations.
The calculations within the $(GG_0)G_0$ and
the $(GG)G_0$ schemes give in a dilute limit very similar results in the
two extreme limits: BCS
and BEC (LP). However, the critical temperature at the unitarity,
determined within the $(GG_0)G_0$ scheme is higher than that obtained within the
$(GG)G_0$ scheme, which is more consistent with the Quantum Monte Carlo
results. 
We have shown that there are large differences between the results obtained
within AHM and those obtained
within the 3D continuum model in the BEC limit. The reason is that in
the LP limit
for AHM, the effective mass of the hard-core bosons increases with $|U|$ and
this behavior is reflected in the results of the T-matrix calculations.

\item We have also obtained the temperature BCS-BEC crossover phase diagrams
for $d=3$, at $h\neq 0$. The critical temperature of the superconducting
transition has been determined within the $(GG_0)G_0$ T-matrix scheme in the
presence of
the Zeeman magnetic field. The interesting result \textcolor{czerwony}{is that} \textcolor{czerwony}{a} spin
polarized superfluid state with a gapless
region for the majority spin species can be stable in the strong coupling
regime, as opposed to the 2D system for $r=1$.

\end{itemize}
  
\section*{Prospects}
Let us finish by pointing to possible directions of extending the work presented
in this thesis:

 \begin{itemize}
\item In this work, we have restricted the analysis to the s-wave and d-wave
pairing only with $\vec{q}=0$. We have also used the strong coupling expansion
to determine the critical
value of particle concentration above which the SC phase can coexist with
charge ordering (CO) at $h=0$ and $r\neq 1$. However, an investigation of the
competition between the superconducting phases, CDW diagonal ordering and
unconventional spin density wave (SDW) \cite{Tamaki}, in particular within the
spin polarized Hubbard model, would be an interesting extension.

\item We have not considered non-homogeneous states such as FFLO, which are
possible in weak to intermediate attraction range, although much more
susceptible to phase fluctuations at finite $T$ in a 2D system \cite{Matsuda,
Shimahara}.

\item In this thesis, we have shown that the BP-2 state is unstable \textcolor{czerwony}{in the ground state} in the whole
range of parameters, within one-band spin polarized AHM, both \textcolor{czerwony}{on} \textcolor{czerwony}{$d=2$ square lattice and $d=3$ simple cubic lattice}. 
\textcolor{czerwony}{Such a phase can be stable in d-wave pairing symmetry case}.  
\textcolor{czerwony}{One can suppose that the Liu-Wilczek phase can be realized in
s-wave case, in two-band model}.

\item \textcolor{czerwony}{Continuation of our study of the AAHM by strong coupling expansion method would be of interest. In particular, this concerns a description of the low density limit in terms of the (hard-core) bosons and fermions mixture.} 

\item Investigation of the influence of \textcolor{czerwony}{a} Zeeman magnetic
field on the
superfluid properties of the model of a mixture of mutually interacting bound
electron pairs and itinerant fermions (boson-fermion model) \cite{MicnasModern,
Micnas2007, Krzyszczak, Krzyszczak2} is an \textcolor{czerwony}{important} extension
of the
analysis presented in this work. The boson-fermion \textcolor{czerwony}{(BF)} model can be applied to the
high-$T_c$ superconductivity and to description of the
superfluidity in ultracold gases near the Feshbach resonance. 
\textcolor{czerwony}{Our preliminary study indicated that 
in a certain limit of parameters and for the dilute case the properties of the BF model in Zeeman field closely follow those of the spin polarized attractive Hubbard model \cite{Micnas+AC}. }

\item In this thesis, the impact of the Zeeman magnetic field on superfluidity
has been investigated. Taking into consideration the orbital pair breaking
mechanism \cite{Maska} would also make an important extension of the
analysis presented in this work.   
   
 \end{itemize}

\newpage
\thispagestyle{empty}
\mbox{}

\chapter*{Streszczenie w języku polskim pracy doktorskiej pt. \emph{Wpływ pola magnetycznego na własności nadprzewodzące i przejście BCS-BEC w układach z lokalnym parowaniem fermionów}}
\addcontentsline{toc}{chapter}{Streszczenie w języku polskim pracy doktorskiej
pt. \emph{Wpływ pola magnetycznego na własności nadprzewodzące i przejście
BCS-BEC w układach z lokalnym parowaniem fermionów}}

Niekonwencjonalne nadprzewodnictwo z nietrywialnym mechanizmem parowania
w u\-kła\-dach silnie skorelowanych elektronów oraz spinowo spolaryzowana
nadciekłość (w kontekście ultrazimnych atomowych gazów fermionowych) są obecnie
intensywnie badane oraz szeroko dyskutowane w wiodącej literaturze światowej.
Stanowią one jedne z najbardziej aktualnych kierunków badań w zakresie fizyki
ciała stałego i ultrazimnych gazów kwantowych. Ogromny postęp technik
eksperymentalnych w ultrazimnych gazach fermionowych z dostrajalnym
oddziaływaniem przyciągającym (poprzez rezonanse Feshbacha) pozwala badać
własności różnych egzotycznych stanów materii realizowanych w tych układach, w
tym fizykę przejścia BCS-BEC. Realizowane są także eksperymenty, w których gazy
fermionowe (lub bozonowe) umieszcza się na sieciach optycznych. Zarówno
głębokość periodycznego potencjału pułapkującego jak i geometria mogą być w
pełni kontrolowane. Dzięki temu można badać układy silnie skorelowane o różnej
geometrii sieci. Gazy atomowe z dostrajalnym oddziaływaniem na sieciach
optycznych umożliwiają nowe eksperymentalne realizacje modeli Hubbarda.

Własności układu fermionów w ekstremalnych granicach BCS oraz BEC są bardzo
różne, zwłaszcza w stanie normalnym.
\\
Reżim BCS charakteryzuje się tym, że:
\begin{itemize}
 \item pary tworzą się i kondensują w tej samej temperaturze ($T_c$),
 \item parowanie zachodzi w przestrzeni odwrotnej, a w tworzeniu par Coopera
uczestniczy tylko niewielka liczba elektronów o energiach bliskich energii
Fermiego,
 \item oddziaływanie pomiędzy fermionami ($|U|$) jest słabe,
 \item rozmiar par w kondensacie jest znacznie większy niż średnia odległość
pomiędzy nimi (stąd silne przekrywanie),
 \item szczelina energetyczna maleje monotonicznie ze wzrostem $T$ i zanika w $T_c$,
 \item temperatura krytyczna oraz termodynamika określone są wzbudzeniami
jednocząstkowymi -- rozerwanymi parami Coopera, z eksponencjalnie małą
szczeliną,
 \item powyżej temperatury krytycznej stan normalny opisywany jest teorią
cieczy Fermiego. 
\end{itemize}

Z kolei granica BEC (lub par lokalnych) charakteryzuje się tym, że:
\begin{itemize}
 \item tworzenie się par następuje w temperaturze $T_p$, znacznie wyższej od
temperatury krytycznej, w której pojawia się dalekozasięgowa koherencja fazowa i
przejście do stanu nadprzewodzącego,
\item parowanie zachodzi w przestrzeni rzeczywistej i sparowane są wszystkie elektrony,
\item oddziaływanie pomiędzy fermionami jest silne,
\item pary są znacznie mniejsze niż średnia odległość pomiędzy nimi,
\item energia wiązania par jest proporcjonalna do $|U|$, pary istnieją powyżej
$T_c$, aż do $T_p$,
\item temperatura krytyczna oraz termodynamika są określone przez mody kolektywne,
\item stan normalny jest opisywany cieczą Bosego silnie związanych i
nieskorelowanych fazowo par (w zakresie $T_c<T<T_p$). 
\end{itemize}

Dla pośrednich sprzężeń stan normalny może wykazywać pseudo-szczelinę (ang.
\emph{pseudo-gap}) i odchylenia od standardowej teorii Landaua cieczy Fermiego.

Efekty przejścia BCS-BEC są bardzo wyraźnie widoczne w zachowaniu potencjału
chemicznego $\mu$ w temperaturze $T=0$.
W granicy słabego sprzężenia $\mu=E_F$ ($E_F$ -- energia Fermiego) i obowiązuje
standardowa teoria BCS.
Jednak przy odpowiednio silnym sprzężeniu potencjał chemiczny zaczyna maleć, aż
w końcu spada do zera i staje się ujemny w granicy BEC.
Punkt $\mu=0$ można uznać za punkt przejścia BCS-BEC, ale ponieważ w jego
okolicach zachowanie układu nie przypomina ani granicy BCS, ani BEC, więc możemy
mówić o całym obszarze przejścia, a nie tylko jednym punkcie.
Gdy $\mu>0$, to w układzie występuje powierzchnia Fermiego i mamy do czynienia z
kondensatem par Coopera, natomiast gdy $\mu$ staje się ujemne, powierzchnia
Fermiego znika i mówimy o kondensacie bozonowym par.

W 2003 roku trzy grupy eksperymentalne zaobserwowały kondensację par atomów
fermionowych w regionie przejścia BCS-BEC. Grupa M. Grimma z Uniwersytetu w
Insbruku oraz W. Ketterlego z MIT zastosowała atomy $^{6}$Li. Grupa D. S. Jin z
Los Angeles schłodziła pułapkowany gaz atomów $^{40}$K do odpowiednio
niskich temperatur rzędu $5\cdot 10^{-8}$K. Następnie, poprzez rezonans
Feshbacha, kontrolowano oddziaływanie międzyatomowe. Na podstawie pomiarów, przy
użyciu spektroskopii radiowej wykazano pojawienie się przerwy energetycznej, rok
później ostatecznie potwierdzono istnienie fazy nadciekłej obecnością wirów.

Kolejne prace grup eksperymentalnych z MIT oraz z Uniwersytetu Rice w roku 2005
zapoczątkowały badania kwantowych gazów fermionowych ($^{6}$Li), o różnej
liczbie fermionów ze ,,spinem w górę'' ($\uparrow$) oraz ,,w dół''
($\downarrow$) (ang. \emph{systems with population imbalance}). W trójwymiarowym
układzie atomów profile gęstości obserwowane w eksperymentach pokazują
niespolaryzowany nadciekły rdzeń w centrum pułapki harmonicznej oraz otaczający
go gaz atomów w stanie normalnym. W tym układzie istnieje również rejon
separacji fazowej pomiędzy niespolaryzowanym stanem nadciekłym i spolaryzowanym
stanem normalnym.

Możliwość realizacji takich silnie skorelowanych układów fermionowych w
laboratorium stanowi motywację zarówno dla grup teoretycznych jak i
eksperymentalnych do prowadzenia badań nad fizyką przejścia BCS-BEC. Umożliwiło
to także badanie wpływu Zeemanowskiego pola magnetycznego na własności
nadciekłe.
 
Obecność pola magnetycznego Zeemana ($h$) powoduje rozszczepienie powierzchni
Fermiego na dwie, odpowiadające cząstkom ze spinem w górę i w dół. Przejściu z
fazy nadprzewodzącej do spinowo spolaryzowanego stanu normalnego towarzyszy skok
parametru porządku i mamy wówczas do czynienia z przejściem pierwszego rodzaju,
zachodzącym dla uniwersalnej wartości krytycznego pola magnetycznego
$h_c=\Delta_0 /\sqrt{2}$ (granica Clogstona-Chandrasekhara), gdzie $\Delta_0$
jest przerwą energetyczną w $T=0$ oraz $h=0$. W granicy słabego sprzężenia, przy
znacznej nierównowadze cząstek, mogą istnieć stany, charakteryzujące się
nietrywialnym mechanizmem parowania. Jednym z nich jest stan FFLO (Fulde-Ferrell
i Larkin-Ovchinnikov), w którym pary Coopera w kondensacie mają niezerowy pęd
całkowity. Istnieją uzasadnione przypuszczenia, że stan FFLO może realizować się
w nadprzewodnikach ciężkofermionowych. Jednak ze względu na niszczący wpływ
efektu orbitalnego na nadprzewodnictwo, zaobserwowanie tego stanu jest niezwykle
trudne w związkach nadprzewodzących. Istnieje jednak szansa na realizację
eksperymentalną fazy FFLO w ultrazimnych gazach fermionowych z dostrajalnym
oddziaływaniem przyciągającym. 

Innym możliwym rodzajem parowania i koherencji fazowej jest jednorodny
przestrzennie spinowo spolaryzowany stan nadciekły z parami o zerowym pędzie
całkowitym (w literaturze określany jako breached pairing (BP) lub faza Sarmy), 
charakteryzujący się bezszczelinowym widmem dla większościowego kierunku spinu. 
 
Głównym celem pracy doktorskiej było zbadanie wpływu pola magnetycznego
Zeemana na własności nadprzewodzące i przejście BCS-BEC w układach z lokalnym
parowaniem fermionów. Przeprowadzono kompletną analizę własności
nadprzewodzących spinowo spolaryzowanego modelu Hubbarda z jednowęzłowym
oddziaływaniem
przyciągającym (symetria parowania typu $s$) w przypadku 2D (sieć
kwadratowa) oraz 3D (sieć prosta kubiczna). Zbadano także wpływ
symetrii parowania typu $d$ na stabilność faz nadciekłych w 2D w
rozszerzonym modelu Hubbarda z oddziaływaniem międzywęzłowym $|W|$, w obecności
pola magnetycznego. Rozważyliśmy też szczegółowo nadprzewodnictwo w układach
ze spinowo zależnymi całkami przeskoku ($t^{\uparrow}\neq t^{\downarrow}$) w
ramach spinowo spolaryzowanego modelu Hubbarda w 2D i 3D.

Główne rezultaty pracy są następujące.
\begin{itemize}
\item Zbadano wpływ pola magnetycznego na własności nadciekłe modelu Hubbarda z
przyciąganiem na węźle, zarówno w granicy słabego ($|U|<2zt$, gdzie $z$ --
liczba sąsiadów, $t$ -- całka przeskoku), jak i silnego sprzężenia ($|U|\gg t$).
W ramach przybliżenia pola średniego (BCS-Stoner) otrzymano diagramy fazowe
stanu podstawowego w przypadku ustalonego potencjału chemicznego oraz ustalonej
liczby cząstek. W tym drugim przypadku, dla koncentracji elektronów ($n$) różnej
od 1, otrzymano na diagramach fazowych dwa krytyczne pola Zeemana, które
wyznaczają obszar separacji fazowej pomiędzy fazą nadprzewodzącą (SC) i stanem
normalnym (NO). 
W granicy słabego sprzężenia, nadprzewodnictwo jest niszczone
poprzez rozrywanie par. 
Z drugiej strony, w granicy silnego sprzężenia oraz w reżimie
przejścia w 2D, przejście od niespolaryzowanego ($P=0$) nadprzewodnictwa
(SC$_0$) do spinowo spolaryzowanego stanu normalnego (NO-I/NO-II --
częściowo ($P\neq 0$)/całkowicie ($P=1$) spolaryzowane stany NO) następuje
poprzez obszar separacji fazowej. Stąd, w przypadku 2D sieci
kwadratowej i spinowo niezależnych całek
przeskoku ($t^{\uparrow}=t^{\downarrow}$), nie istnieje stabilny energetycznie
spolaryzowany stan nadciekły w przypadku silnego przyciągania, lecz
korzystne energetycznie jest utworzenie obszaru sepracji fazowej (przy
ustalonej liczbie cząstek) nawet po stronie LP (BEC).
Skonstruowane bez członu Hartree diagramy fazowe w płaszczyznach ($h-|U|$) i
($P-|U|$), przy niskich koncentracjach elektronów w 2D i 3D bardzo dobrze
zgadzają się z rezultatami dla modelu rozrzedzonego gazu fermionów w
continuum. 
\item Zbadane zostały także diagramy fazowe stanu podstawowego przejścia
BCS-BEC w obecności pola magnetycznego Zeemana dla 3D sieci prostej
kubicznej. W przeciwieństwie do przypadku 2D, dla silnego przyciągania i w
granicy niskiej koncentracji elektronów, realizuje się jednorodny spinowo
spolaryzowany stan nadciekły (SC$_M$ (ang. \emph{breached pairing} (BP))). 
Jest to specyficzny stan nadciekły będący koherentną
mieszaniną silnie związanych par lokalnych (bozonów z twardym rdzeniem) oraz
nadwyżką fermionów ze spinem w górę.
Taki stan może posiadać tylko jedną powierzchnię Fermiego (FS), stąd nazywany
jest BP-1.
Występowanie stanu BP-1 zależy od wymiaru i geometrii sieci, tzn.
dla $t^{\uparrow}=t^{\downarrow}$, faza SC$_M$ jest niestabilna dla 2D, ale
występuje w przypadku sieci 3D.
W układzie zaobserwowano także topologiczne kwantowe przejście fazowe typu
Lifshitza, od niespolaryzowanej fazy nadciekłej do stanu SC$_M$ i wyznaczono na
diagramach fazowych punkty trójkrytyczne.
Przeanalizowano także wpływ członu Hartree na diagramy opisujące przejście
BCS-BEC w polu magnetycznym i pokazano, że
obecność tego członu bardzo ogranicza zakres występowania fazy SC$_M$.
Oszacowano krytyczną koncentrację nośników, powyżej
której spinowo spolaryzowany stan nadciekły staje się niestabilny względem
separacji fazowej (z uwzględnieniem członu Hartree $n=0.0145$, bez członu
Hartree $n=0.154$).
Człon Hartree, który zwykle faworyzuje ferromagnetyzm w modelu Stonera ($U>0$),
w rozważanym przypadku ($U<0$) silnie współzawodniczy z
nadprzewodnictwem.
Przeanalizowano również wpływ różnych geometrii sieci na stabilność fazy SC$_M$
(sieć przestrzennie centrowana (BCC) oraz \textcolor{czerwony}{powierzchniowo} centrowana (FCC)).
Skonstruowano również diagramy fazowe przejścia BCS-BEC w $T=0$ używając
półkolistej (ang. \emph{semicircular}) gęstości stanów, która może być
zrealizowana jako tzw. sieć Bethego. 
Dla tej ostatniej, w przypadku bez członu Hartree, $n_c$ jest najniższa i
wynosi $n_c\approx 0.054$, podczas gdy dla sieci BCC $n_c\approx 0.199$, a
dla sieci FCC jest najwyższa $n_c\approx 0.212$. 

\item Otrzymano temperaturowe diagramy fazowe (zarówno przy ustalonym
$\mu$, jak i $n$) dla 2D i 3D, zarówno z członem Hartree, jak i bez
niego. 
W $d=2$ ($h=0$) przejście ze stanu nadprzewodzącego do normalnego jest
typu Kosterlitza-Thoulessa (KT) ze stanu topologicznie uporządkowanego do stanu
nieuporządkowanego. Poniżej temperatury krytycznej takie uporządkowanie
manifestuje się istnieniem ściśle zwią\-za\-nych par wir-antywir, które są rozrywane
przez fluktuacje termiczne, gdy temperatura wzrasta powyżej temperatury
krytycznej. Przejście fazowe objawia się uniwersalnym skokiem gęstości składowej
nadciekłej ($\rho_s$) w temperaturze $T_c^{KT}$. Uwzględniono \textcolor{czerwony}{częściowo} w
obliczeniach
fluktuacje fazowe dla $h\neq 0$ i oszacowano temperatury $T_c^{KT}$. 

\item Skonstruowano diagramy temperaturowe (w płaszczyźnie ($T-h$) and
($T-P$)) dla modelu Hubbarda z $U<0$ w polu magnetycznym w granicy słabego
sprzężenia na sieci dwu- i trójwymiarowej. W zakresie słabego przyciągania, w 2D
znaleziono następujące stany przy ustalonym $n$: stan kwazi-nadprzewodzący qSC
(poniżej temperatury $T_c^{KT}$), obszar par bez koherencji fazowej (poniżej
temperatury Hartree-Focka), spinowo spolaryzowany obszar z bezszczelinowym
widmem dla cząstek ze spinem w górę, region separacji fazowej oraz stan
normalny. 
Zaobserwowano również występowanie energetycznie stabilnego obszaru diagramów
fazowych z formalnie ujemną gęstością składowej nadciekłej ($\rho_s<0$) w 2D i 3D, na diagramach z
członem Hartree. Sugeruje to istnienie stabilnego stanu FFLO w 3D w granicy
słabego sprzężenia. 
Uwzględnienie członu Hartree prowadzi do podwyższenia
granicy Chan\-drasekhara-Clogstona.

\item Zbadano wpływ pola magnetycznego na własności nadprzewodzące
rozszerzonego modelu Hubbarda z $W<0$ i $t^{\uparrow}=t^{\downarrow}$.
Rozważono przypadek czystej symetrii parowania typu $d$. 
W $T=0$, w obecności pola magnetycznego, stanem podstawowym jest przestrzennie
jednorodny, spinowo spolaryzowany stan nadciekły, który ma bezszczelinowe
widmo dla większościowego kierunku spinu i dwie
powierzchnie Fermiego (BP-2), dla przypadku słabego przyciągania, w
przeciwieństwie do przypadku symetrii parowania typu $s$ w 2D.
Rozszerzono także analizę do skończonych temperatur w 2D, wykorzystując
scenariusz KT. W skończonych temperaturach, w reżimie słabego sprzężenia,
w 2D występują następujące fazy w przypadku ustalonego $\mu$: dla $h=0$ --
faza SC$_0$; w $T=0$, $h\neq 0$ -- spolaryzowany stan nadciekły z
bezszczelinowym widmem dla większościowego kierunku spinu; w $T>0$ -- qSC$(P\neq
0)$ (poniżej $T_c^{KT}$); region
par bez koherencji fazowej (poniżej temperatury Hartree); region PS i
NO. PS kończy się w punkcie trójkrytycznym wyznaczonym w przybliżeniu
pola średniego (na diagramach w płaszczyźnie ($T-P$)).

\item Szczegółowo zbadano także przypadek nierównych mas w spinowo
spolaryzowanym modelu Hubbarda \textcolor{czerwony}{z jednocentrowym przyciąganiem} i spinowo zależnymi całkami
przeskoku ($t^{\uparrow}\neq t^{\downarrow}$, $t^{\uparrow}/t^{\downarrow}\equiv
r$).
Diagramy fazowe przejścia BCS-LP w obecności pola magnetycznego Zeemana
wykazują ciekawe zachowanie. W przeciwieństwie do przypadku $t^{\uparrow}=
t^{\downarrow}$, faza SC$_M$ pojawia się w zerowej temperaturze, w
przypadku silnego przyciągania. W ogólności, ten rodzaj rozwiązań
(typu Sarma, z $\Delta (h)$) pojawia się (dla $r> 1$) kiedy
$h>(\frac{r-1}{r+1})\bar{\mu}+2\Delta
\frac{\sqrt{r}}{r+1}$ (po stronie BCS) lub kiedy 
$h>\sqrt{(\bar{\mu}-\epsilon_0)^2+|\Delta |^2}-D\frac{r-1}{r+1}$ (po stronie
LP). Jeżli $r\neq 1$, faza SC$_M$ (lub BP-2) jest niestabilna w reżimie
słabego sprzeżęnia w $T=0$, ale faza BP-1 może być stabilna w granicy silnego
sprzężenia (LP), zarówno dla 2D, jak i dla 3D.
Zbadano wpływ członu Hartree na stabilność fazy SC$_M$ w przypadku
nierównych mas. Jeśli $t^{\uparrow}\neq t^{\downarrow}$, wyraz Hartree
ogranicza występowanie fazy SC$_M$ do niższych koncentracji nośników.
Jednakże, nierówne masy mogą zmienić to zachowanie nawet dla 2D, ze
względu na polaryzację spinową wynikającą z wyrazu kinetycznego. 

\item Zbadano, że faza BP-2 jest niestabilna w $T=0$ w całym zakresie
parametrów, zarówno w 2D, jak i w 3D jednopasmowym spinowo-spolaryzowanym
modelu Hubbarda \textcolor{czerwony}{z jednocentrowym przyciąganiem}. 
\textcolor{czerwony}{Faza ta może być jednak stabilna w przypadku parowania $d$.}  
\textcolor{czerwony}{Można także} przypuszczać, że faza BP-2 (Liu-Wilczka) może być zrealizowana w modelu dwupasmowym.

\item Metodą \textcolor{czerwony}{transformacji kanonicznej} wyprowadzono hamiltonian efektywny do 2. rzędu w
$t/|U|$ i określono krytyczne wartości koncentracji nośników, powyżej której
faza SC może współistnieć z uporządkowaniem ładunkowym (CO) w $h=0$ i
$r\neq 1$. SC/CO jest regionem separacji fazowej z \textcolor{czerwony}{domeną} SC i \textcolor{czerwony}{domeną} CO
(z $n=1$).

\item  Rozszerzono analizę przejścia BCS-LP do skończonych temperatur w 2D, \textcolor{czerwony}{w ramach spinowo spolaryzowanego modelu Hubbarda},
wykorzystjąc scenariusz KT. Temperatury przejścia KT są znacznie niższe od
temperatur Hartree-Focka. Polaryzacja spinowa ma silny niszczący wpływ na
stan nadciekły KT, który występuje w reżimie słabego sprzężenia, zgodnie z
rezultatami otrzymanymi dla przypadku continuum. W granicy silnego sprzężenia
temperatury krytyczne KT nie zależą od pola magnetycznego (poniżej
$h_{c1}$), ale są zależne od nierównowagi mas nośników -- górna granica na te
temperatury krytyczne:
$k_{B}T_c^{KT}=2\pi\frac{r}{(1+r)^2}\frac{t^2}{|U|}n(2-n)$ ($r>0$). Dla
$k_BT<<|U|$, istnieją jedynie nierozerwane pary lokalne, które mogą stworzyć
niespolaryzowaną fazę qSC poniżej $T_c^{KT}$ lub pozostać bez koherencji
fazowej. Ukłąd taki jest równoważny układowi gazu bozonów z twardym rdzeniem
na sieci. Jeśli $r\neq 1$ w 2D, spinowo spolaryzowany stan nadciekły KT może
być stabilny nawet w \textcolor{czerwony}{w zakresie} średniego i silnego sprzężenia.

\item Zbadano również przejście BCS-BEC w skończonych temperaturach w 3D, w
ramach spinowo spolaryzowanego modelu Hubbarda z $U<0$
($t^{\uparrow}=t^{\downarrow}$), wykraczając poza przybliżenie pola
średniego. Temperatury krytyczne przejścia nadprzewodzącego zostały \textcolor{czerwony}{wyznaczone}
metodą \textcolor{czerwony}{wielociałowej samouzgodnej macierzy T}. Porównano wyniki uzyskane w ramach schematów
$(GG_0)G_0$ i $(GG)G_0$, zarówno dla 3D modelu ciągłego z kontaktowym
oddziaływaniem przyciągającym, jak i dla przyciągającego modelu
Hubbarda przy niskich koncentracjach nośników. W skończonych temperaturach i
zerowym polu, występują następujące fazy w ukłądzie 3D: faza SC/SF
(poniżej $T_c$), region pseudo-szczeliny (powyżej temperatury kondensacji
par $T_c$, ale poniżej temperatury formowania się par $T_p$) i stan normalny
(powyżej $T_p$). W regionie pseudo-szczeliny istnieją długożyjące pary,
będące \textcolor{czerwony}{wzbudzeniami} niekoherentnymi fazowo.
Obliczenia w ramach schematów $(GG_0)G_0$ i $(GG)G_0$ prowadzą do bardzo
podobnych wyników w dwóch skrajnych granicach: BCS i BEC (LP). Jednakże,
krytyczne temperatury w reżimie unitarnym, określone w ramach schematu
$(GG_0)G_0$ są wyższe niż \textcolor{czerwony}{te} otrzymane w ramach schematu $(GG)G_0$,
który jest bardziej zgodny z wynikami symulacji kwantowego Monte Carlo. 
Pokazano, że w granicy BEC istnieją wyraźne różnice pomiędzy rezultatami
otrzymanymi w ramach przyciągającego modelu Hubbarda i w ramach 3D modelu
ciągłego. Wynika to z faktu, że w granicy par lokalnych modelu Hubbarda, masa
efektywna bozonów z twardym rdzeniem rośnie wraz z rosnącym $|U|$ -- ta cecha
modelu znajduje odzwierciedlenie w wynikach obliczeń w ramach T-macierzy.

\item Otrzymano także temperaturowe diagramy przejścia BCS-BEC dla \textcolor{czerwony}{$d=3$}, w
niezerowym polu \textcolor{czerwony}{magnetycznym}. Temperatury krytyczne przejścia nadprzewodzącego zostały
okreslone w ramach schematu $(GG_0)G_0$. Spinowo spolaryzowany stan nadciekły
z bezszczelinowym widmem dla większościowego kierunku spinu może być stabilny
w reżimie silnego sprzężenia, w przeciwieństwie do układu 2D dla $r=1$.
\end{itemize}


\newpage
\thispagestyle{empty}
\mbox{}

\appendix

\chapter{Attraction-repulsion Symmetry}
\label{appendix1}
Let us consider a bipartite lattice\footnote{a lattice which can be decomposed
into two sublattices such that each site of one sublattice is surrounded only by
sites of the other sublattice.}. For $t^{\uparrow}=t^{\downarrow}$ one can
transform the Hamiltonian \eqref{extham} (with $W=0$):
\begin{equation}
\label{hama}
H=\sum_{ij\sigma} (t_{ij}-\mu
\delta_{ij})c_{i\sigma}^{\dag}c_{j\sigma}+U\sum_{i}
n_{i\uparrow}n_{i\downarrow}-h\sum_{i}(n_{i\uparrow}-n_{i\downarrow}),
\end{equation}
with an attractive interaction $|U|$ ($U=-|U|$) into the repulsive Hubbard
model, by means of the \emph{canonical transformation} (particle-hole type)
\cite{MicnasModern, Shiba}:
\begin{equation}
c_{i\downarrow}^{\dag} \rightarrow b_{i\downarrow} e^{i\vec{Q}\cdot \vec{R}_i},
\qquad c_{i\downarrow} \rightarrow b_{i\downarrow}^{\dag}
e^{-i\vec{Q}\cdot\vec{R}_i},
\end{equation}
\begin{equation}
 c_{i\uparrow}^{\dag} \rightarrow b_{i\uparrow}^{\dag}, \qquad c_{i\uparrow}
\rightarrow b_{i\uparrow}.
\end{equation}
The vector $\vec{Q}$ is chosen so that it satisfies
$\textrm{exp}(i\vec{Q}\cdot \vec{R})=-1$ for a translation vector $\vec{R}$
connecting two sublattice sites.
After using the above transformation, the Hamiltonian (\ref{hama}) for
half-filling $\frac{1}{N} \sum_{i\sigma} \langle n_{i\sigma} \rangle =1$, where
$\mu = -|U|/2$ is given by:
\begin{eqnarray}
H&=&\sum_{ij\sigma} t_{ij} b_{i\sigma}^{\dag} b_{j\sigma}+|U|\sum_{i}
n_{i\uparrow}n_{i\downarrow}-\sum_{i\sigma}\Bigg(h+\frac{|U|}{2}\Bigg)n_{i\sigma
}\nonumber\\
&+&\Bigg(h+\frac{|U|}{2}\Bigg)N,
\end{eqnarray}
where: $n_{i\sigma}=b_{i\sigma}^{\dag} b_{i\sigma}$. The magnetization per site
$M=N^{-1}\sum_i \langle n_{i\uparrow}-n_{i\downarrow}\rangle$ in the half-filled
negative $U$ Hubbard model is mapped into a site filling $\langle
n_{i\uparrow}+n_{i\downarrow} \rangle -1 =\langle n \rangle -1$. The
half-filling condition for (\ref{hama}) is transformed to: $M = 0$.

Thus, the half-filled attractive Hubbard model in a Zeeman field has been
transformed into the doped repulsive Hubbard model in which $h+|U|/2$ now plays
the role of the chemical potential. Under this transformation, one can also see
that the FFLO state becomes the striped, charge density ordered and $\pi$-phase
shifted antiferromagnetic state of the doped positive $|U|$ Hubbard model
\cite{moreo}.

The various correspondences between quantities and operators in the attractive
and the repulsive Hubbard model are shown in Tab.~\ref{tab-hubbard}.

\begin{table}
\begin{center}
\caption{Particle-hole transformation mapping the attractive ($U<0$) Hubbard
model onto the repulsive ($U>0$) Hubbard model.}
\label{tab-hubbard}
\vspace{0.5cm}
\begin{tabular}[c]{|c|c|c|}
\hline
Operators and quantities & $U<0$ & $U>0$ \\
\hline\hline
creation operator $\downarrow$ & $c^{\dag}_{i\downarrow}$ & $b_{i\downarrow}
e^{i\vec{Q}\cdot \vec{R}_i}$\\
creation operator $\uparrow$ & $c^{\dag}_{i\uparrow}$ & $b_{i\uparrow}^{\dag}$\\
annihilation operator $\downarrow$ & $c_{i\downarrow}$ & $b_{i\downarrow}^{\dag}
e^{-i\vec{Q}\cdot\vec{R}_i}$\\
annihilation operator $\uparrow$ & $c_{i\uparrow}$ & $b_{i\uparrow}$\\
number of particle $\downarrow$ & $c^{\dag}_{i\downarrow}c_{i\downarrow}$ &
$1-b^{\dag}_{i\downarrow}b_{i\downarrow}$\\
number of particle $\uparrow$ & $c^{\dag}_{i\uparrow}c_{i\uparrow}$ &
$b^{\dag}_{i\uparrow}b_{i\uparrow}$\\
chemical potential & $\mu$ & $\mu=h+|U|/2$\\
external magnetic field & $h$ & $h=\mu+|U|/2$\\
superfluid phase (SF) & $\langle c_{i\downarrow}c_{i\uparrow} \rangle$ &
SDW$_{xy}$: $\langle b^{\dag}_{i\downarrow}b_{i\uparrow}\rangle
e^{-i\vec{Q}\cdot\vec{R}_i}$\\
CDW order & $\langle c^{\dag}_{i\uparrow} c_{i\uparrow} + c^{\dag}_{i\downarrow}
c_{i\downarrow} \rangle e^{i\vec{Q}\cdot \vec{R}_i} $  & SDW$_{z}$:  $(1-\langle
b^{\dag}_{i\downarrow}b_{i\downarrow}-b^{\dag}_{i\uparrow}b_{i\uparrow}\rangle)
e^{i\vec{Q}\cdot\vec{R}_i} \propto $\\
&  &  $\propto \langle S^{z}_{i} \rangle e^{i\vec{Q}\cdot\vec{R}_i}$\\ 
\hline
\end{tabular}
\end{center}
\end{table}

\chapter{Zero-temperature equations}
\label{appendix2}

In Section \ref{HFapprox} the general forms of the mean-field equations for the
case of the finite temperatures were presented.

Here, the zero-temperature equations are shown both for the superconducting and
the normal state explicitly. 

If we set $T=0$, the form of the order parameter equations changes in the
following way:
\begin{equation}
\Delta_{\vec{k}} =
\frac{1}{N}\sum_{\vec{k}}V_{\vec{k}\vec{q}}^s\frac{\Delta_{\vec{k}}}{4\omega_{
\vec{k}}}\Big(\textrm{sgn} (E_{\vec{k} \uparrow})+ \textrm{sgn}(E_{\vec{k}
\downarrow})\Big).
\end{equation}
The particle number equation is:
\begin{equation}
 n=1-\frac{1}{2N} \sum_{\vec{k}}
\frac{-(t^{\uparrow}+t^{\downarrow})\Theta_{\vec{k}}-\bar{\mu}-\textcolor{czerwony}{p\frac{\gamma_{\vec k}}{\gamma_0}W}}{\omega_{\vec{k}}
}\Big(\textrm{sgn} (E_{\vec{k} \uparrow})+ \textrm{sgn}(E_{\vec{k}
\downarrow})\Big).
\end{equation}
The equation for the magnetization takes the form:
\begin{equation}
M=\frac{1}{2N} \sum_{\vec{k}} \Big(\textrm{sgn}(E_{\vec{k} \downarrow}) -
\textrm{sgn}(E_{\vec{k} \uparrow})\Big).
\end{equation}
The Fock parameter equation:
\begin{equation}
p=-\frac{1}{N} \sum_{\vec{k}}\frac{1}{2}\gamma_{\vec{k}} 
\frac{-(t^{\uparrow}+t^{\downarrow})\Theta_{\vec{k}}-\bar{\mu}-\textcolor{czerwony}{p\frac{\gamma_{\vec k}}{\gamma_0}W}}{\omega_{\vec{k}}
} \Big(\textrm{sgn} (E_{\vec{k} \uparrow})+ \textrm{sgn}(E_{\vec{k}
\downarrow})\Big).
\end{equation}
Finally, one obtains the grand canonical potential:
\begin{eqnarray}
\frac{\Omega^{SC}}{N}&=&\frac{1}{4} Un(2-n)-\mu+\frac{1}{4}UM^2+W\gamma_0
n-\frac{1}{2}W\gamma_0 n^2 \nonumber \\
&-&Wp_{\uparrow}^2/2\gamma_0-Wp_{\downarrow}^2/2\gamma_0\nonumber+\frac{1}{N}
\sum_{\vec{k}}\frac{|\Delta_{\vec{k}}|^2}{4\omega_{\vec{k}}}\Big(\textrm{sgn}
(E_{\vec{k} \uparrow})+ \textrm{sgn}(E_{\vec{k} \downarrow})\Big)\\
&-&\frac{1}{2N}\sum_{\vec{k} \sigma}E_{\vec{k} \sigma}\textrm{sgn}
(E_{\vec{k}\sigma}). 
\end{eqnarray}

The corresponding equations for the normal phase ($\Delta = 0$) have the
following forms:
\begin{equation}
n=1-\frac{1}{2N}\sum_{\vec{k}}\Big(\textrm{sgn} (E_{\vec{k} \uparrow}^{NO})+
\textrm{sgn}(E_{\vec{k} \downarrow}^{NO})\Big), 
\end{equation}

\begin{equation}
M=\frac{1}{2N} \sum_{\vec{k}} \Big(\textrm{sgn}(E_{\vec{k} \downarrow}^{NO}) -
\textrm{sgn}(E_{\vec{k} \uparrow}^{NO})\Big),
\end{equation}
\begin{equation}
p=-\frac{1}{N} \sum_{\vec{k}}\frac{1}{2}\gamma_{\vec{k}} \Big(\textrm{sgn}
(E^{NO}_{\vec{k} \uparrow})+ \textrm{sgn}(E^{NO}_{\vec{k} \downarrow})\Big),
\end{equation}
\begin{eqnarray}
\frac{\Omega^{NO}}{N}&=&\frac{1}{4} Un(2-n)-\mu+\frac{1}{4}UM^2+W\gamma_0
n-\frac{1}{2}W\gamma_0 n^2\nonumber\\ 
&-&p_{\uparrow}^2/2\gamma_0-Wp_{\downarrow}^2/2\gamma_0-\frac{1}{N}\sum_{\vec{k}
} E_{\vec{k}}^{NO}\textrm{sgn}(E_{\vec{k}\sigma}^{NO}), 
\end{eqnarray}
where: $E_{\vec{k}\downarrow, \uparrow}^{NO}= \pm
(-t^{\downarrow}+t^{\uparrow})\Theta_{\vec{k}}\pm \frac{UM}{2}\pm 
\frac{1}{2}W(p_{\uparrow}-p_{\downarrow})\frac{\gamma_{\vec{k}}}{\gamma_0}\pm
h+\omega_{\vec{k}}^{NO}$ 
\\
and
$\omega_{\vec{k}}^{NO}=(-t^{\uparrow}-t^{\downarrow})\Theta_{\vec{k}}-\bar{\mu} \textcolor{czerwony}{-p\frac{\gamma_{\vec k}}{\gamma_{0}}W}
$.

\chapter{The analysis of the quasiparticle excitations spectrum: gapless region}
\label{appendix3}

In this Appendix, we determine, by analysis of the quasiparticle excitations
spectrum, the critical magnetic field $h_c$ above which the spin-polarized
superfluidity arises at $T=0$. There has been much work on the possibility of
the existence of the spatially homogeneous spin-polarized superfluidity
(breached pair (BP)) with one or two Fermi surfaces (BP-I or BP-II,
respectively) and a gapless spectrum for the majority spin species. These kinds
of pairing are very interesting not only in the context of superconductivity,
but also in that of trapped unbalanced ultracold Fermi atomic gases and color
superconductivity in quantum chromodynamics.

In our analysis, we take into account the isotropic pairing with the order
parameter: $\Delta=-\frac{U}{N}\sum_{\vec{i}} \langle c_{i \downarrow} c_{i
\uparrow} \rangle=-\frac{U}{N}\sum_{\vec{k}} \langle c_{-\vec{k} \downarrow}
c_{\vec{k} \uparrow} \rangle$. Therefore, the Hamiltonian \eqref{extham} takes
the form:
\begin{equation}
\label{hamb}
H=\sum_{ij\sigma} (t_{ij}^{\sigma}-\mu
\delta_{ij})c_{i\sigma}^{\dag}c_{j\sigma}+U\sum_{i}
n_{i\uparrow}n_{i\downarrow}-h\sum_{i}(n_{i\uparrow}-n_{i\downarrow}).
\end{equation}
Transforming the Hamiltonian (\ref{hamb}) to the reciprocal space and applying
the broken symmetry Hartree approximation, one can obtain two physical branches
of quasiparticle excitations:
\begin{equation}
\label{energy_2}
E_{\vec{k}\downarrow, \uparrow}= \pm
(-t^{\downarrow}+t^{\uparrow})\Theta_{\vec{k}}\pm \frac{UM}{2}\pm
h+\omega_{\vec{k}},
\end{equation}
where:
$\omega_{\vec{k}}=\sqrt{((-t^{\uparrow}-t^{\downarrow})\Theta_{\vec{k}}-\bar{\mu
})^2+|\Delta|^2}$, $\Theta_{\vec{k}}=\sum_{l=1}^{d} \cos(k_l a_l)$ (here $d=2,3$
for two- and three-dimensional lattices, respectively); $a_l$ is the lattice
constant in the $l$-th direction (we set $a_l=1$ in further considerations). 

Let us assume $t^{\uparrow} \geq t^{\downarrow}$ and $h>0$. For clarity, let
us introduce the following notation: $\textcolor{czerwony}{\bar{h}}\equiv h+\frac{UM}{2}$,
$t_{\pm}\equiv t^{\uparrow}\pm t^{\downarrow}$.

Therefore,  the quasiparticle energies \eqref{energy_2} can be rewritten as:
\begin{equation}
 E_{\vec{k}\downarrow, \uparrow}= \pm t_{-}\Theta_{\vec{k}}\pm
\textcolor{czerwony}{\bar{h}}+\sqrt{(-t_{+}\Theta_{\vec{k}}-\bar{\mu})^2+|\Delta|^2}.
\end{equation}

The critical value of $\textcolor{czerwony}{\bar{h}}=\textcolor{czerwony}{\bar{h}_c}$ is determined from the condition
that $E_{\vec{k}\uparrow}$ branch has one zero, i.e. $E_{\vec{k}\uparrow}=0$.
Then, we have the quadratic equation:
\begin{equation}
(t_{-}^2-t_{+}^2)\Theta_{\vec{k}}^2+2(t_{-}\textcolor{czerwony}{\bar{h}}-t_{+}\bar{\mu})\Theta_{\vec
{k}}+\textcolor{czerwony}{\bar{h}}^2-\bar{\mu}^2+|\Delta|^2,
\end{equation}
which can be solved with respect to $\Theta_{\vec{k}}$. In this way:
\begin{equation}
 \Theta_{\vec{k}}^{\pm}=\frac{t_{-}\textcolor{czerwony}{\bar{h}}-t_{+}\bar{\mu}\pm\sqrt{(t_{+}\textcolor{czerwony}{\bar{h}}-t_{-}\bar{\mu})^2-4t^{\uparrow}t^{\downarrow}|\Delta|^2}}{4t^{\uparrow}t^{
\downarrow}}.
\end{equation}
Therefore, above $\textcolor{czerwony}{\bar{h}_c}$, the $E_{\vec{k}\uparrow}$ branch crosses the
zero energy axis at two points.
At the critical value of $\textcolor{czerwony}{\bar{h}}=\textcolor{czerwony}{\bar{h}_c}$, the two Fermi surfaces are
equal $\Theta_{\vec{k}}^{+}=\Theta_{\vec{k}}^{+}\equiv \Theta_{c}$:
\begin{equation}
 \Theta_{c}=\frac{\textcolor{czerwony}{\bar{h}_c}t_{-}-t_{+}\bar{\mu}}{4t^{\uparrow}t^{\downarrow}}.
\end{equation}
In the weak coupling limit, the homogeneous magnetized superconducting phase
(SC$_M$) (Sarma-type solution with $\Delta (h)$) occurs at the critical value of
$\textcolor{czerwony}{\bar{h}}=\textcolor{czerwony}{\bar{h}_c}$:
\begin{equation}
\textcolor{czerwony}{\bar{h}_c}=\Big(\frac{r-1}{r+1}\Big)\bar{\mu}
+2\Delta\frac{\sqrt{r}}{r+1},
\end{equation}
where: $r\equiv t^{\uparrow}/t^{\downarrow}$. If $t^{\uparrow}=t^{\downarrow}$
($r=1$), then $\textcolor{czerwony}{{\bar{h}_c}}=\Delta$, on the BCS side.

On the other hand, in the strong coupling limit (for $\vec{k}=0$), the critical
value of $\textcolor{czerwony}{\bar{h}}$ is:
\begin{equation}
\textcolor{czerwony}{{\bar{h}_c}}=\sqrt{(\bar{\mu}-\epsilon_0)^2+|\Delta|^2}-D\frac{r-1}{r+1},
\end{equation}
where: $D=zt$, $z=2d$ is the coordination number. If
$t^{\uparrow}=t^{\downarrow}$ ($r=1$), then
${\sqrt{(\bar{\mu}-\epsilon_0)^2+|\Delta|^2}}$, on the BEC side.

It is worth mentioning that the transition from the non-polarized superfluid
state (SC$_0$) to the polarized superfluid phase is a topological quantum phase
transition (Lifshitz type \cite{Lifshitz}). As shown above, the quasiparticle
excitation spectrum can change from gapped to gapless one. There is a
change in the electronic structure. Therefore, we can identify the topological
quantum phase transition as the appearance of zero energies in the energy
excitation spectrum. Lifshitz type transitions take place in metals and alloys
\cite{Blanter}. The topological phase transition also appears in non-s-wave
superfluids \cite{Iskin, Iskin-2, Volovik, Duncan, Botelho, TobijaszewskaB}. The SC$_0$
$\rightarrow$ SC$_M$ transition is in some sense unusual, because it concerns
the
s-wave pairing symmetry case and occurs without a change in the order
parameter symmetry \cite{Iskin-2}.

For the BCS state, the momentum distribution in the Brillouin zone is given by
\cite{Tinkham}:
\begin{equation}
 n_{\vec{k}}=|\nu_{\vec{k}}|^2,
\end{equation}
where $|\nu_{\vec{k}}|^2$ \eqref{nu} is a measure of the probability that the
($\vec{k}\uparrow$, $-\vec{k}\downarrow$) state is occupied.

In turn, for the SC$_M$ state:
\begin{equation}
\left\{ \begin{array}{ll}
\textrm{if} \,\, \Theta_{\vec{k}} \in [\Theta_{\vec{k}}^{-},
\Theta_{\vec{k}}^{+} ]: & n_{\vec{k}\uparrow}=1, \,\, n_{\vec{k}\downarrow}=0 \\
   
\textrm{otherwise}: &
n_{\vec{k}\uparrow}=n_{\vec{k}\downarrow}=|\nu_{\vec{k}}|^2.
\end{array} \right.
\end{equation}
The numbers of particles with spin up and spin down are
equal in the BCS state. However, in the presence of the Zeeman magnetic field,
in the SC$_M$ phase, they differ. Therefore, this quantum phase transition is
associated with the change in the Fermi sea topology and it appears at the point
in which the Fermi sea is divided into two regions. 

\newpage
\thispagestyle{empty}
\mbox{}

\chapter{T-matrix approach. Functional derivatives technique}
\label{appendix-Tmatrix}
\section{Functional derivatives technique}
In this appendix a detail derivation of the T-matrix equations for the Hubbard model is presented using the 
functional derivative technique. \textcolor{czerwony}{In adition, we also obtain the mean-field equations derived in Chapter \ref{chapter4}}.
We consider the Hubbard Hamiltonian:
\begin{equation}
\label{HubbardHamiltonian}
H=H_{0}+H_{1}, 
\end{equation}
\begin{equation}
H_{0}=\sum_{i,j,\sigma}\bar{t}_{ij}c_{i\sigma}^{\dagger}c_{j\sigma},\quad
\bar{t}_{ij}=
t_{ij} -\mu\delta_{ij}
\end{equation}
\begin{equation}
H_{1}= U\sum_{i}n_{i\uparrow}n_{i\downarrow}, 
\end{equation} 
Introducing the Nambu representation: ${\hat\Psi}^{\dagger}_{i}=
\left(\begin{array}{cc}c^{\dagger}_{i\uparrow}&c_{i\downarrow}\end{array}
\right)$,
${\hat\Psi}_{i}=
\left(\begin{array}{c}c_{i\uparrow}\\c^{\dagger}_{i\downarrow}\end{array}
\right)$,
we have:
$n_{i\uparrow}=\frac{1}{2}{\hat\Psi}^{\dagger}_{i}(\tau_{3}+\tau_{0}){\hat\Psi}_
{i}$, 
$n_{i\downarrow}=\frac{1}{2}{\hat\Psi}^{\dagger}_{i}(\tau_{3}-\tau_{0}){\hat\Psi
}_{i}+1$,
$c^{\dagger}_{i\uparrow}c^{\dagger}_{i\downarrow}={\hat\Psi}^{\dagger}_{i}\tau_{
+}{\hat\Psi}_{i}$,
$c_{i\downarrow}c_{i\uparrow}={\hat\Psi}^{\dagger}_{i}\tau_{-}{\hat\Psi}_{i}$,
$\tau_{\alpha}$ are the Pauli matrices ($\tau_\pm=\frac{1}{2}(\tau_1\pm
i\tau_2$), $\tau_{0}$ is the identity matrix).

The Hubbard Hamiltonian $H=H_0+H_1$ can be rewritten in the following way:
\begin{equation}
\label{HubbardNambu}
H_{0}=\sum_{i,j}t_{ij}{\hat\Psi}^{\dagger}_{i}\tau_{3}{\hat\Psi}_{j}
-\mu N,
\end{equation}
\begin{equation}
H_{1}=
U\sum_{i}\left({\hat\Psi}^{\dagger}_{i}\tau_{+}{\hat\Psi}_{i}\right)\left({
\hat\Psi}^{\dagger}_{i}\tau_{-}{\hat\Psi}_{i}\right),
\end{equation} 
where $N$ is the total number of sites.

We define the thermal (Matsubara) Green functions using the Nambu field
representation:
\begin{eqnarray}
\hat{\cal{G}}_{ij}(\tau,\tau')= 
-\langle T_{\tau}\hat{\Psi}_{i}(\tau)\hat{\Psi}^{\dagger}_{j}(\tau')\rangle &=&
\left(\begin{array}{cc}-\langle
T_{\tau}c_{i\uparrow}(\tau)c^{\dagger}_{j\uparrow}(\tau')\rangle &
- \langle T_{\tau}c_{i\uparrow}(\tau) c_{j\downarrow}(\tau')\rangle\\
- \langle T_{\tau}c^{\dagger}_{i\downarrow}(\tau)
c^{\dagger}_{j\uparrow}(\tau')\rangle & -\langle
T_{\tau}c^{\dagger}_{i\downarrow}(\tau)c_{j\downarrow}(\tau')\rangle\end{array}
\right)=\nonumber\\
&=&\left(\begin{array}{cc}G^{\uparrow}_{ij}(\tau,\tau') & F_{ij}(\tau,\tau')\\
F_{ij}^{\dagger}(\tau,\tau') &
-G^{\downarrow}_{ji}(\tau',\tau)\end{array}\right)~~,
\end{eqnarray}
The time dependence of the operators is given by $\hat{\Psi}_{i}(\tau)=
e^{\tau H}\hat{\Psi}_{i}e^{-\tau H}, \hat{\Psi}^{\dagger}_{i}(\tau)=
e^{\tau H}\hat{\Psi}^{\dagger}_{i}e^{-\tau H}$. The thermal average is given by:
$\langle\cdot\cdot\cdot\rangle=
\textrm{Tr}[ \cdot\cdot\cdot \exp(-\beta H) ]/\textrm{Tr}[\exp(-\beta H)].$

In order to derive the equations for the Green functions, we use the functional
derivative technique \cite{KB, teubel, Pedersen, RMunpablished}.  We consider
the Green functions
written in the presence of the source fields, as given below:
\begin{eqnarray}
\label{TGF}
\hat{G}_{ij}(\tau,\tau')= 
-\langle T_{\tau}\hat{\Psi}_{i}(\tau)\hat{\Psi}^{\dagger}_{j}(\tau')\rangle_{S}
=
-\frac{\langle
T_{\tau}S\hat{\Psi}_{i}(\tau)\hat{\Psi}^{\dagger}_{j}(\tau')\rangle}
{\langle
T_{\tau}S\rangle}\equiv\langle\langle\hat{\Psi}_{i}(\tau);\hat{\Psi}^{\dagger}_{
j } \rangle\rangle,
\end{eqnarray}
where $T_\tau$ is the chronological ordering operator and the
generating functional $S$ is given by:
\begin{eqnarray}
S&=&\exp\Big(-\int_0^\beta d\tau \sum_{i}[\xi_{i}(\tau)
\hat{\Psi}^{\dagger}_{i}(\tau)\tau_{3} \hat{\Psi}_{i}(\tau) + 
h_{i}(\tau) \hat{\Psi}^{\dagger}_{i}(\tau)\tau_{0}
\hat{\Psi}_{i}(\tau) +\nonumber\\
&+&\eta_{i}(\tau){\hat\Psi}^{\dagger}_{i}(\tau)\tau_{-}{\hat\Psi}_{i}(\tau) +
\eta^{*}_{i}(\tau){\hat\Psi}^{\dagger}_{i}(\tau)\tau_{+}{\hat\Psi}_{i}
(\tau)]\Big) =
\nonumber \\
&=&\exp\Big(-\int d1 [\xi(1) \hat{\Psi}^{\dagger}(1)\tau_{3} \hat{\Psi}(1) + 
h(1) \hat{\Psi}^{\dagger}(1)\tau_{0} \hat{\Psi}(1) + \nonumber\\
&+&\eta(1){\hat\Psi}^{\dagger}(1)\tau_{-}{\hat\Psi}(1) +
\eta^{*}(1){\hat\Psi}^{\dagger}(1)\tau_{+}{\hat\Psi}]\Big),
\end{eqnarray}
where we have introduced convenient shorthand notation:
$1\equiv(\vec{R}_{i},\tau)$, $\int d1\equiv\int_{0}^{\beta} d{\tau}
\sum_{i}$.
This functional of the external sources includes:
$\xi_{i}(\tau)$ -- charge field (that couples to the number of particles
$n_{i\uparrow}+n_{i\downarrow}$),
$h_{i}(\tau)$ -- magnetic field (couples to the magnetization
$n_{i\uparrow}-n_{i\downarrow}$), and $\eta_{i}(\tau), \eta^{*}_{i}(\tau)$ are
the pairing fields (couple to products of operators
$c_{i\downarrow}c_{i\uparrow}$ and
$c^\dagger_{i\uparrow}c^\dagger_{i\downarrow}$, respectively). 
Explicitly, we have:
\begin{eqnarray}
\hat{G}(1,2)=\left(\begin{array}{cc}G^{\uparrow}(1,2)&F(1,2)\\F^{\dagger}(1,
2)&-G^{\downarrow}(2,1)\end{array}\right).
\end{eqnarray}

The equation of motion for $\hat{G}(1,2)$ is the Heisenberg equation:
\begin{equation}
\frac{\partial {\hat\Psi}(1)}{\partial \tau}=[H+H',{\hat \Psi}(1)],
\end{equation}
where $H'$ describes the source term. Explicitly:
\begin{align}
\label{EOM}
&\sum_{m}\left\{\left[-\frac{\partial}{\partial\tau}
-h_{m}(\tau)\right]\tau_{0}\delta_{im}-\left[\bar{t}_{im}+\xi_{m}(\tau)\delta_{
im}\right]\tau_{3}
-\left[\eta_{m}(\tau)\tau_{-}+\eta^{*}_{m}(\tau)\tau_{+}\right]\delta_{im}
\right\}\hat{G}_{mj}(\tau,\tau')\nonumber\\
&=\tau_{0}\delta_{ij}\delta(\tau-\tau') +
U \langle\langle \tau_{+}{\hat
\Psi}(1)\left({\hat\Psi}^{\dagger}(1)\tau_{-}{\hat\Psi}(1)\right); 
 {\hat \Psi}^{\dagger}(2) \rangle\rangle
+U \langle\langle 
\left({\hat\Psi}^{\dagger}(1)\tau_{+}{\hat\Psi}(1)\right)\tau_{-}{\hat \Psi}(1);
 {\hat \Psi}^{\dagger}(2) \rangle\rangle, 
\end{align}
where we have used:
\begin{equation}
\left[{\hat\Psi}(1),\left({\hat\Psi}^{\dagger}(1)\tau_{\pm}{\hat\Psi}
(1)\right)\right]=
\tau_{\pm}{\hat \Psi}(1)
\end{equation}
The first $\langle\langle\cdots\rangle\rangle$ term on the RHS of
Eq.~(\ref{EOM})
can be rewritten  by means of functional derivatives as: 
\begin{eqnarray}
\langle\langle \tau_{+}{\hat
\Psi}(1)\left({\hat\Psi}^{\dagger}(1)\tau_{-}{\hat\Psi}(1)\right);  {\hat
\Psi}^{\dagger}(2)
\rangle\rangle=-\tau_{+}\frac{\delta {\hat G}(1,2)}{\delta \eta(1)}+
\tau_{+}\langle{\hat\Psi}^{\dagger}(1)\tau_{-}{\hat\Psi}(1)\rangle_{S}
{\hat G}(1,2),
\end{eqnarray}
and the  analogous formula holds for the second
$\langle\langle\cdots\rangle\rangle$
term of Eq.(\ref{EOM}).
The RHS of (\ref{EOM}) is the given by
\begin{equation}
U\left[\tau_{+}\langle{\hat\Psi}^{\dagger}(1)\tau_{-}{\hat\Psi}(1)\rangle_{S}+
\tau_{-}\langle{\hat\Psi}^{\dagger}(1)\tau_{+}{\hat\Psi}(1)\rangle_{S}\right]{
\hat
G}(1,2)
-U\left[\tau_{+}\frac{\delta {\hat G}(1,2)}{\delta \eta(1)}
+\tau_{-}\frac{\delta {\hat G}(1,2)}{\eta^{*}(1)}\right].
\end{equation}
Defining the gap function as: 
\begin{eqnarray}
\Delta^{*}(1)=U\langle
c^{\dagger}_{i\uparrow}(\tau)c^{\dagger}_{i\downarrow}(\tau)\rangle_{S},
\end{eqnarray}
the equation of motion takes the form:
\begin{align}
\label{EOM1}
&\sum_{m}\left\{\left[-\frac{\partial}{\partial\tau}
-h_{m}(\tau)\right]\tau_{0}\delta_{im}-\left[\bar{t}_{im}+\xi_{m}(\tau)\delta_{
im}\right]\tau_{3}
-\left[\eta_{m}(\tau)\tau_{-}+\eta^{*}_{m}(\tau)\tau_{+}\right]\delta_{im}
\right\}\hat{G}_{mj}(\tau,\tau')=\nonumber\\
&\tau_{0}\delta_{ij}\delta(\tau-\tau')  +
(\tau_{+}\Delta(1)+\tau_{-}\Delta^{*}(1)){\hat
G}(1,2)-U\left[\tau_{+}\frac{\delta {\hat G}(1,2)}{\delta \eta(1)}
+\tau_{-}\frac{\delta {\hat G}(1,2)}{\eta^{*}(1)}\right].
\end{align}
Let us now introduce the Dyson equation in Nambu space:
\begin{eqnarray}
{\hat G}(1,2)={\hat G}_{0}(1,2)+\int d{\bar 3}\int d{\bar 4}{\hat G}_{0}(
1,{\bar 3}){\hat \Sigma}({\bar 3},{\bar 4}){\hat G}({\bar 4},2).
\end{eqnarray}
The inverse Green function is defined by: 
\begin{eqnarray}
\label{invG}
\int d{\bar 3}{\hat G}(1,{\bar 3}){\hat G}^{-1}({\bar 3},2)=\tau_{0}\delta(1-2).
\end{eqnarray}
In particular, the zeroth order Green function is explicitly given by:
\begin{eqnarray}
\label{G0inverse}
[\hat{G}^{0}_{ij}(\tau,\tau')]^{-1}&=&\Bigg\{\left[-\frac{\partial}{
\partial\tau }
-h_{i}(\tau)\right]\tau_{0}\delta_{ij}-\left[\bar{t}_{ij}+\xi_{i}(\tau)\delta_{
ij}\right]\tau_{3}\nonumber\\
&-&\left[\eta_{i}(\tau)\tau_{-}+\eta^{*}_{i}(\tau)\tau_{+}\right]\delta_{ij}
\Bigg\}\delta(\tau-\tau').
\end{eqnarray}
The Dyson equation can be written as: 
\begin{eqnarray}
\label{Dysoneq}
{\hat G}^{-1}(1,2)={\hat G}^{-1}_{0}(1,2)-{\hat \Sigma}(1,2).
\end{eqnarray}
The self-energy is given by:
\begin{eqnarray}
{\hat
\Sigma}(1,3)&=&\left[\tau_{+}\Delta(1)+\tau_{-}\Delta^{*}(1)\right]
\delta(1-3)\nonumber\\
&-&U\int d{\bar 2} \left[\tau_{+}\frac{\delta {\hat G}(1,{\bar 2})}{\delta
\eta(1)} {\hat G}^{-1}({\bar 2},3)
+\tau_{-}\frac{\delta {\hat G}(1,{\bar 2})}{\eta^{*}(1)}{\hat G}^{-1}({\bar
2},3)\right].
\end{eqnarray}
Using now the identity:
\begin{eqnarray}\label{ID}
\delta {\hat G}=-{\hat G}\delta{\hat G}^{-1} {\hat G}= -{\hat G}\delta{\hat
G_{0}}^{-1}{\hat G}+{\hat G}\delta {\hat \Sigma}{\hat G},
\end{eqnarray}
which follows from the definition of the inverse Green function \eqref{invG}, we
can rewrite the functional derivatives in terms of ${\hat
\Sigma}$.  We get
\begin{eqnarray}
\label{aux}
\frac{\delta {\hat G}(1,2)}{\delta \eta(1^{+})}={\hat G}(1,1^{+})\tau_{-}{\hat
G}(1,2)+\int d{\bar 3}\int d{\bar 4}
{\hat G}(1,{\bar 3})\frac{\delta {\hat \Sigma}({\bar 3},{\bar 4})}{\delta
\eta(1^{+})}{\hat G}({\bar 4},2).
\end{eqnarray}
Substituting (\ref{aux}) and using $\langle n_{\sigma}(1)\rangle_{S}=-\langle
T_{\tau}c(1)c^{\dagger}(1^{+})\rangle_{S}$, we
obtain the equation for the matrix self-energy in the form:
\begin{eqnarray}\label{Fin}
{\hat \Sigma}(1,2)&=&\left(\begin{array}{cc}
U\langle n_{\downarrow}(1)\rangle_{S}&\Delta(1)\\
\Delta^{*}(1)&-U\langle n_{\uparrow}(1)\rangle_{S}
\end{array}\right) \delta(1-2)\nonumber\\
&-&
U\int d{\bar 3}\left[\tau_{+}{\hat G}(1,{\bar 3})\frac{\delta}{\delta \eta(1)}
+\tau_{-}{\hat G}(1,{\bar 3})\frac{\delta}{\delta \eta^{*}(1)}\right]
{\hat \Sigma}({\bar 3},2).
\end{eqnarray}

\textcolor{czerwony}{Or explicitly restoring space and time indices}:
\begin{eqnarray}\label{Fin1}
{\hat \Sigma}_{ij}(\tau,\tau')&=&\left(\begin{array}{cc}
U\langle n_{i\downarrow}(\tau)\rangle_{S}&\Delta_{i}(\tau)\\
\Delta^{*}_{i}\tau )&-U\langle n_{i\uparrow}(\tau)\rangle_{S}
\end{array}\right) \delta_{ij}\delta(\tau-\tau')\\ \nonumber
&-&U\sum_{l}\int_{0}^{\beta} d{\bar \tau} \left[\tau_{+}{\hat G}_{il}(\tau,{\bar
\tau})\frac{\delta}{\delta \eta_{i}(\tau)}
+\tau_{-}{\hat G}_{il}(1,{\bar \tau})\frac{\delta}{\delta \eta^{*}_{i}(\tau)}\right]
{\hat \Sigma}_{lj}({\bar \tau},\tau').
\end{eqnarray}

The compact form of the above equation (\ref{Fin}, \ref{Fin1}) gives an
exact integro-differential
equation for determining the self-energy ${\hat \Sigma}_{ij}(\tau,\tau')$.
This equation can be solved in an iterative way.
The zeroth order self-energy, in which we set $\delta{\hat \Sigma}=0$ 
and the source fields to zero, gives the
Hartree-Fock-Bogoliubov approximation and leads to the BCS theory in the Nambu
formulation. 

Let us consider the case of non-zero Zeeman magnetic field, setting:
$h_{i}(\tau)=-h$,
$\xi_{i}(\tau)\rightarrow 0, 
\eta_{i}(\tau)\rightarrow 0,\eta^{*}_{i}(\tau)\rightarrow 0$ 
and $\delta{\hat \Sigma}=0$. 
Using \eqref{G0inverse} and
\eqref{Fin} in the Dyson equation \eqref{Dysoneq}, the inverse of the full Green
function is given by:
\begin{equation}
\label{full_green_func}
[\hat{G}_{ij}(\tau,\tau')]^{-1}=\left(\begin{array}{cc}\left[-\frac{\partial}{
\partial\tau}
+h-U\langle
n_{i\downarrow}\rangle\right]\delta_{ij}-\bar{t}_{ij}&-\Delta_{i}\delta_{ij}\\
-\Delta^{*}_{i}\delta_{ij}&\left[-\frac{\partial}{\partial\tau}+h-U\langle
n_{i\uparrow}\rangle\right]\delta_{ij}+\hat{t}_{ij}
\end{array}\right)\delta(\tau-\tau').
\end{equation}
Assuming translational invariance, one can introduce the space-time Fourier
transform:
\begin{eqnarray}\label{FT}
\hat{G}_{ij}(\tau,\tau')=\frac{1}{\beta
N}\sum_{\vec{k},\omega_{n}}e^{i\vec{k}\cdot(\vec{R}_{i}-\vec{R}_{j})-i\omega_{n}
(\tau-\tau')}
{\hat G}(\vec{k}, i\omega_{n}).   
\end{eqnarray}
Considering only homogeneous superconducting  ordering and  setting $\langle
n_{i\sigma}\rangle=n_{\sigma}, \Delta_{i}=\Delta$ one gets 
for the  Green function:
\begin{eqnarray}
{\hat G}(\vec{k},i\omega_{n})&=&
\left(\begin{array}{cc} G^{\uparrow}(\vec{k},i\omega_{n})
&F(\vec{k},i\omega_{n})\\
F^{\dagger}(\vec{k},i\omega_{n})&-G^{\downarrow}(-\vec{k},-i\omega_{n})
\end{array}\right)=\nonumber\\
&=&
\left(\begin{array}{cc}
i\omega_{n}+h-\epsilon_{\vec{k}}+\mu-Un_{\downarrow}&-\Delta\\
-\Delta^{*}&i\omega_{n}+h+\epsilon_{\vec{k}}-\mu+Un_{\uparrow}
\end{array}\right)^{-1}.
\end{eqnarray}
This yields explicit solutions in the following form:
\begin{eqnarray}\label{BCS}
G^{\uparrow}(\vec{k},i\omega_{n})=
\frac{ i\omega_{n}+{\bar h}+{\bar \epsilon}_{\vec{k}} }
{(i\omega_{n}+{\bar h}-{\bar \epsilon}_{\vec{k}})
(i\omega_{n}+{\bar h}+{\bar \epsilon}_{\vec{k}} )-|\Delta|^2},\nonumber \\
G^{\downarrow}(\vec{k},i\omega_{n})=
\frac{ i\omega_{n}-{\bar h}+{\bar \epsilon}_{\vec{k}} }
{(i\omega_{n}-{\bar h}-{\bar \epsilon}_{\vec{k}})
(i\omega_{n}-{\bar h}+{\bar \epsilon}_{\vec{k}} )-|\Delta|^2},\nonumber \\
F(\vec{k},i\omega_{n})=
\frac{ \Delta}
{(i\omega_{n}+{\bar h}-{\bar \epsilon}_{\vec{k}})
(i\omega_{n}+{\bar h}+{\bar \epsilon}_{\vec{k}} )-|\Delta|^2},\nonumber\\
F^{*}(\vec{k},i\omega_{n})=
\frac{ \Delta^{*} }
{(i\omega_{n}-{\bar h}+{\bar \epsilon}_{\vec{k}})
(i\omega_{n}-{\bar h}-{\bar \epsilon}_{\vec{k}} )-|\Delta|^2},
\end{eqnarray}
where:
${\bar h}=h+UM/2$, ${\bar \epsilon}_{\vec{k}}=\epsilon_{\vec{k}}- 
{\bar \mu}$, ${\bar \mu} =\mu-Un/2$, $n=n_{\uparrow}+n_{\downarrow}$,
$M=n_{\uparrow}-n_{\downarrow}$,
${\epsilon}_{-\vec{k}}={\epsilon}_{\vec{k}}$ and
$F^{\dagger}(\vec{k},i\omega_{n})=F^{*}(-\vec{k},-i\omega_{n})$.  
The quasiparticle spectrum is obtained upon analytic continuation $i\omega_n \rightarrow E+i\eta$ and is given by the four poles:
\begin{eqnarray}
E^{\pm}_{\vec{k}\uparrow}=-{\bar h}\pm E_{\vec{k}}\nonumber\\
E^{\pm}_{\vec{k}\downarrow}={\bar h}\pm E_{\vec{k}}\nonumber\\
E_{\vec{k}}=\sqrt{{\bar \epsilon}^{2}_{\vec{k}}+|\Delta|^2},
\end{eqnarray}
among which it is sufficient to choose only two:
\begin{eqnarray}
E_{\vec{k}\uparrow}=-{\bar h}+E_{\vec{k}}\nonumber\\
E_{\vec{k}\downarrow}={\bar h}+E_{\vec{k}}.
\end{eqnarray}

The equation (\ref{BCS}) can be written as follows:
\begin{eqnarray}
G^{\uparrow}(\vec{k},i\omega_{n})=
\frac{u^{2}_{\vec{k}}}{i\omega_{n}-E_{\vec{k}\uparrow}}
+\frac{ v^{2}_{\vec{k}}}{i\omega_{n}+E_{\vec{k}\downarrow}},\nonumber\\
G^{\downarrow}(\vec{k},i\omega_{n})=
\frac{u^{2}_{\vec{k}}}{i\omega_{n}-E_{\vec{k}\downarrow}}
+\frac{ v^{2}_{\vec{k}}}{i\omega_{n}+E_{\vec{k}\uparrow}},\nonumber\\
F(\vec{k},i\omega_{n})=
\frac{ \Delta}{(i\omega_{n}-E_{\vec{k}\uparrow})
(i\omega_{n}+E_{\vec{k}\downarrow})},
\end{eqnarray}
where $u^{2}_{\vec{k}}+v^{2}_{\vec{k}}=1,
u^{2}_{\vec{k}}
=\frac{1}{2}\left(1+\frac{{\bar\epsilon}_{\vec{k}}}{E_{\vec{k}}}\right)$ 
are the usual BCS coherence factors.
In this way, one obtains the gap equation:
\begin{eqnarray}
\label{gapnambu}
\Delta=U\frac{1}{\beta N}\sum_{\vec{k},\omega_{n}}F(\vec{k},i\omega_{n})=
-U\frac{1}{N}\sum_{\vec{k}}\frac{\Delta}{2E_{\vec{k}}}
\left[1-f(E_{\vec{k}\uparrow})
-f(E_{\vec{k}\downarrow})\right]
\end{eqnarray}
and the particle numbers are:
\begin{eqnarray}
n_{\vec{k}\sigma}=\frac{1}{\beta}\sum_{\omega_{n}}
G^{\sigma}(\vec{k},i\omega_{n})e^{i\omega_{n}0^{+}}=
u^{2}_{\vec{k}}f(E_{\vec{k}\sigma})
+v^{2}_{\vec{k}}f(-E_{\vec{k},-\sigma}),
\end{eqnarray}
where $f(x)=\left[e^x+1\right]^{-1}$ is the Fermi function.
Thus, the particle number equation and the magnetization equation are given by:
\begin{eqnarray}
\label{n_nambu}
n=\frac{1}{N}\sum_{\vec{k}}\left[1-
\frac{{\bar\epsilon}_{\vec{k}}}{E_{\vec{k}}}
\left[1-f(E_{\vec{k}\uparrow})
-f(E_{\vec{k}\downarrow})\right]\right],\\
M=\frac{1}{N}\sum_{\vec{k}}\left[f(E_{\vec{k}\uparrow})-
f(E_{\vec{k}\downarrow})\right],
\end{eqnarray}
in complete agreement with the standard treatment
(Eqs.~\eqref{del2},~\eqref{particle_eq} and \eqref{Magn_s} for the gap
parameter, particle number and magnetization, respectively).

The above scheme can be generalized to the hopping imbalance ($t^{\uparrow}\neq
t^{\downarrow}$) case. The Hubbard Hamiltonian with asymmetric hopping takes the
form:
\begin{equation}
\label{HamiltonianAsymmetric}
H=H_{0}+H_{1},
\end{equation}
where:
\begin{equation}
H_{0}=\sum_{i,j,\sigma}({t}^{\sigma}_{ij}
-\mu\delta_{ij})c_{i\sigma}^{\dagger}c_{j\sigma},
\end{equation}
\begin{equation}
\label{oddzialywanie}
H_{1}= U\sum_{i}n_{i\uparrow}n_{i\downarrow}, 
\end{equation}
${t}^{\sigma}_{ij}={t}^{\sigma}_{ji}, {t}^{\sigma}_{ii}=0.$

After introducing the Nambu representation, the Hamiltonian
\eqref{HamiltonianAsymmetric}-\eqref{oddzialywanie} can be rewritten:
\begin{equation}
H_{0}=\sum_{i,j}\left[{\bar t}_{ij}{\hat\Psi}^{\dagger}_{i}\tau_{3}{\hat\Psi}_{j}
-t^{-}_{ij}{\hat\Psi}^{\dagger}_{i}\tau_{0}{\hat\Psi}_{j}\right]
-\mu N,
\end{equation}
\begin{equation}
H_{1}=
U\sum_{i}\left({\hat\Psi}^{\dagger}_{i}\tau_{+}{\hat\Psi}_{i}\right)\left({
\hat\Psi}^{\dagger}_{i}\tau_{-}{\hat\Psi}_{i}\right),
\end{equation} 
where: $\bar{t}_{ij}=t_{ij}-\mu \delta_{ij}$, $t_{ij}=\frac{1}{2}(t^{\uparrow}_{ij}+t^{\downarrow}_{ij})$, 
$t^{-}_{ij}=\frac{1}{2}(t^{\downarrow}_{ij}-t^{\uparrow}_{ij})$. 

The inverse of the full Green
function given by Eq.~\eqref{full_green_func} in Hartree-Fock-Bogoliubov approximation for the asymmetric hopping case
takes the form:

\begin{equation}
[\hat{G}_{ij}(\tau,\tau')]^{-1}=\left(\begin{array}{cc}\left[-\frac{\partial}{
\partial\tau}
+h-U\langle
n_{i\downarrow}\rangle +\mu \right]\delta_{ij}-t^{\uparrow}_{ij}&-\Delta_{i}\delta_{ij
}\\
-\Delta^{*}_{i}\delta_{ij}&\left[-\frac{\partial}{\partial\tau}+h-U\langle
n_{i\uparrow}\rangle -\mu \right]\delta_{ij}+t^{\downarrow}_{ij}
\end{array}\right)\delta(\tau-\tau').
\end{equation}
After the use of the space-time Fourier transform \eqref{FT} and considering
only homogeneous superconducting  ordering,  one gets for the  Green function:
\begin{eqnarray}
{\hat G}(\vec{k},i\omega_{n})&=&
\left(\begin{array}{cc} G^{\uparrow}(\vec{k},i\omega_{n})
&F(\vec{k},i\omega_{n})\\
F^{\dagger}(\vec{k},i\omega_{n})&-G^{\downarrow}(-\vec{k},-i\omega_{n})
\end{array}\right)=\nonumber\\
&=&
\left(\begin{array}{cc}
i\omega_{n}+h-\epsilon_{\vec{k}\uparrow}+\mu-Un_{\downarrow}&-\Delta\\
-\Delta^{*}&i\omega_{n}+h+\epsilon_{\vec{k}\downarrow}-\mu+Un_{\uparrow}
\end{array}\right)^{-1},
\end{eqnarray}
where: $\epsilon_{\vec{k}\sigma}=-2t^{\sigma}\Theta_{\vec{k}}$. The solutions
are:
\begin{eqnarray}\label{BCS_asymmetric}
G^{\uparrow}(\vec{k},i\omega_{n})=
\frac{ i\omega_{n}+{\bar h}+{\bar \epsilon}_{\vec{k}\downarrow} }
{(i\omega_{n}+{\bar h}-{\bar \epsilon}_{\vec{k}\uparrow})
(i\omega_{n}+{\bar h}+{\bar \epsilon}_{\vec{k}\downarrow}
)-|\Delta|^2},\nonumber \\
G^{\downarrow}(\vec{k},i\omega_{n})=
\frac{ i\omega_{n}-{\bar h}+{\bar \epsilon}_{\vec{k}\uparrow} }
{(i\omega_{n}-{\bar h}-{\bar \epsilon}_{\vec{k}\downarrow})
(i\omega_{n}-{\bar h}+{\bar \epsilon}_{\vec{k}\uparrow} )-|\Delta|^2},\nonumber
\\
F(\vec{k},i\omega_{n})=
\frac{ \Delta}
{(i\omega_{n}+{\bar h}-{\bar \epsilon}_{\vec{k}\uparrow})
(i\omega_{n}+{\bar h}+{\bar \epsilon}_{\vec{k}\downarrow}
)-|\Delta|^2},\nonumber\\
F^{*}(\vec{k},i\omega_{n})=
\frac{ \Delta^{*} }
{(i\omega_{n}-{\bar h}+{\bar \epsilon}_{\vec{k}\uparrow})
(i\omega_{n}-{\bar h}-{\bar \epsilon}_{\vec{k}\downarrow} )-|\Delta|^2},
\end{eqnarray}
where:
${\bar h}=\textcolor{czerwony}{h}+UM/2$, ${\bar \epsilon}_{\vec{k}\sigma}=\epsilon_{\vec{k}\sigma}- 
{\bar \mu}$, ${\bar \mu} =\mu-Un/2$, $n=n_{\uparrow}+n_{\downarrow}$,
$M=n_{\uparrow}-n_{\downarrow}$,
${\epsilon}_{-\vec{k}\sigma}={\epsilon}_{\vec{k}\sigma}$ and
$F^{\dagger}(\vec{k},i\omega_{n})=F^{*}(-\vec{k},-i\omega_{n})$.

The quasiparticle spectrum is given by the four poles:
\begin{eqnarray}
E^{\pm}_{\vec{k}\uparrow}=-{\bar h}+\bar{\epsilon}^{-}_{\vec{k}}\pm
E_{\vec{k}}\nonumber\\
E^{\pm}_{\vec{k}\downarrow}={\bar h}-\bar{\epsilon}^{-}_{\vec{k}} \pm E_{\vec{k}}\nonumber\\
E_{\vec{k}}=\sqrt{({\bar \epsilon}^{+}_{\vec{k}})^2+|\Delta|^2},
\end{eqnarray}
where: ${\bar \epsilon}^{\pm}_{\vec{k}}=
\frac{{\bar \epsilon}_{\vec{k}\uparrow}\pm{\bar
\epsilon}_{\vec{k}\downarrow}}{2},
{\bar \epsilon}^{+}_{\vec{k}}={\bar \epsilon}_{\vec{k}}=
\epsilon_{\vec{k}}-\bar{\mu},
\epsilon_{\vec{k}}=-2t\Theta_{\vec{k}},
t=\frac{t^{\uparrow}+t^{\downarrow}}{2}$,
among which it is sufficient to choose only two:
\begin{eqnarray}
E_{\vec{k}\uparrow}=-{\bar
h}+\bar{\epsilon}^{-}_{\vec{k}}+E_{\vec{k}}\nonumber\\
E_{\vec{k}\downarrow}={\bar h}-\bar{\epsilon}^{-}_{\vec{k}}+E_{\vec{k}}.
\end{eqnarray}

The equation (\ref{BCS_asymmetric}) can be written analogously to the equal
hopping integral case (see: Eq.~\eqref{BCS}) as follows:
\begin{eqnarray}
\label{D46}
G^{\uparrow}(\vec{k},i\omega_{n})=
\frac{u^{2}_{\vec{k}}}{i\omega_{n}-E_{\vec{k}\uparrow}}
+\frac{ v^{2}_{\vec{k}}}{i\omega_{n}+E_{\vec{k}\downarrow}},\nonumber\\
G^{\downarrow}(\vec{k},i\omega_{n})=
\frac{u^{2}_{\vec{k}}}{i\omega_{n}-E_{\vec{k}\downarrow}}
+\frac{ v^{2}_{\vec{k}}}{i\omega_{n}+E_{\vec{k}\uparrow}},\nonumber\\
F(\vec{k},i\omega_{n})=
\frac{ \Delta}{(i\omega_{n}-E_{\vec{k}\uparrow})
(i\omega_{n}+E_{\vec{k}\downarrow})},
\end{eqnarray}
where $u^{2}_{\vec{k}}+v^{2}_{\vec{k}}=1,
u^{2}_{\vec{k}}
=\frac{1}{2}\left(1+\frac{{\bar\epsilon}_{\vec{k}}}{E_{\vec{k}}}\right)$ 
are the usual BCS coherence factors.

One can also derive the spin-dependent superconducting density of states,
given by the formula:
\begin{equation}
 g_{\sigma}(E)=\frac{-1}{\pi N}\sum_{\vec k}\textrm{Im}\,G^{\sigma}({\vec k},
i\omega_{n}\rightarrow E+ i0^{+}),
\end{equation} 
which upon the use of \eqref{D46} yields:
\begin{equation}
\label{app-DOS}
 g_{\sigma}(E)=\frac{1}{N} \sum_{\vec{k}}\Big[u_{\vec{k}}^2\delta
(E-E_{\vec{k}\sigma}) + \nu_{\vec{k}}^2\delta (E+E_{\vec{k}\,-\sigma})\Big],
\end{equation}
i.e. Eq.~\eqref{DOS}.

After simple transformations, one can obtain a set of self-consistent equations,
in analogy to Eqs.~\eqref{gapnambu}-\eqref{n_nambu}.   

\section{T-matrix scheme}
Let us write the exact equation for the self-energy (\ref{Fin}) in the form:
\begin{equation}
{\hat \Sigma}(1,2)={\hat \Sigma}^{0}(1,2) +{\hat \Sigma}^{1}(1,2),
\end{equation}
\begin{equation}
{\hat \Sigma}^{0}(1,2)=\left(\begin{array}{cc}U\langle
n_{\downarrow}(1)\rangle_{S}&\Delta(1)
\\ \Delta^{*}(1)&-U\langle n_{\uparrow}(1)\rangle_{S}\end{array}\right)
\delta(1-2),
\end{equation}
\begin{equation}
{\hat \Sigma}^{1}(1,2)=
-\int d{\bar 3}\left[\tau_{+}{\hat G}(1,{\bar 3}){\hat \Gamma}_{1}({\bar 3},2|1)
+\tau_{-}{\hat G}(1,{\bar 3}){\hat \Gamma}_{2}({\bar
3},2|1)\right],\label{SELF1}
\end{equation}
where the vertex functions ${\hat \Gamma}_{s}({\bar 3},2|1)$, $s=1,2$ are:
\begin{eqnarray}\label{VF}
{\hat \Gamma}_{1}({\bar 3},2|1)=U\frac{\delta{\hat \Sigma}({\bar 3},2)}{\delta
\eta(1)}, \\
\label{VF2}
{\hat \Gamma}_{2}({\bar 3},2|1)=U\frac{\delta{\hat \Sigma}({\bar 3},2)}{\delta
\eta^{*}(1)},
\end{eqnarray}
The equation for the vertex function can be obtained by substituting again the
Dyson equation:
${\hat \Sigma}({\bar 3},2)={\hat G}^{-1}({\bar 3},2)-{\hat G}^{-1}_{0}({\bar
3},2)$
in (\ref{VF}) and (\ref{VF2}). 
To a first iteration, we get:
\begin{eqnarray}\label{1It}
{\hat \Gamma}_{1}({\bar 3},2|1)=
\left(\begin{array}{cc}
U^{2} \frac{\delta \langle n_{\downarrow}(2)\rangle_{S}}
{\delta \eta(1)}& U \frac{\delta \Delta(2)}{\delta \eta(1)}\\
U \frac{\delta \Delta^{*}(2)}{\delta \eta(1)}& 
-U^{2} \frac{\delta \langle n_{\uparrow}(2)\rangle_{S}}
{\delta \eta(1)} \end{array}\right)\delta(2-{\bar3}),
\end{eqnarray}
and analogously for ${\hat \Gamma}_{2}({\bar 3},2|1)$.
Substituting (\ref {1It}) to (\ref {SELF1}) we have 
\begin{align}\label{SELF1It}
&{\hat \Sigma}^{1}(1,2)=\\
&\left(\begin{array}{cc}
-U^{2} F^{+}(1,2)\frac{\delta \langle n_{\downarrow}(2)\rangle_{S}}
{\delta \eta(1)}+UG_{\downarrow}(2,1)\frac{\delta \Delta^{*}(2)}{\delta \eta(1)}
& -U^{2}G_{\downarrow}(2,1) \frac{\delta \langle n_{\downarrow}(2)\rangle_{S}}
{\delta\eta(1)}-UF^{\dagger}(1,2)\frac{\delta \Delta(2)}{\delta \eta^(1)}\\
-U^{2}G_{\uparrow}(1,2) \frac{\delta \langle
n_{\downarrow}(2)\rangle_{S}}{\delta
\eta^{*}(1)}-U F(1,2)\frac{\delta \Delta^{*}(2)}{\delta \eta^{*}(1)}&
U^{2} F(1,2)\frac{\delta \langle n_{\uparrow}(2)\rangle_{S}}{\delta \eta^{*}(1)}
-UG_{\uparrow}(1,2)\frac{\delta \Delta(2)}{\delta \eta^{*}(1)}\nonumber
\end{array}\right).
\end{align}
In this way, in this approach we included corrections up to second
order in $U$.

From now on we consider only the normal state and set all the sources to zero \textcolor{czerwony}{as well as $t^{\uparrow}=t^{\downarrow}$}.
(At this stage, we also do not consider the effect of the Zeeman field).
The self energy to a first iteration is then diagonal and given by:
\begin{eqnarray}\label{SELF1ItN}
{\hat \Sigma}(1,2)=\left(\begin{array}{cc}\Sigma_{11}(1,2)&0\\
0& \Sigma_{22}(1,2) \end{array}\right),
\end{eqnarray}
\begin{eqnarray}\label{SELF1Ita}
\Sigma_{11}(1,2)=U\langle
n_{\downarrow}(1)\rangle\delta(1-2)+UG_{\downarrow}(2,1)\left.\frac{\delta
\Delta^{*}(2)}{\delta \eta(1)}\right|_{0}~~,\\\label{SELF1Itb}
\Sigma_{22}(1,2)=-U\langle
n_{\uparrow}(1)\rangle\delta(1-2)-UG_{\uparrow}(1,2)\left.\frac{\delta\Delta(2)}
{\delta
\eta^{*(1)}}\right|_{0}~~.
\end{eqnarray}
In the above, the terms  given by functional derivatives 
generate the Cooper pair fluctuations.
These terms are calculated from:
\begin{eqnarray}
\frac{\delta\Delta^{*}(2)}{\delta \eta(1)}=U\frac{\delta
F^{\dagger}(2,2^{+})}{\delta \eta(1)}
=U\left. \frac{\delta {\hat G}(2,2^{+})}{\delta \eta(1)}\right|_{21}~~,
\end{eqnarray}
by substituting the self-energy through (\ref{ID}). 
Keeping the terms involving the Cooper pair fluctuations to the same order of
approximation, we have:
\begin{eqnarray}
\left.\frac{\delta\Delta^{*}(2)}{\delta\eta(1)}\right|_{0}=-UG_{\downarrow}(1,
2)G_{\uparrow}(1,2)-
U\int d{\bar 4}G_{\downarrow}({\bar 4},2)\left.\frac{\delta\Delta^{*}({\bar
4})}{\delta
\eta(1)}\right|_{0}G_{\uparrow}({\bar4},2),\\ \label{Eq1}
\left.\frac{\delta\Delta(2)}{\delta
\eta^{*}(1)}\right|_{0}=-UG_{\uparrow}(2,1)G_{\downarrow}(2,1)-U\int d{\bar
4}G_{\uparrow}({2,\bar 4})\left.\frac{\delta\Delta({\bar 4})}{\delta
\eta^{*}(1)}\right|_{0}G_{\downarrow}({2,\bar4}). \label{Eq2}
\end{eqnarray}
On the other hand:  
\begin{eqnarray}
\left.\frac{\delta\Delta^{*}(2)}{\delta\eta(1)}\right|_{0}=-UG_{2}(1,2), \quad
\left.\frac{\delta\Delta(2)}{\delta\eta^{*}(1)}\right|_{0}=-UG_{2}(2,1),
\end{eqnarray}
where $G_{2}(1,2)=\langle T_{\tau}\rho^{-}(1)\rho^{+}(2)\rangle$ is the
two-particle Green function,
$\rho^{-}(1)=c_{\downarrow}(1)c_{\uparrow}(1),\rho^{+}(2)=
c^{\dagger}_{\uparrow}(2)c^{\dagger}_{\downarrow}(2).$
Using (\ref{Eq1}), we obtain the equation satisfied by $G_{2}(1,2)$:
\begin{eqnarray}\label{G2}
G_{2}(1,2)=G_{\downarrow}(1,2)G_{\uparrow}(1,2) -
U\int d{\bar 4}G_{\downarrow}({\bar 4},2)G_{\uparrow}({\bar4},2)G_{2}(1,{\bar
4}).
\end{eqnarray}
We now define the T-matrix:
\begin{eqnarray}
T_{\downarrow\uparrow}(1,2)=U\delta(1-2)+U\left.\frac{\delta\Delta^{*}(2)}{
\delta\eta(1)}\right|_{0},\\
T_{\uparrow\downarrow}(2,1)=U\delta(2-1)+U\left.\frac{\delta\Delta(2)}{
\delta\eta^{*}(1)}\right|_{0},
\end{eqnarray}
which leads to:
\begin{eqnarray}
\left.\frac{\delta\Delta^{*}(2)}{\delta\eta(1)}\right|_{0}=
-\int d{\bar 4}G_{\downarrow}({\bar 4},2)G_{\uparrow}({\bar
4},2)T_{\downarrow\uparrow}(1,{\bar 4}),\\
\left.\frac{\delta\Delta(2)}{\delta
\eta^{*}(1)}\right|_{0}=
-\int d{\bar 4}G_{\uparrow}(2,{\bar 4})G_{\downarrow}(2,{\bar
4})T_{\uparrow\downarrow}({\bar 4},1).\\
\end{eqnarray}
The T-matrix satisfies the equation:
\begin{eqnarray}
T_{\downarrow\uparrow}(1,2)=U\delta(1-2)-U\int d{\bar 4}
G_{\downarrow}({\bar 4},2)G_{\uparrow}({\bar 4},2)T_{\downarrow\uparrow}(1,{\bar
4}),\\
T_{\uparrow\downarrow}(2,1)=U\delta(1-2)-U\int d{\bar 4}
G_{\uparrow}(2,{\bar 4})G_{\downarrow}(2,{\bar 4})T_{\uparrow\downarrow}({\bar
4},1).
\end{eqnarray}
The self-energies in the T-matrix scheme are obtained from (\ref{SELF1ItN}) and
given by:
\begin{eqnarray}\label{TS1}
\Sigma_{11}(1,2)=G_{\downarrow}(2,1)\left[U\delta(1-2)
+U\left. \frac{\delta\Delta^{*}(2)}{\delta\eta(1)}\right|_{0}\right]=
T_{\downarrow\uparrow}(1,2)G_{\downarrow}(2,1),\\\label{TS2}
\Sigma_{22}(1,2)=-G_{\uparrow}(1,2)\left[U\delta(1-2)+U\left.\frac{
\delta\Delta(2)}{\delta\eta^{*}(1)}\right|_{0}\right]
=- T_{\uparrow\downarrow}(2,1)G_{\uparrow}(1,2),
\end{eqnarray}
where we substituted $\langle n_{\sigma}(1)\rangle=G_{\sigma}(1,2)\delta(1-2)$
in (\ref{SELF1Ita}-\ref{SELF1Itb}).
Since in the absence of magnetic field, the \textcolor{czerwony}{normal state} Green functions are 
\textcolor{czerwony}{spin symmetric},
 we simply set 
$T_{\downarrow\uparrow}(1,2)=T(1,2).$
Thus,   for the electron Green function $G(1,2)=-\langle
T_{\tau}c_{i\sigma}(\tau)c^{\dagger}_{j\sigma}(\tau')\rangle$ in the 
T-matrix approximation, we have the following set
of self-consistent equations:
\begin{equation}
G^{-1}(1,2)=G^{-1}_{0}(1,2)-\Sigma(1,2),\\
\end{equation}
\begin{equation}
\Sigma(1,2)=T(1,2)G(2,1),\\
\end{equation}
\begin{equation}
T(1,2)=U\delta(1-2)-U\int d{\bar 3}G({\bar 3},2)G({\bar 3},2)T(1,{\bar 3}),\\
\end{equation}
\begin{equation}
G^{-1}_{0}(1,2)=\left(-\frac{\partial}{\partial\tau}\delta_{ij}-{\bar
t}_{ij}\right)\delta(\tau-\tau'),
\end{equation}
and the self-energy is expressed by fully dressed Green functions $G$.
For the translationally invariant system, we go to the space-time Fourier
representation using ({\ref {FT}) and: 
\begin{eqnarray}
\Sigma_{ij}(\tau,\tau')=\frac{1}{\beta
N}\sum_{\vec{k},\omega_{n}}e^{i\vec{k}\cdot(\vec{R}_{i}-\vec{R}_{j})-i\omega_{n}
(\tau-\tau')}
\Sigma(\vec{k}, i\omega_{n}),\\
T_{ij}(\tau,\tau')=\frac{1}{\beta
N}\sum_{\vec{q},\nu_{m}}e^{i\vec{q}\cdot(\vec{R}_{i}-\vec{R}_{j})-i\nu_{m}
(\tau-\tau')}T(\vec{q}, i\nu_{m}),
\end{eqnarray}
to obtain:
\begin{equation}\label{TM}
T(\vec{q}, i\nu_{m})=\frac{U}{1+U\chi(\vec{q}, i\nu_{m})},\\
\end{equation}
\begin{equation}
\label{SELFTM}
\Sigma(\vec{k}, i\omega_{n})
=\frac{1}{\beta N}\sum_{\vec{q}, \nu_{m}}T(\vec{q}, i\nu_{m}) 
G(\vec{q}-\vec{k},i\nu_{m}-i\omega_{n}),\\ 
\end{equation}
\begin{equation}
\label{PSU}
\chi(\vec{q}, i\nu_{m})=\frac{1}{\beta N}\sum_{\vec{k}, 
\omega_{n}}G(\vec{k},i\omega_{n})G(\vec{q}-
\vec{k},i\nu_{m}-i\omega_{n}),
\end{equation}
where $\nu_{m}=2\pi m/\beta$, $m$ integer, is the bosonic Matsubara frequency,
$\chi(\vec{q}, i\nu_{m})$ is the pairing susceptibility.
The free Green function is given by 
$G_{0}(\vec{k},i\omega_{n})=\left[i\omega_{n}-\epsilon_{\vec{k}}+\mu\right]^{
-1}$ and has no Hartree term.
In diagrammatic interpretation, the T-matrix approximation is equivalent to
the summation of repeated particle-particle scatterings 
described by the ladder diagrams \cite{FetterWalecka}. For the attractive
interaction, it describes particle-particle
correlations in the Cooper channel.
It can also be considered as a ladder approximation to
the full Bethe-Salpeter equation for the two-particle Green functions (compare
(\ref{G2})).
The  $G_{2}(q)$ is simply related to the T-matrix:
\begin{eqnarray}
G_{2}(q)=\frac{\chi(q)T(q)}{U}=\frac{\chi(q)}{1+U\chi(q)}.
\end{eqnarray}

The so called $(GG_{0})G_{0}$ T-matrix scheme follows as an 
approximation to the above equations (\ref{SELFTM}), (\ref{PSU}) by replacing on
the RHS one dressed $G$ by $G_{0}$.

The T-matrix scheme goes beyond the 
standard
mean-field, since includes the pairing fluctuations (effect of noncondensed
pairs with ${\vec q}\neq 0$) and allows for description of the BEC regime of
the crossover.In general, the method works best in a low density regime.

\section{T-matrix approach for $h\neq 0$}
\label{appD3}
We shall now consider the case of the Hubbard model in a Zeeman magnetic field
and in the normal state.
The matrix Green function is diagonal with the components given by: 
\begin{eqnarray}
G^{-1}_{11}(1,2)=\left[G^{0}_{11}(1,2)\right]^{-1}-\Sigma_{11}(1,2)\\
G^{-1}_{22}(1,2)=\left[G^{0}_{22}(1,2)\right]^{-1}-\Sigma_{22}(1,2),
\end{eqnarray}
where the self-energies are given by (\ref{TS1}) and (\ref{TS2}), i.e.:
\begin{eqnarray}
\Sigma_{11}(1,2)=
T_{\downarrow\uparrow}(1,2)G_{\downarrow}(2,1),\\
\Sigma_{22}(1,2)=- T_{\uparrow\downarrow}(2,1)G_{\uparrow}(1,2), 
\end{eqnarray}
and 
$\left[G_{0}(1,2)\right]^{-1}=\left\{\left[
-\frac{\partial}{\partial\tau}-h\right]\tau_{0}\delta_{ij}-
{\bar t}_{ij}\tau_{3}\right\}\delta(\tau-\tau').$
We can also write:
\begin{eqnarray}
G^{-1}_{\uparrow}(1,2)=\left[G_{0\uparrow}(1,2)\right]^{-1}-\Sigma_{11}(1,2)\\
G^{-1}_{\downarrow}(1,2)=\left[G_{0\downarrow}(1,2)\right]^{-1}+\Sigma_{22}(2,
1).
\end{eqnarray}
Therefore, one has the Dyson equation for both $\sigma$: 
\begin{eqnarray}
G^{-1}_{\sigma}(1,2)=\left[G_{0\sigma}(1,2)\right]^{-1}-\Sigma_{\sigma}(1,2),
\end{eqnarray}
where:
\begin{eqnarray}
\Sigma_{\uparrow}(1,2)=
T_{\downarrow\uparrow}(1,2)G_{\downarrow}(2,1),\\
\Sigma_{\downarrow}(1,2)= T_{\uparrow\downarrow}(1,2)G_{\uparrow}(2,1).
\end{eqnarray}
Performing the Fourier transform, we obtain:
\begin{equation}\label{HDE}
G^{-1}_{\sigma}(k)=G^{-1}_{0\sigma}(k)-\Sigma_{\sigma}(k),
\end{equation}
\begin{equation}
\label{HSE1}
\Sigma_{\uparrow}(k)=\sum_{q} T_{\downarrow\uparrow}(q)G_{\downarrow}(q-k),
\end{equation}
\begin{equation}
\label{HSE2}
\Sigma_{\downarrow}(k)=\sum_{q} T_{\uparrow\downarrow}(q)G_{\uparrow}(q-k),
\end{equation}
where  we have used the four-vector notation: $k=(\vec{k},i\omega_{n}), 
q=(\vec{q},i\nu_{m}),
\sum_{q}=\frac{1}{N}\sum_{\vec{q}}\frac{1}{\beta}\sum_{\nu_{m}}$.\\ 
The free Green function is  given by: 
$G^{-1}_{0\sigma}(k)=i\omega_{n}-(\epsilon_{\vec{k}}-\mu)+h\sigma$.

The corresponding T-matrix satisfies the equations:
\begin{eqnarray}\label{HTM1}
T^{-1}_{\downarrow\uparrow}(q)=\frac{1}{U}+\chi_{\downarrow\uparrow}(q),\\
\label{HTM2}
T^{-1}_{\uparrow\downarrow}(q)=\frac{1}{U}+\chi_{\downarrow\uparrow}(q),
\end{eqnarray}
where the pairing susceptibilites are given by: 
\begin{eqnarray}\label{HPS}
\chi_{\sigma\bar{\sigma}}(q)=\sum_{k}G_{\sigma}(k)G_{\bar{\sigma}}(q-k),~~
\bar{\sigma}=-\sigma.
\end{eqnarray}
One can check, by the change of summation variables, that
$\chi_{\downarrow\uparrow}(q)=\chi_{\uparrow\downarrow}(q)$ and therefore we
have $T^{-1}_{\downarrow\uparrow}(q)=T^{-1}_{\uparrow\downarrow}(q)$, and thus
only one T-matrix for both spin directions.
The effects of superconducting correlations in the normal state are 
described by the above spin-dependent self-energies.
The above equations (\ref{HDE}-\ref{HTM2}), supplemented by the number equation:
\begin{eqnarray}
n=\sum_{k,\sigma}G_{\sigma}(k)e^{i\omega_{n}0^{+}},
\end{eqnarray}
form the required set for the normal state.

For calculations of the critical temperature, one resorts to
 the Thouless criterion of the divergent T-matrix: 
\begin{eqnarray} 
\frac{1}{U}+\chi_{\downarrow\uparrow}({\vec{q}},0,T_{c})=0,
\end{eqnarray}
which in general can yield instability of the normal state 
for $\vec{q}=0$ or $\vec{q}\neq 0$, 
if $h\neq 0$.
For the free case, one gets for the pairing susceptibilities:
\begin{eqnarray}
\chi^{0}_{\downarrow\uparrow}(q)=\sum_{k}G_{0\downarrow}(k)G_{0\uparrow}(q-k)=
\frac{1}{N}\sum_{\vec{k}}\frac{1-f(\epsilon_{\vec{k}}-\mu+h)-
f(\epsilon_{\vec{q}-\vec{k}}-\mu-h)}
{\epsilon_{\vec{k}}+\epsilon_{\vec{q}-\vec{k}}-2\mu-i\nu_{m}},\\
\chi^{0}_{\uparrow\downarrow}(q)=
\sum_{k}G_{0\uparrow}(k)G_{0\downarrow}(q-k)=
\frac{1}{N}\sum_{\vec{k}}\frac{1-f(\epsilon_{\vec{k}}-\mu-h)-
f(\epsilon_{\vec{q}-\vec{k}}-\mu +h)}
{\epsilon_{\vec{k}}+\epsilon_{\vec{q}-\vec{k}}-2\mu-i\nu_{m}},
\end{eqnarray}
and indeed $\chi^{0}_{\downarrow\uparrow}(q)=\chi^{0}_{\uparrow\downarrow}(q)$,
as
it is seen,  by a change of the variable in the momentum summation. 
The unique solution which follows from the Thouless criterion is in
agreement with the expression for $T_{c}$ in a simple BCS theory:
\begin{eqnarray}
0&=&\frac{1}{U}+\chi^{0}_{\downarrow\uparrow}(\vec{0},0,T_{c})=\frac{1}{U}+\
\frac{1}{N}\sum_{\vec{k}}\frac{1-f(\epsilon_{\vec{k}}-\mu+h)-
f(\epsilon_{\vec{k}}-\mu-h)}
{2(\epsilon_{\vec{k}}-\mu)}=\nonumber\\
&=&\frac{1}{U}+\frac{1}{N}\sum_{\vec{k}}\frac{1-
2{\bar f}(\epsilon_{\vec{k}}-\mu)}{2(\epsilon_{\vec{k}}-\mu)},
\end{eqnarray}
where $\bar {f}(x)$ is the symmetrized Fermi function:
\begin{eqnarray}
\bar {f}(x)=\frac{1}{2}\left[f(x+h)+f(x-h)\right].
\end{eqnarray}

For approximate solutions, one can use the $(GG_{0})G_{0}$ scheme, where we
replace one dressed $G_{\bar{\sigma}}(q-k)$ by $G_{0\bar{\sigma}}(q-k)$ in 
the self-energies (\ref{HSE1}), (\ref{HSE2}) and pairing
susceptibilities (\ref{HPS}). However, for consistency one has to use  a
T-matrix which is the same for both spin directions. A solution 
has been suggested by the Levin group and consists in the use of the
symmetrized pairing susceptibility:
\begin{eqnarray}\label{HPSS}
{\bar\chi}(q)=\frac{1}{2}\left[\chi_{\downarrow\uparrow}(q)+\chi_{
\uparrow\downarrow}(q)\right].
\end{eqnarray}
The corresponding  T-matrix is, according to (\ref{HTM1}-\ref{HTM2}):
\begin{eqnarray}\label{HTSS}
T^{-1}(q)=\frac{1}{U}+
{\bar\chi}(q).
\end{eqnarray}     

Finally, in the $(GG_{0})G_{0}$ scheme, one has the following equations:
\begin{equation} 
\label{GG0G0}
G^{-1}_{\sigma}(k)=G^{-1}_{0\sigma}(k)-\Sigma_{\sigma}(k),
\end{equation}
\begin{equation}
\Sigma_{\sigma}(k)=\sum_{q} T(q)G_{0\bar{\sigma}}(q-k),
\end{equation}
\begin{equation}
T^{-1}(q)=\frac{1}{U}+{\bar\chi}(q),
\end{equation}
\begin{equation}
{\bar\chi}(q)=\frac{1}{2}\left[\chi_{\downarrow\uparrow}(q)+\chi_{
\uparrow\downarrow}(q)\right],
\end{equation}
\begin{equation}
\chi_{\sigma,\bar {\sigma}}(q)=\sum_{k}G_{\sigma}(k)G_{0\bar{\sigma}}(q-k).
\end{equation}
It is clear that this scheme arises as an approximation to the
fully self-consistent T-matrix approach.
An attractive feature of the above formulation is 
that the structure of equations is much similar to
the case without magnetic field (only the Fermi function is replaced by
the symmetrized Fermi function) and numerical computations are feasible.

\chapter{Density of states}
\label{appendix4}
In this Appendix, the densities of states \cite{rjj} which have been
used for numerical calculations in this thesis are shown.

\begin{itemize}
 \item 2D square lattice:
\begin{equation}
 \rho_{2D}=\left\{ \begin{array}{ll} 
 \frac{2}{4\pi^2} K(1-(\epsilon/4t)^2) & \,\, |\epsilon/t|\leq 4 \\ 
 0 & \,\, |\epsilon/t|> 4
\end{array} \right.
\end{equation}
The density of states for the 2D square lattice is expressed by the complete
elliptic integral of the first kind
$K(x)=\int_{0}^{\pi/2}\frac{dt}{\sqrt{1-x^2\sin^2t}}$. 
\end{itemize}

\begin{figure}[h!]
\begin{center}
\includegraphics[width=0.34\textwidth,angle=270]{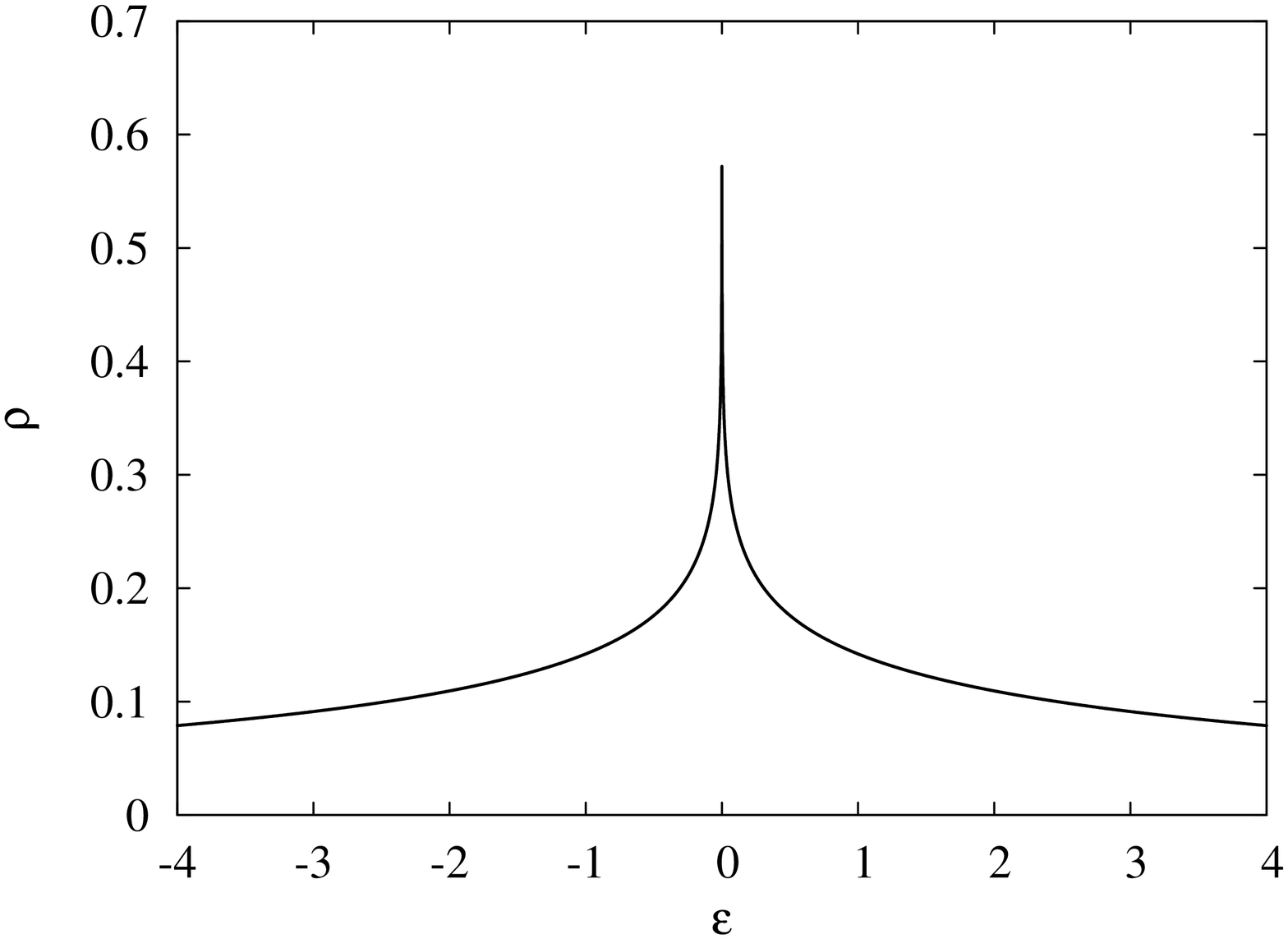}
\end{center}
\end{figure}
\begin{itemize}
\item simple cubic lattice (sc):
\end{itemize}
\begin{equation}
\rho_{sc}=\frac{3}{8\pi^3} \left\{ \begin{array}{ll} 
 70.7801+1.0053|3\epsilon/6t|^2 & \,\, |\epsilon/6t|< 1/3 \\ 
\sqrt{3-3|\epsilon/6t|}(80.3702-16.3846(3-3|\epsilon/6t|)\\
+0.78978(3-3|\epsilon/6t|)^2+\sqrt{3|\epsilon/6t|-1}(-44.2639\\
+3.66394(3-3|\epsilon/6t|)-0.17248(3-3|\epsilon/6t|)^2))& \,\,
1\geq|\epsilon/6t|\geq
1/3\\
  0 & \,\, \textrm{otherwise}
\end{array} \right.
\end{equation}
\begin{figure}[t!]
\begin{center}
\includegraphics[width=0.34\textwidth,angle=270]{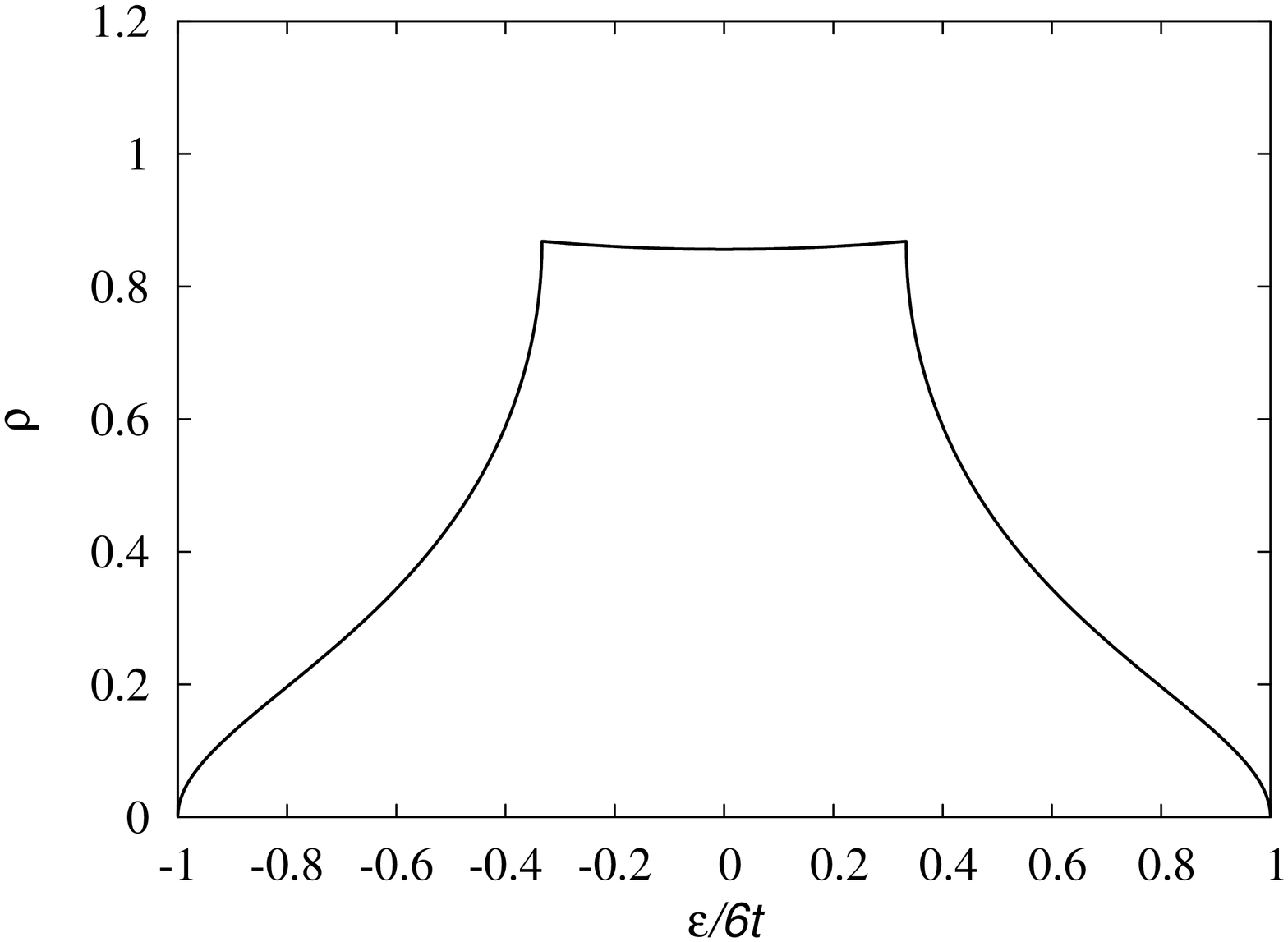}
\end{center}
\end{figure}

\begin{itemize}
 \item semicircular DOS (s-circ):
\begin{equation}
 \rho_{s-circ}=\frac{2}{\pi D}\sqrt{(\epsilon/D+1)(2-(\epsilon/D+1))},
\end{equation}
where $D=10.80732\,t$ is the half bandwidth.
\end{itemize}

\begin{figure}[h!]
\begin{center}
\includegraphics[width=0.34\textwidth,angle=270]{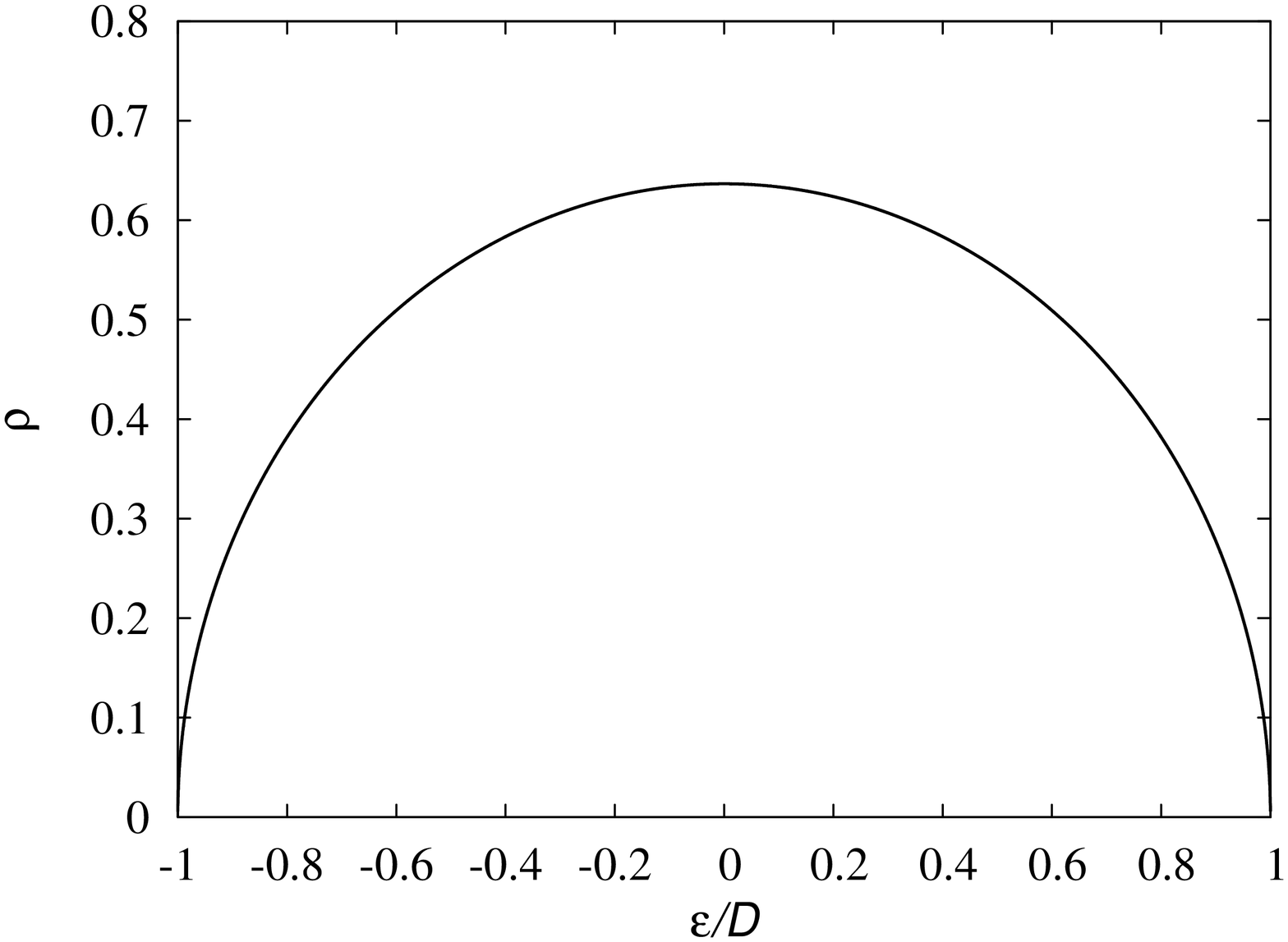}
\end{center}
\end{figure}
\begin{itemize}
 \item  body-centered cubic lattice (BCC):
\end{itemize}
\begin{equation}
 \rho_{BCC}=\left\{ \begin{array}{ll} 
 \frac{1}{8t}(2\sqrt{1-|\epsilon/8t|}\ln^2(5.845/|\epsilon/8t|)(16.6791\\ 
+ 3.636|\epsilon/8t|+2.4880|\epsilon/8t|^2)) & \,\, |\epsilon/8t|<1 \\ 
 0 & \,\, \textrm{otherwise}
\end{array} \right.
\end{equation}
\begin{figure}[t!]
\begin{center}
\includegraphics[width=0.34\textwidth,angle=270]{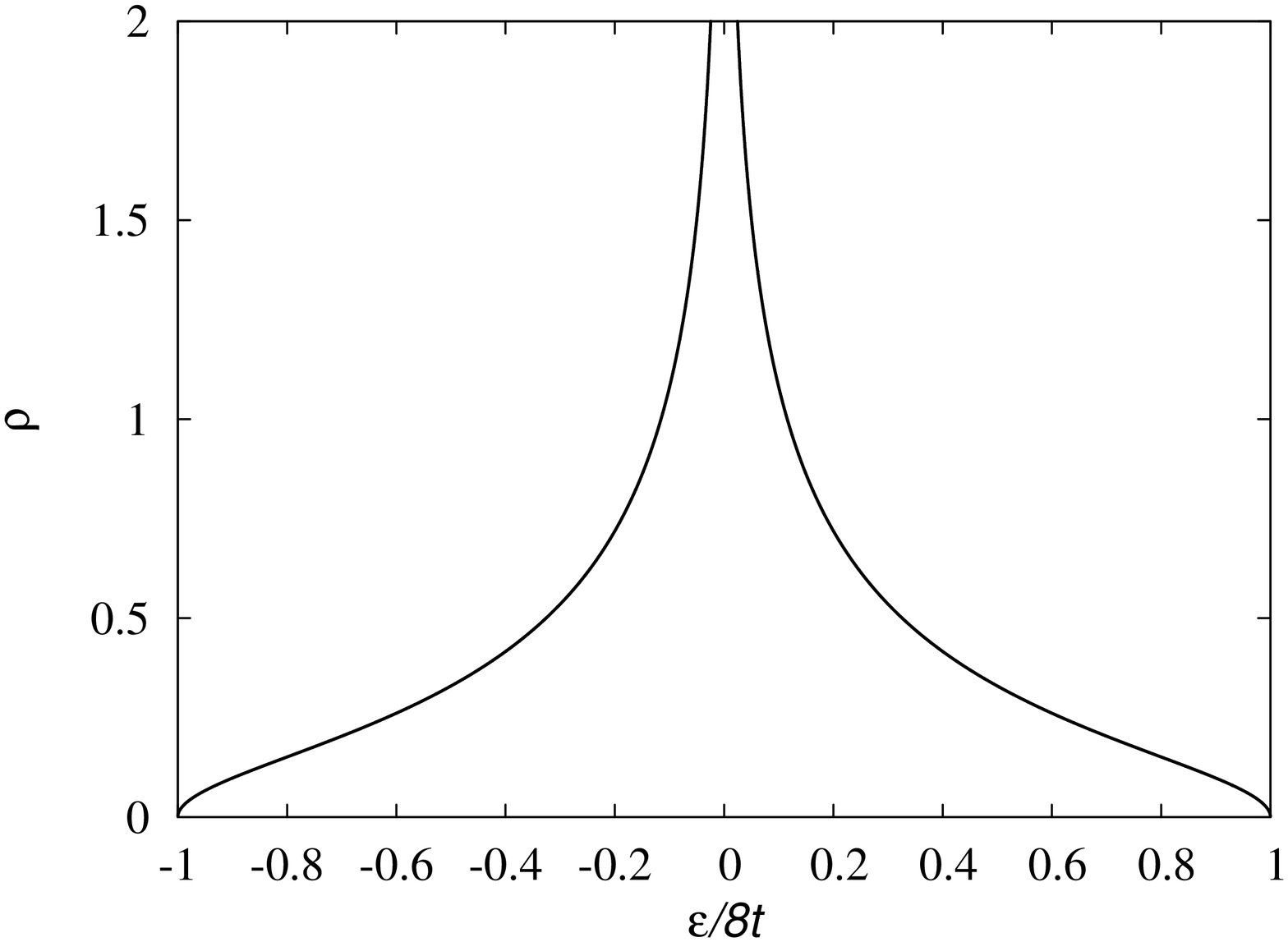}
\end{center}
\end{figure}

\begin{itemize}
 \item  face-centered cubic lattice (FCC):
\end{itemize}
\begin{equation}
 \rho_{FCC}= \frac{2}{8\pi^3}\left\{ \begin{array}{ll} 
 \frac{1}{4t}(4(122.595-19.4100(2\epsilon/8t-1)\\
+1.76011((2\epsilon/8t-1)^2& \,\,
0.5<\epsilon/8t\leq 1\\ 
+(-44.8100+7.18628(2\epsilon/8t-1))\ln(1-(2\epsilon/8t-1))))  \\
\frac{1}{4t}
(4\sqrt{3+(2\epsilon/8t-1)}((-85.9325\\
+101.103(3+(2\epsilon/8t-1))\\
-16.2885(3+(2\epsilon/8t-1))^2\\
+(56.8683-47.1215(3+(2\epsilon/8t-1))\\
+2.9045(3+(2\epsilon/8t-1))^2)\sqrt{|2\epsilon/8t-1|}))) & \,\, -1\leq
\epsilon/8t \leq 0.5\\
 0 & \,\, \textrm{otherwise}
\end{array} \right.
\end{equation}
\begin{figure}[h!]
\begin{center}
\includegraphics[width=0.34\textwidth,angle=270]{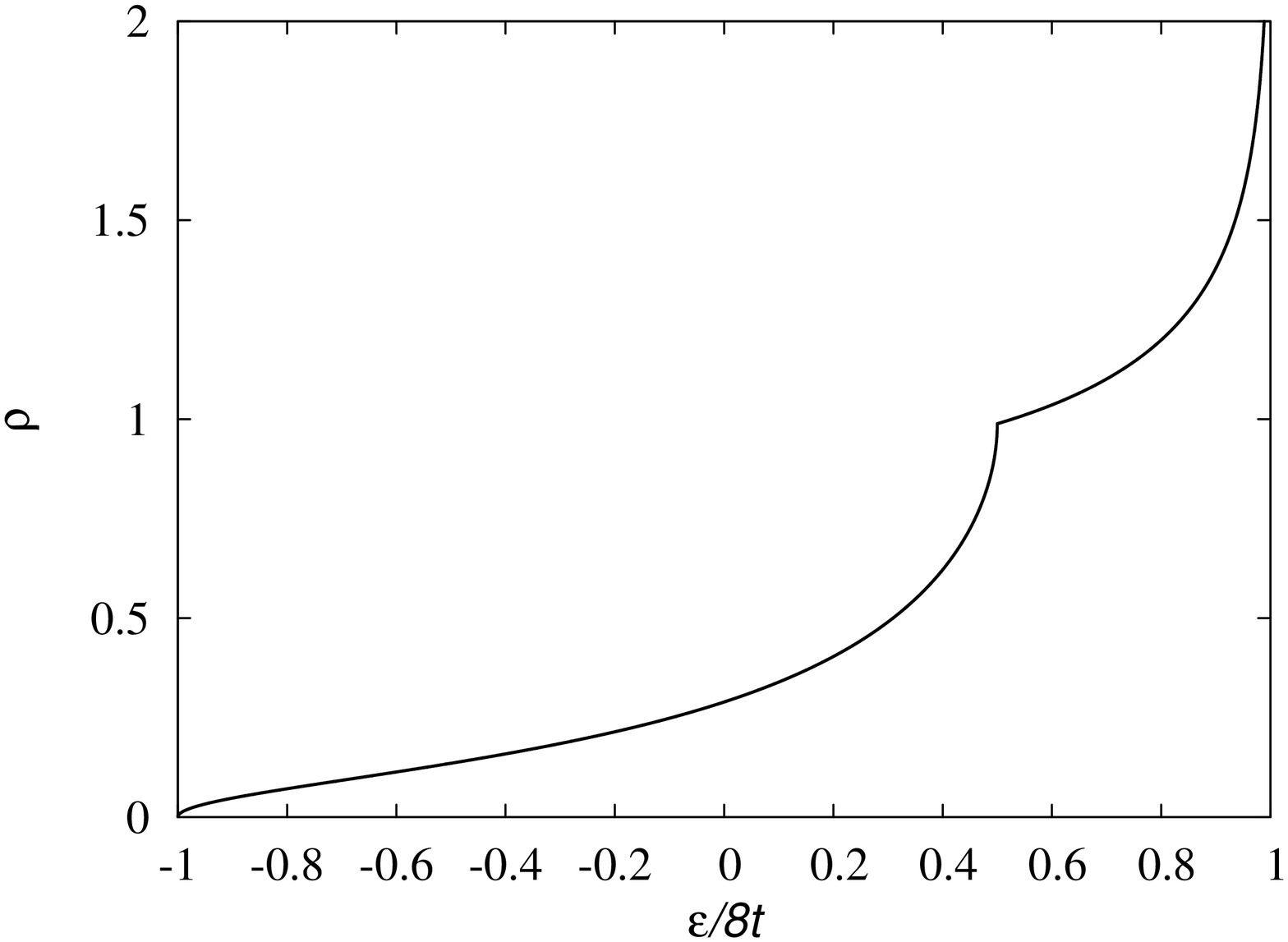}
\end{center}
\end{figure}

\newpage
\thispagestyle{empty}
\mbox{}

\listoftables
\newpage
\thispagestyle{empty}
\mbox{}
\listoffigures

\chapter*{List of publications}
\addcontentsline{toc}{chapter}{List of publications}
\begin{footnotesize}
\begin{enumerate}
\item 
A.~Kujawa,
``The BCS-BEC Crossover and Superconductivity in a Lattice Fermion Model with Hard Core Repulsion,''
Acta Phys.\ Polon.\ A {\bf 111} (2007) 745
[arXiv:0808.1374 [cond-mat.str-el]].

\item
K.~Cichy, J.~Gonzalez Lopez, K.~Jansen, A.~Kujawa and A.~Shindler,
  ``Cutoff effects for Wilson twisted mass fermions at tree-level of perturbation theory,''
  PoS LAT {\bf 2007} (2007) 098
  [arXiv: 0710.2036 [hep-lat]].

\item  
A.~Kujawa, R.~Micnas,
``On the Phase Diagram of the Spin-Polarized Attractive Hubbard Model: Weak Coupling Limit,''
Acta Phys.\ Polon.\ A {\bf 114} (2008) 43
[arXiv:0808.1578 [cond-mat.str-el]].

\item 
K.~Cichy, J.~Gonzalez Lopez, K.~Jansen, A.~Kujawa and A.~Shindler,
  ``Twisted Mass, Overlap and Creutz Fermions: Cut-off Effects at Tree-level of Perturbation Theory,''
  Nucl.\ Phys.\ B {\bf 800} (2008) 94
  [arXiv:0802.3637 [hep-lat]].

\item
A.~Kujawa, R.~Micnas,
``Some Properties of the Spin-Polarized Attractive Hubbard Model,''
Acta Phys.\ Polon.\ A {\bf 115} (2009) 138
[arXiv:0809.1420 [cond-mat.str-el]].

\item
K.~Cichy, J.~Gonzalez Lopez and A.~Kujawa,
``A Comparison of the cut-off effects for Twisted Mass, Overlap and Creutz fermions at tree-level of Perturbation Theory,''
  Acta Phys.\ Polon.\ B {\bf 39} (2008) 3463
  [arXiv:0811.0572 [hep-lat]].

\item
A.~Kujawa-Cichy,
``On the BCS–BEC Crossover in the Spin-Polarized Attractive Hubbard Model at T = 0,''
Acta Phys.\ Polon.\ A {\bf 118} (2010) 423.

\item
A.~Kujawa,
``Teoria superstrun jako zunifikowana teoria fizyki,''
in: Antoni Szczuciński, Zdzisław Błaszczak (eds.),
``Wokół ewolucjonizmu. Dylematy filozofów i fizyków,''
Oficyna Wydawnicza Batik, Poznań 2010.

\item
A.~Kujawa-Cichy and R.~Micnas,
  ``Stability of superfluid phases in the 2D spin-polarized attractive Hubbar d model,''
  Europhys.\ Lett.\  {\bf 95} (2011) 37003.

\item
A.~Kujawa-Cichy,
``On the Imbalanced d-wave Superfluids within the Spin Polarized Extended Hubbard Model: Weak Coupling Limit,''
Acta Phys.\ Polon.\ A {\bf 121} (2012) 824.

\item
A.~Kujawa-Cichy,
``On the BCS-BEC crossover in the 2D Asymmetric Attractive Hubbard Model,''
Acta Phys.\ Polon.\ A {\bf 121} (2012), 1066. 
\end{enumerate}
\end{footnotesize}

\chapter*{List of conferences and schools}
\addcontentsline{toc}{chapter}{List of conferences and schools}
\begin{footnotesize}
\begin{enumerate}
\item 
Physics of Magnetism 2005, Poznań, Poland, June 2005,
poster: ``The quantum spin-1/2 Heisenberg antiferromagnet on a triangular lattice with some bonds removed in a translationally invariant manner - ground state and lowest excitations,''
in collaboration with: K. Cichy, J. Richter and P. Tomczak.

\item
XII National School on Superconductivity, Ustroń, Poland, September 2006,
poster: ``BCS-BEC crossover in a lattice fermion model. The 3D case with infinite on-site repulsion''.

\item
Cosmological Dilemmas of Physics and Philosophy, Adam Mickiewicz University, Poznań, Poland, November 2006,
presentation: ``Superstring theory as a unified theory of physics''

\item
XXV International Symposium on Lattice Field Theory LATTICE 2007, Regensburg, Germany, July 2007,
poster: ``Cutoff effects for Wilson twisted mass fermions at tree-level of perturbation theory,''
in collaboration with: K. Cichy, J. Gonzalez Lopez, K. Jansen and A. Shindler.

\item
Les Houches Predoctoral School in Statistical Physics, Les Houches, France, 26 August -- 7 September 2007.

\item
XIII National School on Superconductivity, Lądek Zdrój, Poland, November 2007,
poster: ``Phase Diagram of Spin-Polarized Attractive Hubbard Model.''

\item
Physics Of Magnetism 2008, Poznań, Poland , June 2008,
poster: ``Superconducting Properties of the Spin-Polarized Attractive Hubbard Model,''
in collaboration with: R. Micnas.

\item
Physics Of Magnetism 2008, Poznań, Poland , June 2008,
poster: ``Quantum Monte Carlo study of the repulsive Hubbard model on a Sierpinski gasket'',
in collaboration with:  K. Cichy, P. Tomczak.

\item
Lattice Practices 2008, Zeuthen, Germany, October 2008.

\item
XIV National School on Superconductivity, Ostrów Wlkp., Poland, October 2009,
poster and presentation: ``On the BCS-BEC crossover in the Spin-Polarized Attractive Hubbard Model,''
in collaboration with: R. Micnas.

\item
Modern perspectives in lattice QCD: Quantum field theory and high performance computing,
Les Houches Summer School, Les Houches, France, 3-28 August 2010.

\item
XXXIV International Conference of Theoretical Physics ``Correlations and coherence at different scales'', Ustroń, Poland, September 2010,
poster: ``BCS-BEC crossover in the Spin-Polarized Attractive Hubbard Model.''

\item
Workshop on Frontiers in Ultracold Fermi Gases, Abdus Salam International  Centre  for  Theoretical  Physics, Trieste, Italy, June 2011,
poster: ``Superfluid phases in the 2D Spin-Polarized Attractive Hubbard Model.''

\item
Physics Of Magnetism 2011, Poznań, Poland, June 2011,
poster: ``Superfluid phases  in the 2D Spin-Polarized Attractive Hubbard Model.''

\item
XIV National School on Superconductivity, Kazimierz Dolny, Poland, October 2011,
poster: ``On the Imbalanced d-wave Superfluids within  the Spin-Polarized Extended Hubbard Model: Weak Coupling Limit.''
\end{enumerate}
\end{footnotesize}
\end{document}